\documentclass[12pt,preprint]{emulateapj}
  
\usepackage{apjfonts}
\usepackage{amsmath}
\usepackage{epsf}
\usepackage{color}
\shorttitle{Galaxy Evolution at $6.3 < z \leq 8.6$}
\shortauthors{Finkelstein et al.}

\newcommand{\sol}{$_{\odot}$}
\newcommand{\lya}{Ly$\alpha$}
\newcommand{\zb}{$z^{\prime}_{850}$}
\newcommand{\yb}{$Y_{105}$}
\newcommand{\jb}{$J_{125}$}
\newcommand{\hb}{$H_{160}$}

\newcommand{\fig}[1]{Figure~\ref{#1}}
\newcommand{\authorspace}{\vspace{-10pt}}

\newcommand{\acsb}{\hbox{$B_{435}$}}
\newcommand{\acsv}{\hbox{$V_{606}$}}
\newcommand{\acsi}{\hbox{$i_{775}$}}
\newcommand{\acsz}{\hbox{$z_{850}$}}
\newcommand{\wfcy}{\hbox{$Y_{105}$}}
\newcommand{\wfcj}{\hbox{$J_{125}$}}
\newcommand{\wfch}{\hbox{$H_{160}$}}

\newcommand{\msol}{\hbox{$M_\odot$}}
\newcommand{\Zsol}{\hbox{$Z_\odot$}}
\newcommand{\zsol}{\hbox{$Z_\odot$}}

\newcommand{\mone}{\hbox{$[3.6]$}}
\newcommand{\mtwo}{\hbox{$[4.5]$}}

\newcommand{\spitzer}{\textit{Spitzer}}

\newcommand{\gsim}{\gtrsim}
\newcommand{\degree}{\hbox{$^\circ$}}

\def\arcs{\hbox{$^{\prime\prime}$}}

\begin{document}
\slugcomment{Accepted to the Astrophysical Journal}
\title{On the Stellar Populations and Evolution of Star--Forming Galaxies at 6.3 $<$ $\MakeLowercase{z}$ $\leq$ 8.6}

\author{\sc Steven L. Finkelstein\altaffilmark{1} \& Casey Papovich}
\affil{George P. and Cynthia Woods Mitchell Institute for Fundamental Physics and Astronomy, \\ Department of Physics and Astronomy, Texas A\&M University, 4242 TAMU, College Station, TX 77843}
\author{\sc \authorspace Mauro Giavalisco}
\affil{University of Massachusetts, 710 North Pleasant Street, Amherst, MA 01003}
\author{\sc \authorspace Naveen A. Reddy\altaffilmark{2}}
\affil{National Optical Astronomy Observatory, 950 N. Cherry Avenue, Tucson, AZ 85719}
\author{\sc \authorspace Henry C. Ferguson \& Anton M. Koekemoer}
\affil{Space Telescope Science Institute, 3700 San Martin Drive, Baltimore, MD 21218}
\and
\author{\sc Mark Dickinson}
\affil{National Optical Astronomy Observatory, 950 N. Cherry Avenue, Tucson, AZ 85719}
\altaffiltext{1}{stevenf@physics.tamu.edu}
\altaffiltext{2}{Hubble Fellow}

\begin{abstract}
Observations of very distant galaxies probe both the formation and evolution of galaxies, and also the nature of the sources responsible for reionizing the intergalactic medium (IGM).  Here, we study the physical characteristics of galaxies at $6.3 < z \leq 8.6$, selected from deep near--infrared imaging with the Wide Field Camera 3 (WFC3) on board the {\it Hubble Space Telescope}.  We investigate the rest--frame ultraviolet (UV) colors, stellar masses, ages, metallicities and dust extinction of this galaxy sample.  Accounting for the photometric scatter using simulations, galaxies at $z \sim$ 7 have bluer UV colors compared to typical local starburst galaxies at $>$ 4 $\sigma$ confidence.  Although the blue colors of galaxies at these redshifts necessitate young ages ($<$100 Myr), low or zero dust attenuation, and low metallicities, these are explicable by normal (albeit unreddened) stellar populations, with no evidence for near-zero metallicities and/or top-heavy initial mass functions.  Most of these galaxies are undetected in deep {\it Spitzer} IRAC imaging.  However, the age of the Universe at these redshifts limits the amount of stellar mass in late-type populations, and the WFC3 photometry implies galaxy stellar masses $\sim 10^8 - 10^9$ M\sol\ for Salpeter initial mass functions to a limiting magnitude of $M_{1500} \sim -18$.  The masses of ``characteristic'' (L$^{\ast}$) z $>$ 7 galaxies are smaller than those of L$^{\ast}$ Lyman break galaxies (LBGs) at lower redshifts, and are comparable to less evolved galaxies selected on the basis of their Lyman~$\alpha$~emission at 3 $<$ z $<$ 6, implying that the $6.3 < z \leq 8.6$ galaxies are the progenitors of more evolved galaxies at lower redshifts.  We estimate that \lya\ emission is able to contribute to the observed WFC3 colors of galaxies at these redshifts, with an estimated typical line flux of $\approx$ 10$^{-18}$ erg s$^{-1}$ cm$^{-2}$, roughly a factor of four below currently planned surveys.  The integrated UV specific luminosity for the detected galaxies at $z \sim$ 7 and $z \sim$ 8 is within factors of a few of that required to reionize the IGM assuming low clumping factors, even with no correction for luminosity incompleteness.  This implies that in order to reionize the Universe galaxies at these redshifts have a high ($\sim$ 50\%) escape fraction of Lyman continuum photons, possibly substantiated by the very blue colors of this population.
\end{abstract}


\section{Introduction}

Galaxies evolve very strongly as one observes them during earlier cosmic epochs.  At some point in the past we should begin to witness the periods during which galaxies are very young, likely having formed no more than a few generations of stars.  Strong evolution is observed in the integrated galaxy star--formation rate (SFR) density, which increases by roughly an order of magnitude over the redshift range $z\sim 0$ to 1, and reaches a peak at some point between $1.5 \lesssim z \lesssim 3$ \citep[e.g.,][]{lilly96, steidel99, dickinson03, hopkins04, reddy09}.  There is a marked decline in the ultraviolet (UV) luminosity density from $z \sim 3$ to $z \sim 6$, implying a lower SFR density when the Universe was $<$ 10\% its current age \citep[e.g.,][]{sawicki06, bouwens07, reddy08, reddy09, bouwens09b}.

Star--forming galaxies are readily identified at $z > 2$ as the redshifted Lyman--break (and the increasing opacity in the \lya~forest) moves through the optical passbands for $2 < z < 6$ \citep[e.g.,][]{steidel93}.  The broad--band, multi--wavelength photometry of typical (``L$^\ast$'') Lyman break galaxies (LBGs) at $z \sim 2 - 3$ shows that they have rest--frame UV colors consistent with local starburst galaxies, whose light is dominated by young stellar populations (on the order of 10~Myr to 1~Gyr) with modest amounts of dust attenuation \citep[e.g.,][and references therein]{meurer99, papovich01, shapley01, shapley05, erb06, reddy04, reddy05, reddy06, reddy08, reddy09, bouwens09b}.  

Star--forming galaxies have implied stellar masses of $10^{10}$~M\sol~at $z \sim$ 2--6 \citep[e.g.,][]{sawicki98, papovich01, shapley01, shapley05, yan05, yan06, erb06b, fontana06, reddy06, overzier09}, with some LBGs having stellar masses as high as $\sim 10^{11}$M\sol~\citep[e.g.,][]{shapley05, huang07}.  At higher redshifts, $4 < z < 6$, the LBG population shifts to bluer rest--frame UV colors \citep[e.g.,][]{papovich04, overzier09, bouwens09b}.  While some LBGs at these higher redshifts have implied stellar masses as high as $\sim 10^{11}$~\msol\ \citep[e.g.,][]{yan05, eyles05, yan06, eyles07, stark09}, the stellar masses of L$^{\ast}$ galaxies decline with redshift to values of several times $10^9$~\msol\ at $z \sim 6 - 7$ \citep[e.g.,][]{stark09, gonzalez09, labbe09}.

While our knowledge of the properties of galaxies at $2 < z < 6$ has grown substantially, at $z > 6.5 $ the combined Ly$\alpha$ and Lyman continuum break moves beyond $\sim$ 1~$\mu$m, into the near--infrared, where the terrestrial background limits astronomical surveys to only the brightest objects.  While several recent ground--based surveys have searched for $z > 7$ candidate galaxies using deep $Y$--band imaging at 1.0~$\mu$m \citep[e.g.,][]{castellano09, ouchi09}, these have produced few high redshift objects.  Even the deepest NICMOS exposures are limited in sample size due to the relatively large point-spread-function \citep{bouwens08, gonzalez09}.

The recent installment of the Wide Field Camera 3 (WFC3) on the \textit{Hubble Space Telescope} provides the ability to obtain deep images in broad--bands from 1--2 \micron.  Recent deep WFC3 imaging in the Hubble Ultra Deep Field \citep[HUDF;][]{beckwith06} allows for the identification and study of galaxies selected primarily as LBGs at $z \sim$ 7--9.  Several groups have capitalized on these new WFC3 images, and have studied the luminosity densities, average stellar masses, morphologies, and inferred ionizing properties of galaxies at $6 \lesssim z \lesssim 10$ \citep[e.g.,][]{bouwens09, oesch09, mclure09, bunker09, yan09}.

Here, we use a new sample of galaxies at 6.3 $<$ z $\leq$ 8.6 selected from an updated reduction of the WFC3 imaging combined with existing imaging from 0.4--1~\micron\ from the Advanced Camera for Surveys (ACS) on board \textit{HST} and deep {\it Spitzer Space Telescope} observations (3.6--4.5 \micron).  In \S 2 we describe the data and galaxy sample selection.  In \S 3 we compare our sample to those obtained in previous studies of these WFC3 data.  In \S 4 we investigate the UV colors of these objects.  In \S 5 we compare the full spectral energy distributions (SEDs) of these objects to stellar population models, putting robust constraints on their stellar masses.  In \S 6 we discuss the effect of \lya~emission on the galaxies' SEDs of our sample, and in \S 7 we discuss the implications of this sample on cosmic reionization.  In \S 8 we present our conclusions.  We denote photometric magnitudes measured in the ACS F435W, F606W, F775W, and F850LP filters as $B_{435}$, $V_{606}$, $i_{775}$, $z_{850}$, respectively.  Similarly, we denote magnitudes measured in the WFC3 F105W, F125W, and F160W filters as $Y_{105}$, $J_{125}$, and $H_{160}$, respectively.  Where applicable, we assume a concordance cosmology with H$_\mathrm{o}$ = 70 km s$^{-1}$ Mpc$^{-1}$, $\Omega_m$ = 0.3 and $\Omega_\lambda$ = 0.7.  All magnitudes are reported in the AB system, where $m_\mathrm{AB} = 31.4 - 2.5\log(f_\nu / 1\, \mathrm{nJy})$ \citep{oke83}.

\section{Data and Sample Selection}

\subsection{WFC3 Observations and Reduction}\label{section:data}
The HUDF was observed by WFC3 from 2009 August 26 to 2009 September 6 as part of program ID 11563 (PI: G.~Illingworth) in three broad-band filters, F105W (16 orbits), F125W (16 orbits), and F160W (25 orbits), with central wavelengths $\lambda_{c}$ $\sim$ 1.05~$\mu$m, 1.25~$\mu$m, and 1.60~$\mu$m, respectively.  The WFC3 pointing is centered at $3^h32^m38.5^s$, $-27\degree 47^\prime 00\farcs$ (J2000) and covers approximately a single WFC3 field of view. The total area surveyed is 4.7 arcmin$^2$ with maximum depth, taking into account dithered offsets between images.
\begin{figure}
\epsscale{1.2}
\plotone{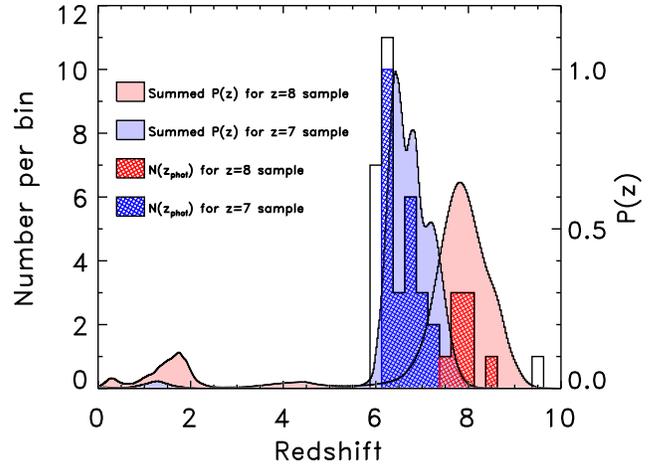}
\caption{Photometric redshift distribution of the high--redshift sample.  All galaxies in this sample have S/N(\wfcj) $>$ 3.5 and S/N(\wfch) $>$ 3.5 and $\mathcal{P}_6 > 0.6$, where $\mathcal{P}_6$ is the photometric redshift probability distribution function integrated from $z = 6$ to 11.  The black, blue and red histograms show the distribution of best--fit photometric redshifts for all 31 objects, the 23 objects in the $z \sim$ 7 sample and the 8 objects in the $z \sim$ 8 sample, respectively.  The blue and red curves are the summed photometric redshift probability distribution function of all galaxies in the $z \sim$ 7 and $z \sim$ 8 samples, respectively.\label{zhist}}
\end{figure}

We processed the WFC3 data (described in more detail in Koekemoer et al.\ 2010, in prep), first reducing the data from the archive using the STSDAS PyRAF task \texttt{calwfc3}, which applies steps to remove the dark current, and corrects for the flat field and detector linearity. We used on--orbit calibration files for the dark current and flat--field rather than the pre-launch calibration files \citep[c.f.,][]{bouwens09, oesch09}.   Several of the F105W images suffered from persistence effects from a previously--observed program.  We reduced the WFC3 images with and without these images, and found that it affected the \wfcy\ photometry by as much as several tenths of a magnitude.  We therefore excluded these data to prevent the persistence--laden images from affecting our photometry.  We combined the images using MultiDrizzle \citep{koekemoer02}, aligning each image by their fractional-pixel dithers, and weighting by the inverse variance of each pixel.  Each WFC3 image was aligned to the original ACS HUDF images to determine the unique astrometric and geometric transformation.  Other reduction steps are similar to those described in \citet{mclure09}. The final mosaic has $0\farcs03$ pixels in each band.  We measure a full--width at half--maximum (FWHM) of $0\farcs176$ in \hb\ for point sources identified in the field.

We use the latest zeropoint information available from the STScI/WFC3 instrument science team\footnote[1]{http://www.stsci.edu/hst/wfc3/phot$\_$zp$\_$lbn}, which were 26.27, 26.25, and 25.96 AB mag for the F105W, F125W, and F160W images, respectively.  These are identical to those employed by McLure et al.\ (2009).  However, they are offset from those used in Bunker et al.\ by 0.10--0.15~mag.  This offset is typically smaller than the statistical uncertainty in object $Y_{105} - J_{125}$ and $J_{125} - H_{160}$ colors, however it is a systematic offset in colors that can effect the interpretation of object colors, photometric--redshift estimation, and stellar-population properties (see \S~3 below). We measure the limiting flux depth of each image by measuring the distribution of fluxes in 10$^{4}$ randomly placed $0\farcs4$--diameter apertures.  We derive $5\sigma$ limiting magnitudes of $Y_{105} = 28.98$~mag, $J_{125} = 29.17$~mag, and $H_{160} = 29.21$~mag.  These are consistent with respect to the findings of other groups.

We convolved the ACS data with a kernel to match the PSF of the WFC3 data.  Our tests indicated that this PSF convolution matches point-source photometry between the ACS and WFC3 bands to better than 5\% for apertures larger that $0\farcs5$-diameter.    Our same tests indicated that we recover the same fraction light  to better than 5\% for unconvolved versions of the three WFC3 images (F105W, F125W, F160W) in apertures larger than $0\farcs5$-diameter.  Because the photometry measured in the native angular resolution of these bands is already well matched we make no adjustment to the FWHM of the WFC3 images.  We use the PSF-matched versions of the ACS data for our photometric catalogs.

\begin{figure*}
\epsscale{0.18}
\vspace{2mm}
\plotone{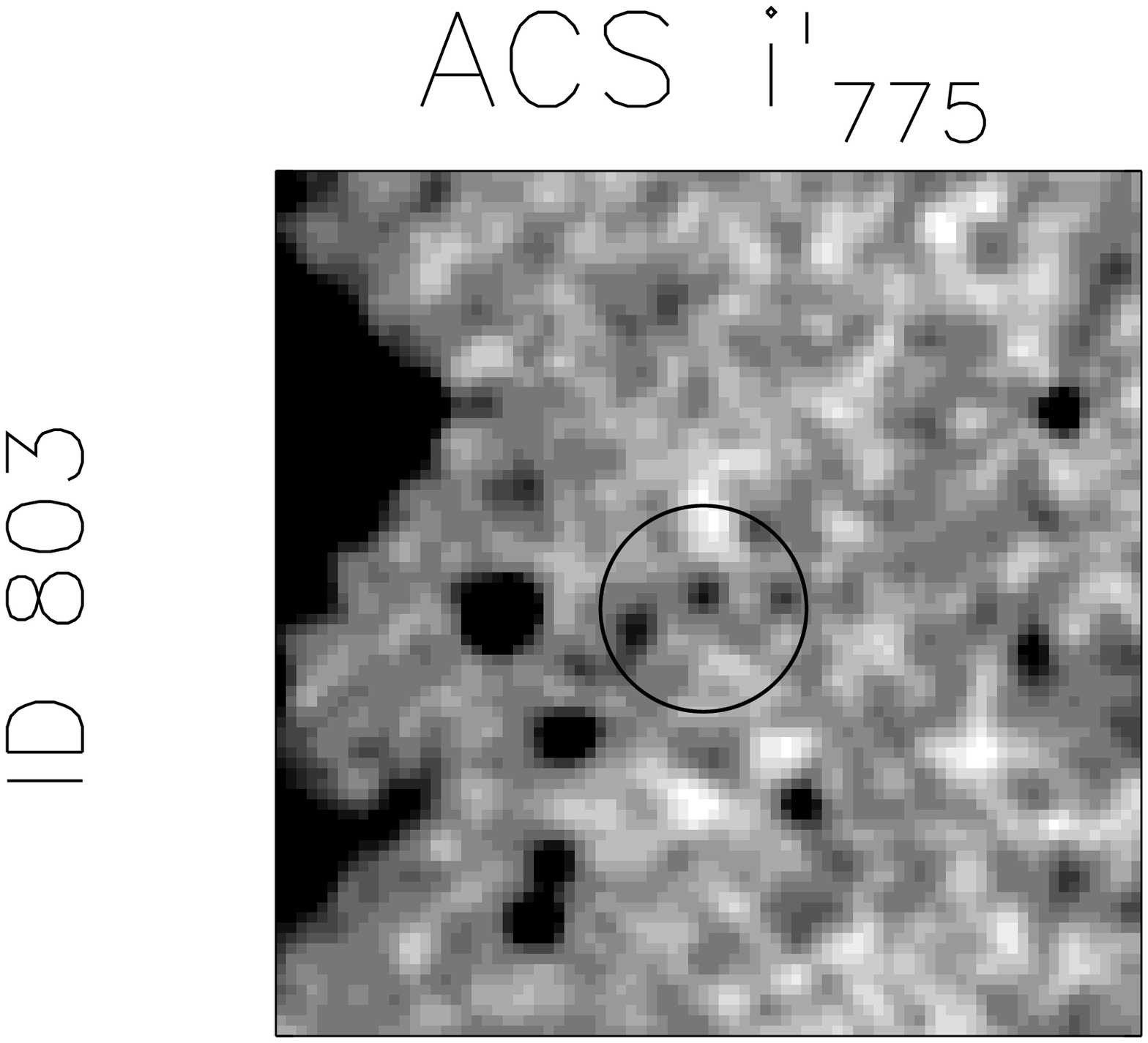}
\hspace{-10mm}
\plotone{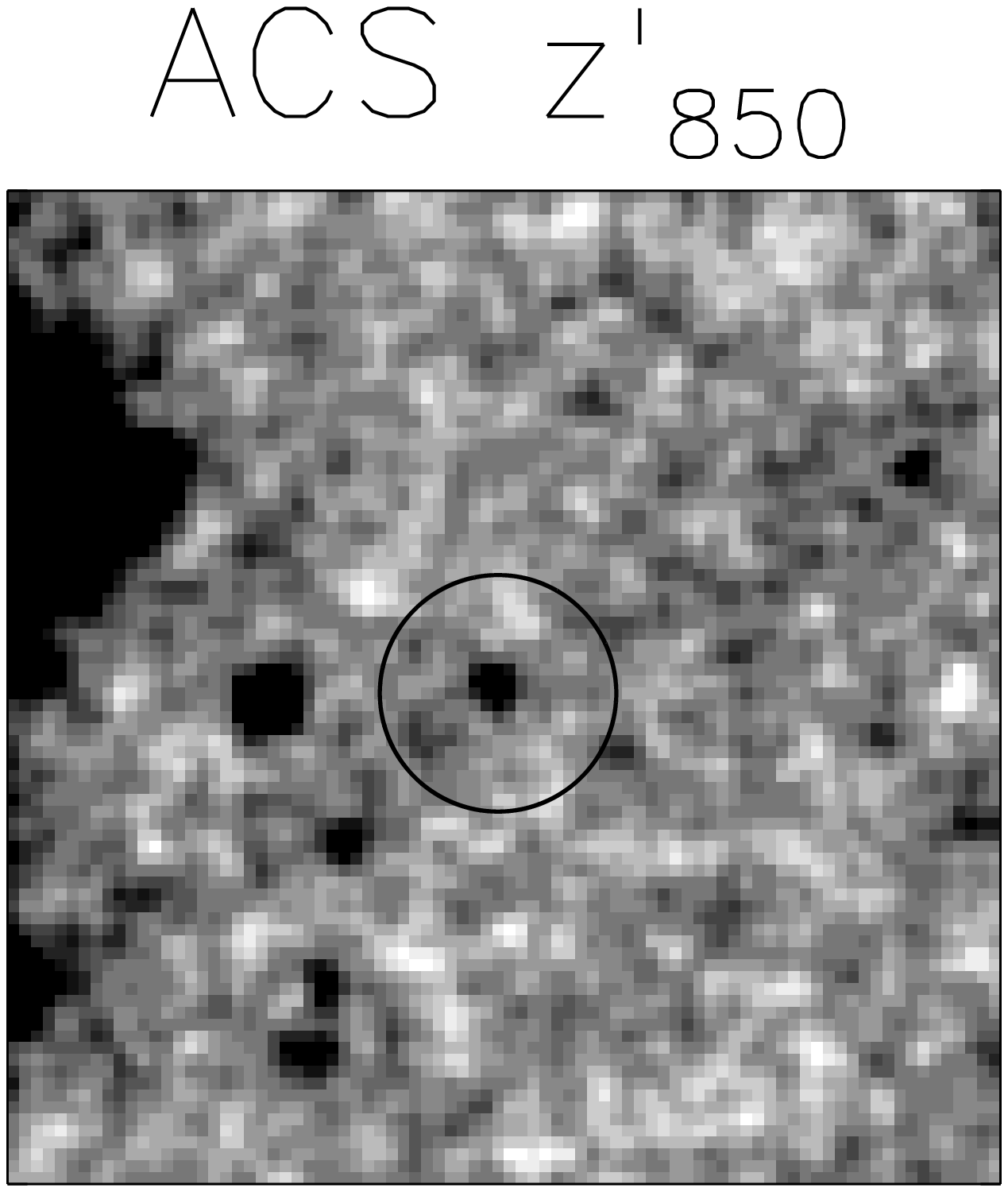}
\hspace{-10mm}
\plotone{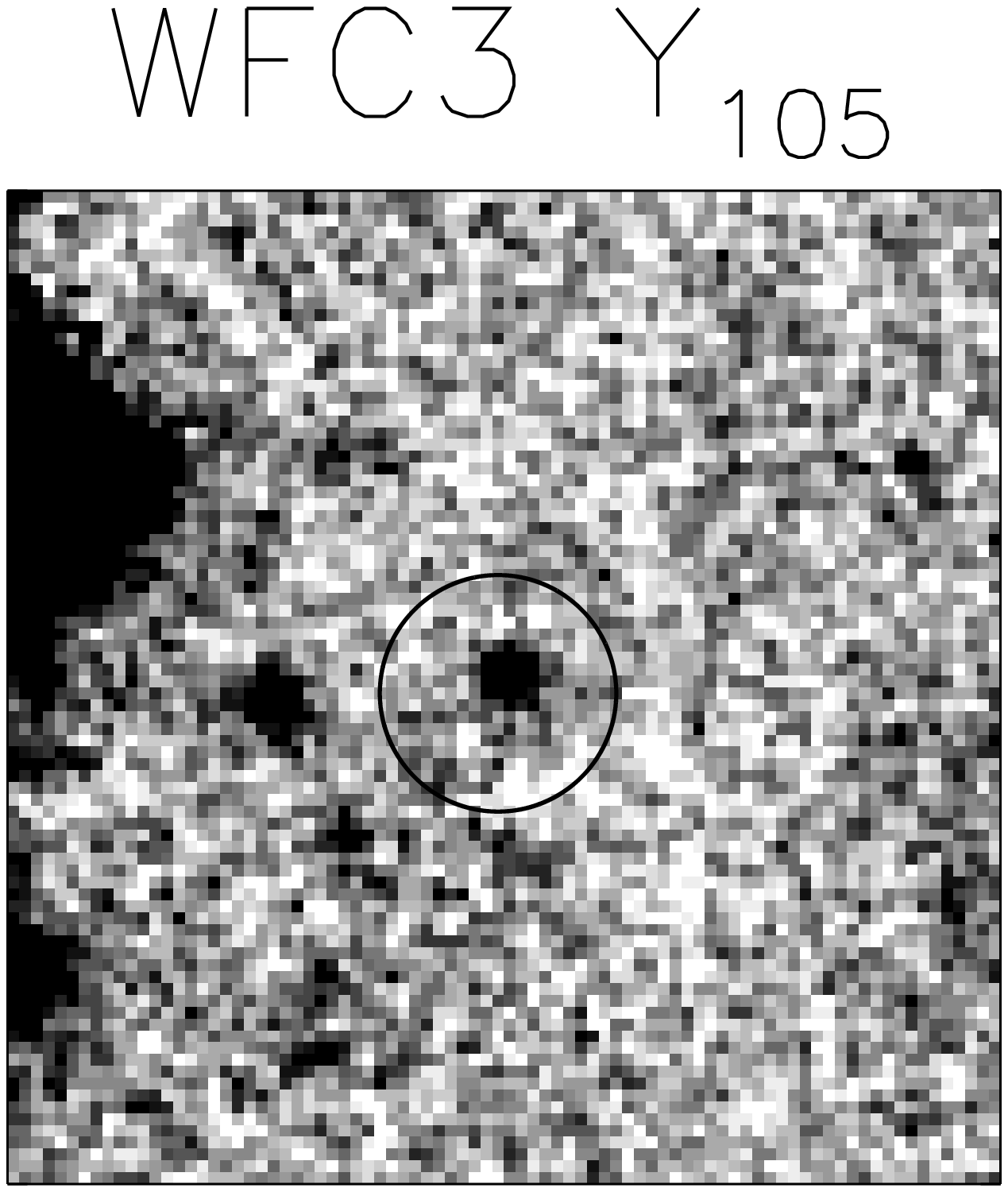}
\hspace{-10mm}
\plotone{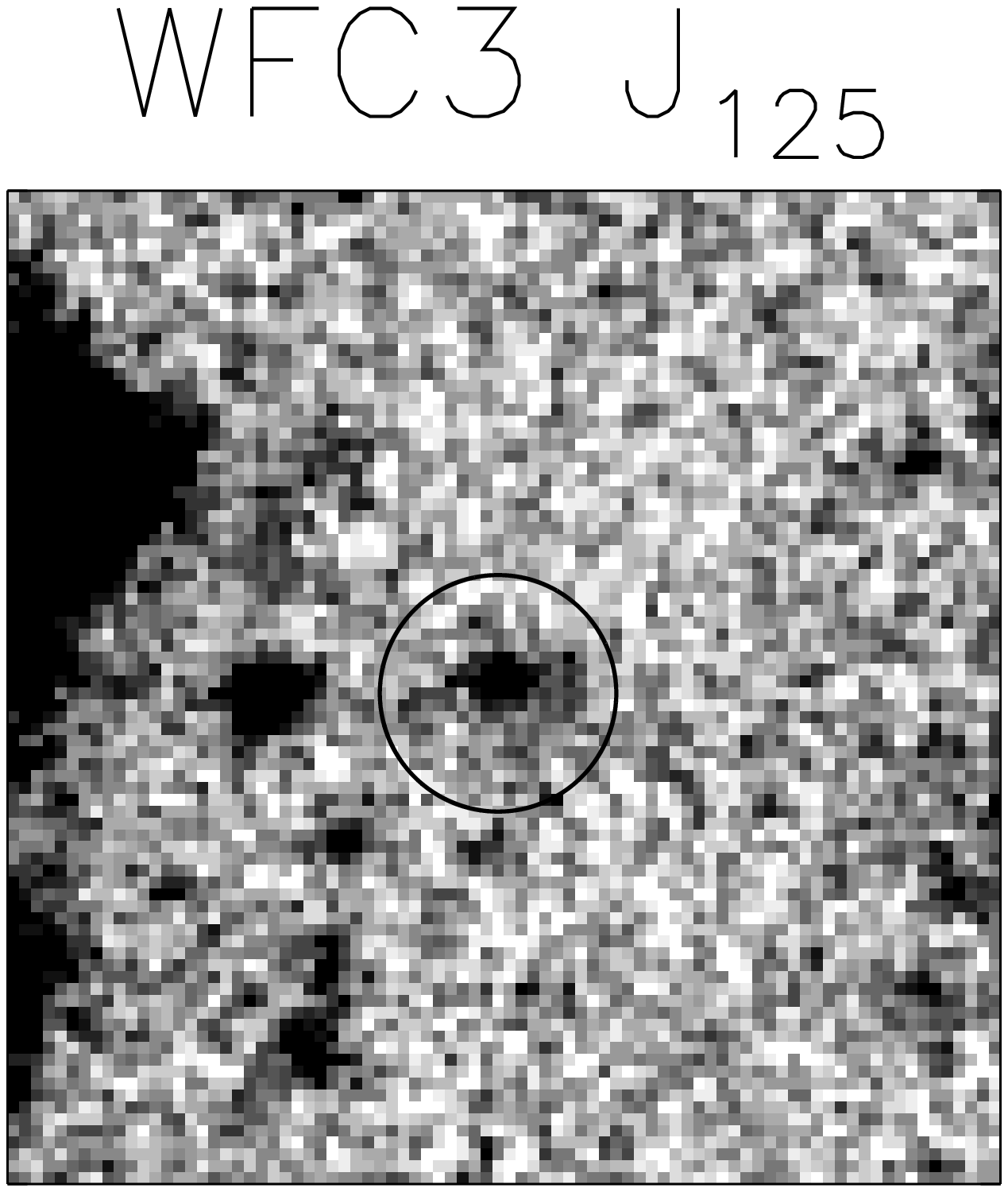}
\hspace{-10mm}
\plotone{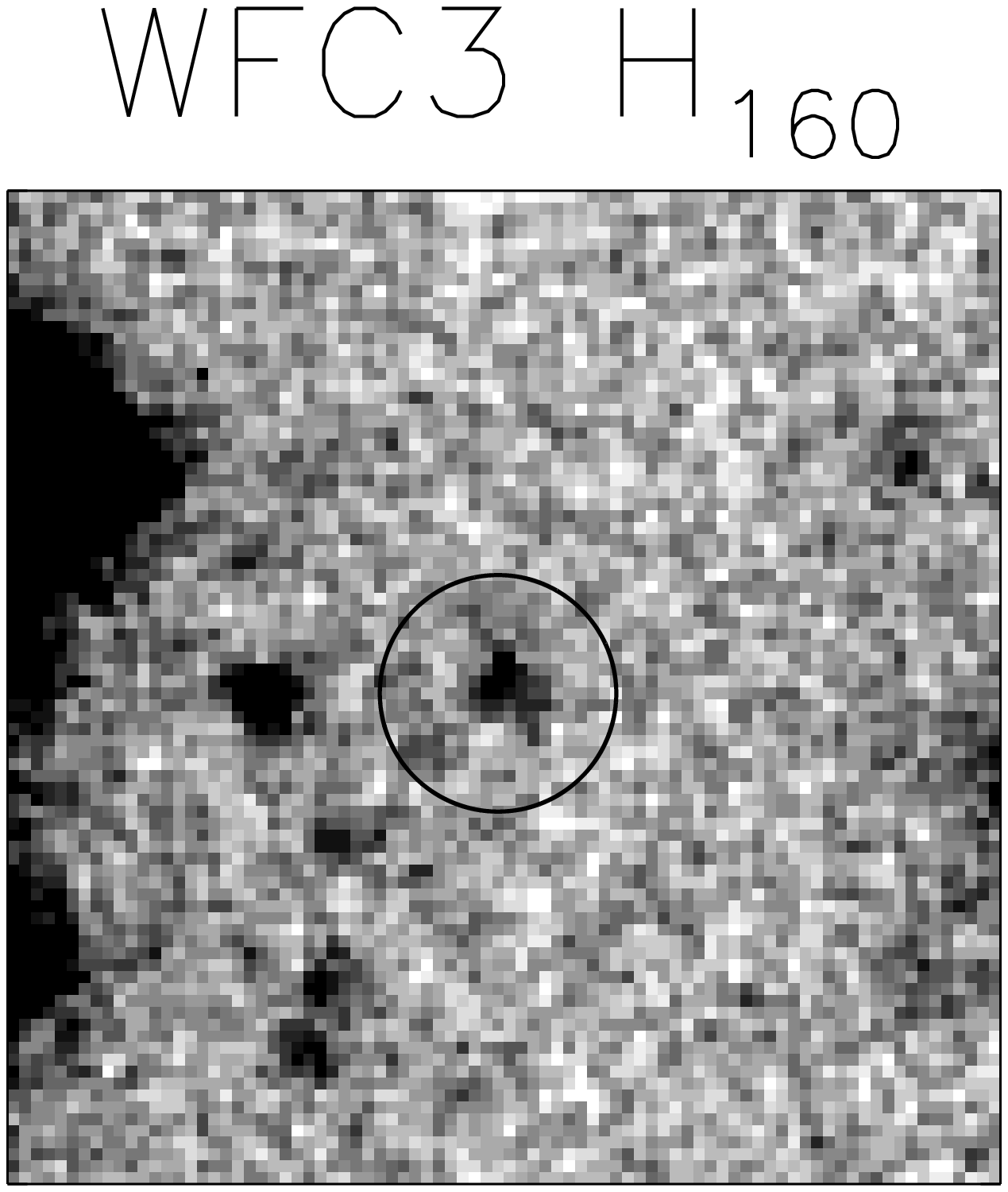}
\hspace{-10mm}
\plotone{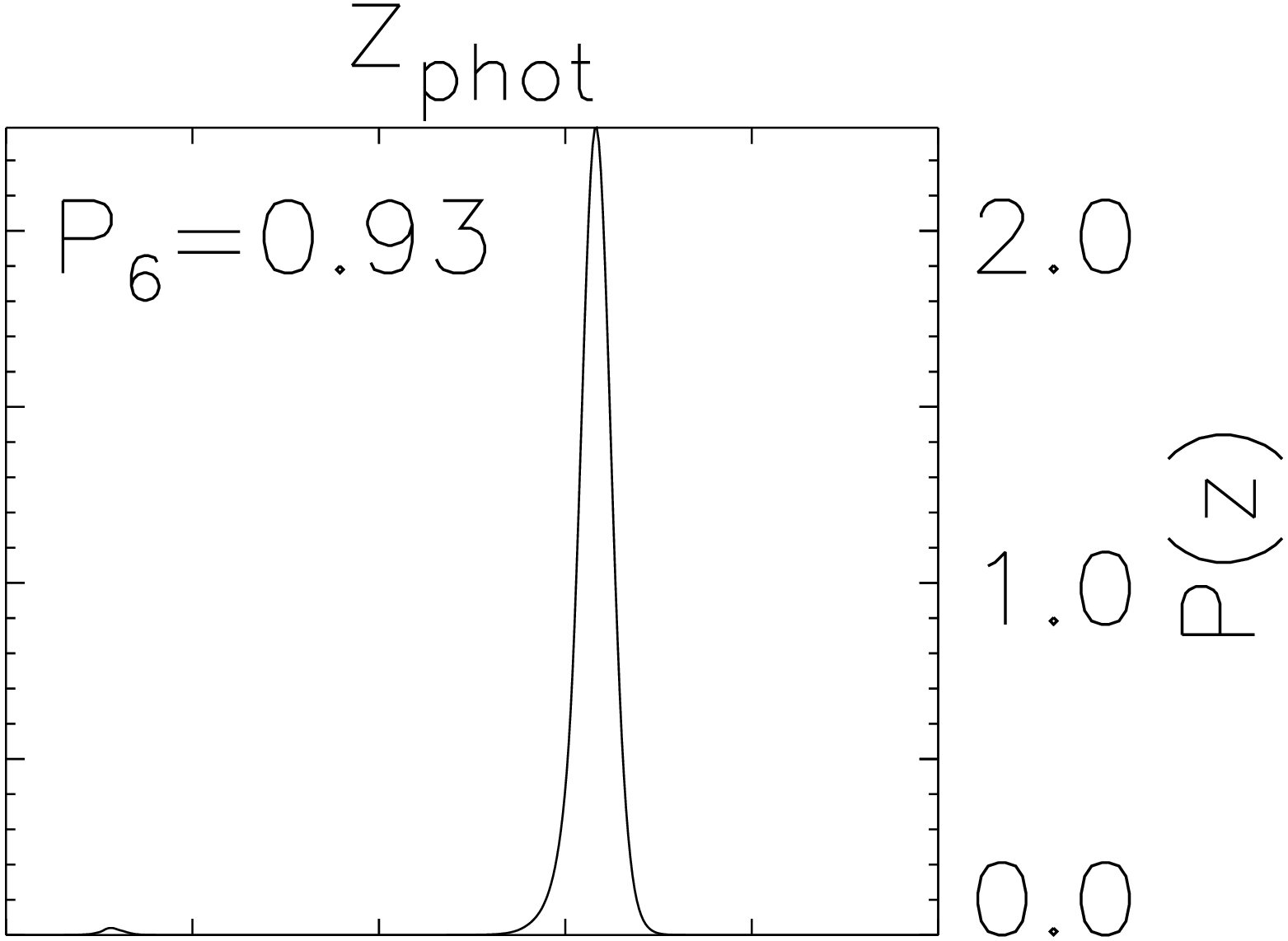}
\vspace{0.5mm}

\plotone{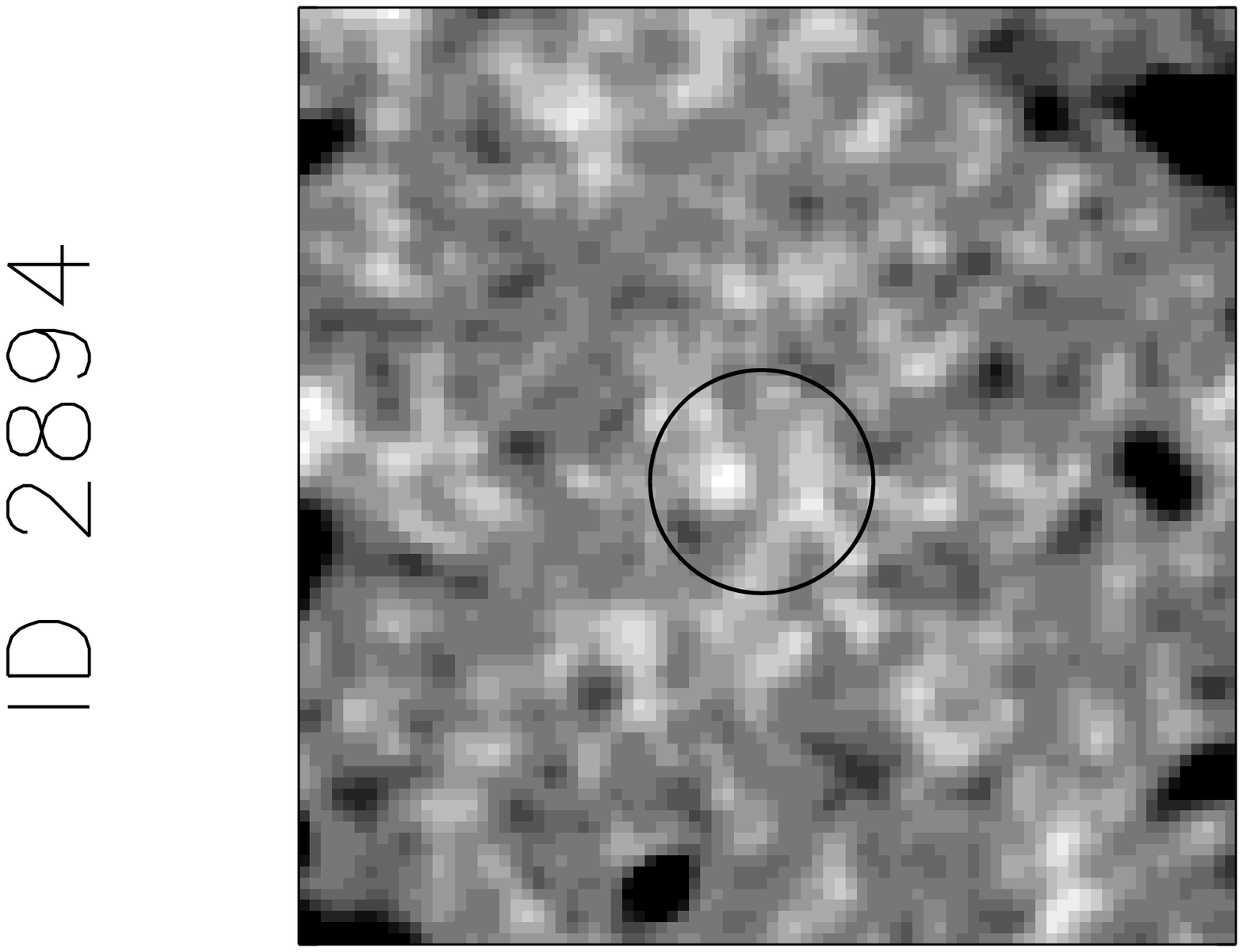}
\hspace{-10mm}
\plotone{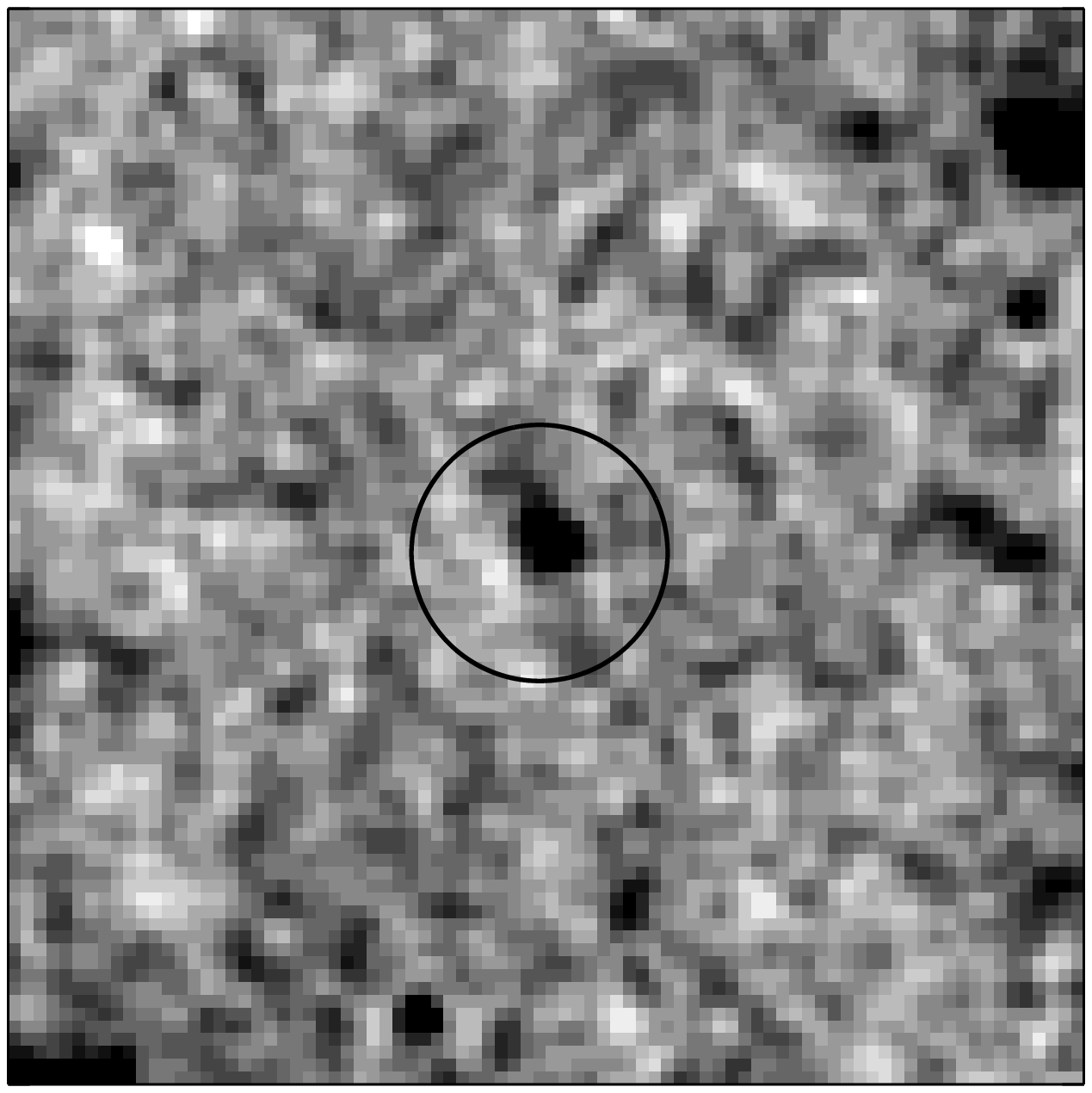}
\hspace{-10mm}
\plotone{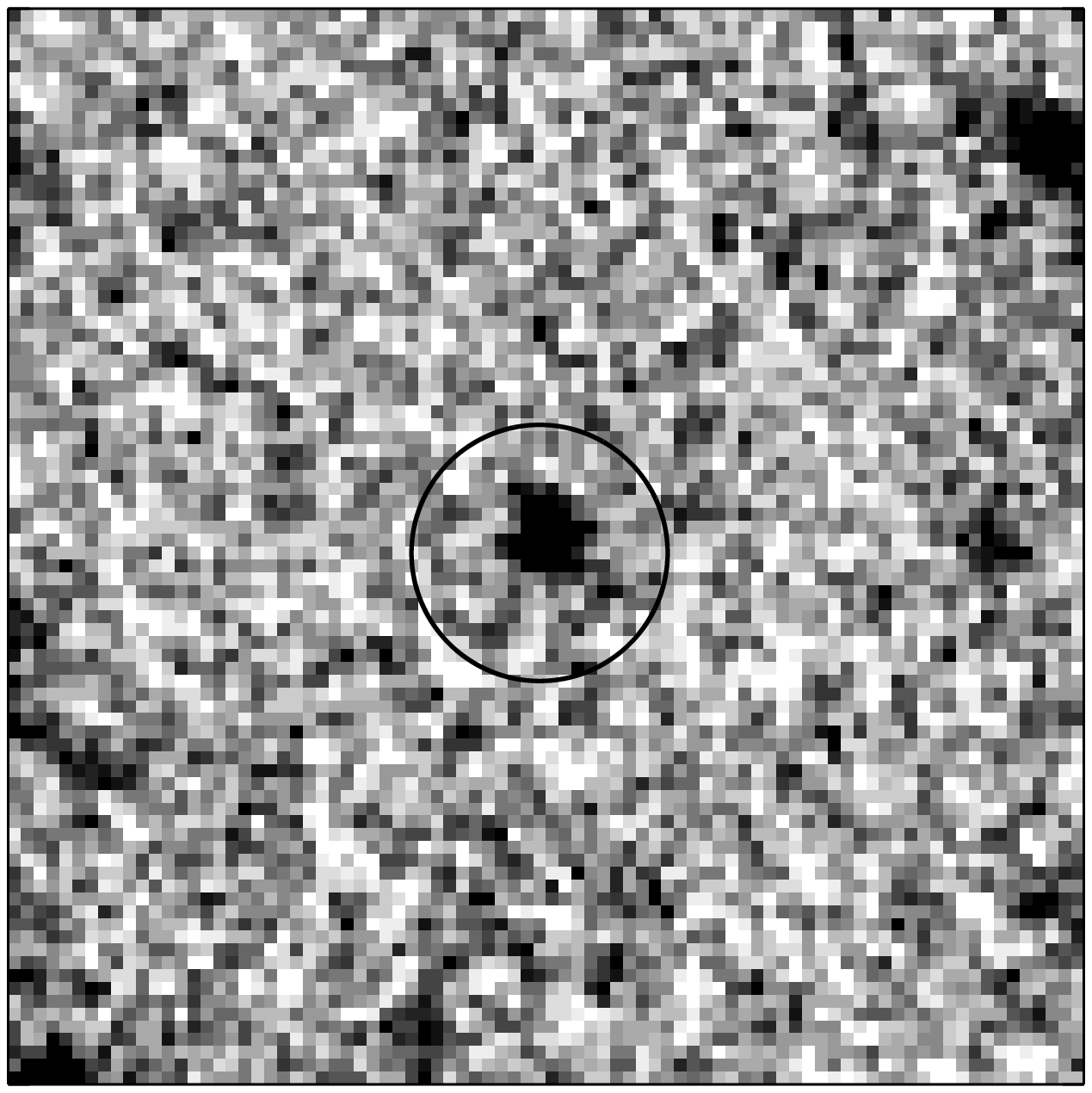}
\hspace{-10mm}
\plotone{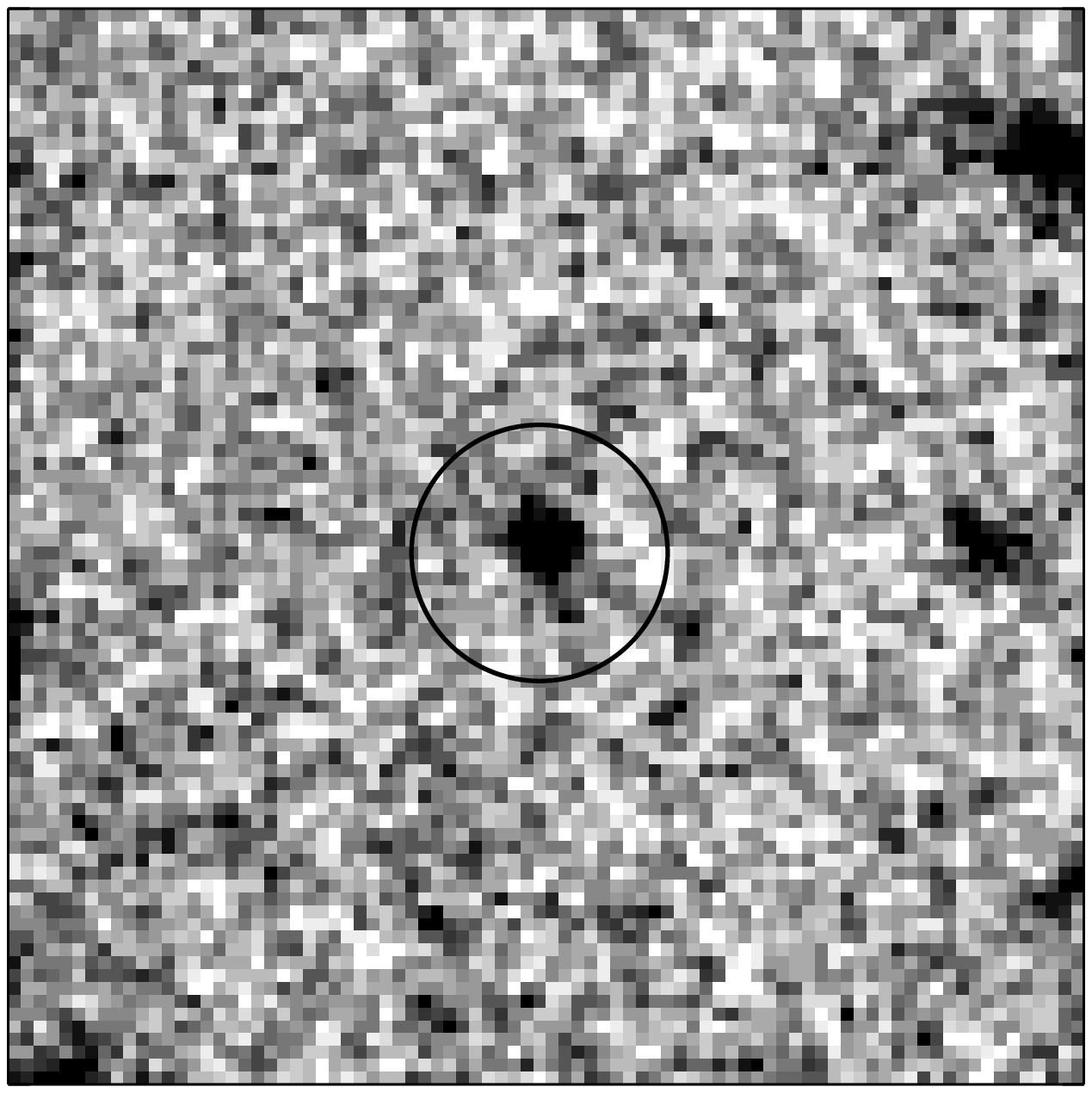}
\hspace{-10mm}
\plotone{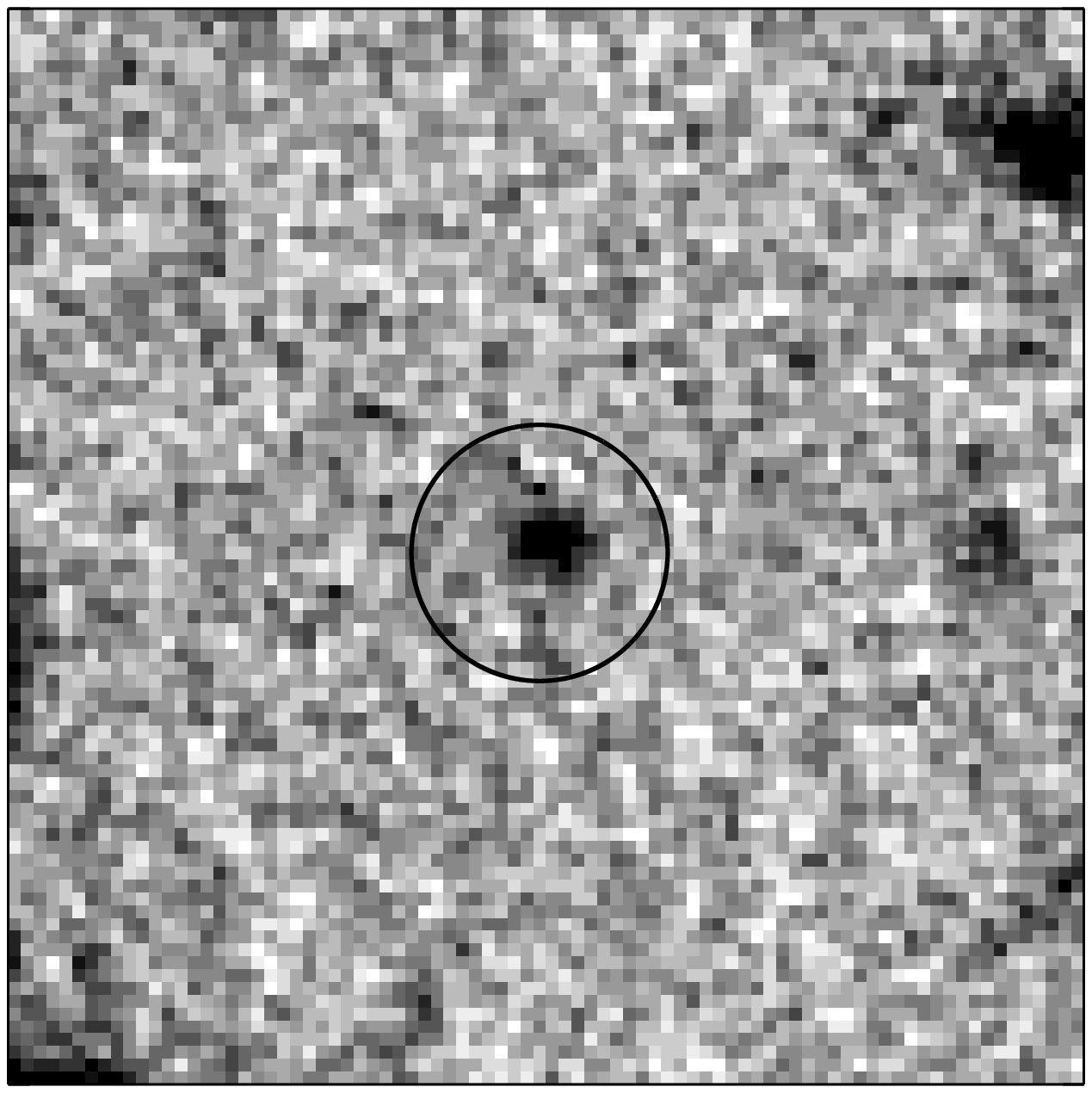}
\hspace{-10mm}
\plotone{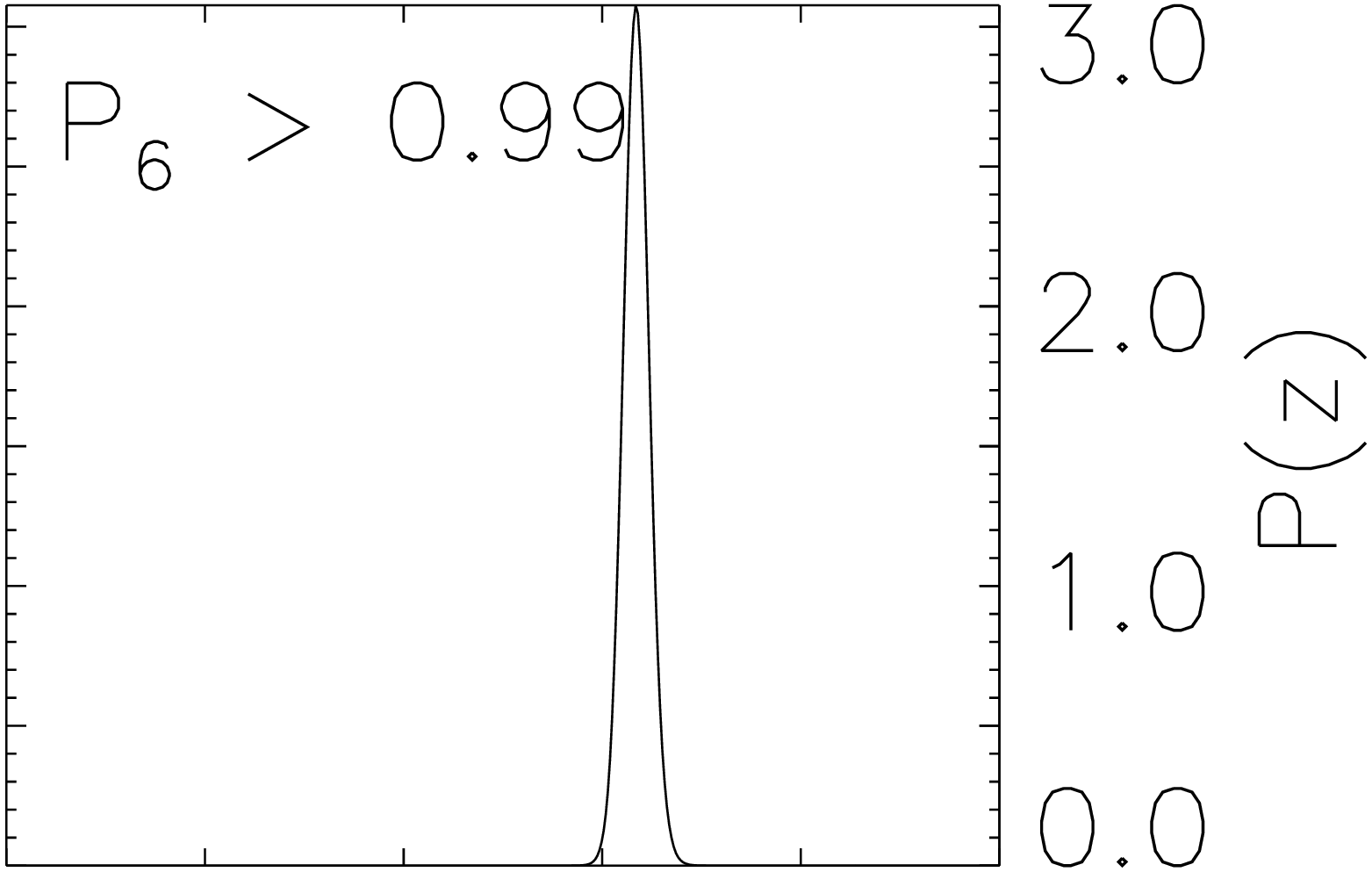}
\vspace{0.5mm}

\plotone{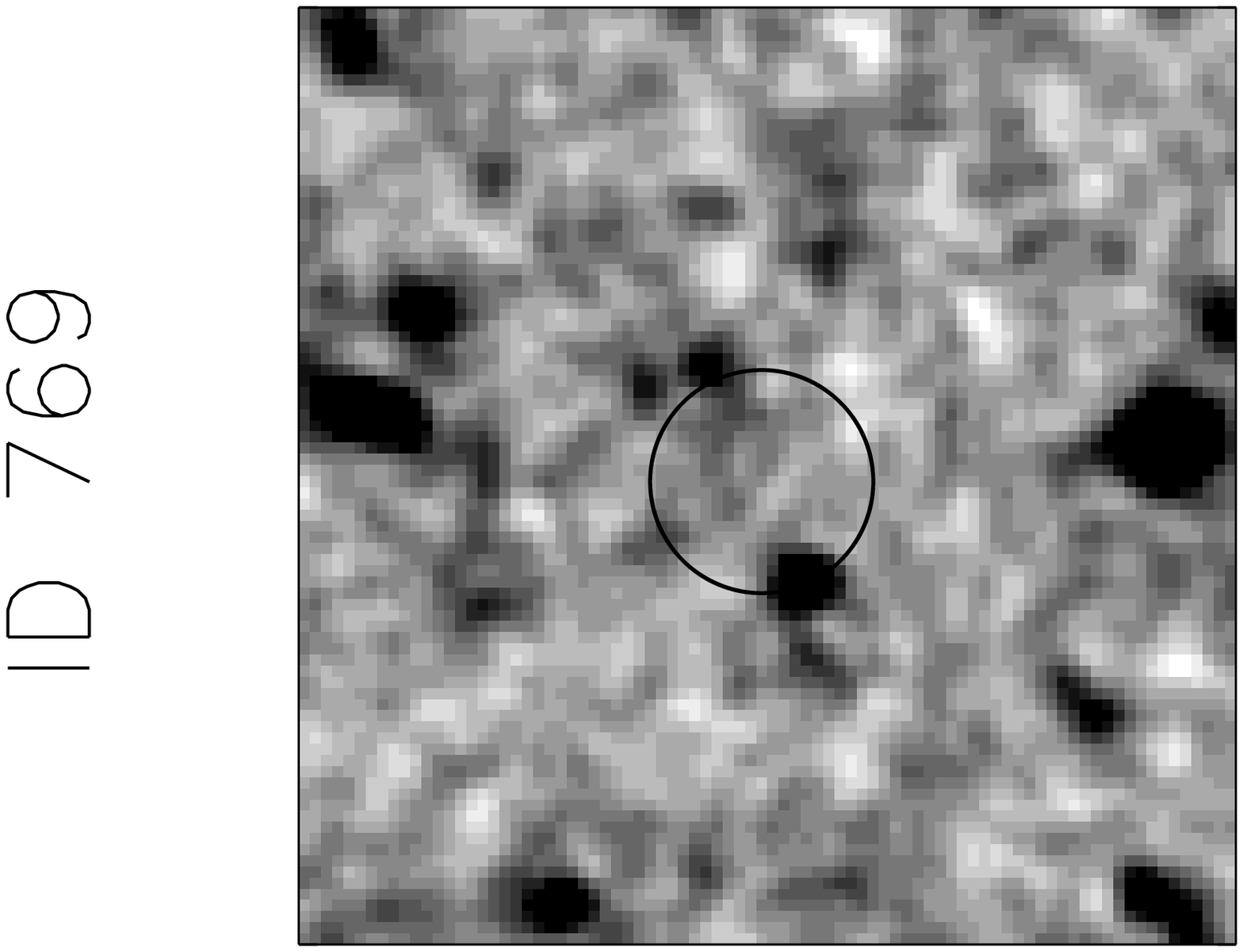}
\hspace{-10mm}
\plotone{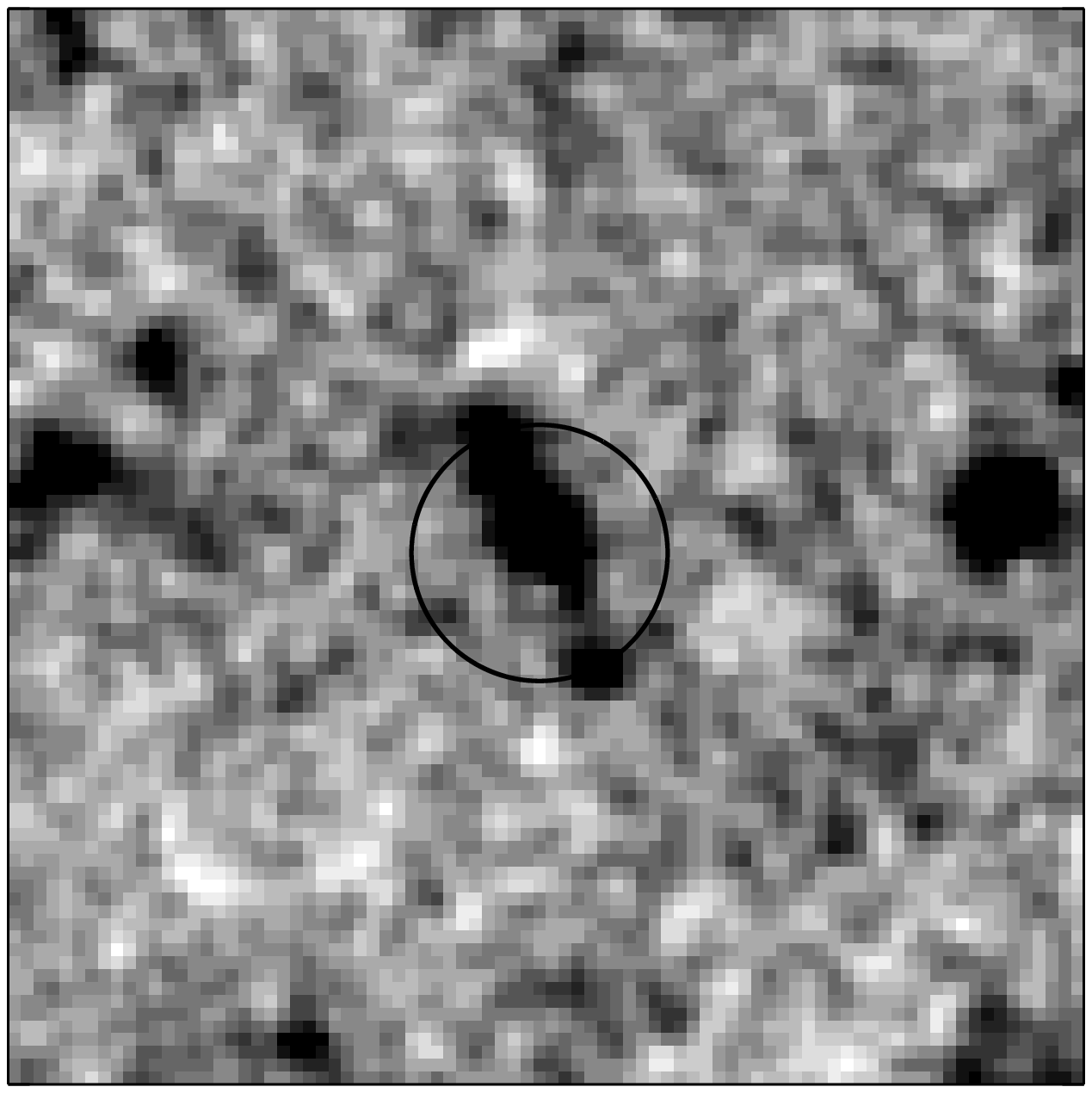}
\hspace{-10mm}
\plotone{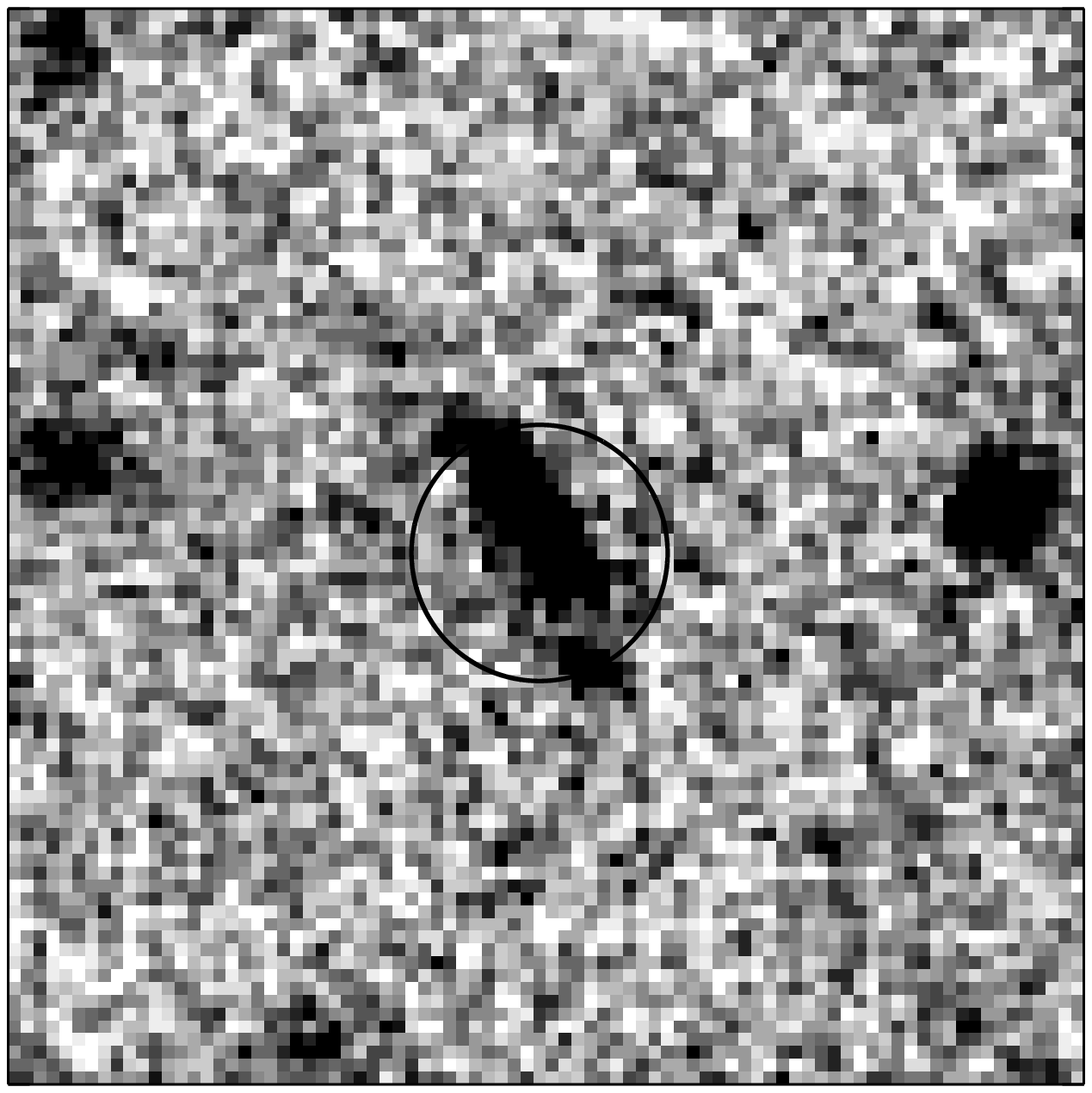}
\hspace{-10mm}
\plotone{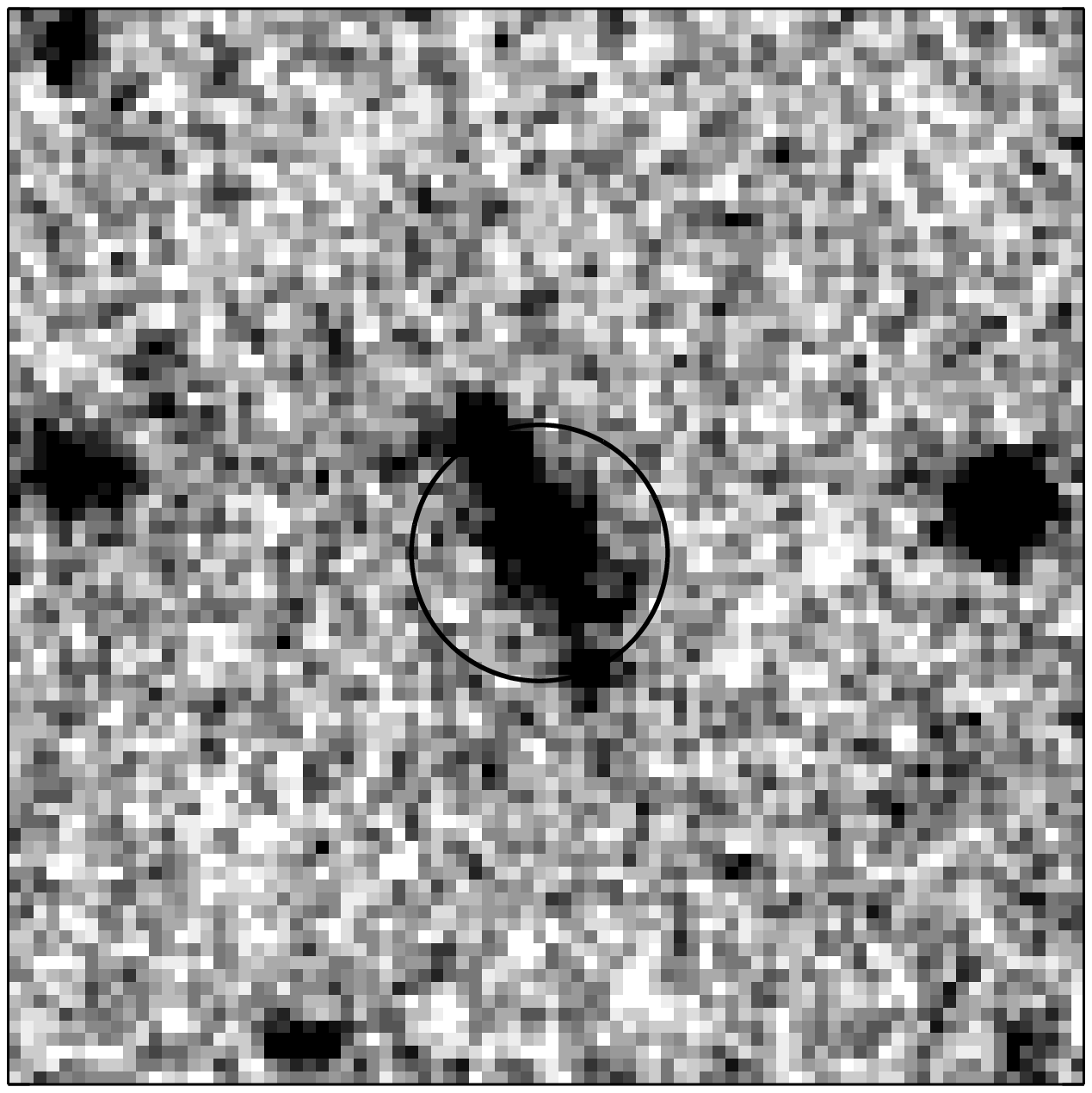}
\hspace{-10mm}
\plotone{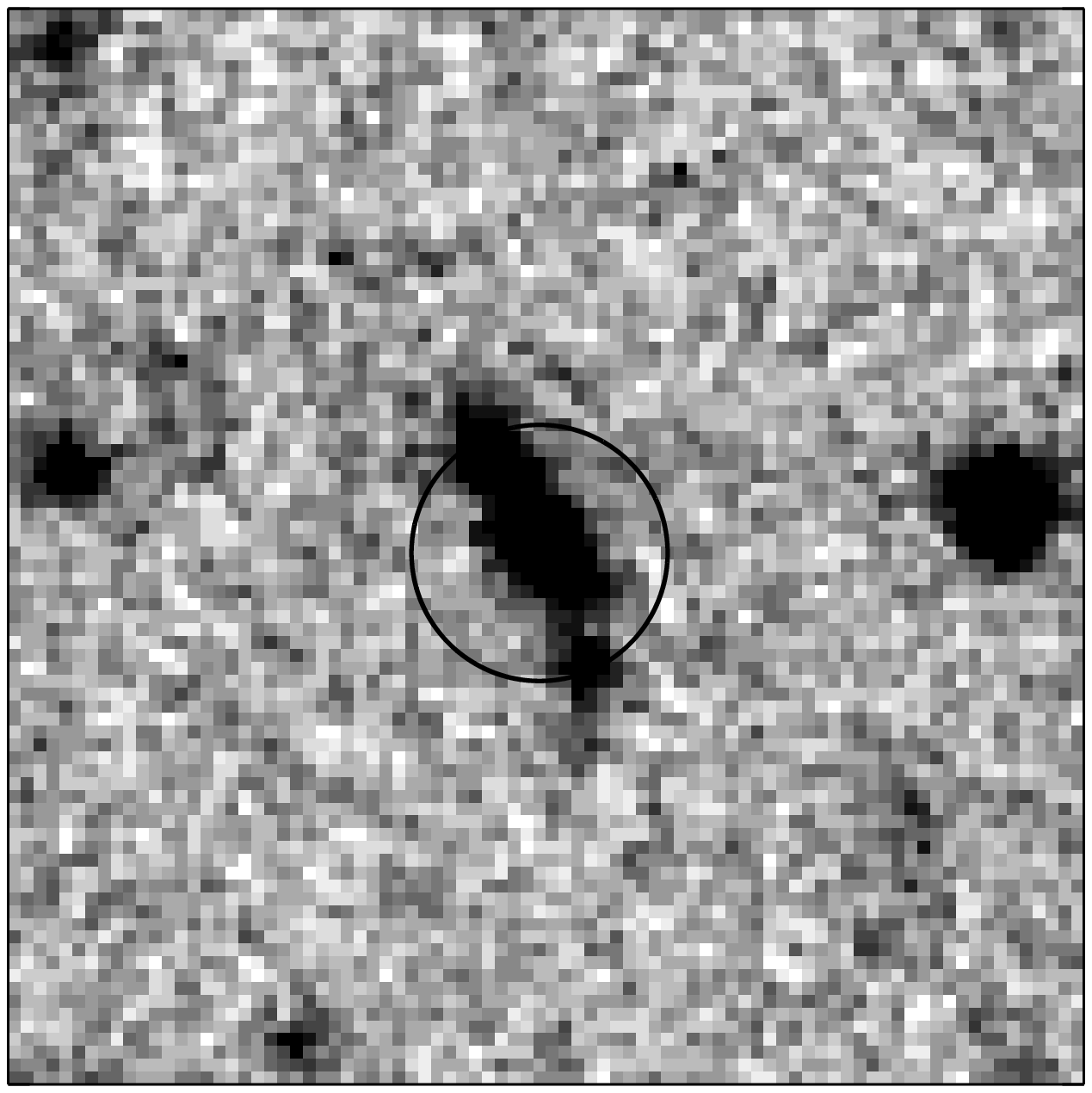}
\hspace{-10mm}
\plotone{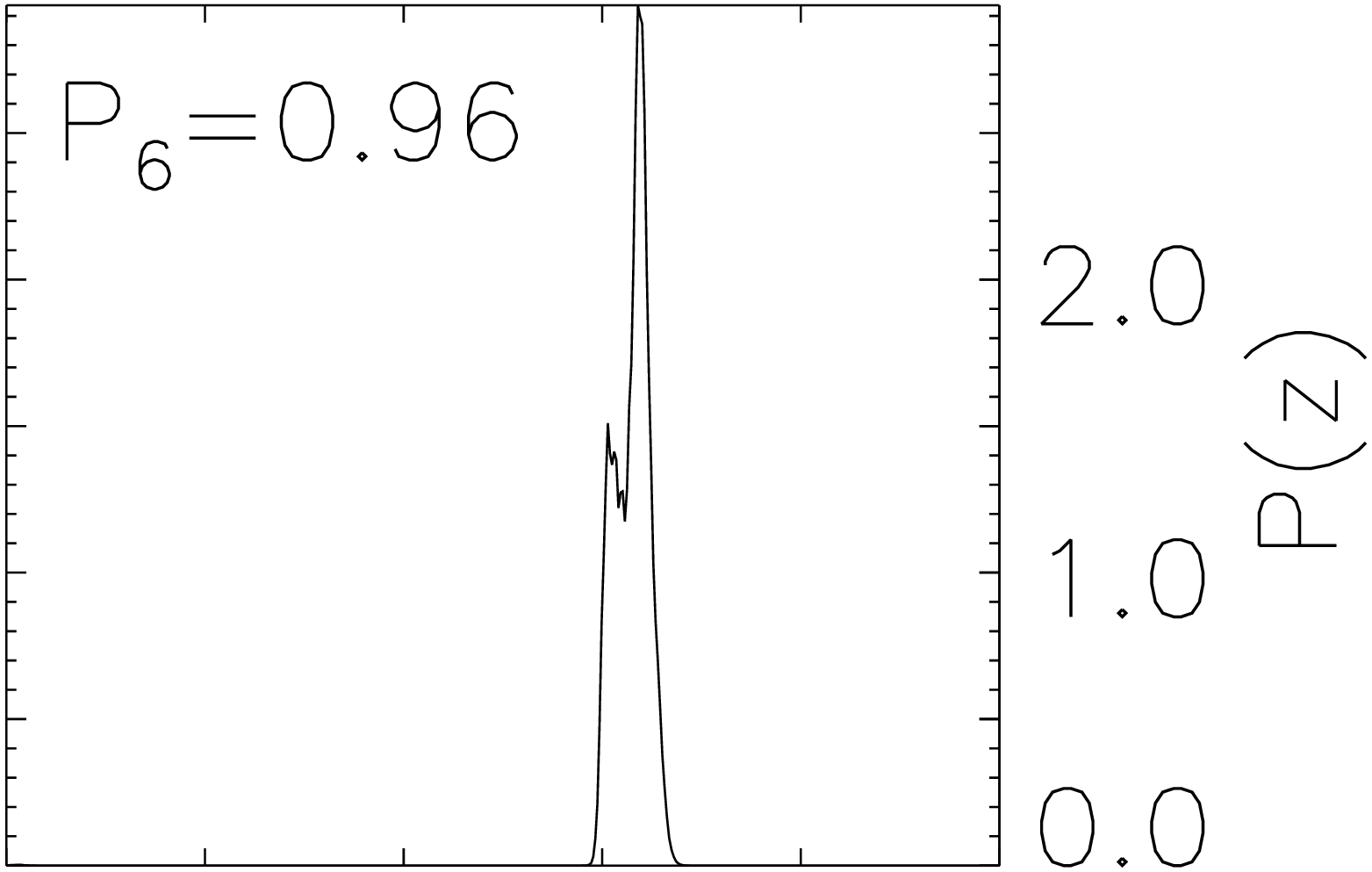}
\vspace{0.5mm}

\plotone{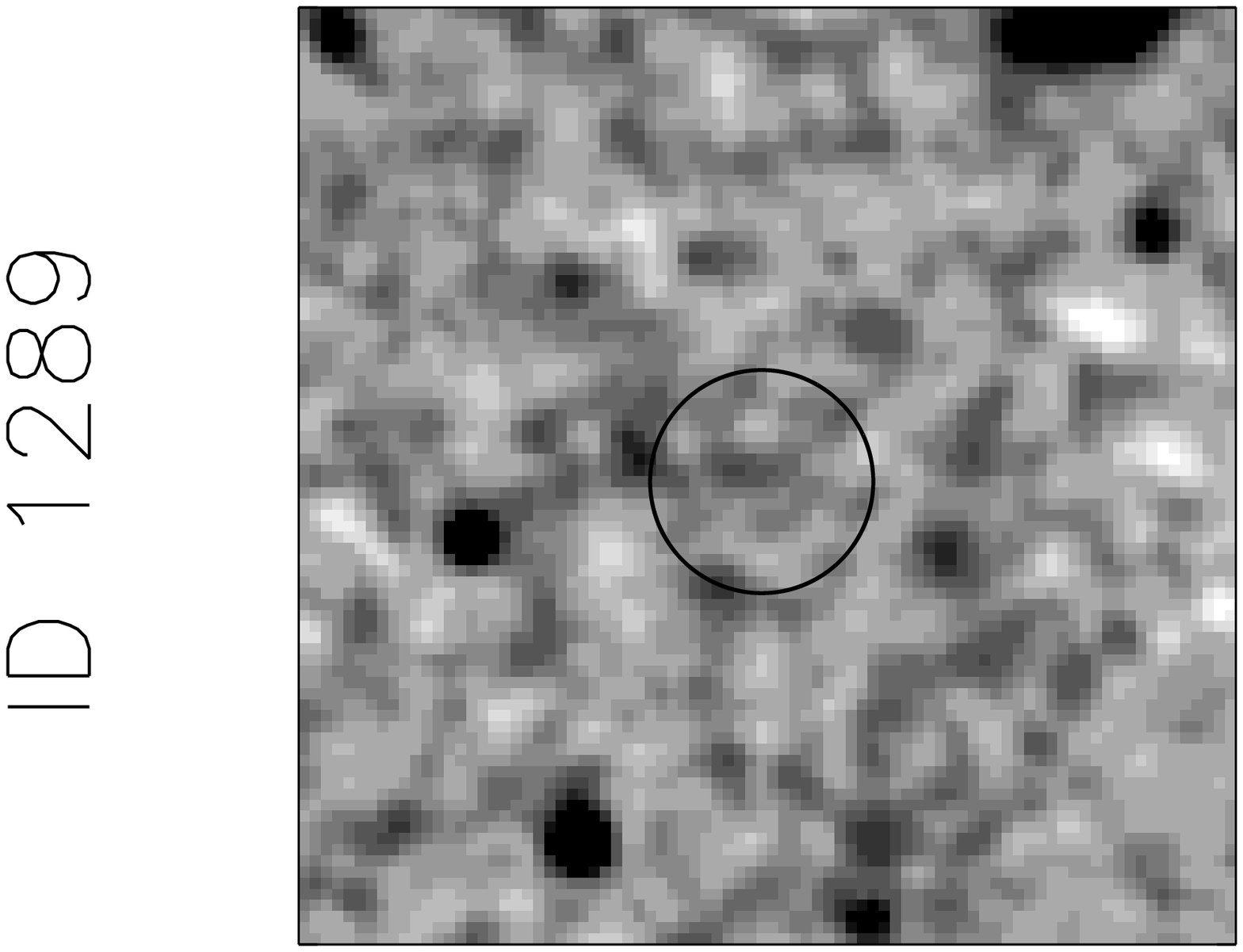}
\hspace{-10mm}
\plotone{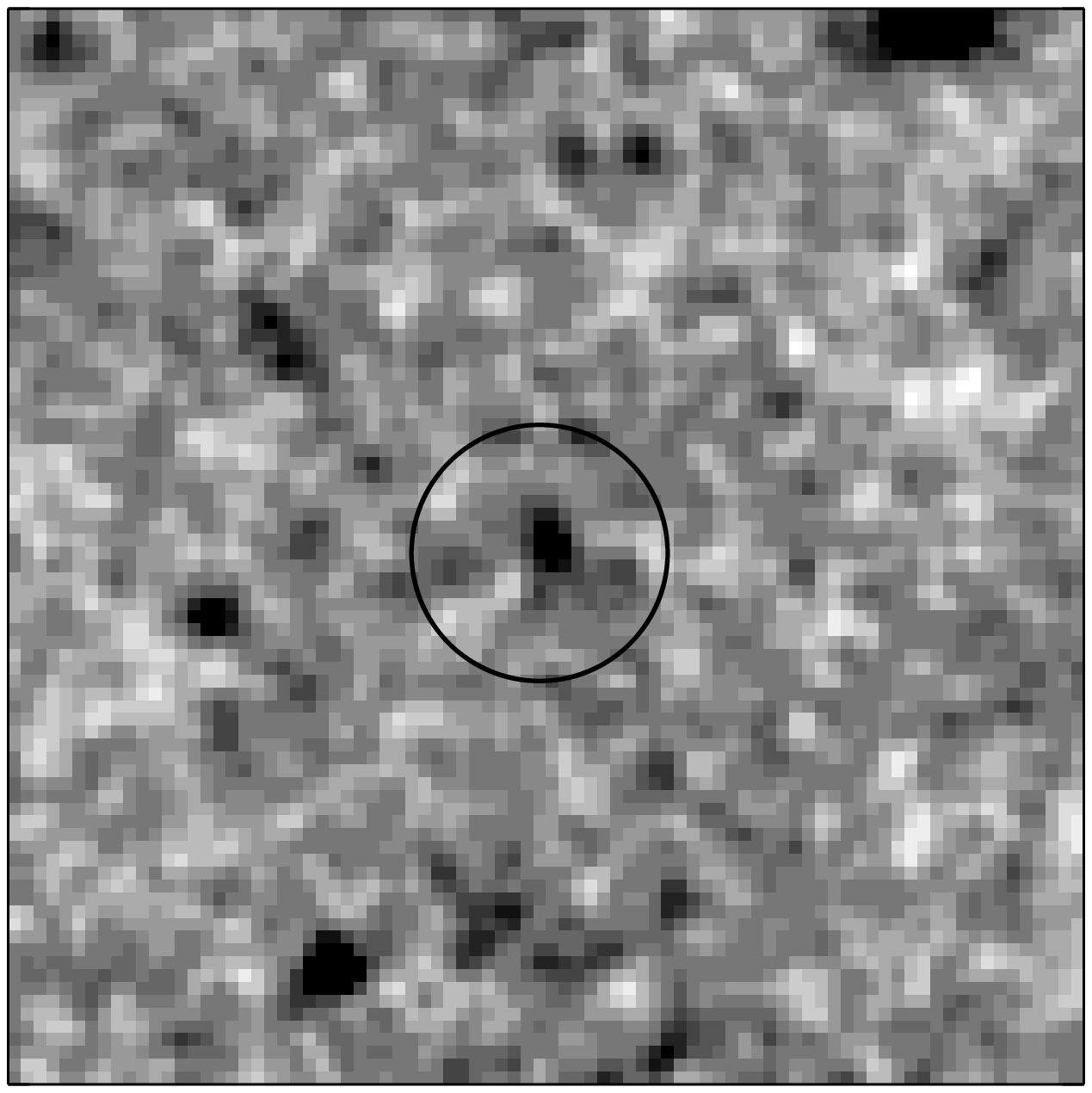}
\hspace{-10mm}
\plotone{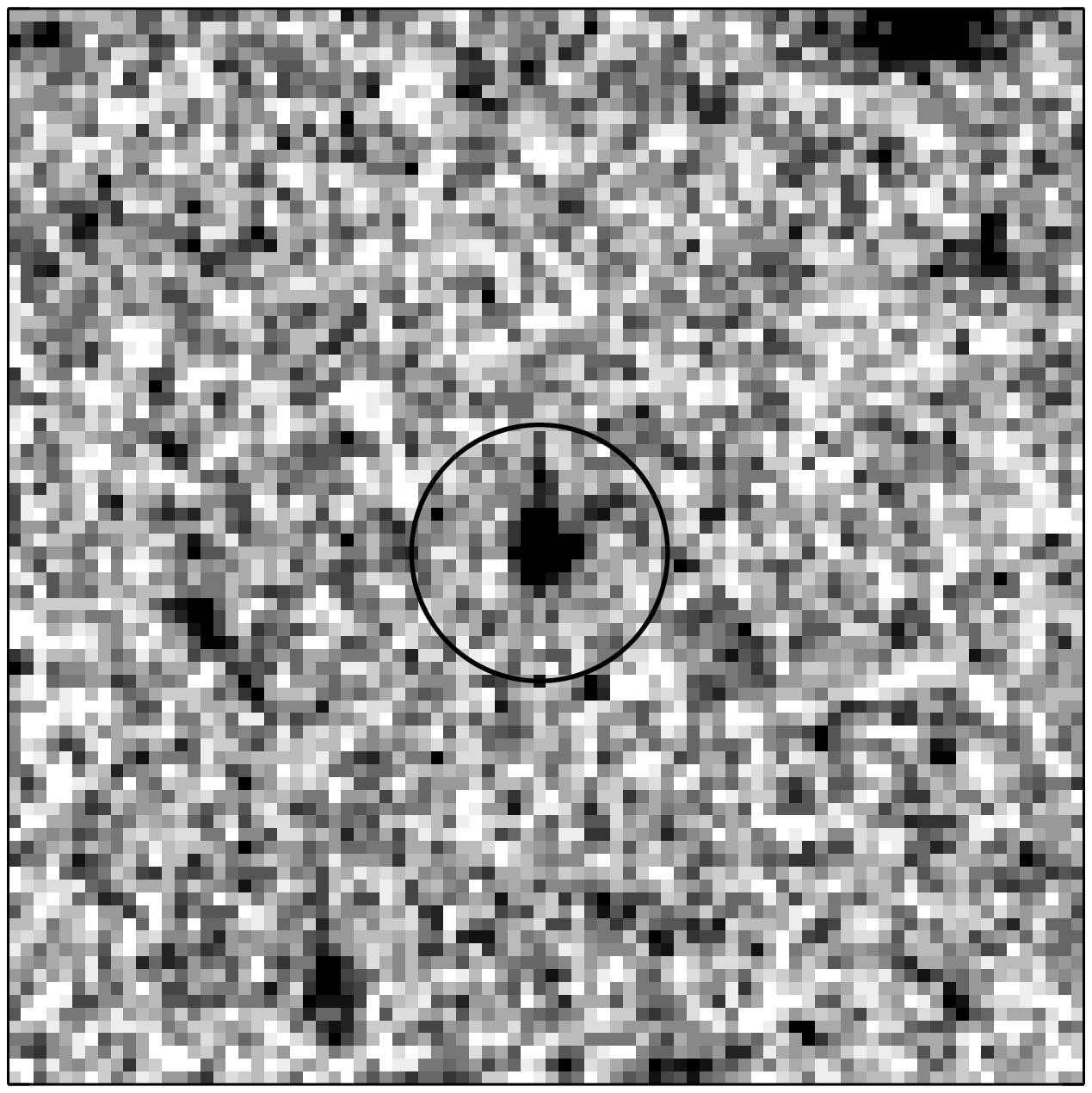}
\hspace{-10mm}
\plotone{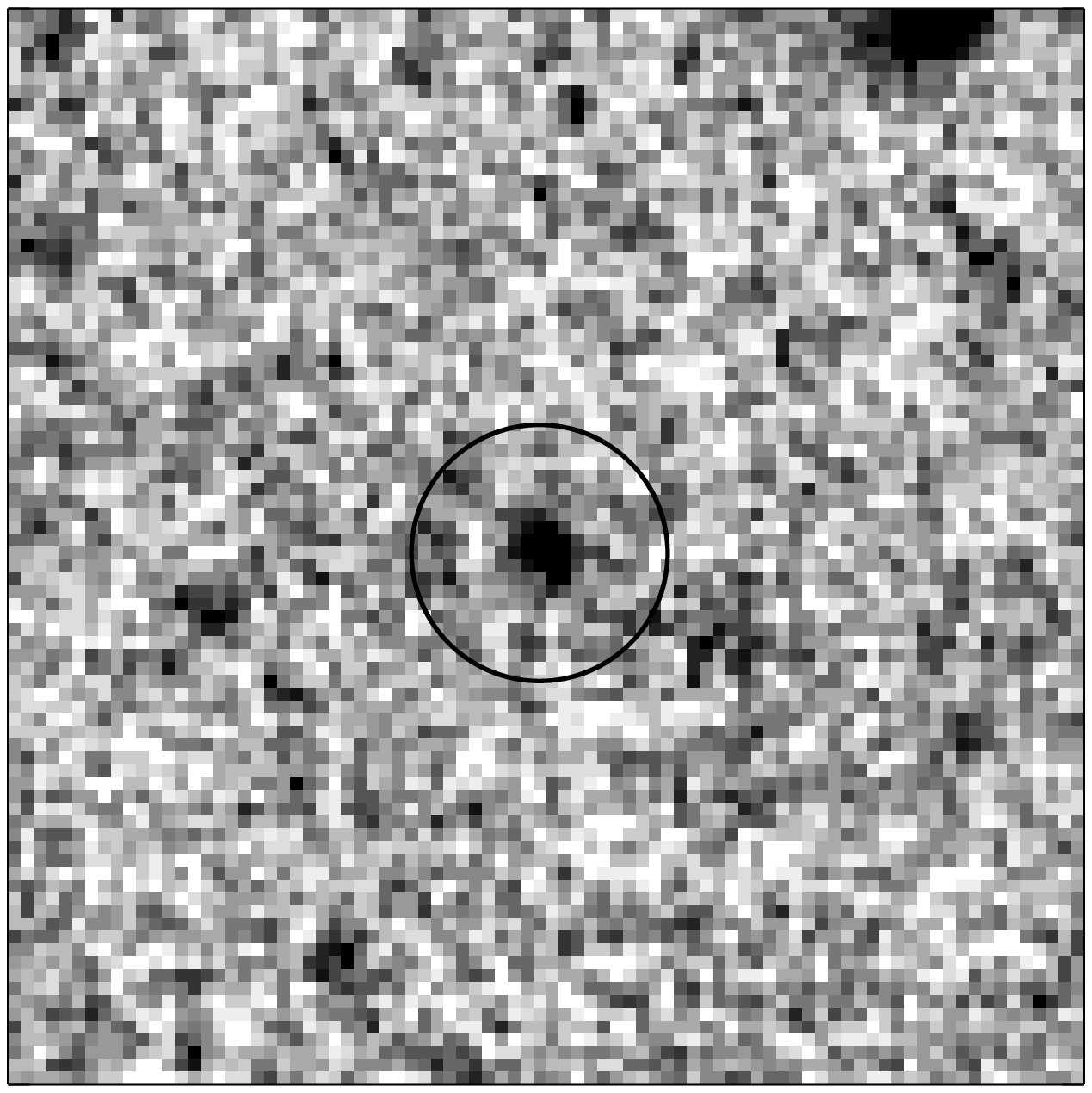}
\hspace{-10mm}
\plotone{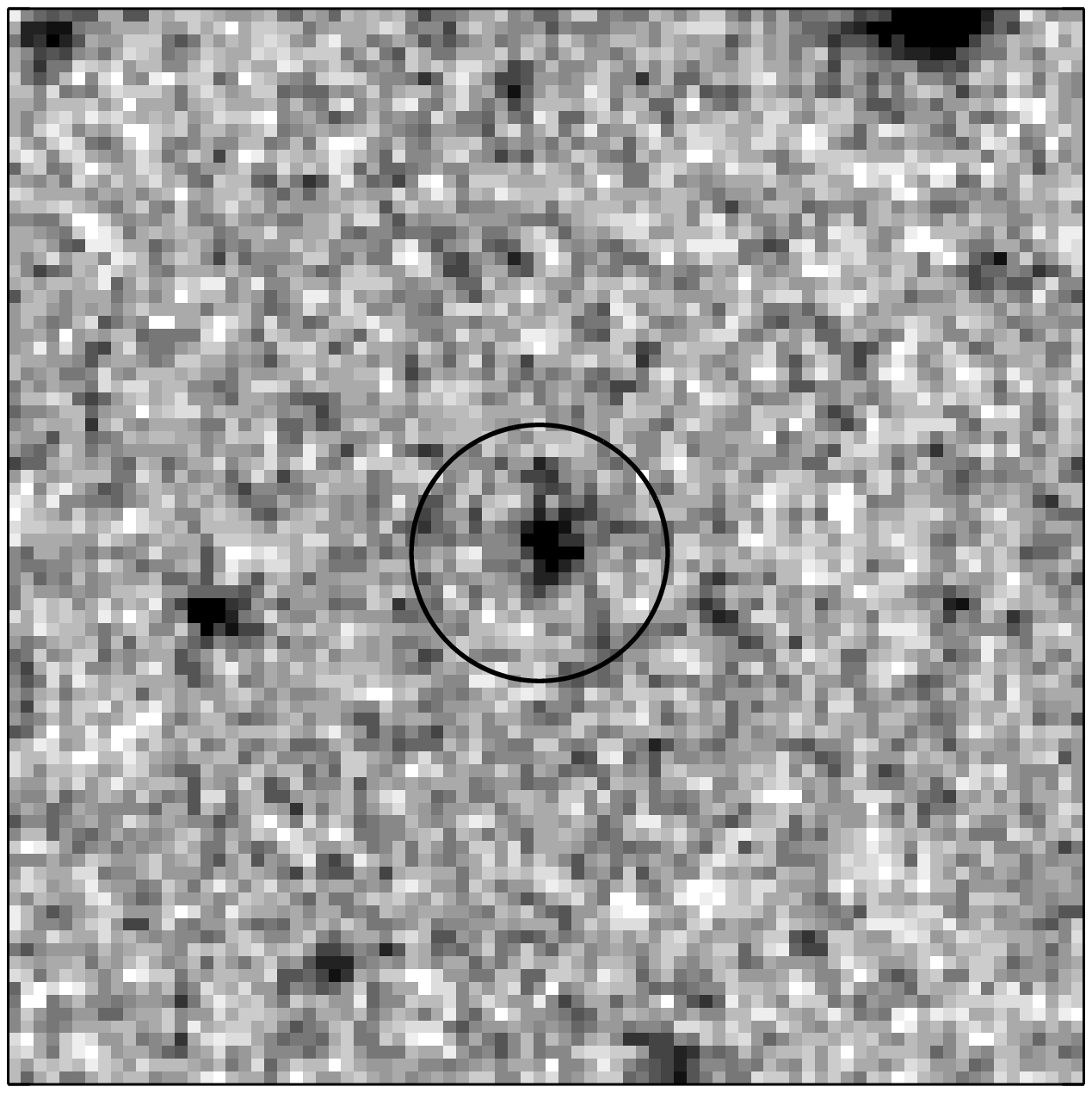}
\hspace{-10mm}
\plotone{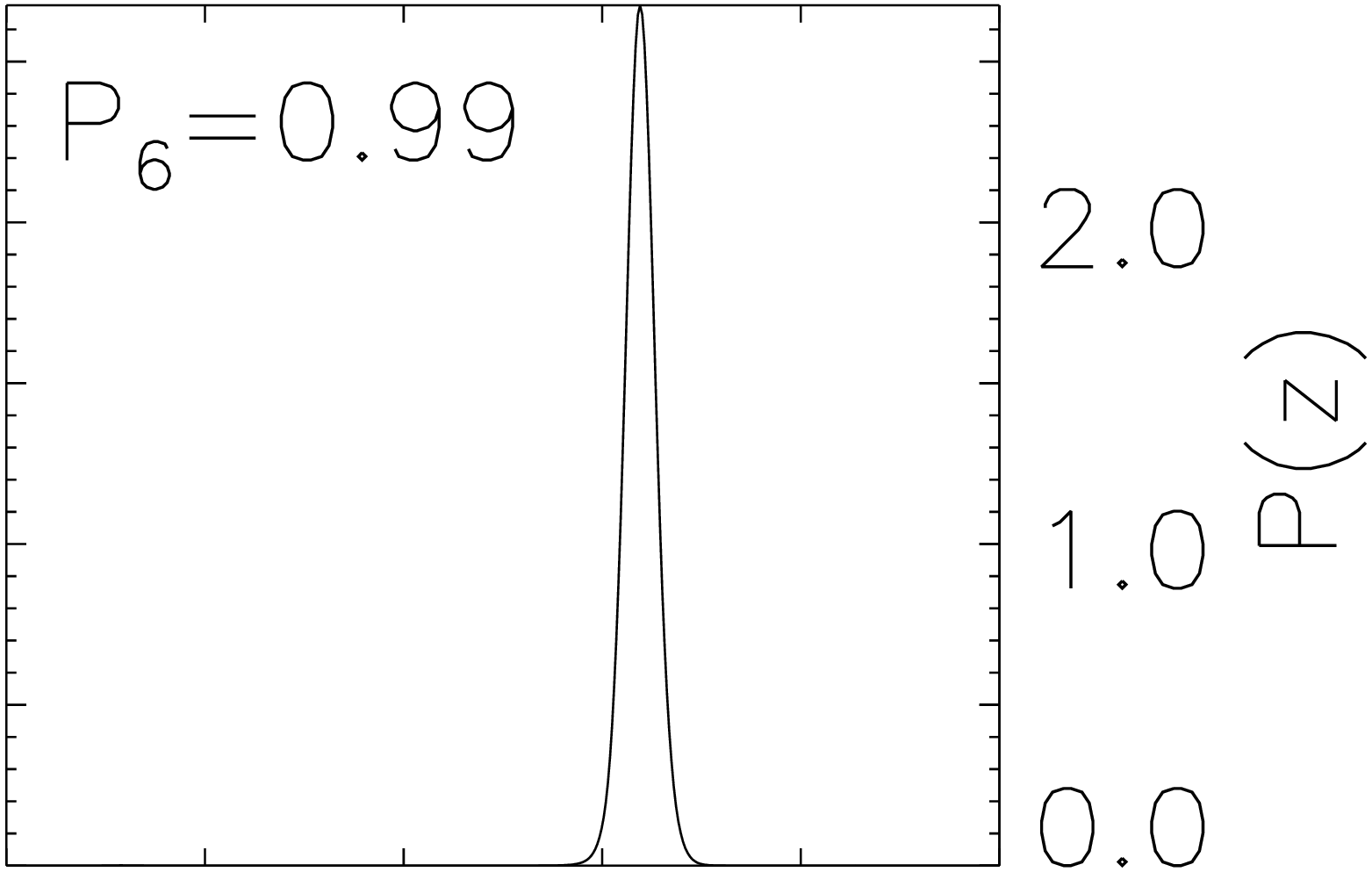}
\vspace{0.5mm}

\plotone{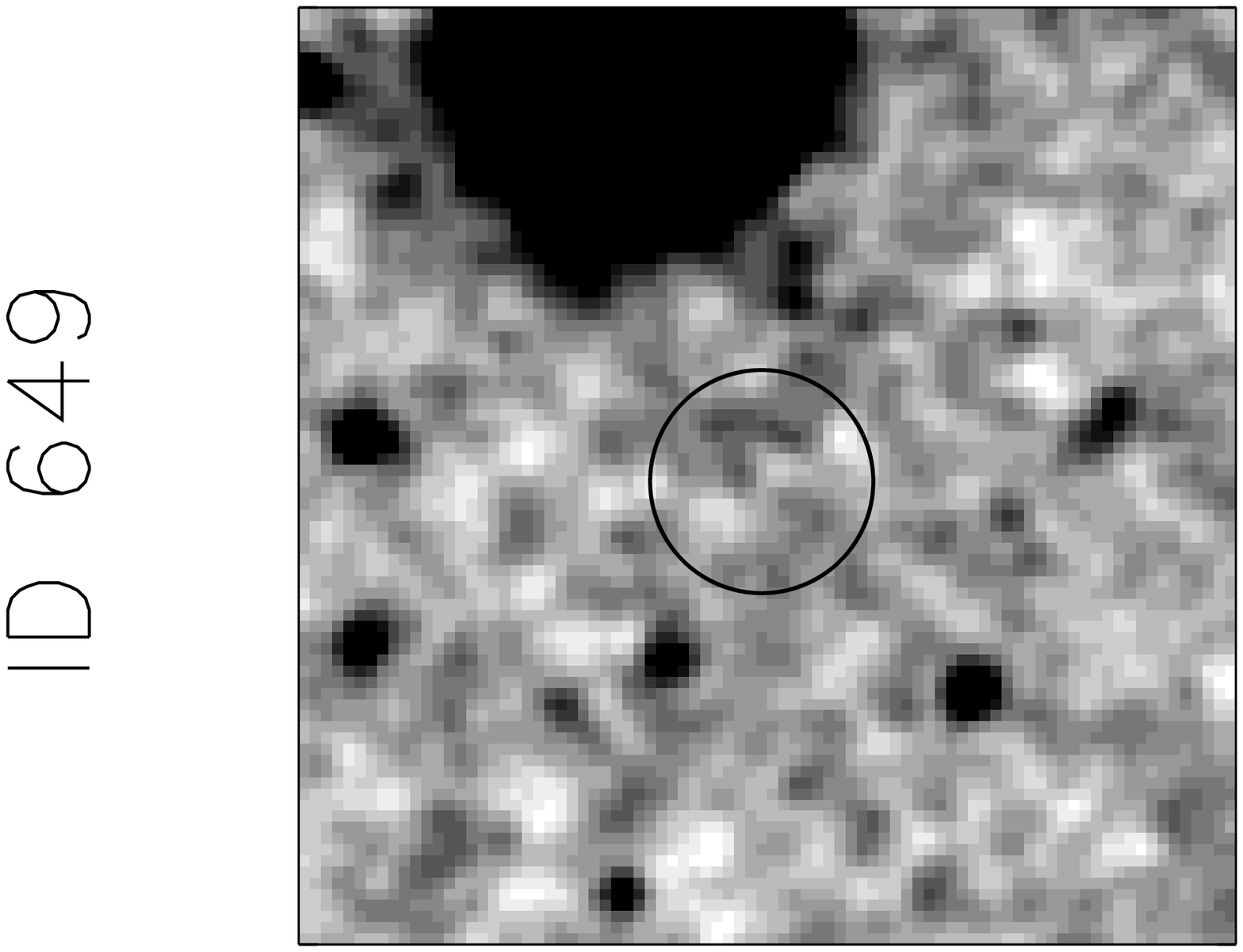}
\hspace{-10mm}
\plotone{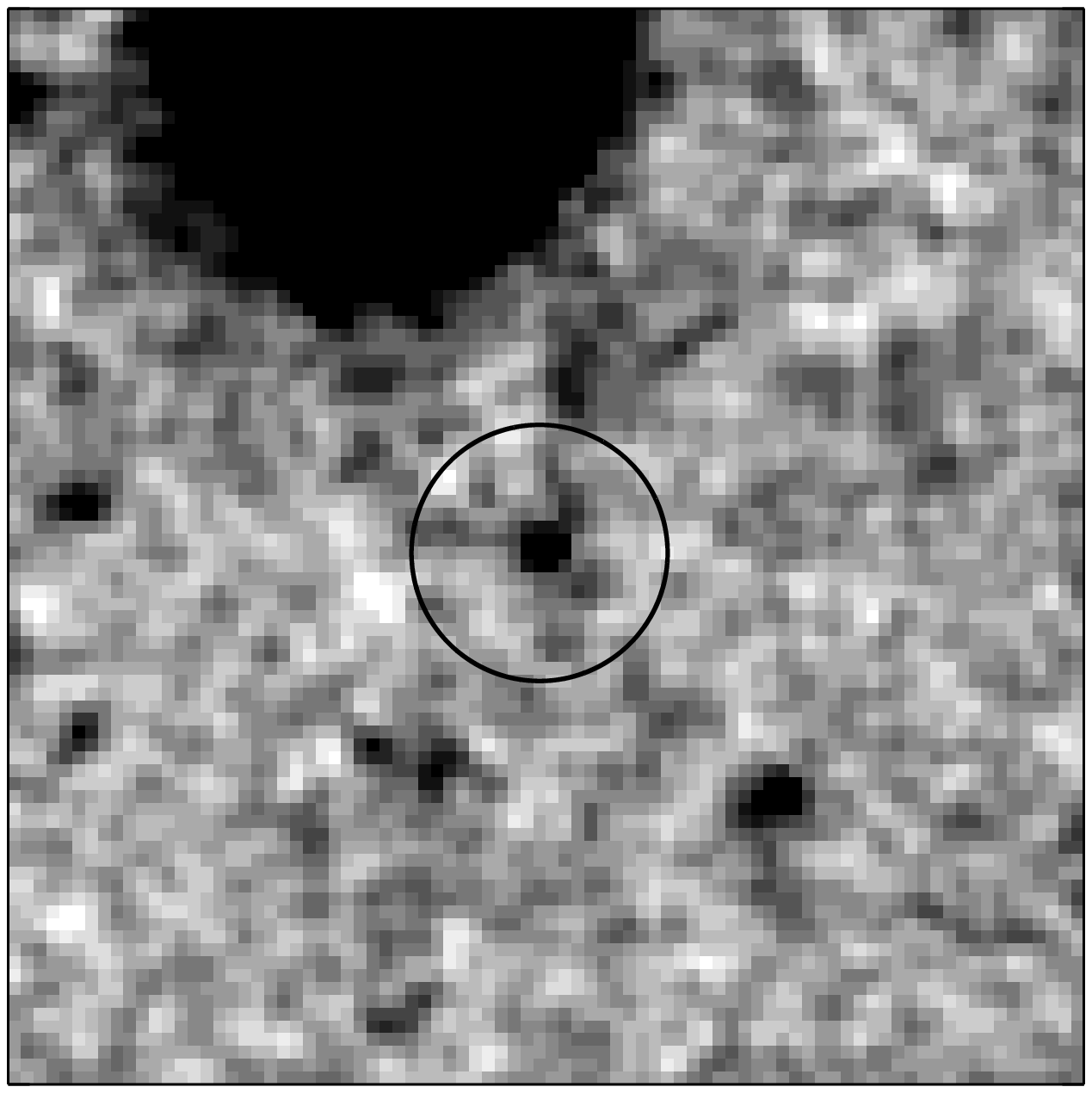}
\hspace{-10mm}
\plotone{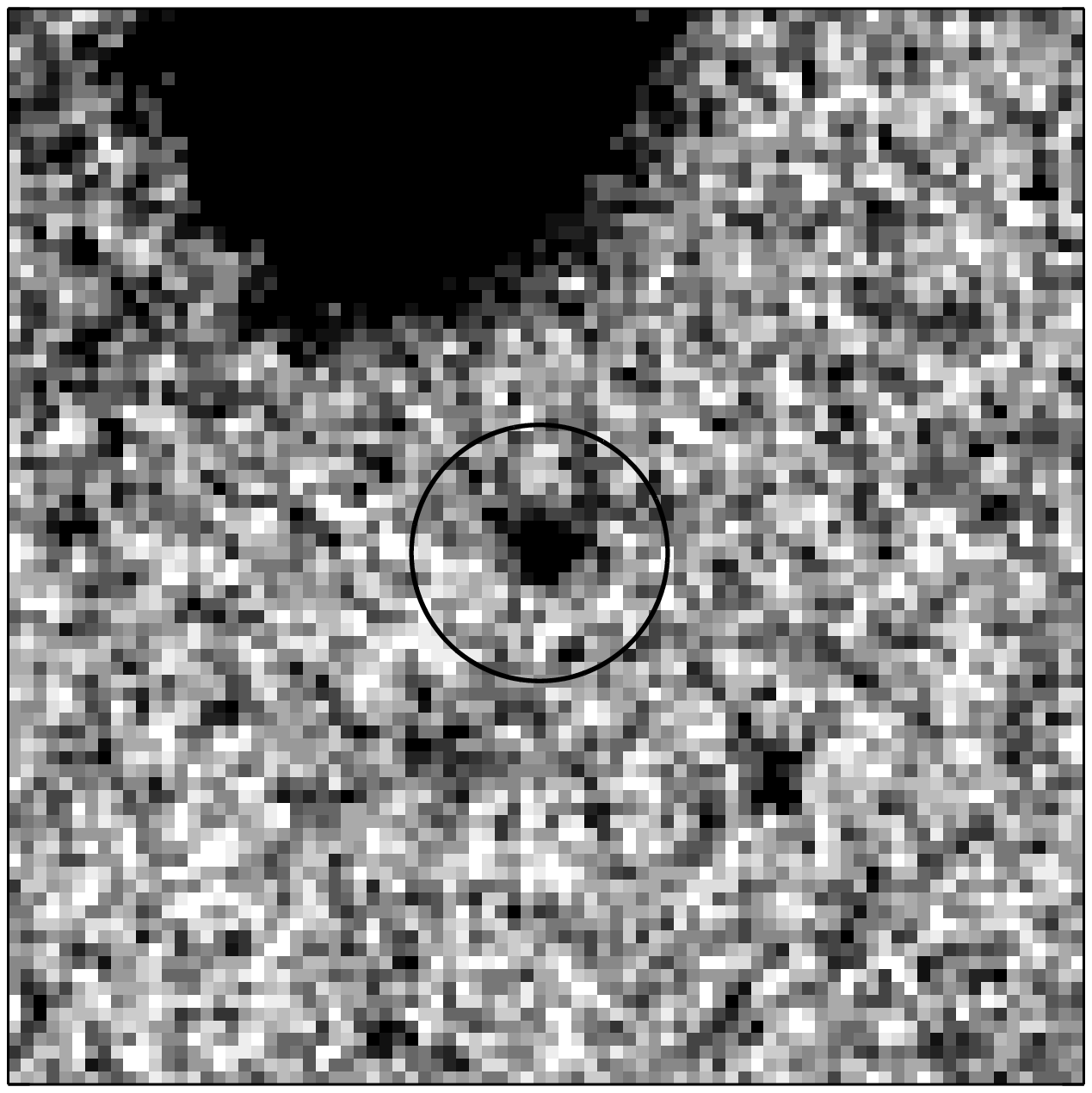}
\hspace{-10mm}
\plotone{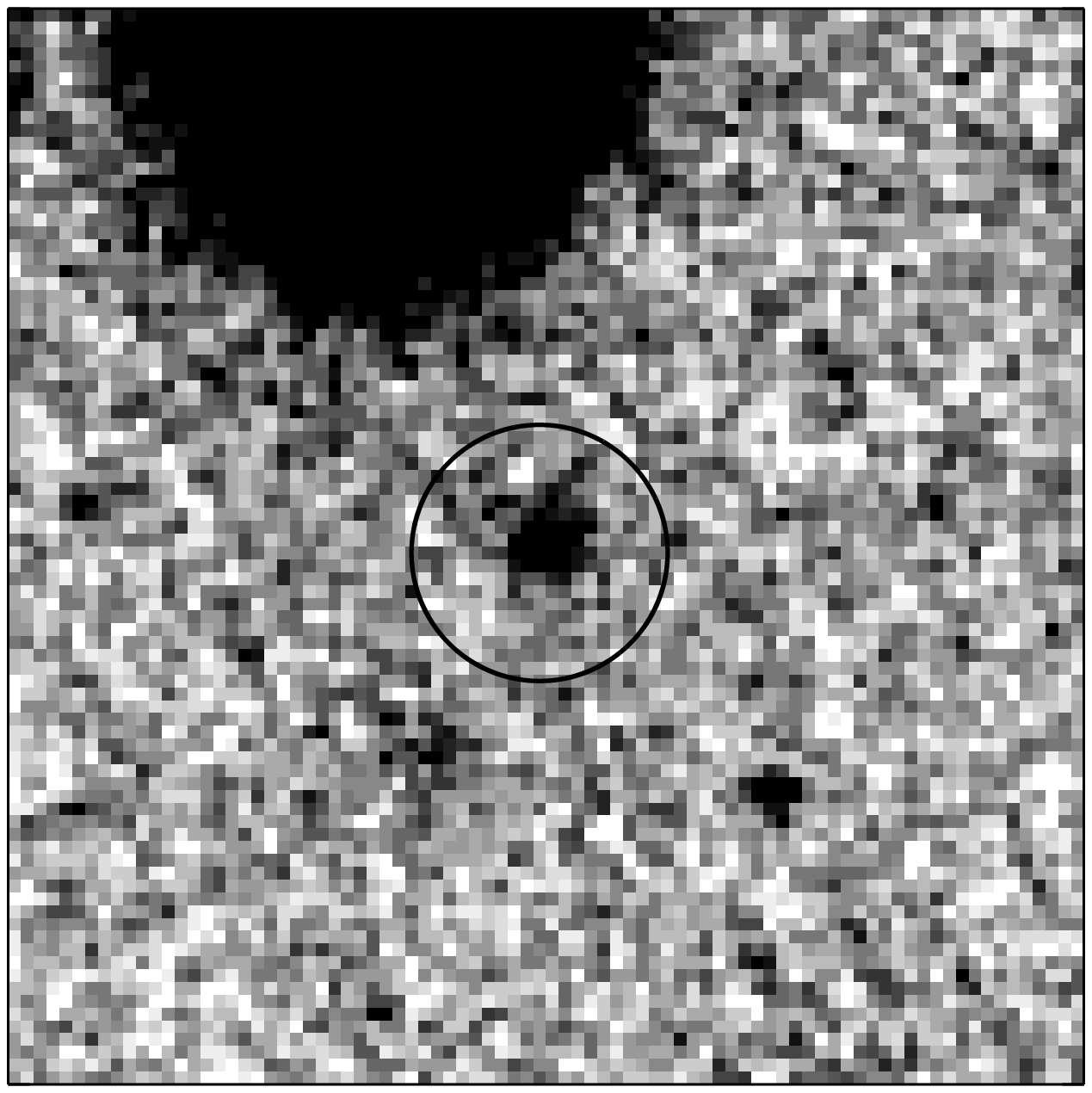}
\hspace{-10mm}
\plotone{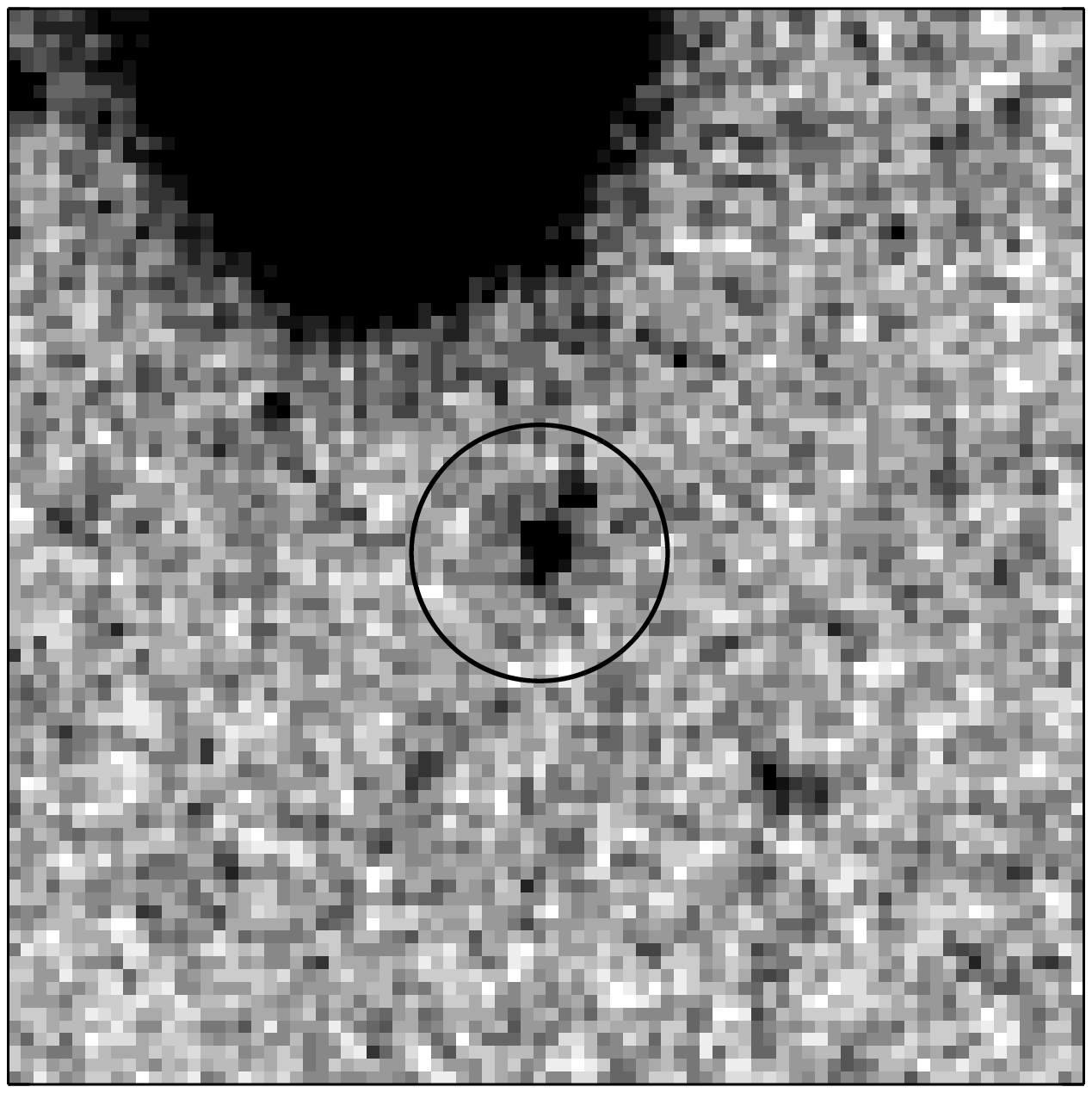}
\hspace{-10mm}
\plotone{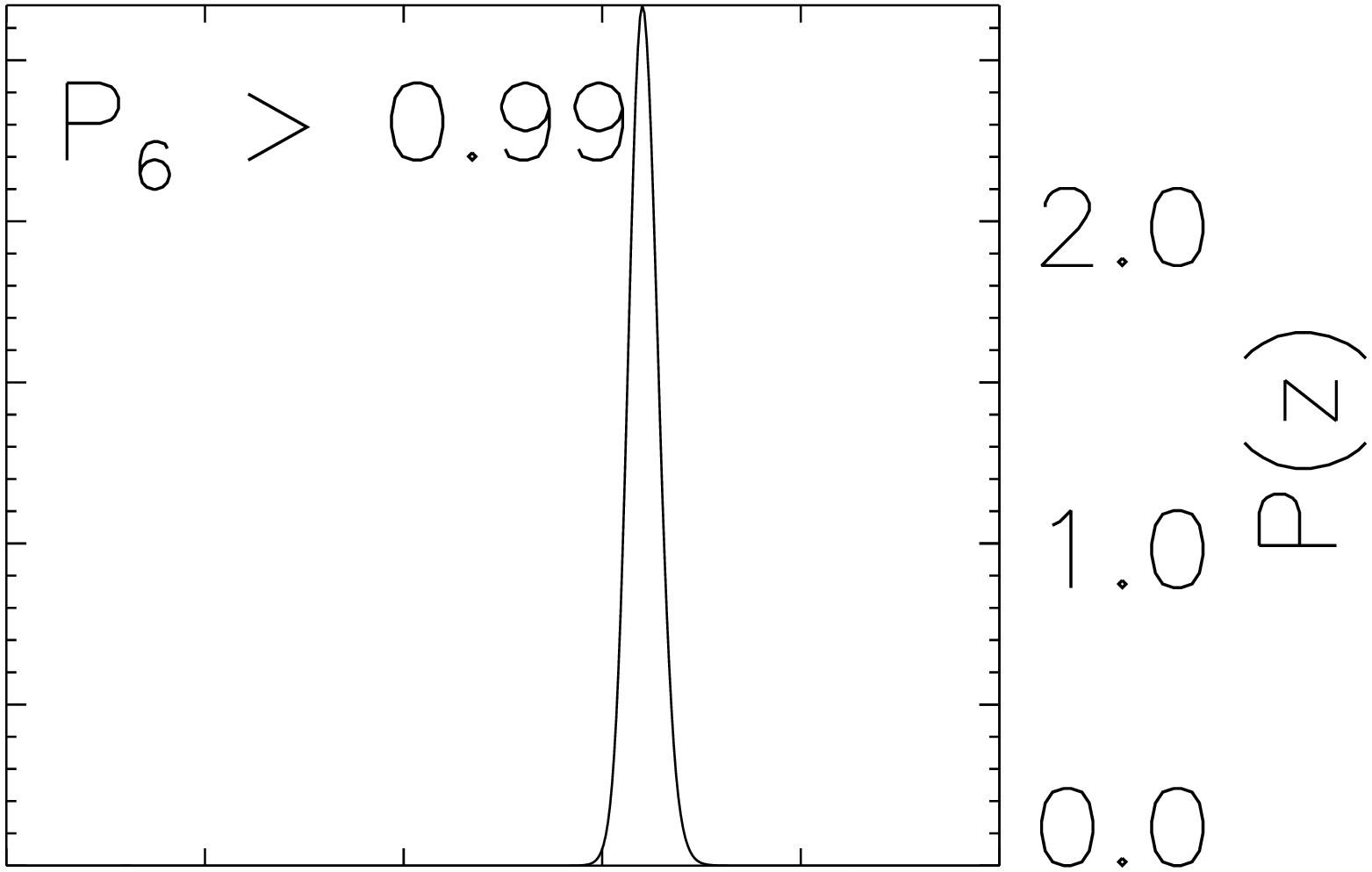}
\vspace{0.5mm}

\plotone{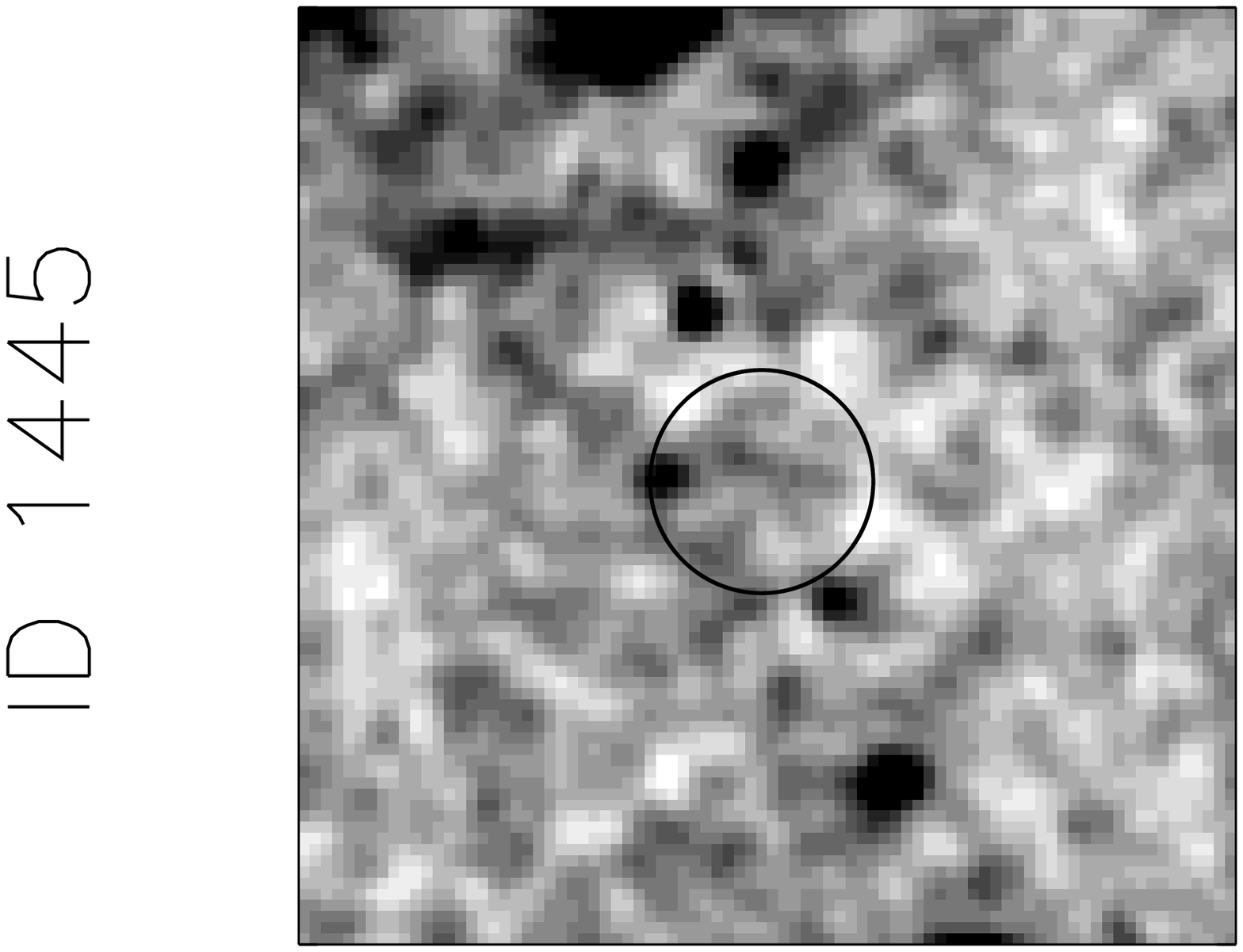}
\hspace{-10mm}
\plotone{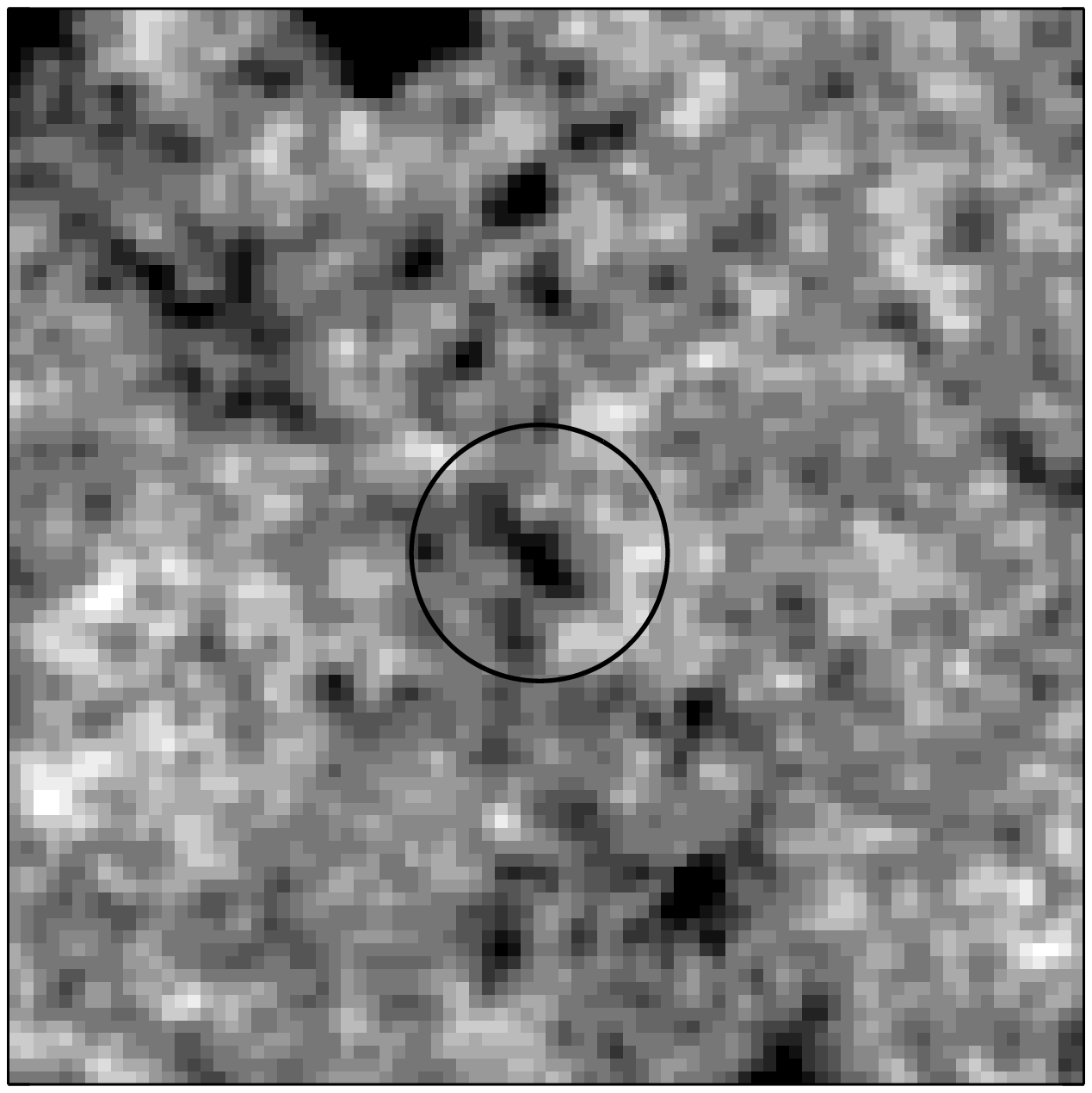}
\hspace{-10mm}
\plotone{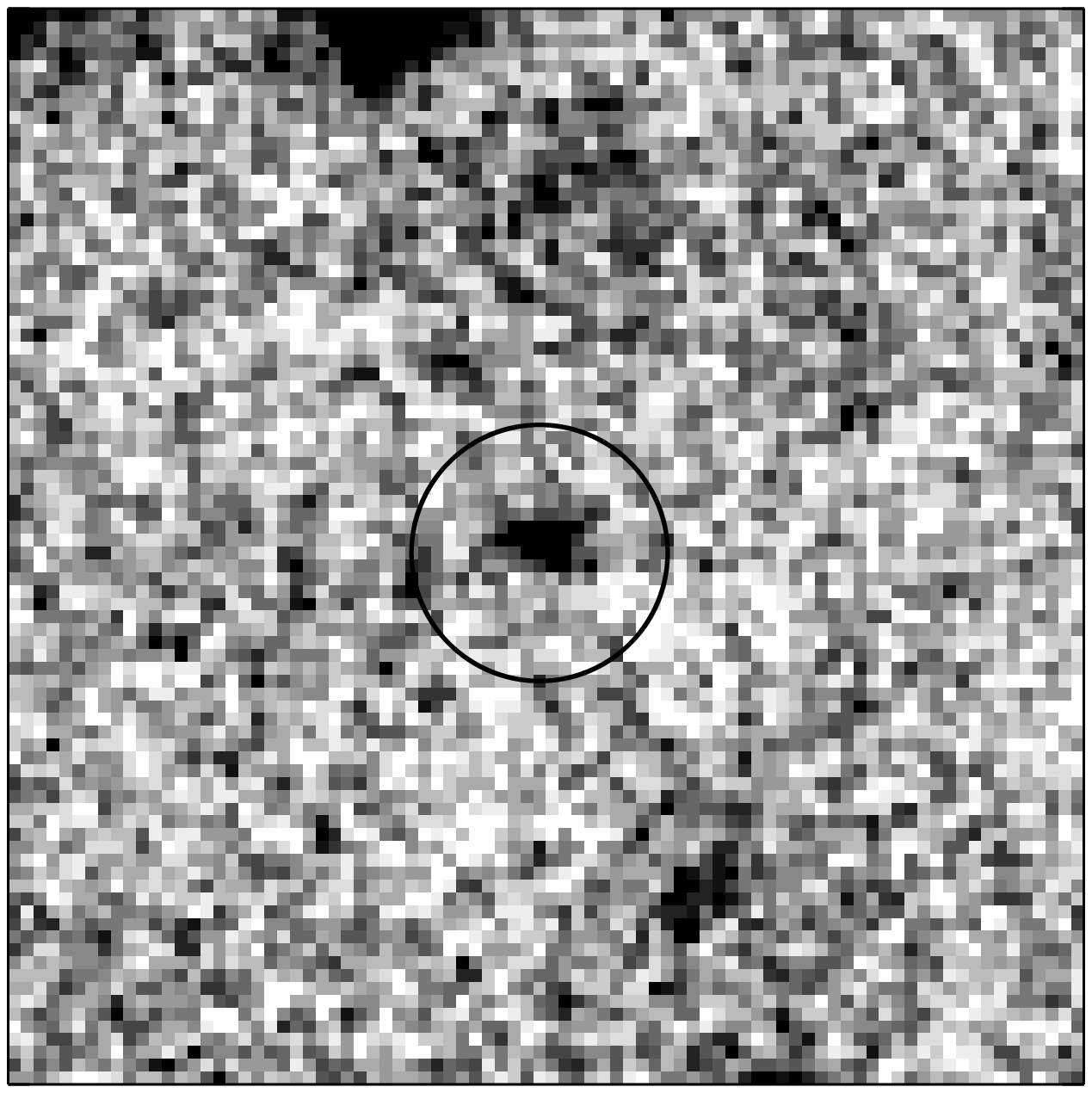}
\hspace{-10mm}
\plotone{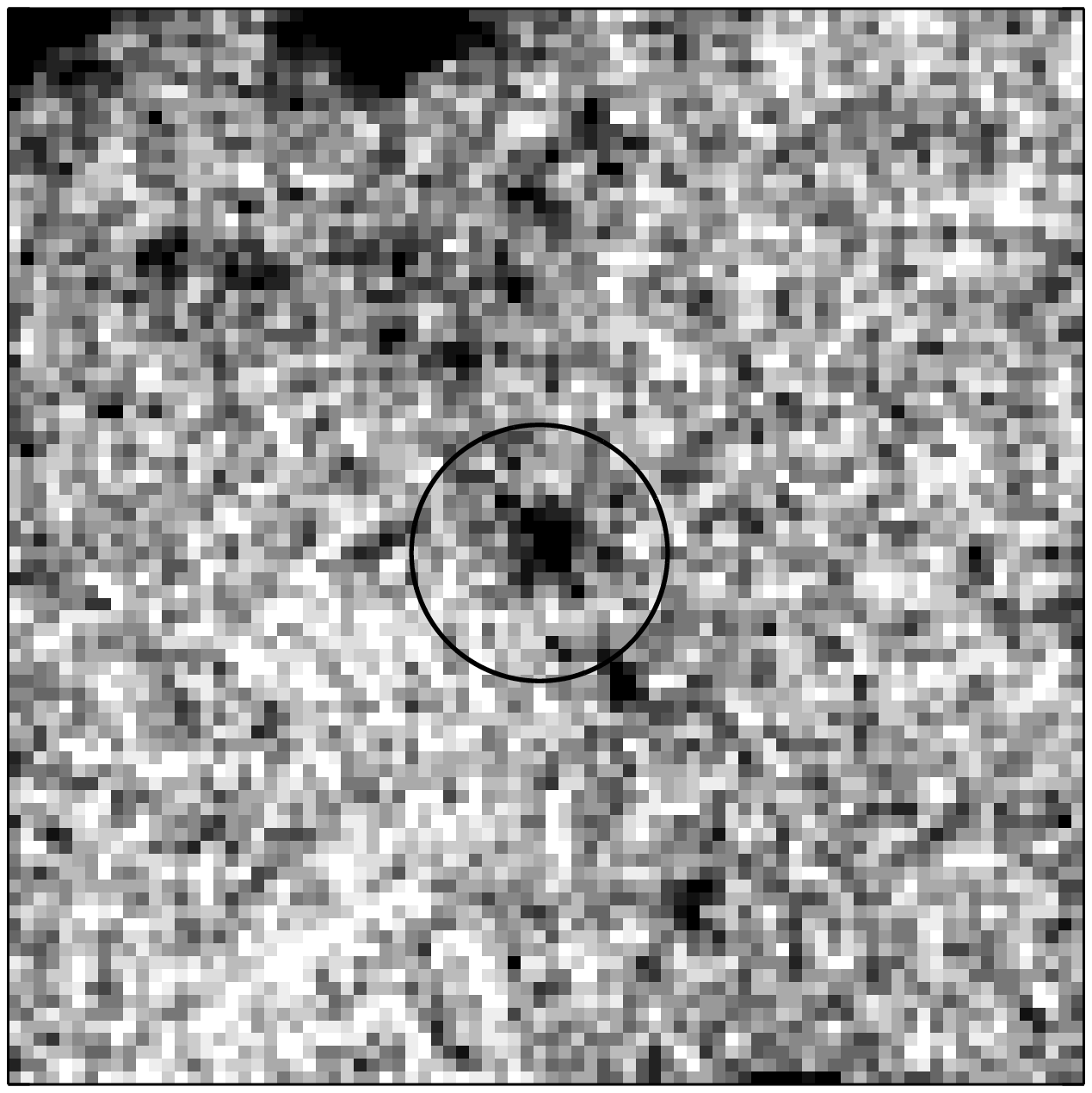}
\hspace{-10mm}
\plotone{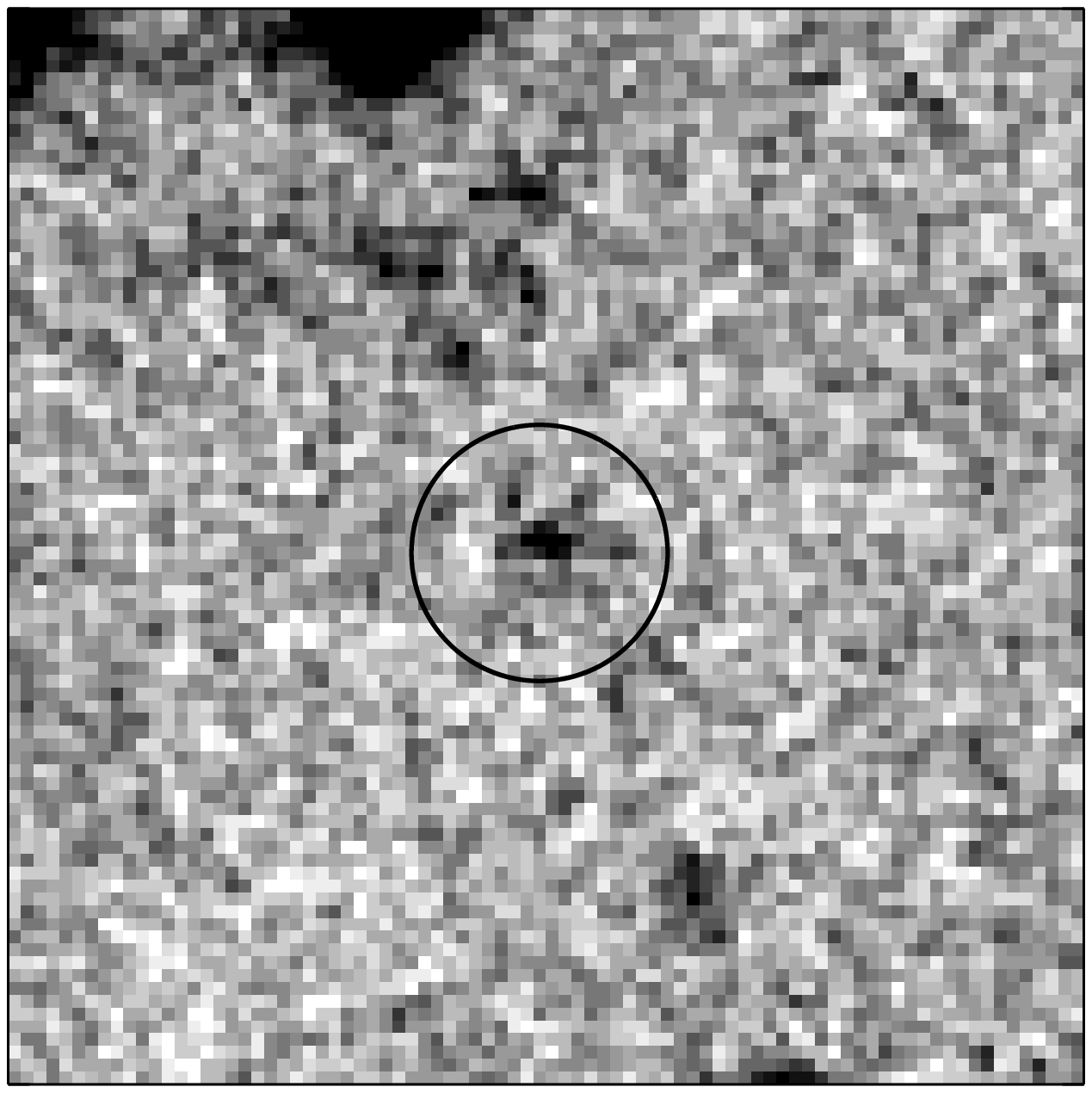}
\hspace{-10mm}
\plotone{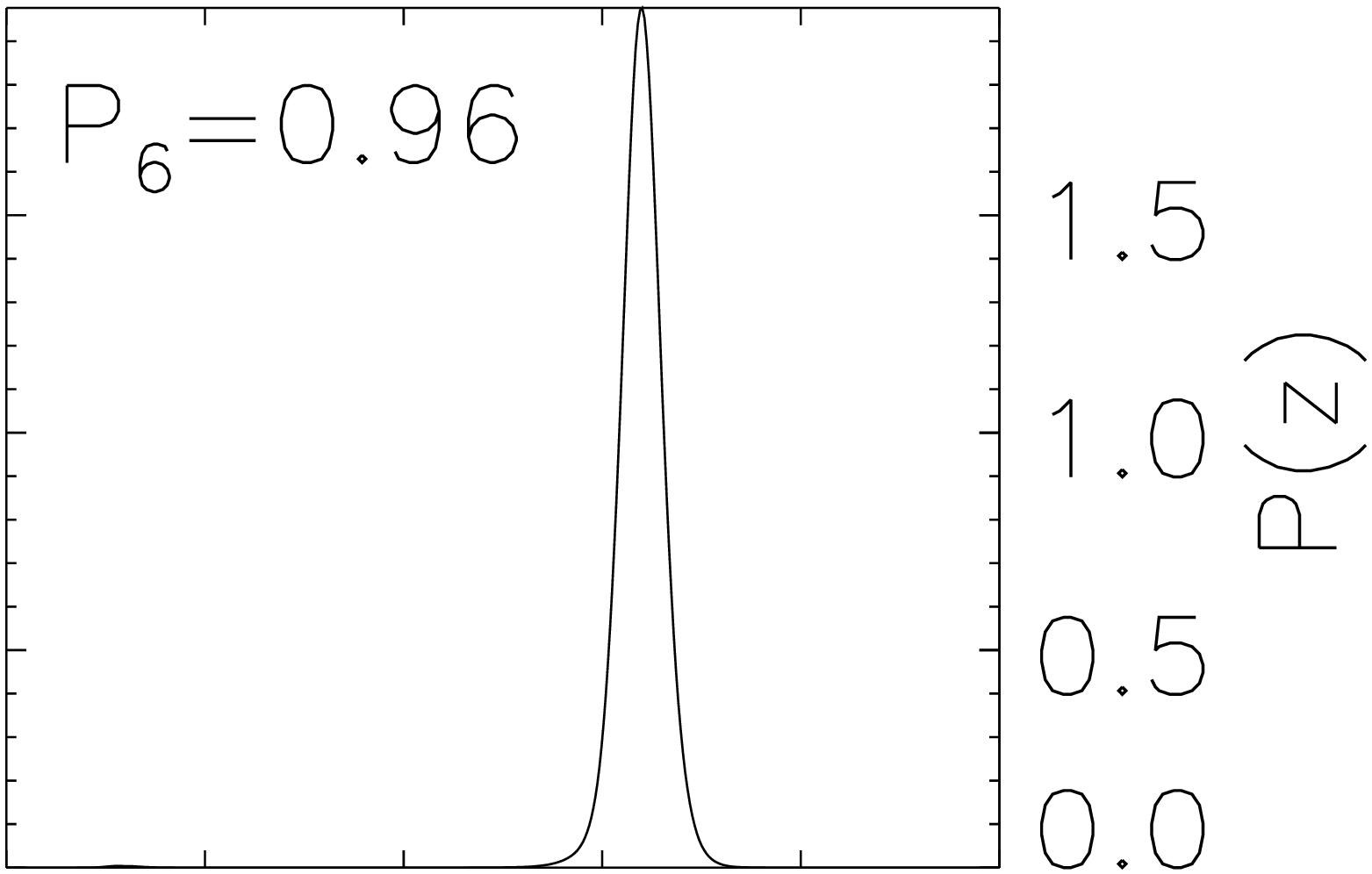}
\vspace{0.5mm}

\plotone{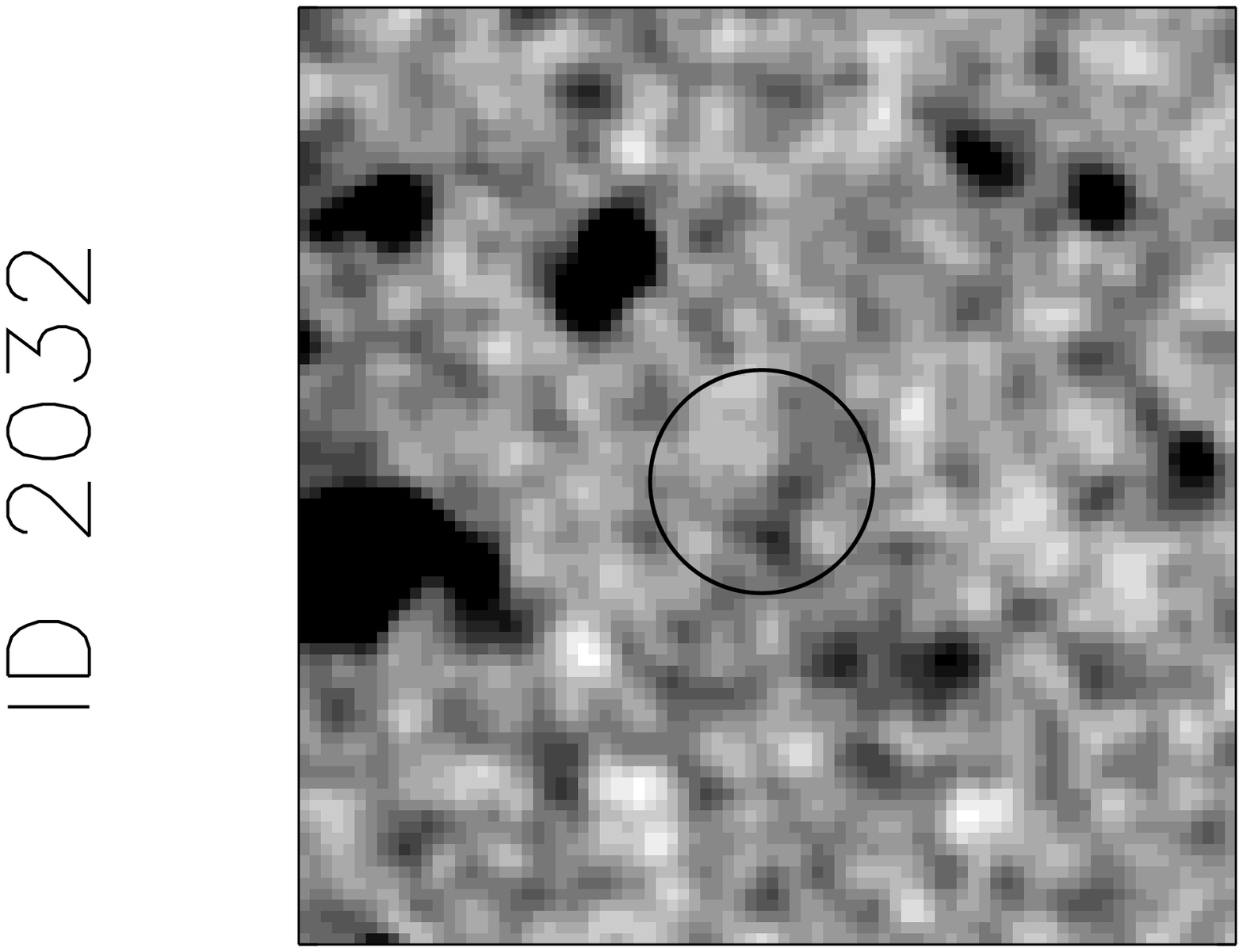}
\hspace{-10mm}
\plotone{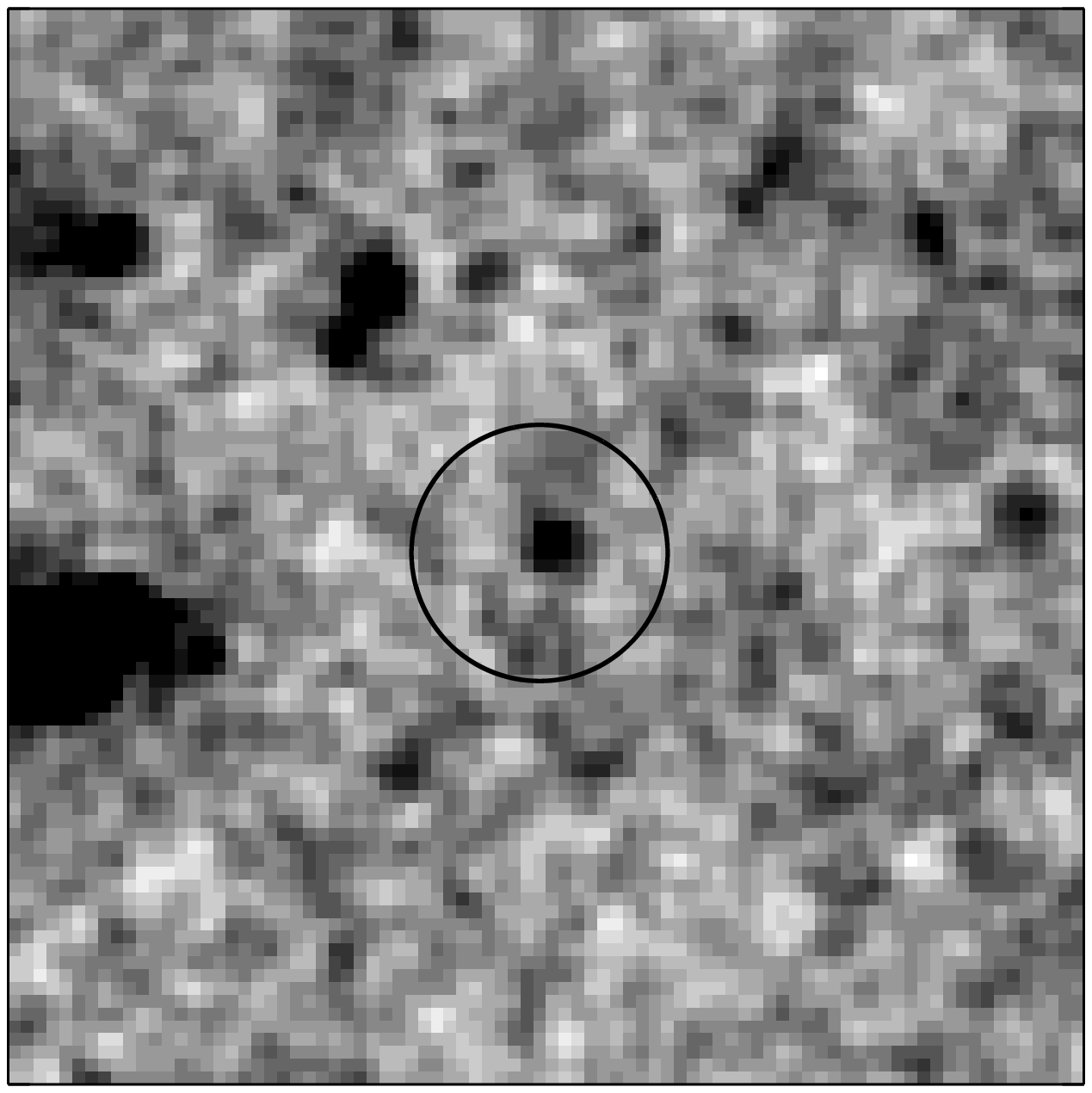}
\hspace{-10mm}
\plotone{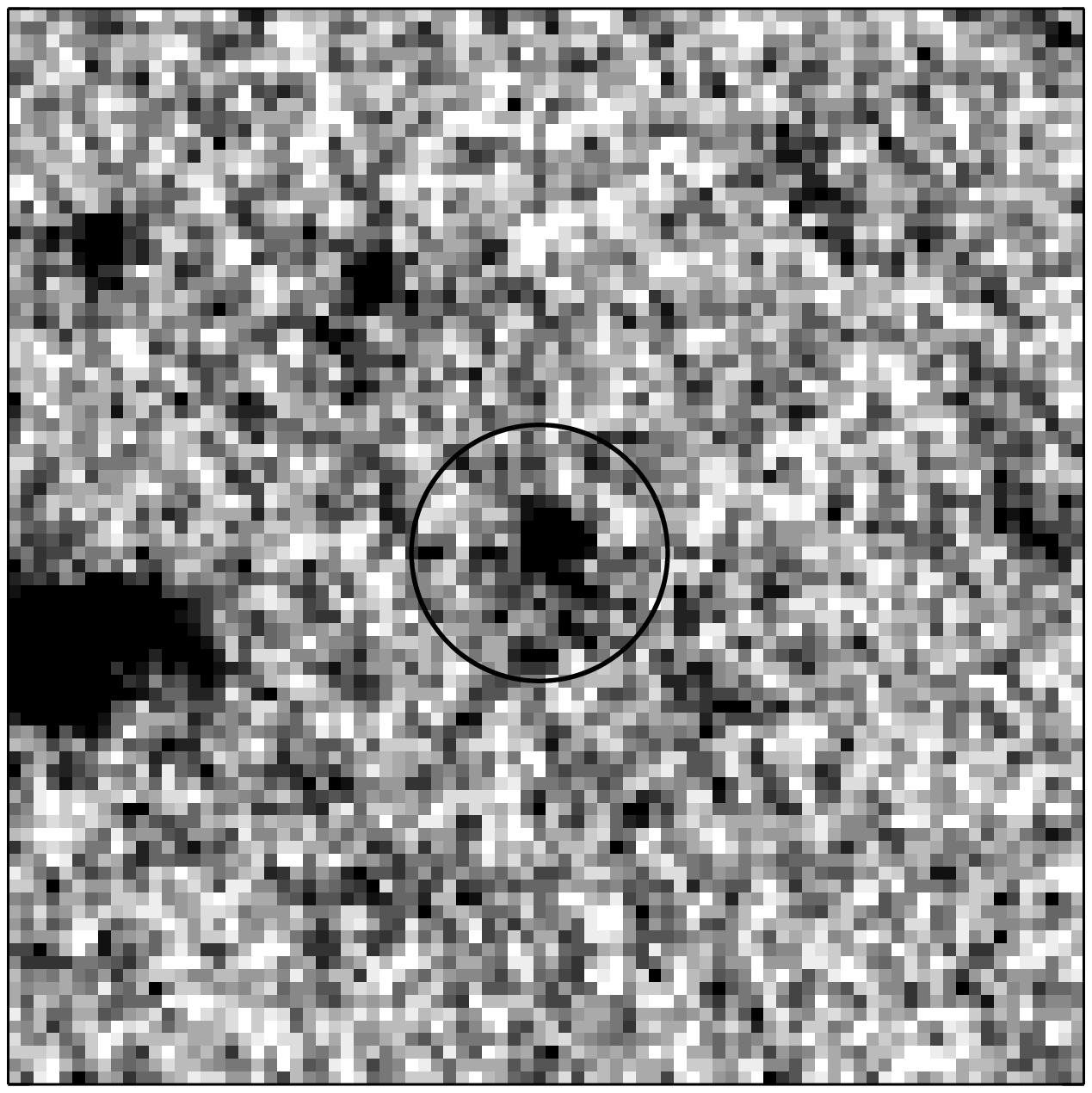}
\hspace{-10mm}
\plotone{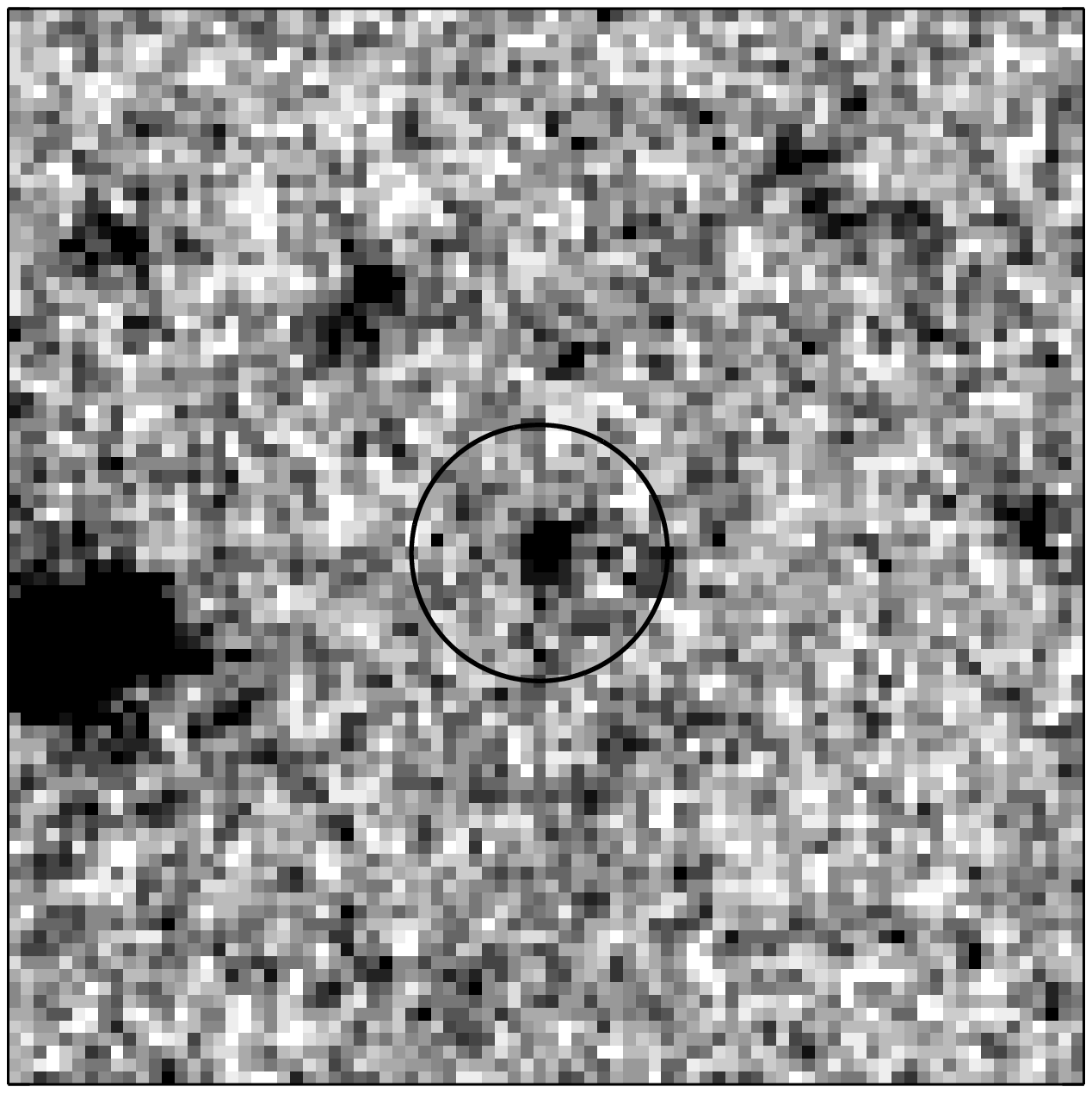}
\hspace{-10mm}
\plotone{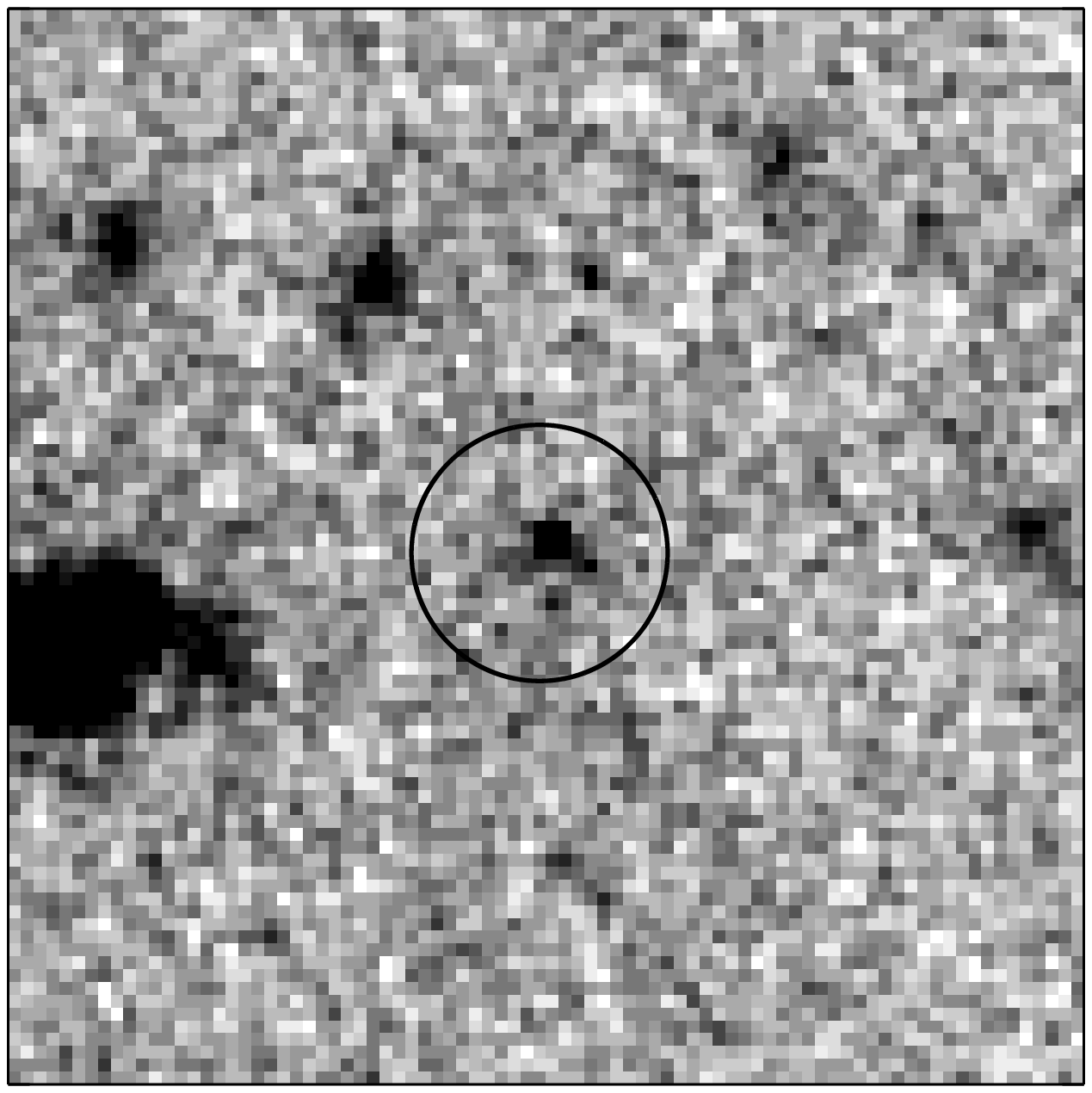}
\hspace{-10mm}
\plotone{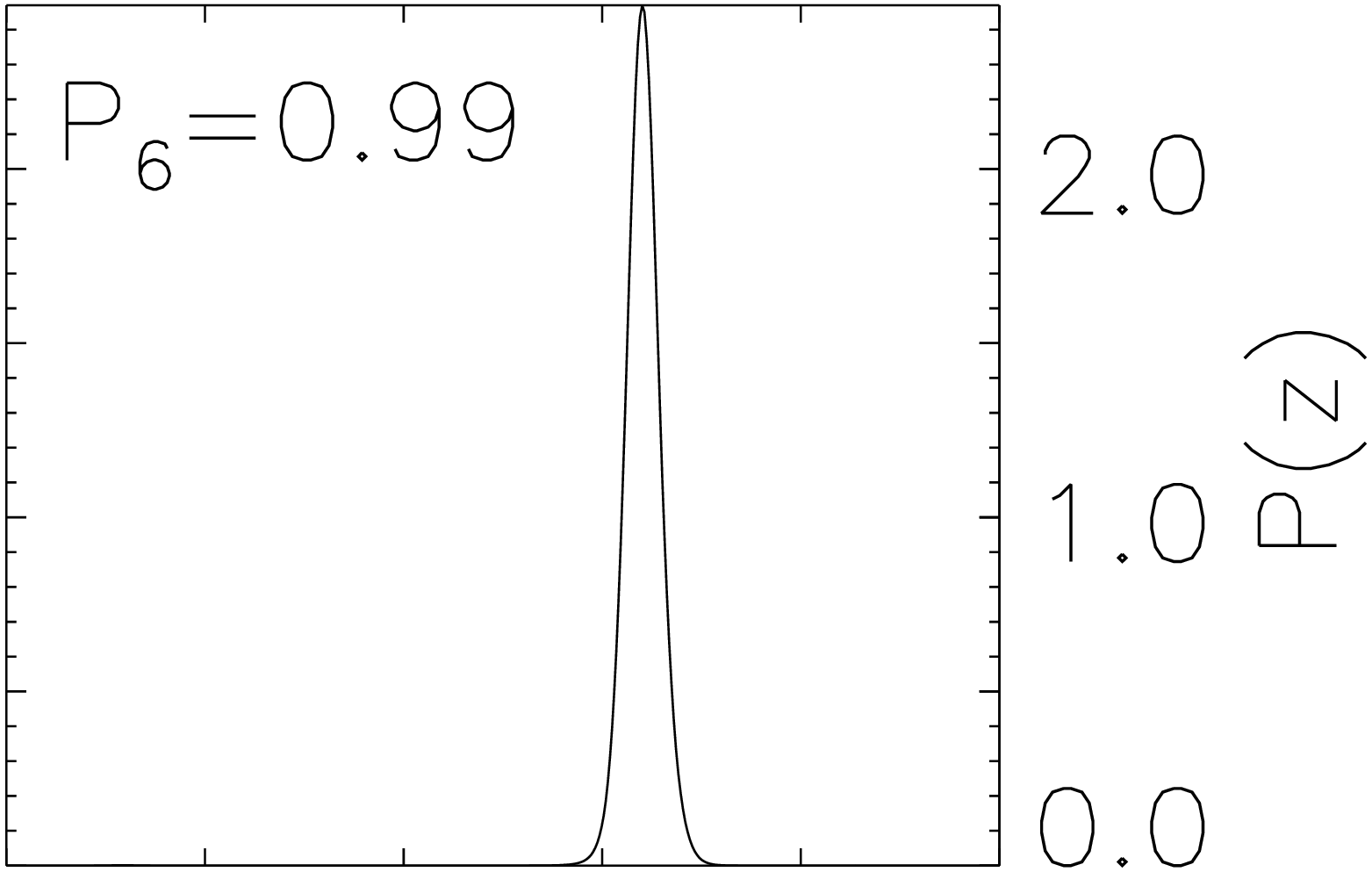}
\vspace{0.5mm}

\plotone{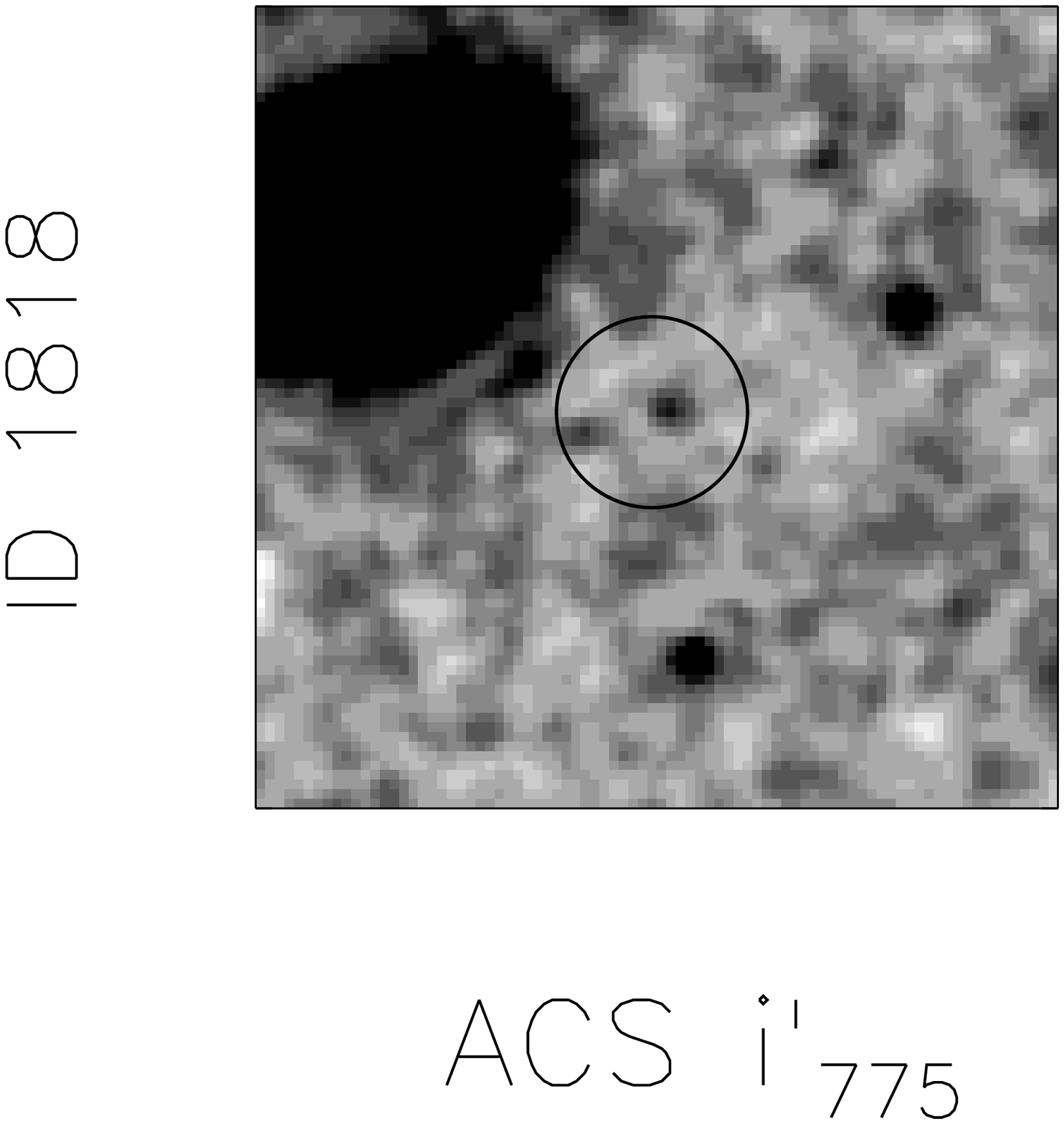}
\hspace{-10mm}
\plotone{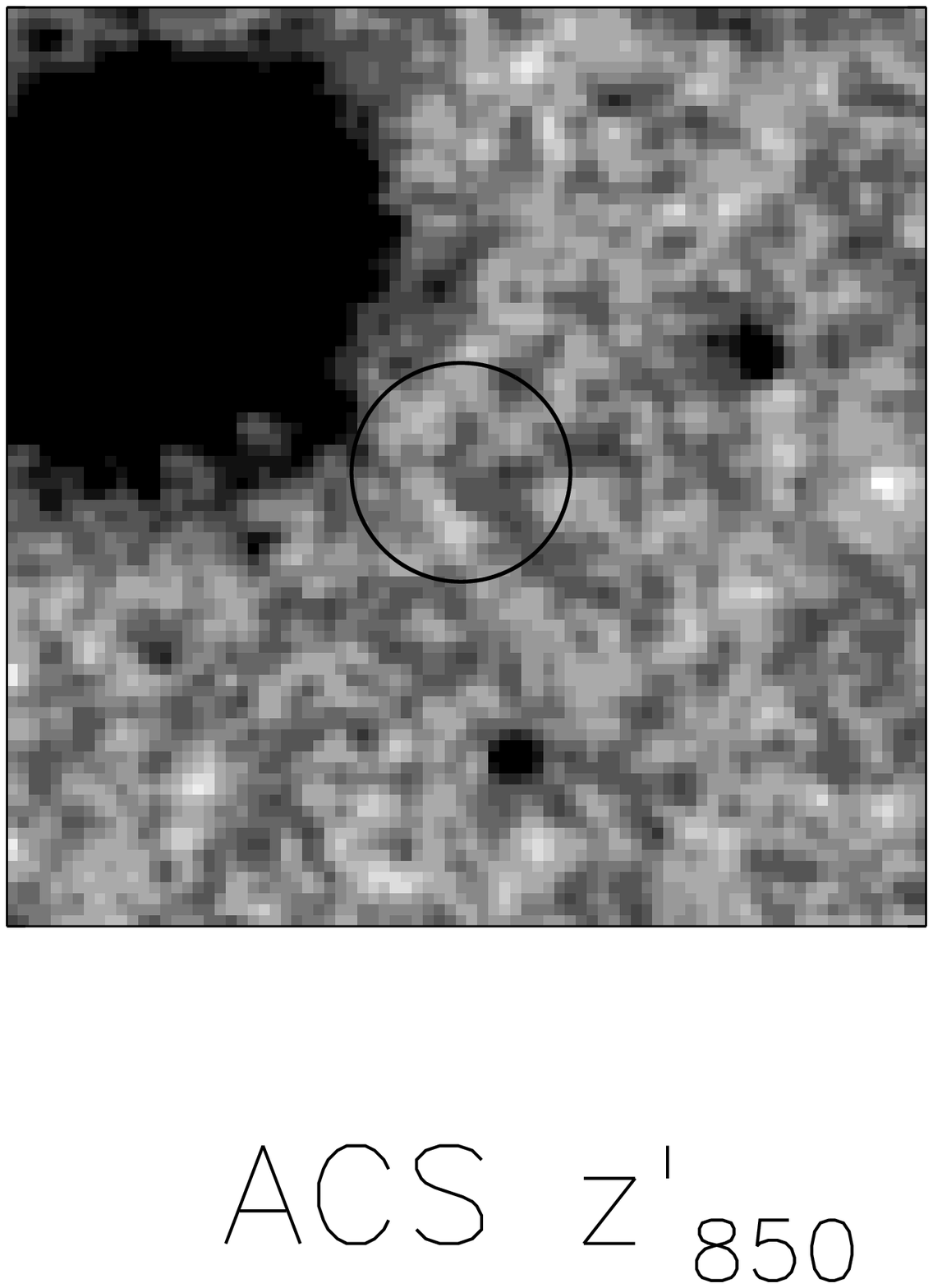}
\hspace{-10mm}
\plotone{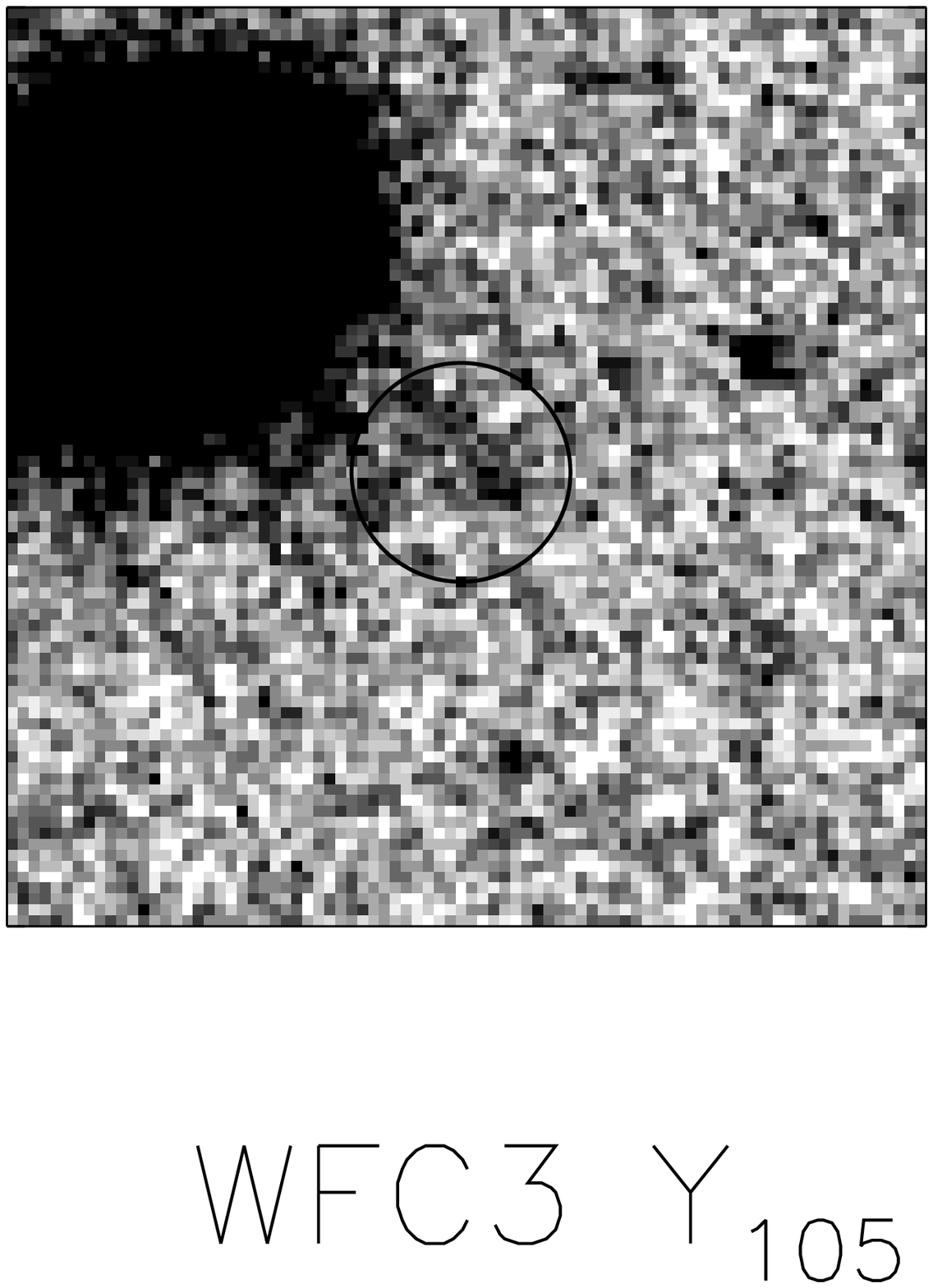}
\hspace{-10mm}
\plotone{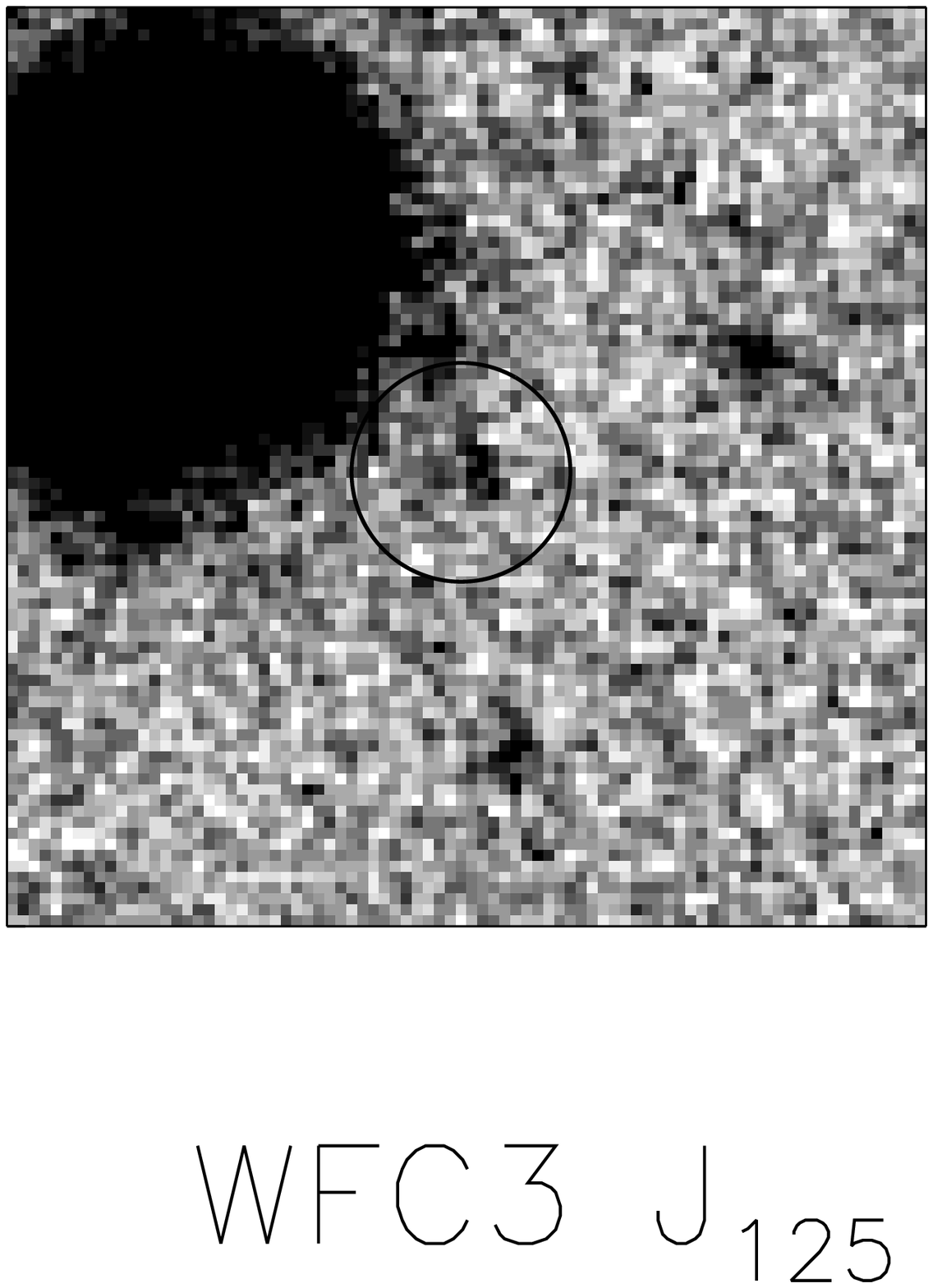}
\hspace{-10mm}
\plotone{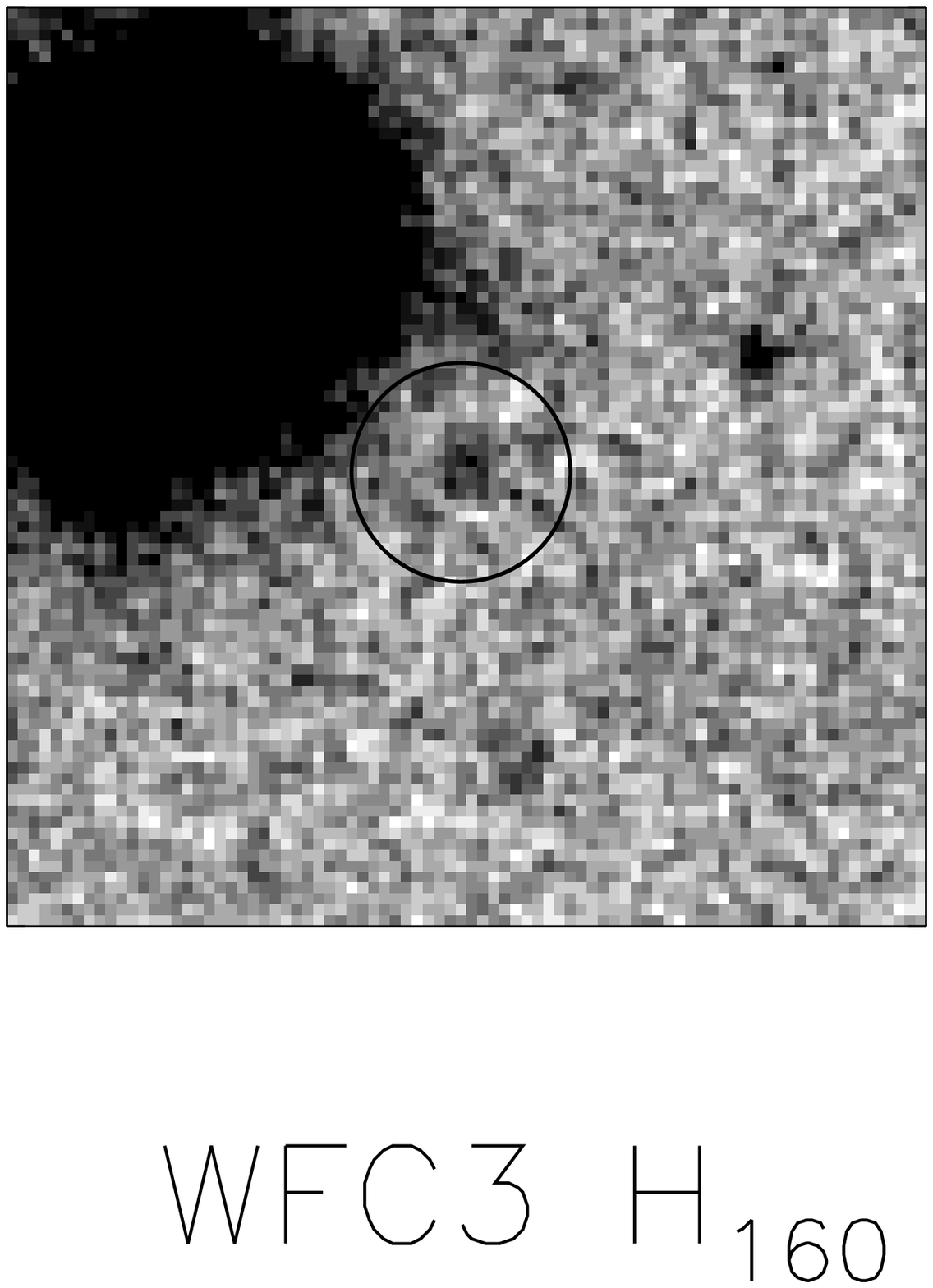}
\hspace{-10mm}
\plotone{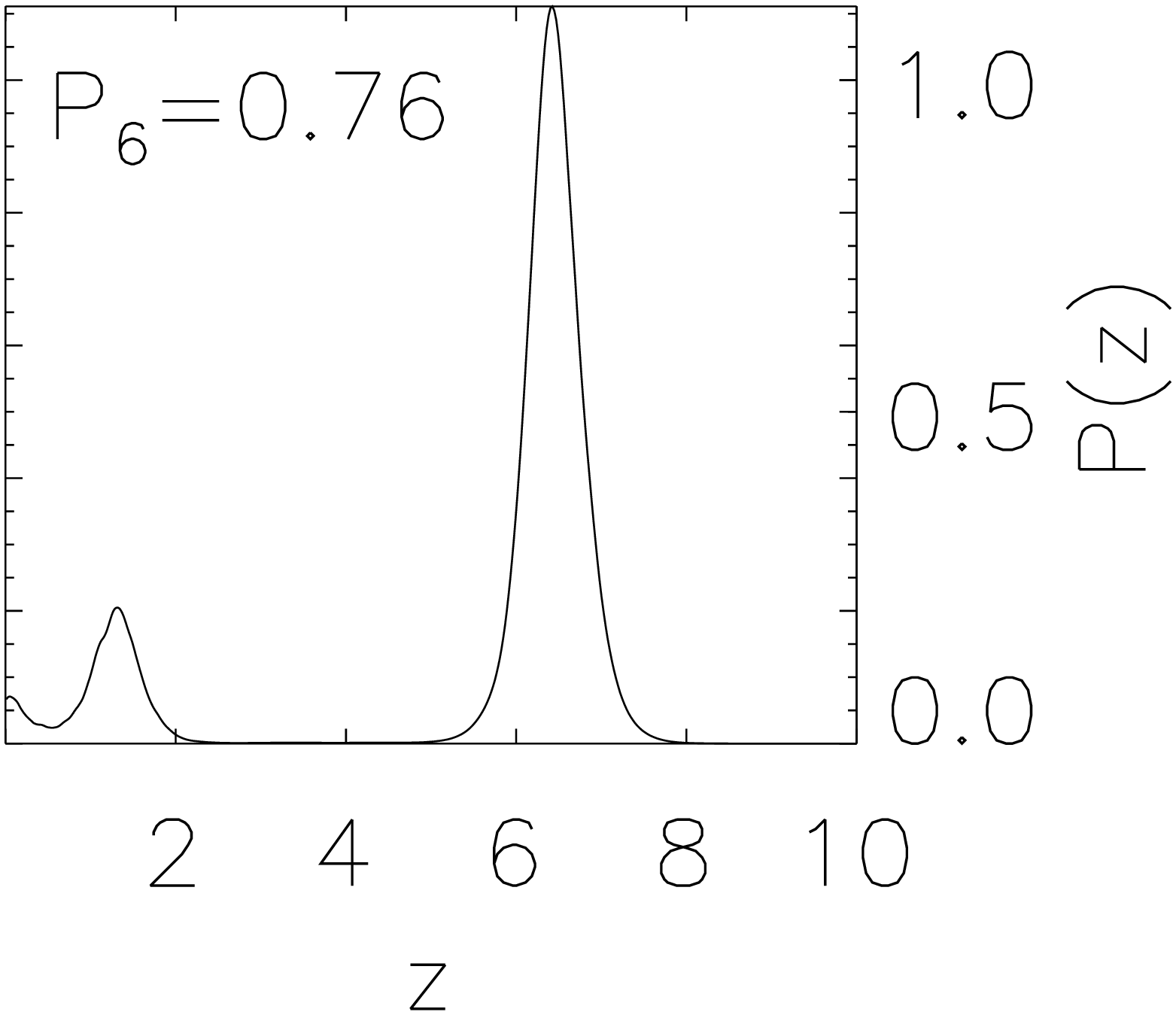}
\vspace{7mm}
\caption{5\arcs cutouts of our 31 candidate z $>$ 6.3 galaxies, from left-to-right: (smoothed) ACS i$^{\prime}_{775}$ and z$^{\prime}_{850}$, and WFC3 \yb, \jb~and \hb.  The objects are shown in order of increasing photometric redshift.  The circles are centered on the objects and have a radius of 1\arcs.  The last column shows the probability distribution function for the photometric redshifts, and prints the integrated probability over 6 $\leq$ z $\leq$ 11 ($\mathcal{P}_6$).}\label{stamps}
\end{figure*}
\addtocounter{figure}{-1}

\begin{figure*}
\epsscale{0.18}
\vspace{2mm}
\plotone{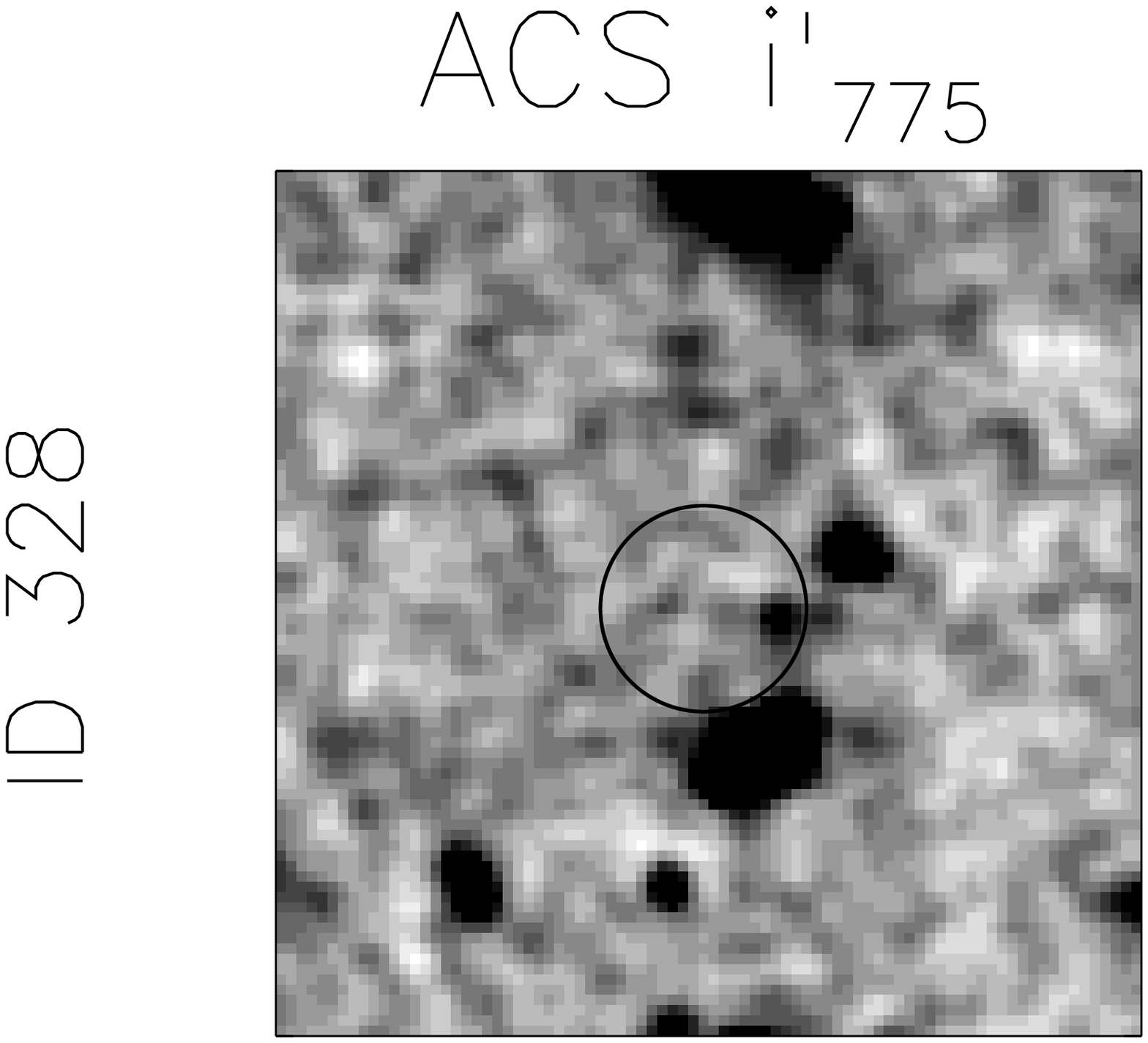}
\hspace{-10mm}
\plotone{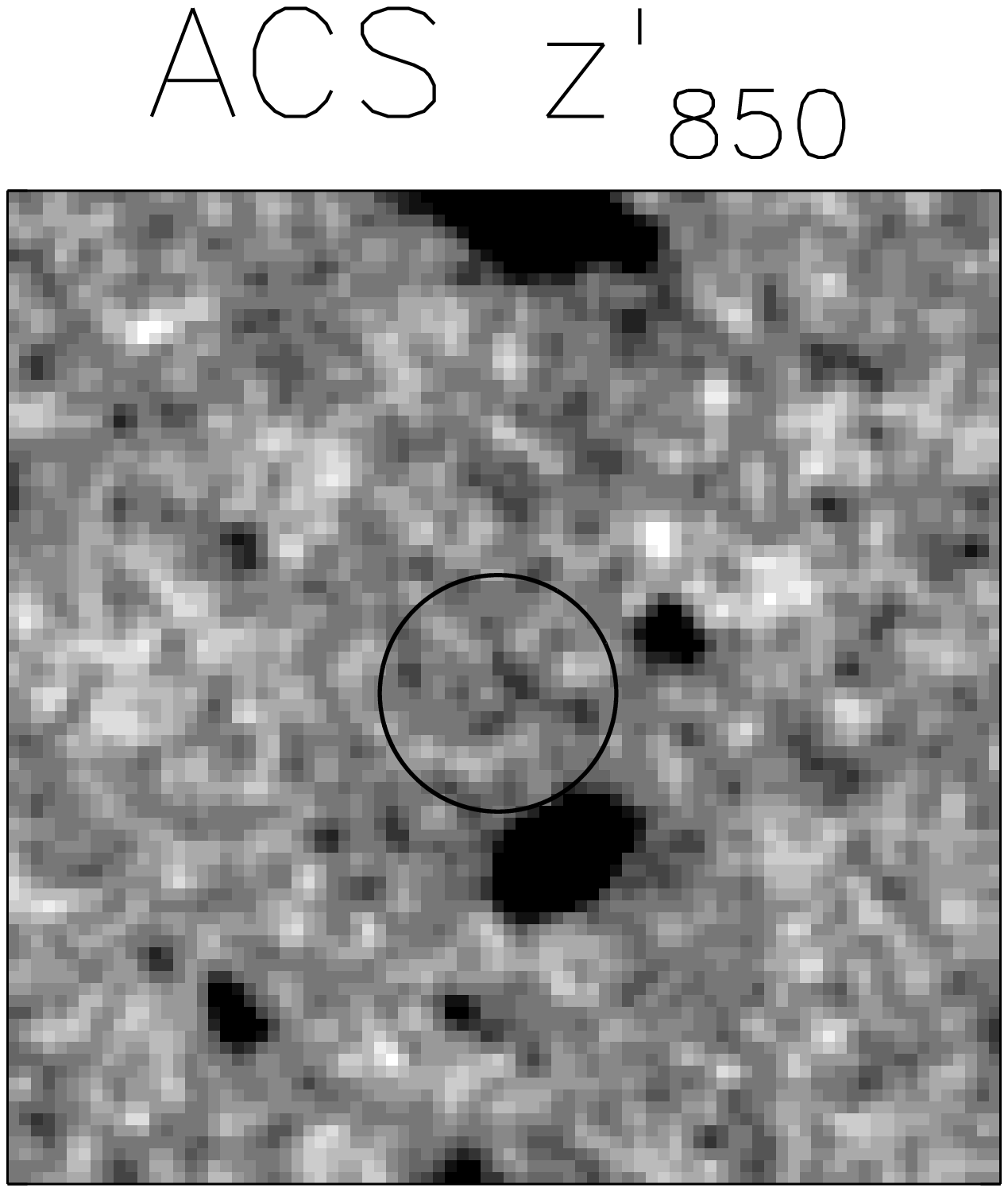}
\hspace{-10mm}
\plotone{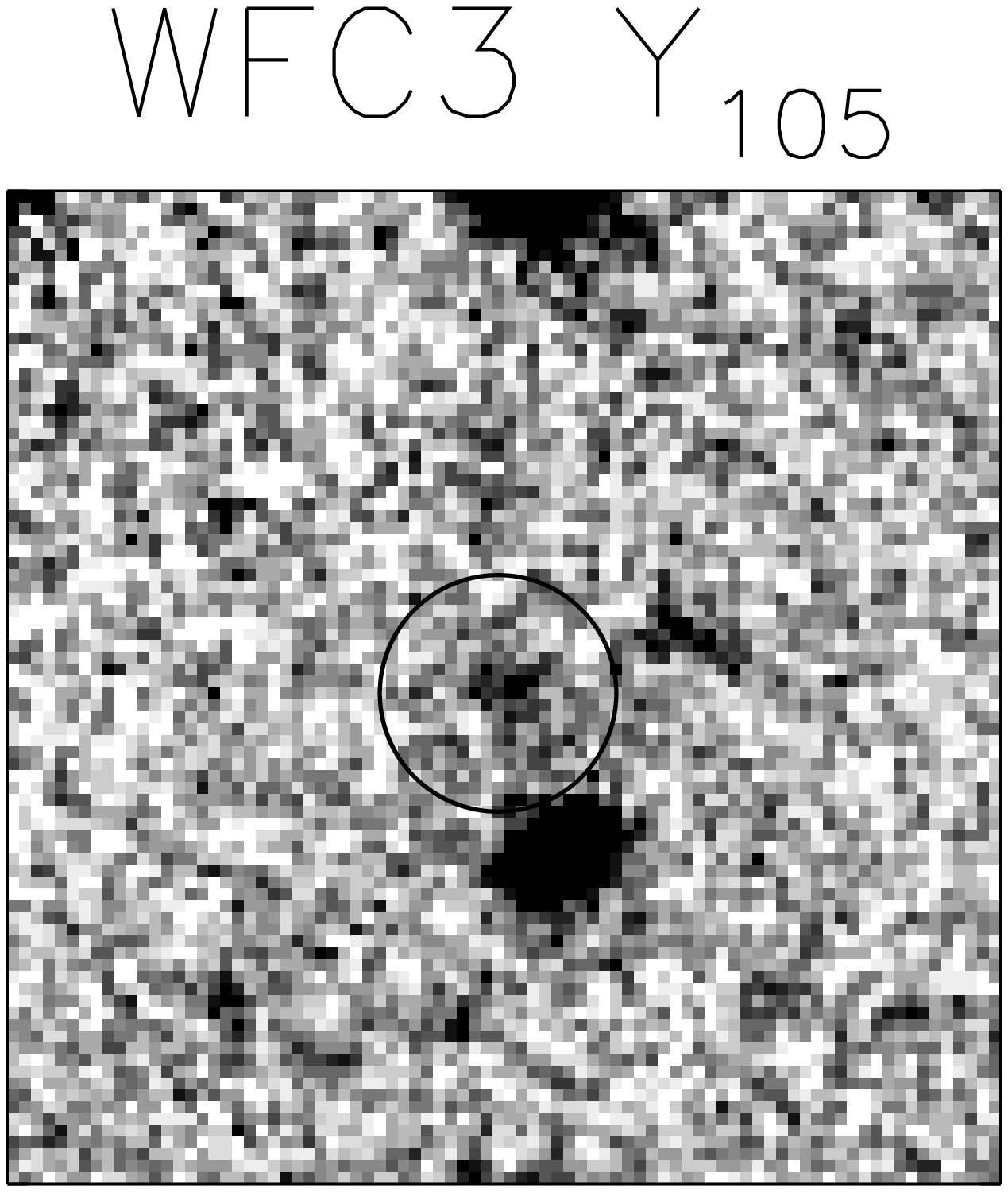}
\hspace{-10mm}
\plotone{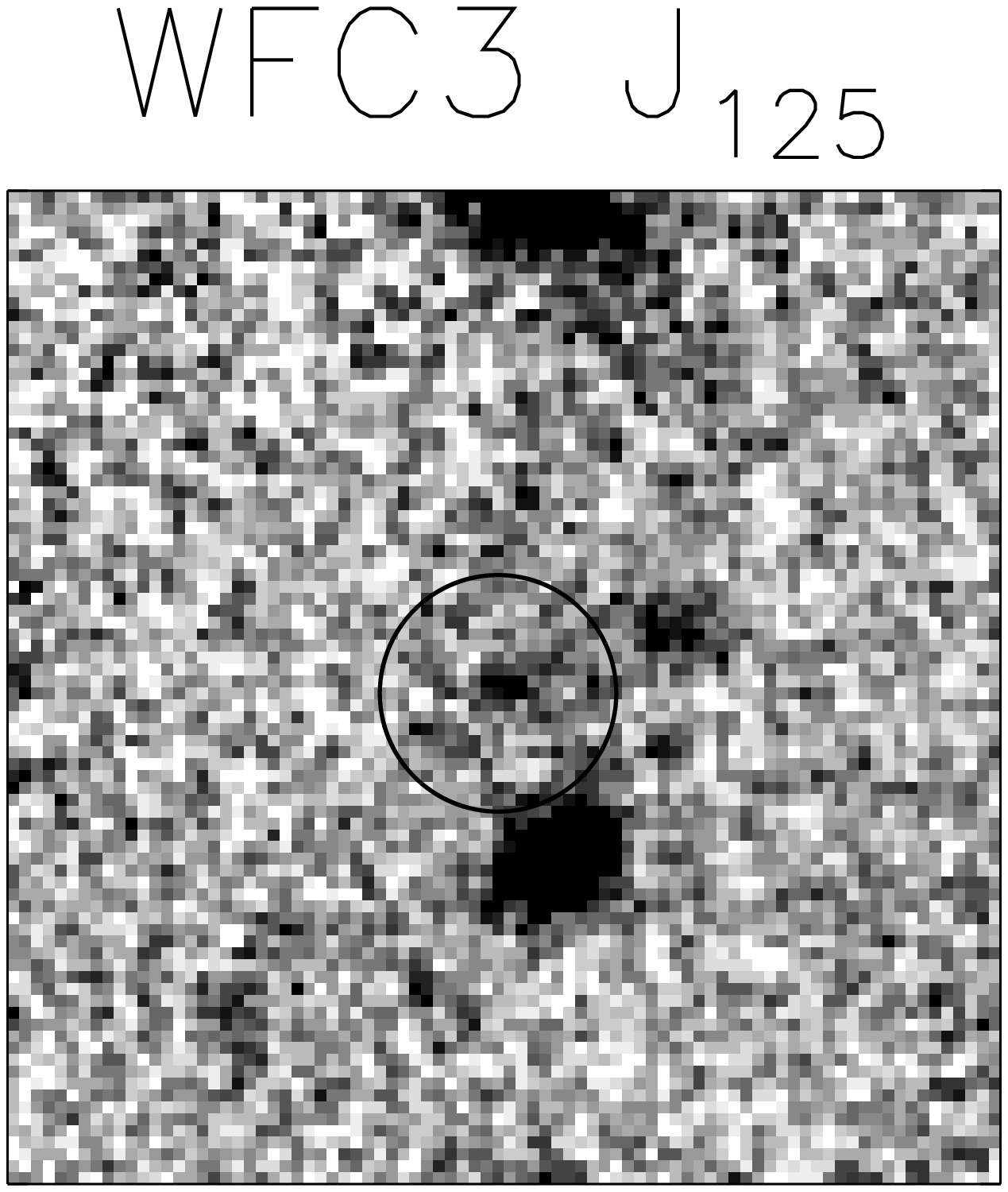}
\hspace{-10mm}
\plotone{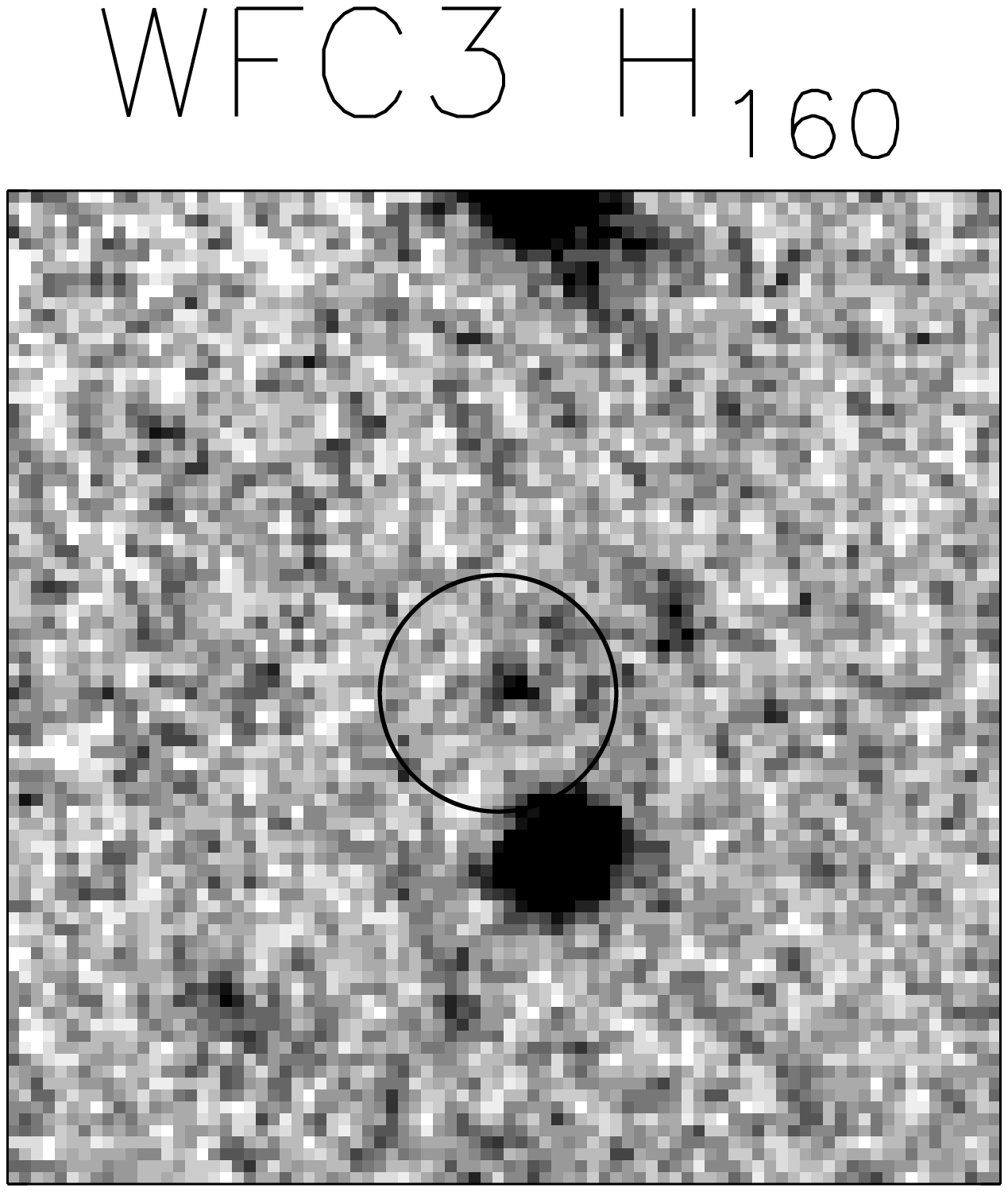}
\hspace{-10mm}
\plotone{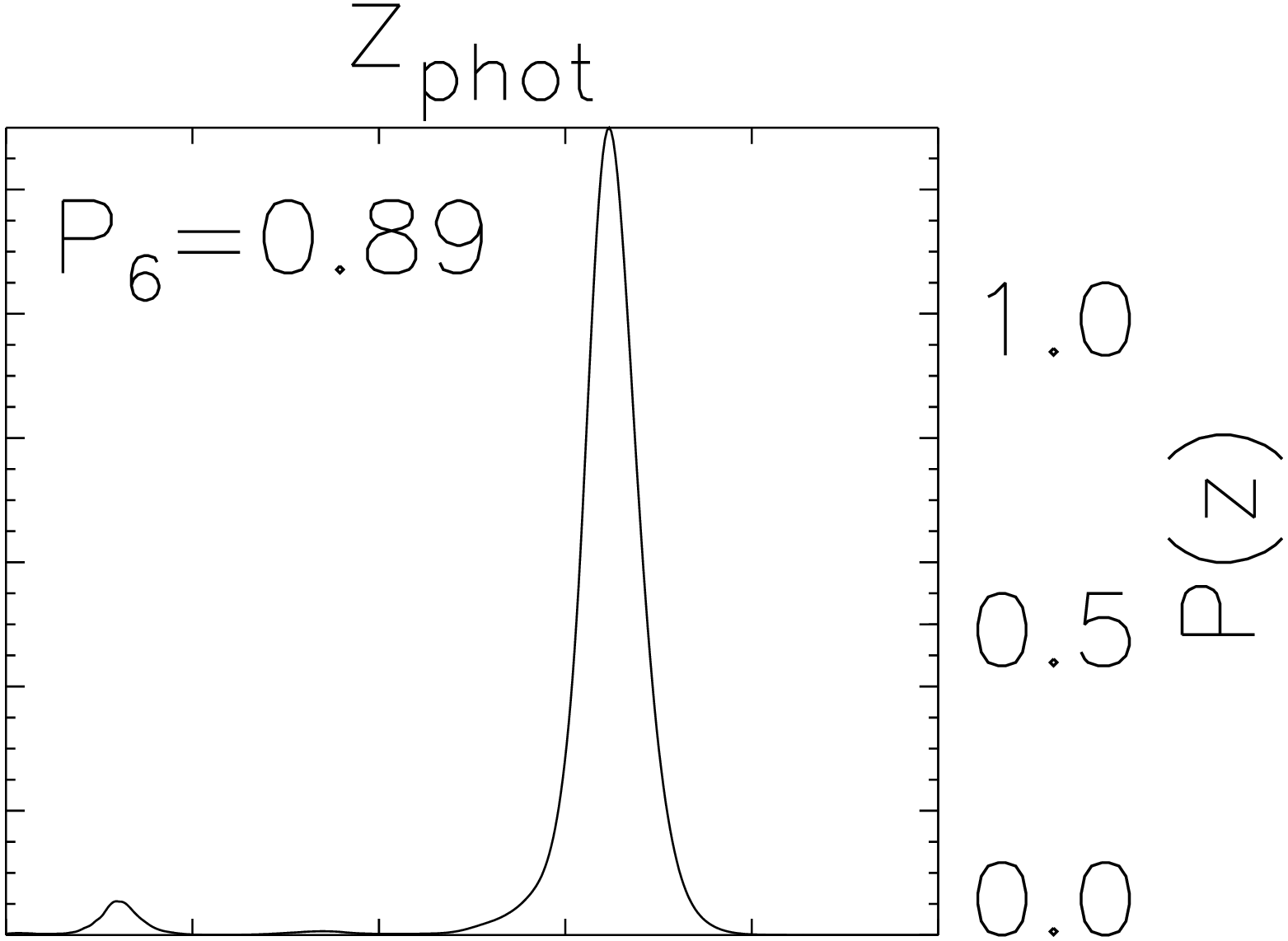}
\vspace{0.5mm}

\plotone{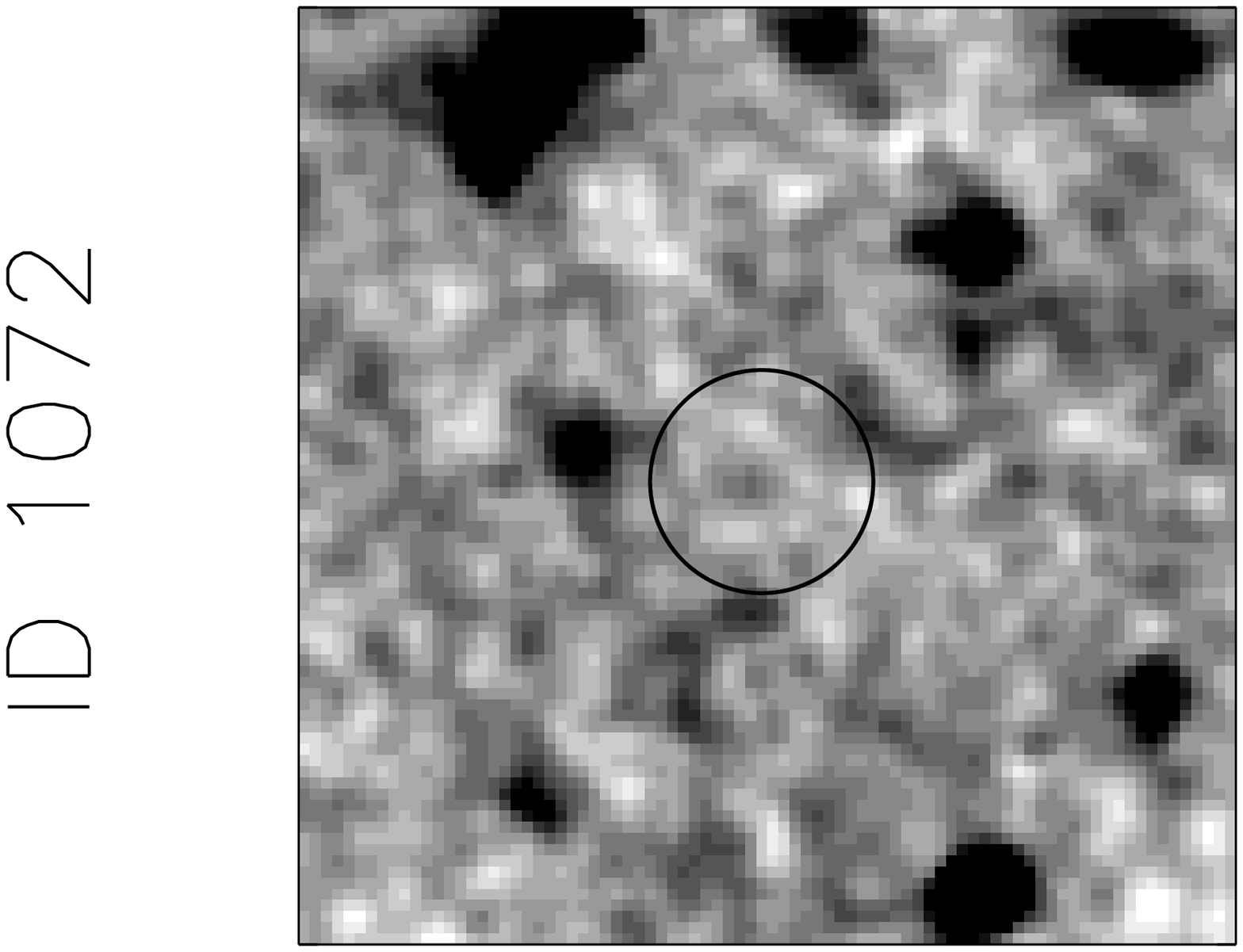}
\hspace{-10mm}
\plotone{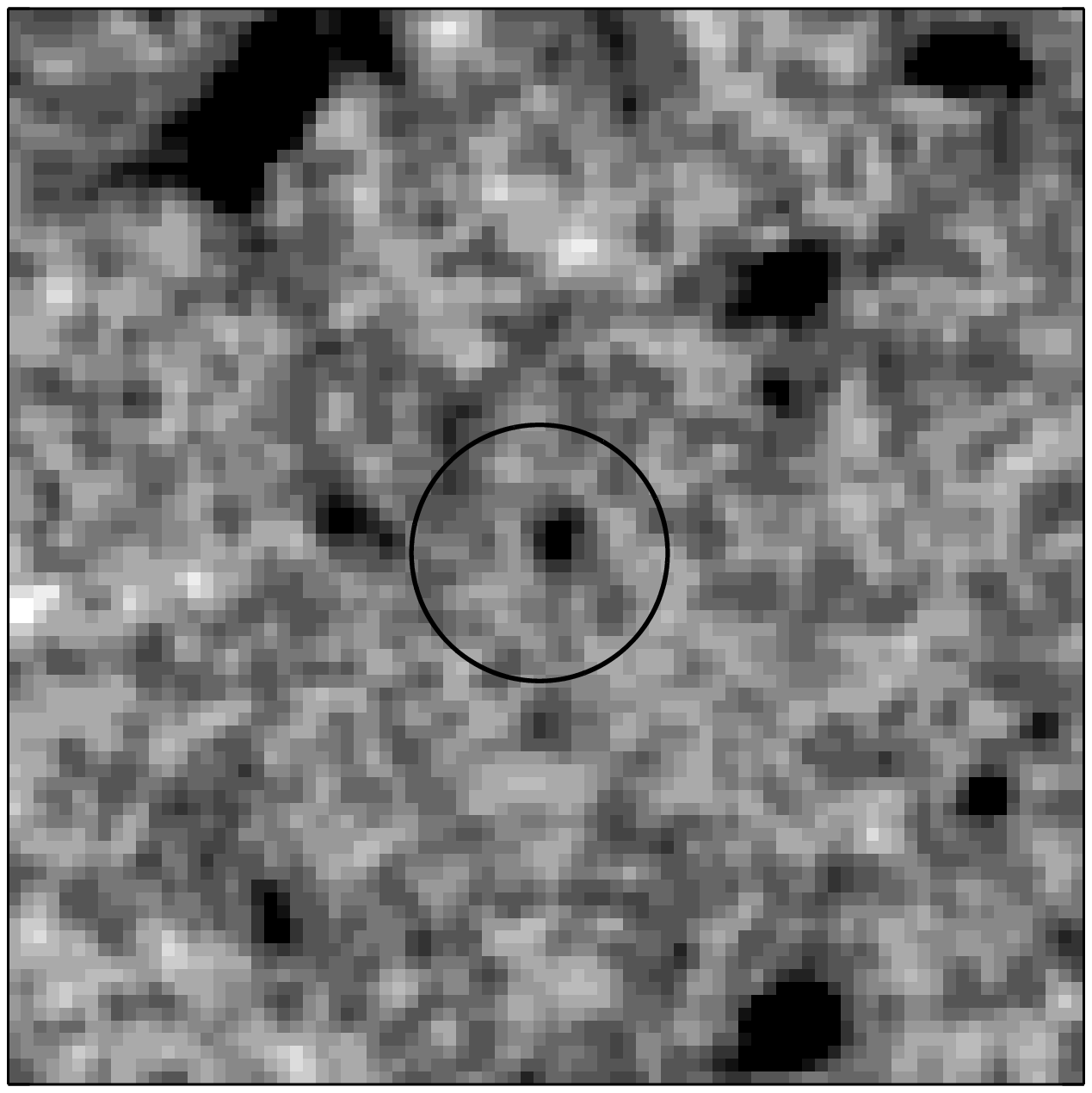}
\hspace{-10mm}
\plotone{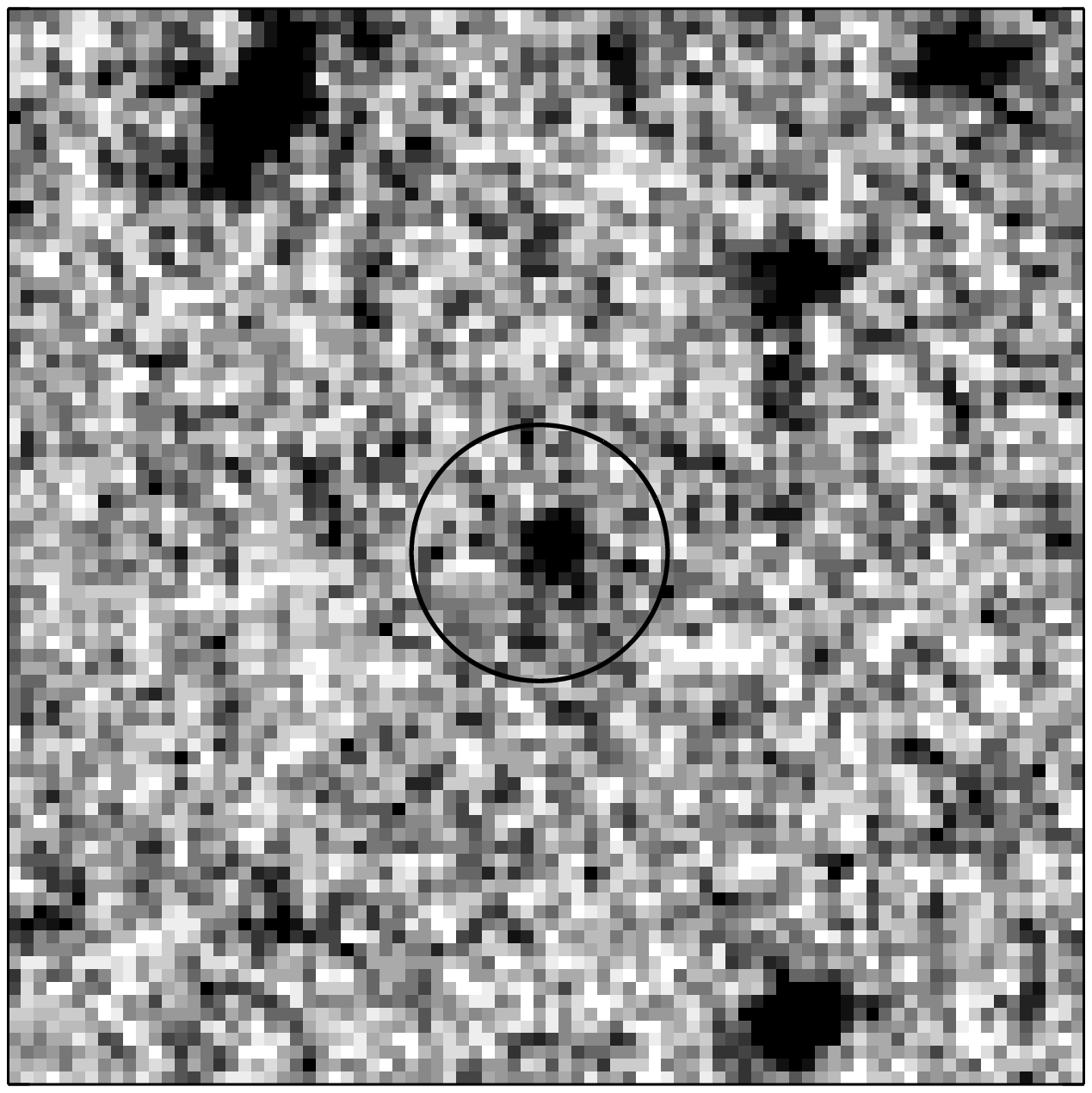}
\hspace{-10mm}
\plotone{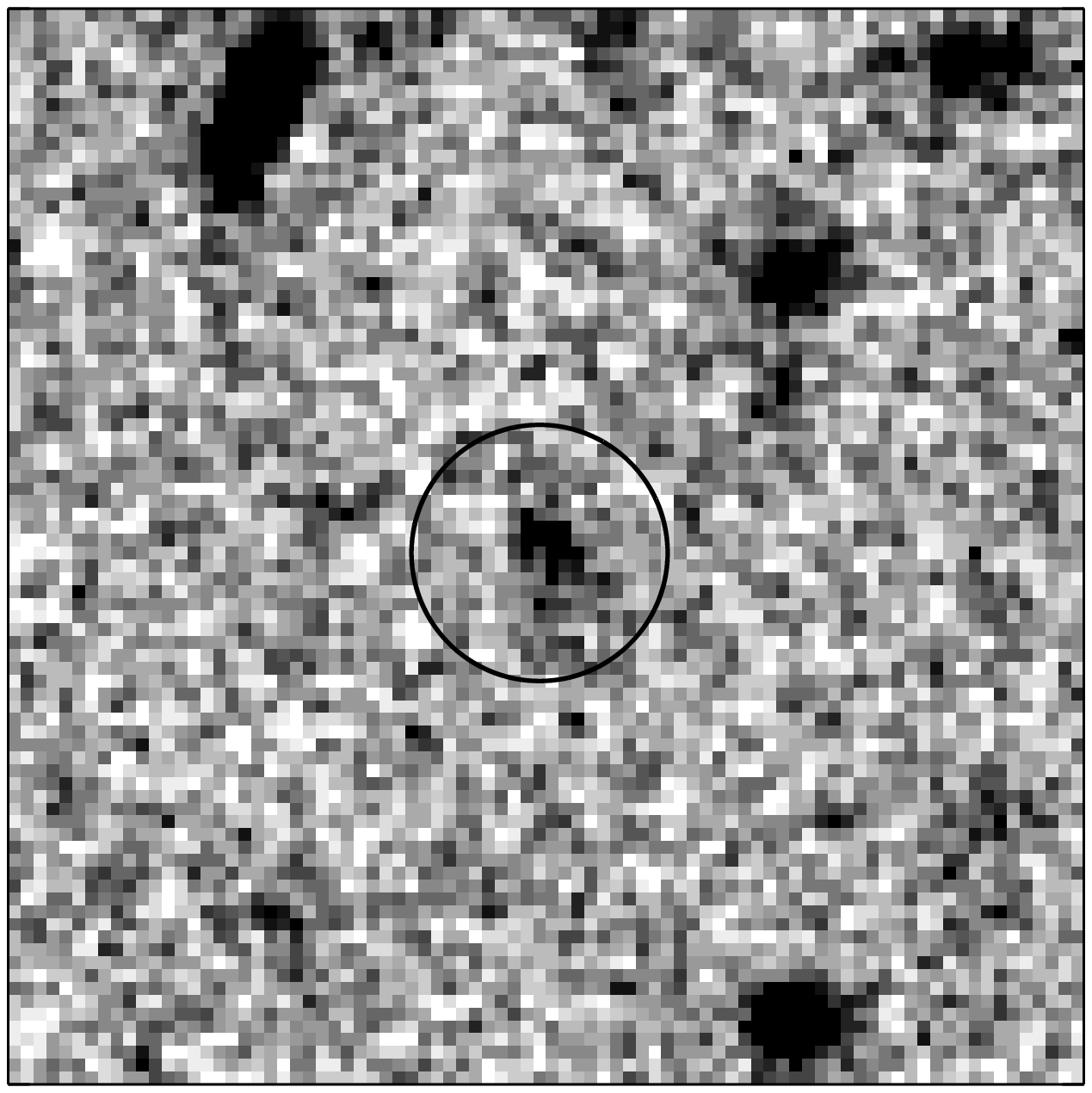}
\hspace{-10mm}
\plotone{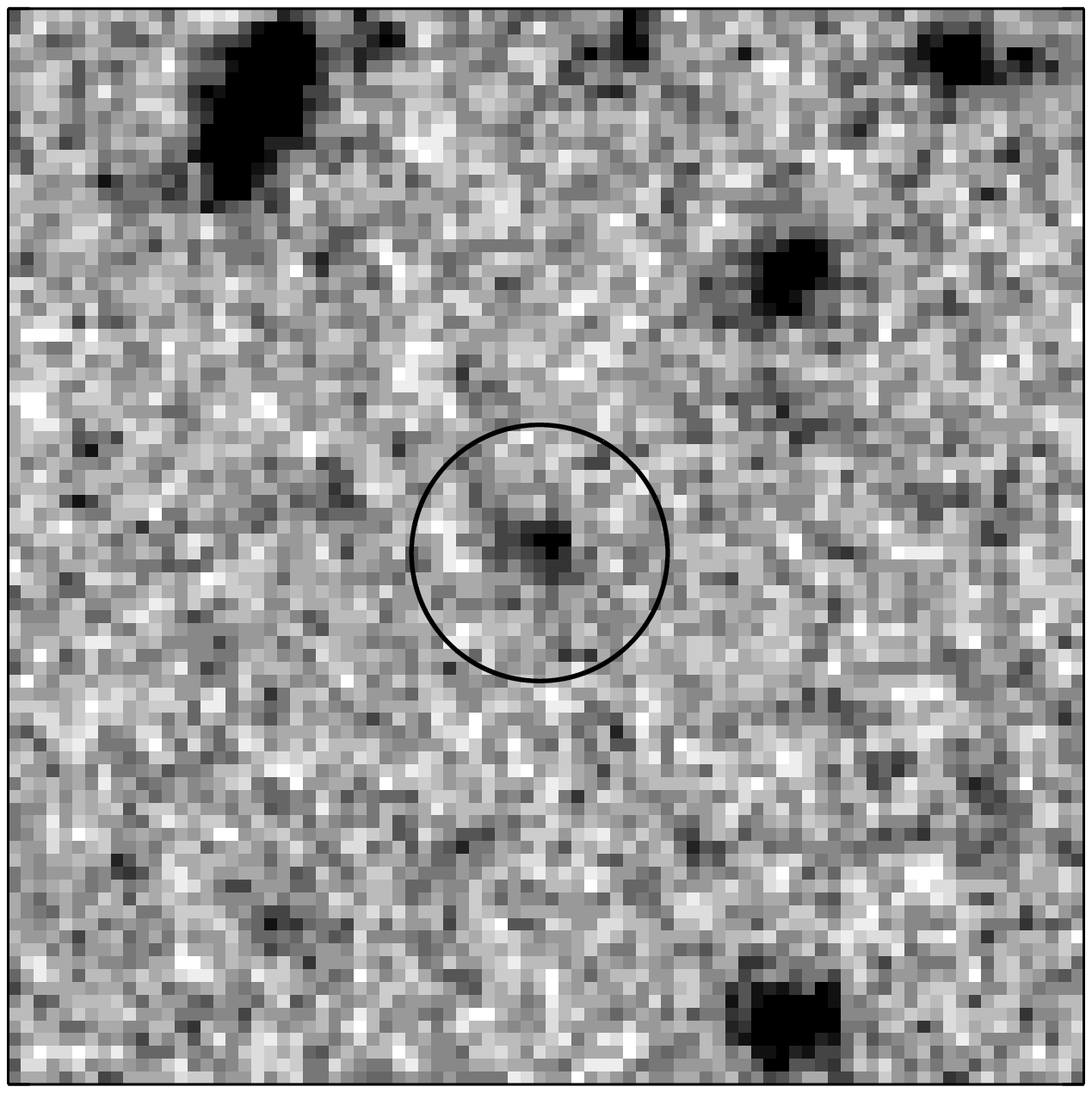}
\hspace{-10mm}
\plotone{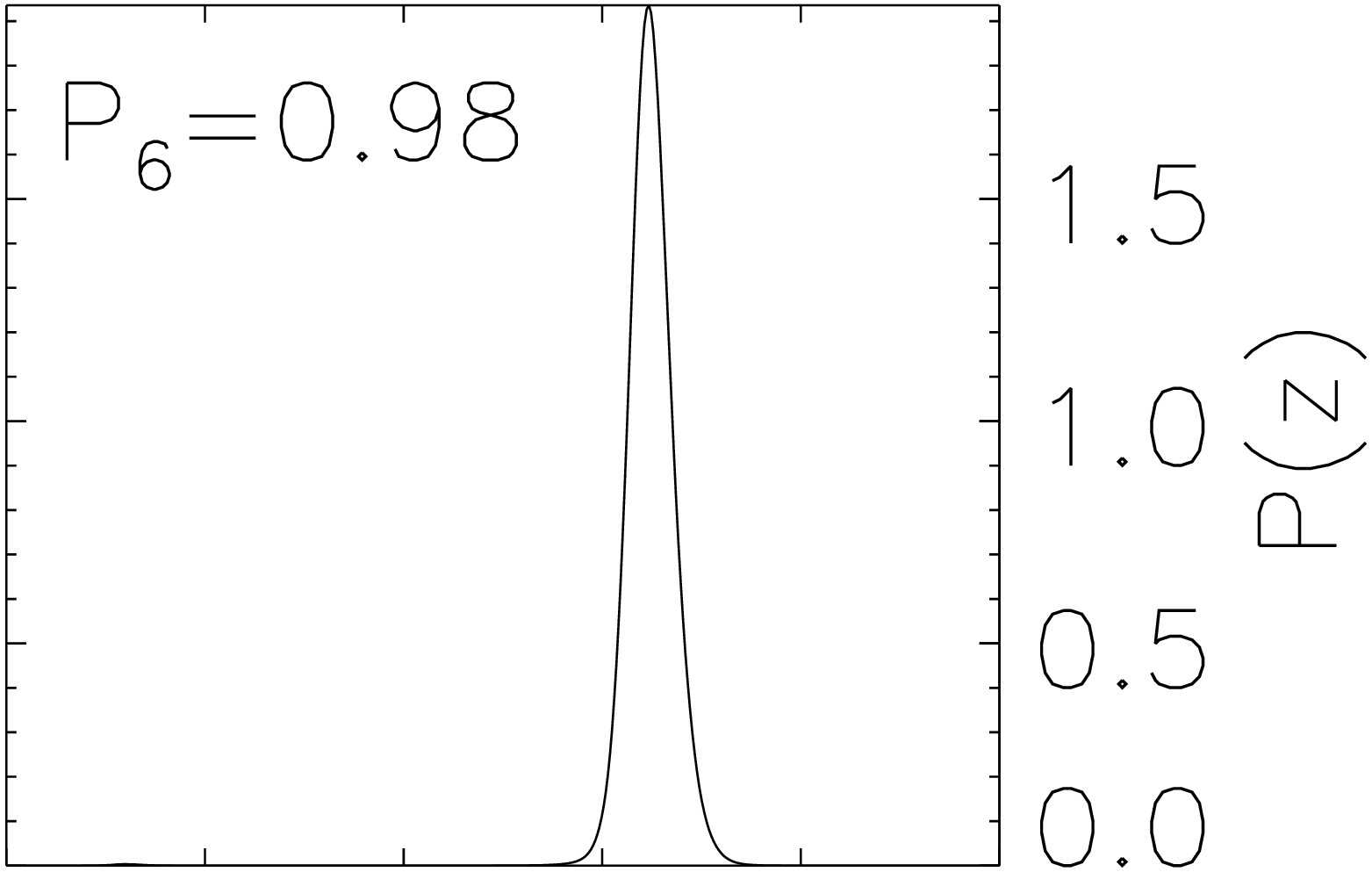}
\vspace{0.5mm}

\plotone{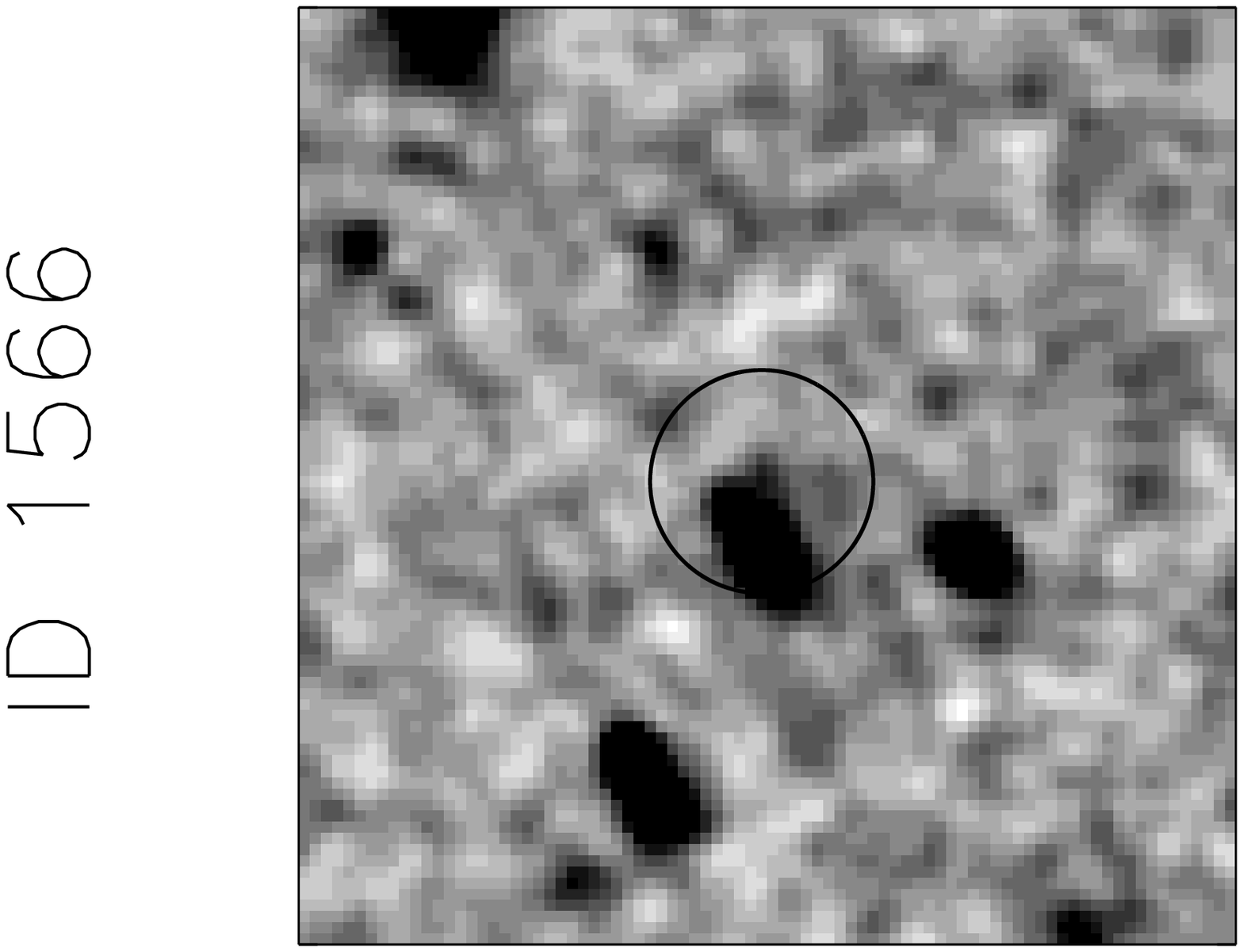}
\hspace{-10mm}
\plotone{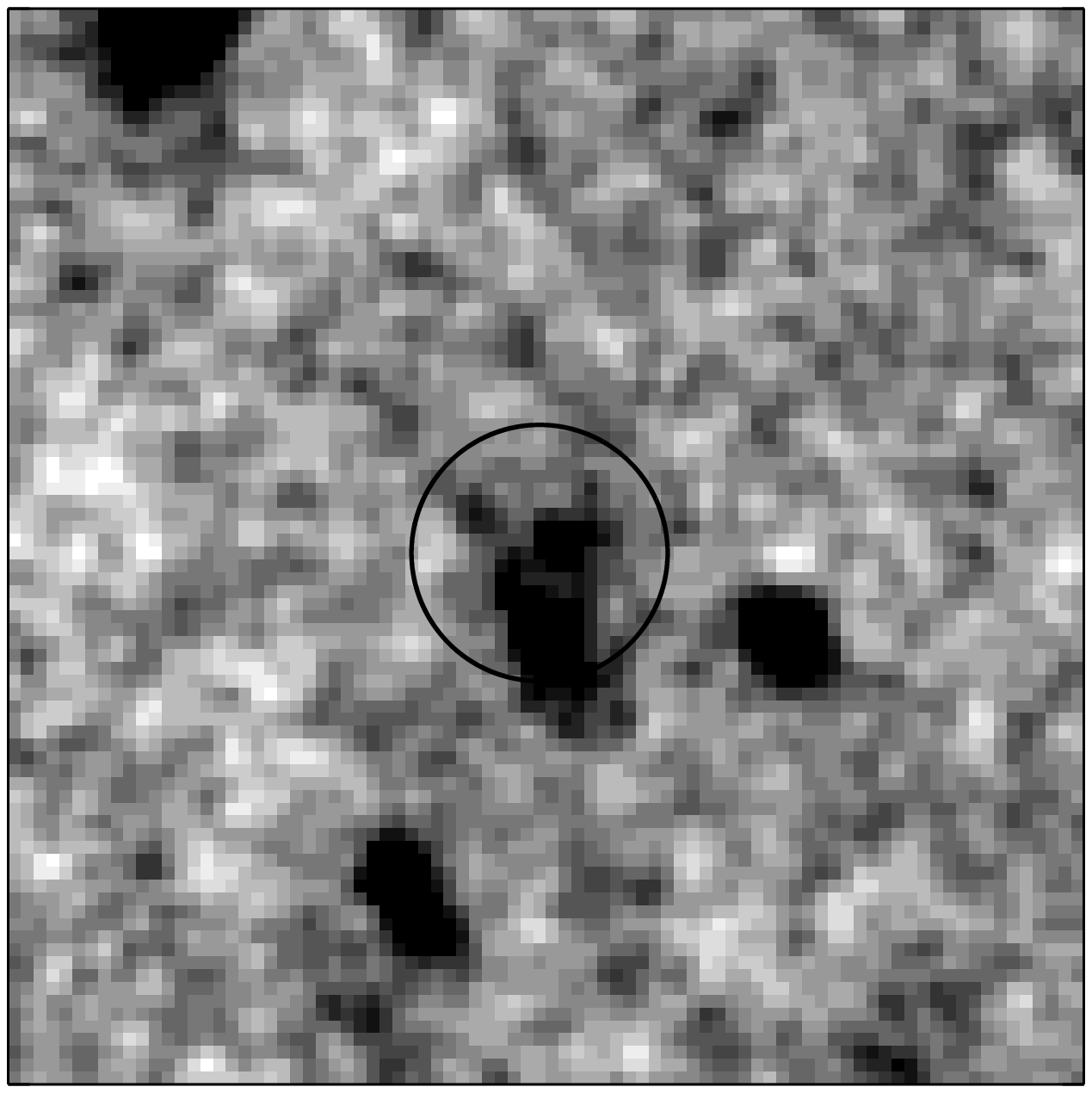}
\hspace{-10mm}
\plotone{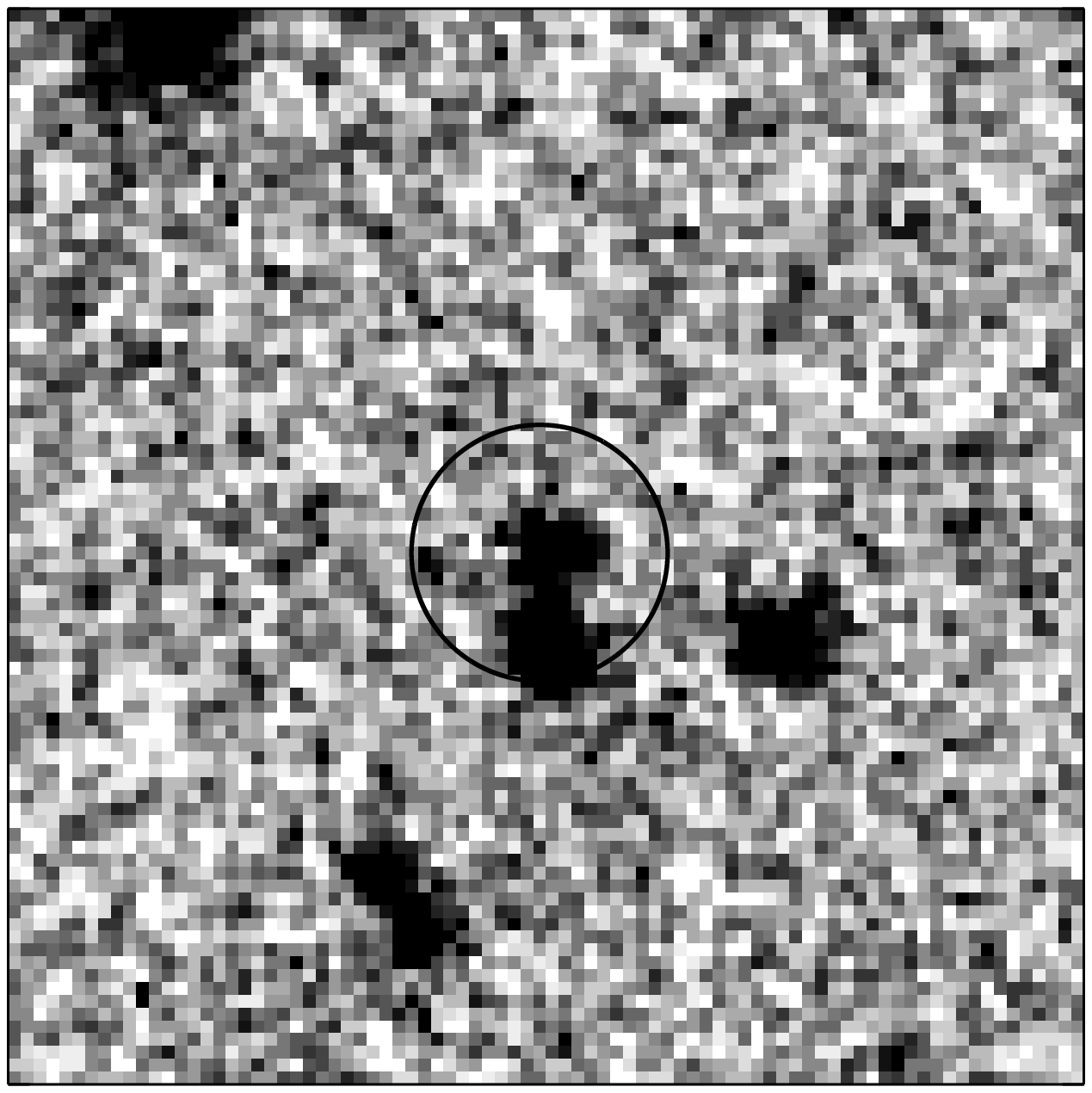}
\hspace{-10mm}
\plotone{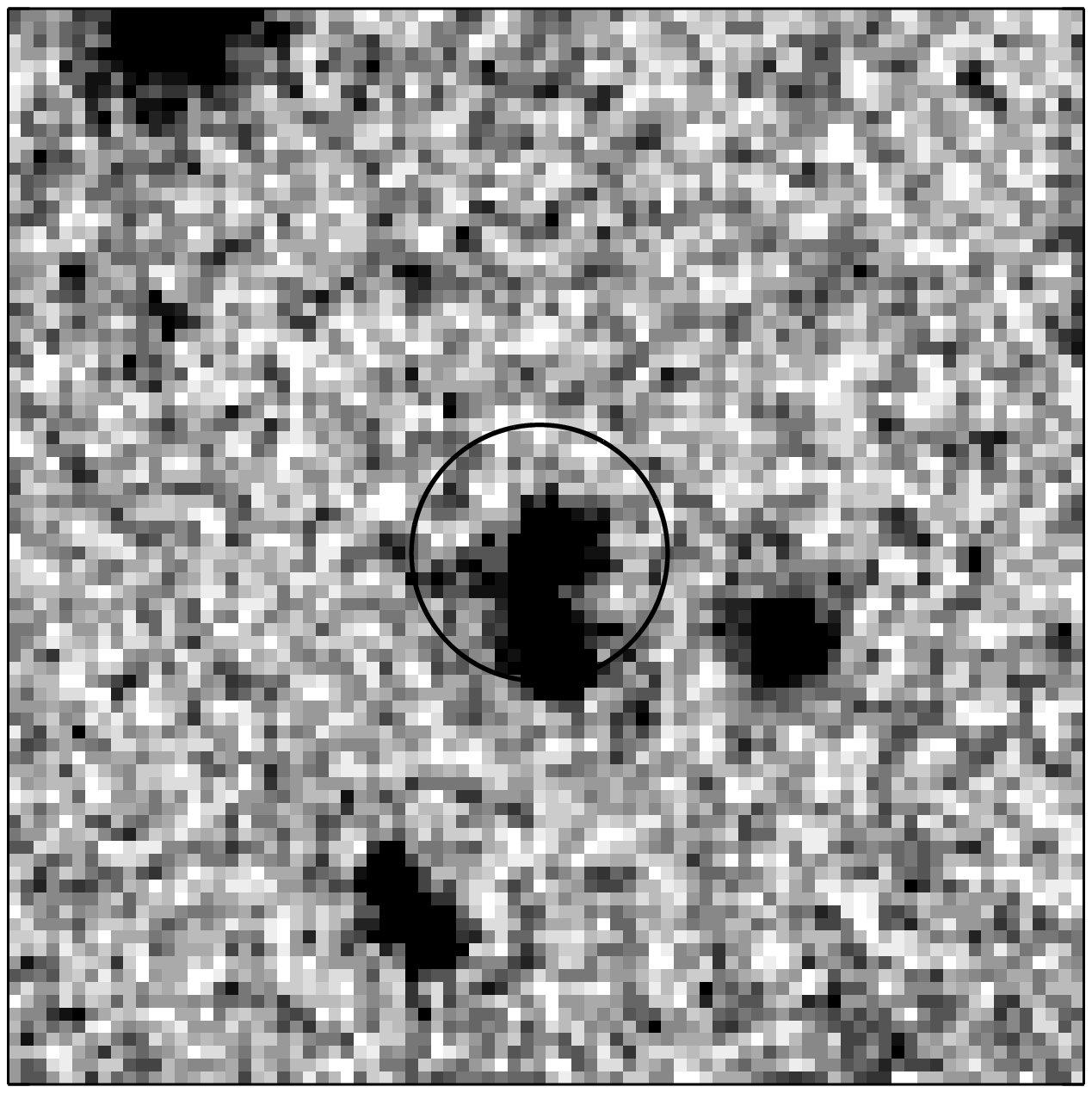}
\hspace{-10mm}
\plotone{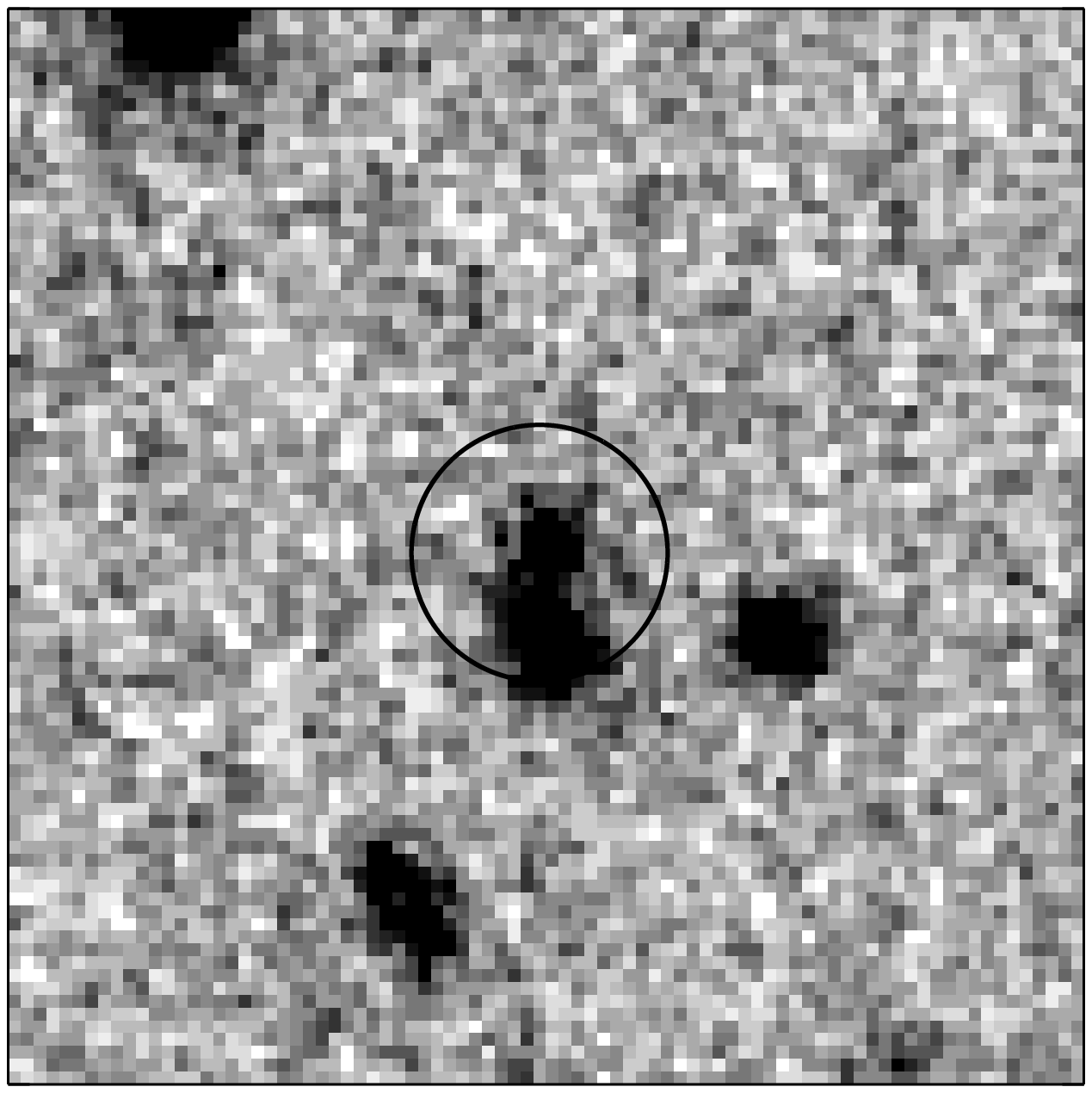}
\hspace{-10mm}
\plotone{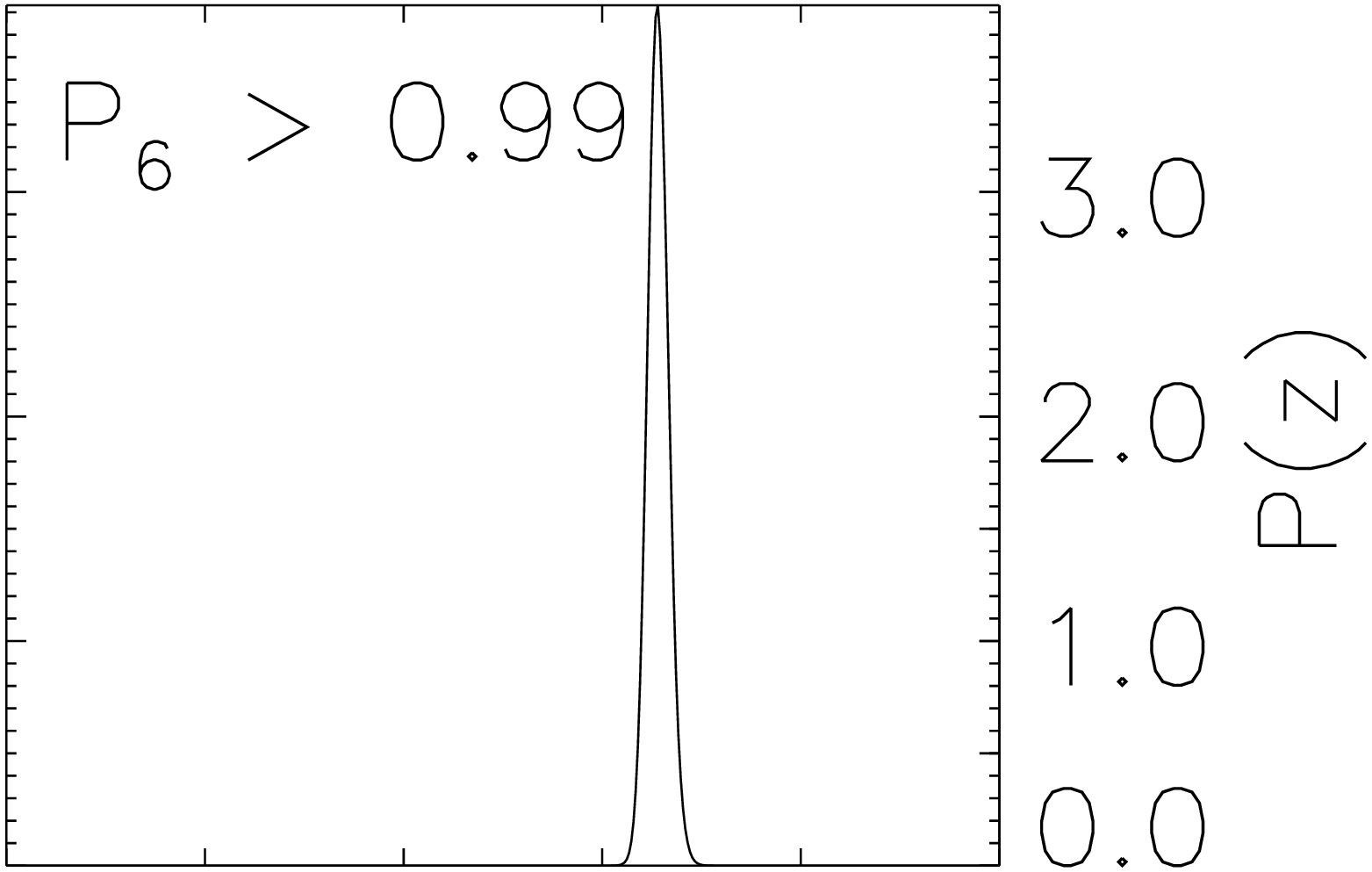}
\vspace{0.5mm}

\plotone{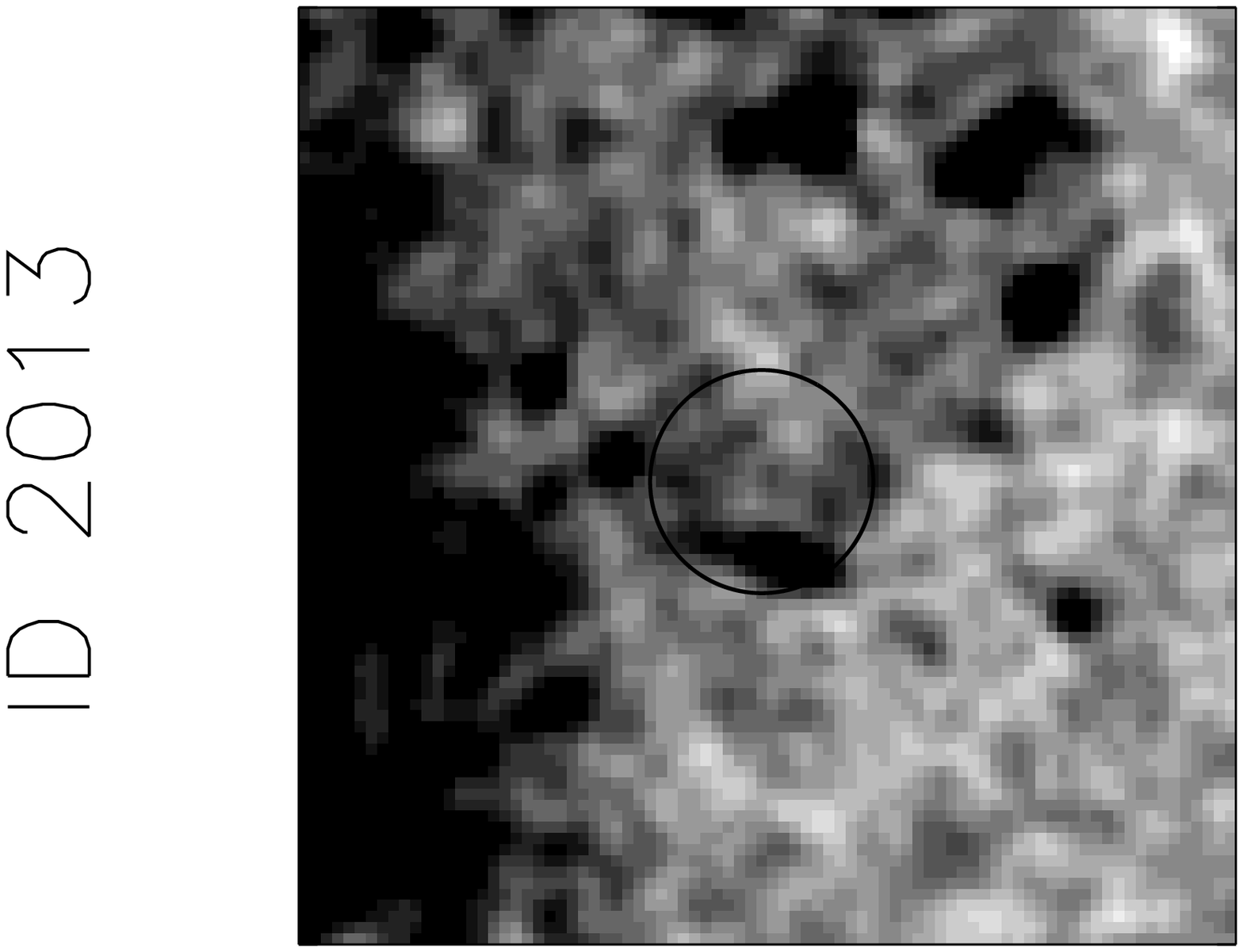}
\hspace{-10mm}
\plotone{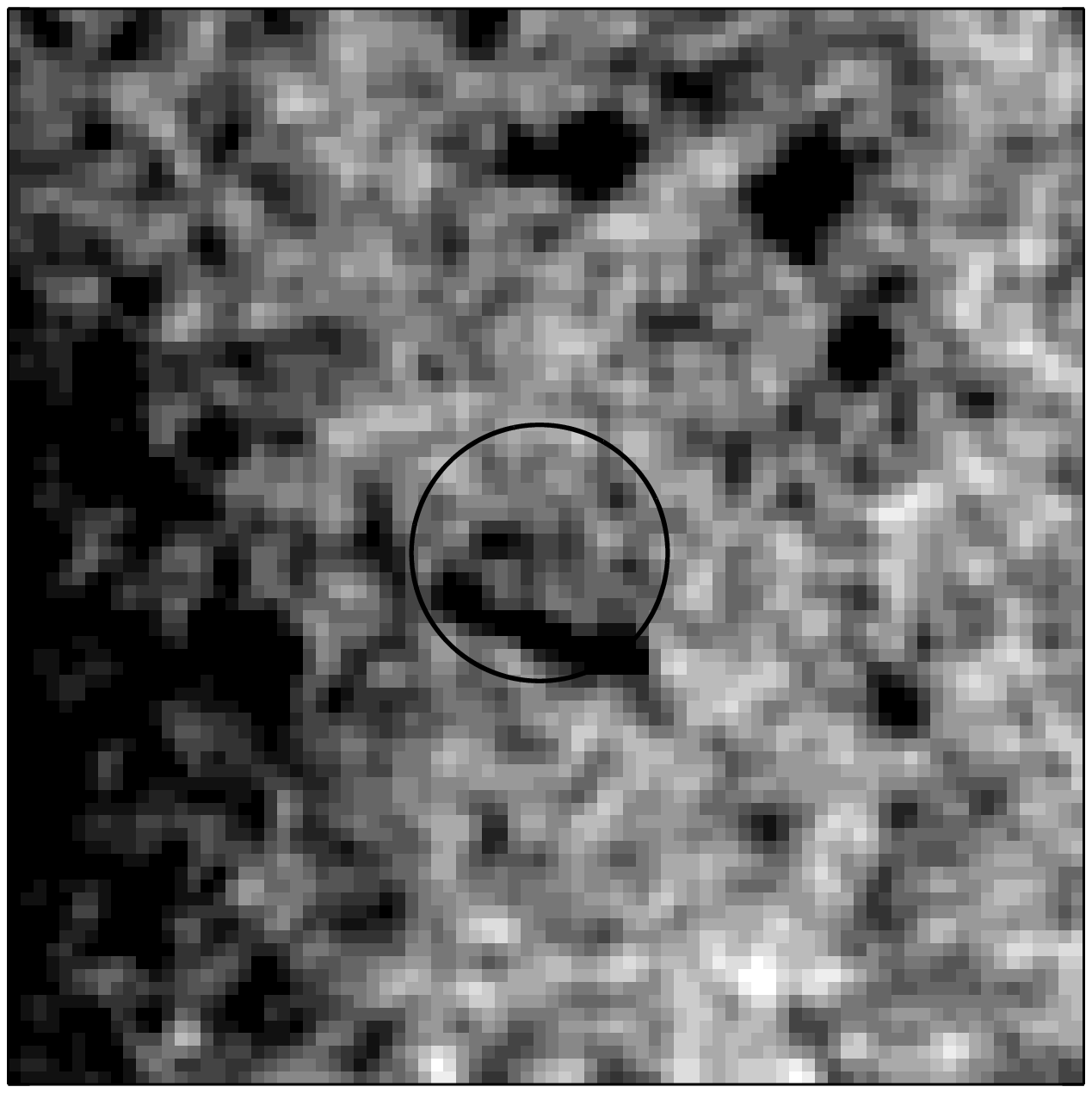}
\hspace{-10mm}
\plotone{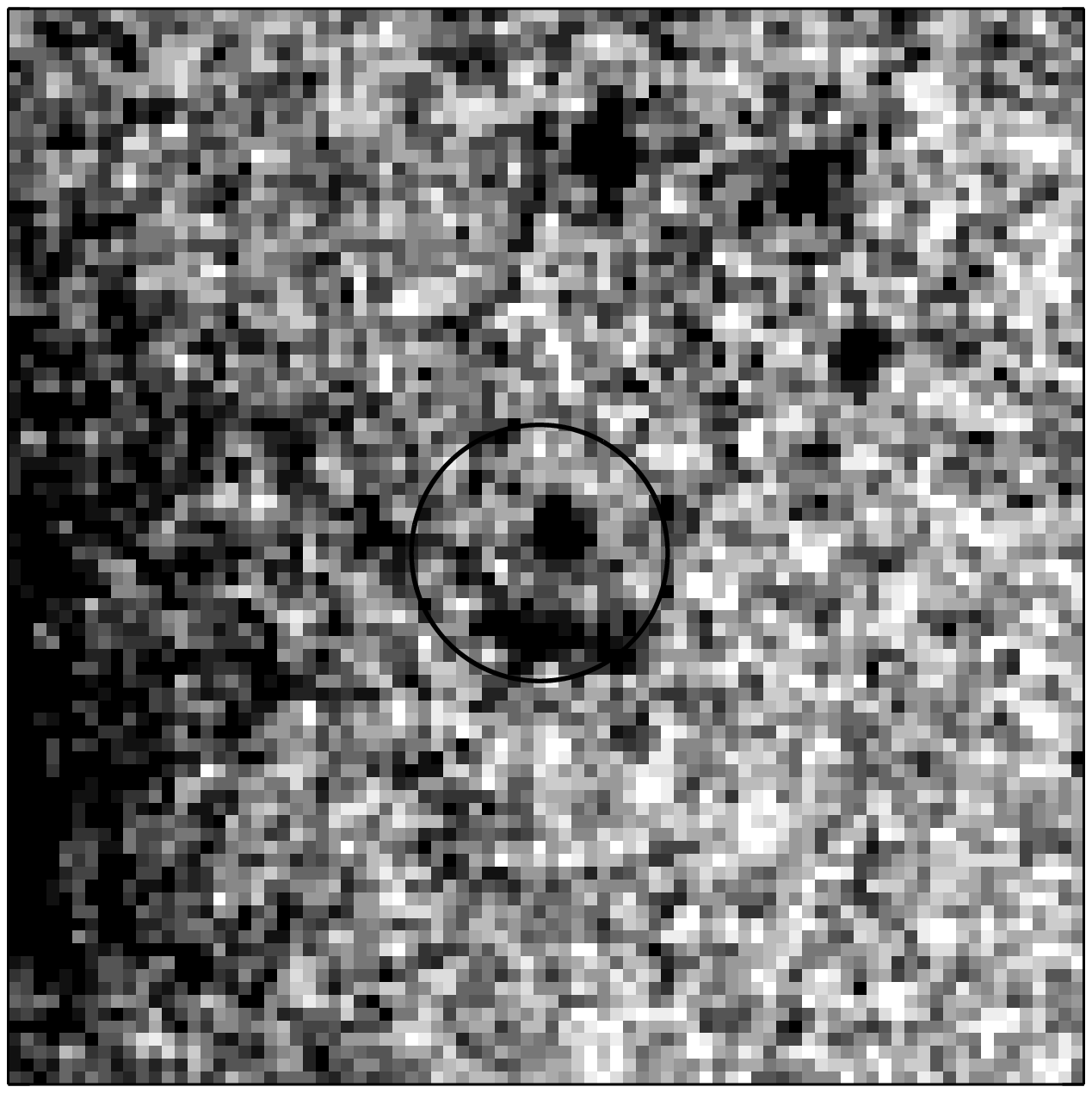}
\hspace{-10mm}
\plotone{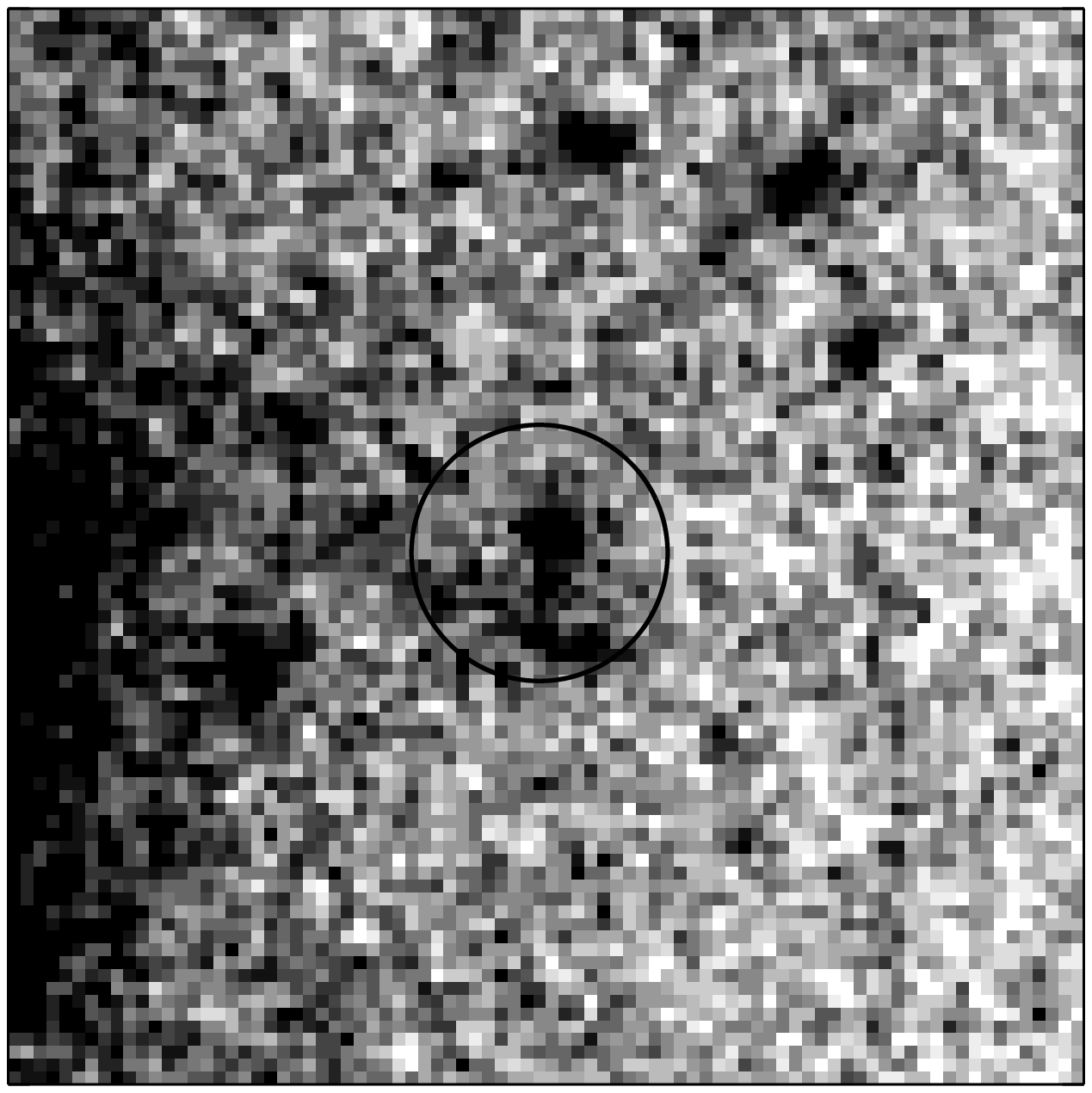}
\hspace{-10mm}
\plotone{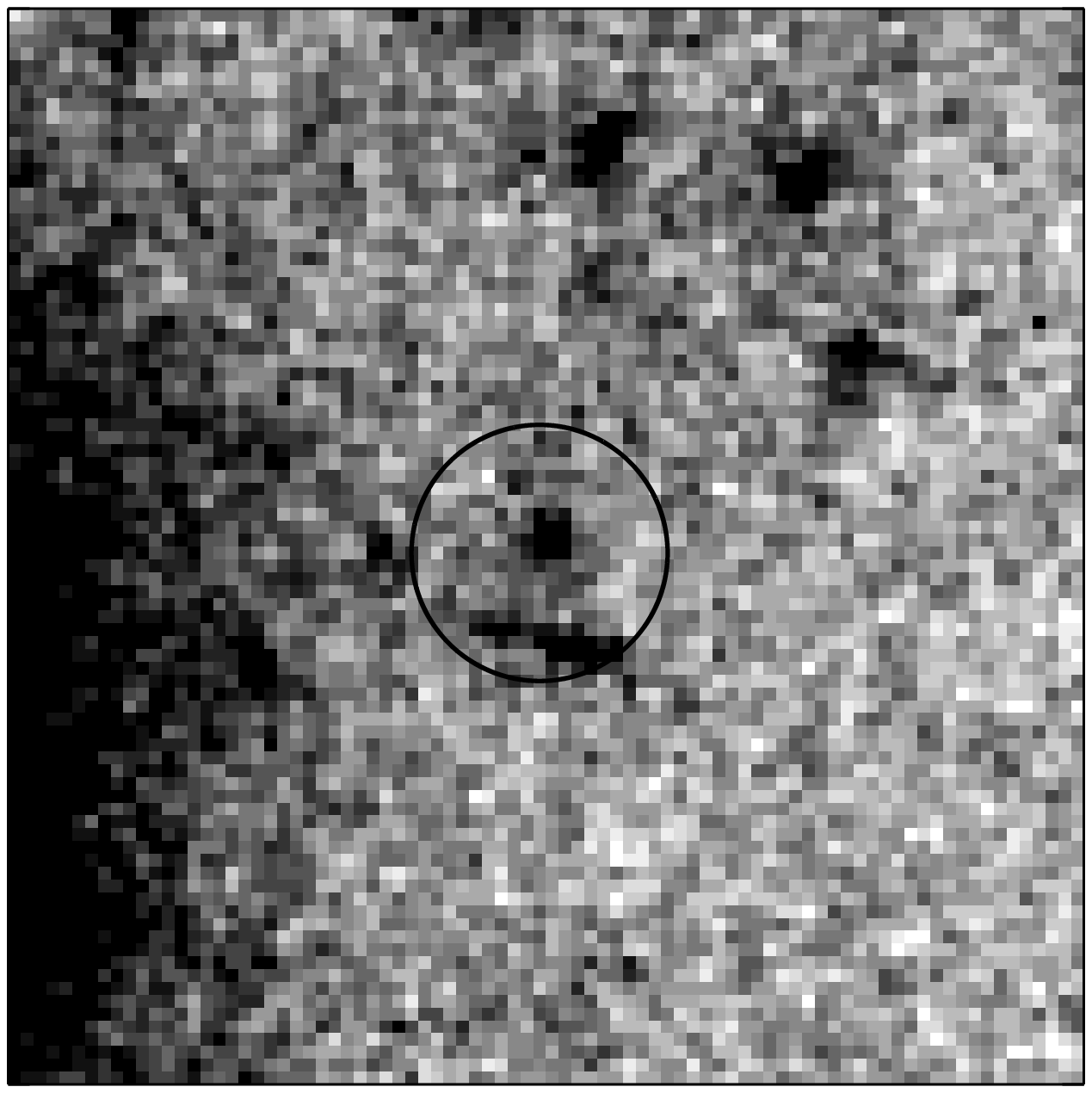}
\hspace{-10mm}
\plotone{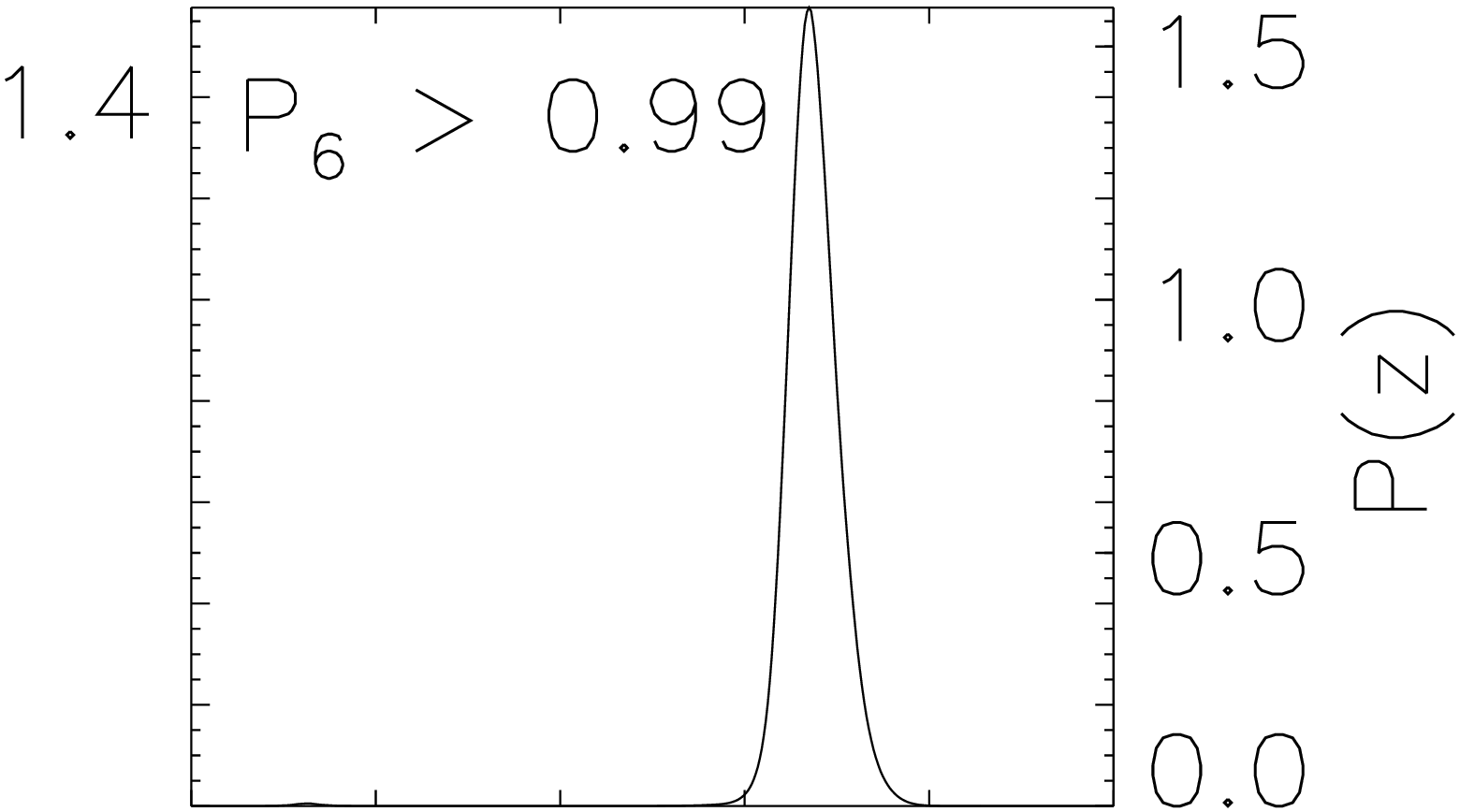}
\vspace{0.5mm}

\plotone{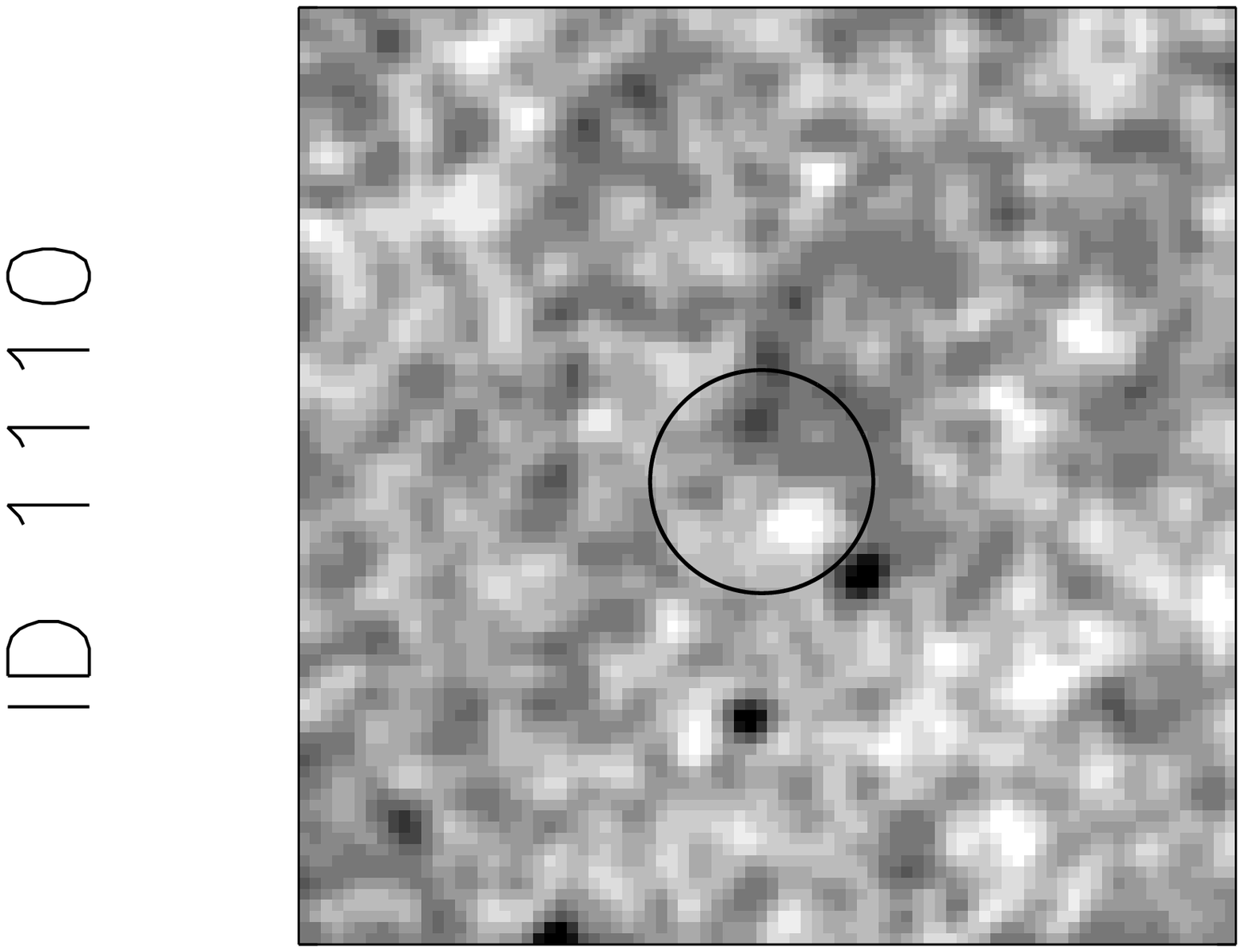}
\hspace{-10mm}
\plotone{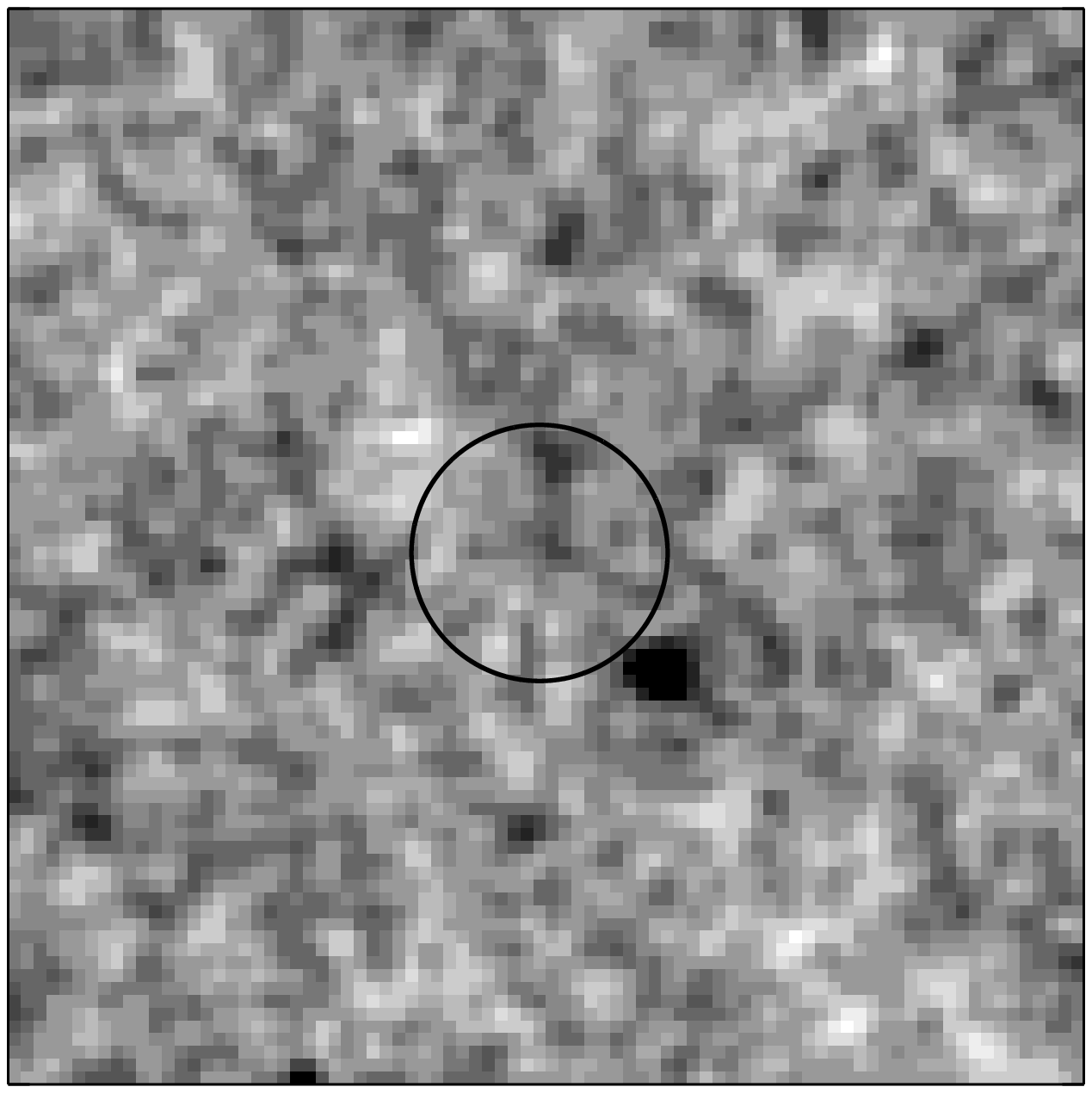}
\hspace{-10mm}
\plotone{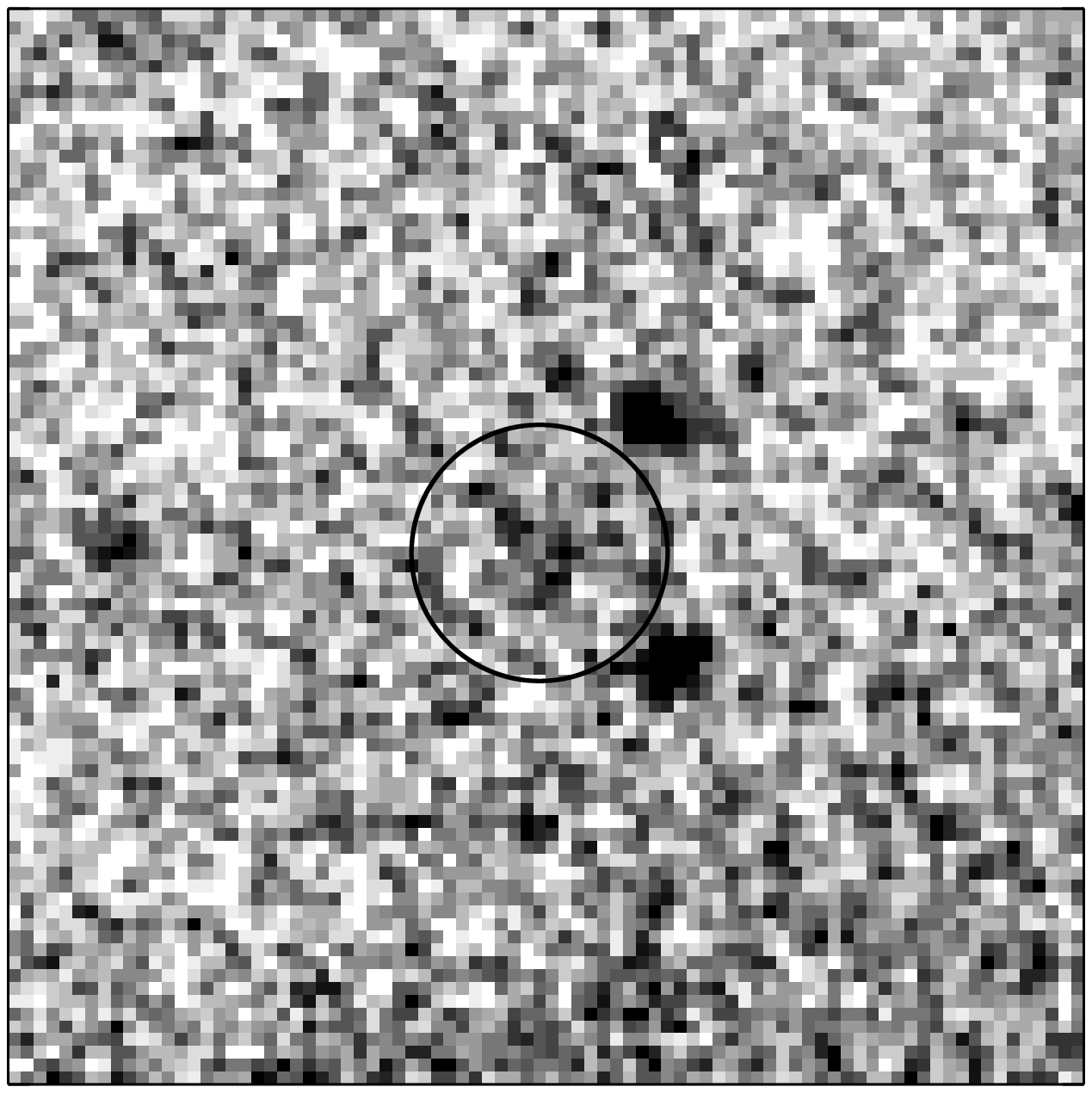}
\hspace{-10mm}
\plotone{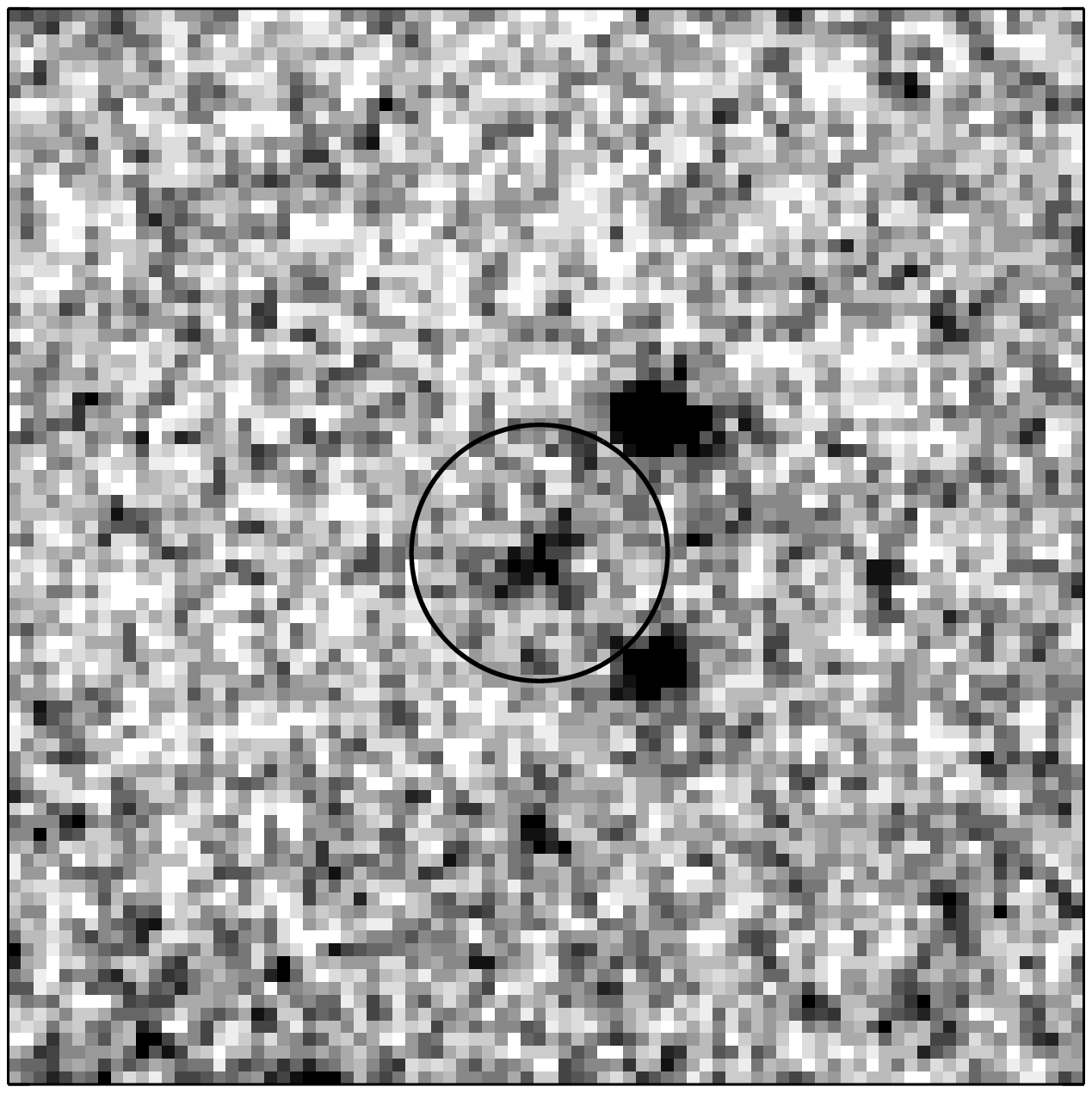}
\hspace{-10mm}
\plotone{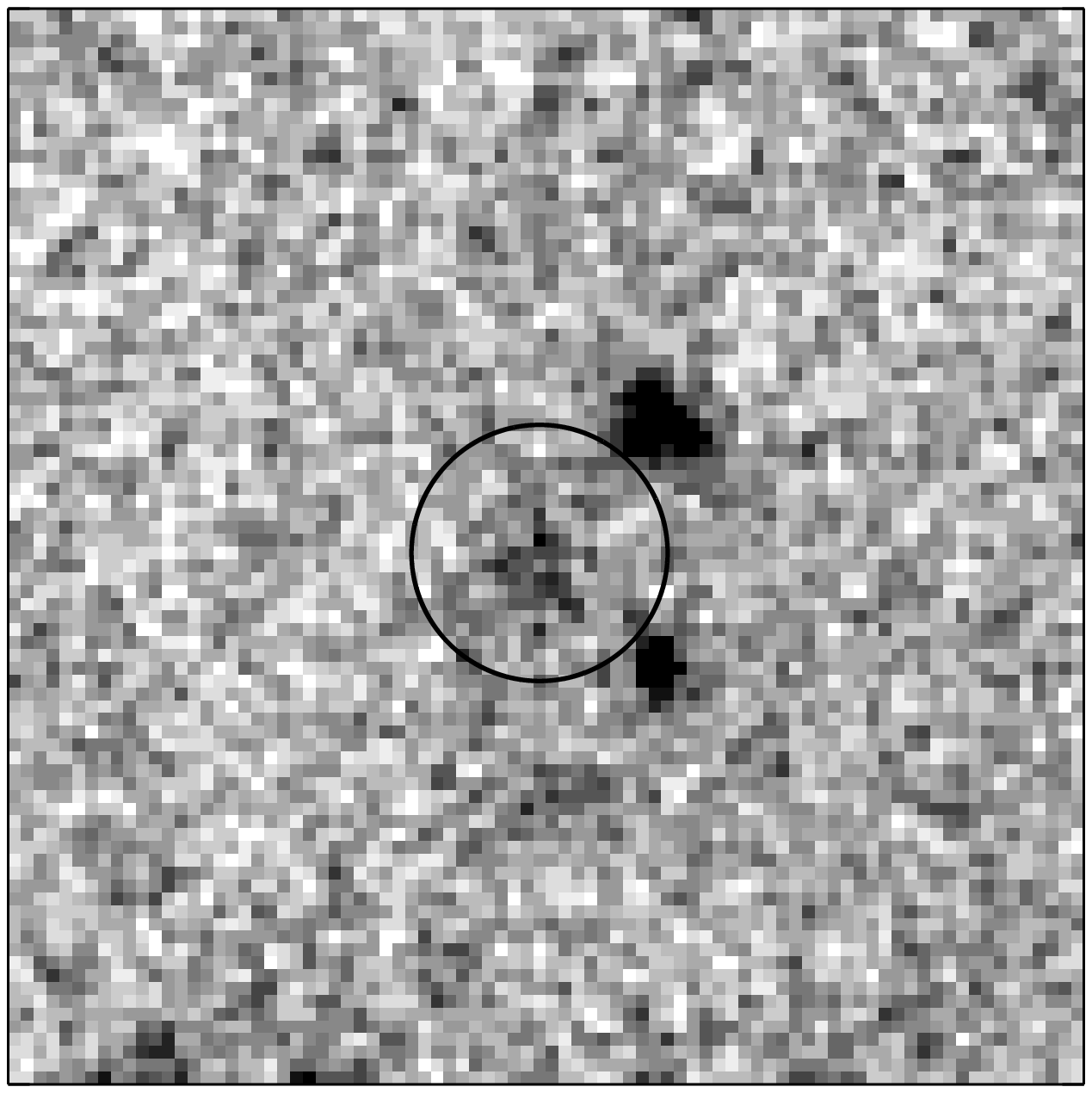}
\hspace{-10mm}
\plotone{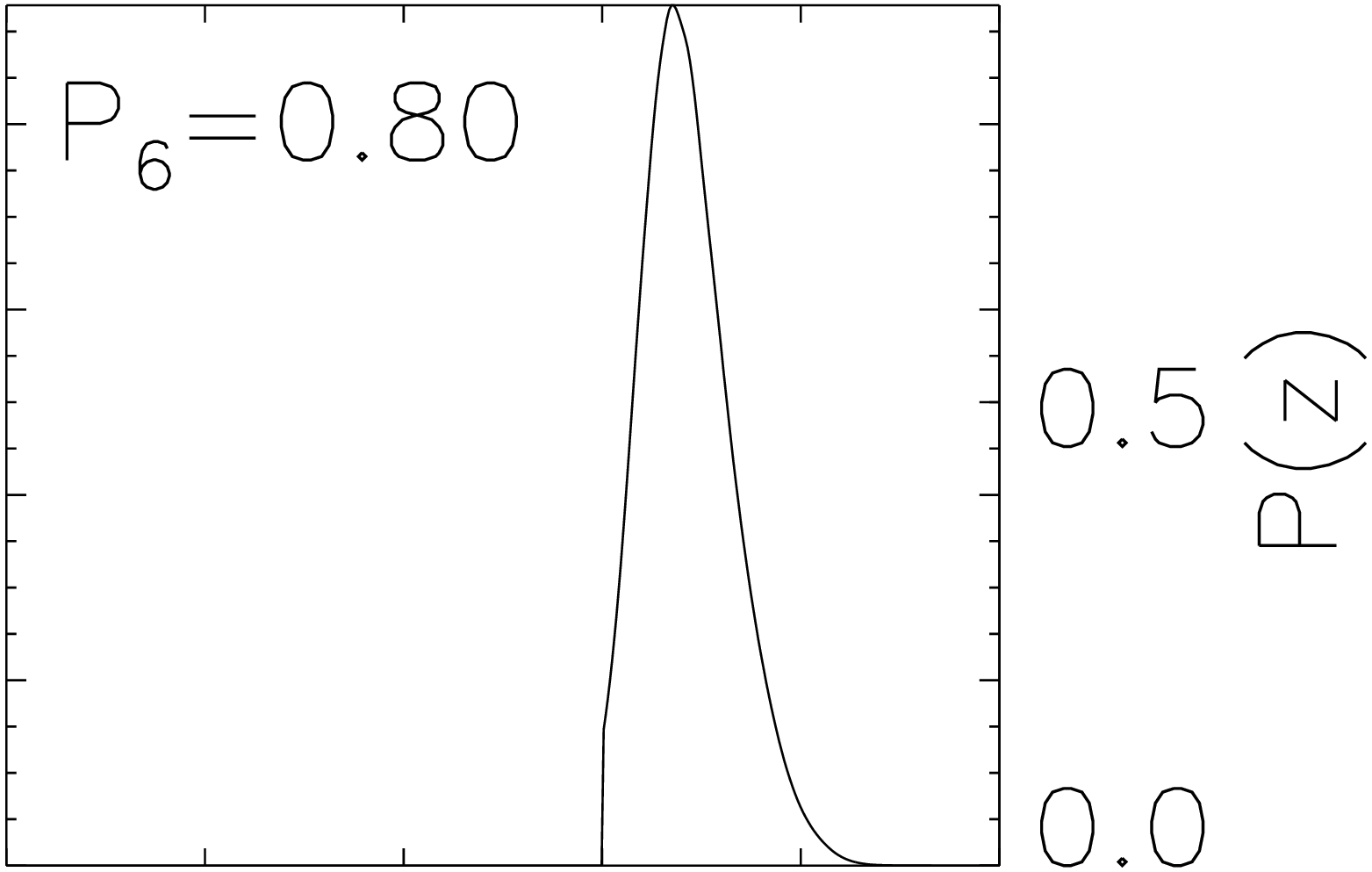}
\vspace{0.5mm}

\plotone{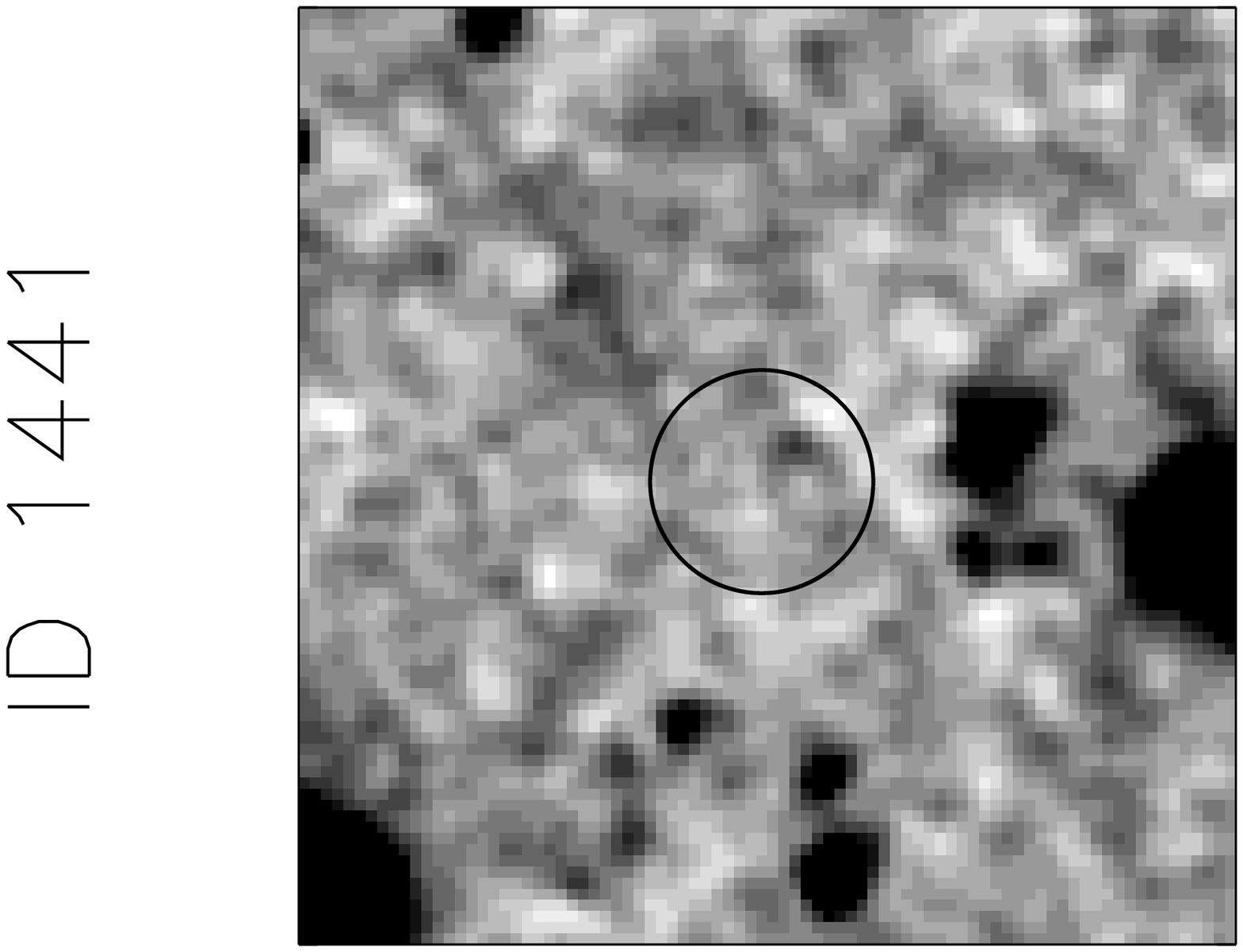}
\hspace{-10mm}
\plotone{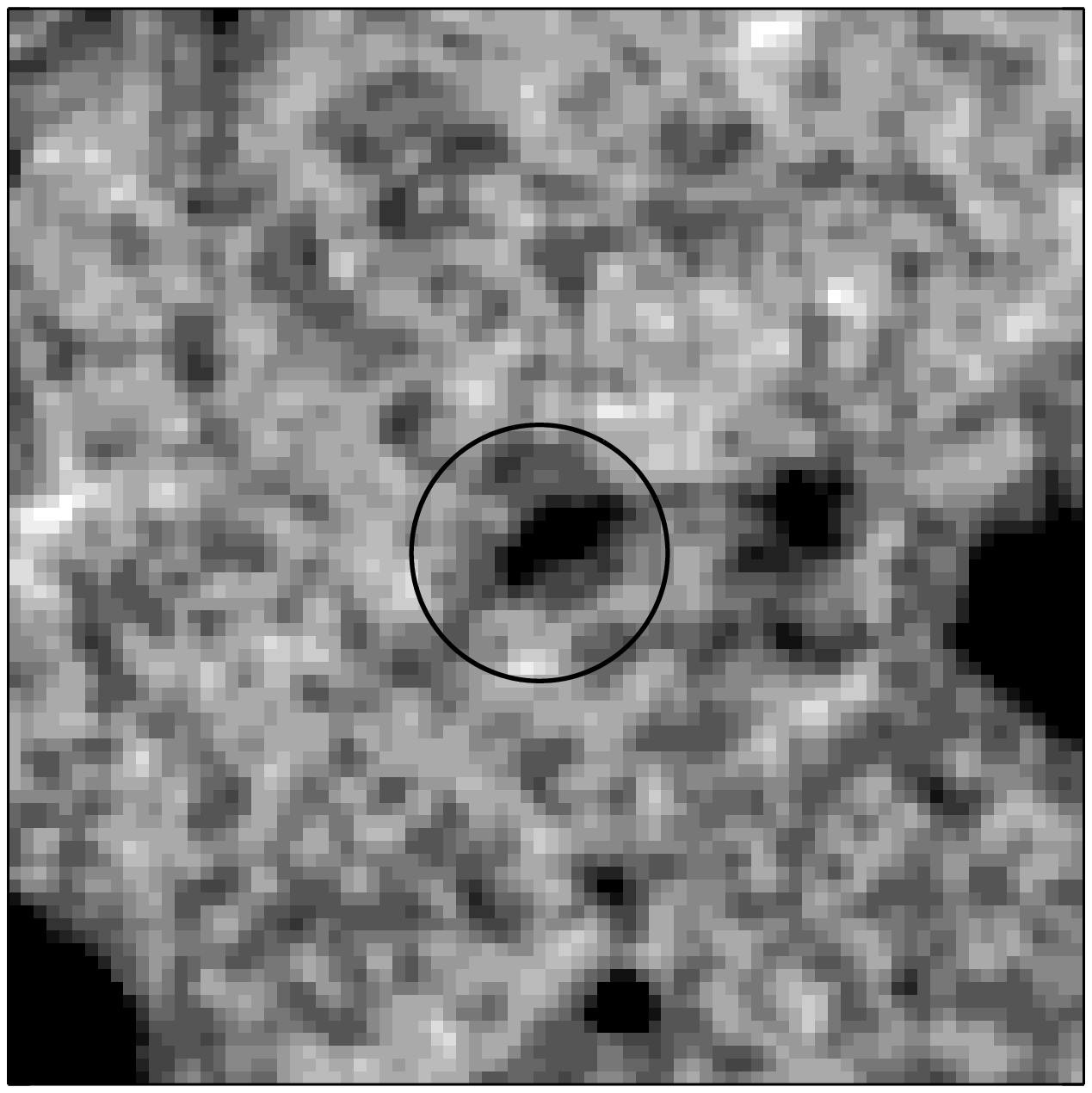}
\hspace{-10mm}
\plotone{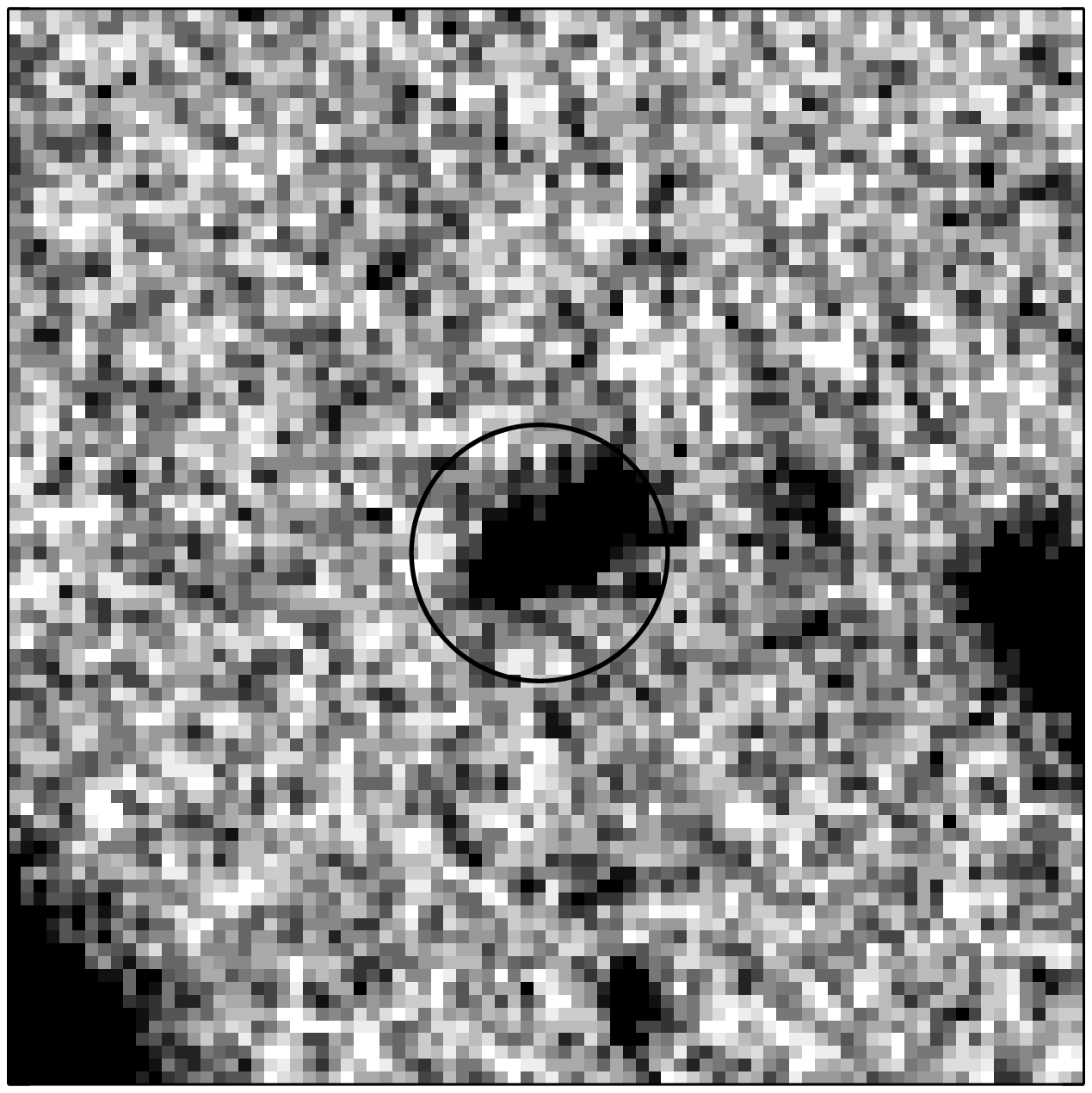}
\hspace{-10mm}
\plotone{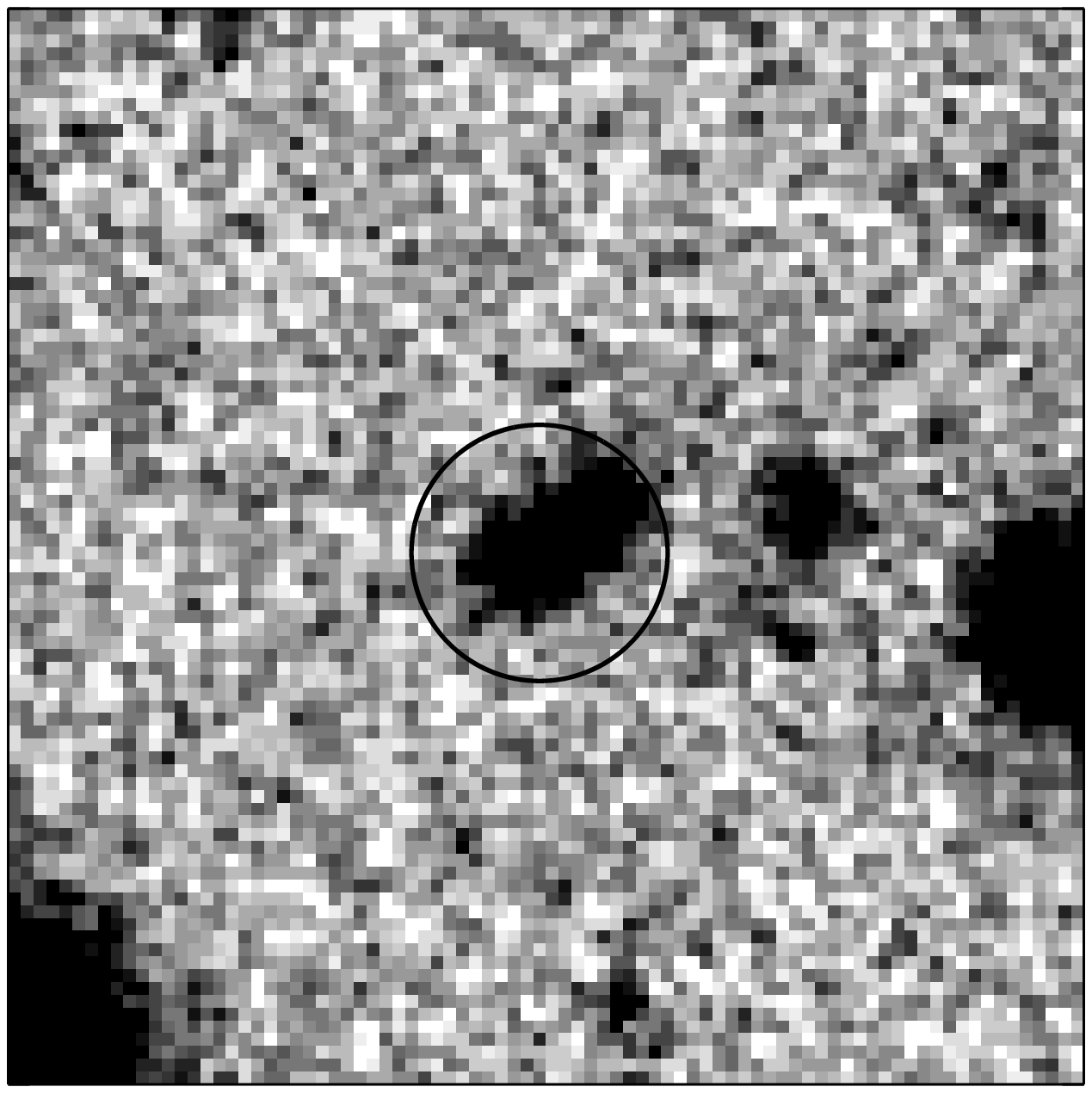}
\hspace{-10mm}
\plotone{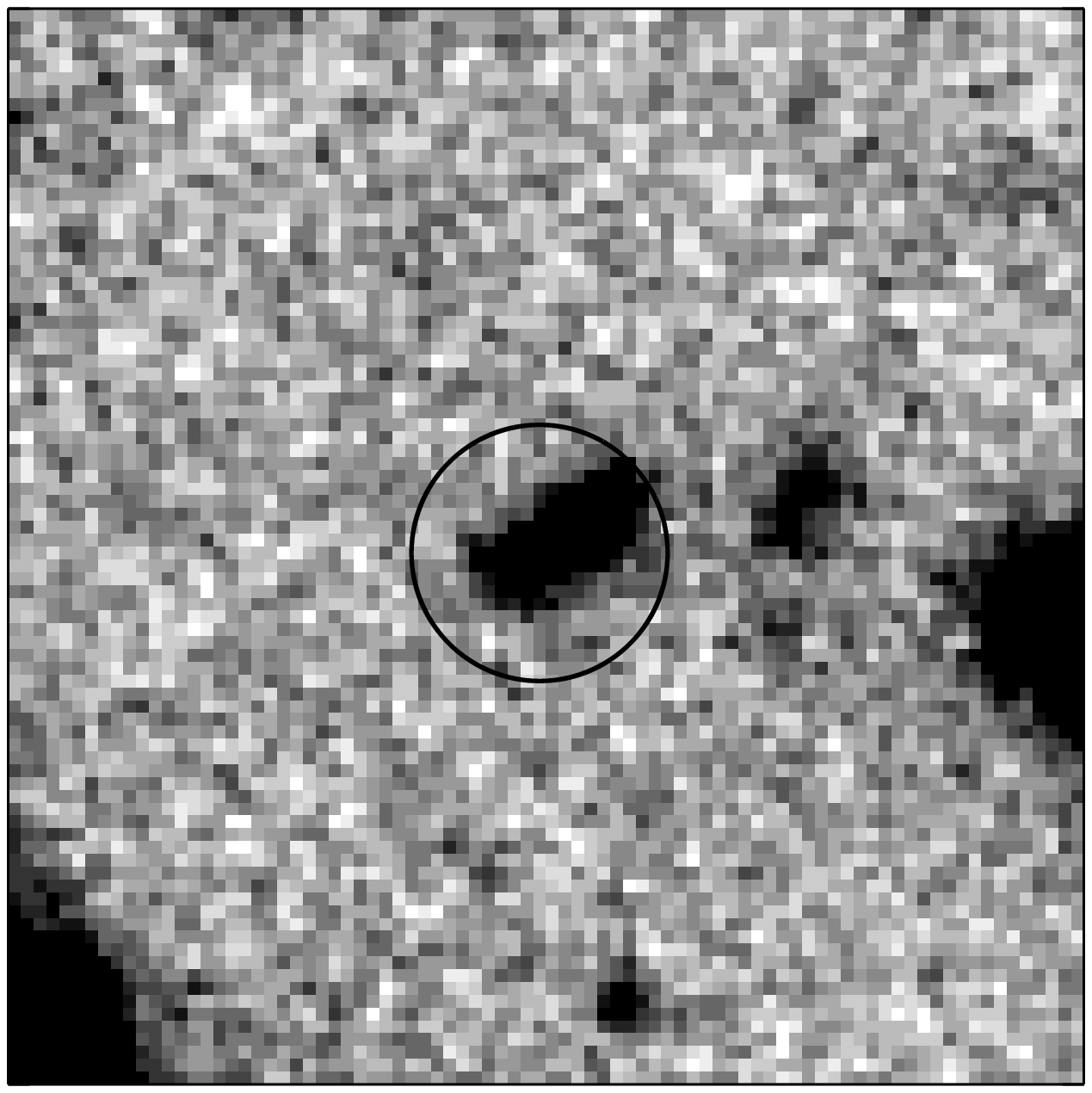}
\hspace{-10mm}
\plotone{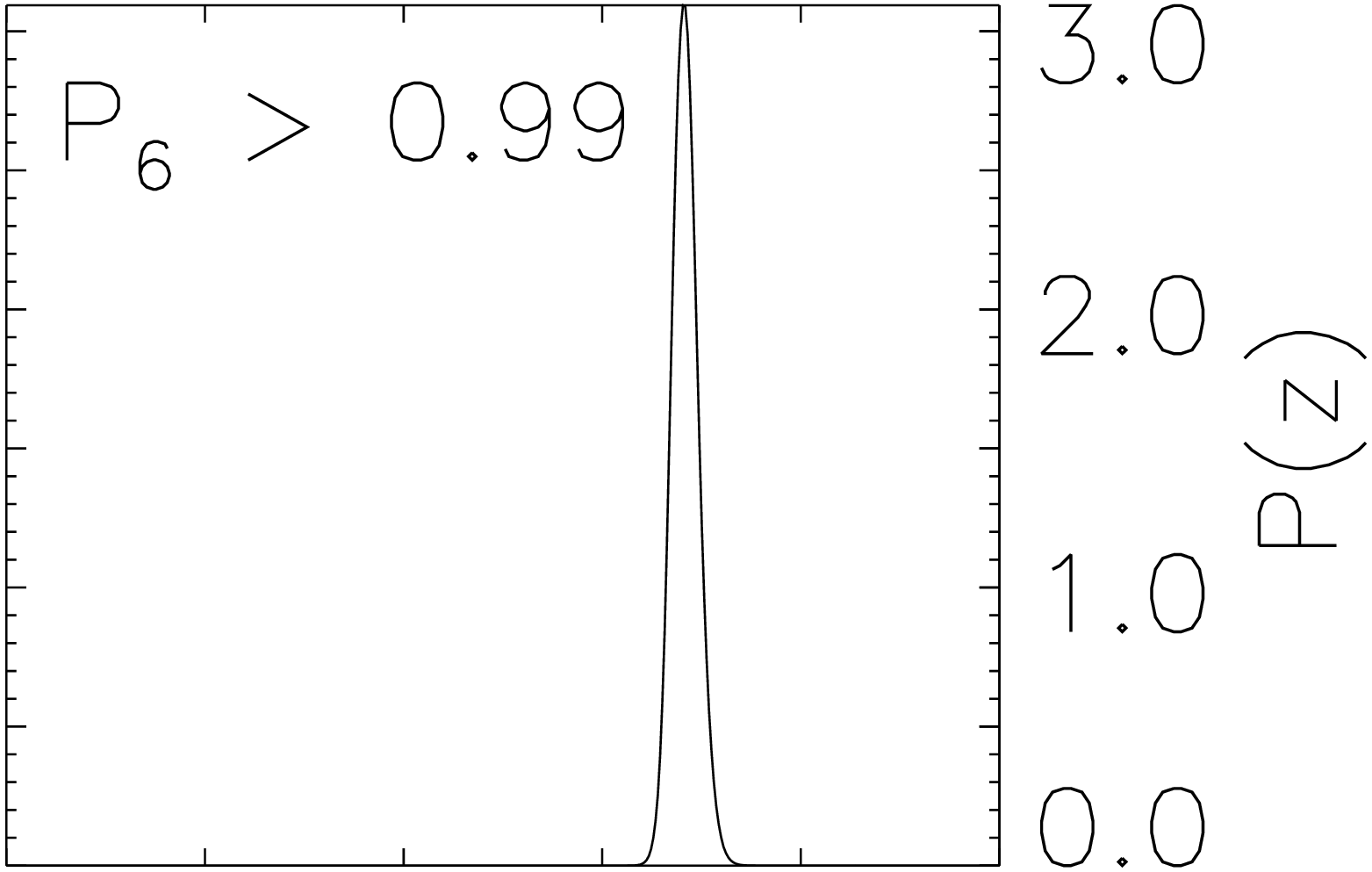}
\vspace{0.5mm}

\plotone{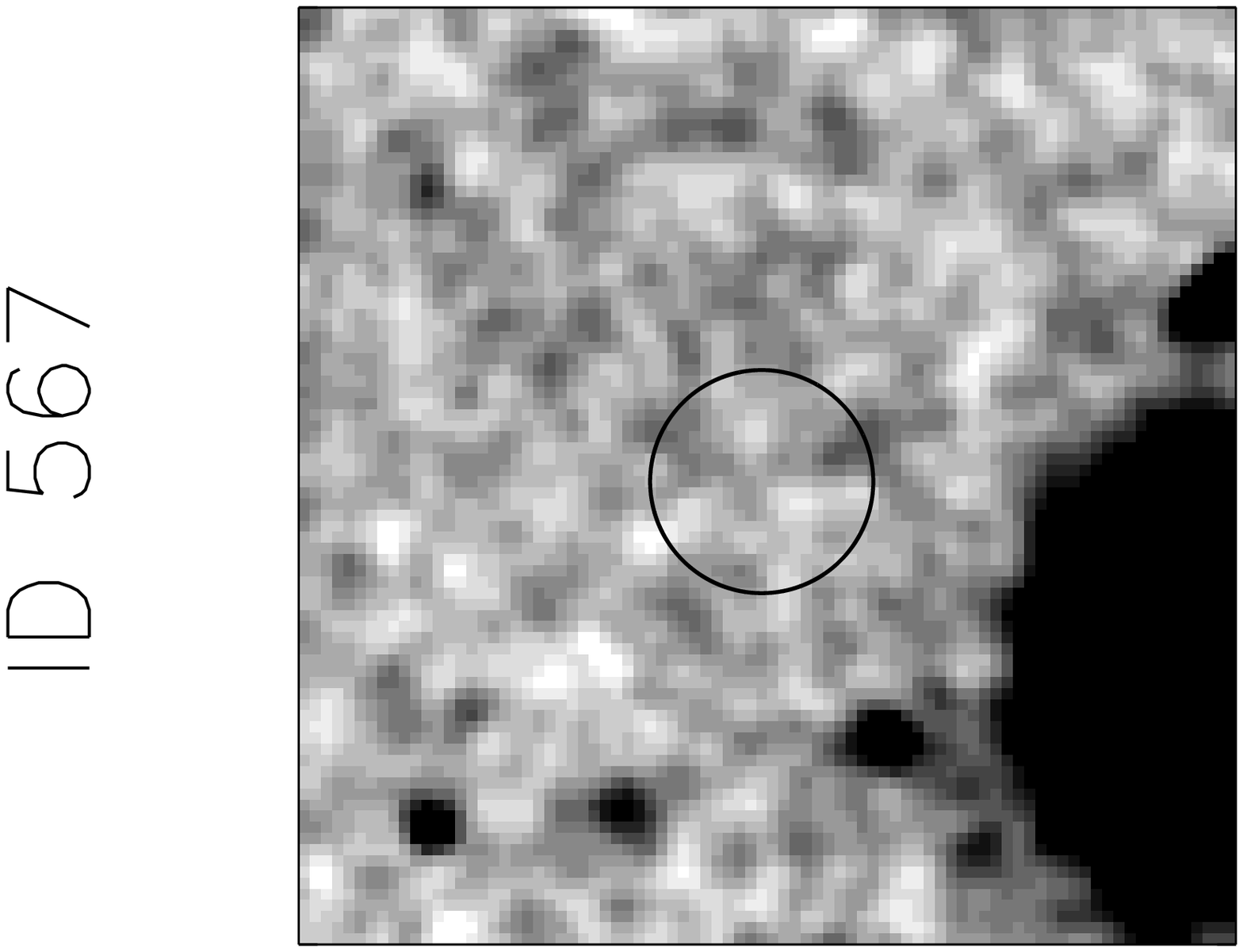}
\hspace{-10mm}
\plotone{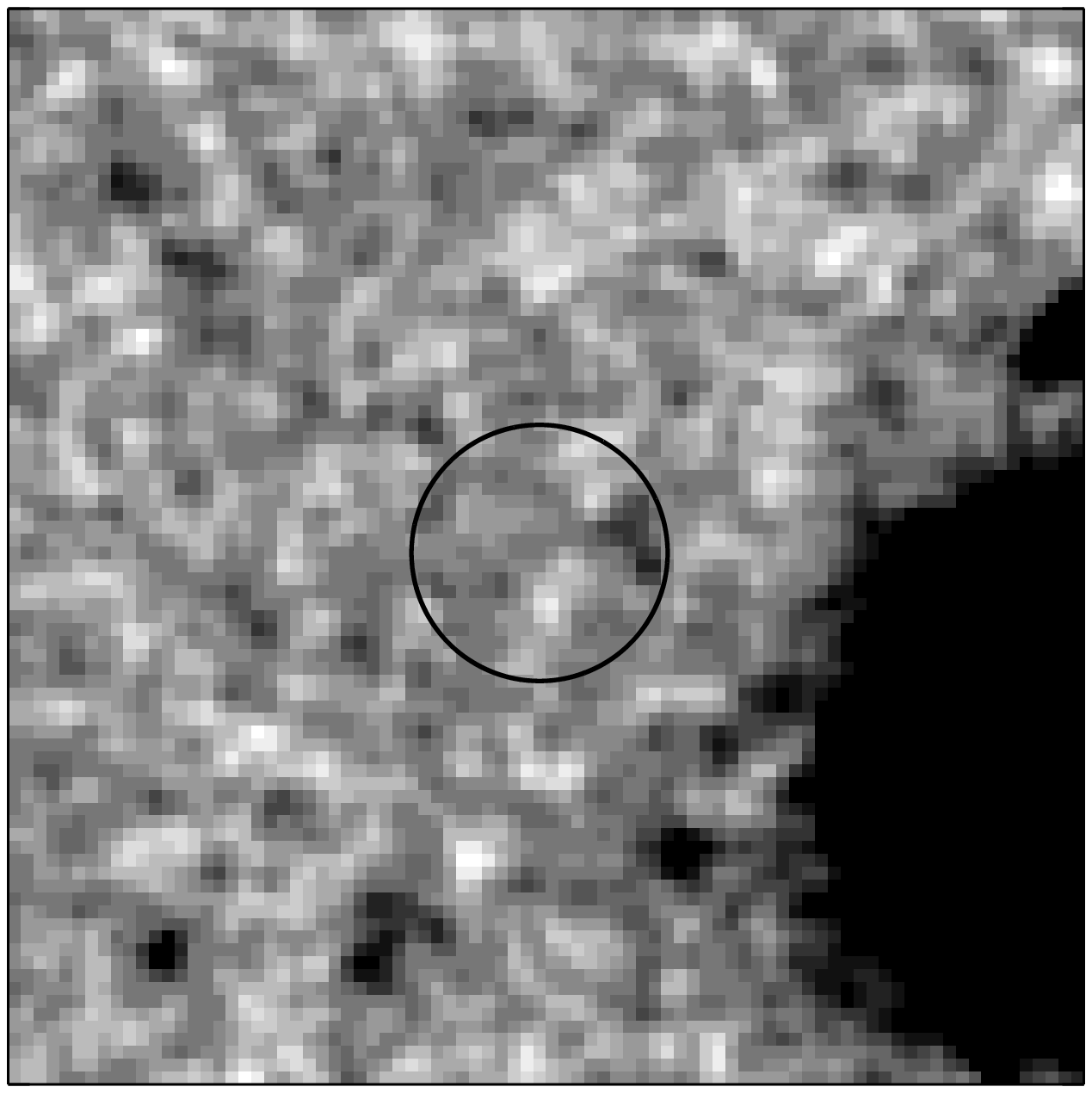}
\hspace{-10mm}
\plotone{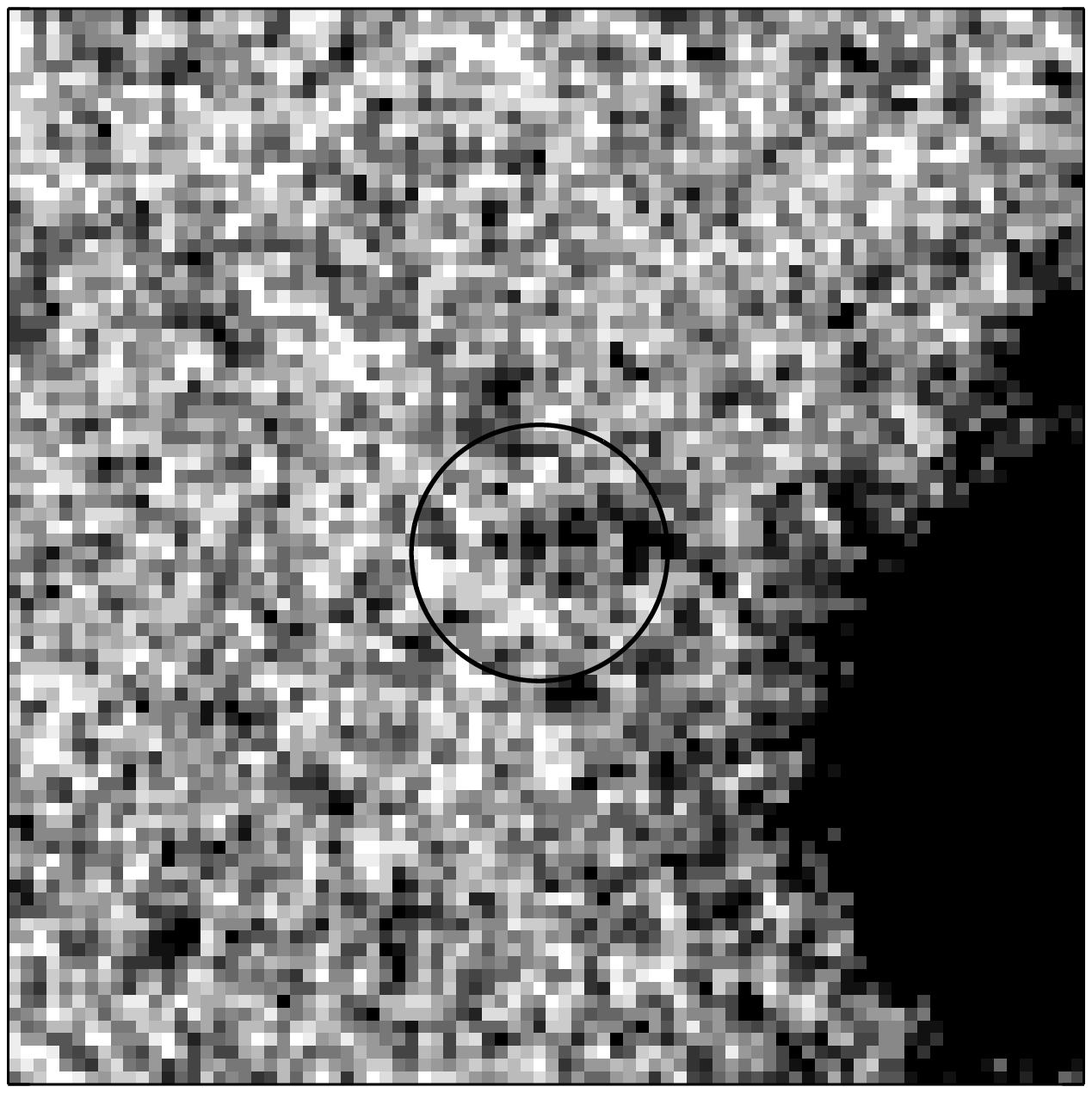}
\hspace{-10mm}
\plotone{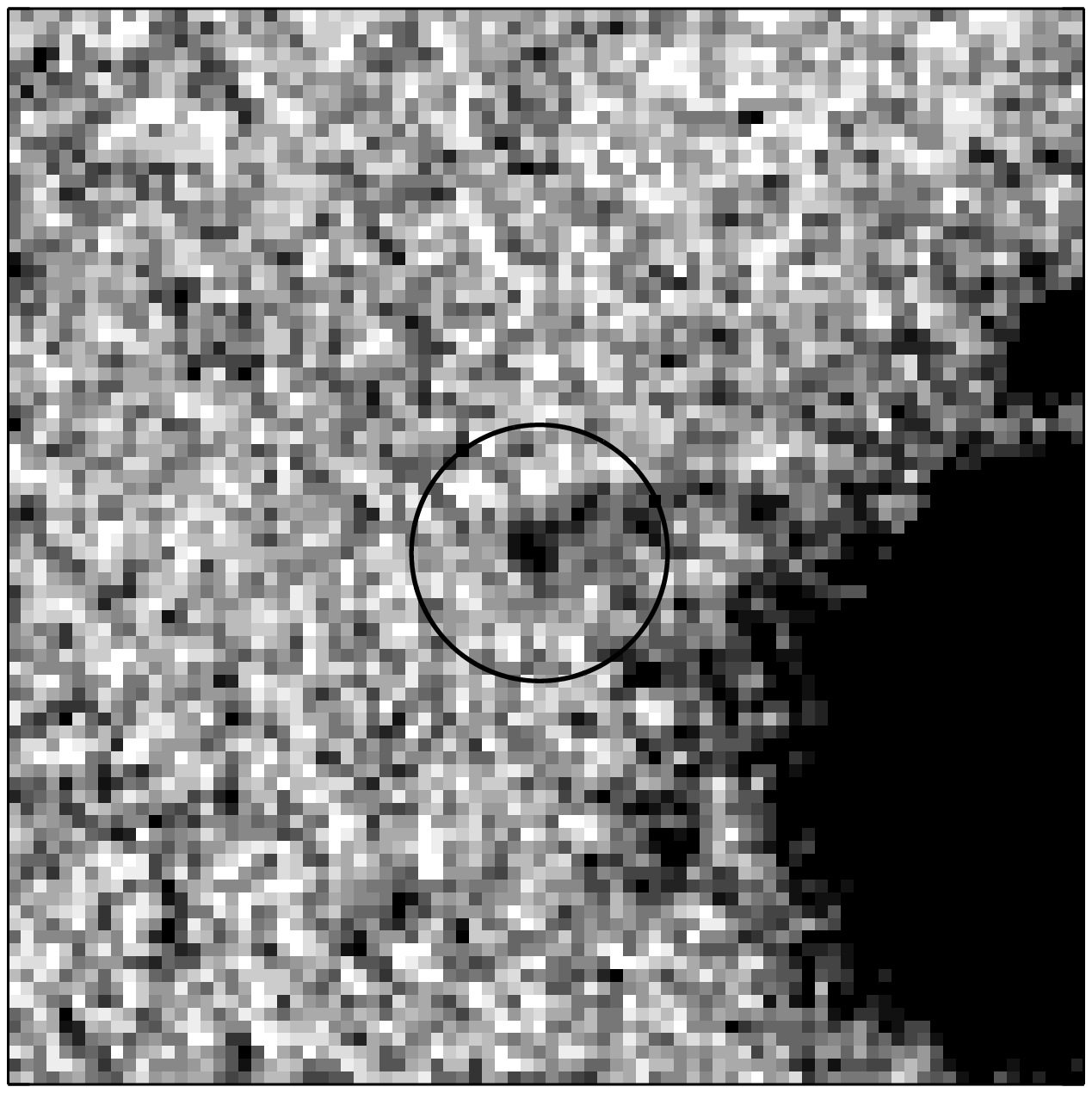}
\hspace{-10mm}
\plotone{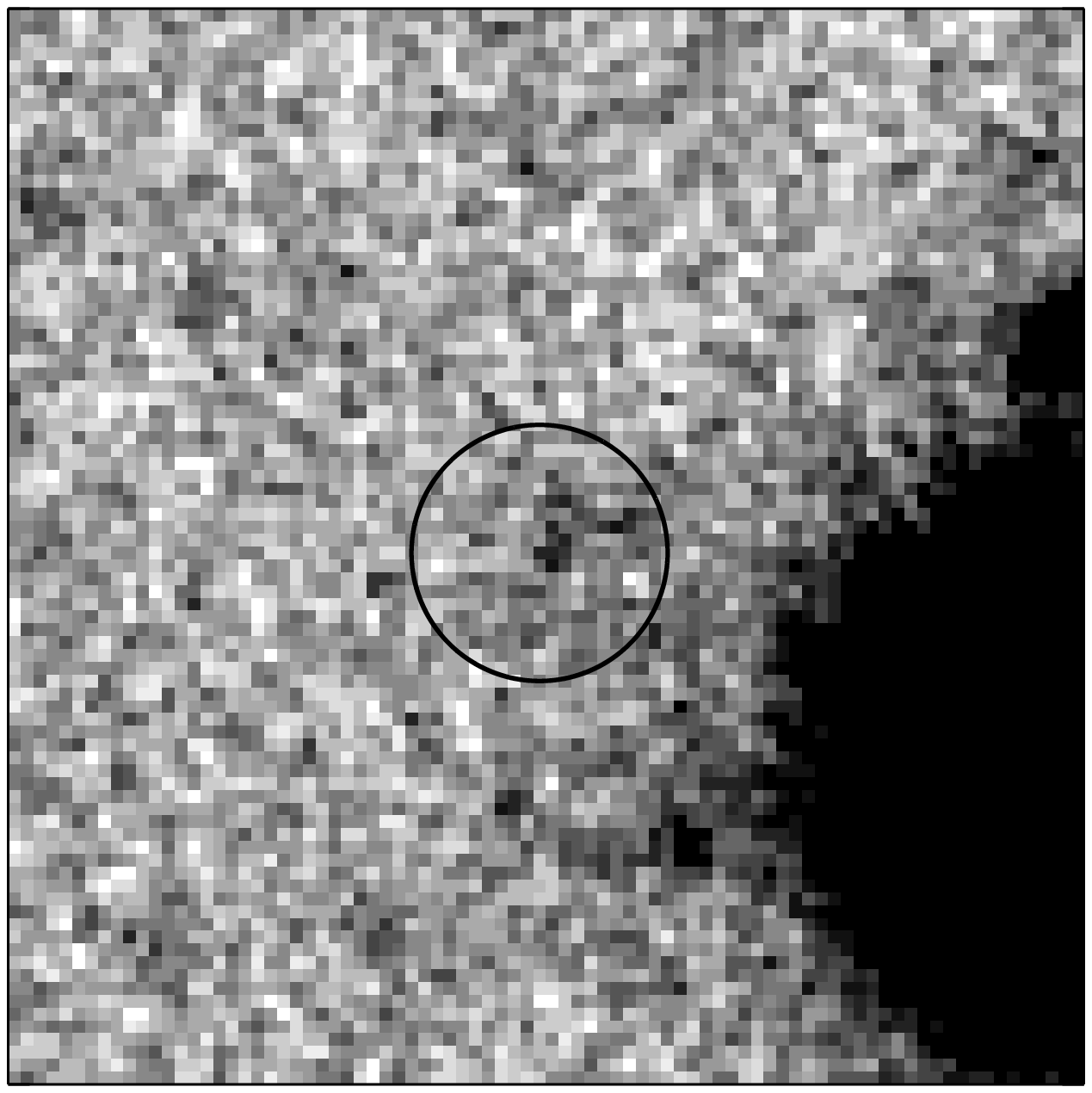}
\hspace{-10mm}
\plotone{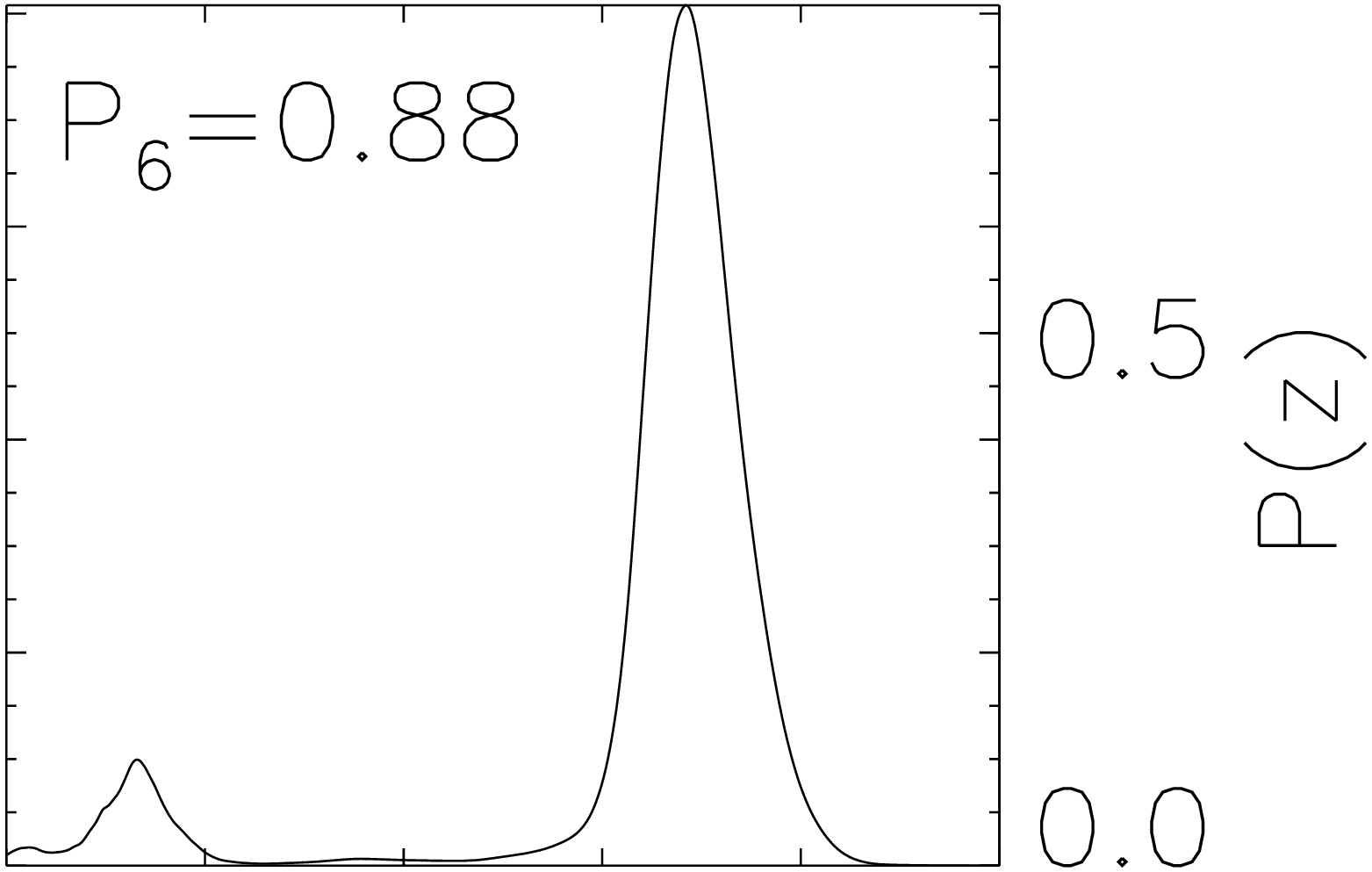}
\vspace{0.5mm}

\plotone{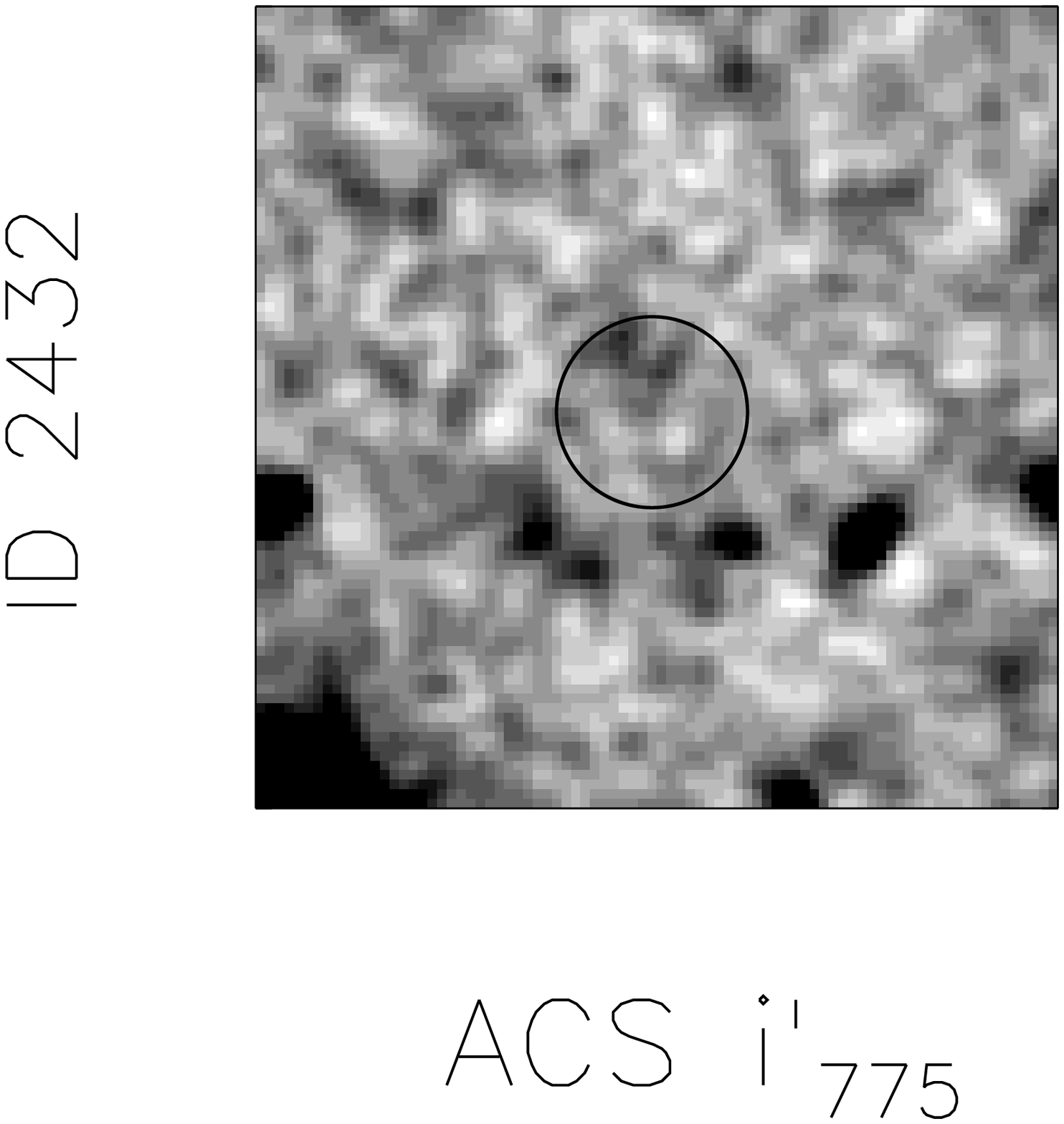}
\hspace{-10mm}
\plotone{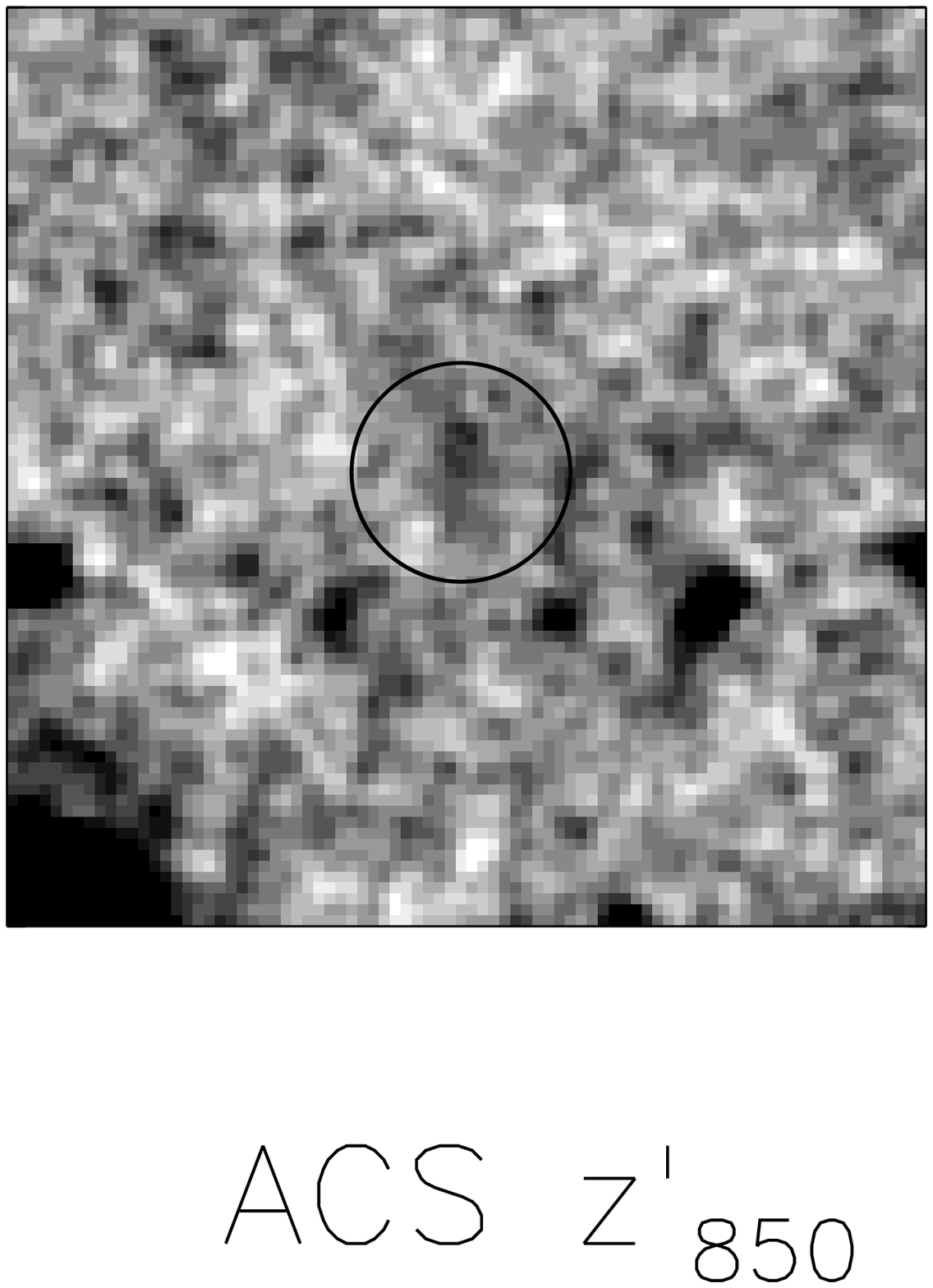}
\hspace{-10mm}
\plotone{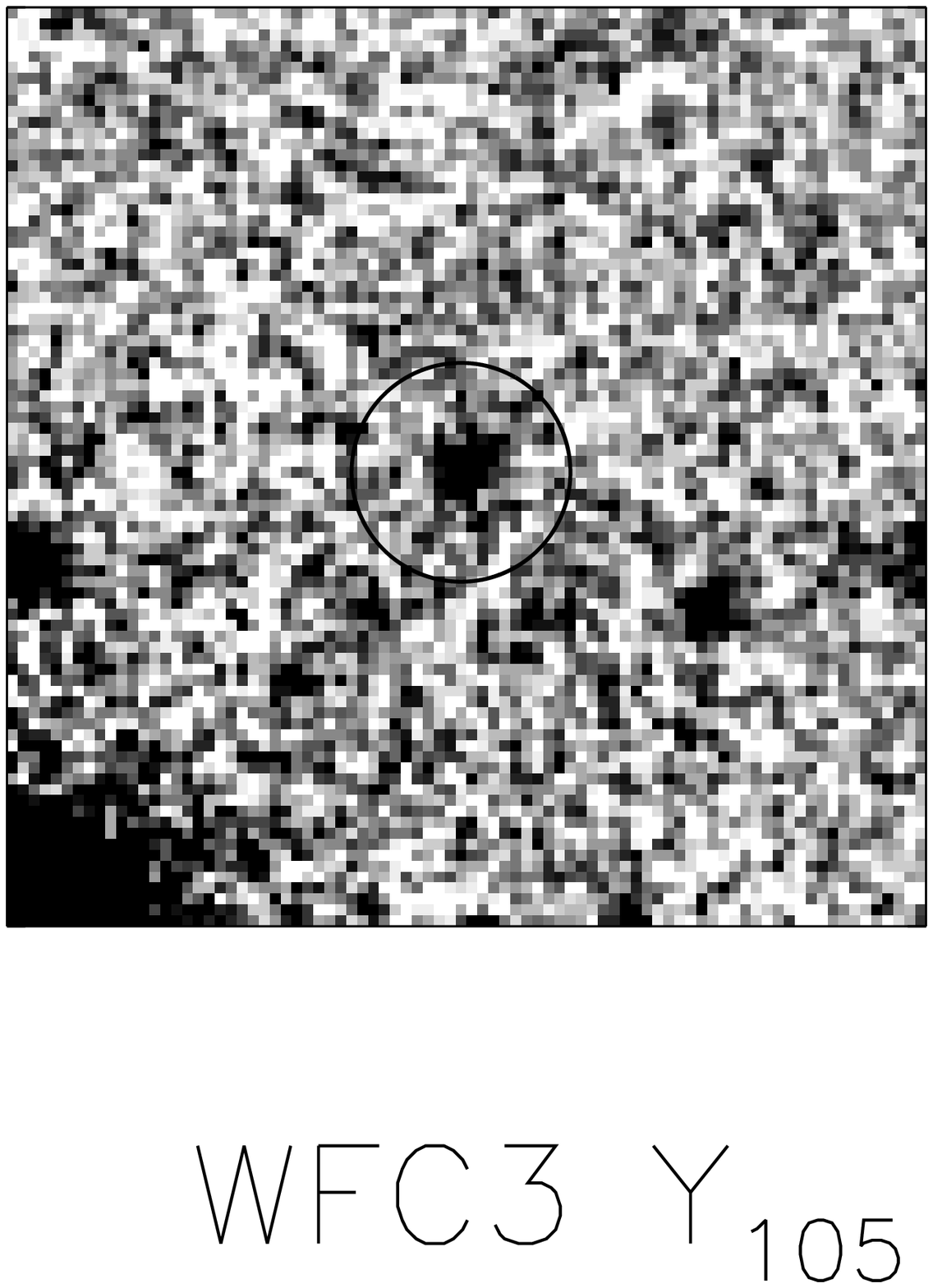}
\hspace{-10mm}
\plotone{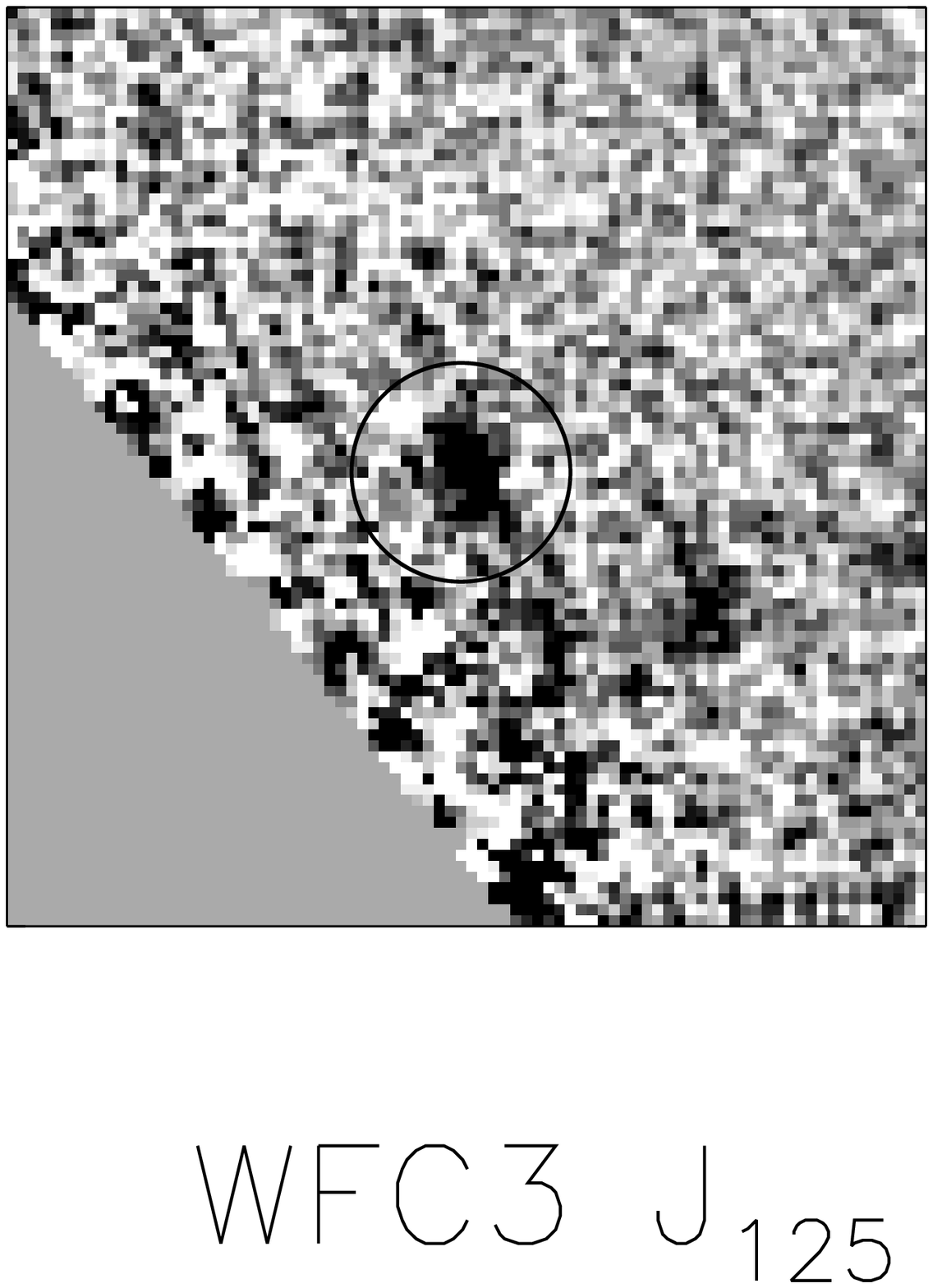}
\hspace{-10mm}
\plotone{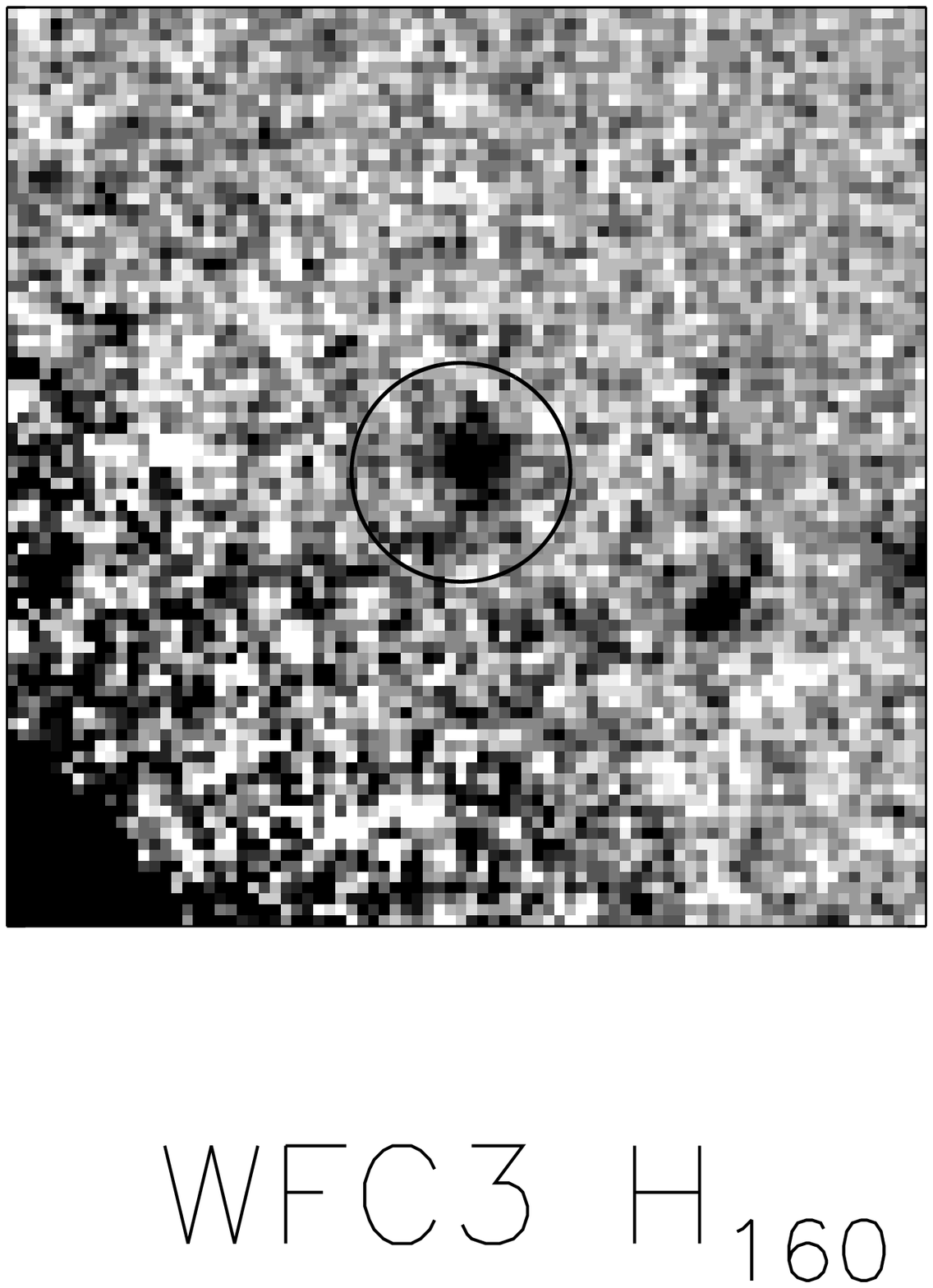}
\hspace{-10mm}
\plotone{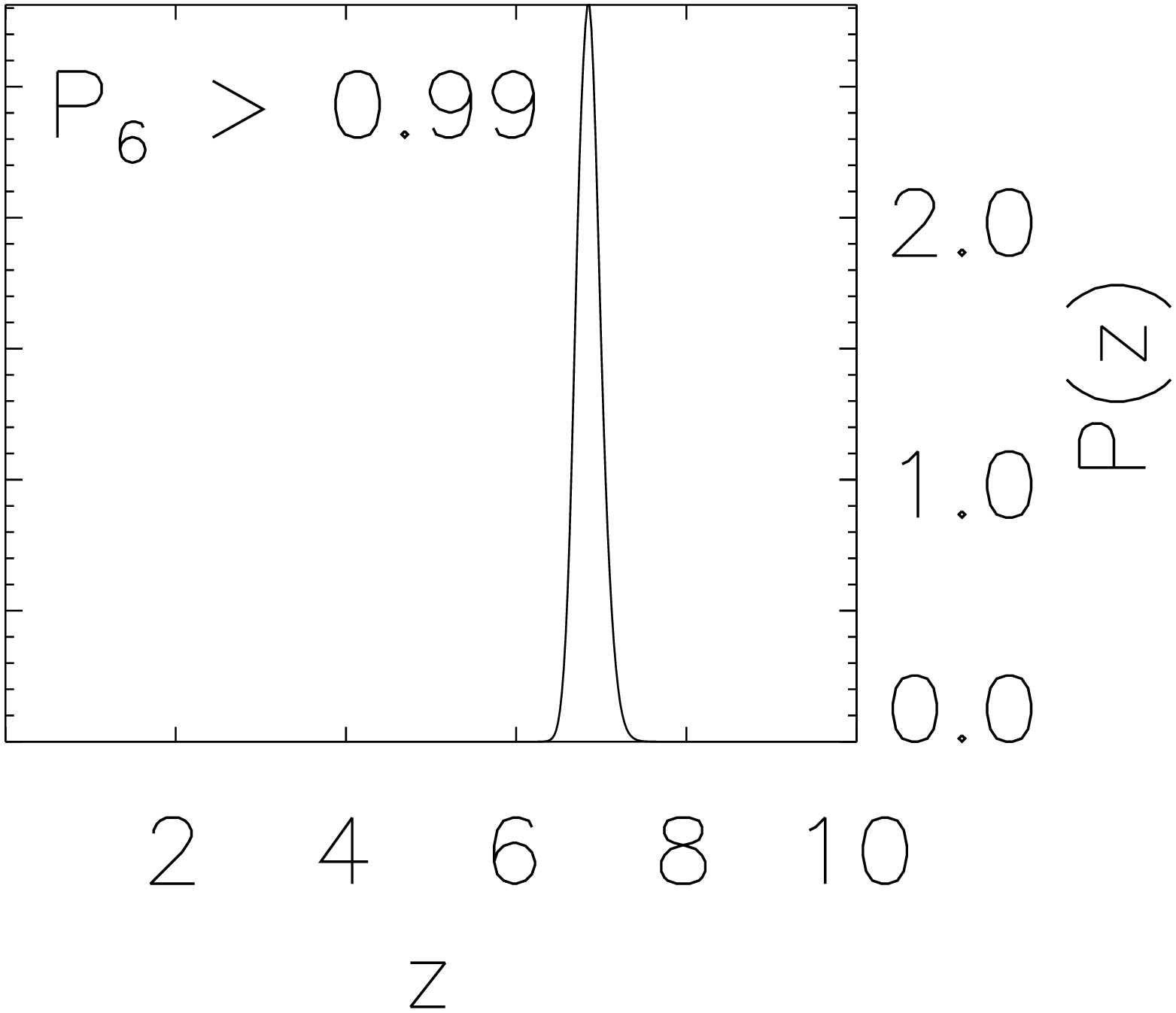}
\vspace{5mm}
\caption{Continued.}\label{stamps2}
\end{figure*}
\addtocounter{figure}{-1}

\begin{figure*}
\epsscale{0.18}
\vspace{2mm}
\plotone{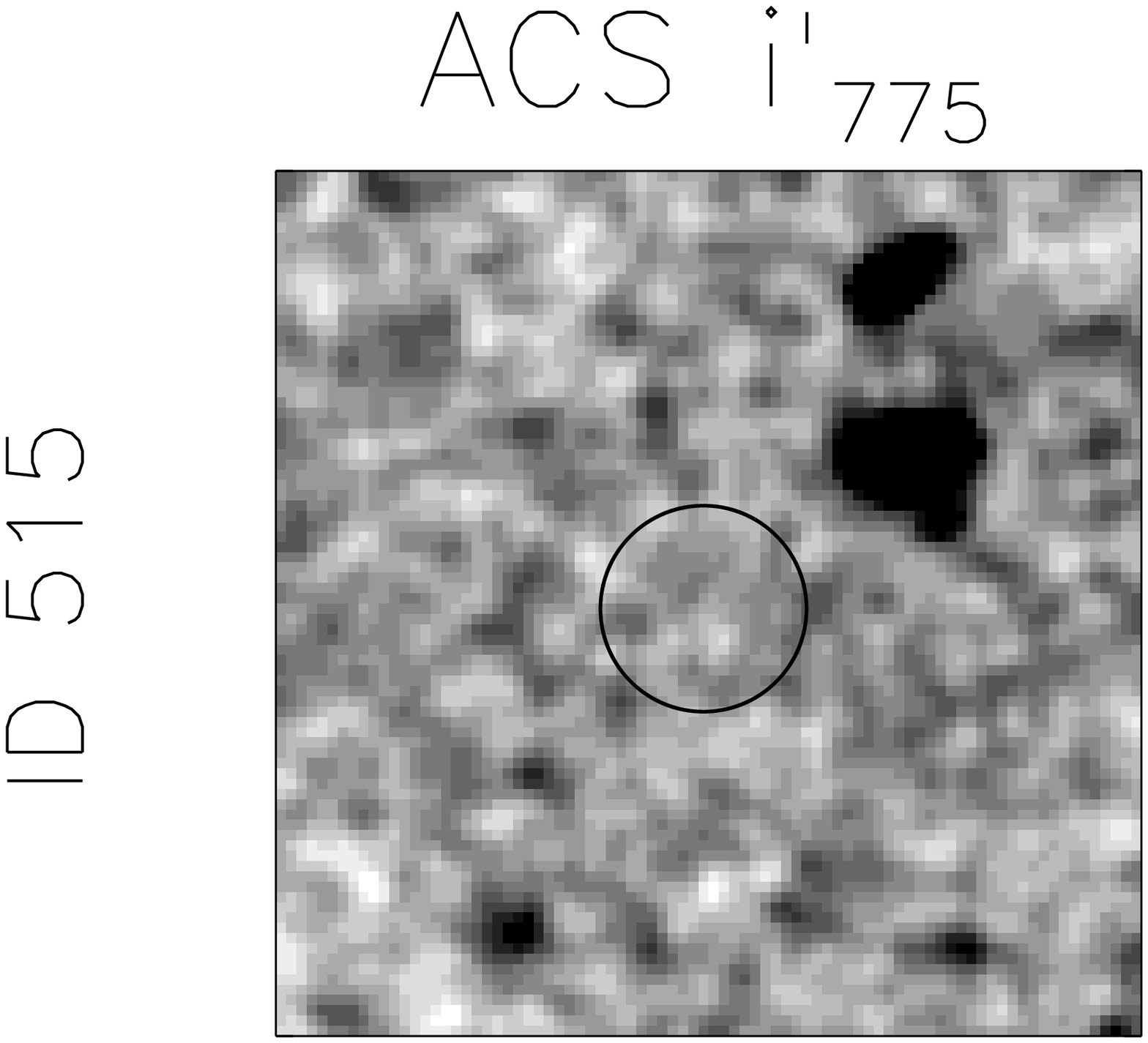}
\hspace{-10mm}
\plotone{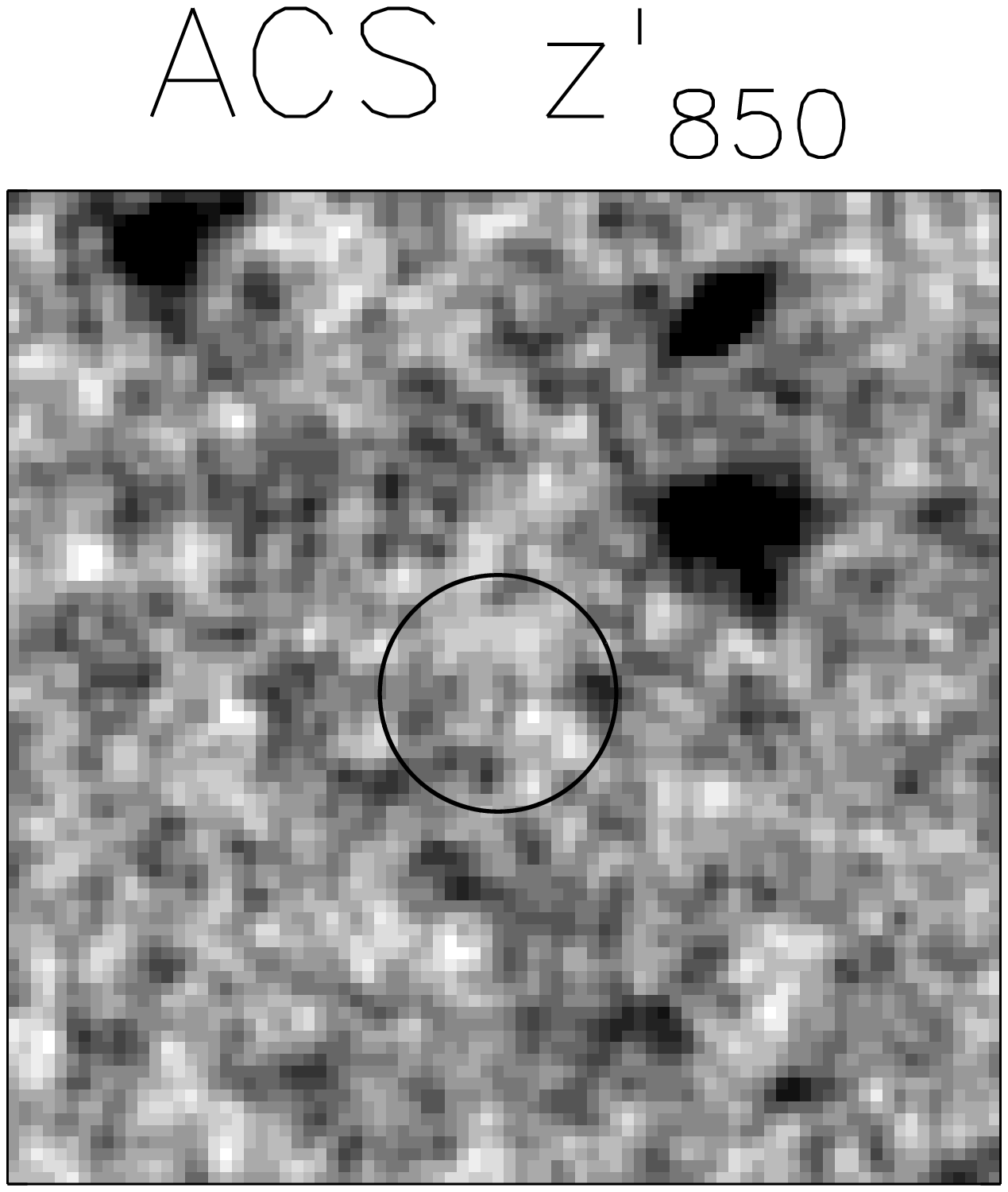}
\hspace{-10mm}
\plotone{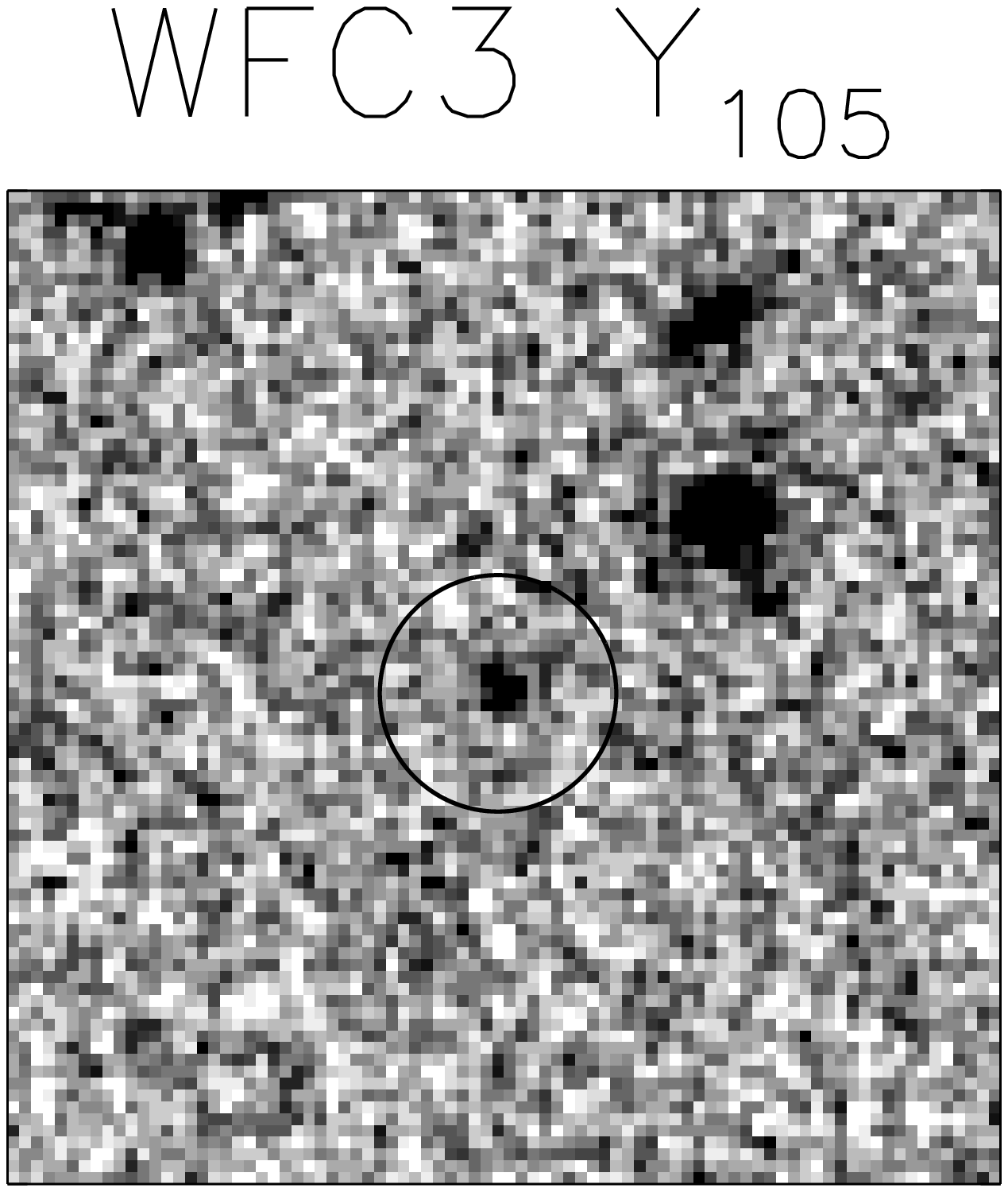}
\hspace{-10mm}
\plotone{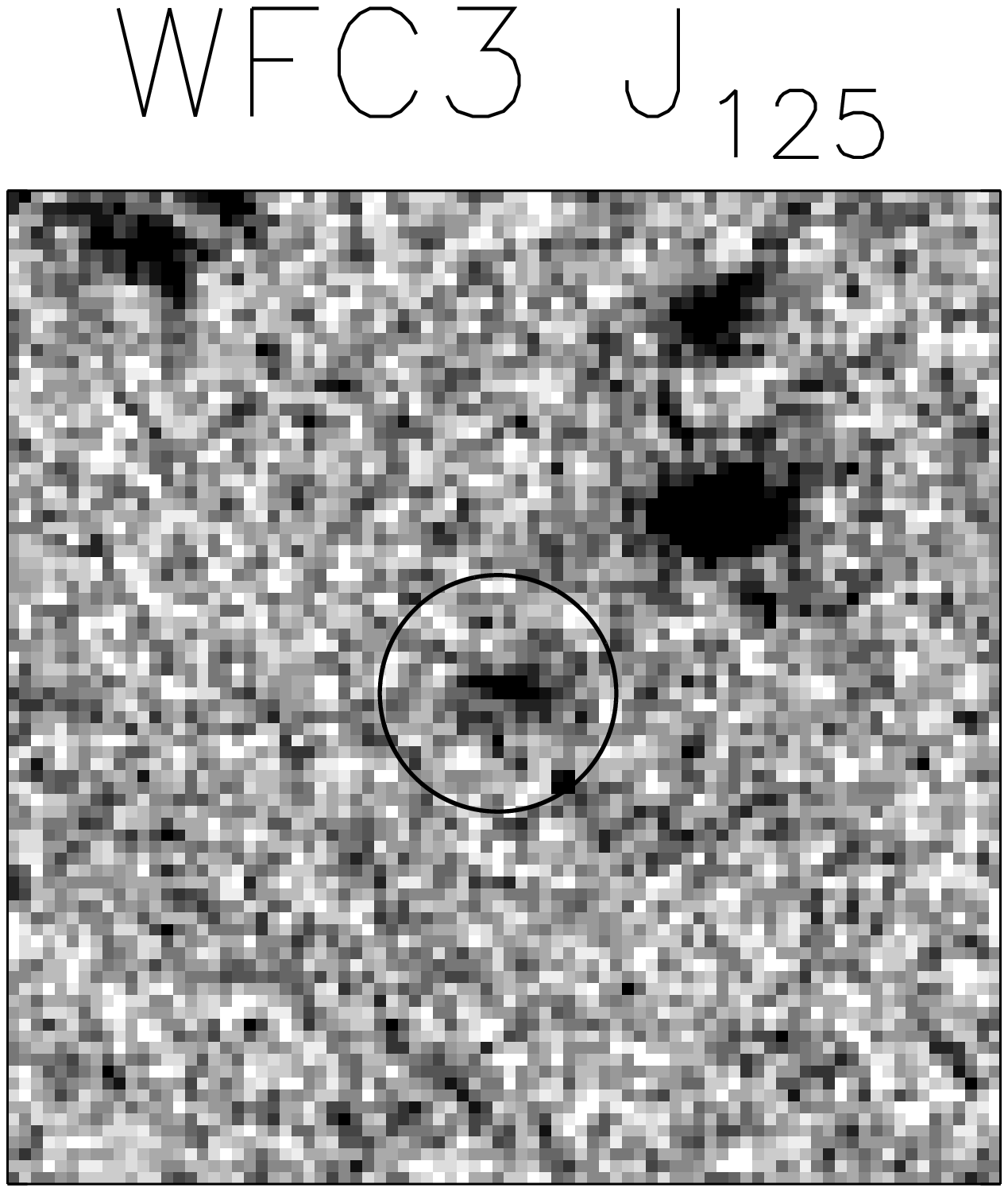}
\hspace{-10mm}
\plotone{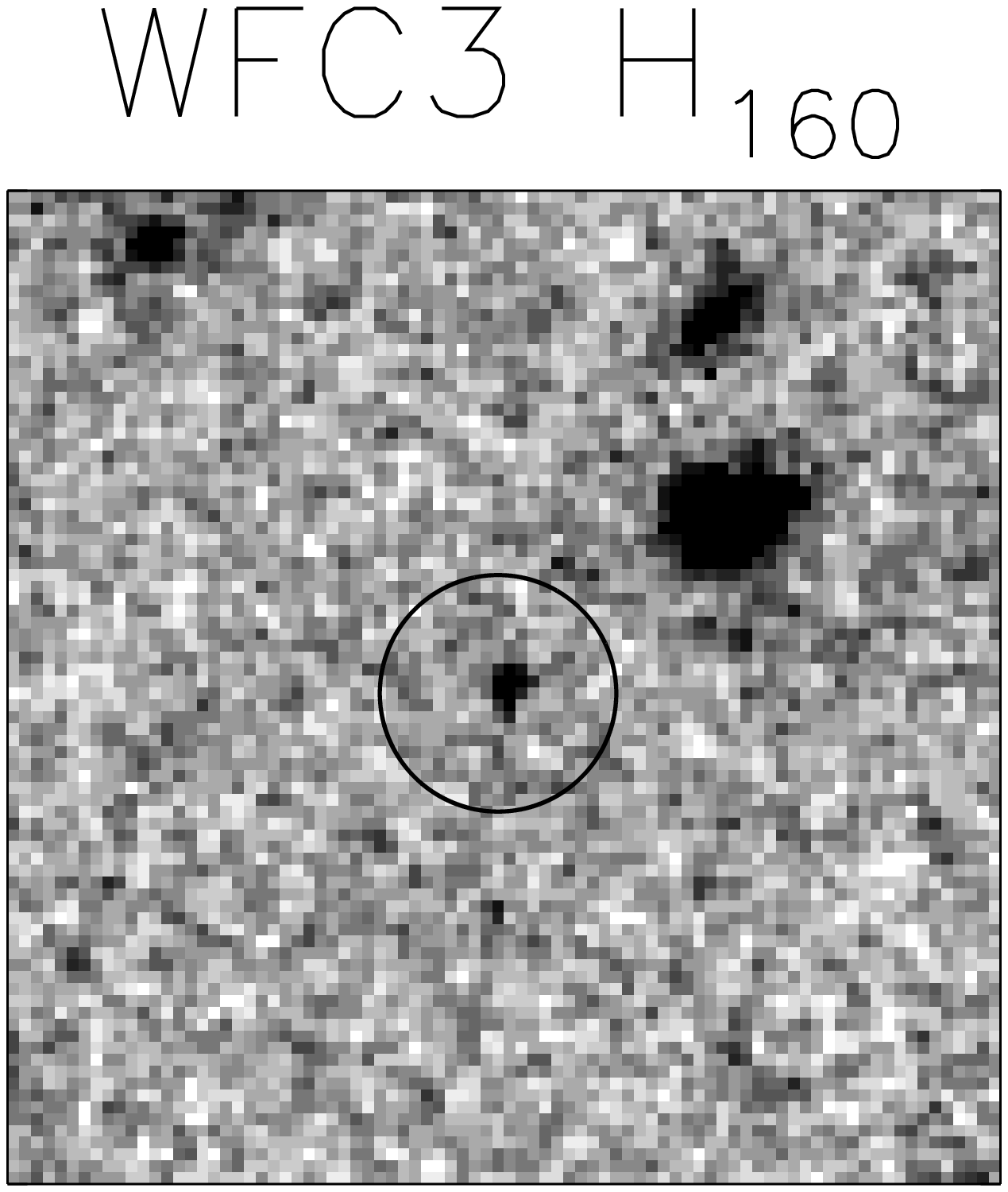}
\hspace{-10mm}
\plotone{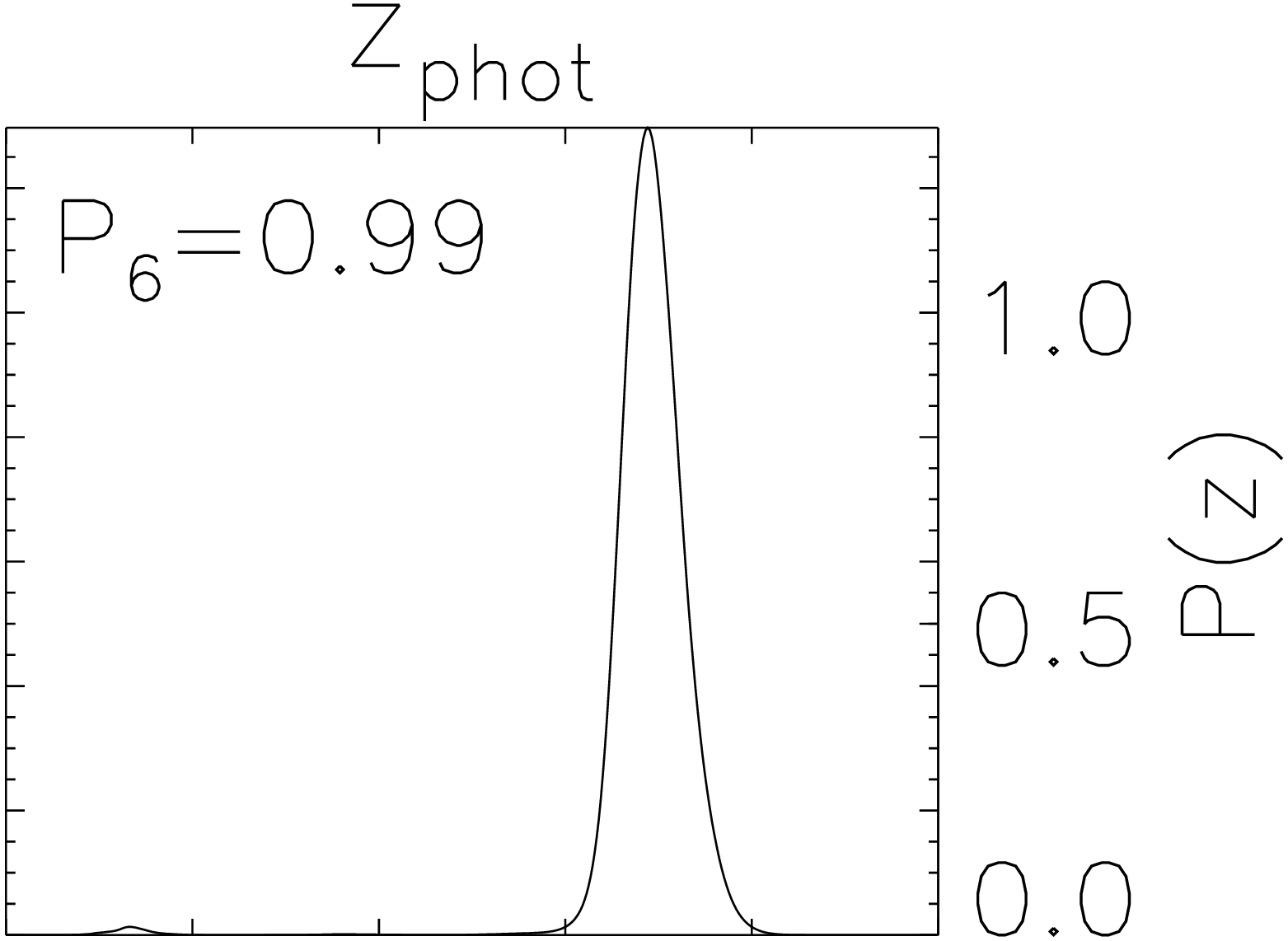}
\vspace{0.5mm}

\plotone{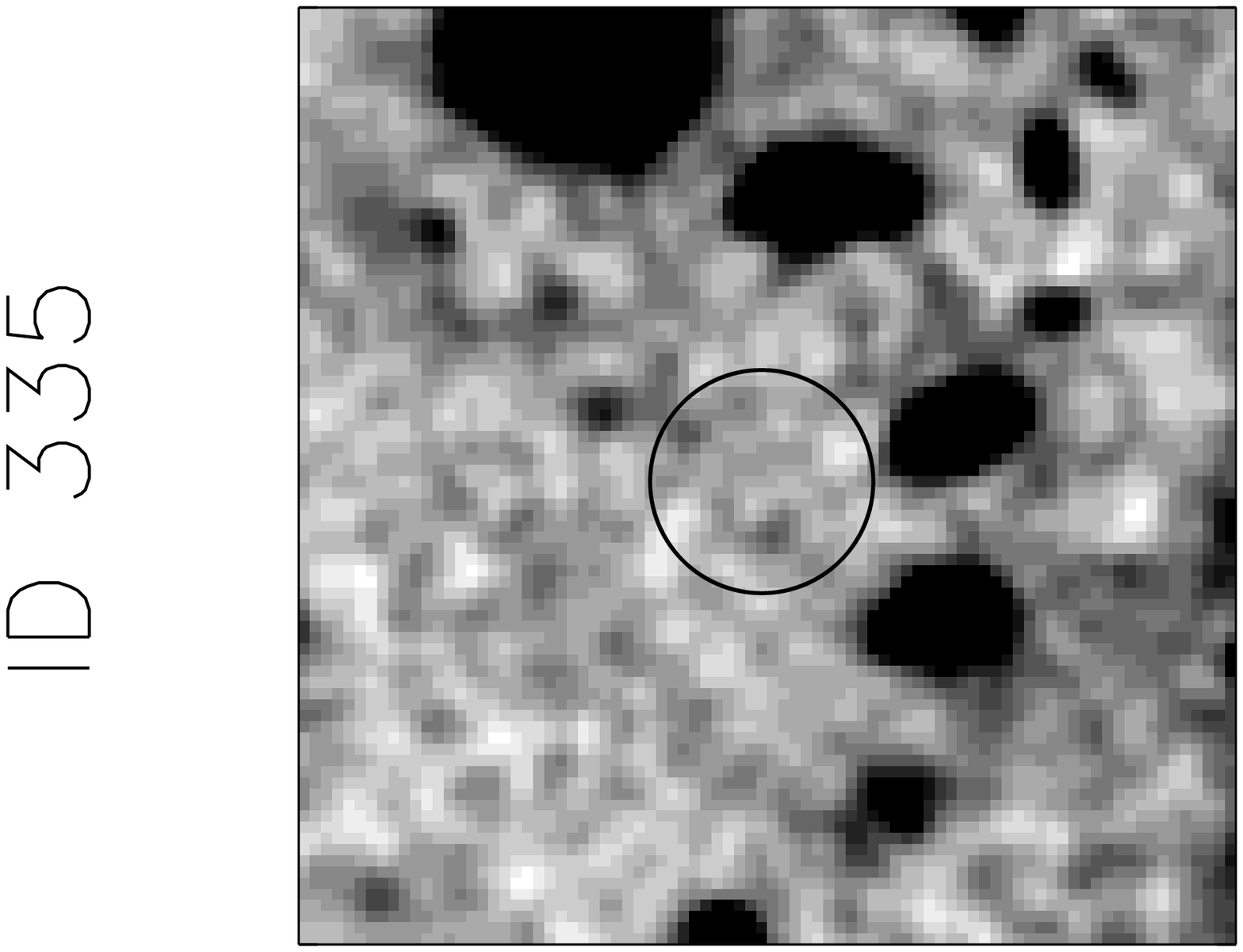}
\hspace{-10mm}
\plotone{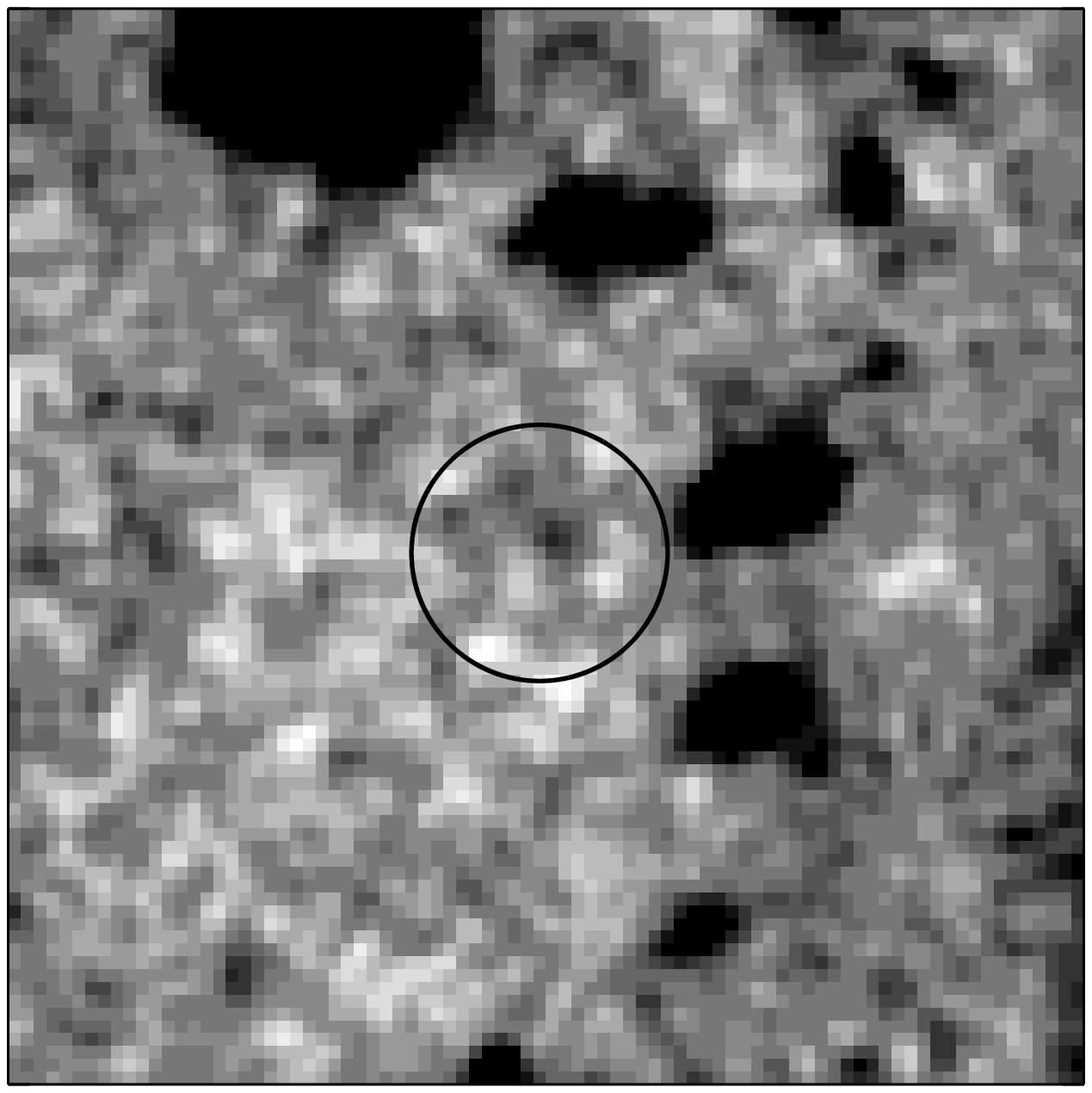}
\hspace{-10mm}
\plotone{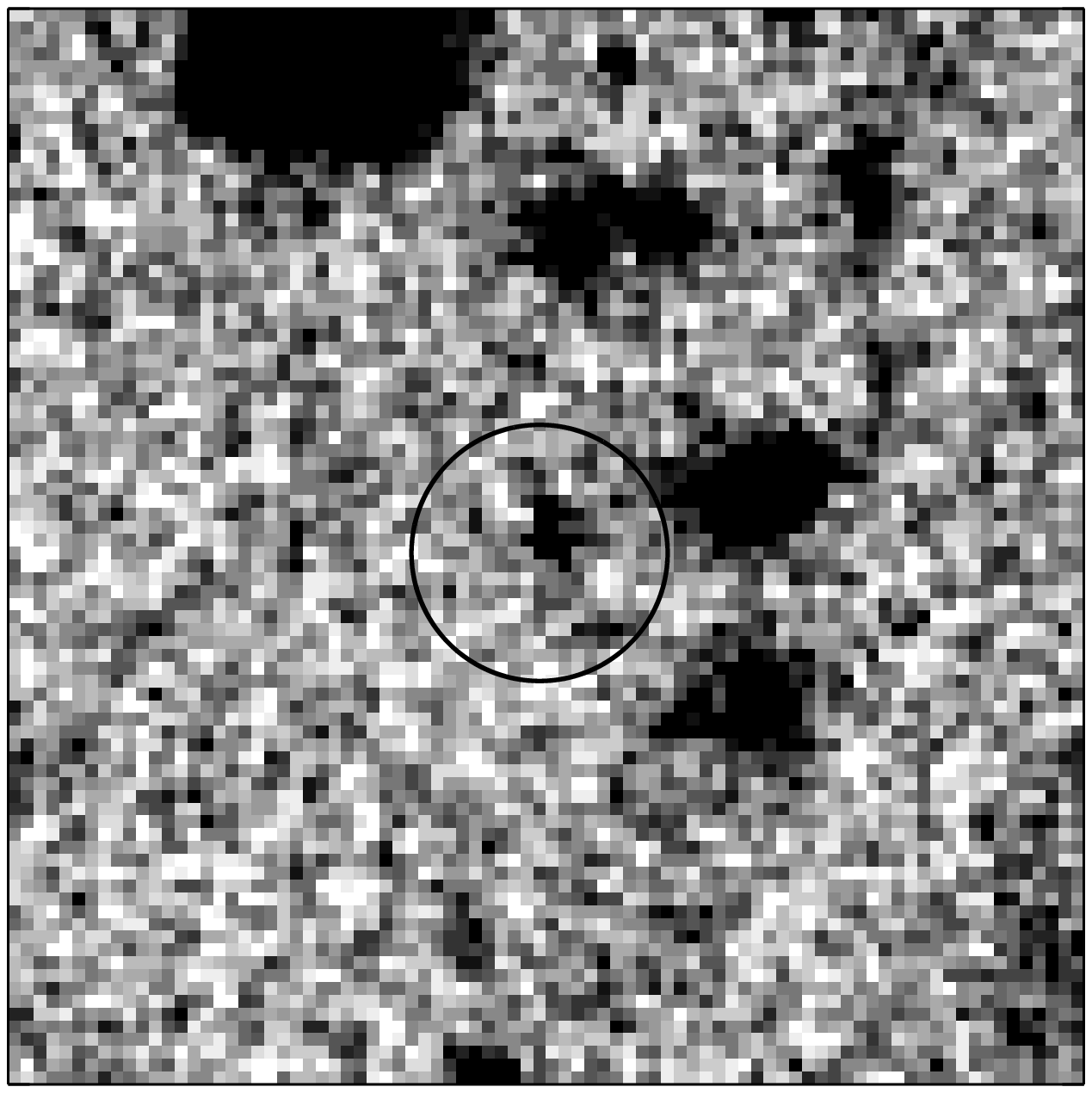}
\hspace{-10mm}
\plotone{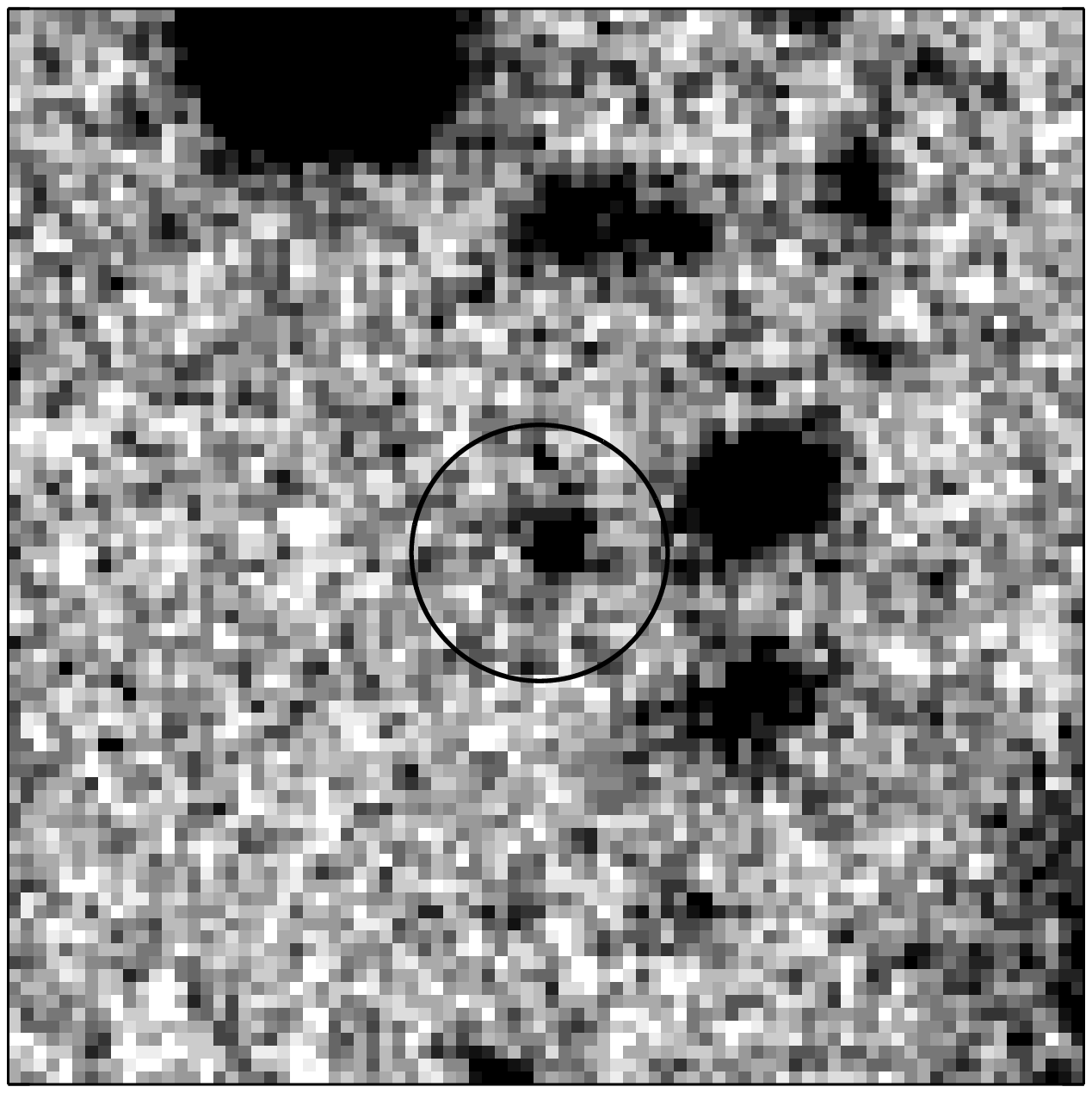}
\hspace{-10mm}
\plotone{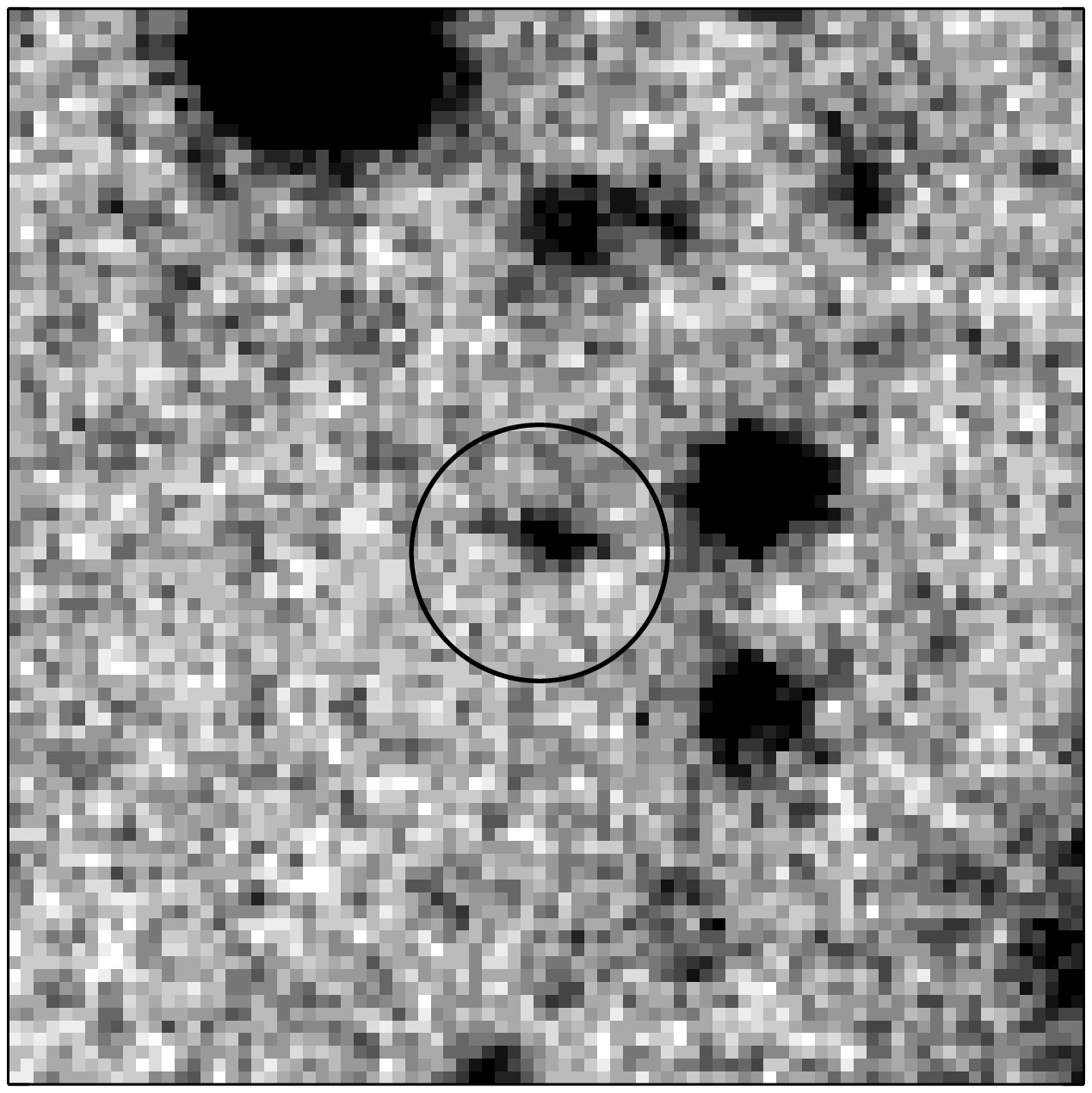}
\hspace{-10mm}
\plotone{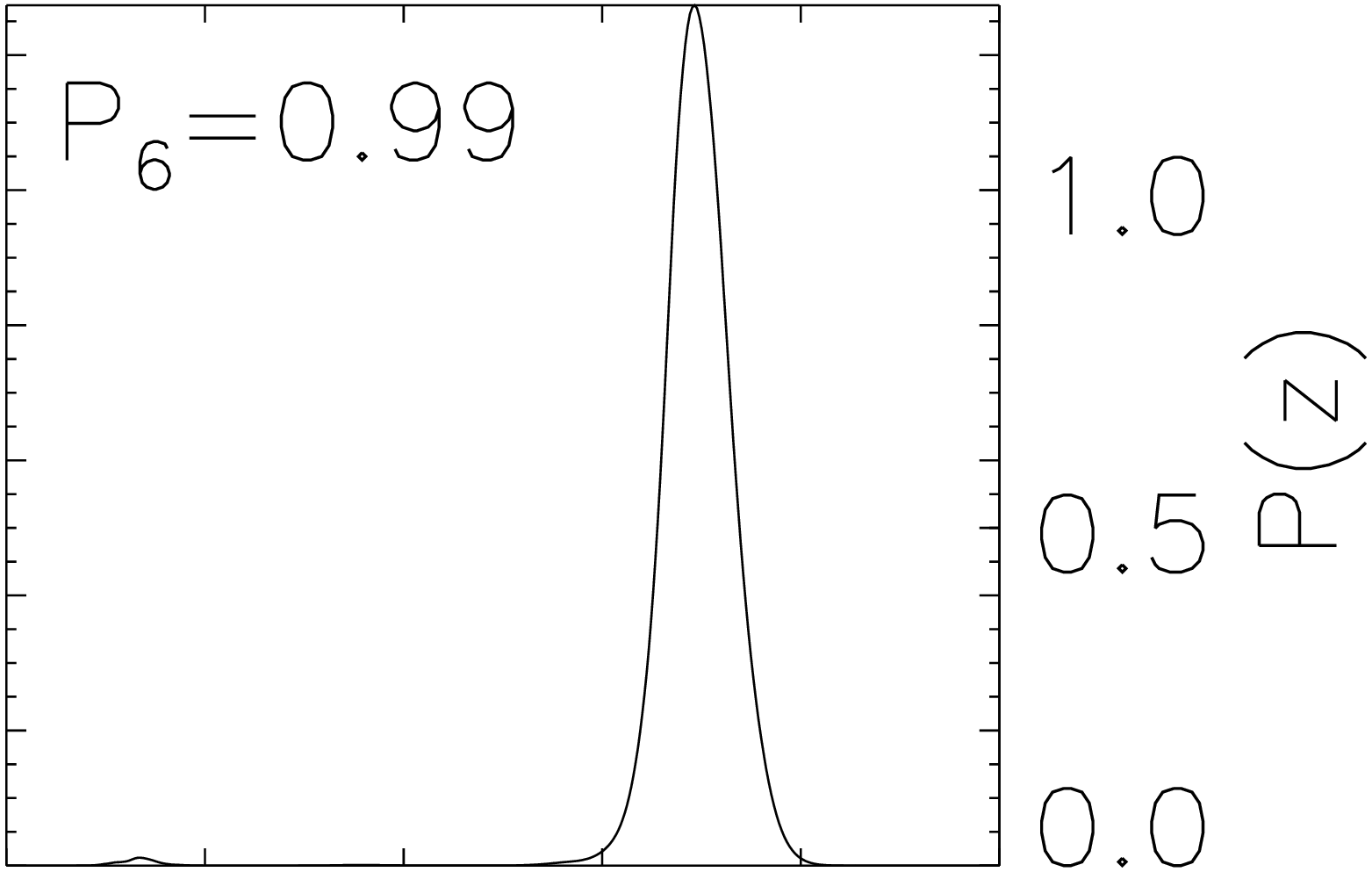}
\vspace{0.5mm}

\plotone{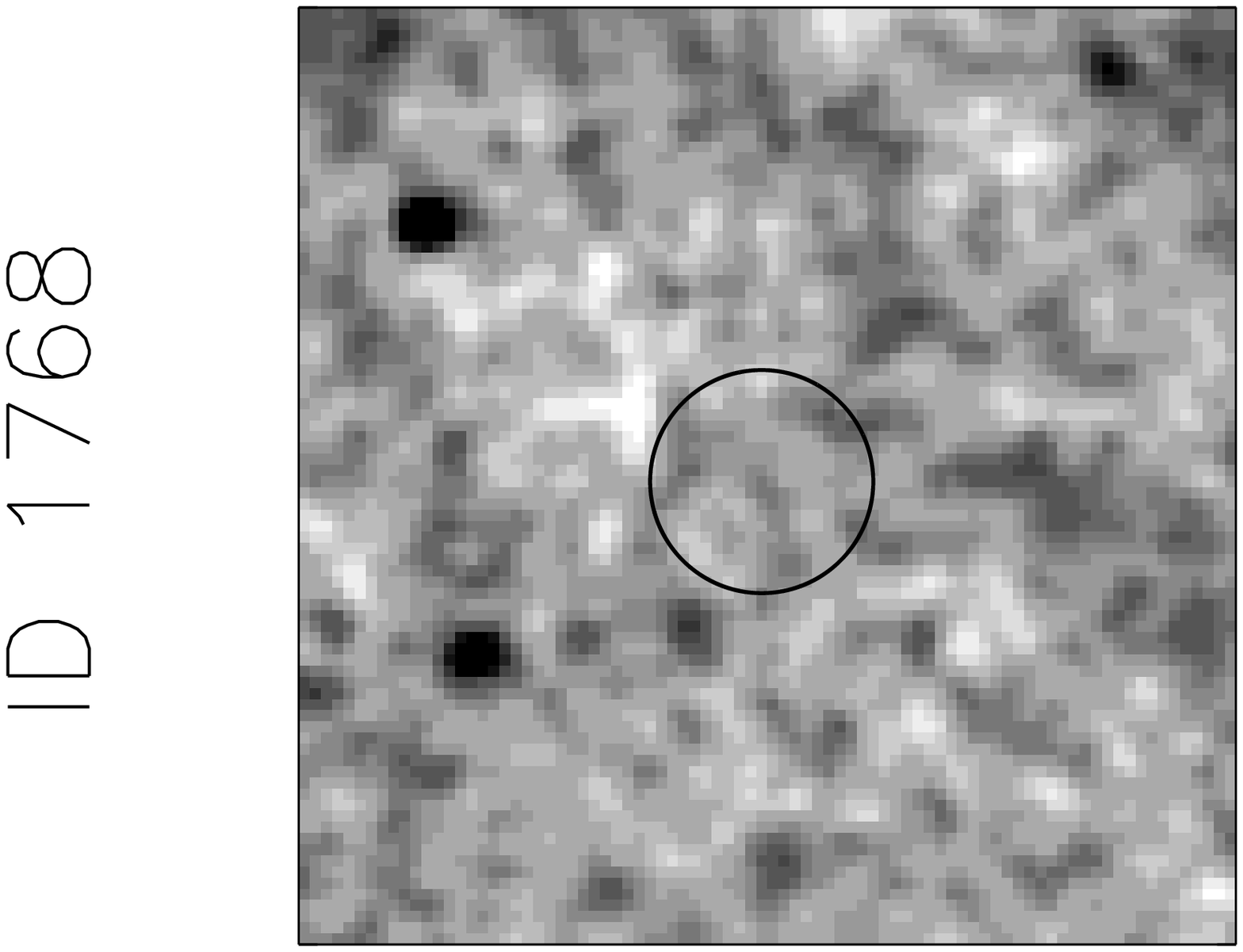}
\hspace{-10mm}
\plotone{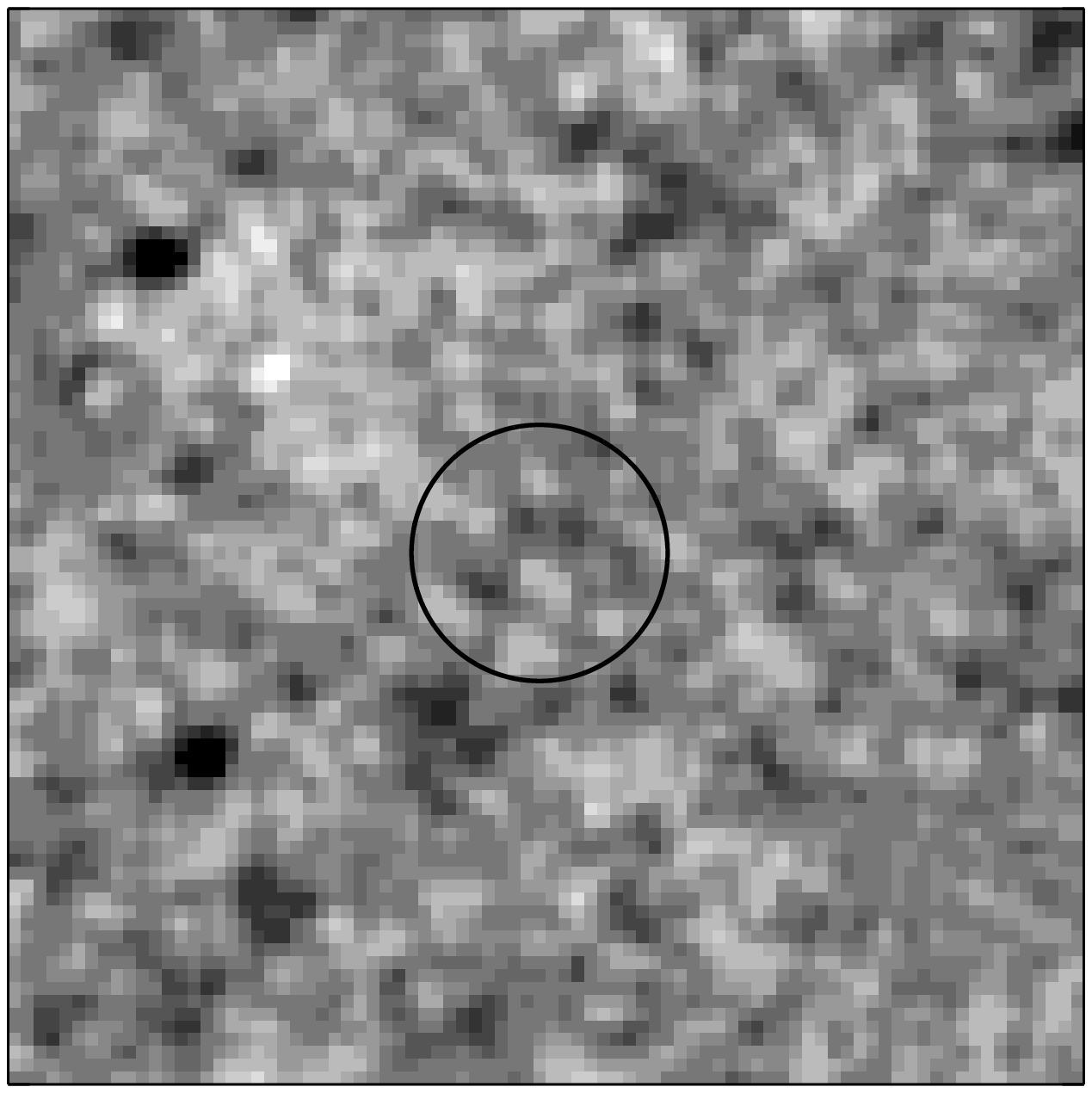}
\hspace{-10mm}
\plotone{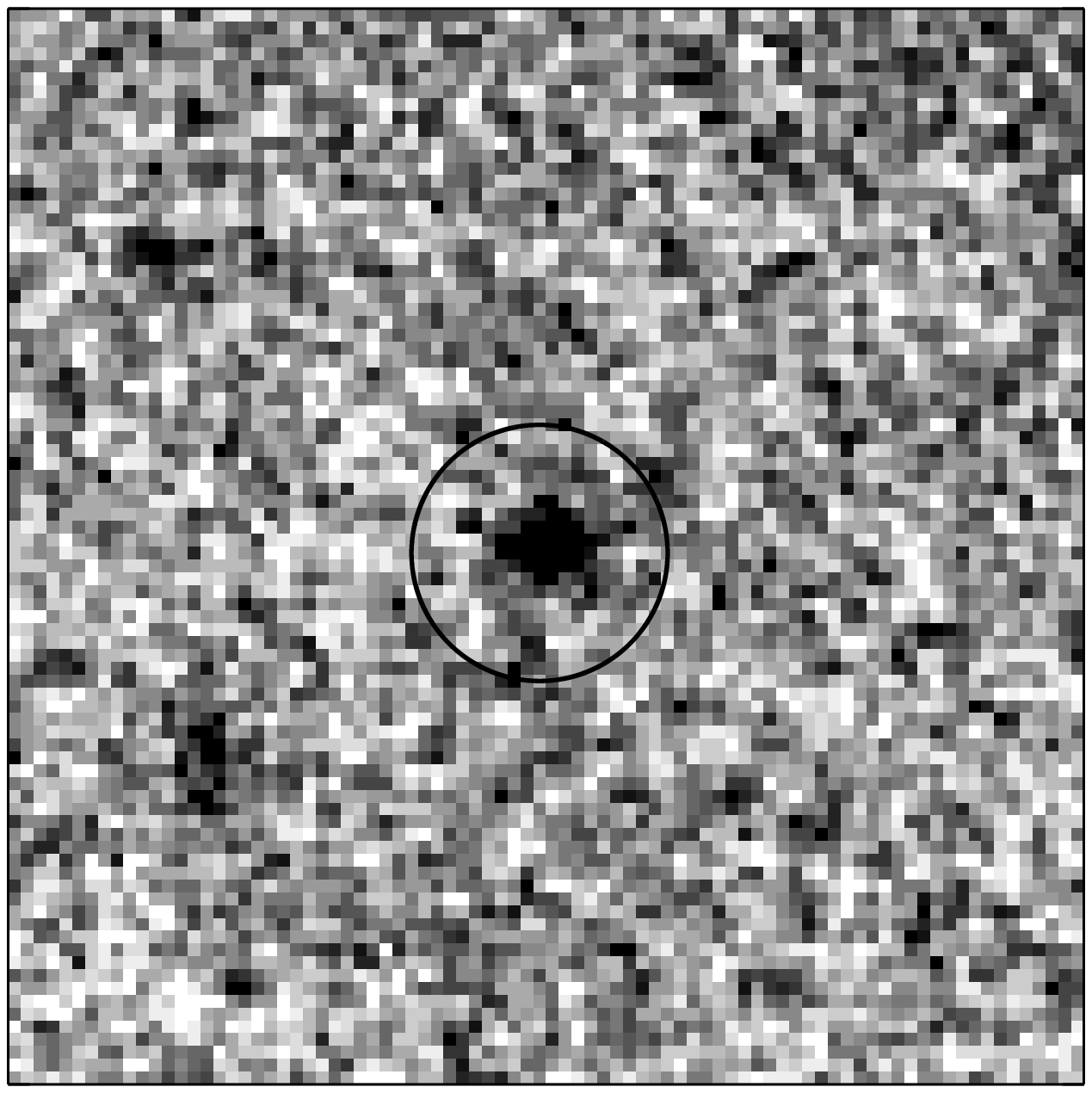}
\hspace{-10mm}
\plotone{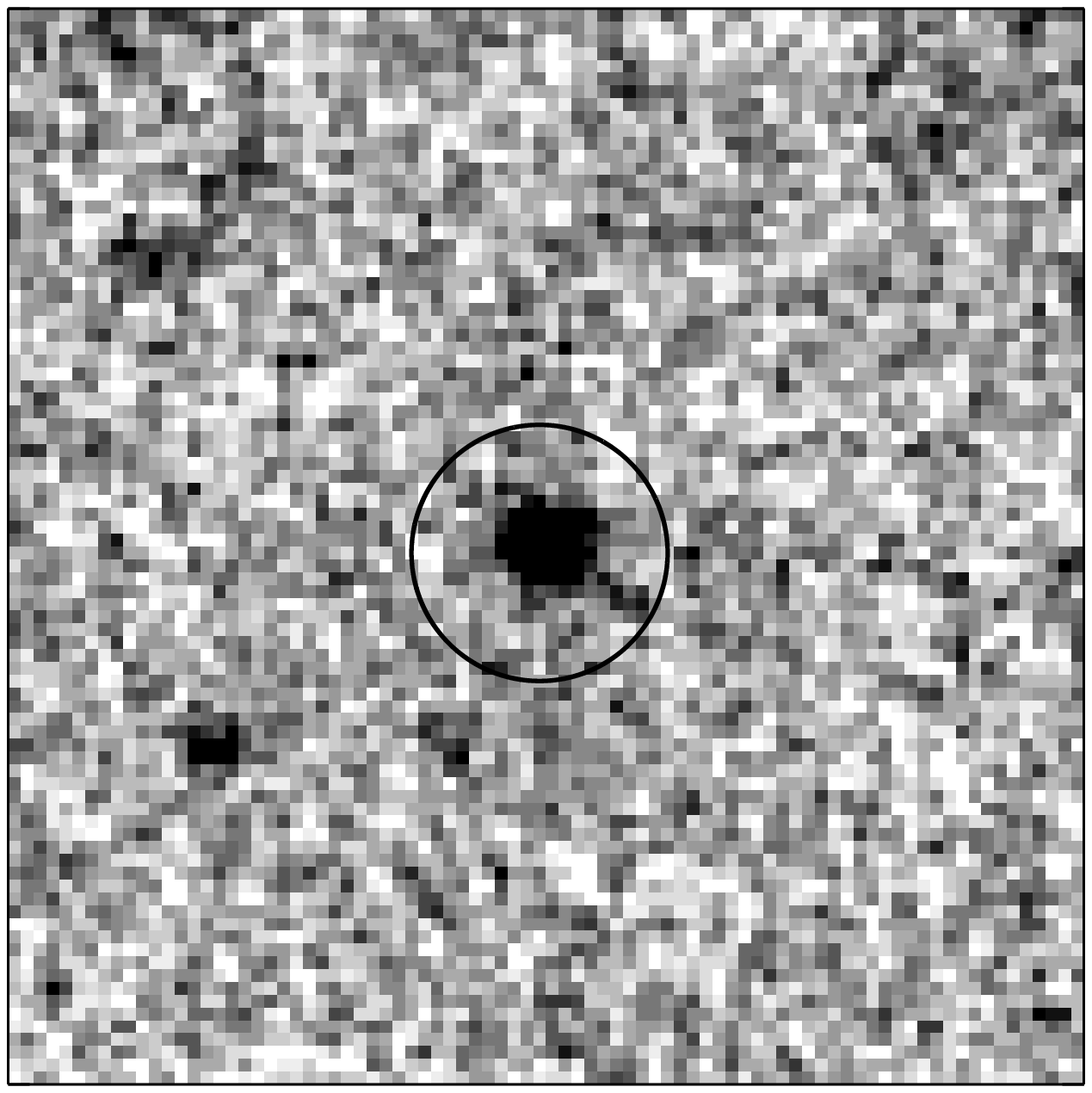}
\hspace{-10mm}
\plotone{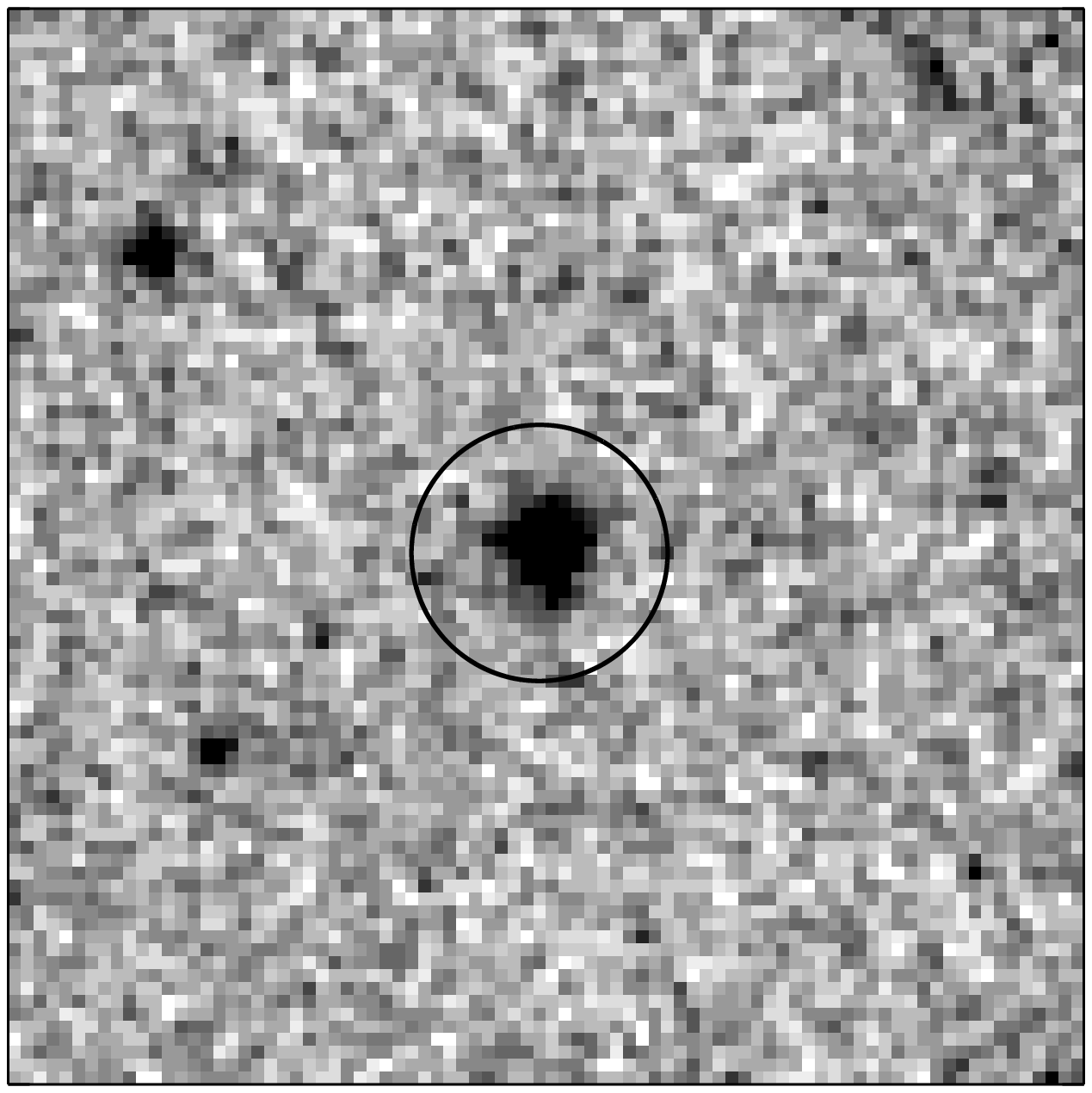}
\hspace{-10mm}
\plotone{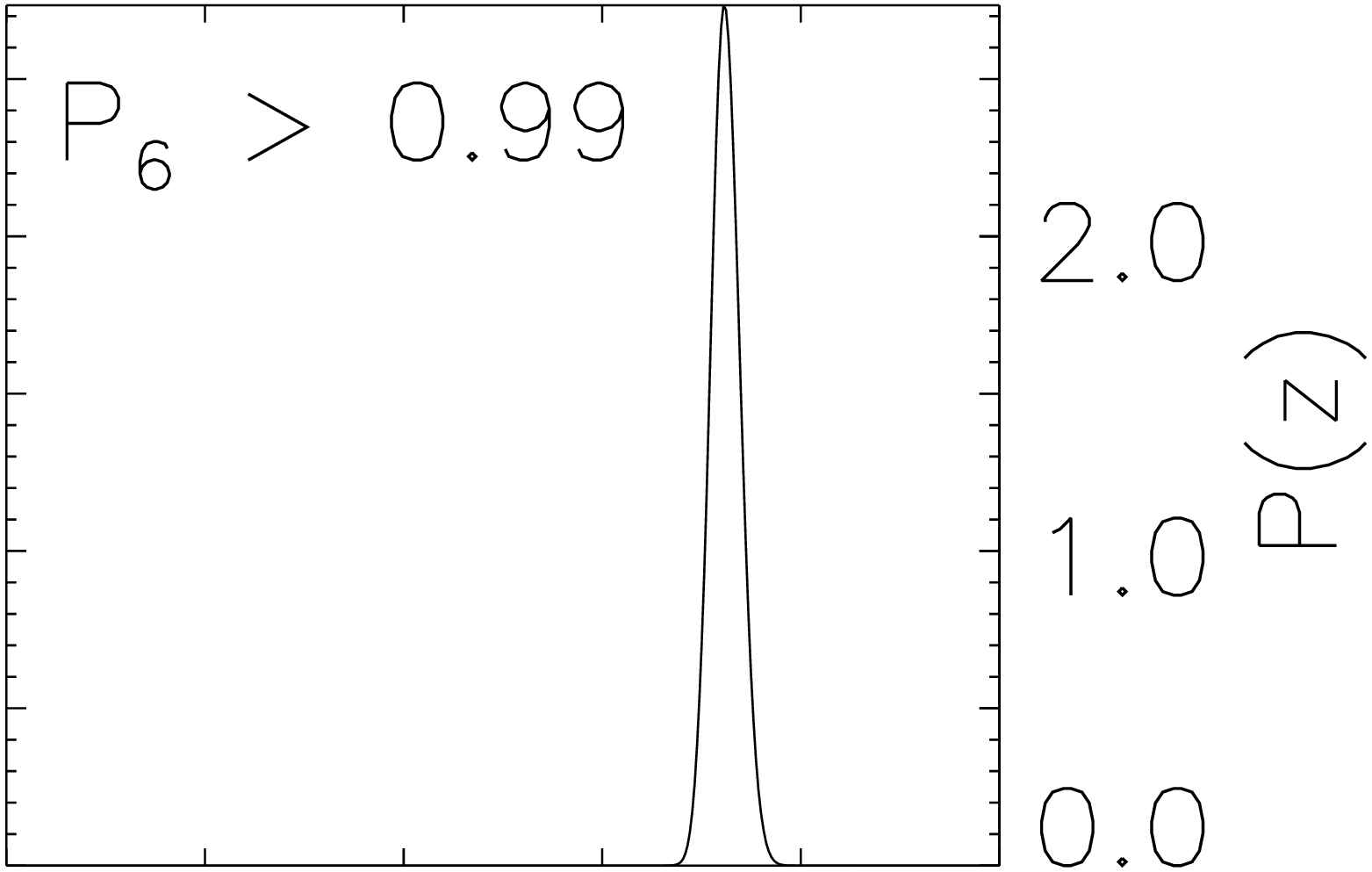}
\vspace{0.5mm}

\plotone{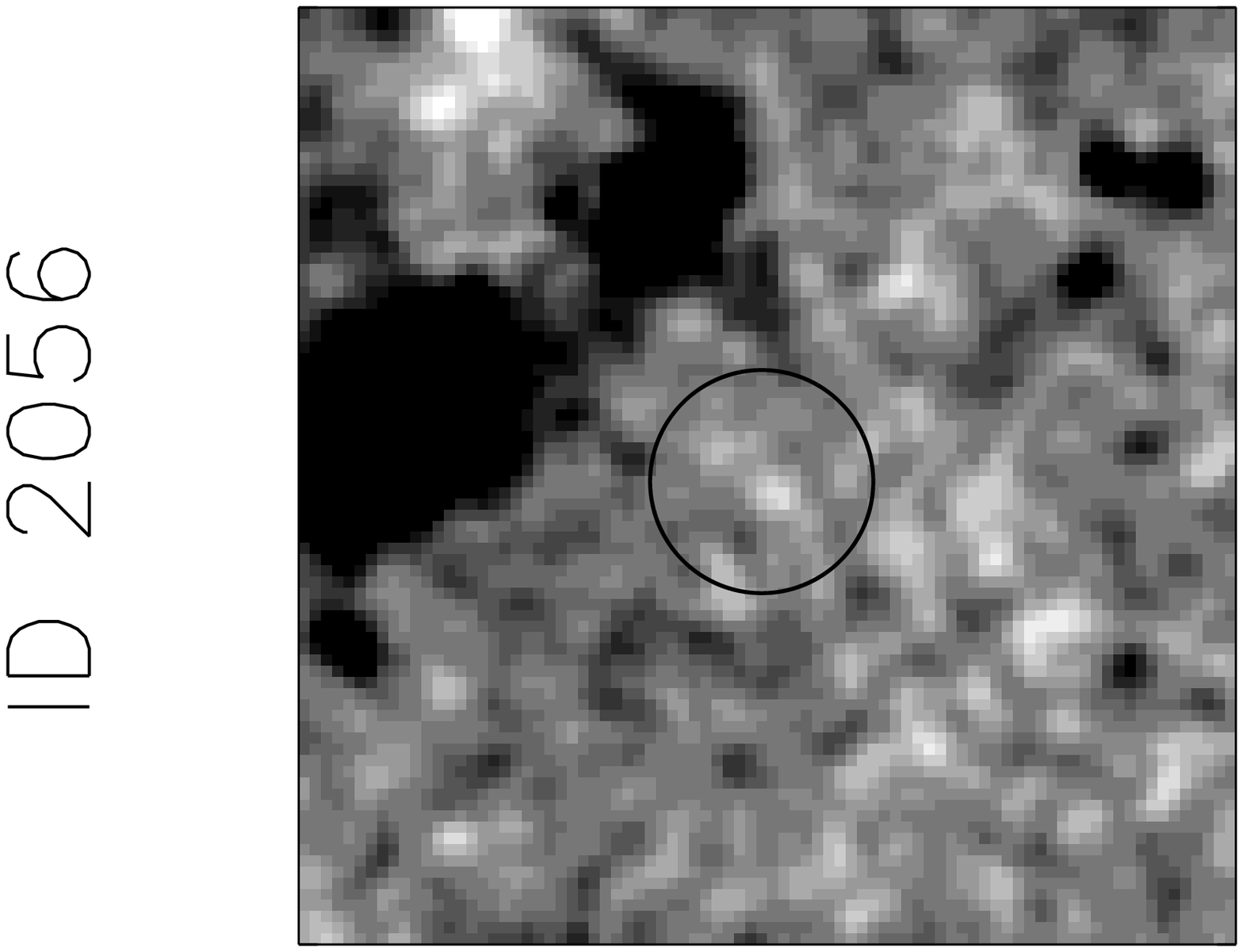}
\hspace{-10mm}
\plotone{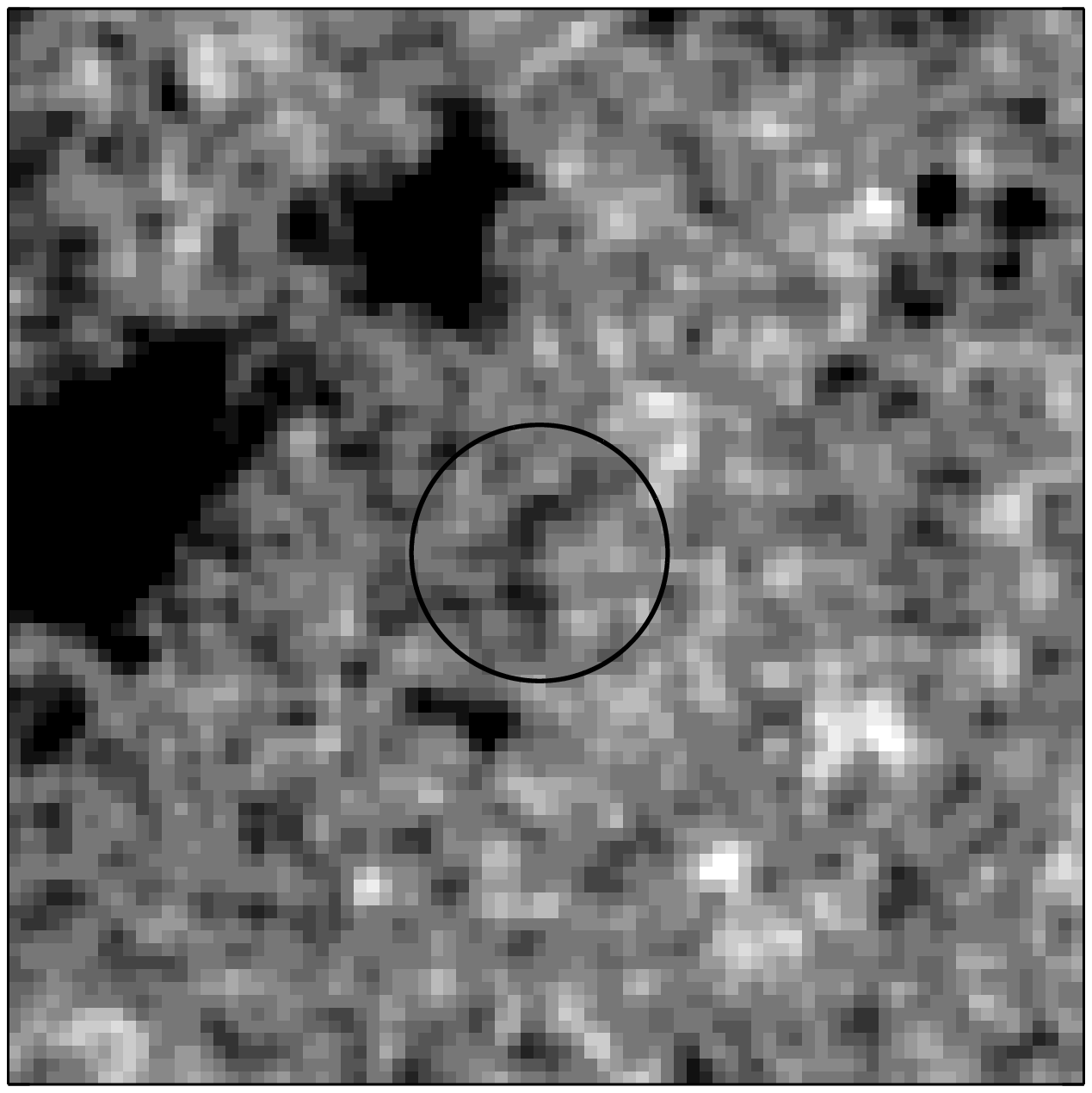}
\hspace{-10mm}
\plotone{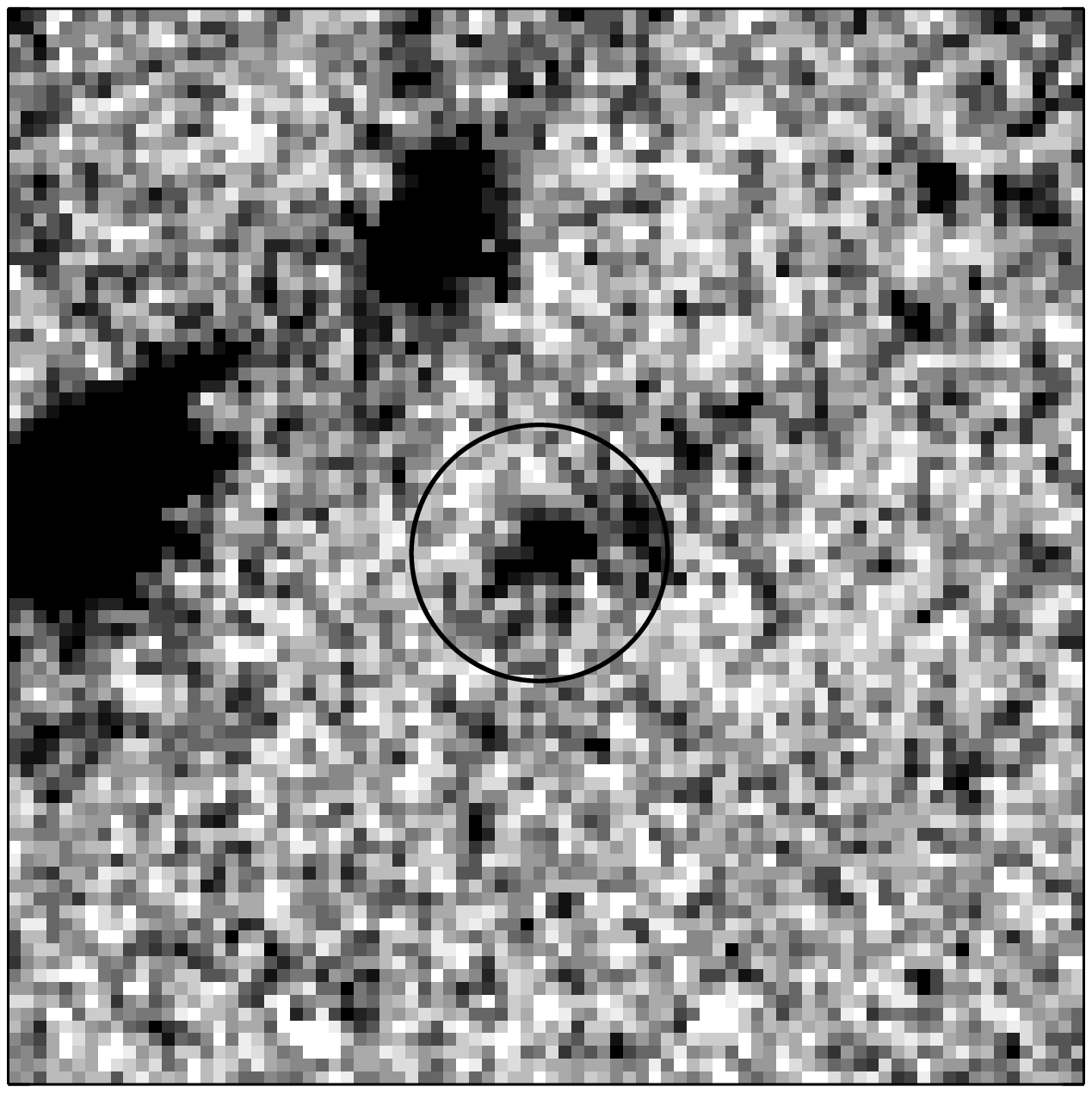}
\hspace{-10mm}
\plotone{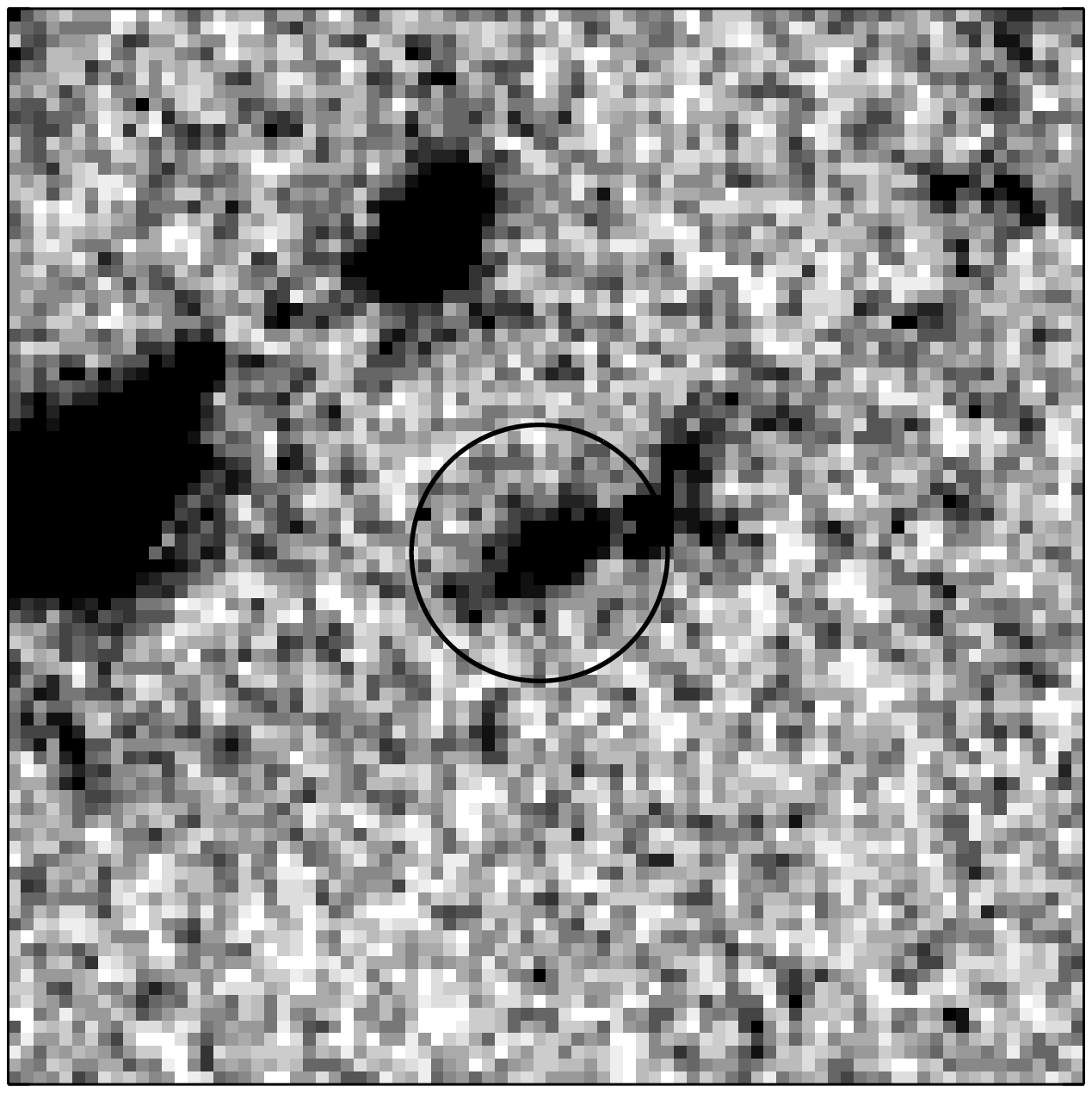}
\hspace{-10mm}
\plotone{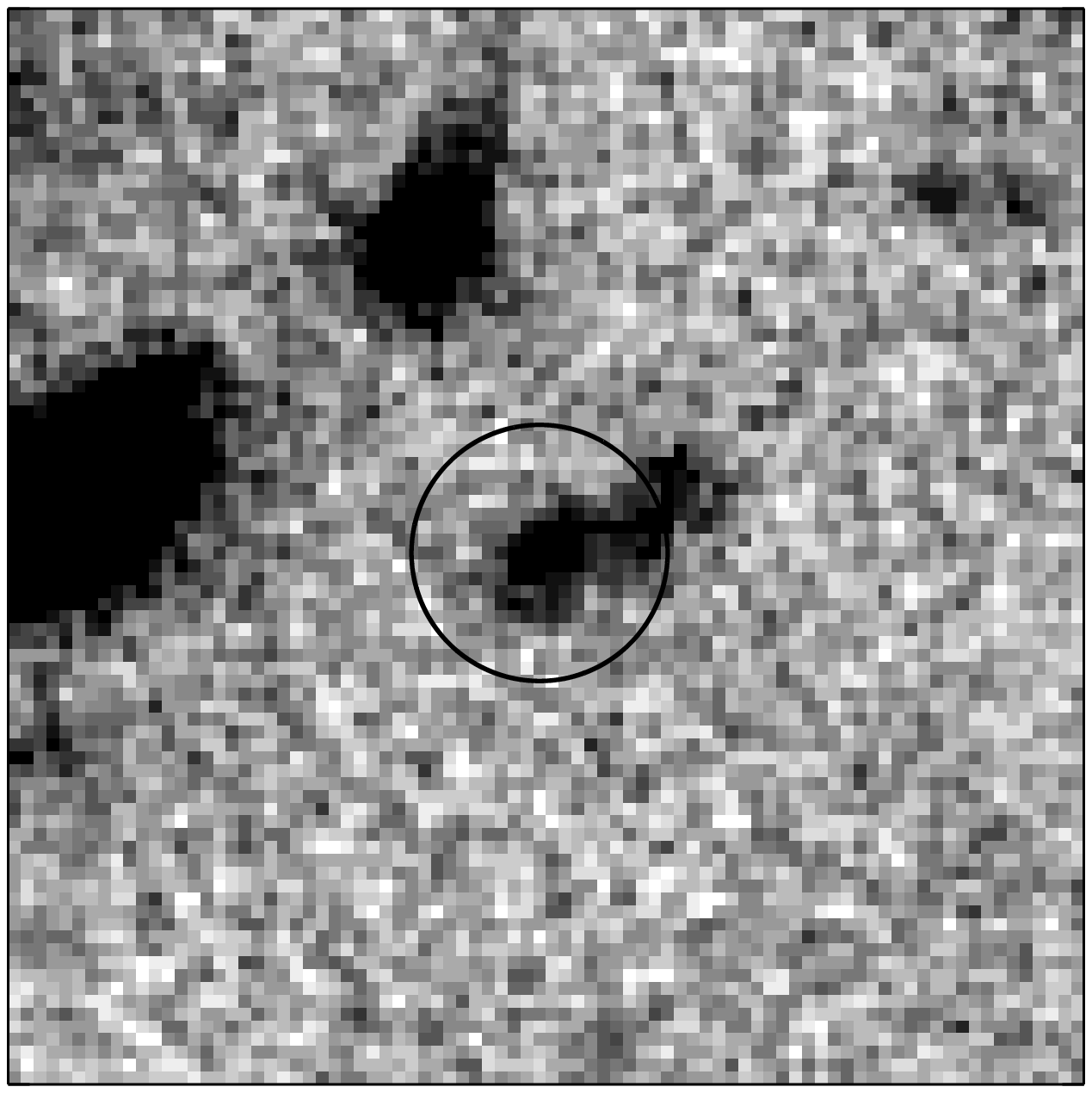}
\hspace{-10mm}
\plotone{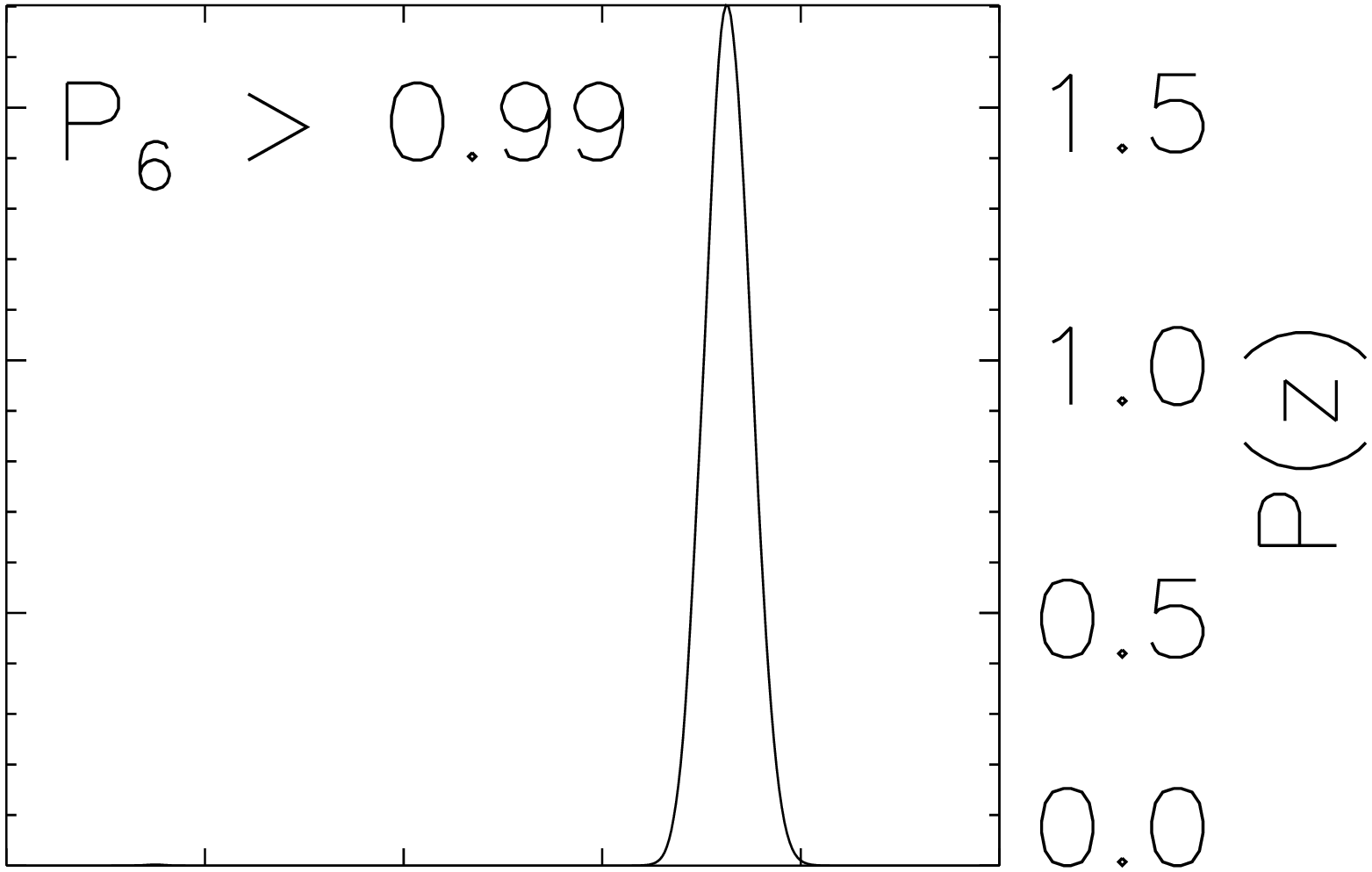}
\vspace{0.5mm}

\plotone{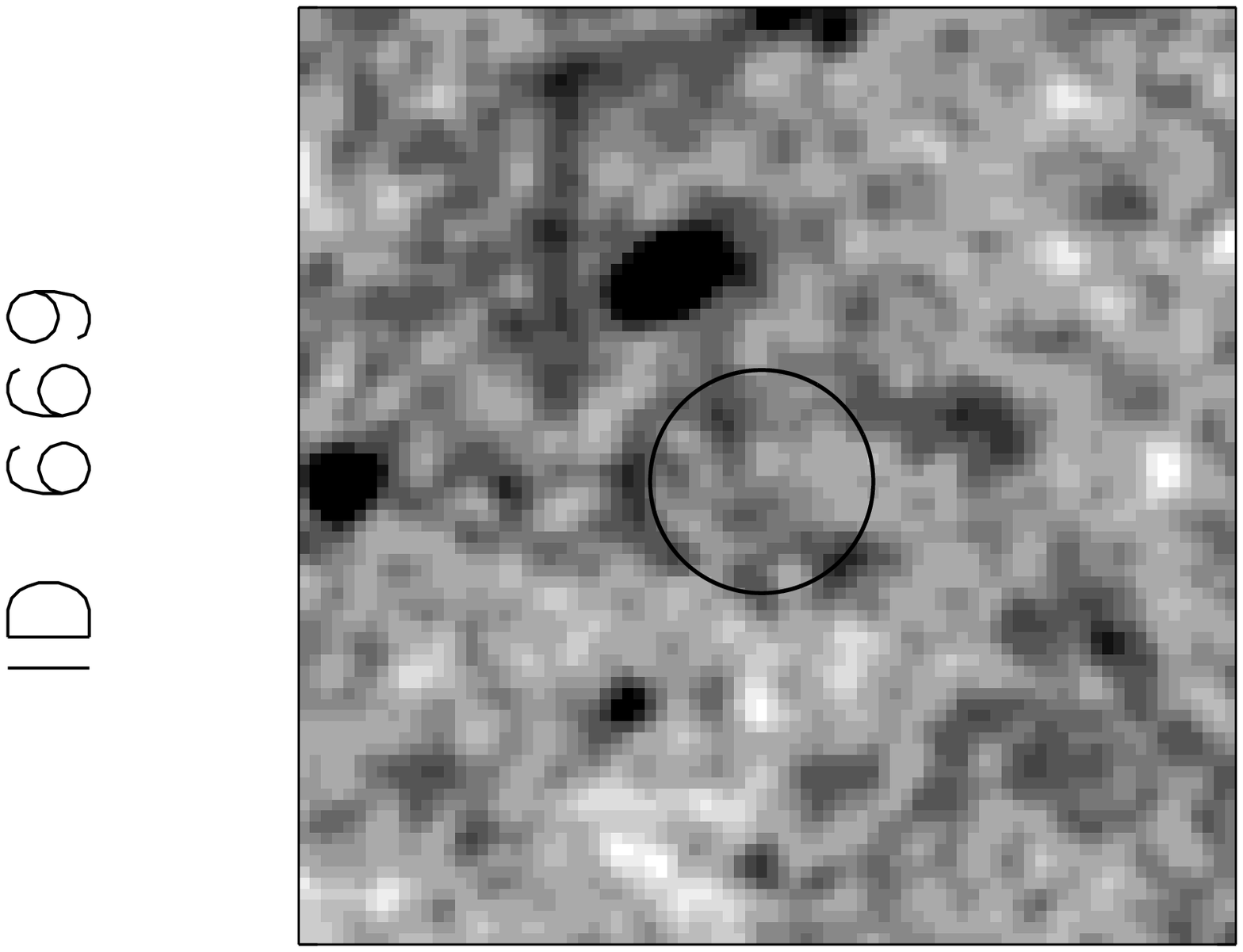}
\hspace{-10mm}
\plotone{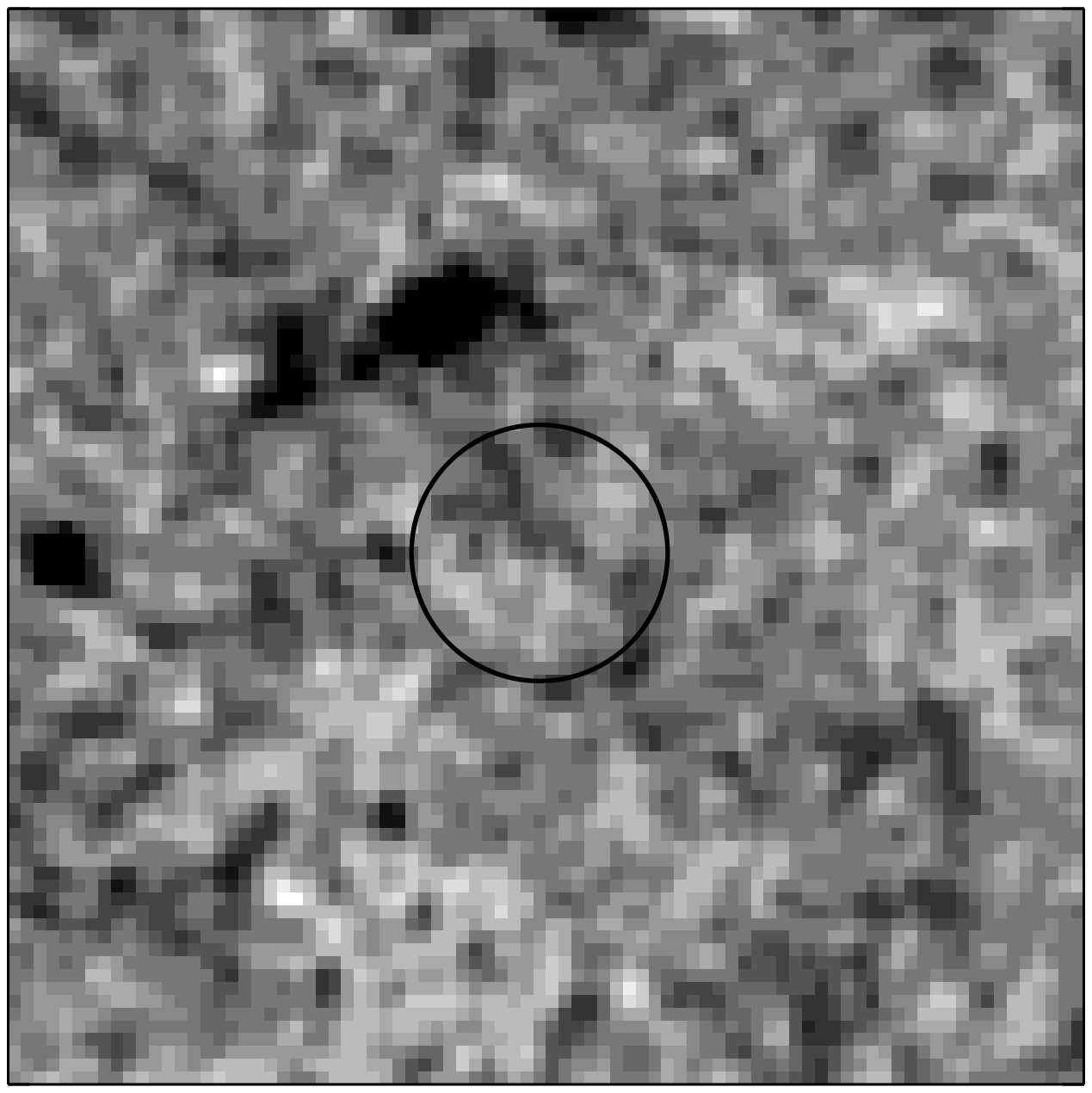}
\hspace{-10mm}
\plotone{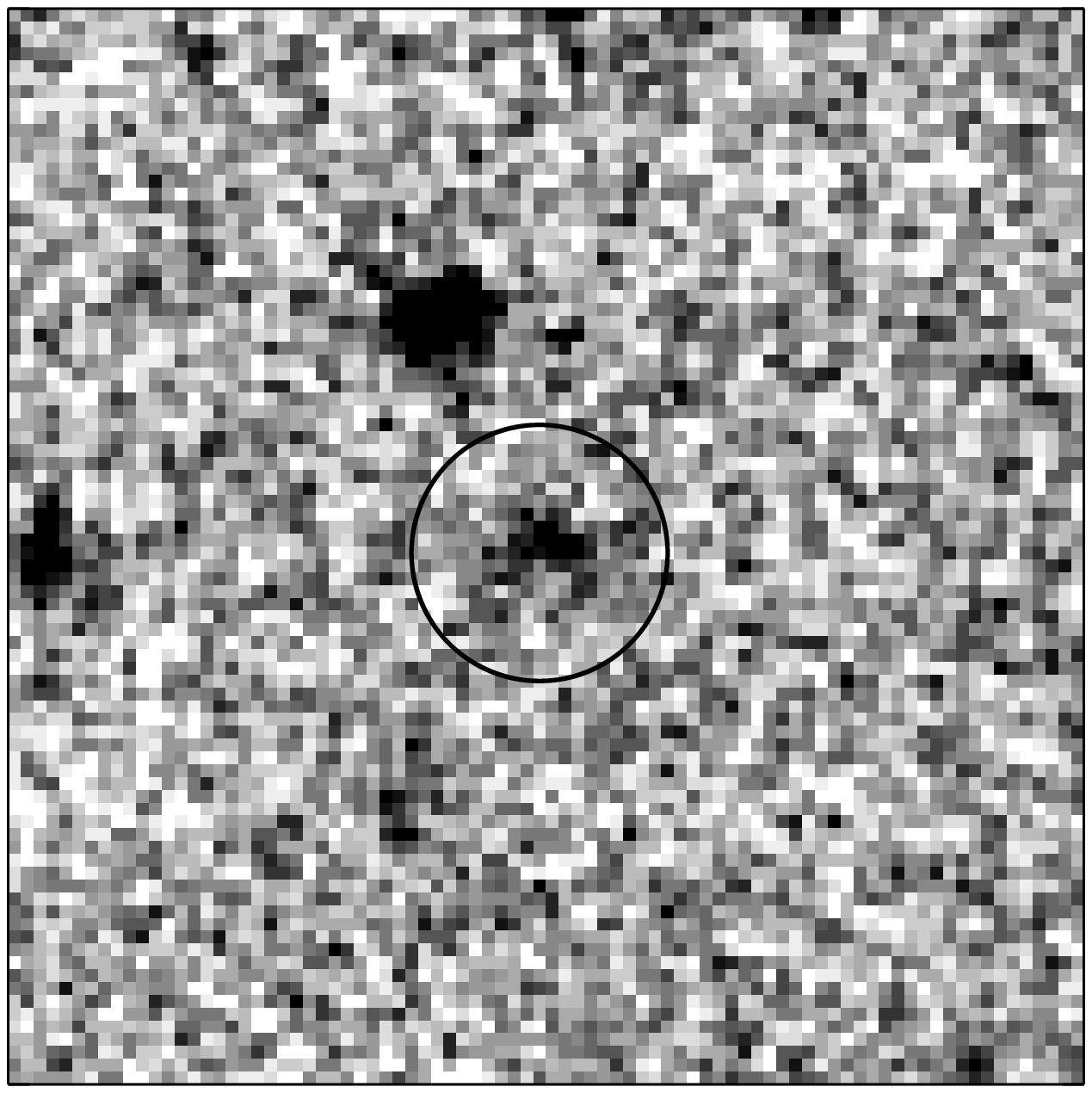}
\hspace{-10mm}
\plotone{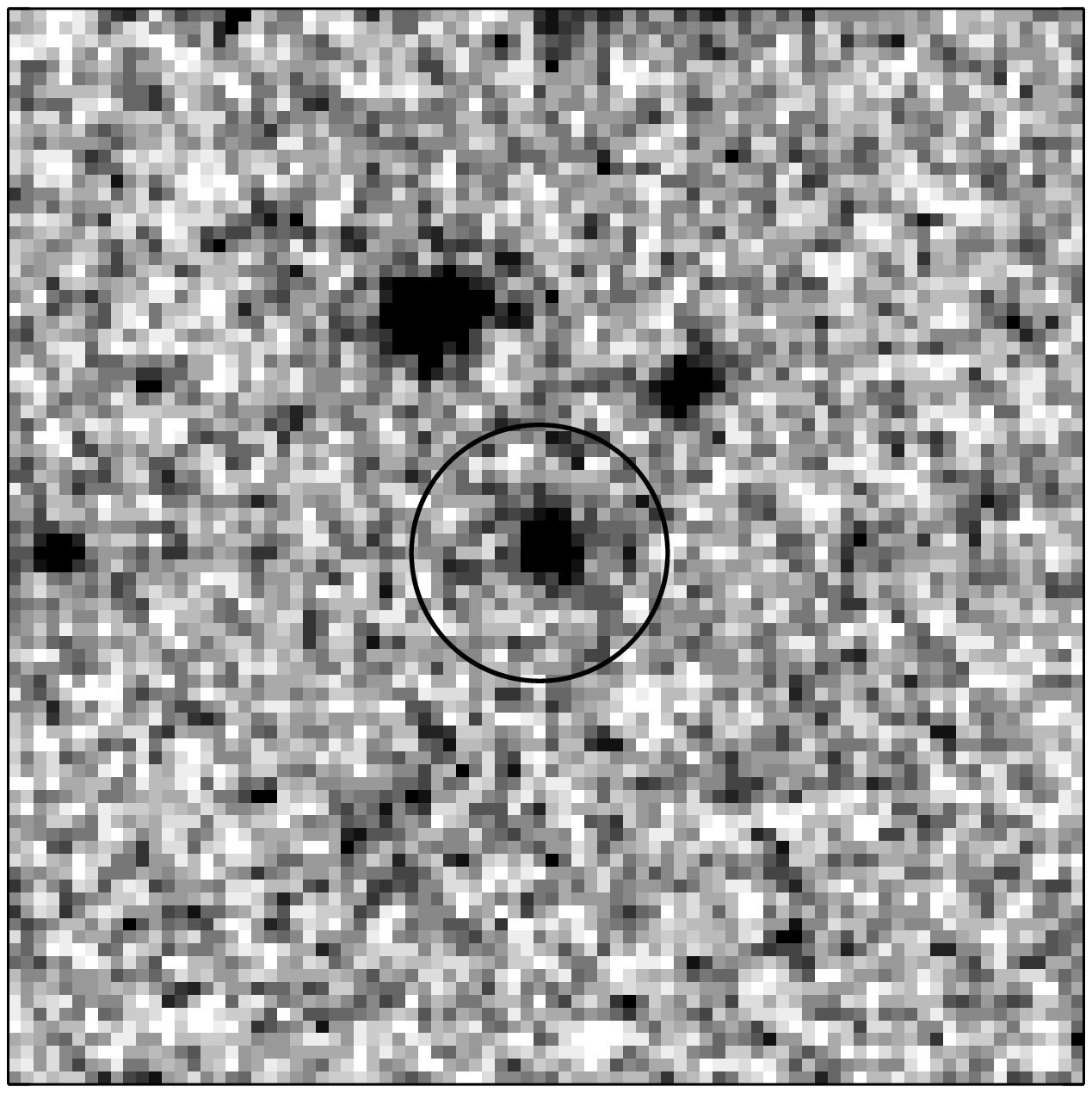}
\hspace{-10mm}
\plotone{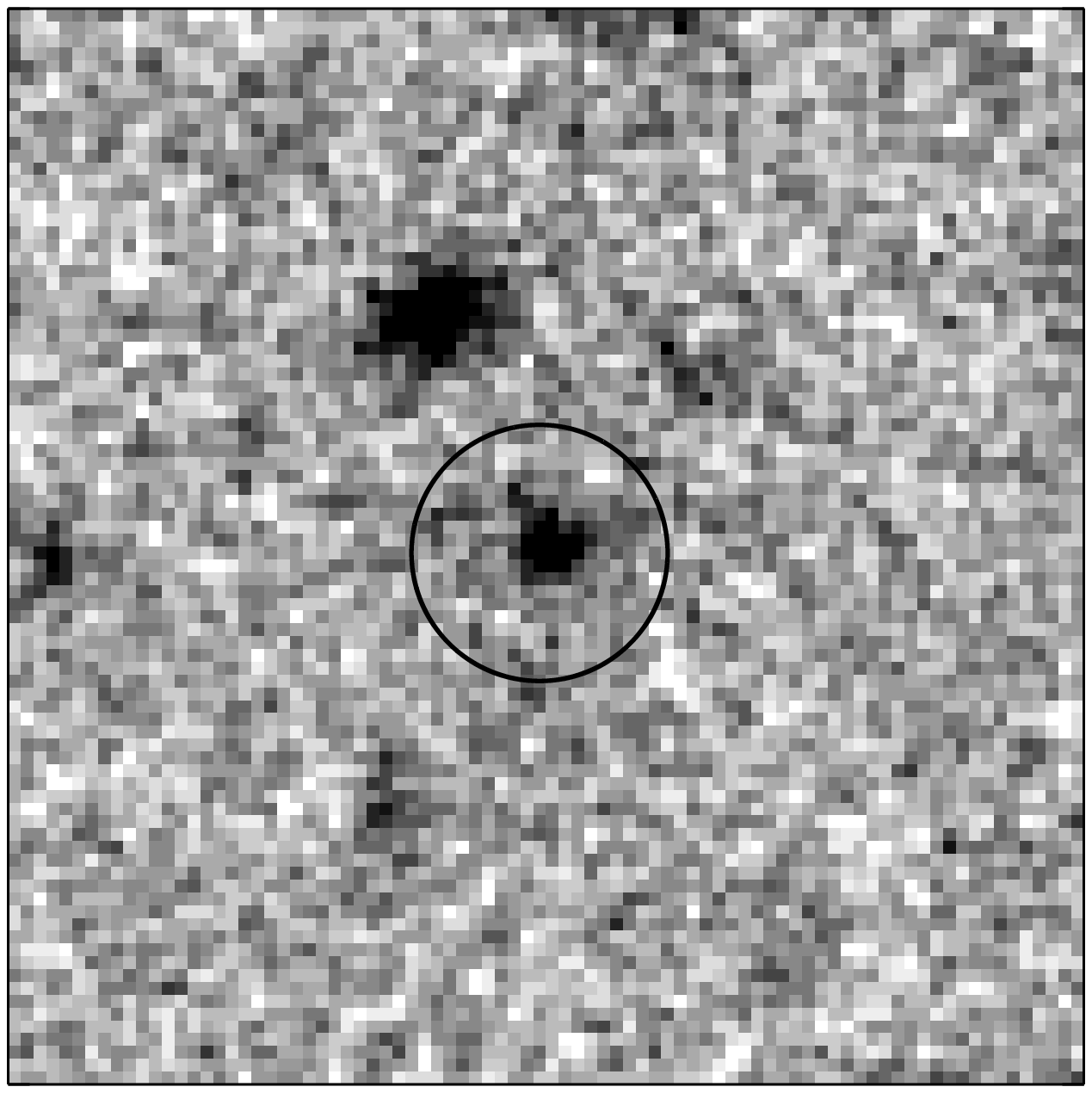}
\hspace{-10mm}
\plotone{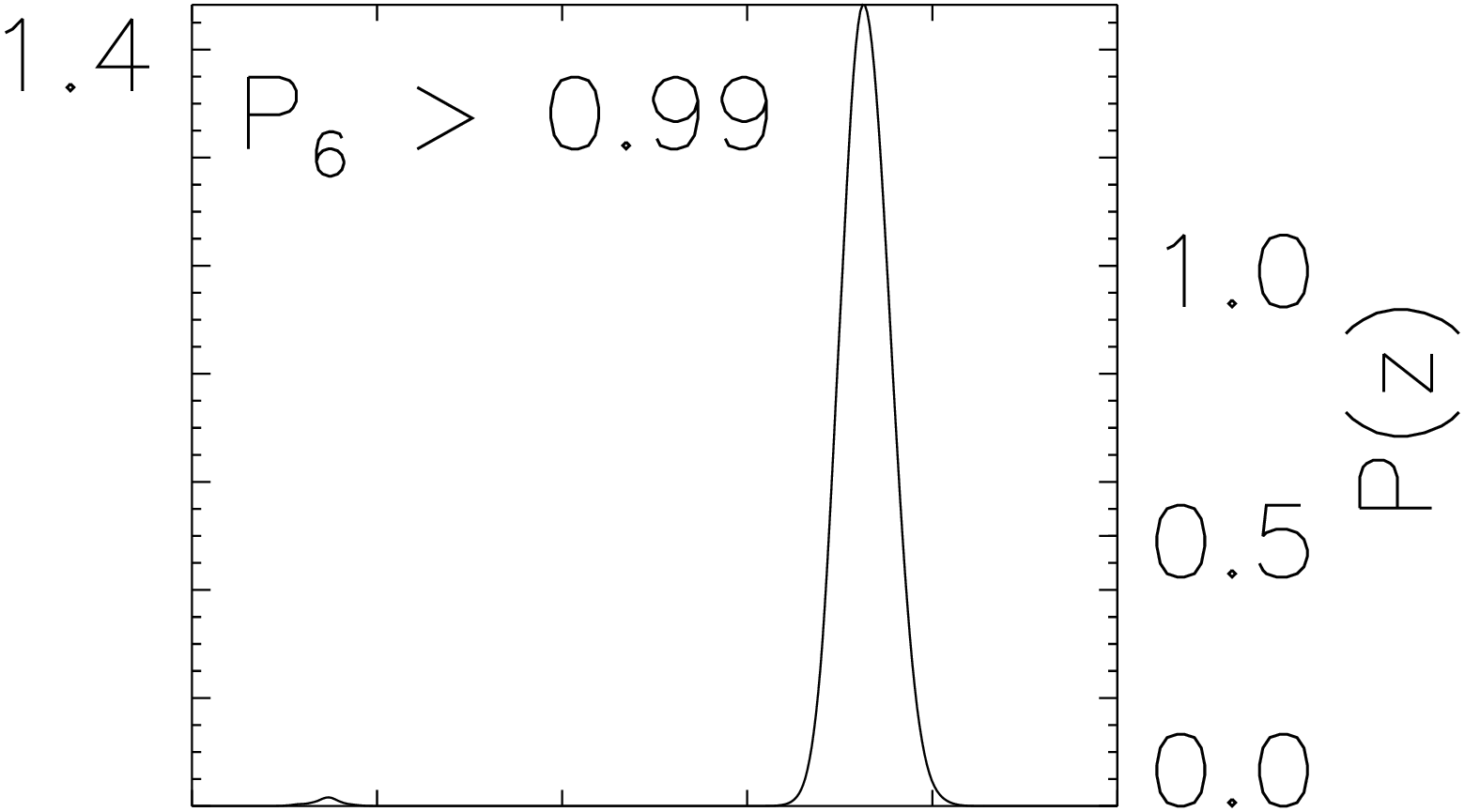}

\plotone{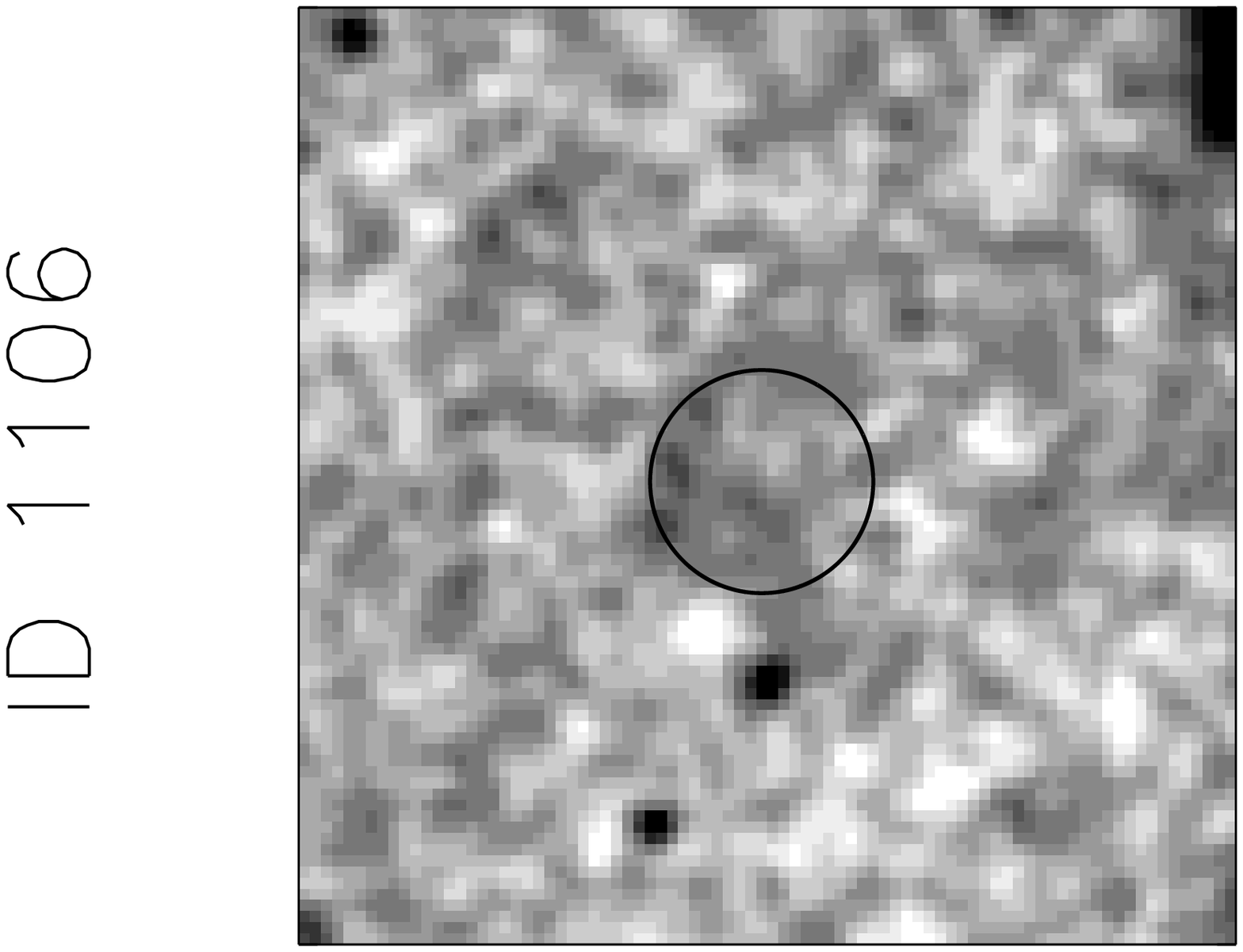}
\hspace{-10mm}
\plotone{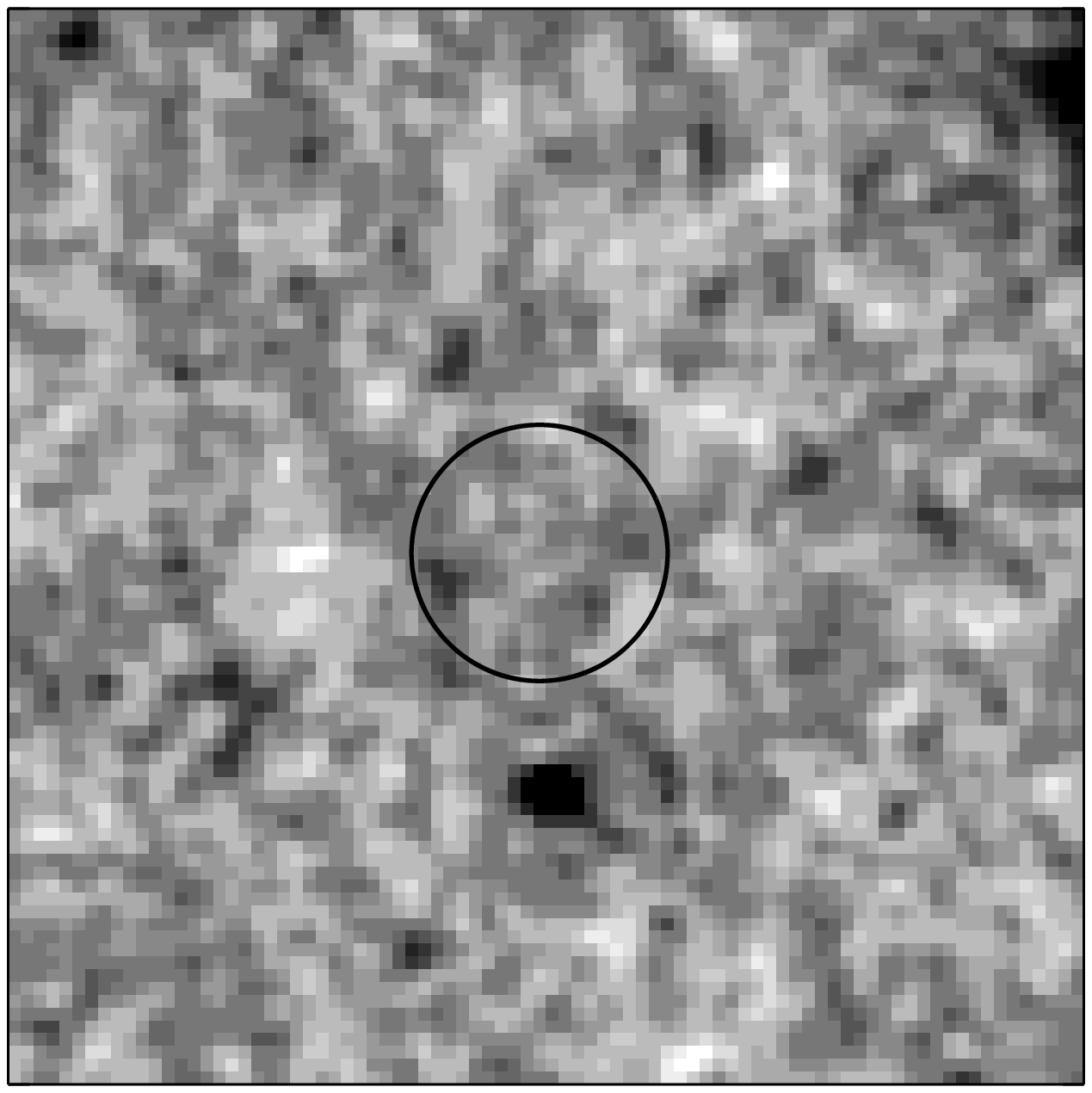}
\hspace{-10mm}
\plotone{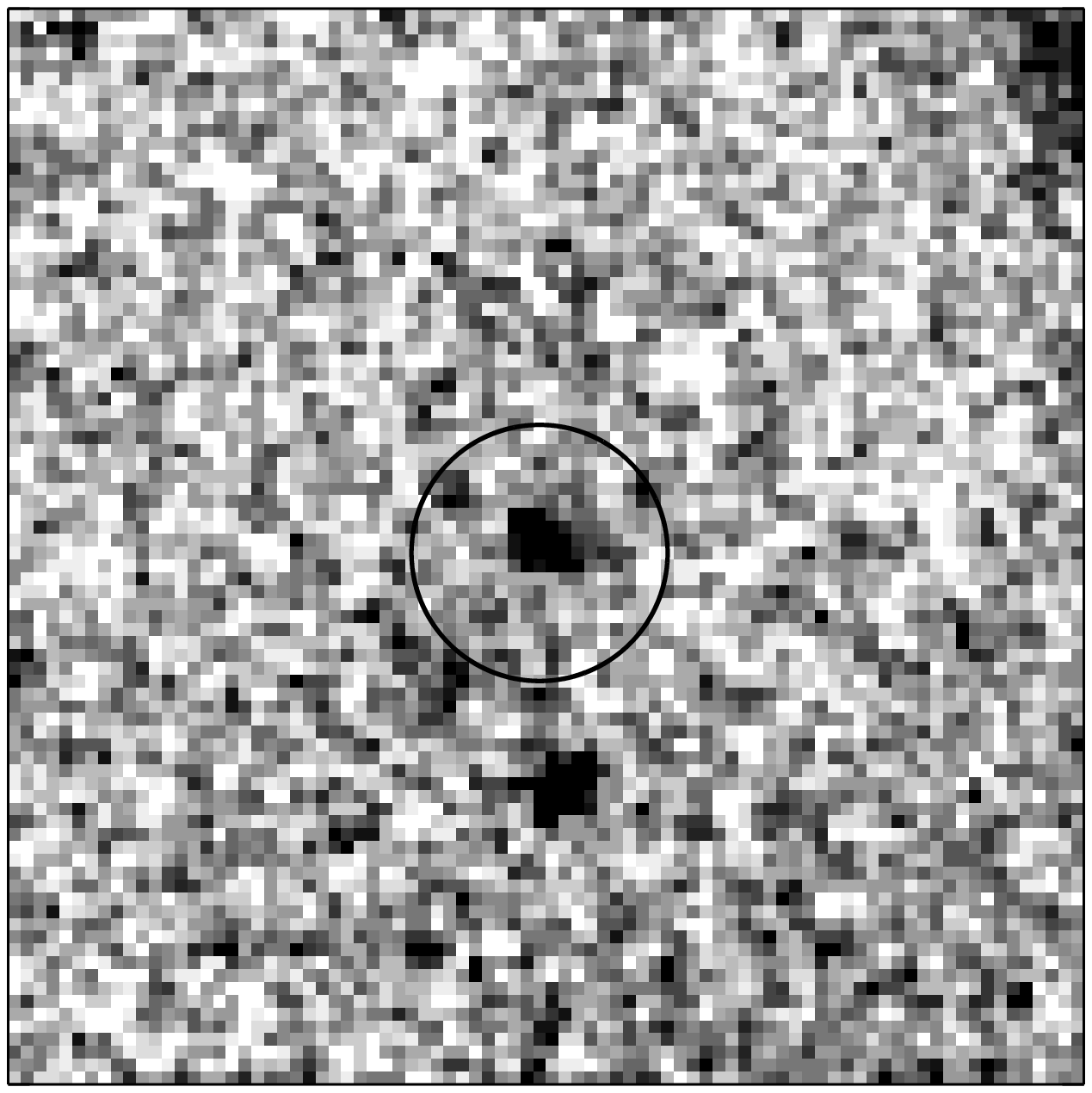}
\hspace{-10mm}
\plotone{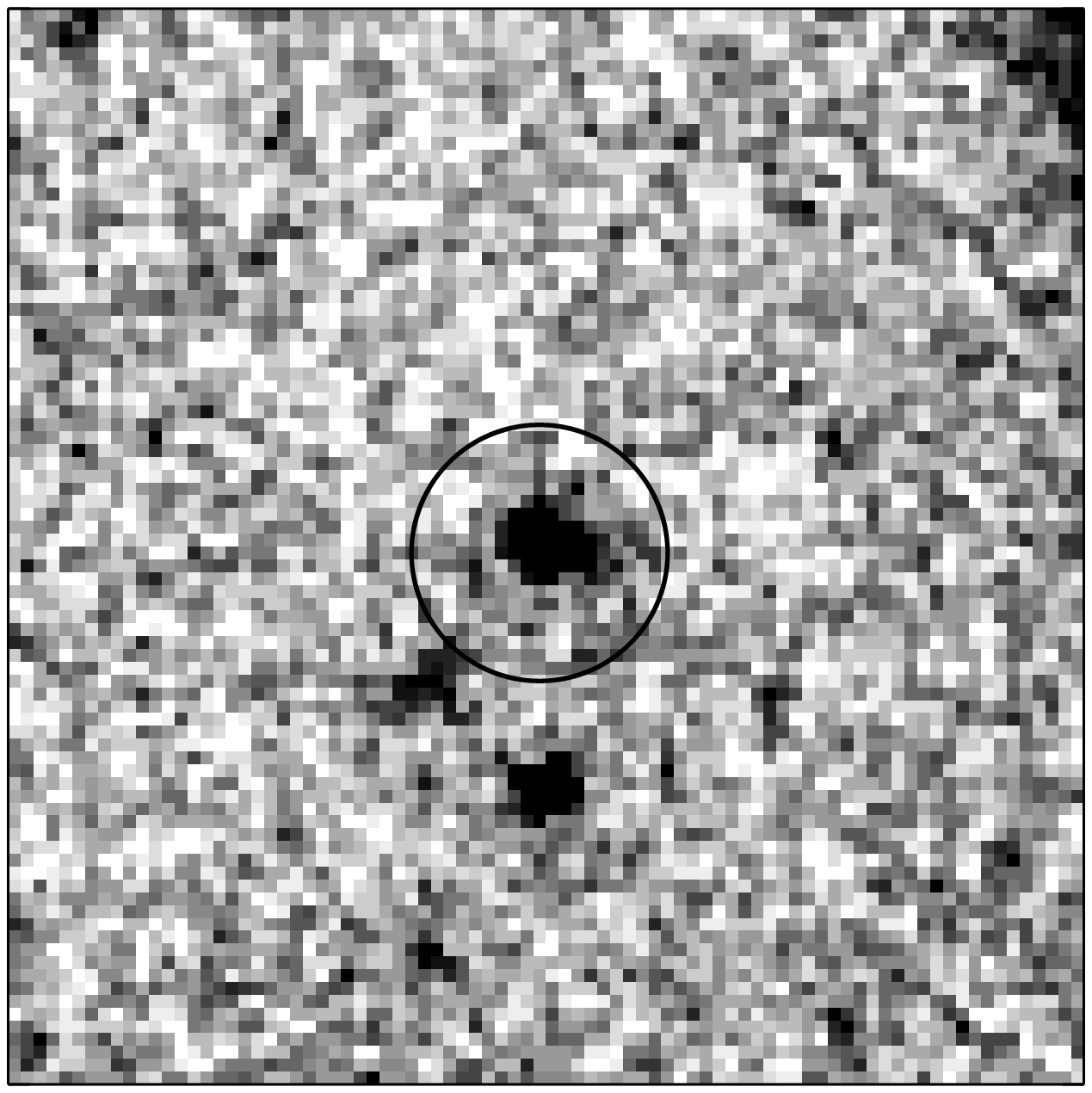}
\hspace{-10mm}
\plotone{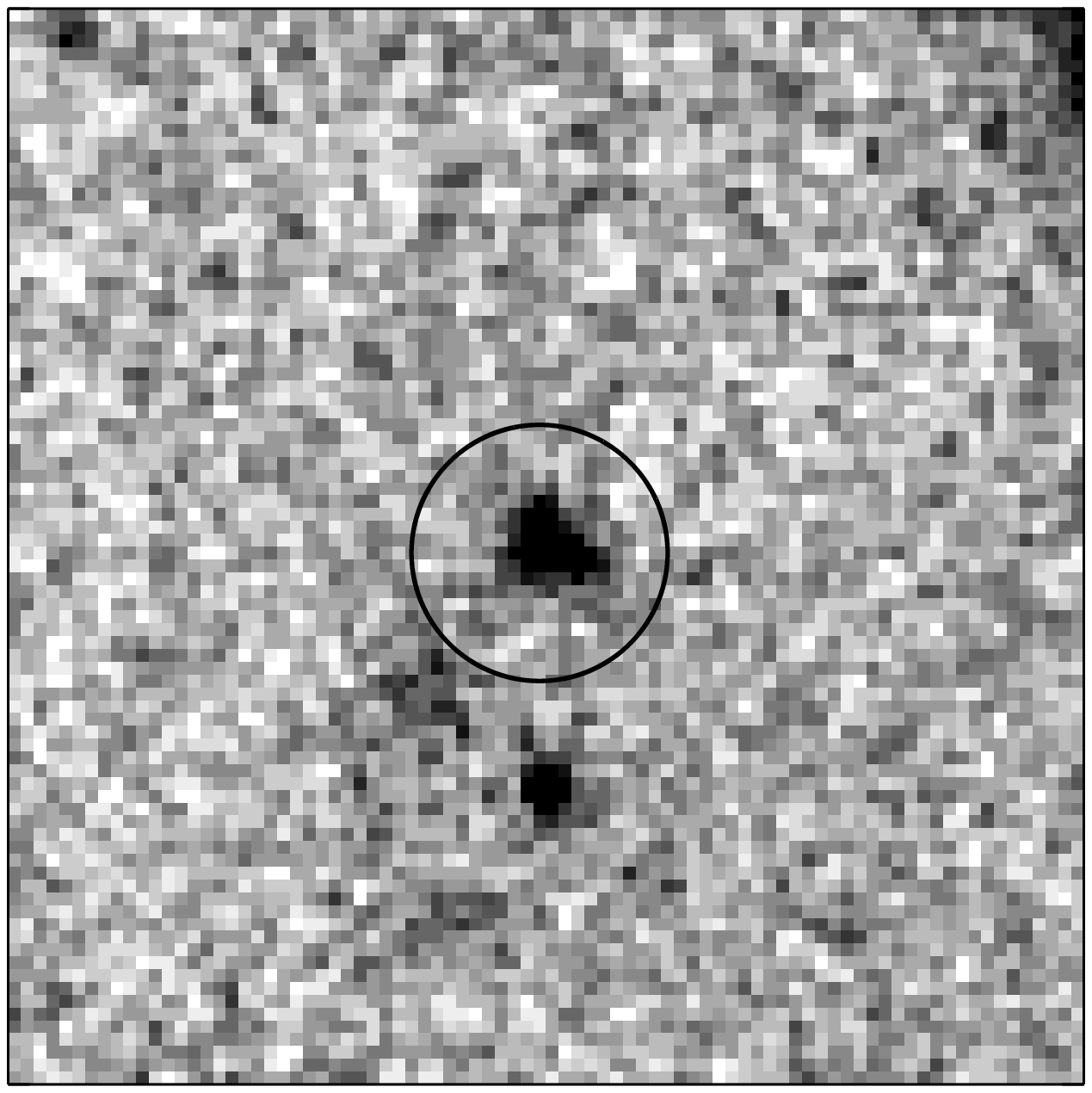}
\hspace{-10mm}
\plotone{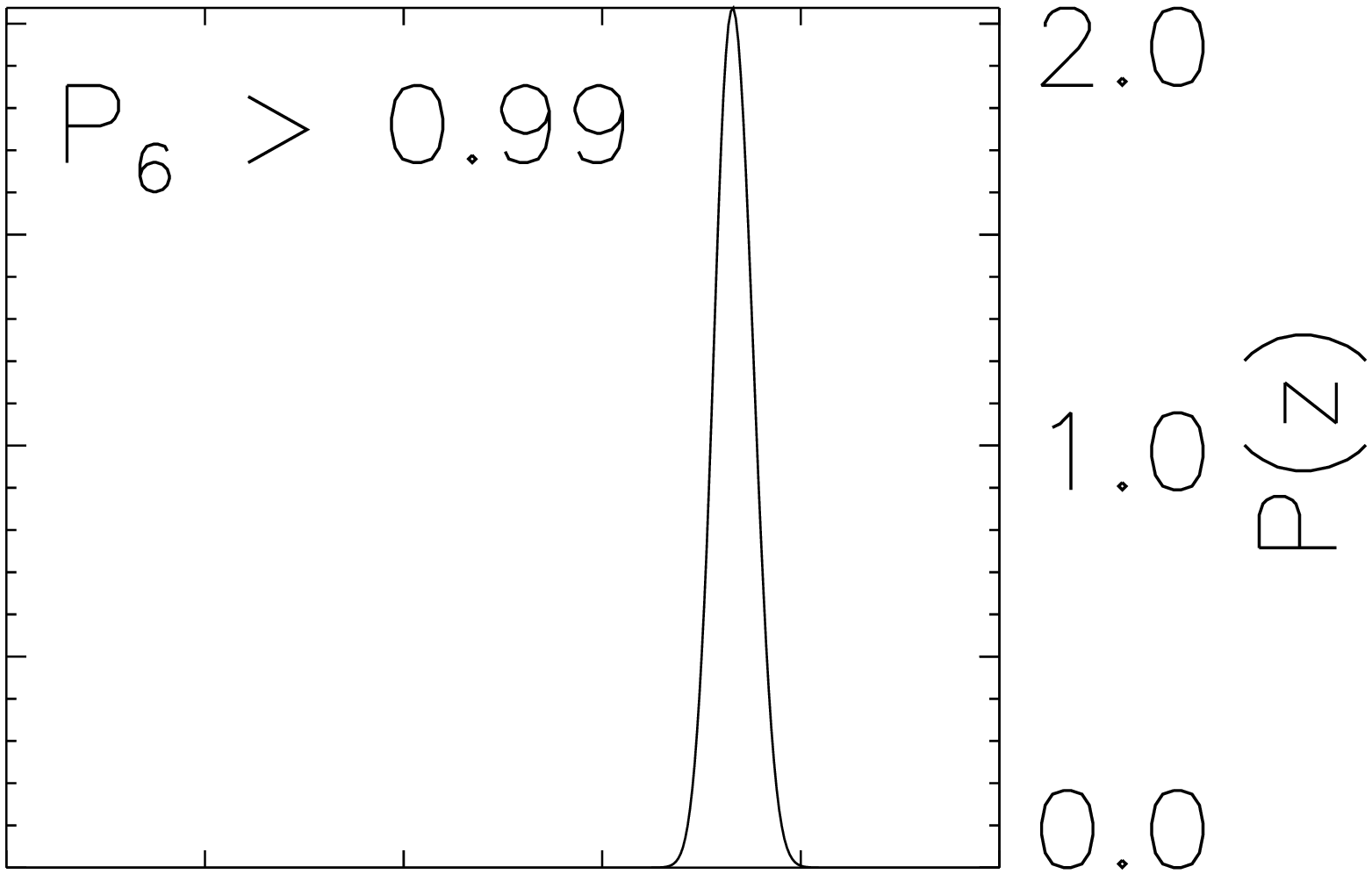}
\vspace{0.5mm}

\plotone{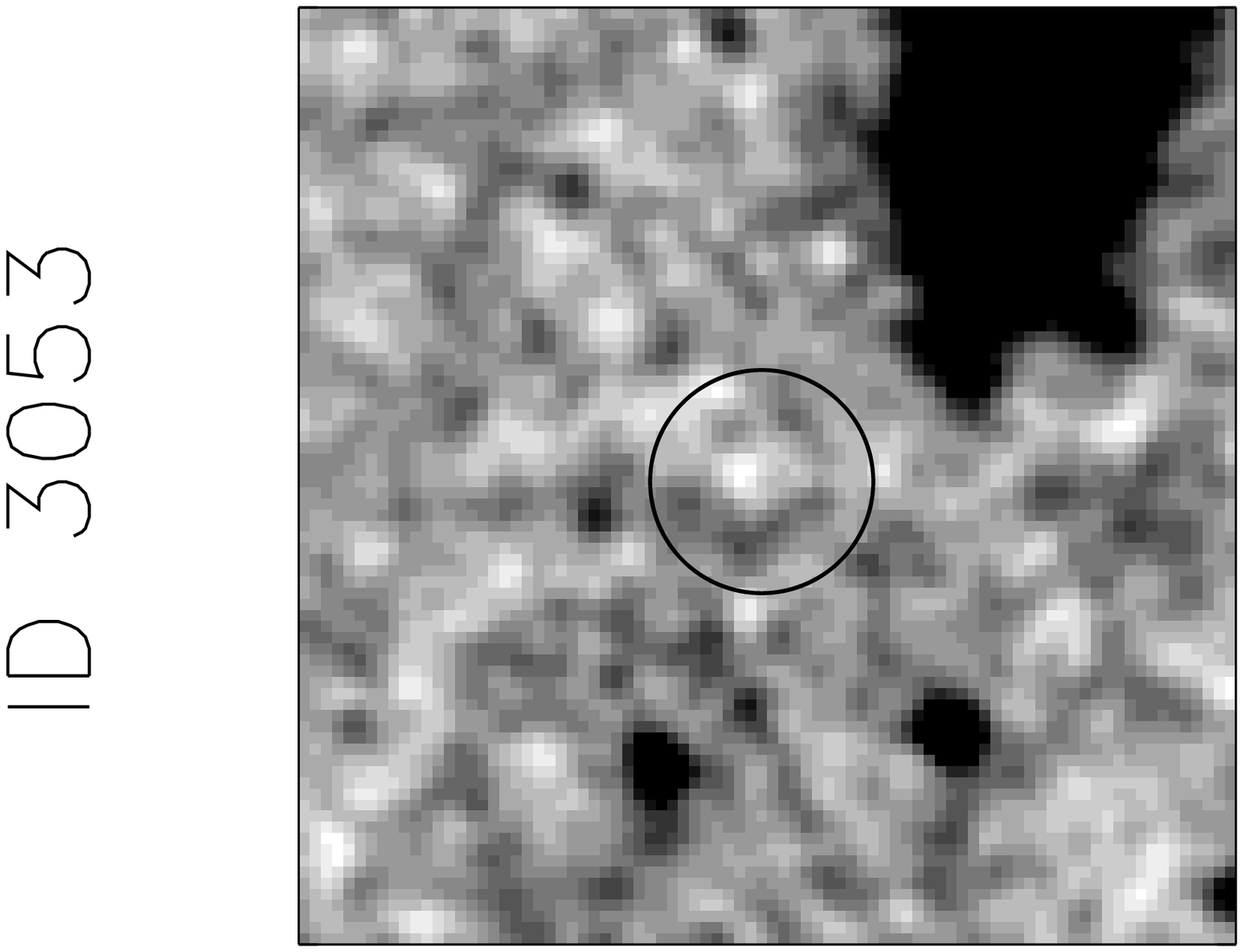}
\hspace{-10mm}
\plotone{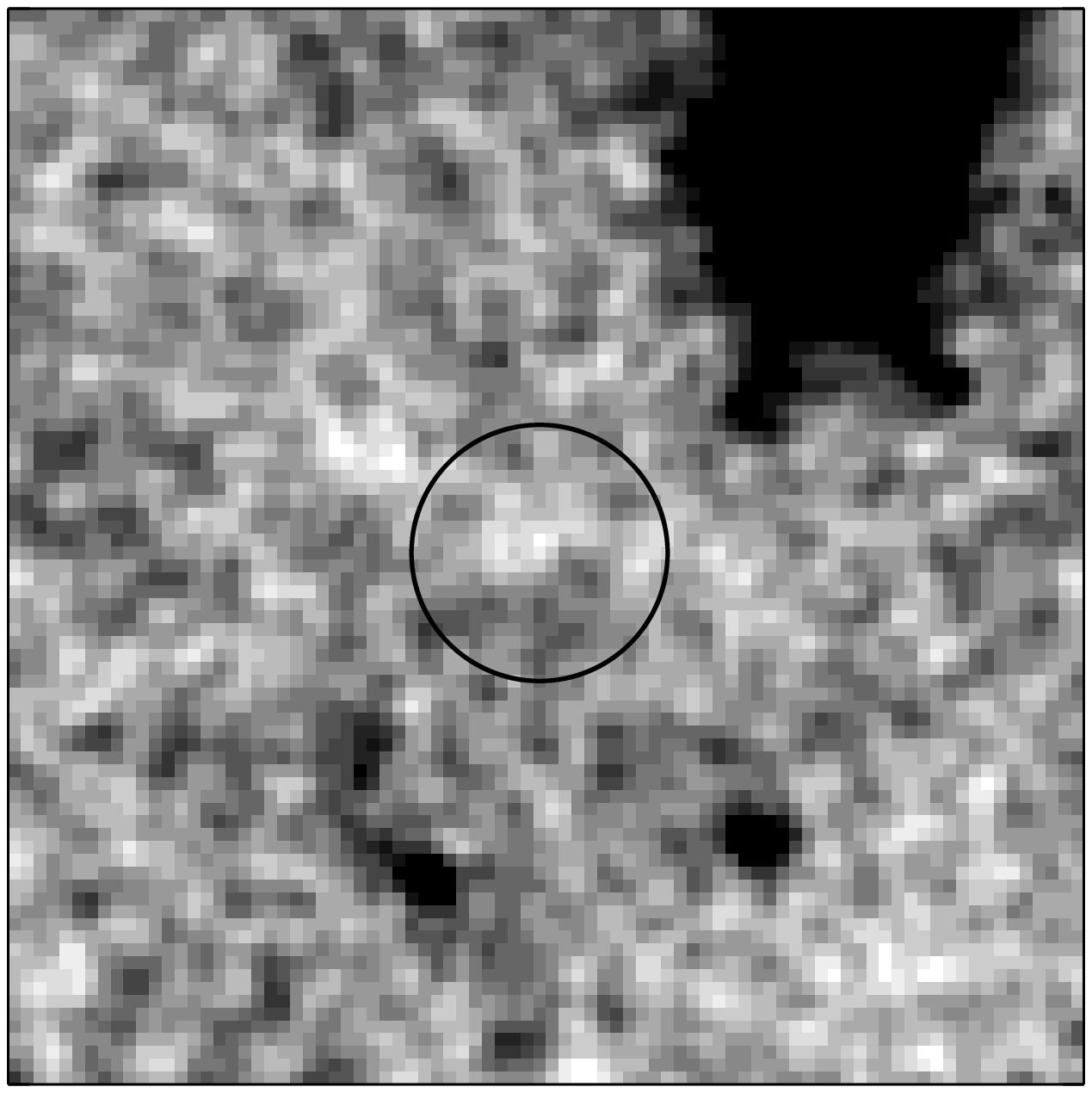}
\hspace{-10mm}
\plotone{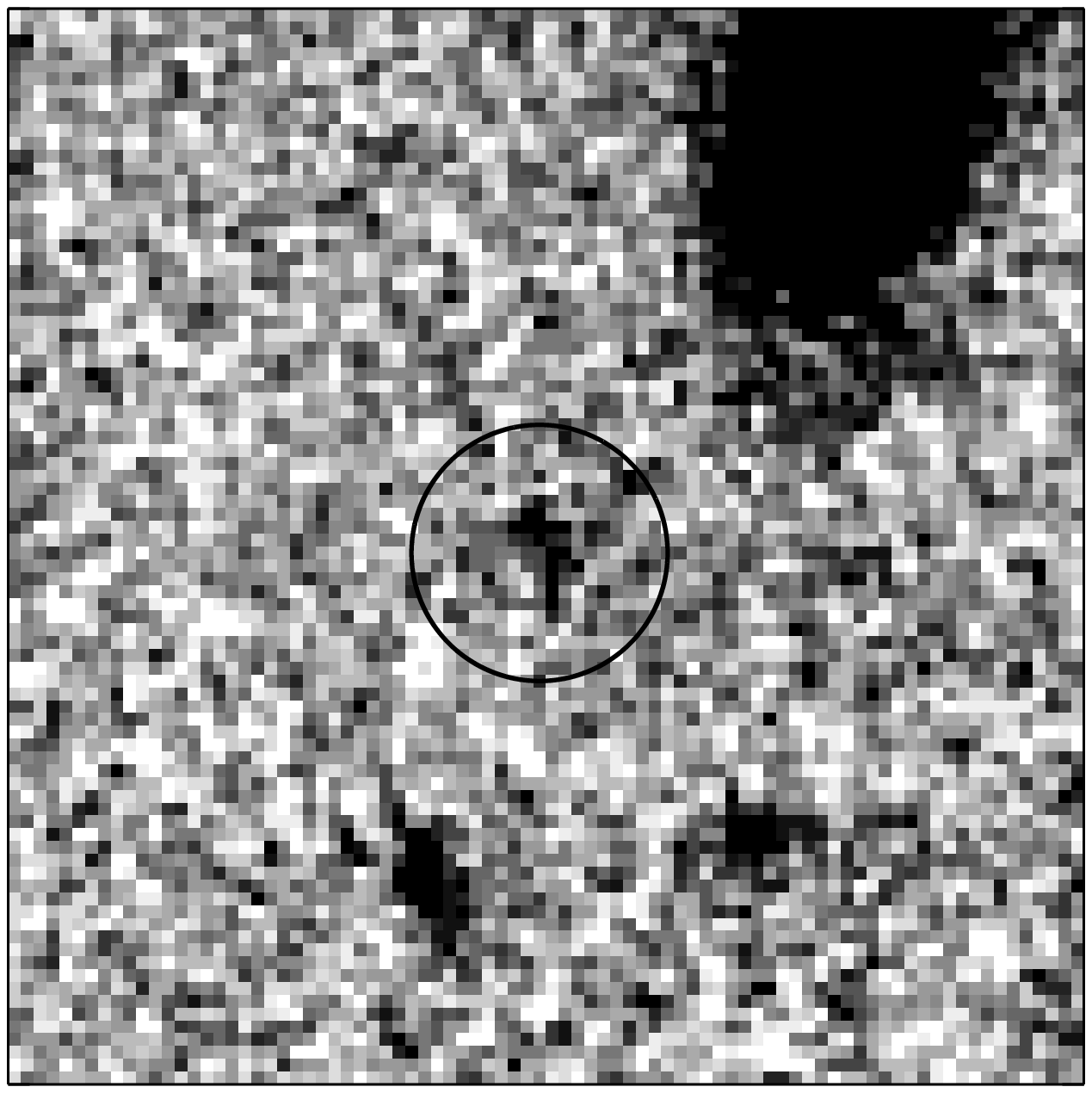}
\hspace{-10mm}
\plotone{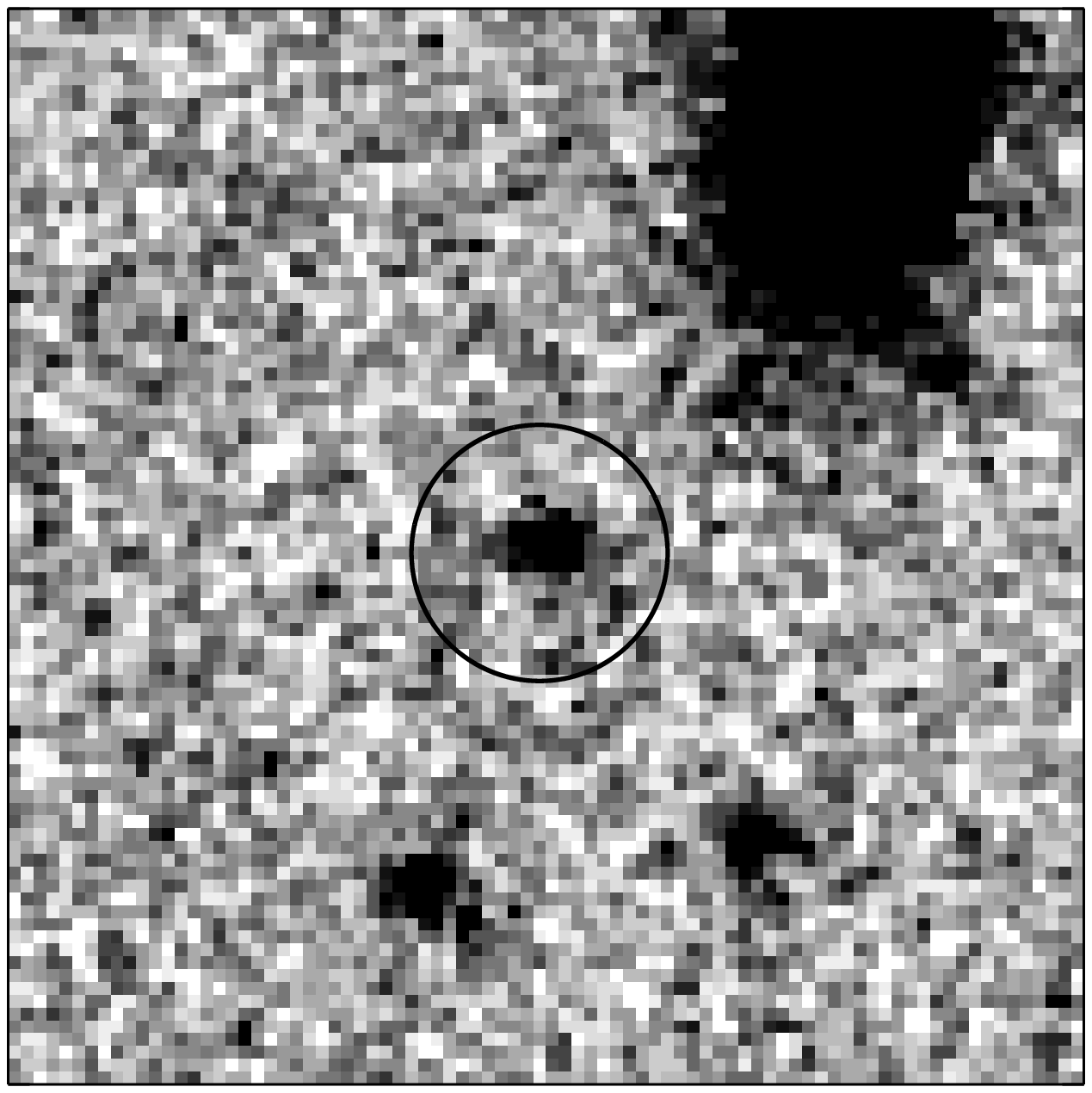}
\hspace{-10mm}
\plotone{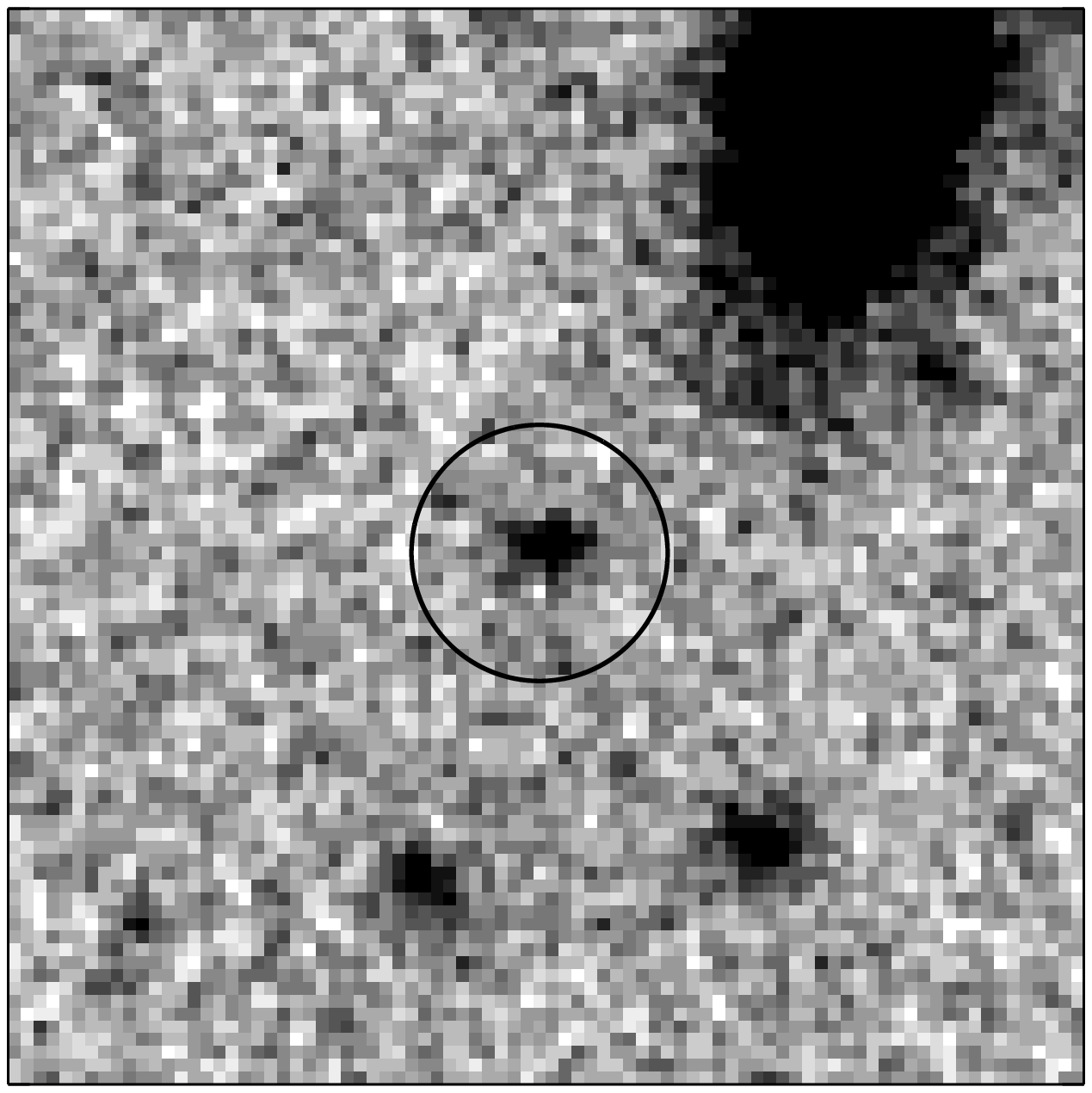}
\hspace{-10mm}
\plotone{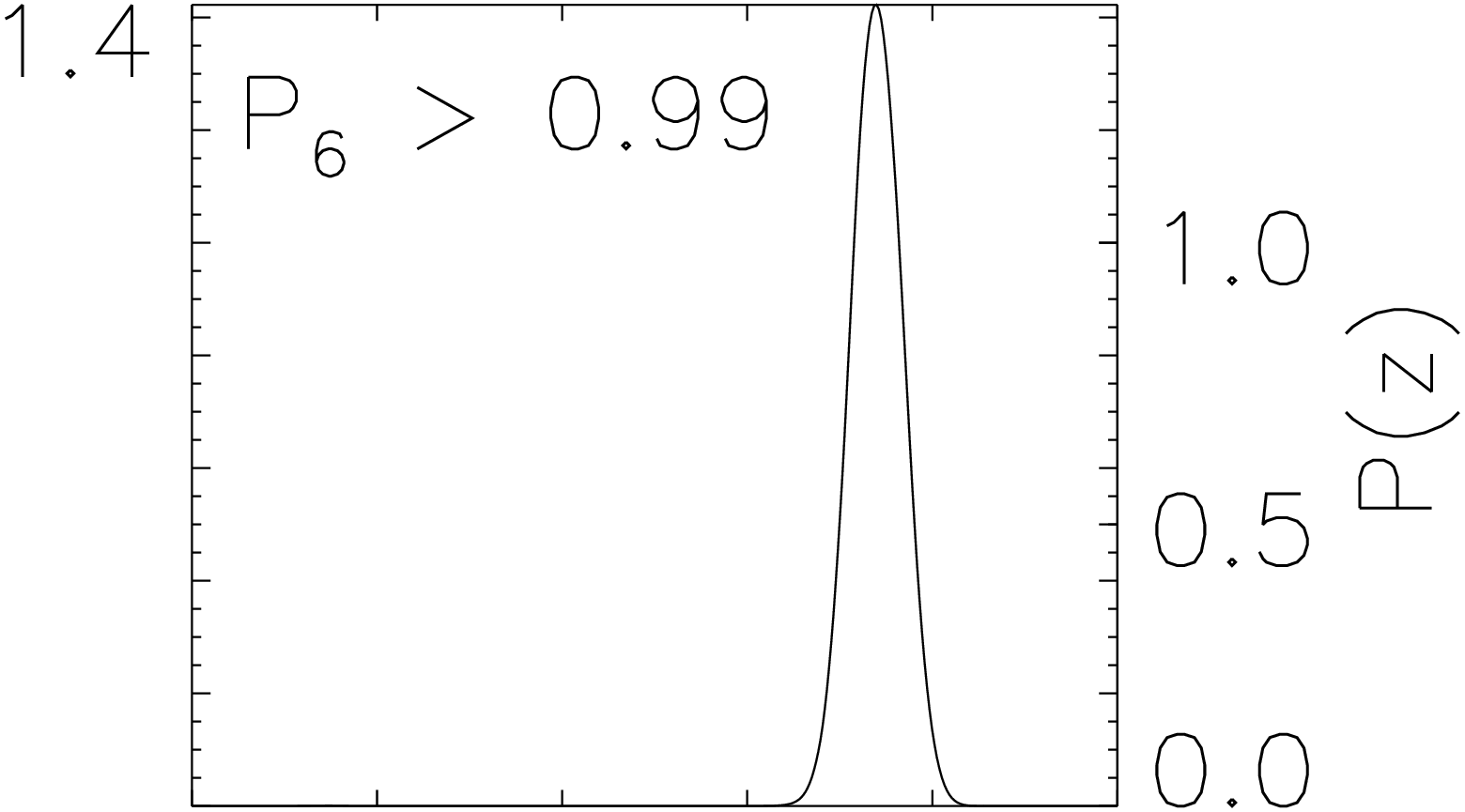}
\vspace{0.5mm}

\plotone{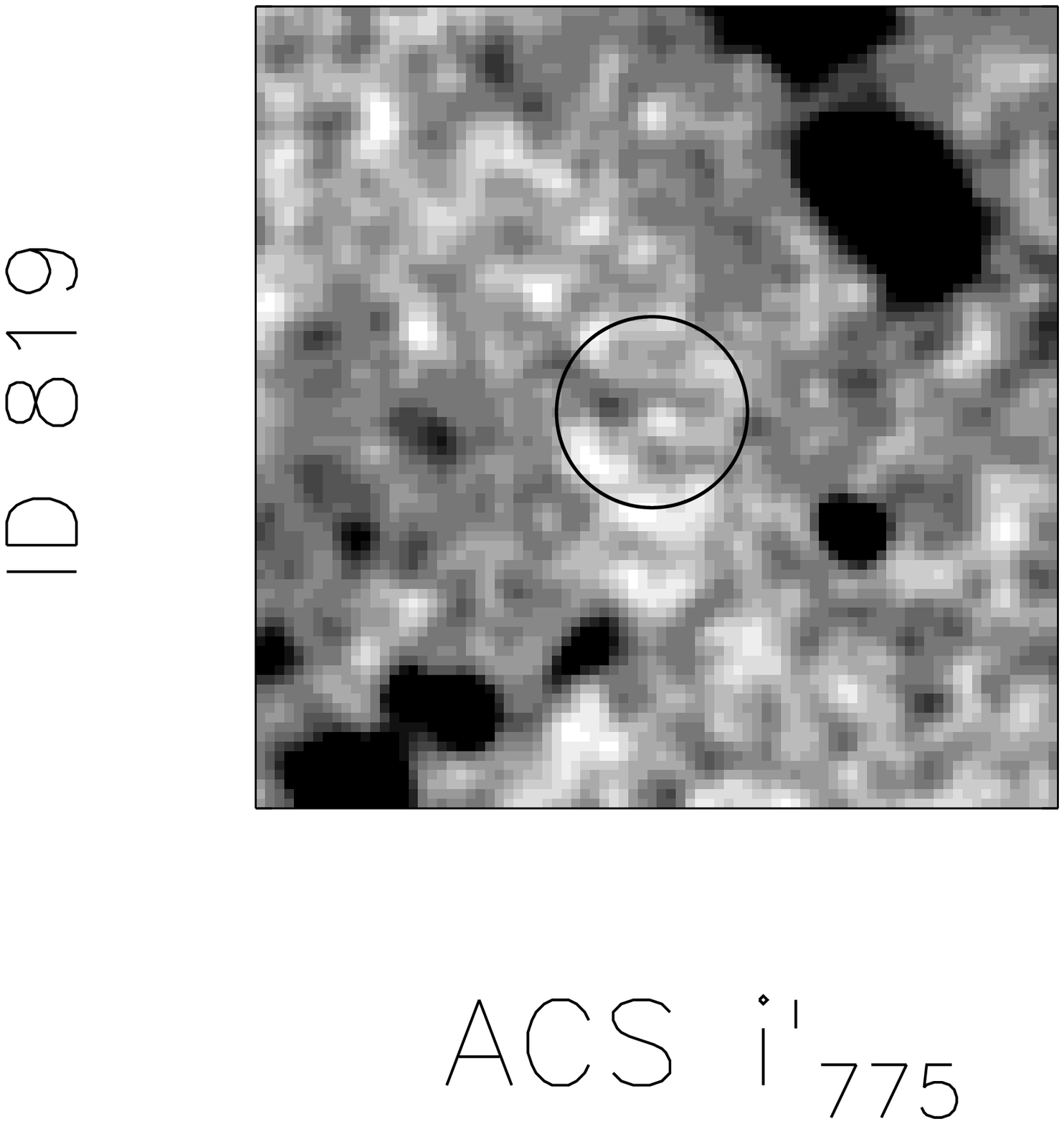}
\hspace{-10mm}
\plotone{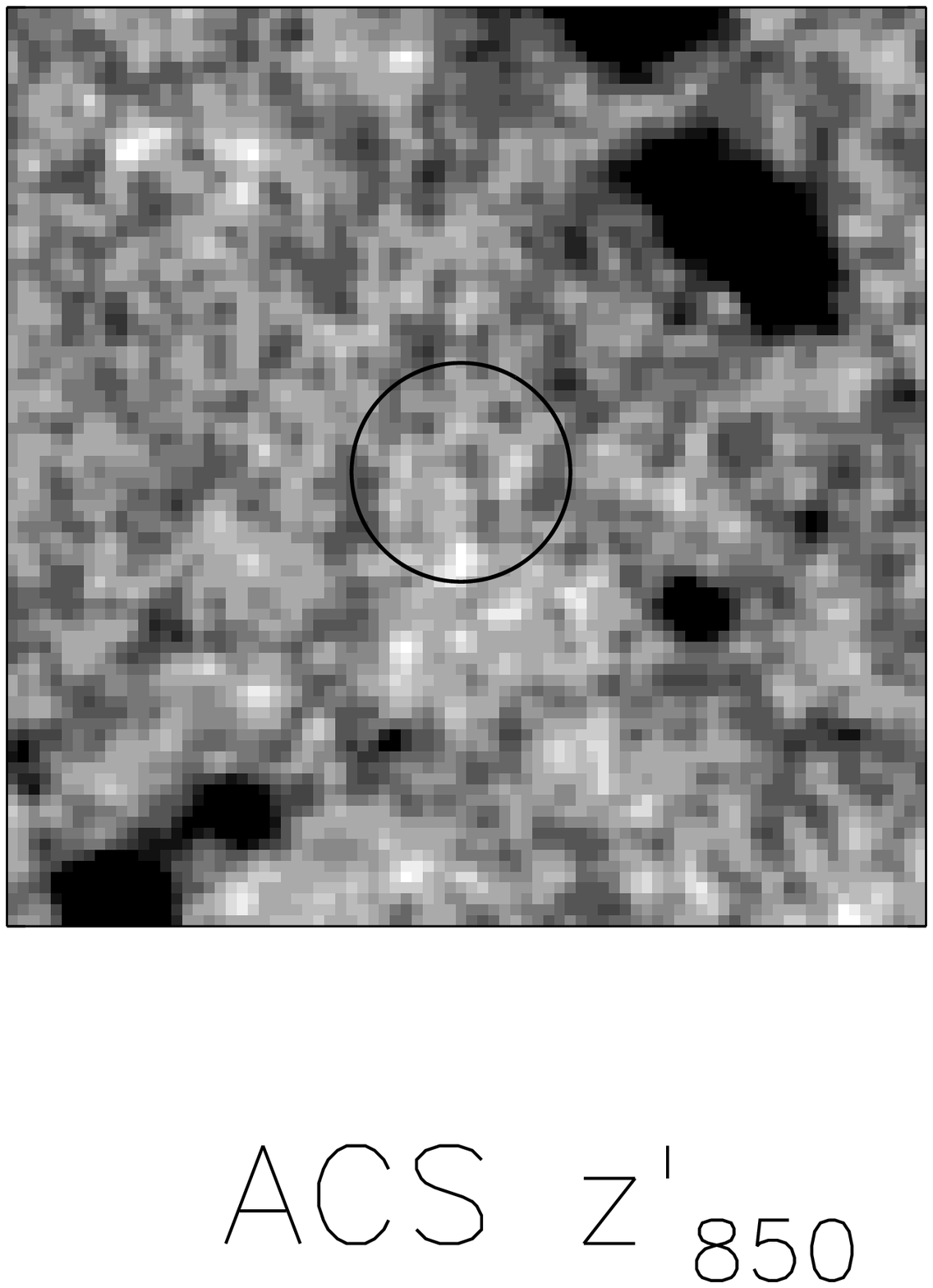}
\hspace{-10mm}
\plotone{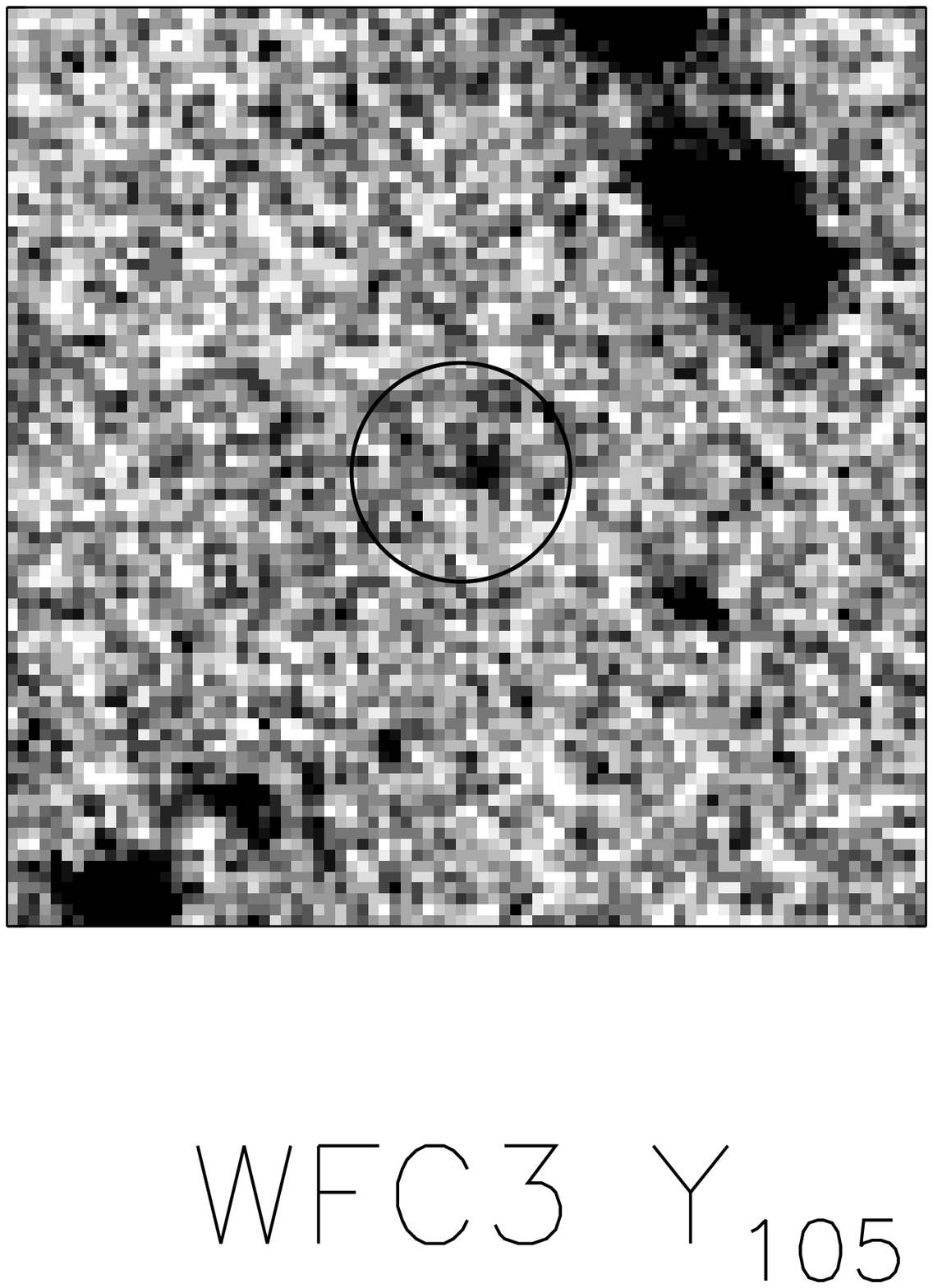}
\hspace{-10mm}
\plotone{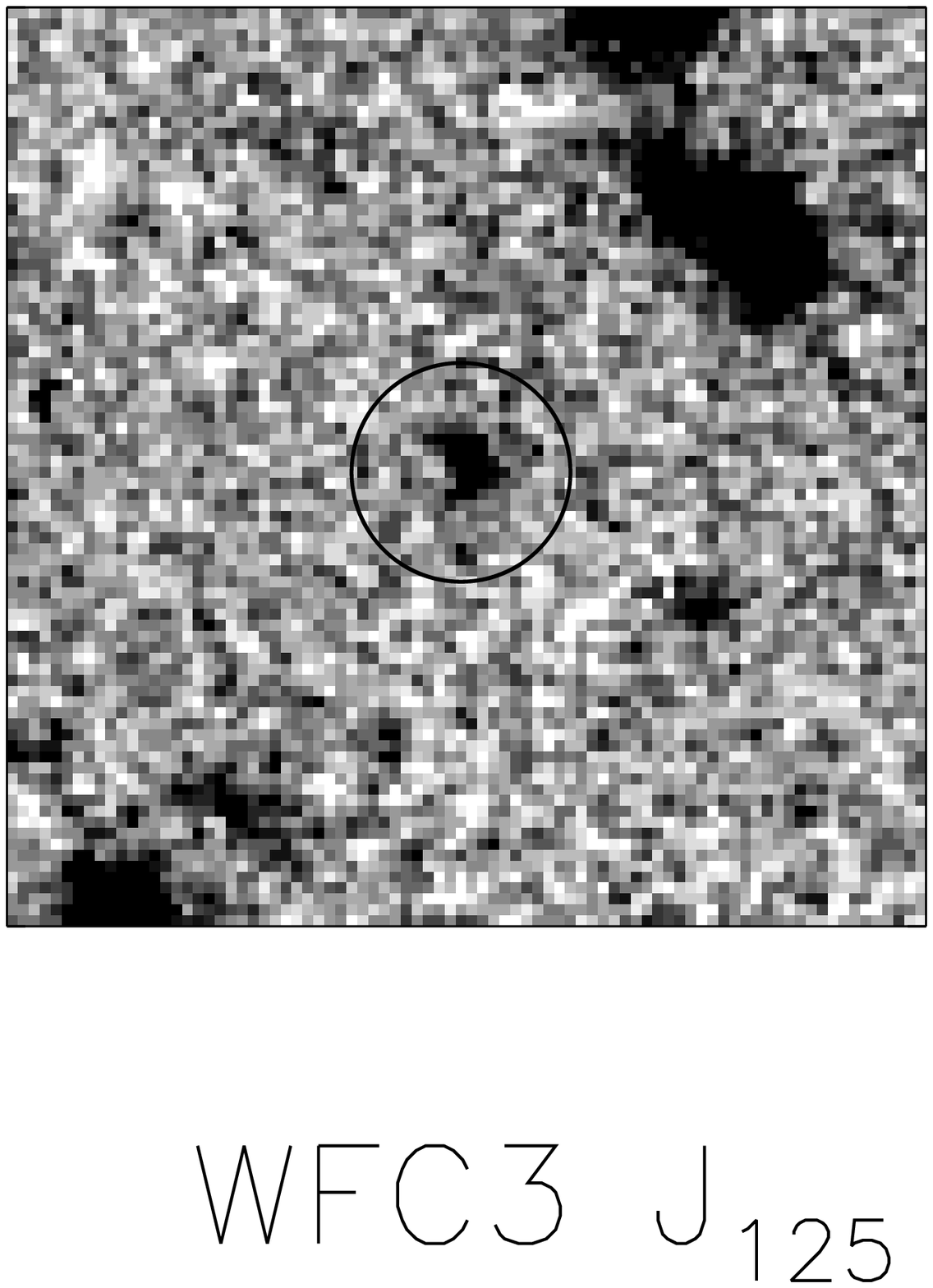}
\hspace{-10mm}
\plotone{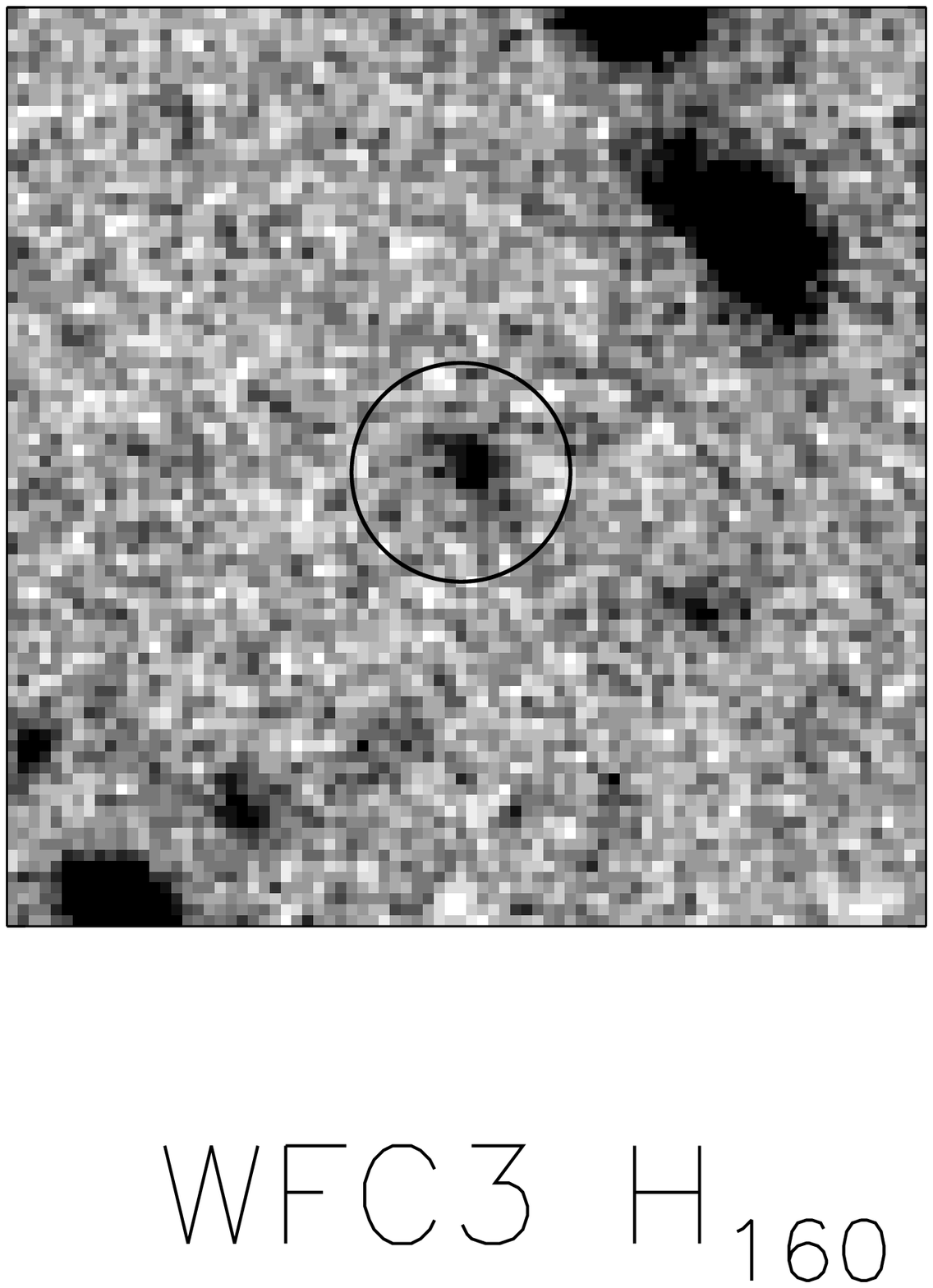}
\hspace{-10mm}
\plotone{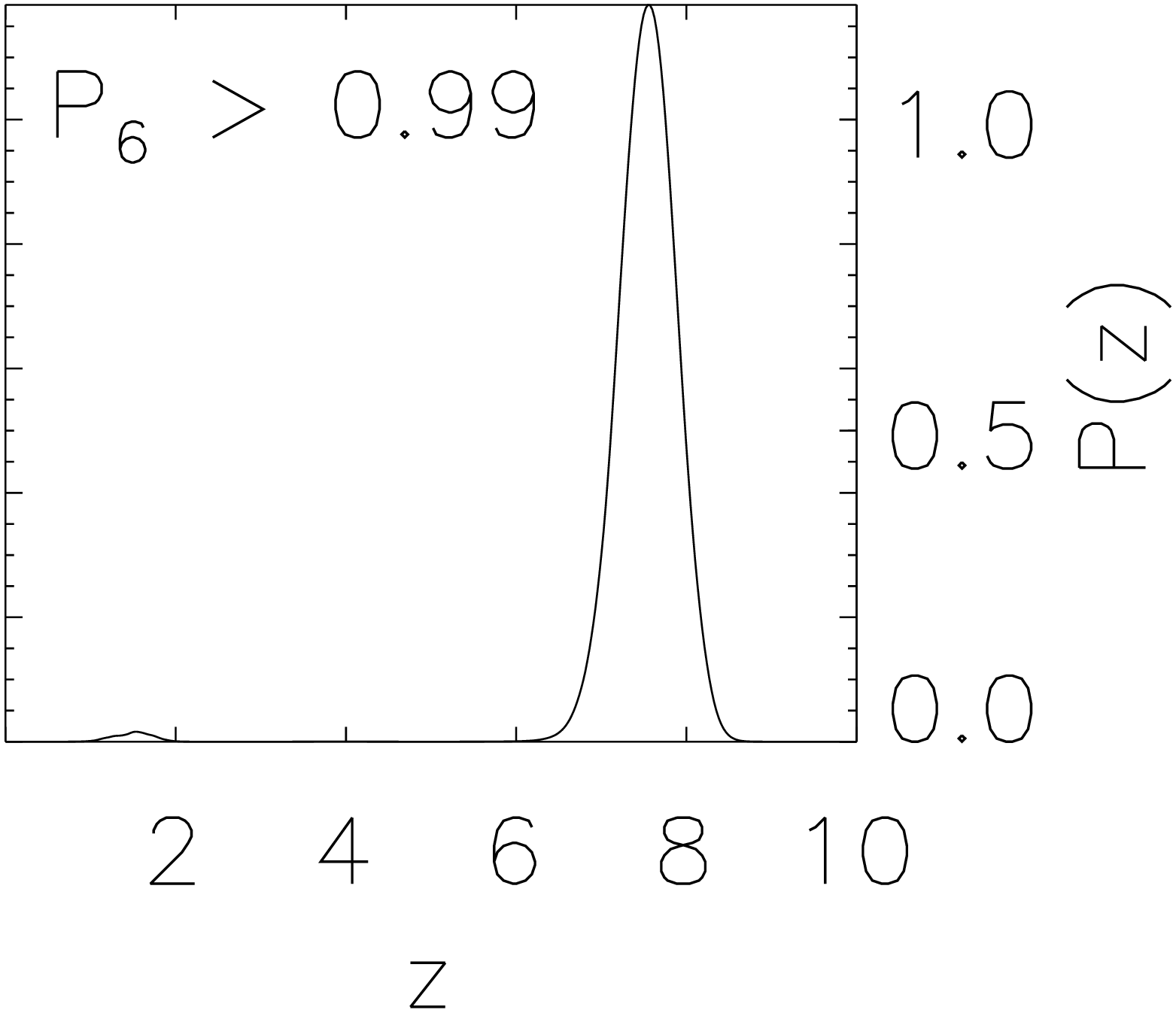}
\vspace{5mm}
\caption{Continued.}\label{stamps3}
\end{figure*}
\addtocounter{figure}{-1}

\begin{figure*}
\epsscale{0.18}
\vspace{2mm}
\plotone{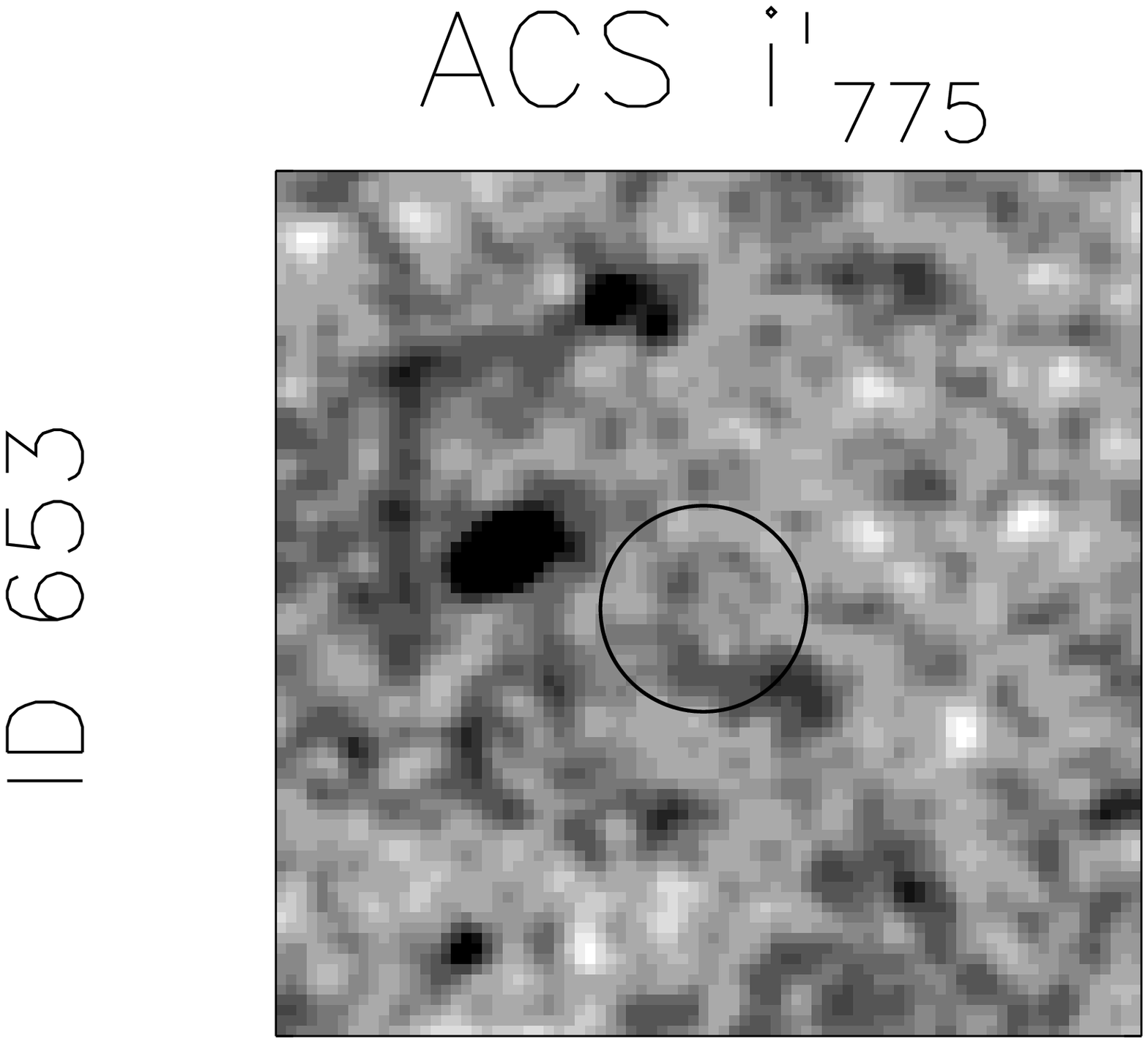}
\hspace{-10mm}
\plotone{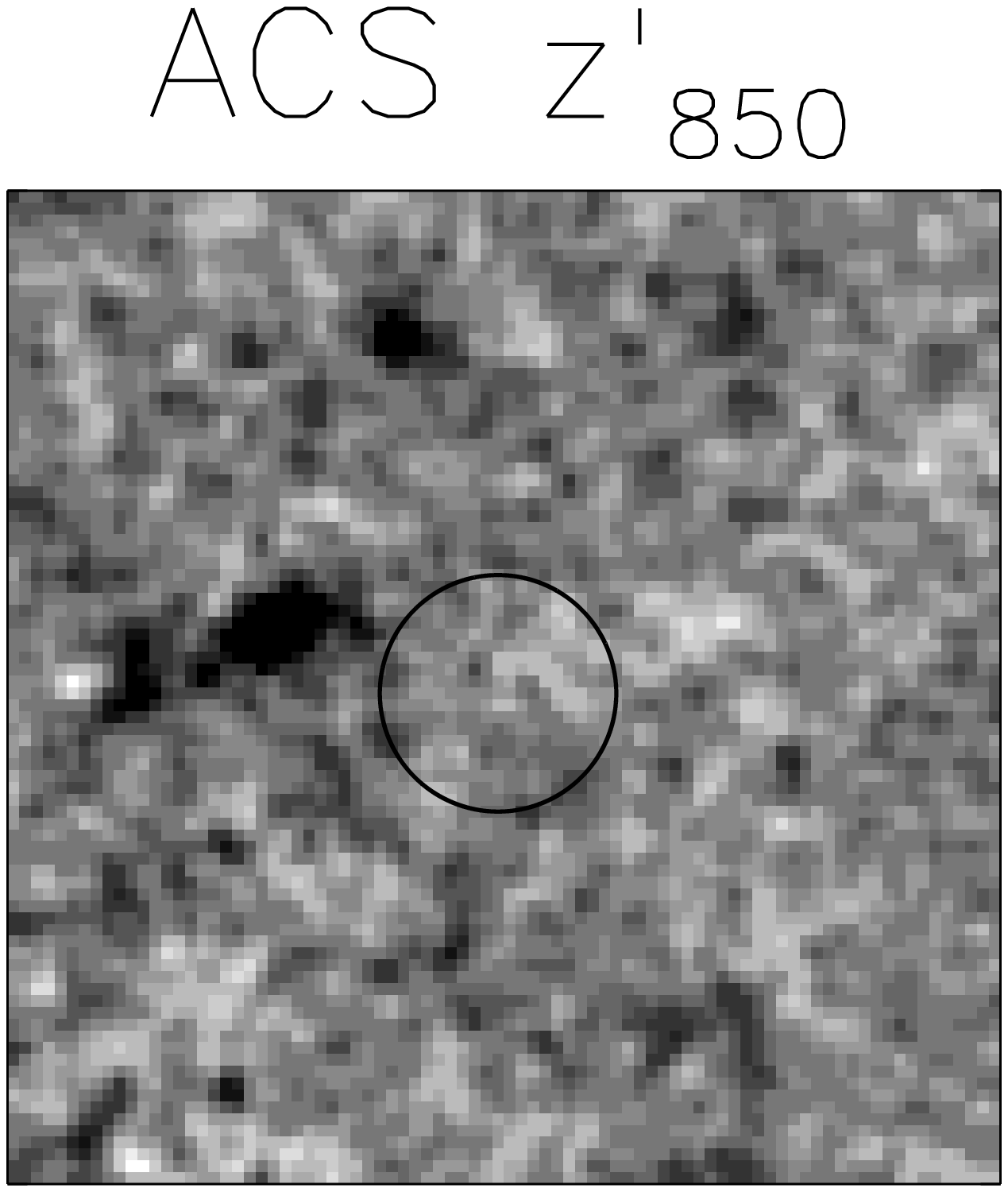}
\hspace{-10mm}
\plotone{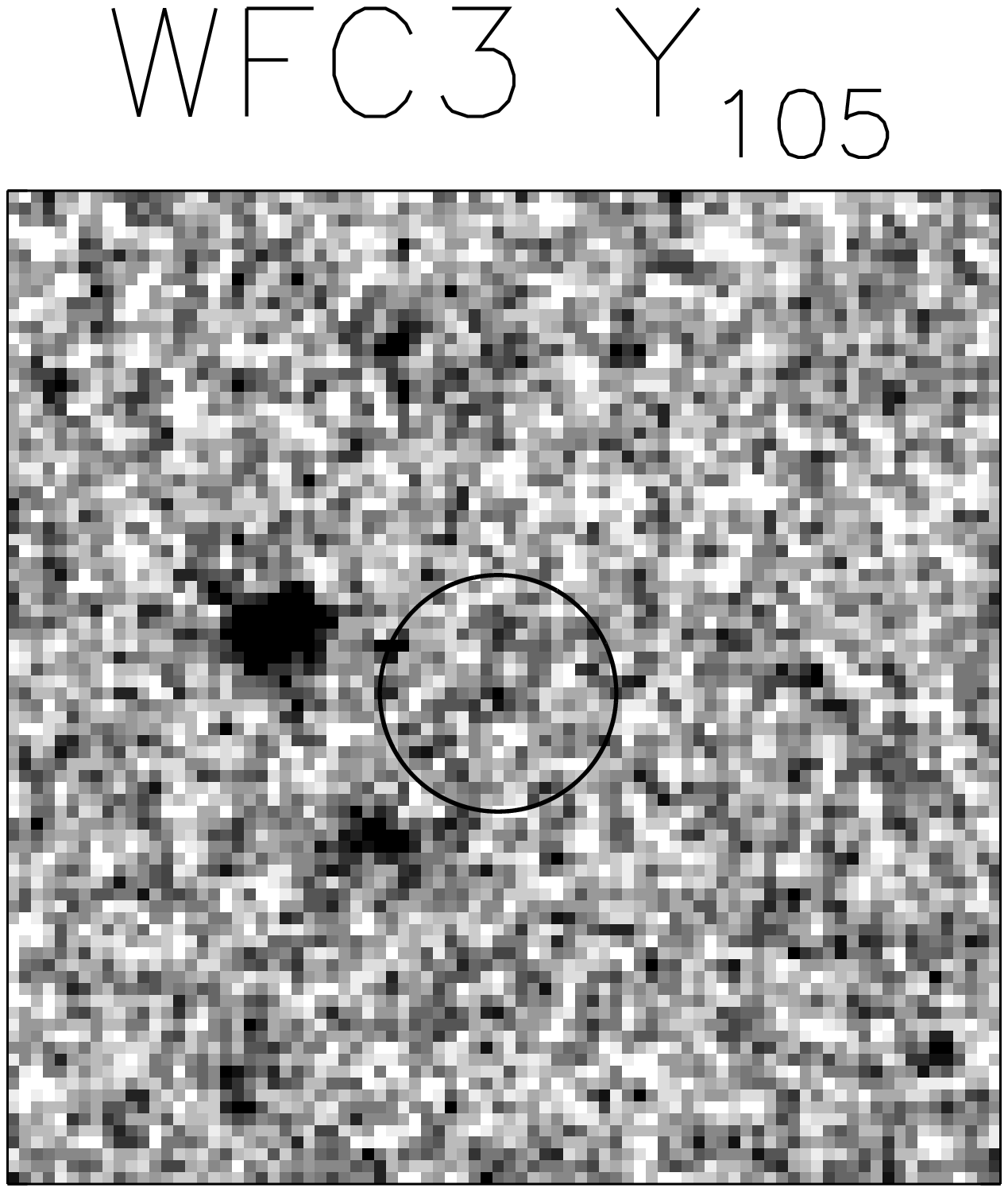}
\hspace{-10mm}
\plotone{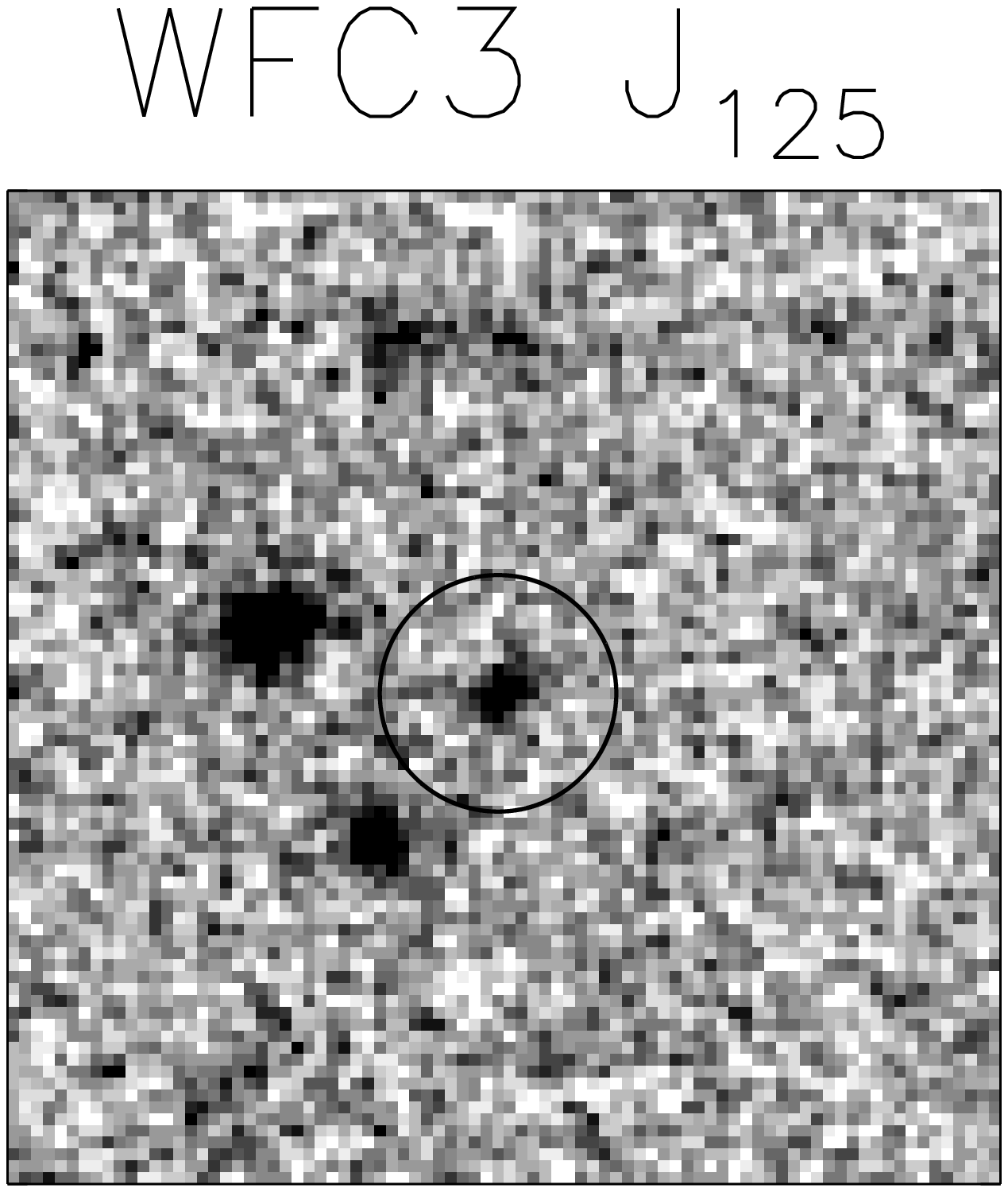}
\hspace{-10mm}
\plotone{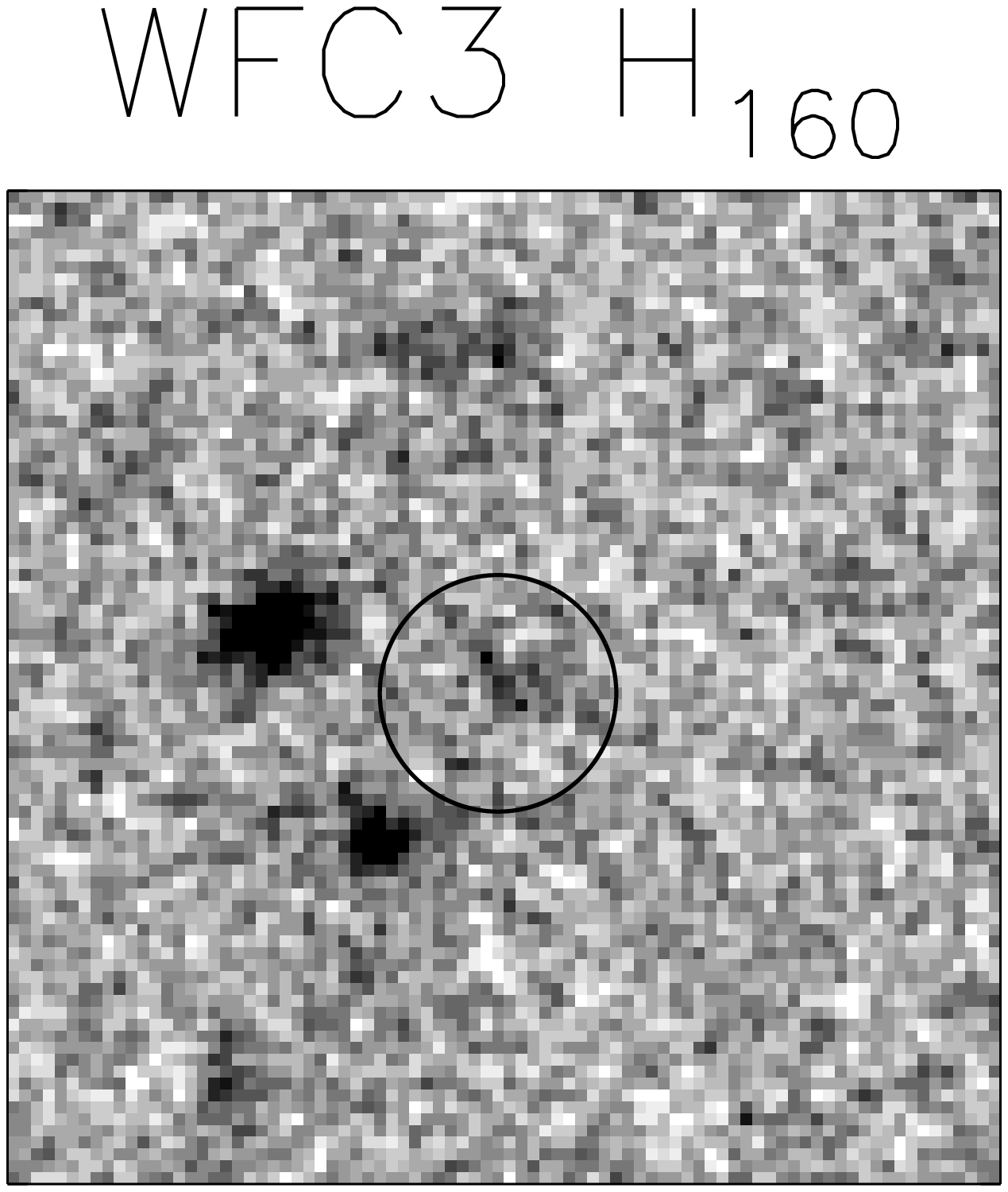}
\hspace{-10mm}
\plotone{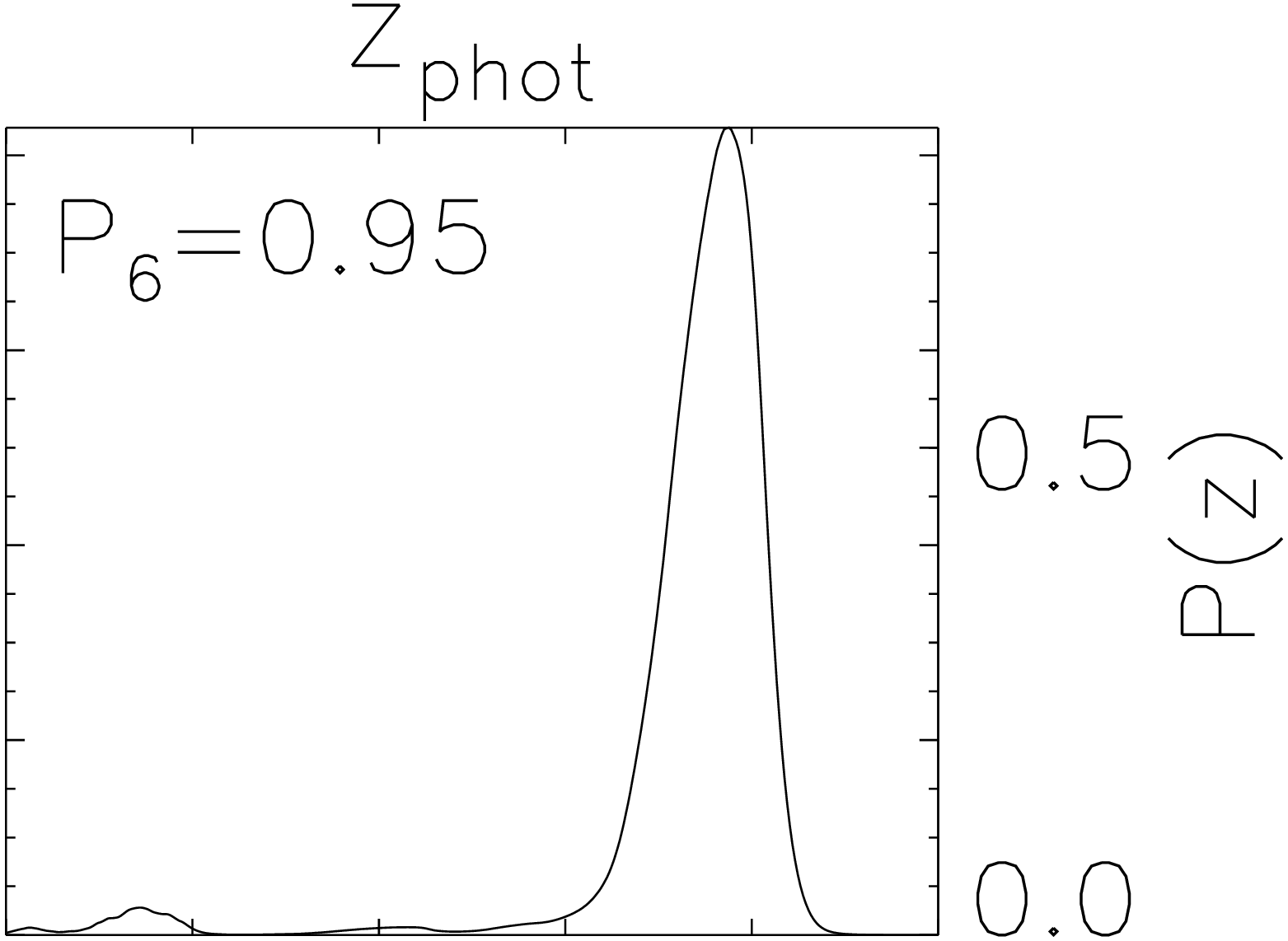}
\vspace{0.5mm}

\plotone{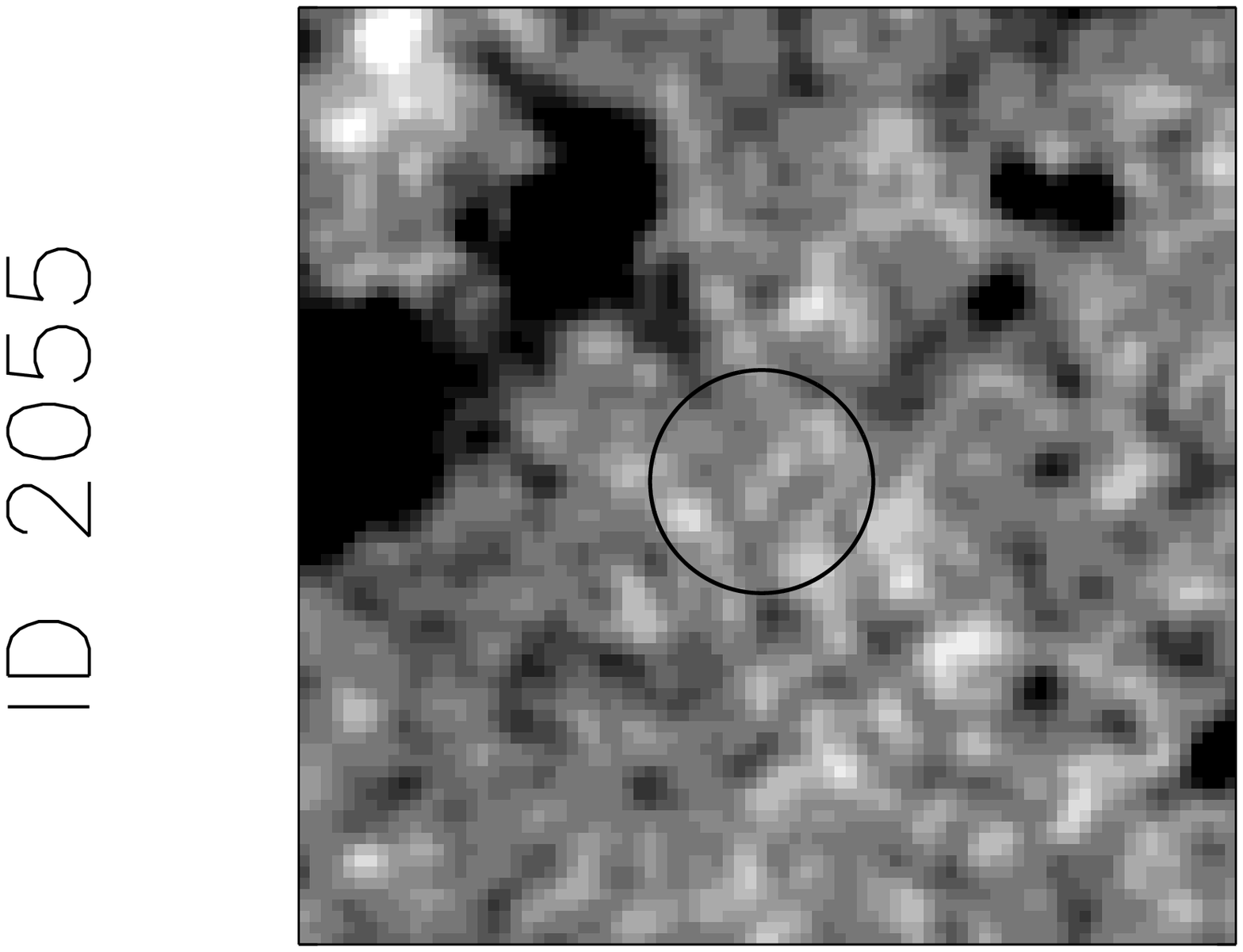}
\hspace{-10mm}
\plotone{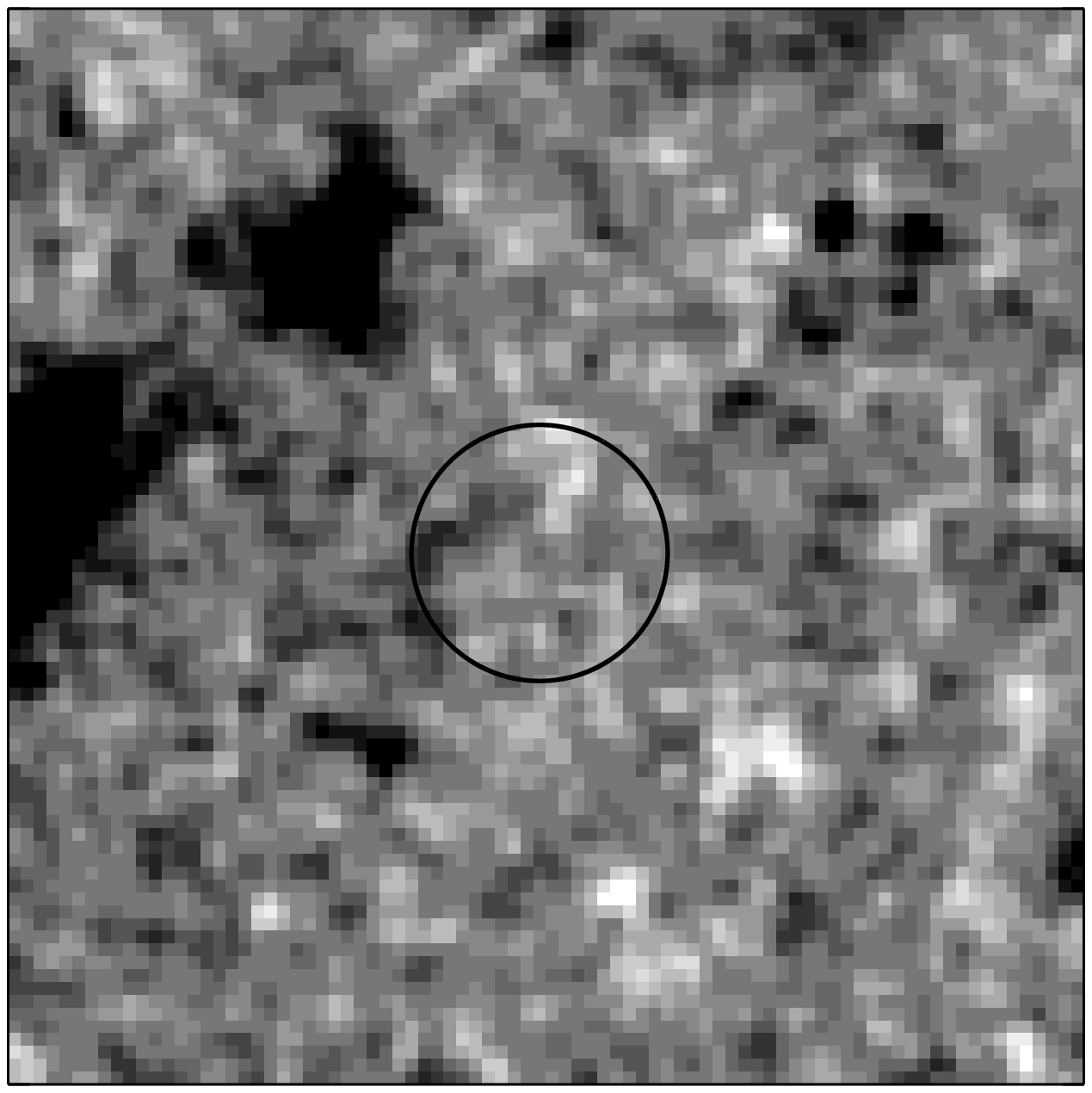}
\hspace{-10mm}
\plotone{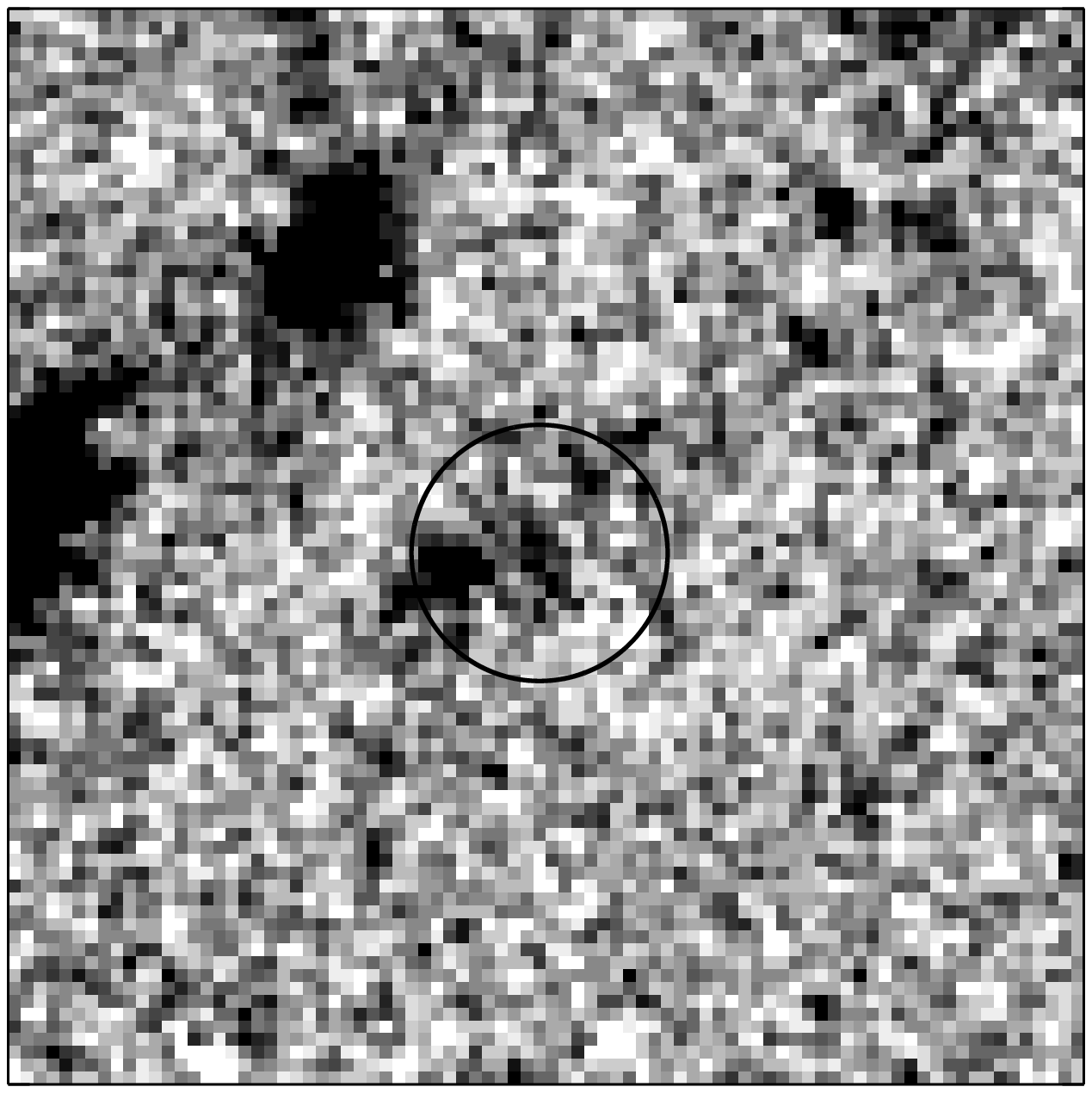}
\hspace{-10mm}
\plotone{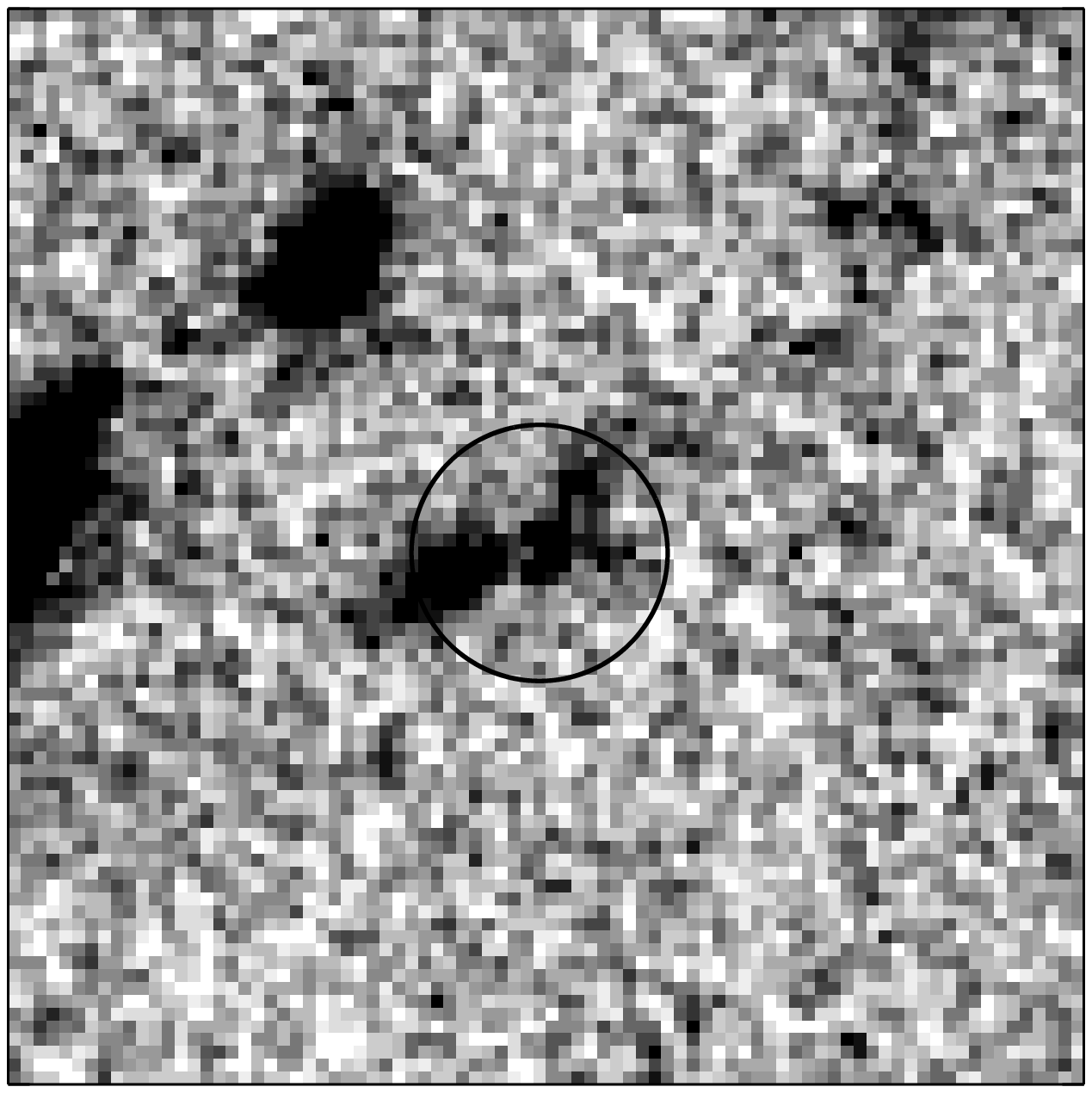}
\hspace{-10mm}
\plotone{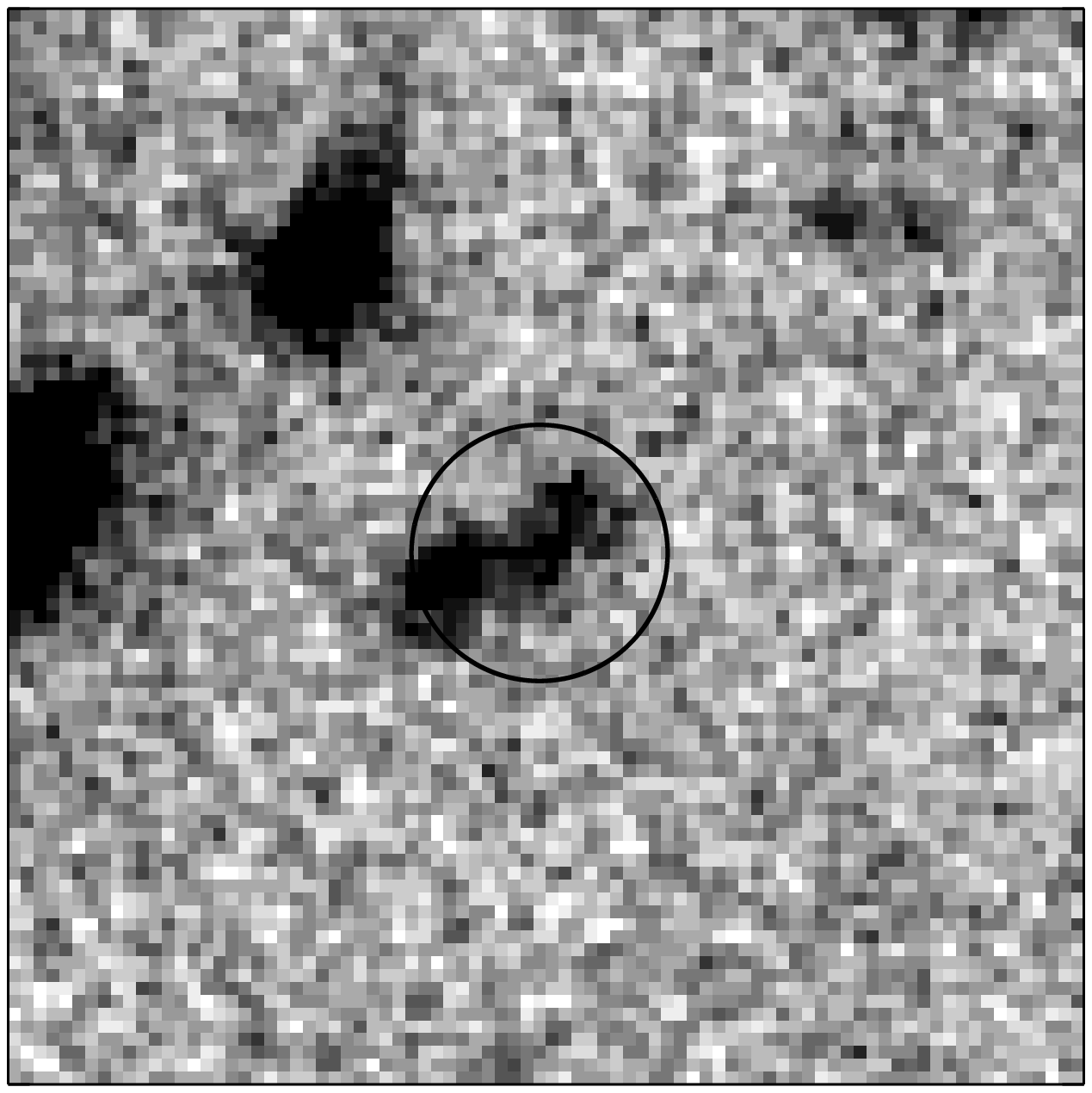}
\hspace{-10mm}
\plotone{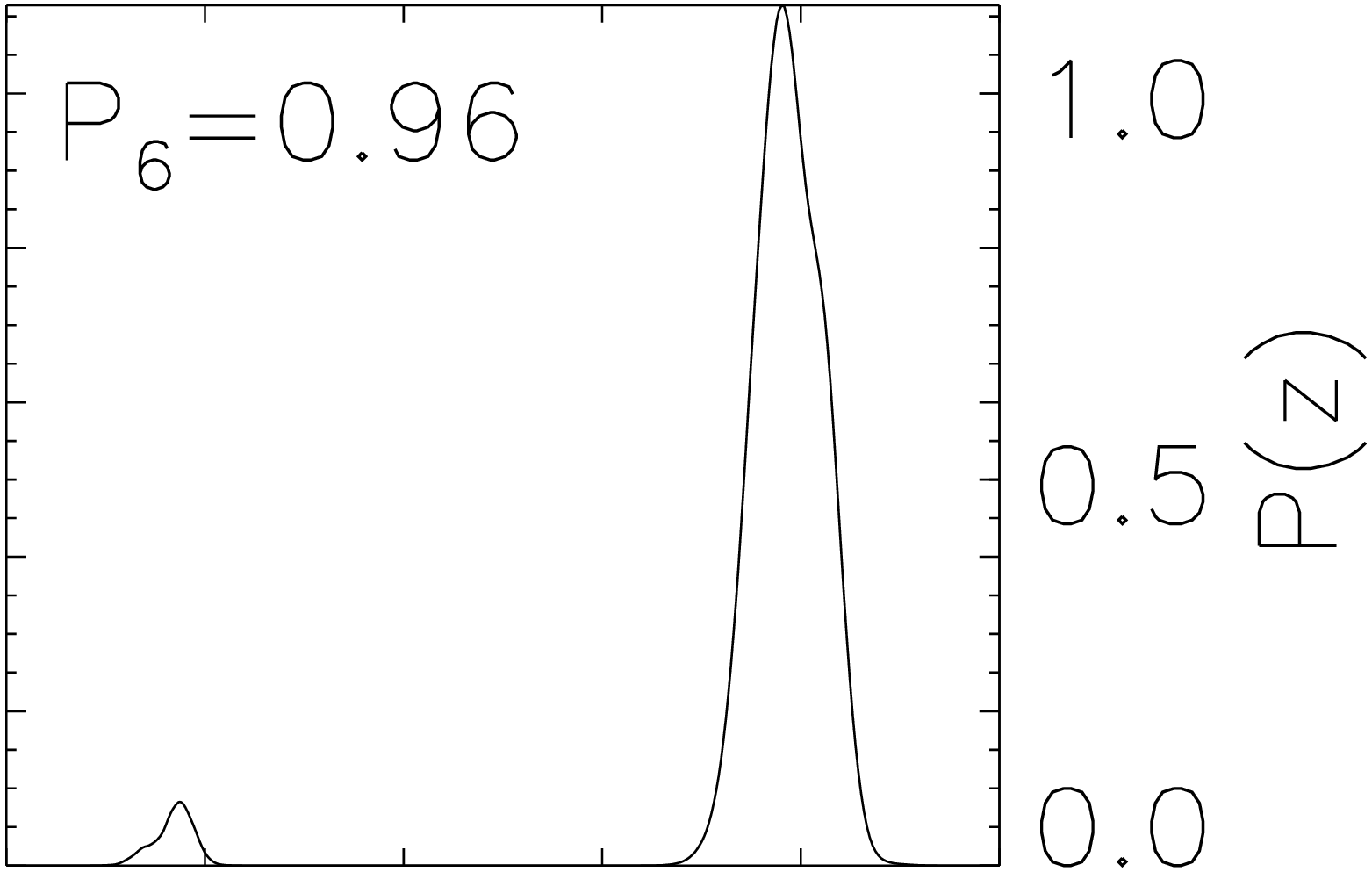}
\vspace{0.5mm}

\plotone{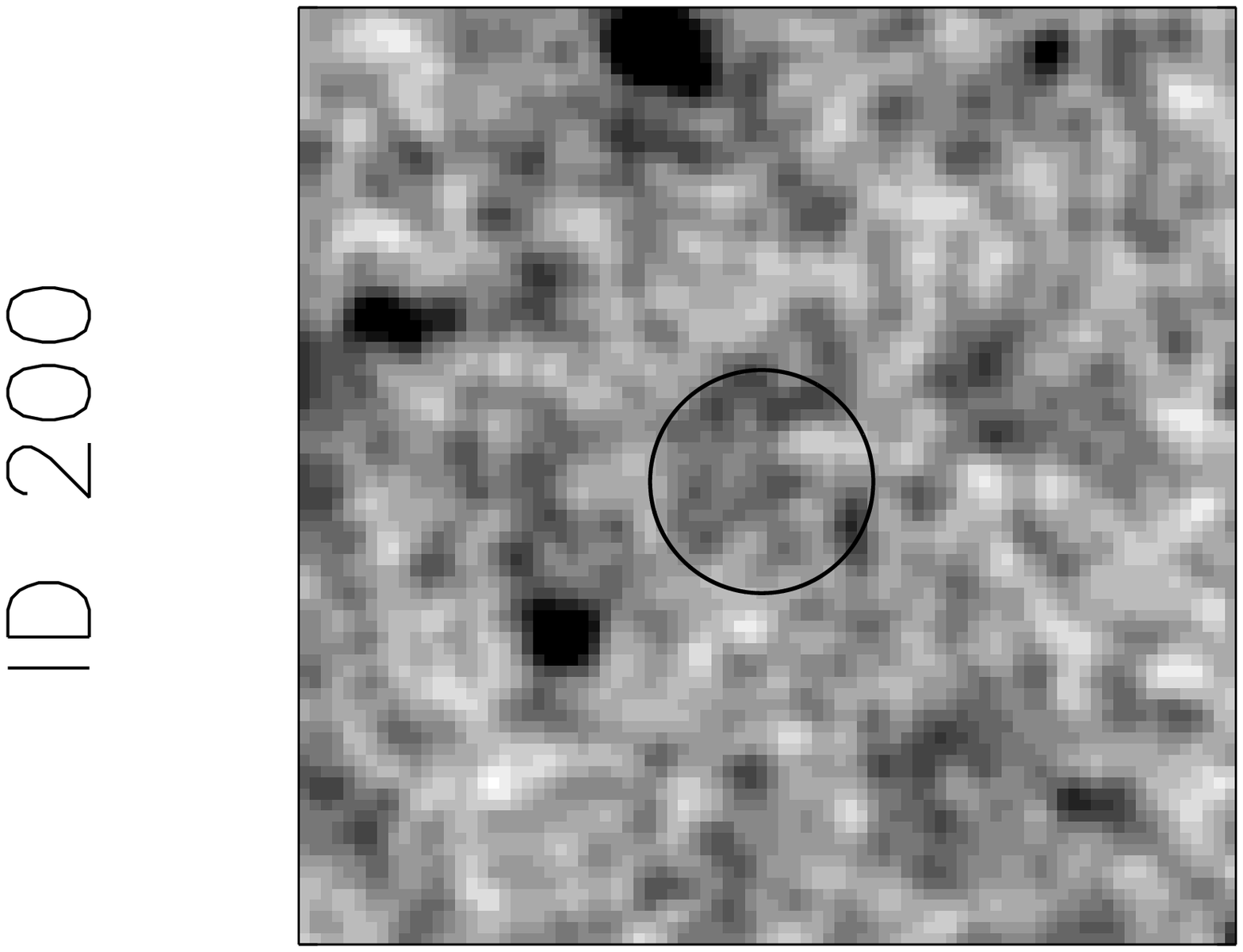}
\hspace{-10mm}
\plotone{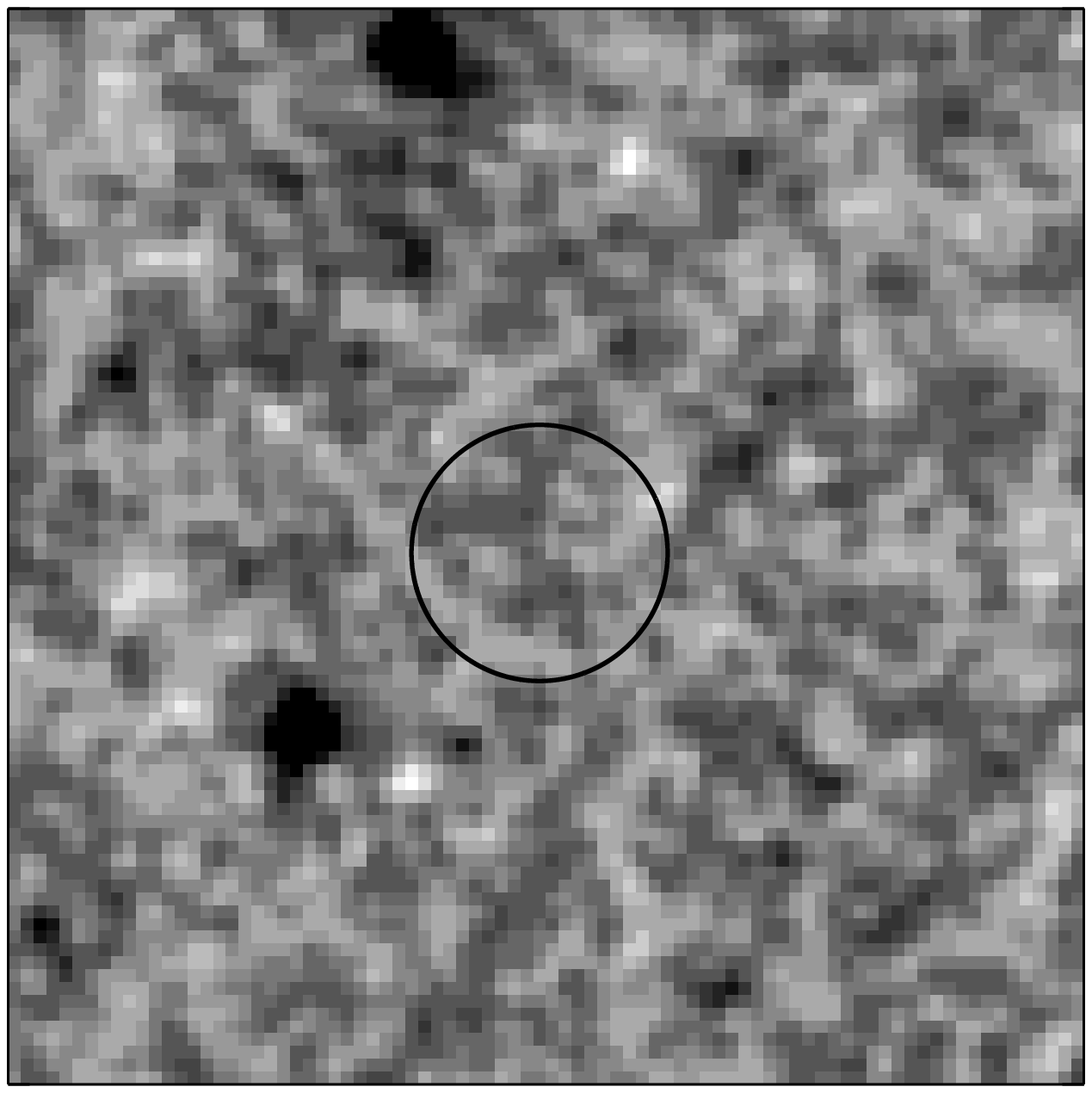}
\hspace{-10mm}
\plotone{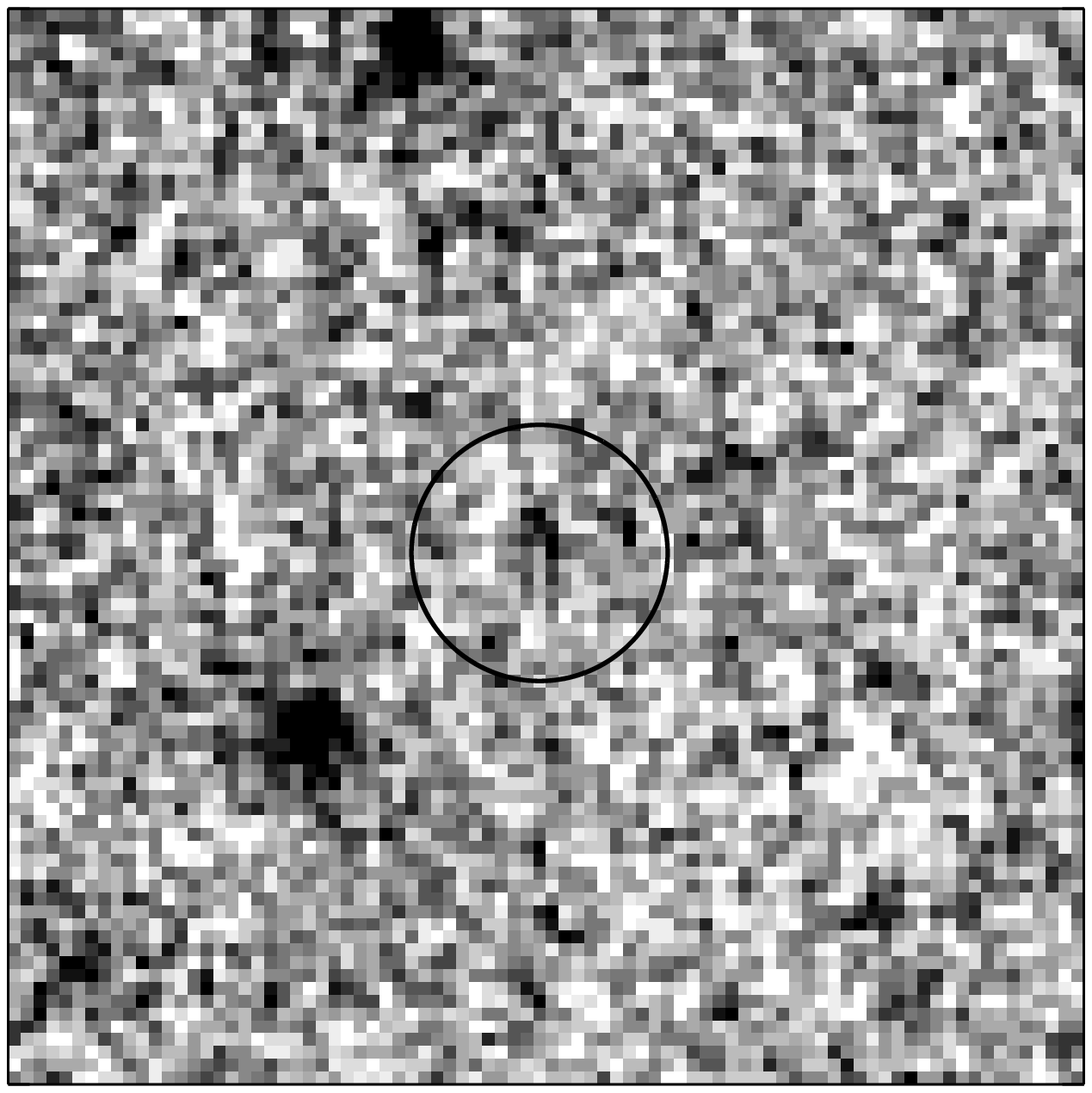}
\hspace{-10mm}
\plotone{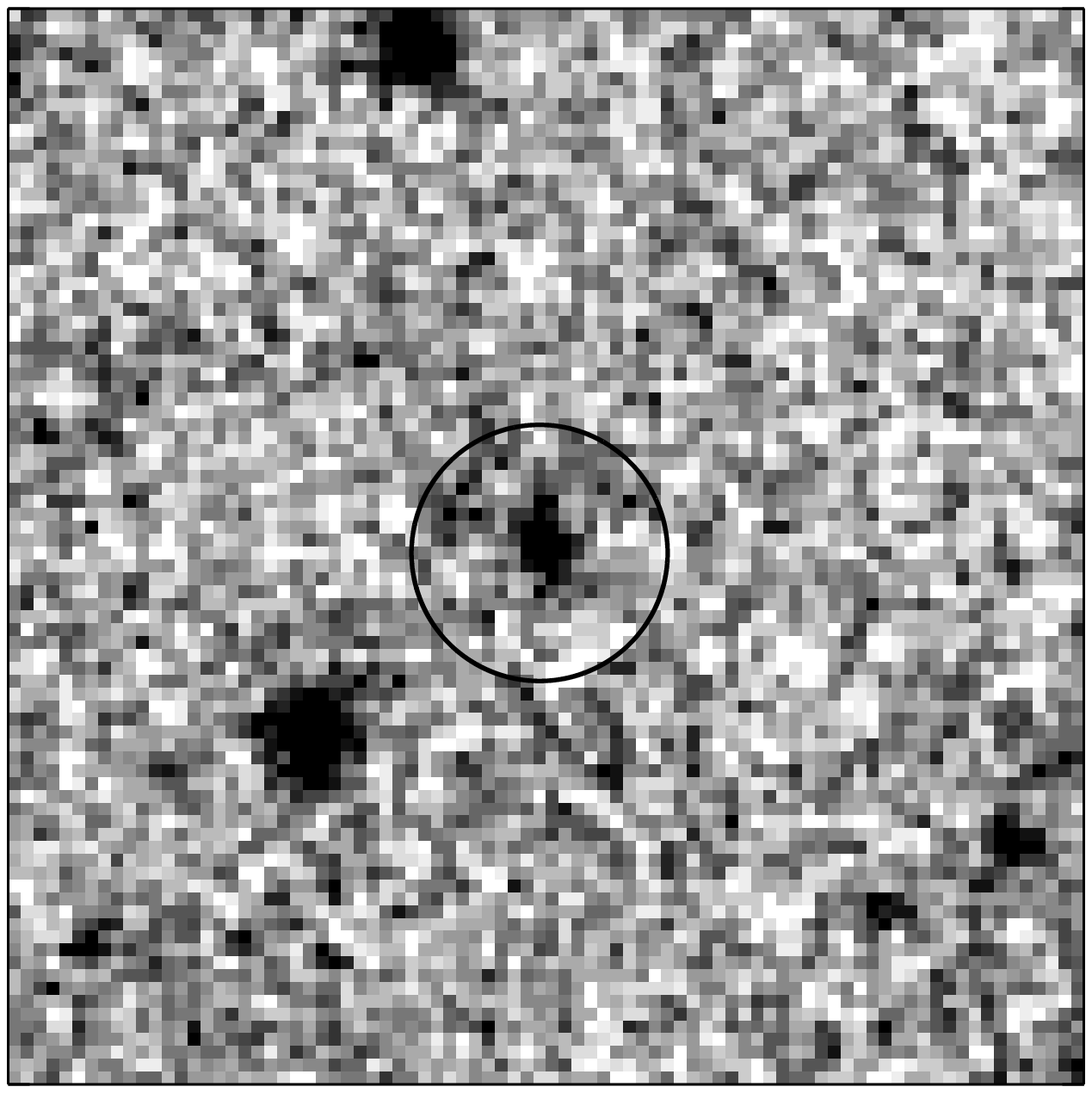}
\hspace{-10mm}
\plotone{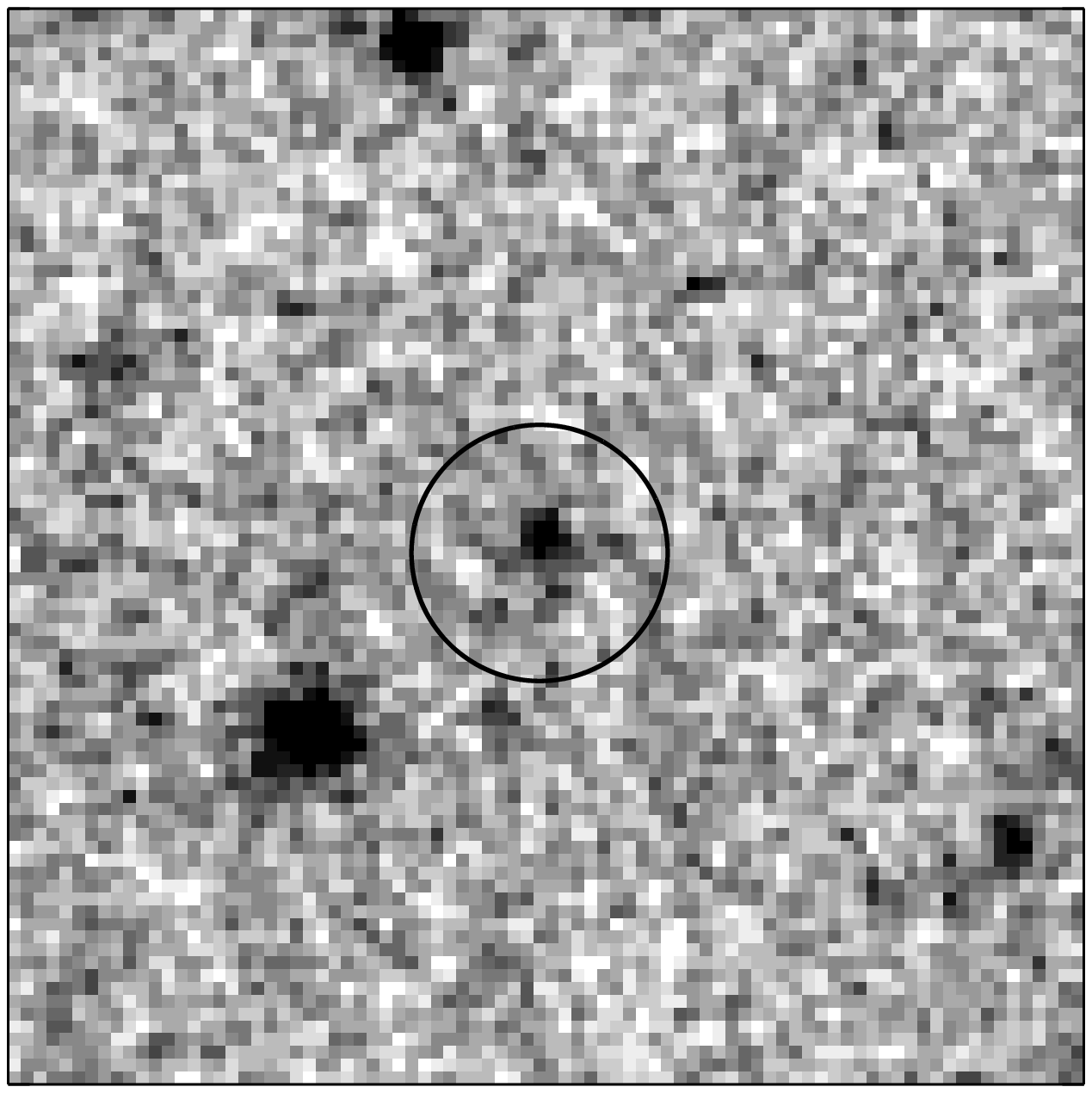}
\hspace{-10mm}
\plotone{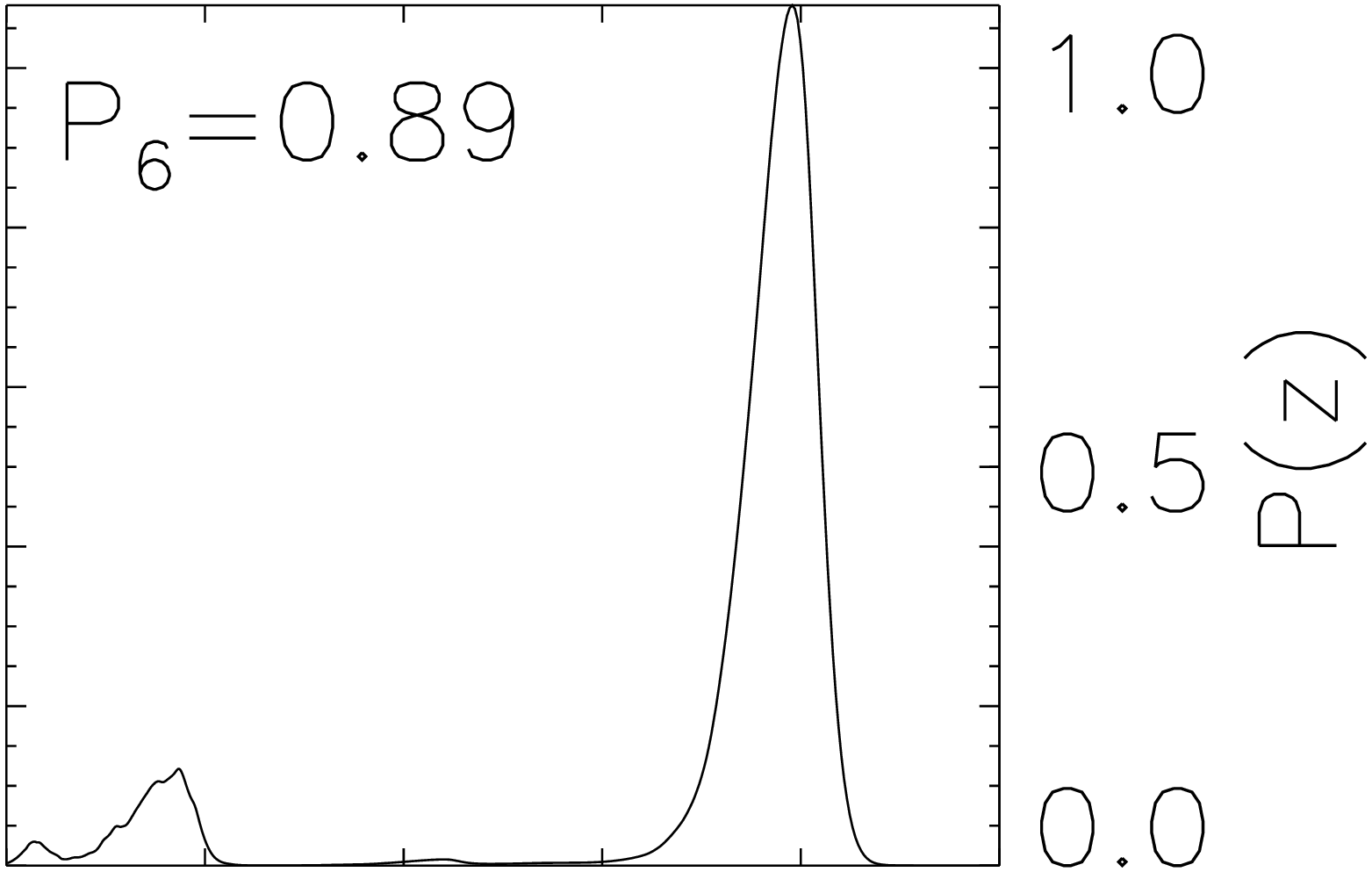}
\vspace{0.5mm}

\plotone{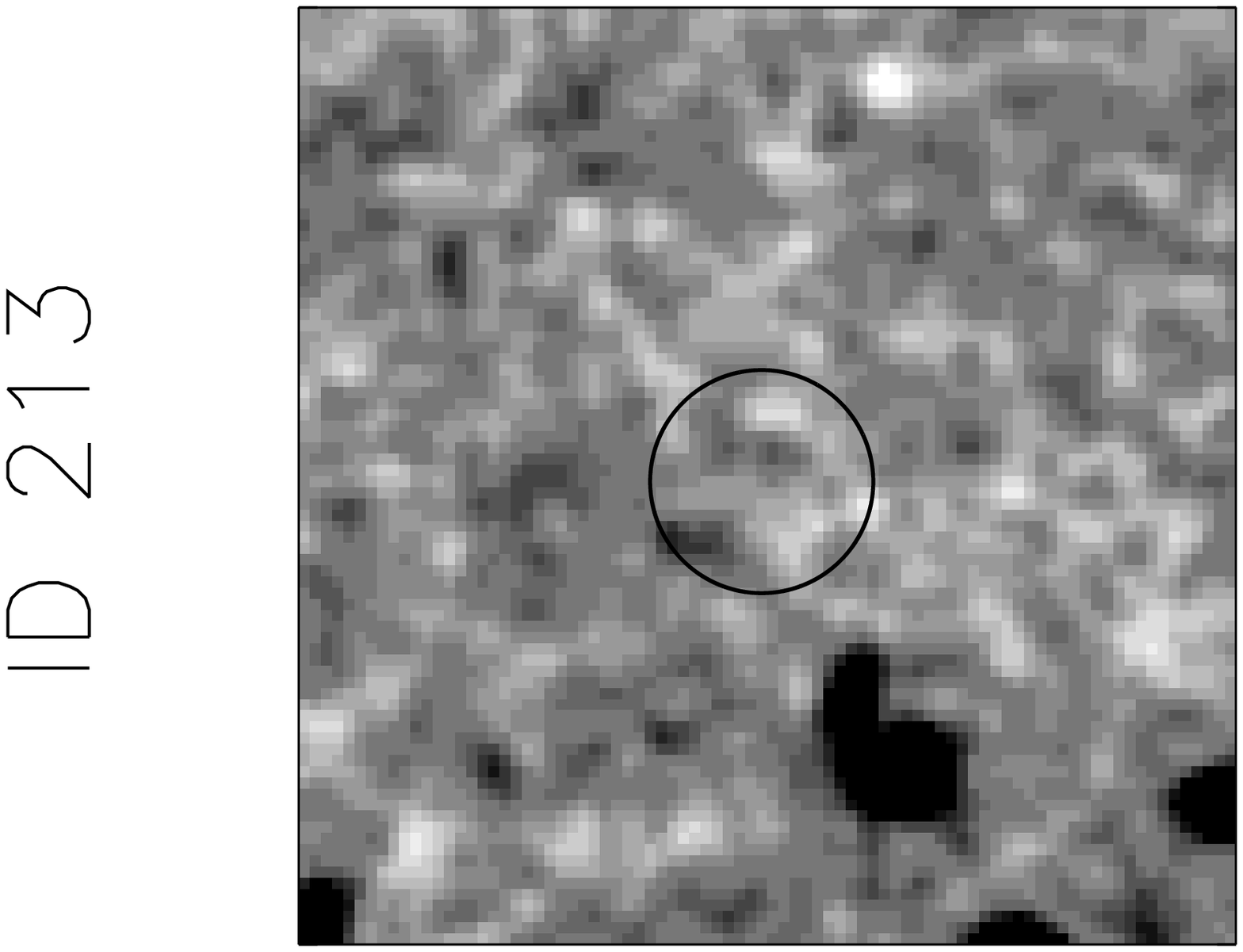}
\hspace{-10mm}
\plotone{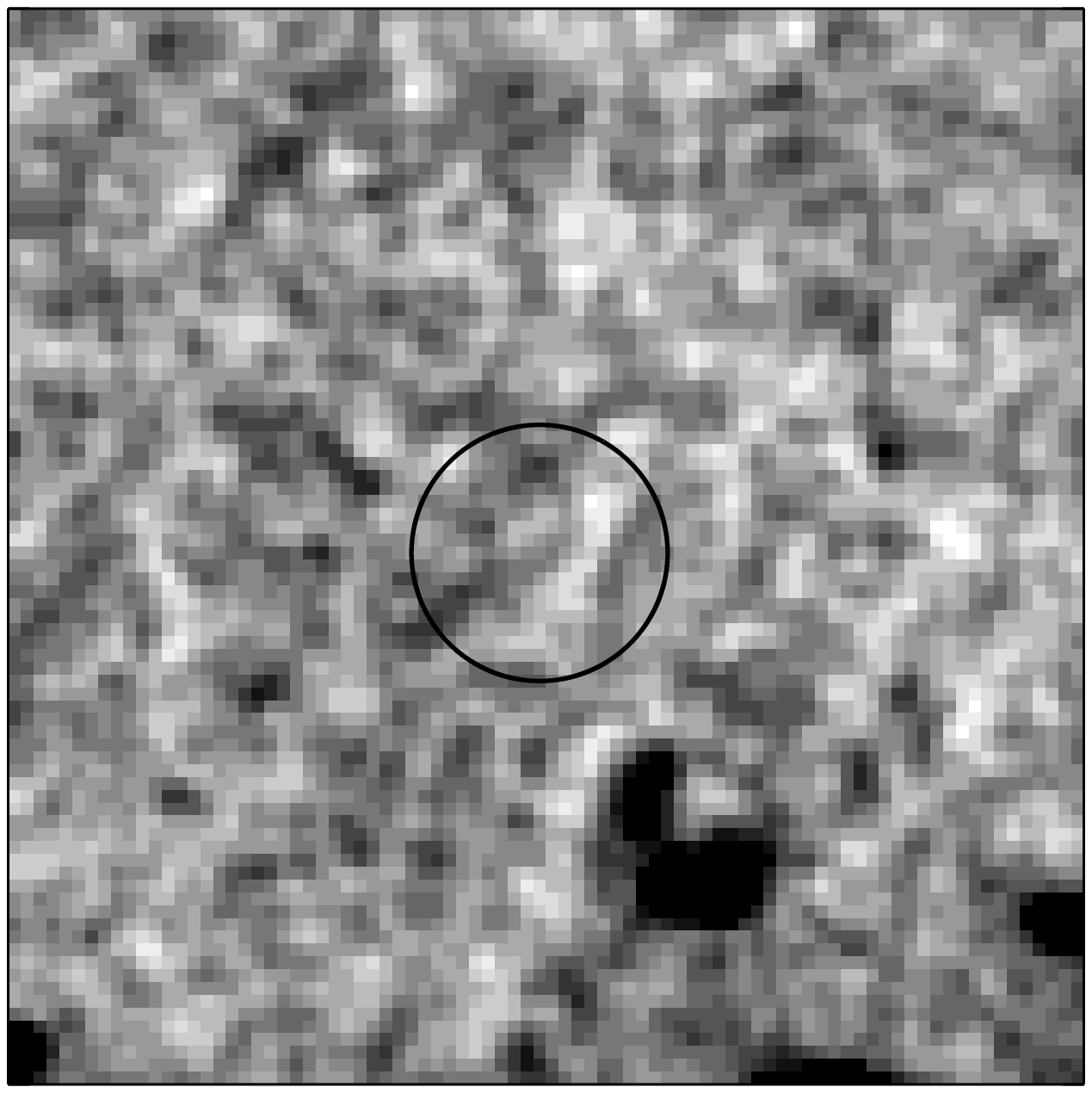}
\hspace{-10mm}
\plotone{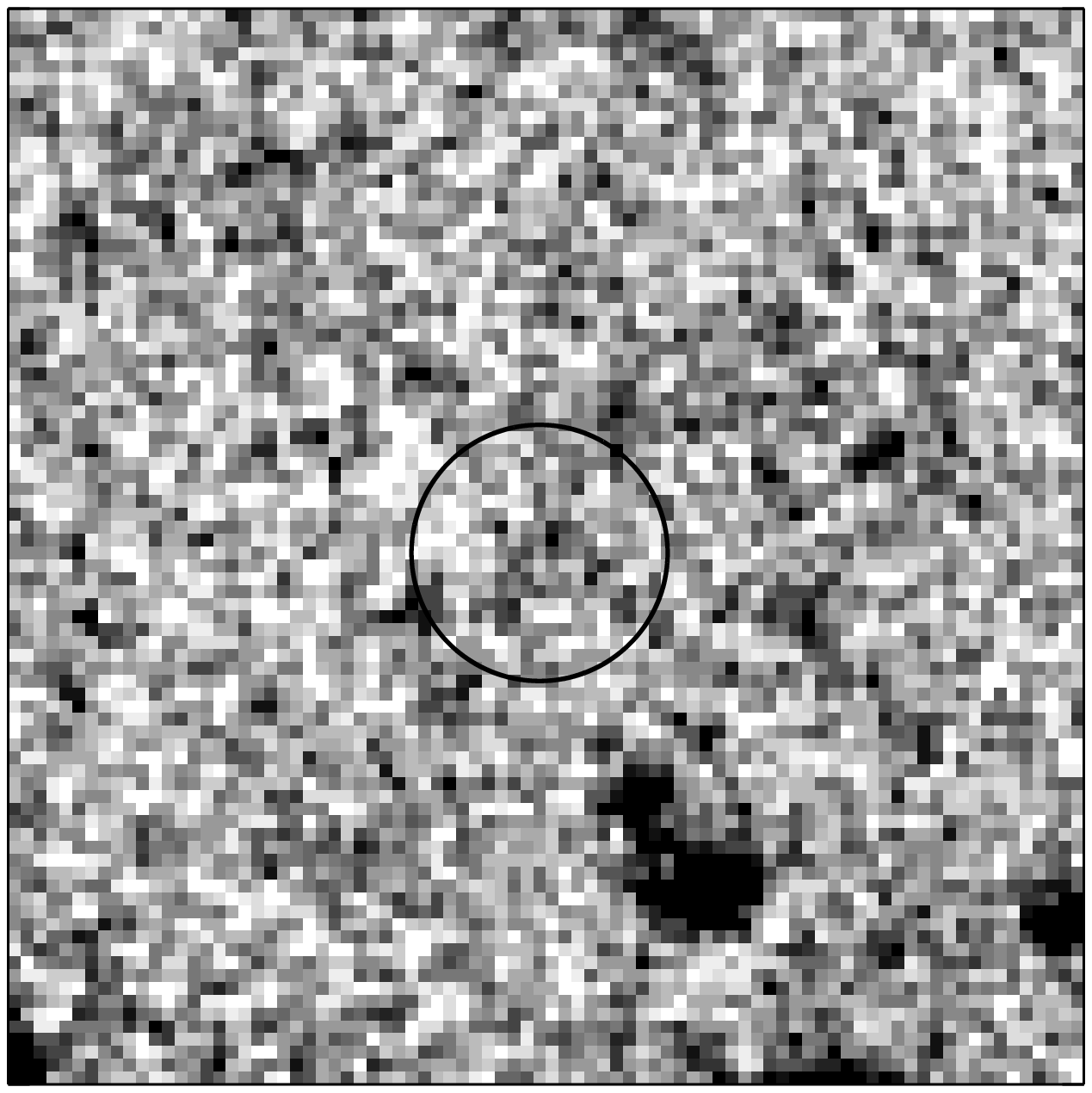}
\hspace{-10mm}
\plotone{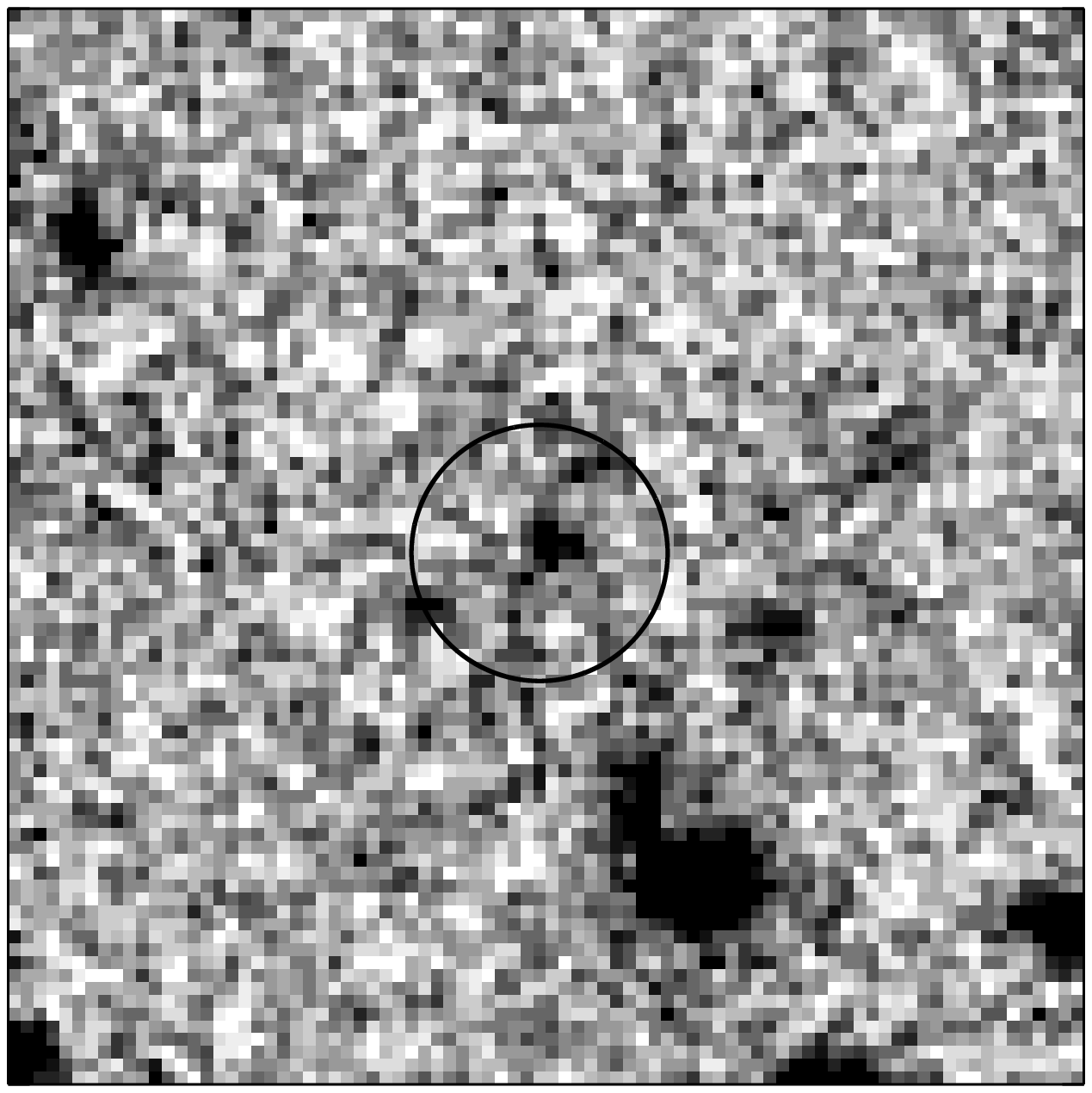}
\hspace{-10mm}
\plotone{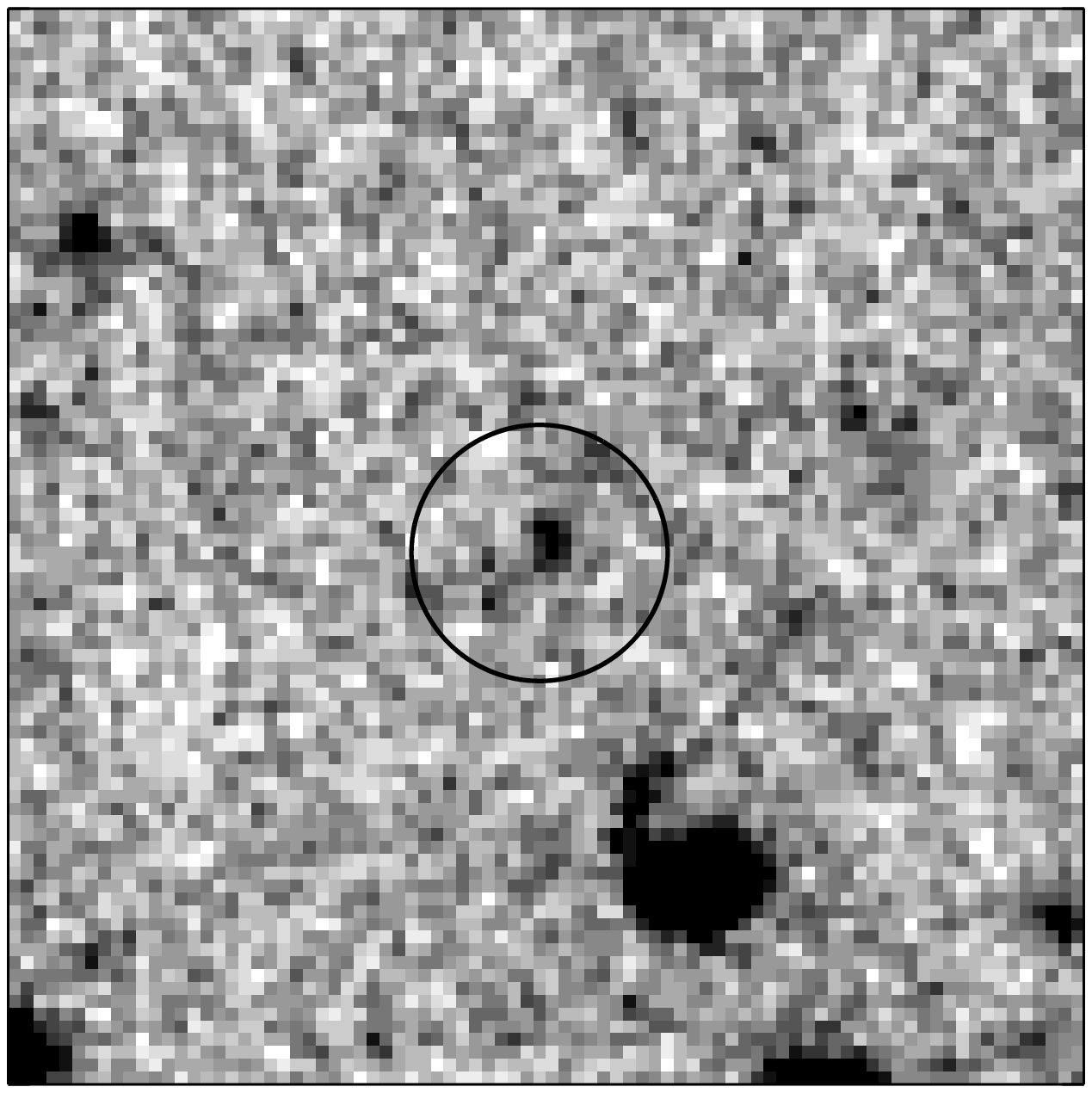}
\hspace{-10mm}
\plotone{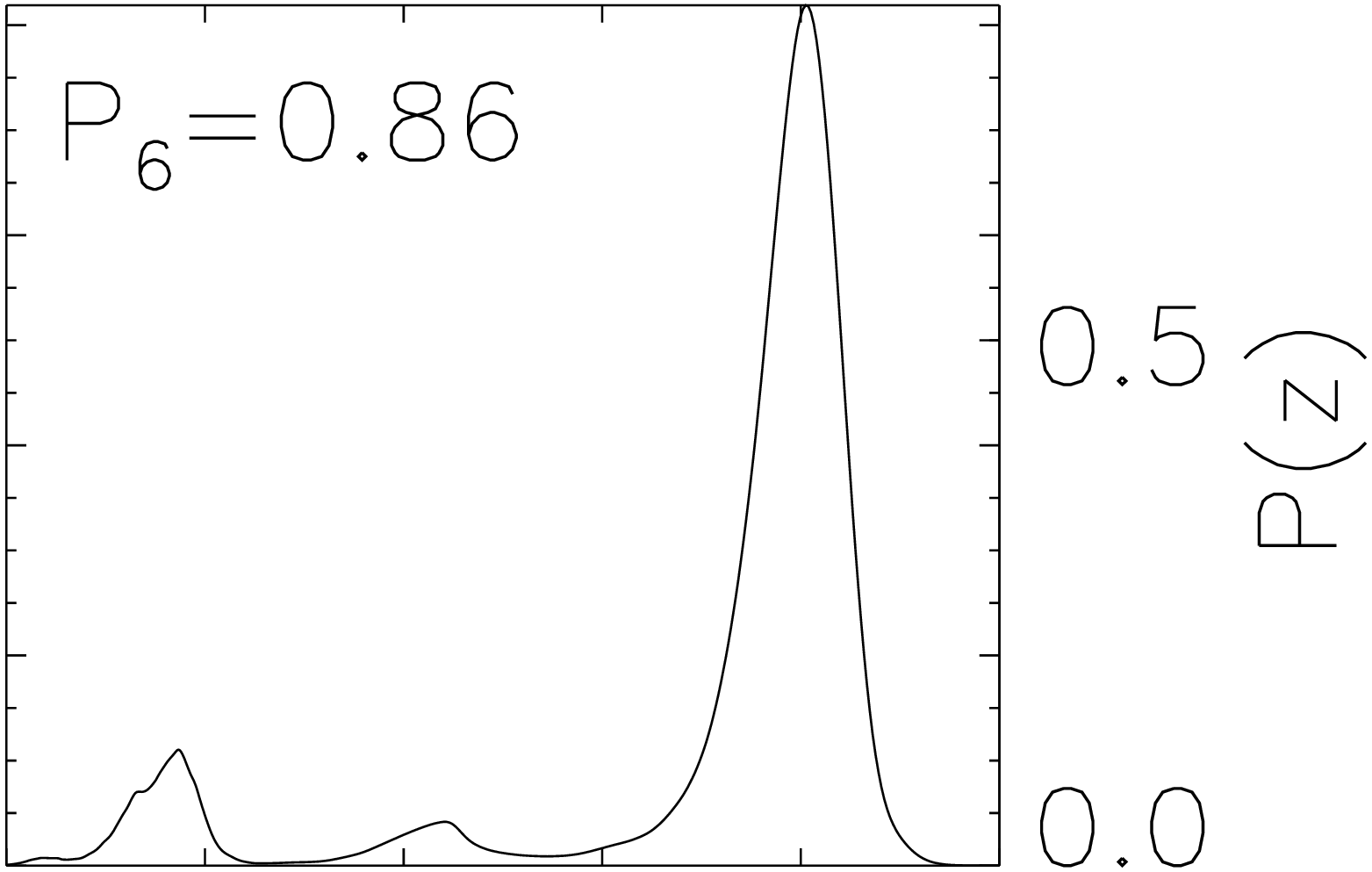}
\vspace{0.5mm}

\plotone{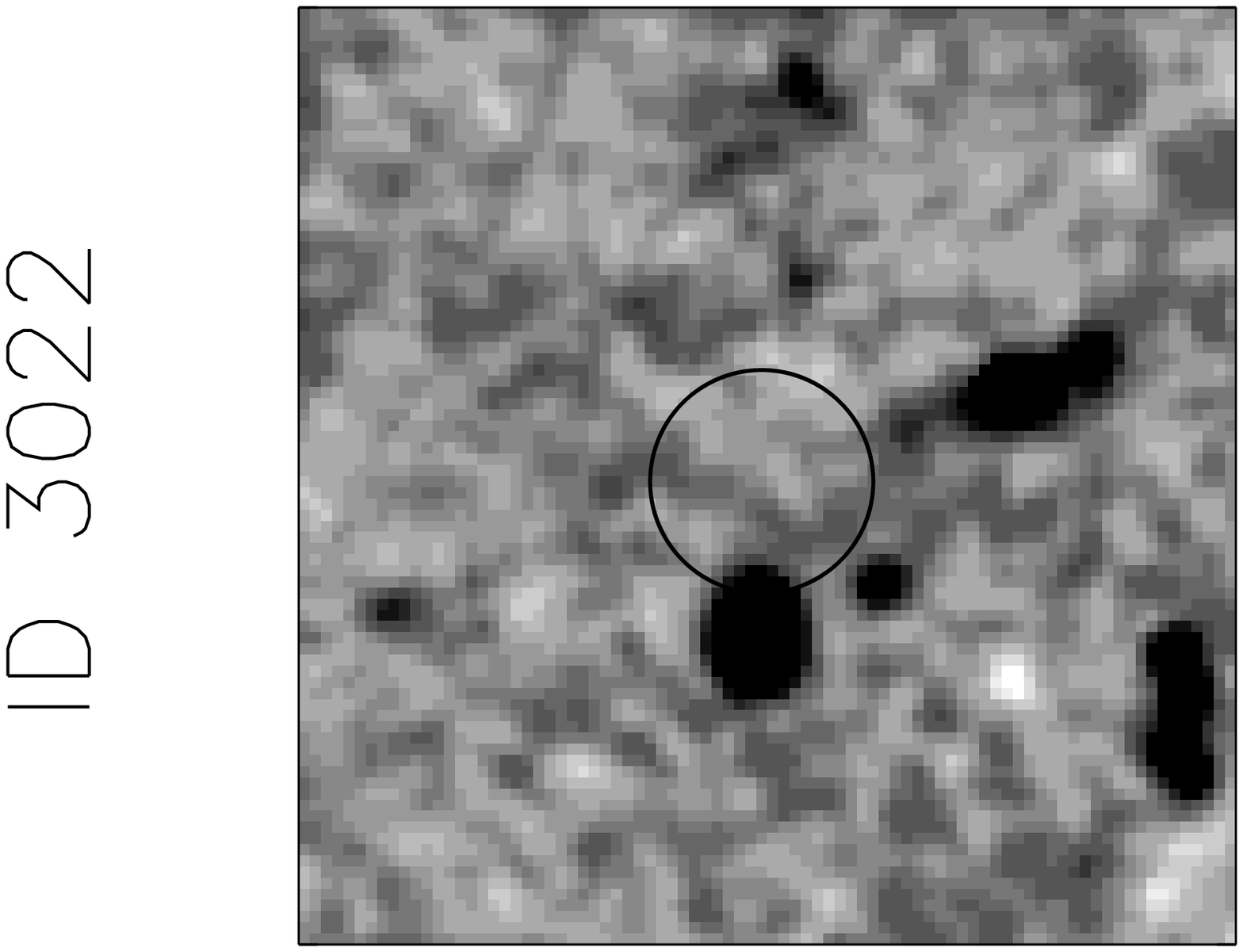}
\hspace{-10mm}
\plotone{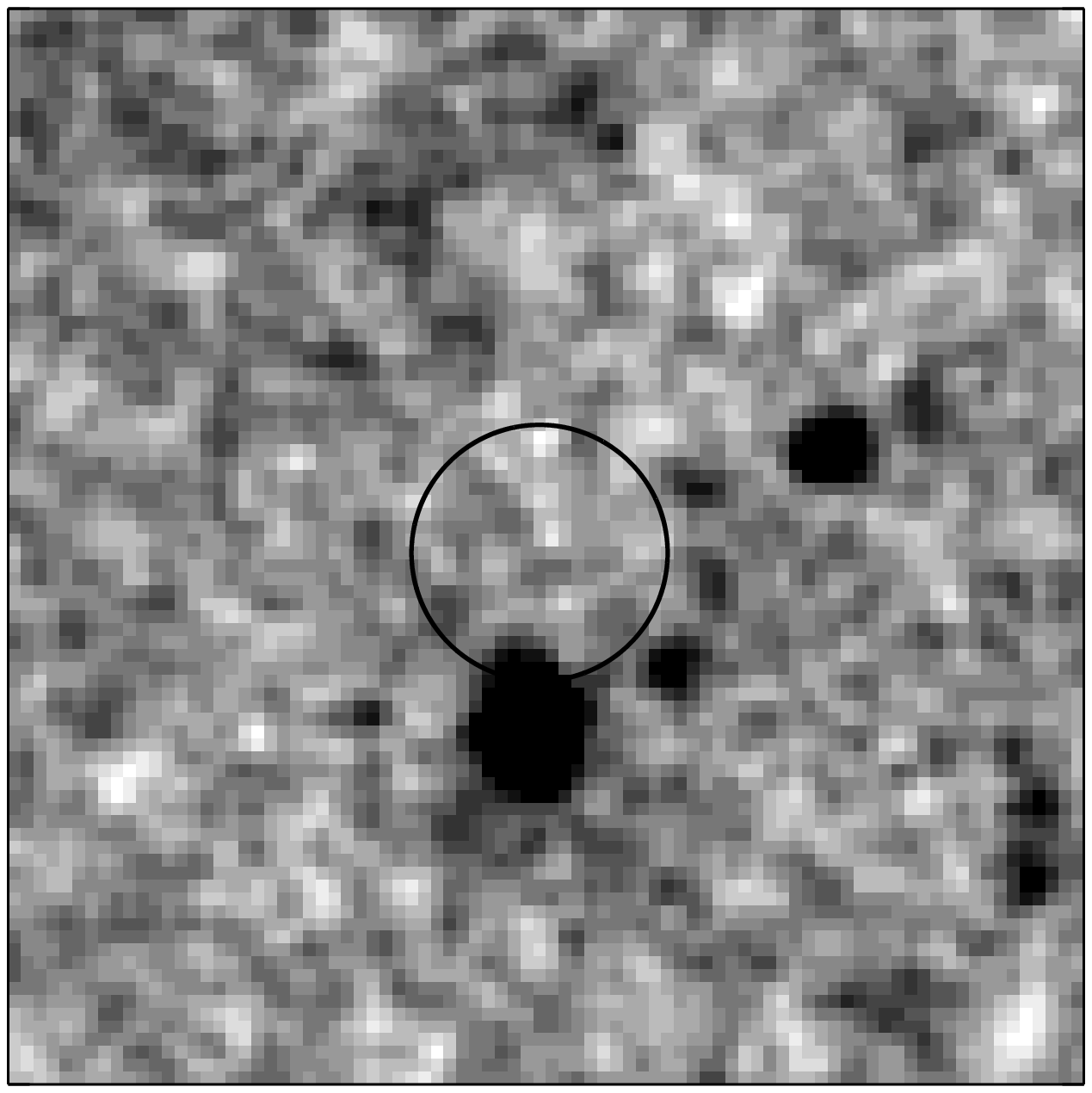}
\hspace{-10mm}
\plotone{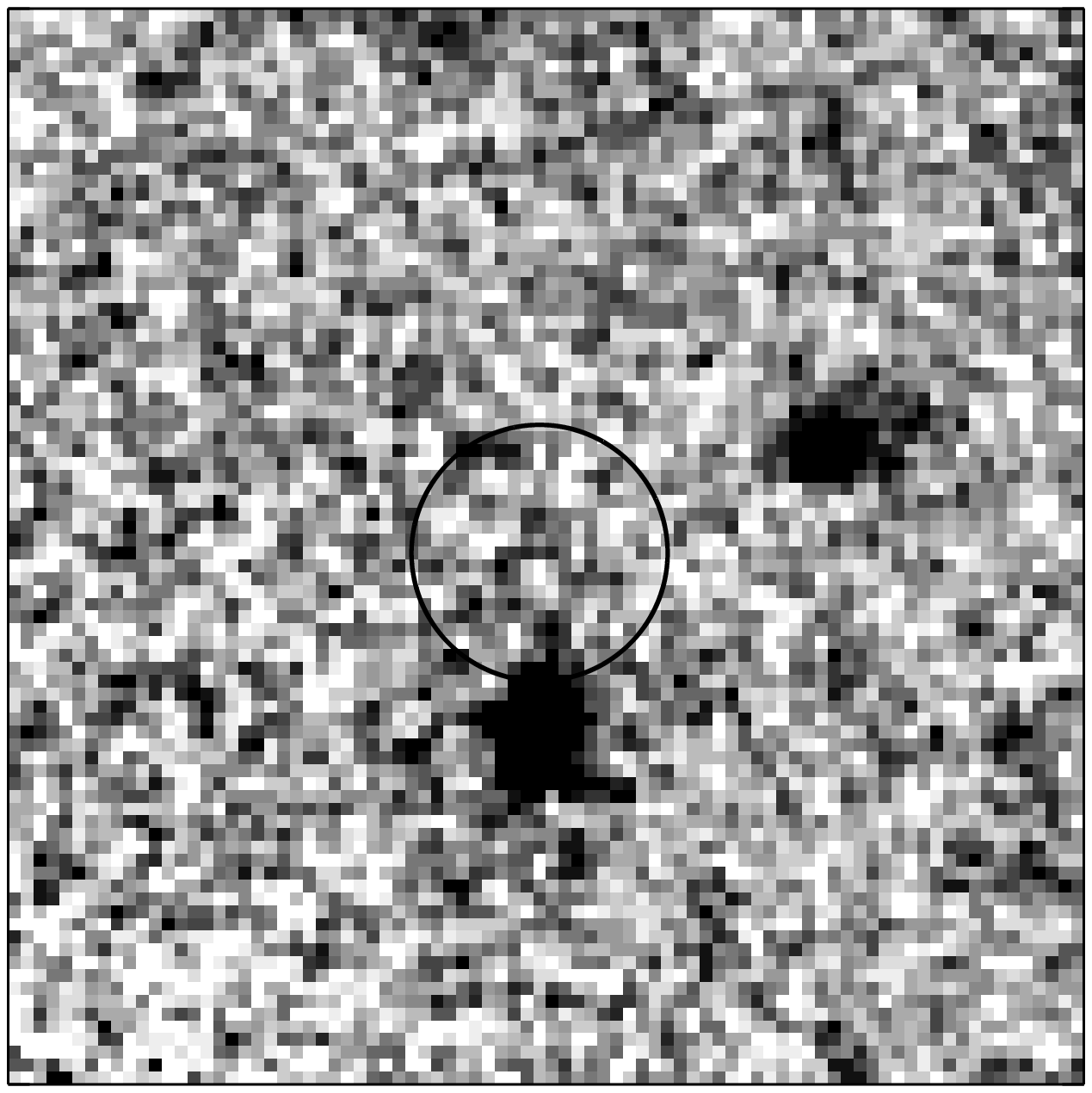}
\hspace{-10mm}
\plotone{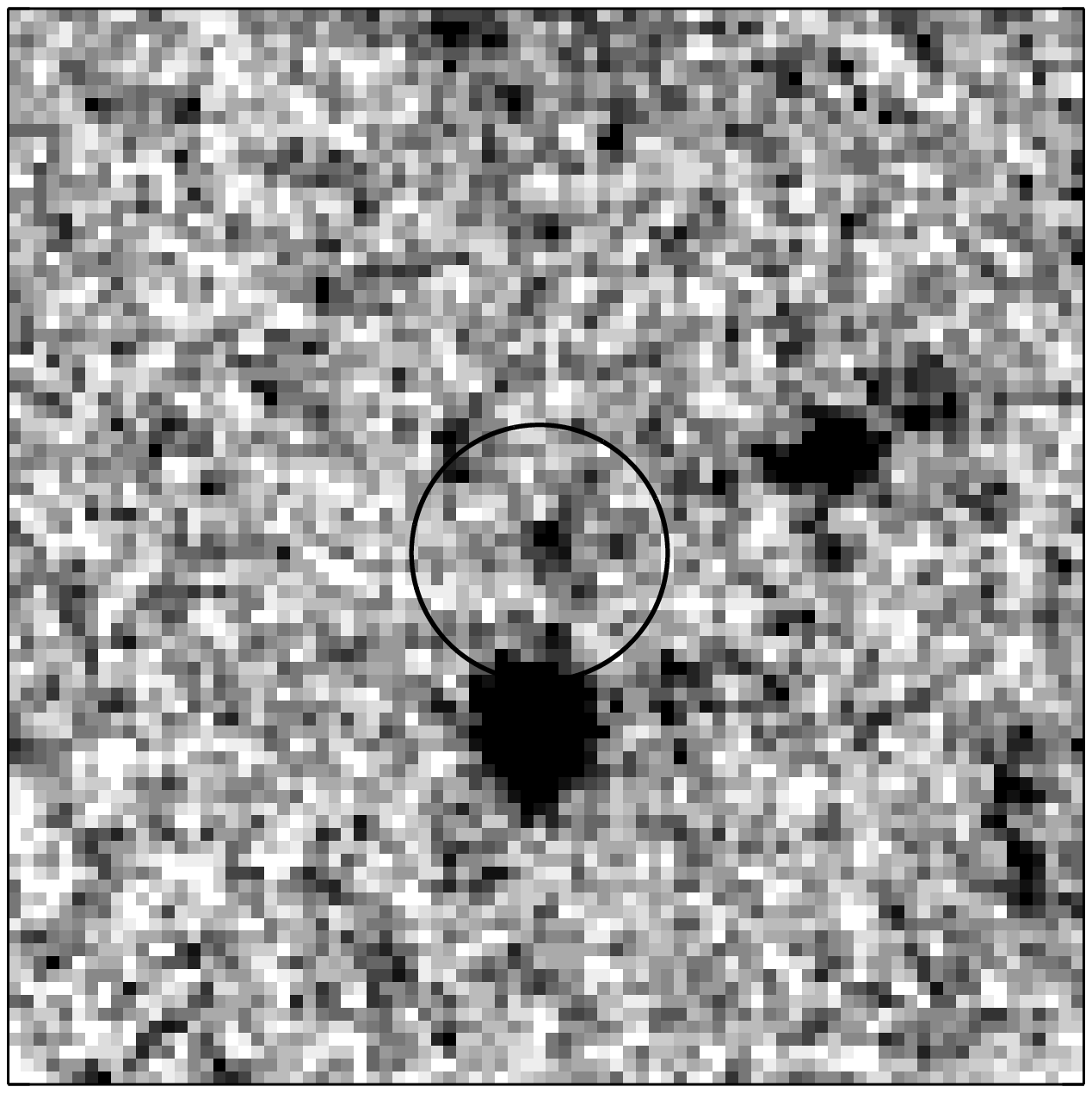}
\hspace{-10mm}
\plotone{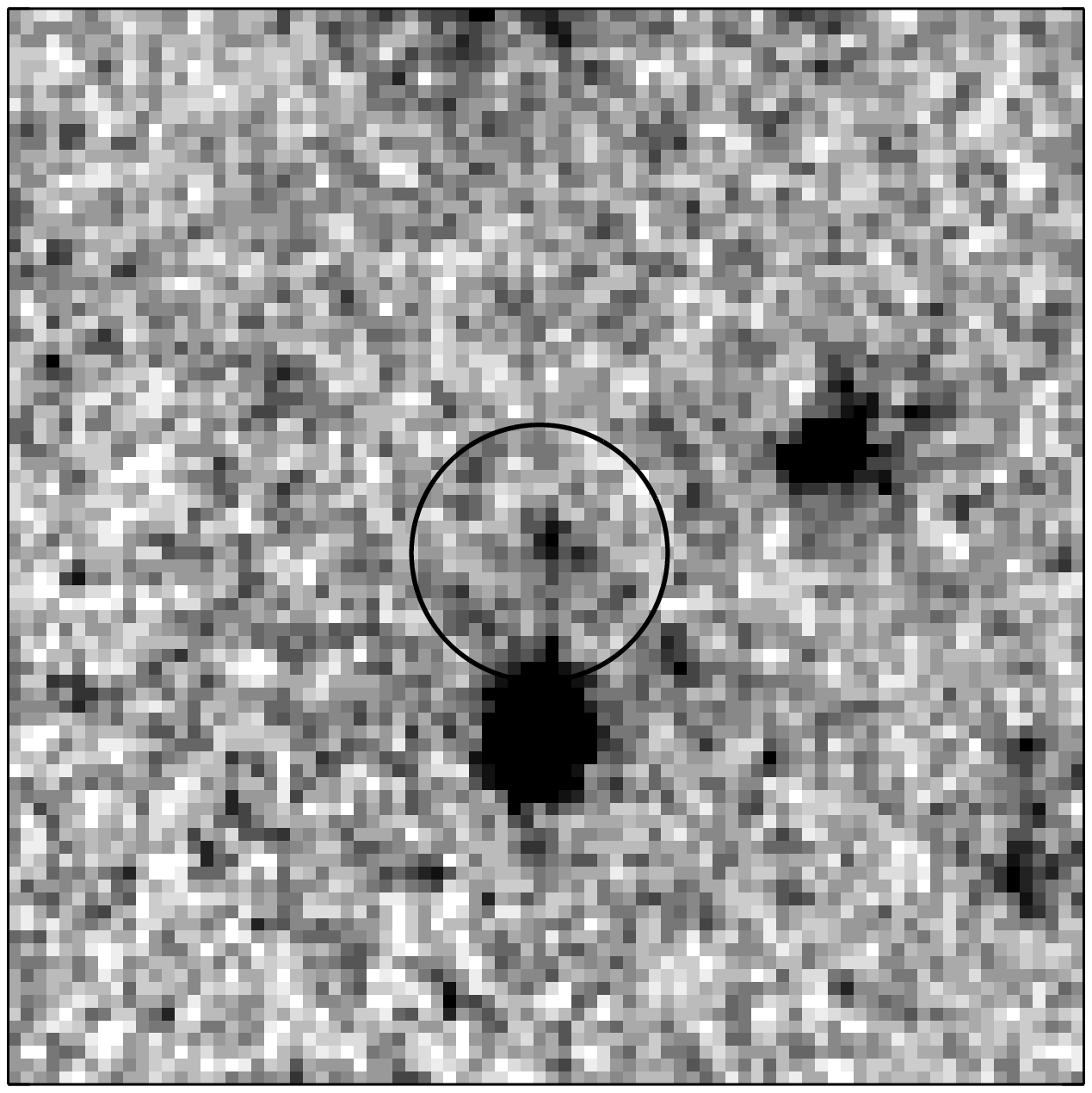}
\hspace{-10mm}
\plotone{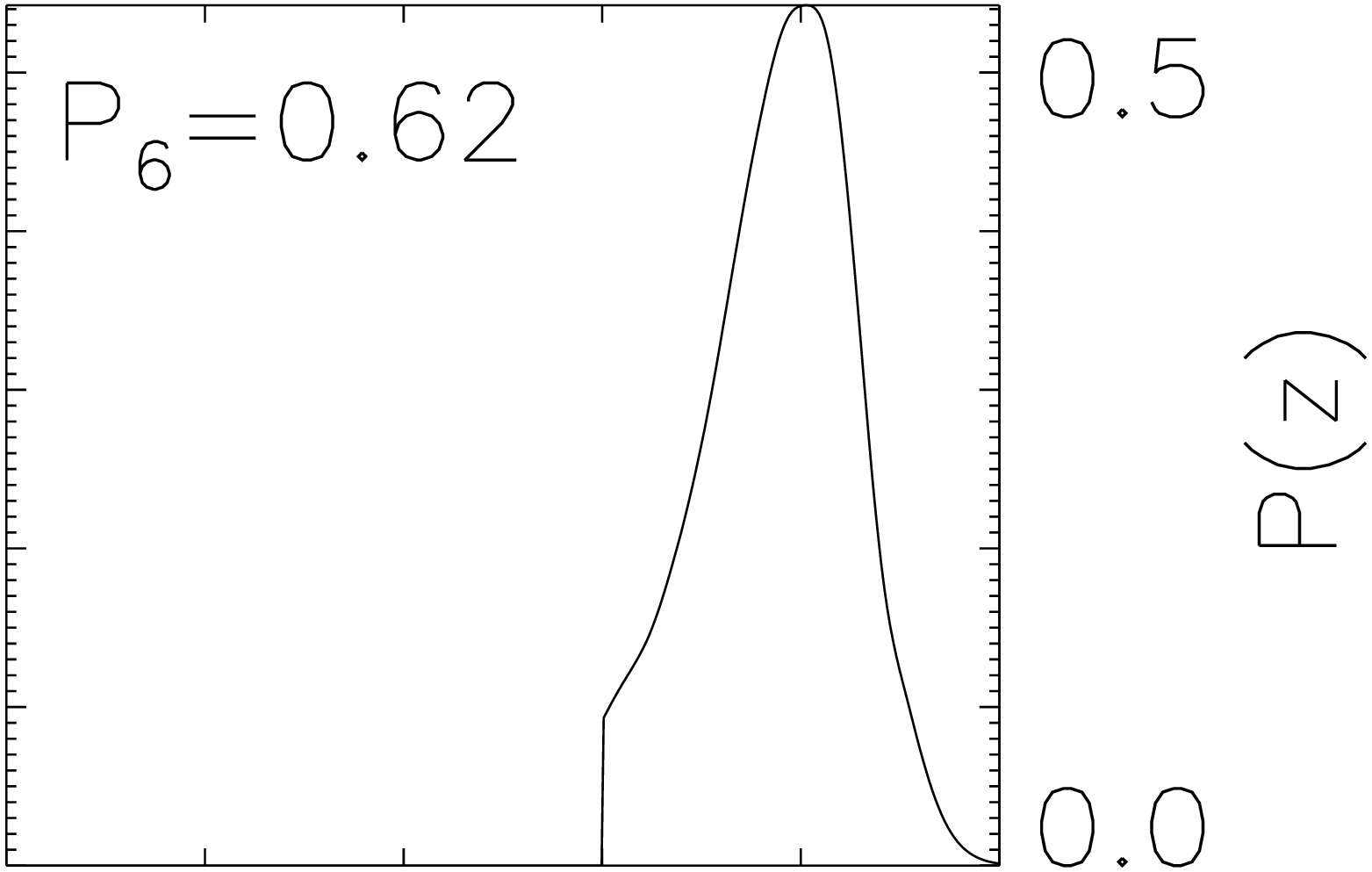}
\vspace{0.5mm}

\plotone{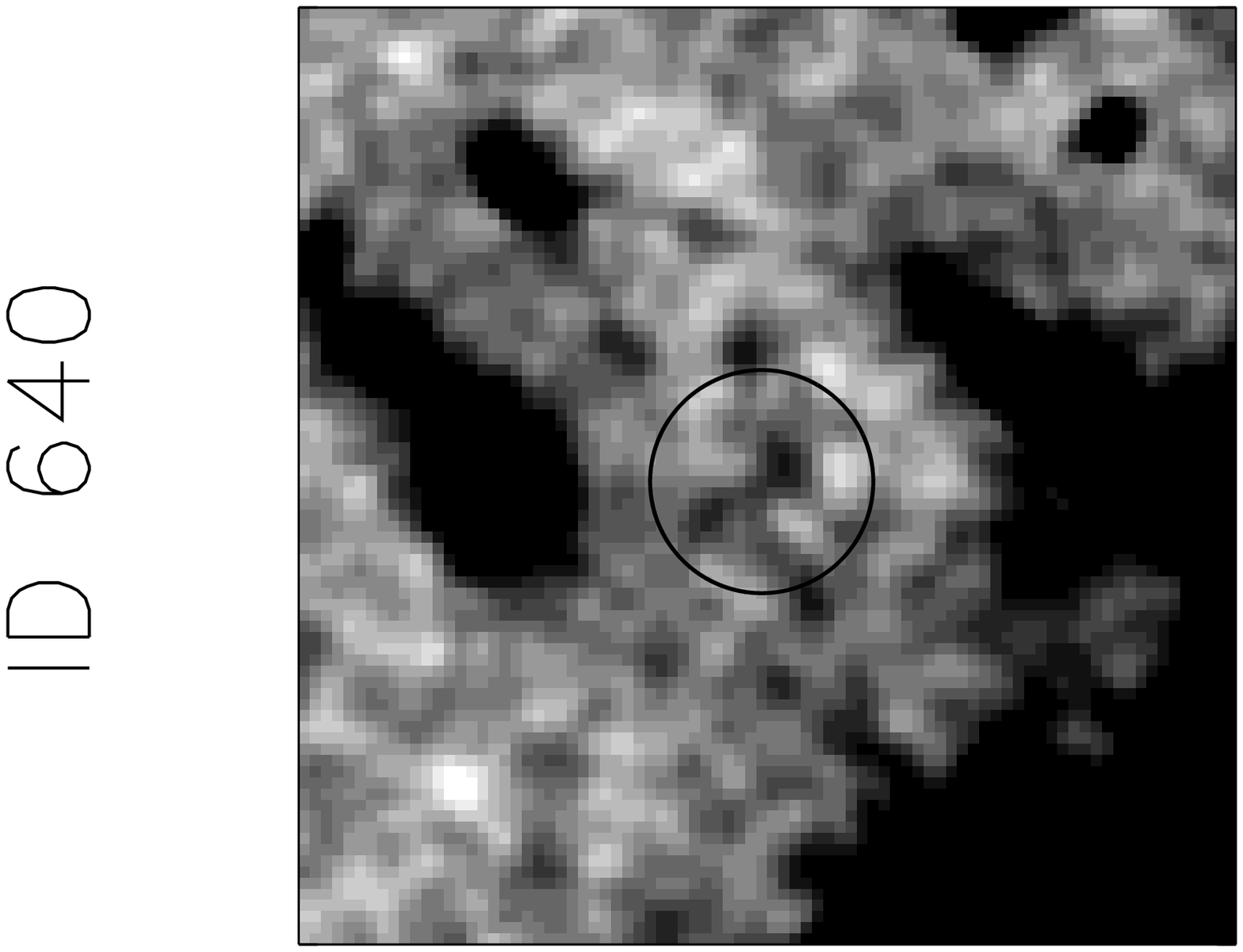}
\hspace{-10mm}
\plotone{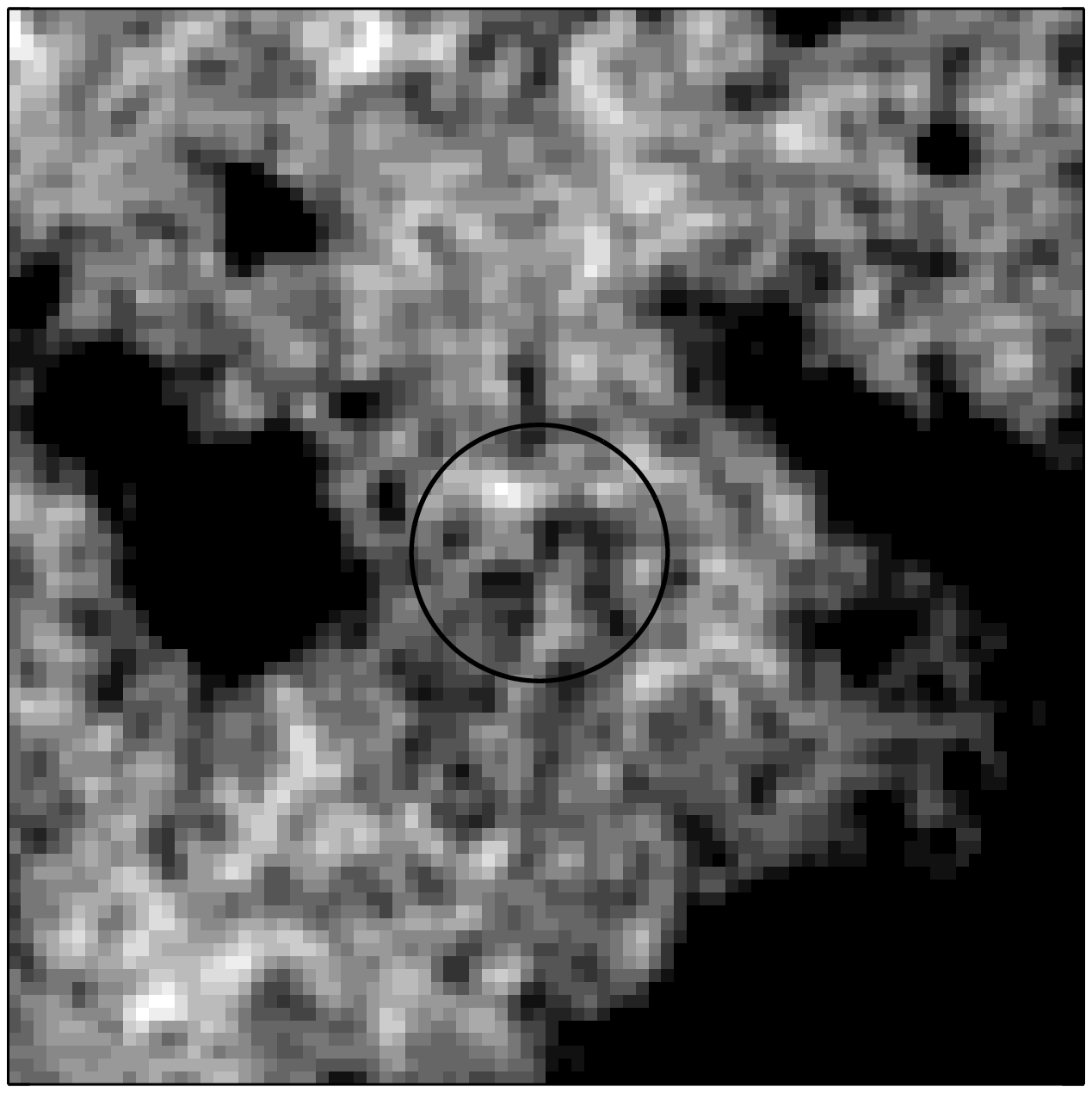}
\hspace{-10mm}
\plotone{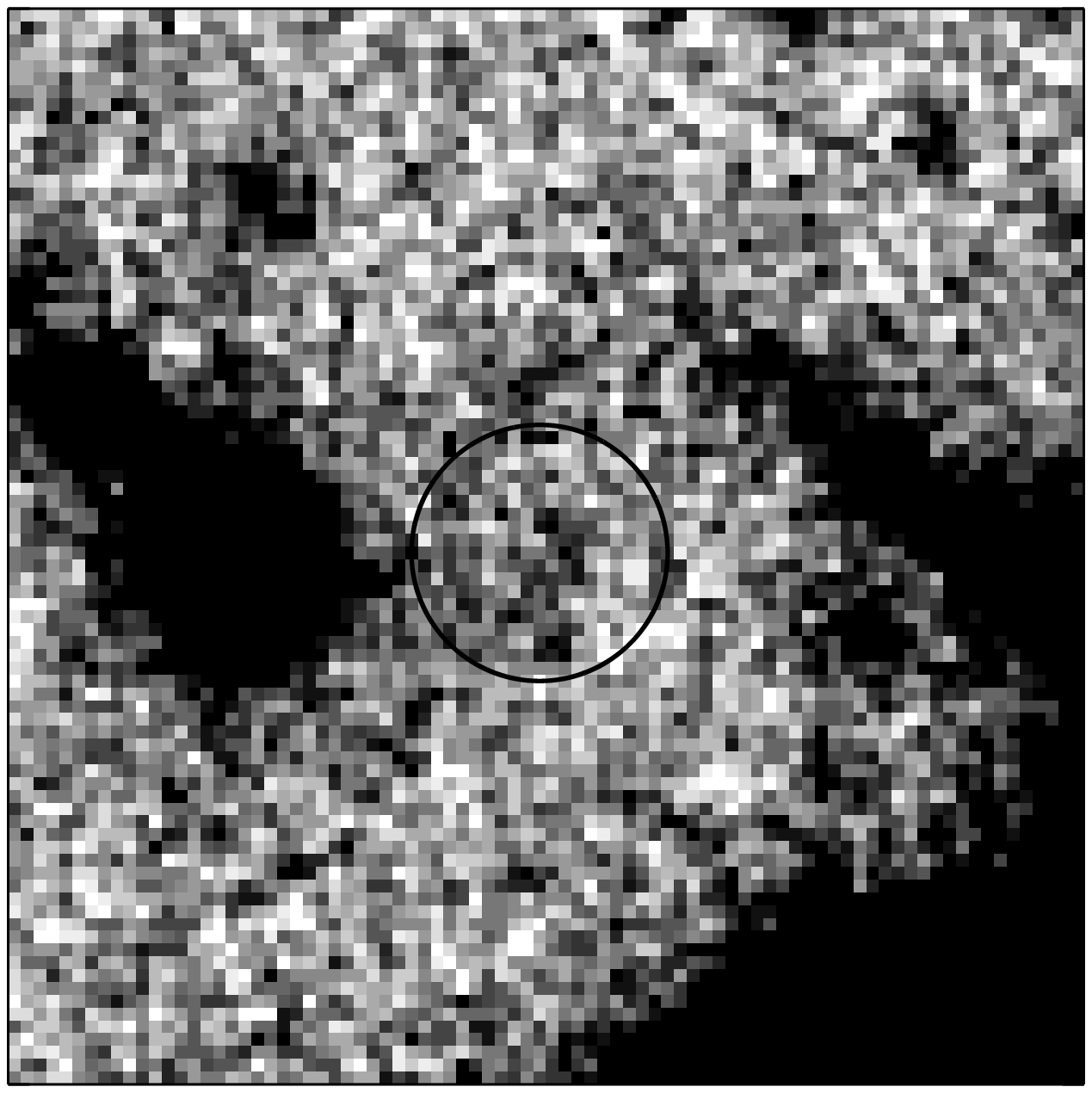}
\hspace{-10mm}
\plotone{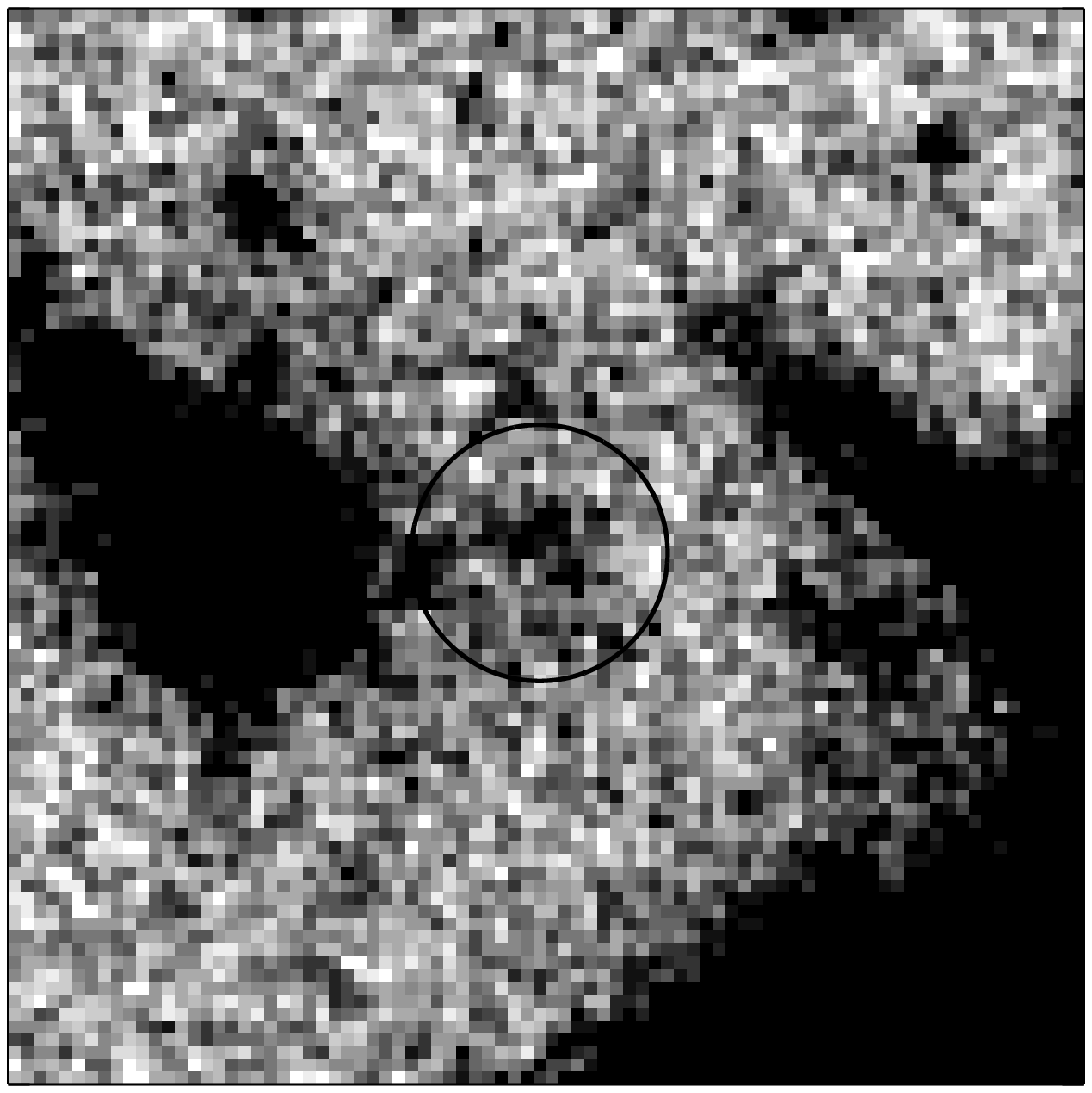}
\hspace{-10mm}
\plotone{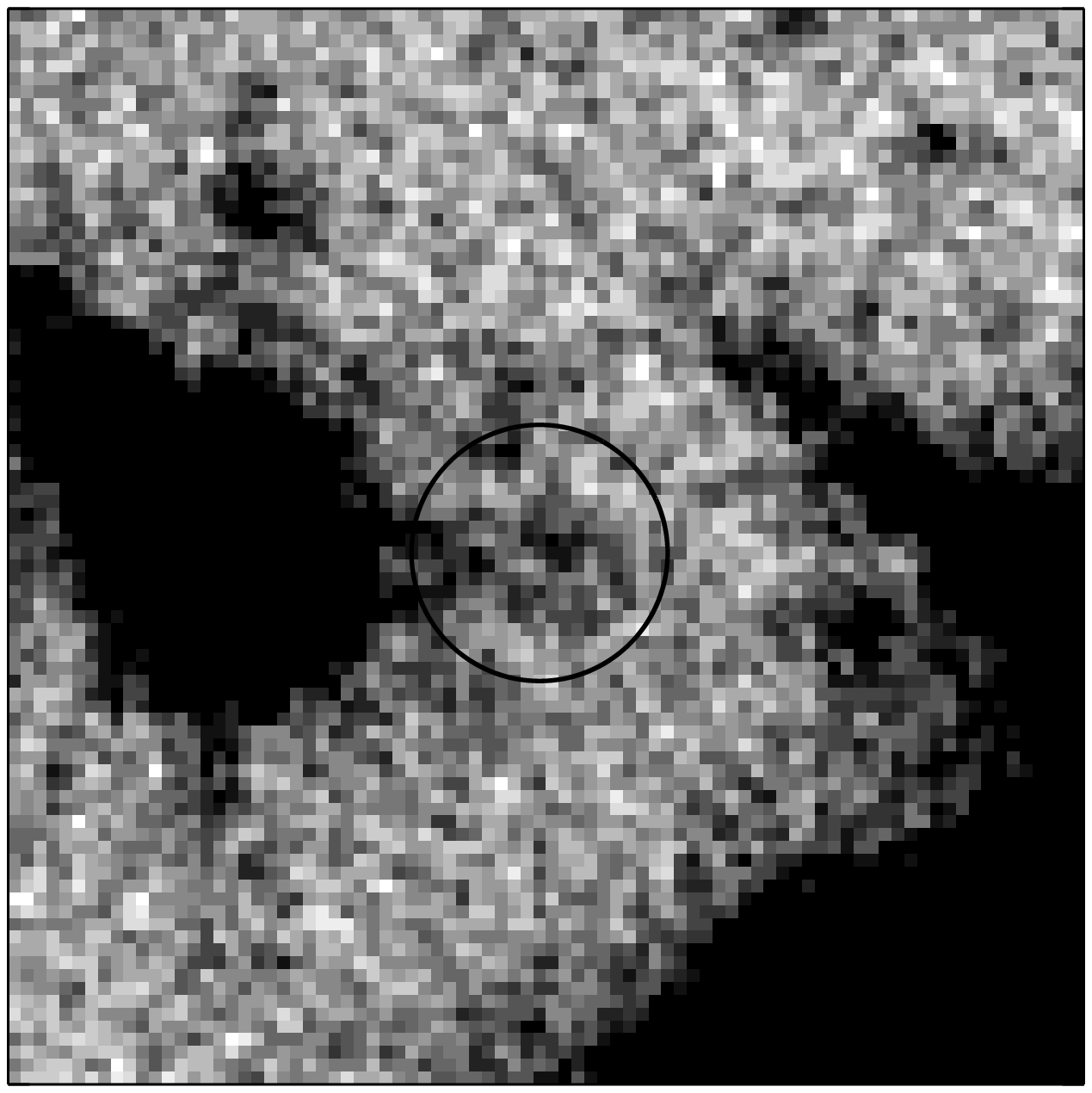}
\hspace{-10mm}
\plotone{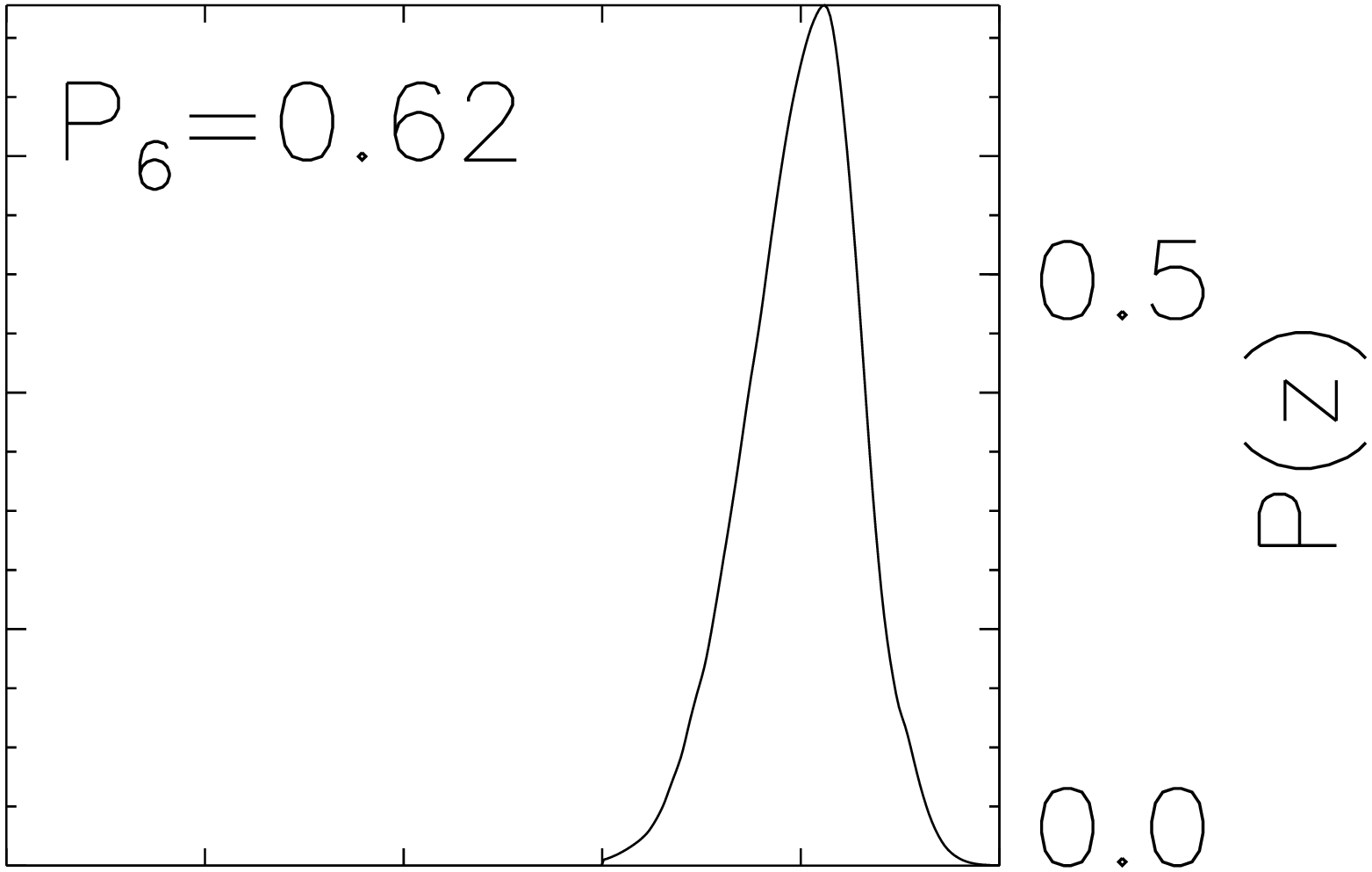}
\vspace{0.5mm}

\plotone{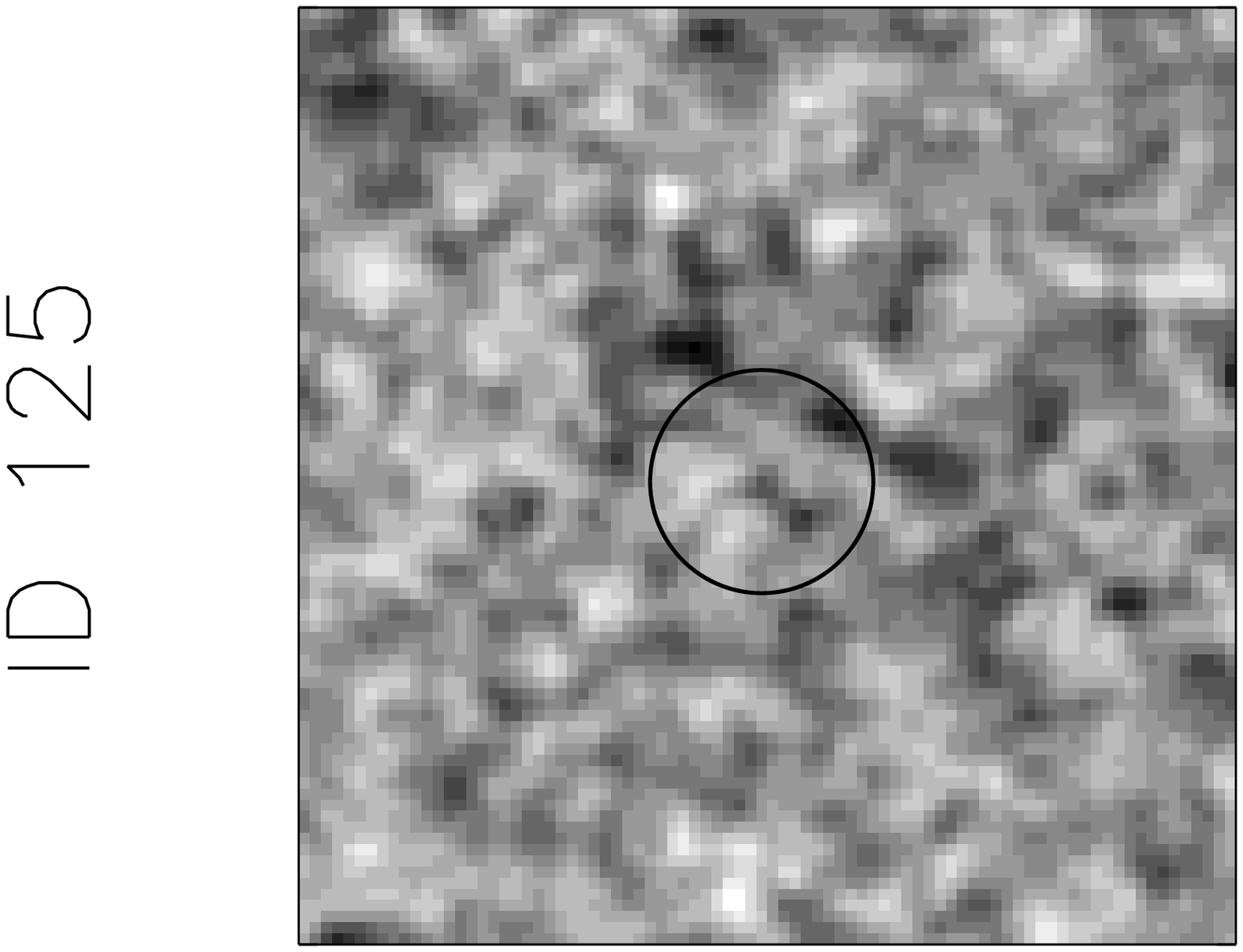}
\hspace{-10mm}
\plotone{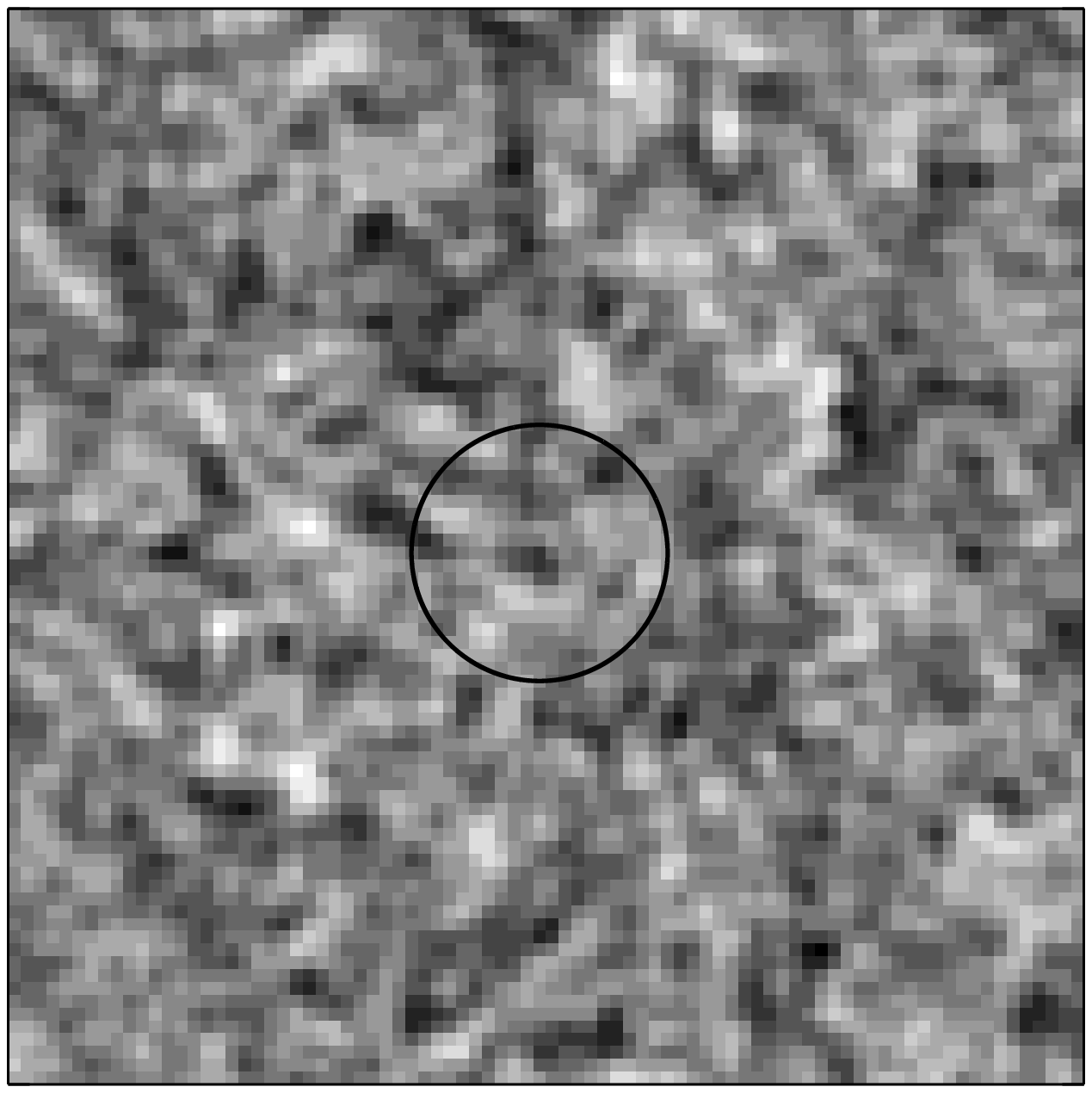}
\hspace{-10mm}
\plotone{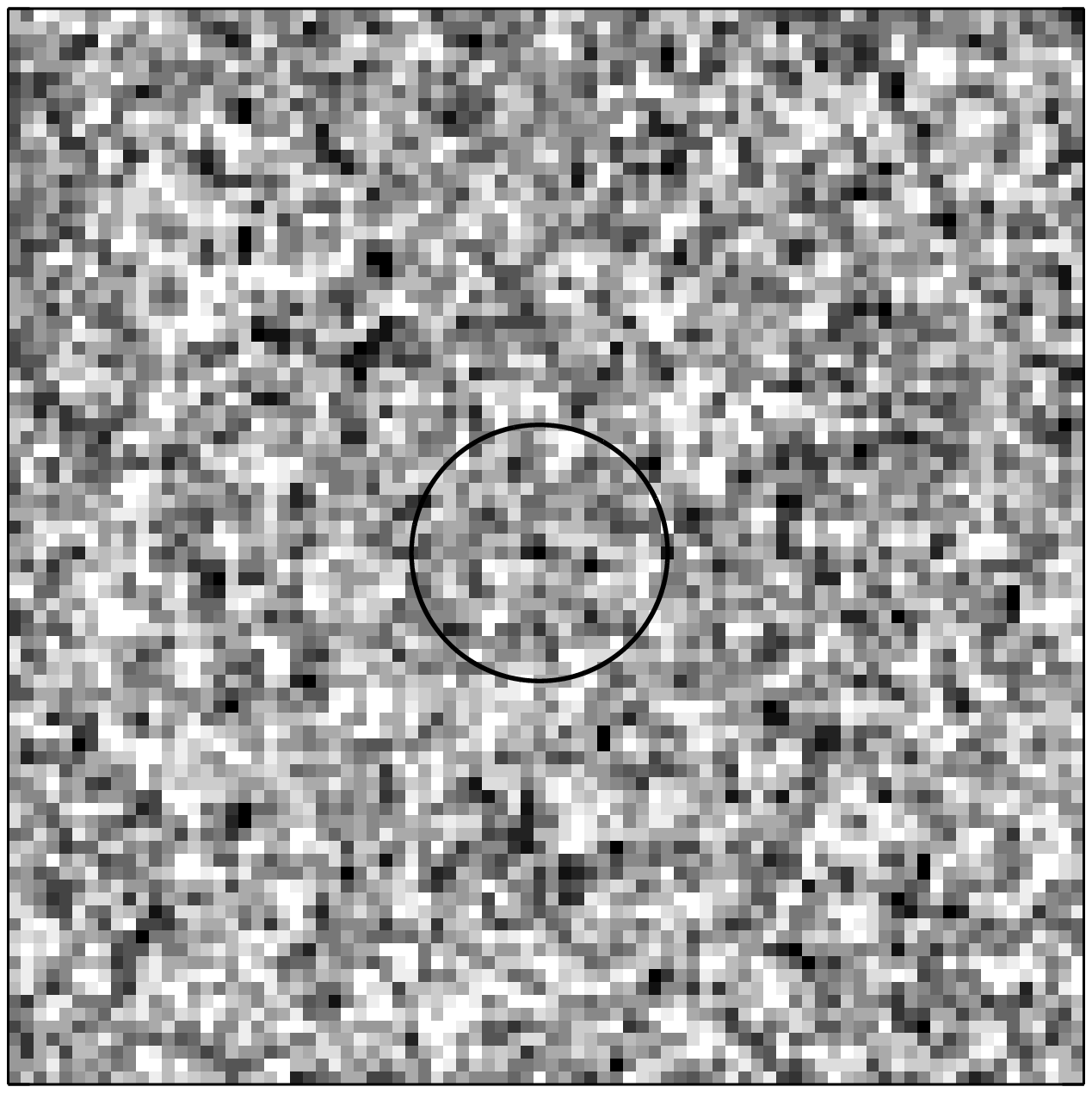}
\hspace{-10mm}
\plotone{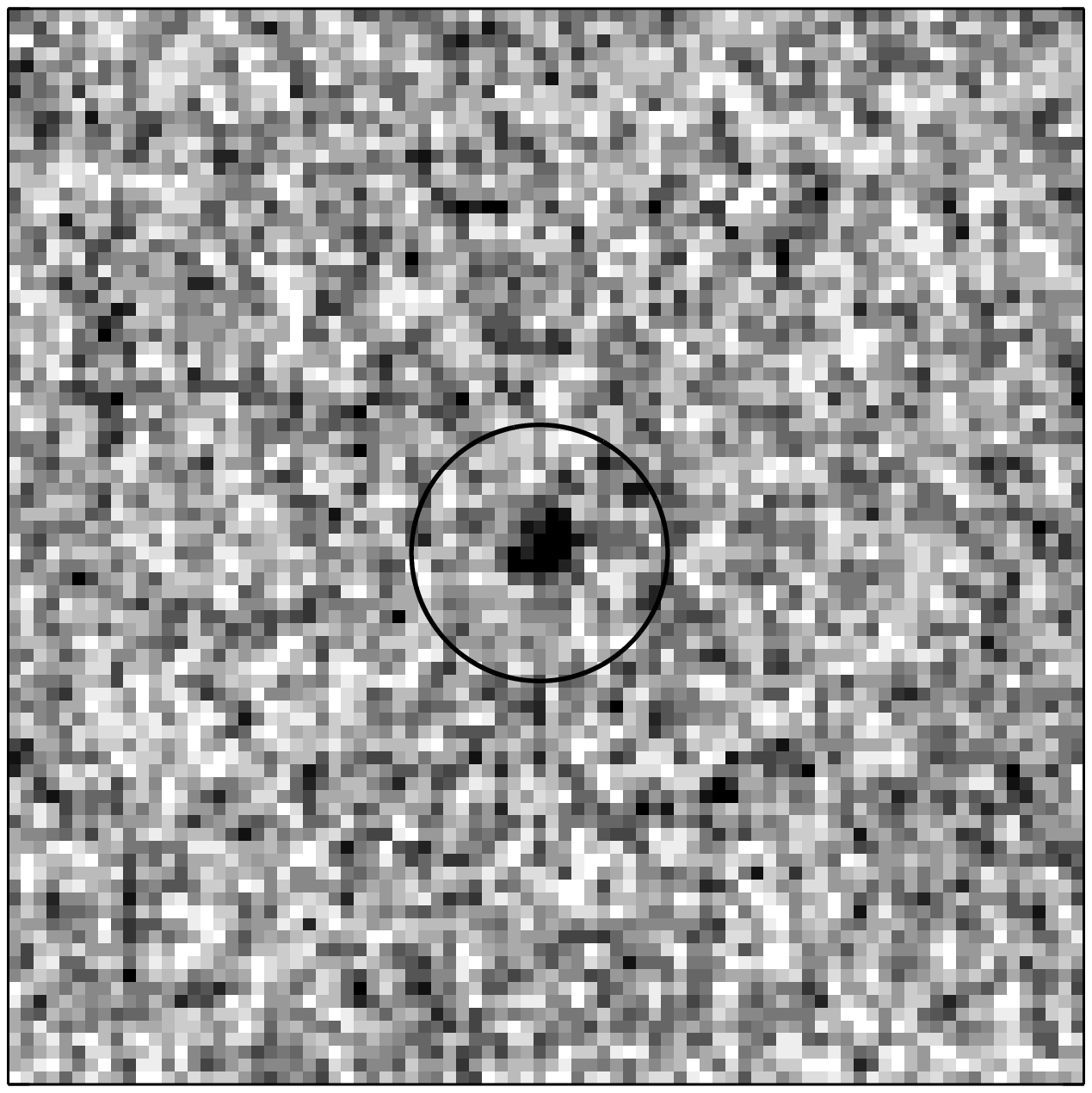}
\hspace{-10mm}
\plotone{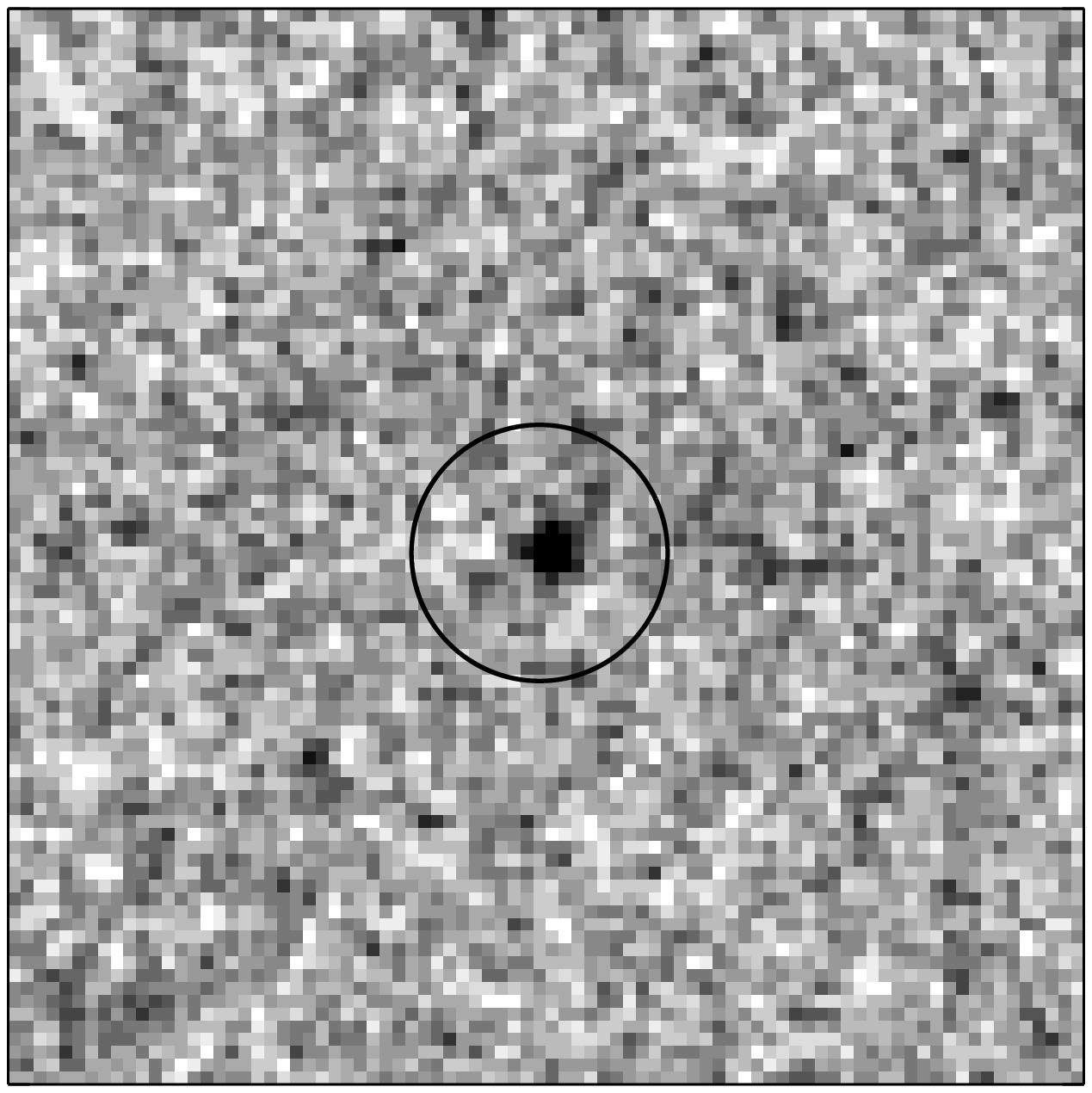}
\hspace{-10mm}
\plotone{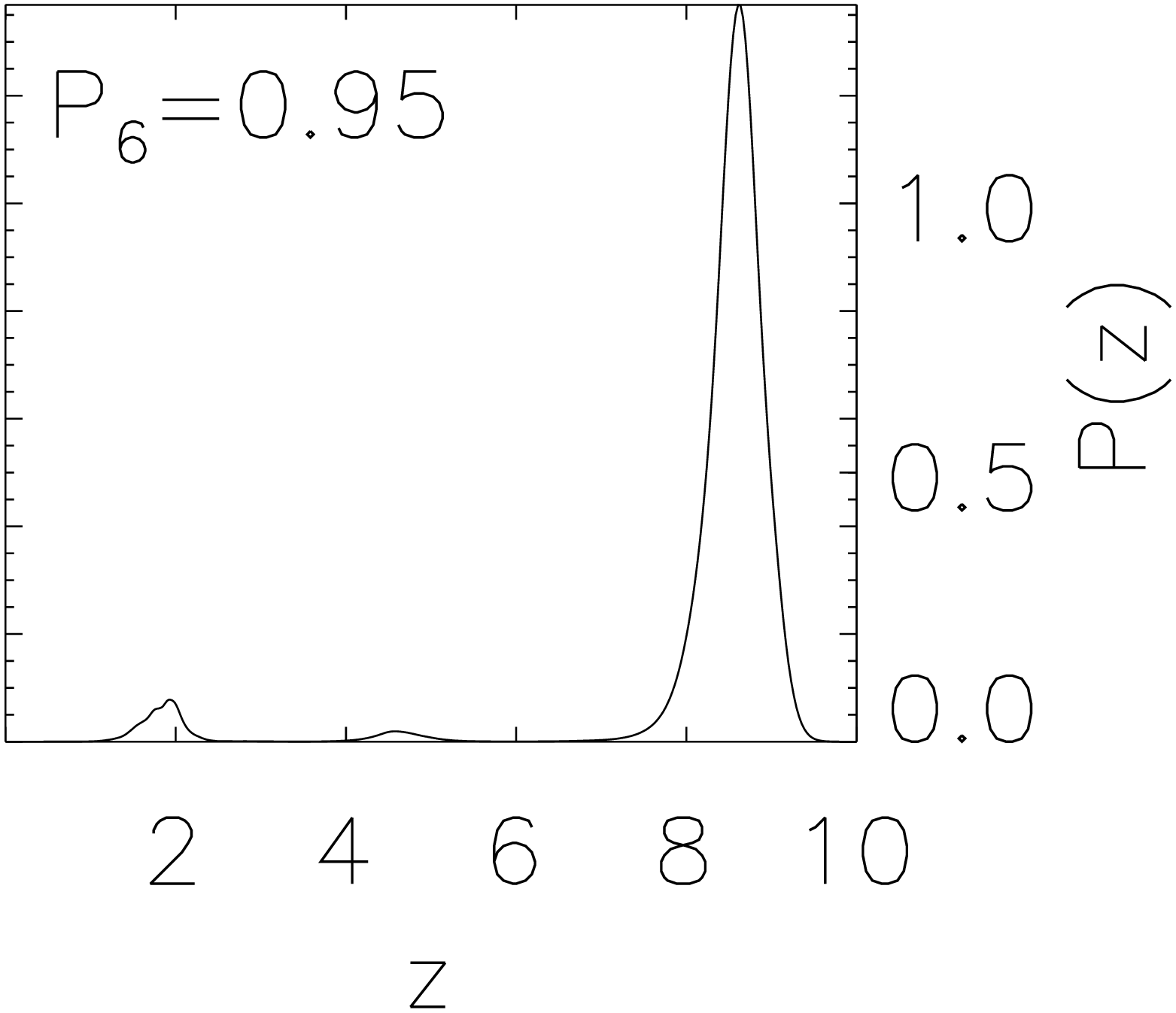}
\vspace{0.5mm}
\vspace{5mm}
\caption{Continued.}\label{f4}
\end{figure*}

\subsection{\spitzer\ IRAC Data}\label{section:irac}
The HUDF was observed with \spitzer\ at 3.6, 4.5, 5.8, and 8.0~\micron\ with the Infrared Array Camera (IRAC, \citet{fazio04}) as part of the Great Observatories Origins Deep Survey (GOODS; Dickinson et al., in prep).  The GOODS IRAC data are substantially deeper at 3.6 and 4.5~\micron\ than at longer wavelengths, and we use only these data here.  We measure $1\sigma$ limiting magnitudes in 3.0\arcs~diameter apertures of m$_{AB}$ = 26.94 and 26.88 mag in the 3.6 and 4.5 $\mu$m bands, respectively, using the same method as with the WFC3 data.

Given the faint nature of the sources of interest in our sample, they will be near the limit of the IRAC data if detected.  Owing to the fact that the IRAC point--response function is substantially larger than that of \textit{HST}/WFC3 ($\approx 0\farcs$2 FWHM versus $1\farcs6$ FWHM), much of the IRAC image is ``contaminated'' by other (foreground) galaxies.  Therefore, when we measure IRAC flux densities and upper limits in our data, we used a version of the IRAC image in which all sources detected in the GOODS v1.9 ACS F850LP image were removed using TFIT \citep{laidler07}.

\subsection{Object Detection and Photometry}\label{section:catalog}

We used the \textit{Source Extraction} (SExtractor) software package v2.5.0 \citep{bertin96} in two--image mode to construct an object catalog and perform photometry.  We constructed a detection image using a sum of the F125W and F160W images weighted by the image inverse variance filtered by a 3--pixel Gaussian kernel.  Objects were detected at $> 2 \sigma$ significance with a minimum area of 12 pixels.  Objects were then photometered on each of the ACS F435W, F606W, F775W, F850LP, and WFC3 F105W, F125W, and F160W images.  We measure object photometric colors using elliptical apertures that scale with the size of each source defined following \citet{kron80}, with a factor (\texttt{KRON\_FACT}) of 1.2 and a minimum size (\texttt{MIN\_RADIUS}) of 1.7.  Our tests have shown that for the faint galaxies in our sample ($H_{160} \gtrsim 28.5$~mag) these Kron--like apertures yield fluxes comparable to those measured in either isophotal apertures or $0\farcs4$--diameter apertures.  For brighter galaxies in our sample ($H \lesssim 28.0$~mag), these Kron--like apertures contain substantially more flux, in some cases by as much as 1 magnitude.  Furthermore, we find that the signal--to--noise ratios (S/N) in these Kron--like apertures are comparable to those measured in isophotal apertures, but because they are based on the light profile of each galaxy they produce higher S/N than that measured in circular apertures by 10--30\% for all galaxy magnitudes.  Therefore we adopt the Kron--like apertures to measure galaxy colors as they maximize the S/N while containing an optimal light fraction.  To measure galaxy total fluxes, we apply an aperture correction to each band constructed from the weighted--summed $J_{125} + H_{160}$ detection image.  The aperture corrections are defined on a source--by--source basis as the difference between the light in the Kron--like aperture defined above to a larger Kron--like aperture, with a Kron factor of 2.5 and a minimum radius of 3.5.

\subsection{Photometric Redshifts}\label{section:photoz}

We derive redshift estimates for each galaxy in our WFC3--based catalog using the full ACS \acsb, \acsv, \acsi, \acsz\ and WFC3 \wfcy, \wfcj, \wfch\ photometry and detections or upper limits from IRAC \mone, \mtwo.  For the ACS and WFC3 photometry, we use flux densities measured for each galaxy in the Kron--like elliptical aperture defined in \S~\ref{section:catalog}.  We scale these to total magnitudes using the aperture corrections defined in \S~\ref{section:catalog}.  For the IRAC data, we use the measured flux densities from the GOODS IRAC catalog (Dickinson et al., in prep) for those objects with detections.  For the remaining sources, we use upper limits measured from the image ``cleaned'' of sources detected in the GOODS \acsz\ image, as described in \S~\ref{section:irac}.  We correct the IRAC fluxes for light falling outside the circular apertures by applying corrections of 0.36~mag and 0.31~mag (derived for point sources) to the 3.6~\micron\ and 4.5~\micron\ data, respectively.  The upper limits are handled identically to detections, i.e. the measured flux in the aperture is used in the fitting, and the 1 $\sigma$ error is the noise.  The only exception is when the measured flux is formally negative -- in these cases, the flux is set to zero, and the 1 $\sigma$ error is still used as the noise.

We use the photometric--redshift code EAZY \citep{brammer08} to estimate photometric redshifts and derive redshift probability distribution functions, $P(z)$, for the sources in our WFC3 catalog.  The measured photometry for each galaxy was fit with non--negative linear combinations of a suite of optimized templates provided with EAZY, based on the P\'EGASE stellar--synthesis models \citep{fioc97}.  We also include the composite rest--frame UV SED for $z\sim 3$ LBGs from \citet{shapley03} as this may best represent the galaxies of interest in our sample and it contains Lyman $\alpha$ emission.  We included the effects of absorption from intervening \ion{H}{1} in the intergalactic medium (IGM) along the line of sight following the prescription of \citet{madau95}, though we note the \lya~forest is effectively opaque at these redshifts.

We fit photometric redshifts to all objects in our $(\wfcj + \wfch)$--selected catalog, including the 4 ACS bands, 3 WFC3 bands, and the 2 bands from the IRAC data.  For the purposes of photometric--redshift analysis, we included those galaxies in our catalog with $\wfcj \le 29.55$~mag and $\wfch \le 29.60$~mag, which are equivalent to the $3.5\sigma$--detection limits for sources in $0\farcs4$--diameter apertures.  For each galaxy we calculate the photometric redshift with the lowest $\chi^2$, as well as the full photometric redshift probability distribution function, $P(z)$, defined as $P(z) \propto \exp(-\chi^2/2)$, normalized such that the integral of $P(z)$ over all redshifts is equal to unity.

\subsection{Selection of Galaxies at $6.3 < \MakeLowercase{z} < 8.6$}

We used the full redshift probability distribution function, $P(z)$, defined above to select those targets with a high likelihood of having redshift $z > 6$.  We define the ``integrated probability'', $\mathcal{P}_6$, as the integral of the redshift probability distribution function from $6 \le z \le 11$, that is
\begin{equation}
\mathcal{P}_6 = \int_6^{11}\, P(z)\, dz.
\end{equation}  
For our sample, we selected all objects satisfying $\mathcal{P}_{6} > 0.6$, thus each of these galaxies has $>$60\% of its probability distribution function above redshift $z = 6$.  This method excludes no information on the galaxies.  We make no other requirements on any of the galaxy colors nor on the limits (non--detections) in any of the ACS bands.  This is in contrast to other typical UV--``dropout'' techniques targeting the redshifted Lyman break \citep[see][ and discussion below]{bouwens09, oesch09, bunker09, yan09}.  Instead, our photometric redshifts use the full set of photometry from all bandpasses including all of the photometric uncertainties to construct the likelihood that a galaxy lies at a given redshift, although this method primarily keys off of the Lyman break feature.

Our sample of galaxies satisfying $\wfcj \le 29.55$~mag, $\wfch \le 29.60$~mag and $\mathcal{P}_6 > 0.6$ includes 45 galaxies.  Upon visual inspection of these sources, we exclude 3 objects as obvious spurious sources (stellar diffraction spikes, oversplit areas of bright foreground galaxies).  We also exclude one additional source with $\wfcj=26.2$~mag and $\wfch=26.1$~mag.  This object has been noted by others \citep{oesch09, mclure09, yan09} as it is bright, yet undetected in existing NICMOS F110W and F160W imaging of the HUDF, which implies it is a transient source. Furthermore, this object is unresolved at the WFC3 resolution, with a measured FWHM=$0\farcs18$ (compared to $0\farcs176$ PSF FWHM for WFC3, see also discussion below).  Our final sample of high--redshift objects includes 41 galaxies.

\fig{zhist} shows the number distribution of the best--fit photometric--redshifts, $z_\mathrm{phot}$, of the galaxies in this sample, which have $P(z)$ distributions that are sharply peaked above $z = 6$.  From this parent sample, we construct two redshift samples for study here denoted as the ``Sample at $z \sim$ 7'' and ``Sample at $z \sim$ 8'' for convenience.  These samples are defined as,
\begin{equation}
\begin{split}
\mathrm{Sample\,\,at}\,\, z \sim 7&: 6.3 < z_\mathrm{phot} < 7.5 \\
\mathrm{Sample\,\,at}\,\,z \sim 8&: 7.5 \leq z_\mathrm{phot} \leq 8.6.
\end{split}
\end{equation} 

The $z \sim$ 7 sample includes 23 galaxies\footnote[2]{One of the galaxies in the $z = 7$ sample (ID 1441) was split into two separate objects by SExtractor.  Based on the similarity in the object colors and photometric redshift probability distribution functions, we have merged these objects by hand in our sample.} with a mean redshift $\langle z\rangle = 6.73$ and the $z \sim$ 8 sample includes 8 galaxies with $\langle z \rangle = 8.00$.  For our analysis, we will focus primarily on these 31 galaxies with 6.3 $<$ z$_{\mathrm phot}$ $\leq$ 8.6.  \fig{zhist} shows also the joint redshift probability distribution functions for each sample, defined as $P_\mathrm{joint}(z) = \Sigma P_i(z)$, where the $P_i(z)$ are the redshift probability distribution functions for the $i$th galaxy in each sample.  We have tested that the $z \sim$ 7 and $z \sim$ 8 subsamples are not subject to our arbitrary definition of the integrated probability distribution function, $\mathcal{P}_6$.  Lowering the integration limits on $\mathcal{P}$ from $z_\mathrm{lower}=6$ to $z_\mathrm{lower}=5$ has no change on the objects in these samples.  Furthermore, the majority of the objects in these samples have high $\mathcal{P}_6$ .  Of the 31 objects, 27 (23) have $\mathcal{P}_6$ $>$ 0.8 (0.9).  We chose the limit at $\mathcal{P}_6$ $>$ 0.6 to maximize our sample size while minimizing the contamination rate, which we found becomes higher at $\mathcal{P}_6$ $<$ 0.6 based on sources not satisfying the LBG photometric criteria (see \S 3.1).  Out of our 31 galaxies, 4 have $z <$ 5 contained within the 68\% confidence range of their photometric redshift probability distributions, yet 3/4 satisfy the $\acsz$- or $\wfcy$-dropout criteria \citep{oesch09, bouwens09}.  To keep the interpretation physical, we applied a Bayesian prior restricting the redshift to $>$ 6.5 or 7.5, respectively, for these three objects.  The remaining object, ID 1818, retains its large redshift uncertainty.

Table~1 provides the astrometric coordinates, photometry, and photometric--redshift information of the 23 objects in the $z \sim$ 7 sample and the 8 objects in the $z \sim$ 8 sample.  \fig{stamps}~shows postage--stamp images in the ACS F775W, F850LP and WFC3 F105W, F125W, and F160W bands, and the $P(z)$ distribution, for each of the galaxies in the $z \sim$ 7 and $z \sim$ 8 samples.  Generically, we will designate the galaxies in these samples as LBGs.  Although they were not selected strictly based on the strength of their Lyman breaks, this is the key feature driving their photometric redshifts, and as a result there is substantial overlap with LBG--selected samples (see \S~\ref{section:comparison}).   Each galaxy in our sample is formally undetected in the \acsb, \acsv, and \acsi\ bands (with S/N$<$2.0 in each band, with the exception of ID 1566, which has S/N$_{i}$ = 2.7\footnote[3]{Due to image convolution, some flux from an adjacent galaxy bleeds into this object's \acsi--band aperture.}), and they each exhibit large color indexes between either the $\acsz - \wfcy$, or $\wfcy - \wfcj$ bands, indicative of the Lyman--break redshifted into these bandpasses.
\begin{deluxetable*}{cccccccccccc}
\tabletypesize{\small}
\tablecaption{Candidate Galaxies at z $\geq$ 6.3}\label{tab:final}
\tablewidth{0pt}
\tablehead{
\colhead{ID} & \colhead{RA} & \colhead{Dec} & \colhead{z$_{phot}$} & \colhead{$\mathcal{P}_6$}  & \colhead{$z^{\prime}_{850}$} & \colhead{$Y_{105}$} & \colhead{$J_{125}$} & \colhead{$H_{160}$}  & \colhead{$M_{1500}$}  & \colhead{SFR$_{UV}$} & \colhead{Flags}\\
\colhead{$ $} & \colhead{(J2000)} & \colhead{(J2000)} & \colhead{$ $} & \colhead{$ $} & \colhead{(m$_{AB}$)} & \colhead{(m$_{AB}$)} & \colhead{(m$_{AB}$)} & \colhead{(m$_{AB}$)}  & \colhead{(mag)}  & \colhead{(M\sol\ yr$^{-1}$)} & \colhead{$ $}\\
}
\startdata
\phantom{0}803&53.18242&-27.77603&6.34$^{+0.17}_{-0.22}$&\phantom{< }0.93&28.96 $\pm$ 0.29&28.04 $\pm$ 0.13&28.15 $\pm$ 0.12&28.17 $\pm$ 0.11&-18.71&\phantom{1}13.7&3\\
2894&53.15322&-27.79822&6.34$^{+0.15}_{-0.13}$&> 0.99&28.61 $\pm$ 0.21&27.76 $\pm$ 0.10&27.85 $\pm$ 0.09&27.96 $\pm$ 0.09&-19.00&\phantom{1}17.8&2,3\\
\phantom{0}769&53.15610&-27.77578&6.36$^{+0.11}_{-0.27}$&\phantom{< }0.96&27.38 $\pm$ 0.09&26.68 $\pm$ 0.05&26.64 $\pm$ 0.04&26.47 $\pm$ 0.04&-20.12&\phantom{1}49.9&3\\
1289&53.15601&-27.78091&6.38$^{+0.17}_{-0.16}$&\phantom{< }0.99&28.89 $\pm$ 0.29&27.81 $\pm$ 0.11&27.98 $\pm$ 0.11&28.05 $\pm$ 0.11&-18.94&\phantom{1}17.0&2,3,5\\
\phantom{0}649&53.17579&-27.77440&6.40$^{+0.18}_{-0.15}$&> 0.99&28.53 $\pm$ 0.24&27.69 $\pm$ 0.11&27.53 $\pm$ 0.08&27.73 $\pm$ 0.09&-19.16&\phantom{1}20.8&3\\
1445&53.16491&-27.78235&6.40$^{+0.21}_{-0.23}$&\phantom{< }0.96&\phantom{0000}> 29.24&28.23 $\pm$ 0.15&28.23 $\pm$ 0.12&28.68 $\pm$ 0.18&-18.57&\phantom{1}12.1&2,3,5\\
2032&53.15158&-27.78784&6.40$^{+0.18}_{-0.17}$&\phantom{< }0.99&28.97 $\pm$ 0.36&27.74 $\pm$ 0.12&28.00 $\pm$ 0.13&28.05 $\pm$ 0.13&-18.99&\phantom{1}17.8&2,3,4,5\\
1818&53.17601&-27.78576&6.43$^{+0.31}_{-4.54}$&\phantom{< }0.76&\phantom{0000}> 28.28&27.66 $\pm$ 0.21&28.07 $\pm$ 0.25&28.02 $\pm$ 0.23&-19.03&\phantom{1}18.5&--\\
\phantom{0}328&53.17427&-27.76979&6.47$^{+0.34}_{-0.36}$&\phantom{< }0.89&\phantom{0000}> 29.46&28.85 $\pm$ 0.21&28.89 $\pm$ 0.18&29.05 $\pm$ 0.21&-17.95&\phantom{10}6.9&5\\
1072&53.16903&-27.77876&6.47$^{+0.25}_{-0.20}$&\phantom{< }0.98&\phantom{0000}> 29.11&28.15 $\pm$ 0.15&28.29 $\pm$ 0.15&28.47 $\pm$ 0.16&-18.61&\phantom{1}12.7&2,3,4,5\\
1566&53.14554&-27.78372&6.56$^{+0.12}_{-0.11}$&> 0.99&28.53 $\pm$ 0.21&27.27 $\pm$ 0.07&27.17 $\pm$ 0.05&27.31 $\pm$ 0.06&-19.61&\phantom{1}32.2&3\\
2013&53.17104&-27.78767&6.70$^{+0.33}_{-0.22}$&> 0.99&\phantom{0000}> 29.18&28.42 $\pm$ 0.18&28.05 $\pm$ 0.11&28.49 $\pm$ 0.16&-18.66&\phantom{1}13.7&2,3,5\\
1110&53.18642&-27.77913&6.72$^{+0.60}_{-0.30}$&\phantom{< }0.80&\phantom{0000}> 28.39&28.20 $\pm$ 0.31&27.94 $\pm$ 0.20&27.93 $\pm$ 0.19&-18.82&\phantom{1}15.9&3,5\\
1441&53.17735&-27.78240&6.83$^{+0.15}_{-0.13}$&> 0.99&27.77 $\pm$ 0.16&26.13 $\pm$ 0.04&25.95 $\pm$ 0.03&25.95 $\pm$ 0.02&-20.89&108.1&2,3,4,5\\
\phantom{0}567&53.17677&-27.77342&6.84$^{+0.54}_{-0.58}$&\phantom{< }0.88&\phantom{0000}> 28.63&28.43 $\pm$ 0.30&28.41 $\pm$ 0.25&28.57 $\pm$ 0.28&-18.48&\phantom{1}11.8&5\\
2432&53.17734&-27.79206&6.86$^{+0.15}_{-0.16}$&> 0.99&\phantom{0000}> 28.93&27.38 $\pm$ 0.09&27.17 $\pm$ 0.06&27.25 $\pm$ 0.06&-19.66&\phantom{1}35.2&2,3,4,5\\
\phantom{0}515&53.16553&-27.77259&6.88$^{+0.38}_{-0.27}$&\phantom{< }0.99&\phantom{0000}> 29.15&28.56 $\pm$ 0.21&28.46 $\pm$ 0.16&28.70 $\pm$ 0.19&-18.42&\phantom{1}11.2&2,3,4,5\\
\phantom{0}335&53.15984&-27.76997&6.93$^{+0.37}_{-0.29}$&\phantom{< }0.99&\phantom{0000}> 29.55&28.90 $\pm$ 0.20&28.47 $\pm$ 0.11&28.65 $\pm$ 0.13&-18.34&\phantom{1}10.5&4,5\\
1768&53.16169&-27.78532&7.22$^{+0.18}_{-0.14}$&> 0.99&\phantom{0000}> 29.16&27.50 $\pm$ 0.08&27.07 $\pm$ 0.05&27.01 $\pm$ 0.04&-19.89&\phantom{1}45.2&2,3,4,5\\
2056&53.16481&-27.78819&7.25$^{+0.27}_{-0.23}$&> 0.99&\phantom{0000}> 28.71&27.72 $\pm$ 0.14&27.15 $\pm$ 0.07&27.05 $\pm$ 0.06&-19.82&\phantom{1}42.5&2,3,4,5\\
\phantom{0}669&53.17973&-27.77457&7.25$^{+0.32}_{-0.25}$&> 0.99&\phantom{0000}> 28.76&28.03 $\pm$ 0.19&27.58 $\pm$ 0.10&27.50 $\pm$ 0.09&-19.39&\phantom{1}28.8&2,3,4,5\\
1106&53.18626&-27.77897&7.32$^{+0.21}_{-0.20}$&> 0.99&\phantom{0000}> 29.01&27.87 $\pm$ 0.13&27.27 $\pm$ 0.06&27.45 $\pm$ 0.07&-19.73&\phantom{1}39.7&2,3,4,5\\
3053&53.15506&-27.80171&7.40$^{+0.30}_{-0.28}$&> 0.99&\phantom{0000}> 29.04&28.45 $\pm$ 0.21&27.77 $\pm$ 0.09&27.96 $\pm$ 0.11&-19.24&\phantom{1}25.5&2,3,4,5\\
\phantom{0}819&53.17866&-27.77625&7.55$^{+0.34}_{-0.36}$&> 0.99&\phantom{0000}> 29.10&28.82 $\pm$ 0.28&28.09 $\pm$ 0.12&28.11 $\pm$ 0.12&-18.96&\phantom{1}20.1&1,3,4,5\\
\phantom{0}653&53.17952&-27.77436&7.76$^{+0.27}_{-0.76}$&\phantom{< }0.95&\phantom{0000}> 29.06&\phantom{0000}> 29.04&28.31 $\pm$ 0.15&28.92 $\pm$ 0.26&-18.68&\phantom{1}15.8&1,3,5\\
2055&53.16466&-27.78815&7.81$^{+0.43}_{-0.33}$&\phantom{< }0.96&\phantom{0000}> 28.64&28.41 $\pm$ 0.30&27.58 $\pm$ 0.11&27.21 $\pm$ 0.08&-19.63&\phantom{1}38.3&2,3,4,5\\
\phantom{0}200&53.15749&-27.76670&7.91$^{+0.19}_{-0.80}$&\phantom{< }0.89&\phantom{0000}> 29.06&\phantom{0000}> 29.05&28.02 $\pm$ 0.12&28.29 $\pm$ 0.14&-19.03&\phantom{1}22.2&1,3,4,5\\
\phantom{0}213&53.15681&-27.76709&8.05$^{+0.29}_{-1.41}$&\phantom{< }0.86&\phantom{0000}> 29.32&\phantom{0000}> 29.32&28.70 $\pm$ 0.17&28.77 $\pm$ 0.18&-18.43&\phantom{1}12.9&1,3,4,5\\
3022&53.16806&-27.80073&8.05$^{+0.44}_{-1.05}$&\phantom{< }0.62&\phantom{0000}> 29.36&\phantom{0000}> 29.35&29.15 $\pm$ 0.26&28.97 $\pm$ 0.21&-18.07&\phantom{10}9.3&--\\
\phantom{0}640&53.15334&-27.77457&8.24$^{+0.31}_{-0.82}$&\phantom{< }0.62&\phantom{0000}> 27.67&\phantom{0000}> 27.62&27.23 $\pm$ 0.20&26.96 $\pm$ 0.15&-20.10&\phantom{1}61.6&--\\
\phantom{0}125&53.15890&-27.76500&8.61$^{+0.28}_{-0.42}$&\phantom{< }0.95&\phantom{0000}> 29.48&\phantom{0000}> 29.47&28.61 $\pm$ 0.14&28.31 $\pm$ 0.11&-18.85&\phantom{1}20.4&1,3,4,5\\

\enddata
\tablecomments{The objects are listed in order of increasing photometric redshift, and the magnitude upper limits shown are 3 $\sigma$.  The photometric redshift errors are 1 $\sigma$.  The column $\mathcal{P}_6$ is the integrated probability from 6 $\leq$ z $\leq$ 11, where only objects with $\mathcal{P}_6$ $>$ 0.6 were included in our sample.  M$_{1500}$ denotes the absolute magnitude at rest--frame 1500 \AA, which was computed for each object by converting the WFC3 fluxes to rest--frame (with $z = z_\mathrm{phot}$), and interpolating to 1500 \AA.  The SFRs were computed using M$_{1500}$ along with the Kennicutt (1998) UV luminosity density -- SFR relation.  The flag column denotes which objects were previously discovered by other studies, with the numbers representing: 1) \citet{bouwens09}; 2) \citet{oesch09}; 3) \citet{mclure09}; 4) \citet{bunker09} \& 5) \citet{yan09}.  Only ID 2056 was detected at $>$ 3 $\sigma$ in either of the IRAC bands, with m$_{[4.5]}$ = 25.53 $\pm$ 0.31.  The remaining objects have 3 $\sigma$ upper limits of $>$ 25.75 in [3.6] and $>$ 25.69 in [4.5].}
\end{deluxetable*}

\subsection{IRAC Photometry of Candidate High-Redshift Galaxies}\label{section:irac2}

In order to measure the IRAC flux at the position of each of our candidates, we constructed a simulated detection image using the IRAF\footnote[4]{The Image Reduction and Analysis Facility (IRAF) is distributed by the National Optical Astronomy Observatory (NOAO), which is operated by the Association of Universities for Research in Astronomy, Inc.\ (AURA) under cooperative agreement with the National Science Foundation.} task {\tt mkobjects} in the {\tt artdata} package, populated with sources with Gaussian profiles inserted at the positions of the high--redshift sources in our sample.  We then performed photometry in 3.0\arcs~diameter circular apertures on the source--subtracted IRAC images using SExtractor in two-image mode.

While \citet{labbe09b} detect $z \sim$ 7 galaxies with IRAC, their objects were selected from the shallower WFC3 Early Release Science (ERS; Windhorst et al.\ 2010) data, and are thus intrinsically brighter.  In contrast, only one object in our high-redshift sample (ID 2013) is detected in the IRAC 3.6~\micron\ data at $>$ 3 $\sigma$ significance (3.29 $\sigma$).  Upon visual inspection of this source, we found it to be strongly blended with an adjacent galaxy, with large residuals in the TFIT--subtracted image.  We therefore discard this flux measurement and replace its IRAC flux with zero in our subsequent analysis, using the measured 3.6~\micron\ magnitude limit as the $1\sigma$ error.

Four objects in our high-redshift sample are detected formally in the IRAC 4.5~\micron\ image, with 2.6--3.5 $\sigma$ significance.  None of these sources were detected in the IRAC 3.6~\micron\ image.  Two of the objects with 4.5~\micron--detections (ID 335 \& 1818) are blended with light from galaxies that are poorly subtracted by TFIT, and we set their 4.5 $\mu$m flux to zero, similar to how we handled the 3.6~\micron\ source discussed in the previous paragraph.

The other two objects, source IDs 2055 and 2056, are detected at 4.5~\micron\ with 3.1$\sigma$ and 3.5$\sigma$ significance, respectively.  Based on the photometric redshift estimates of these objects (see \S~\ref{section:photoz}) it is possible that the IRAC 4.5~\micron\ is contaminated by emission from [\ion{O}{3}] $\lambda\lambda$4959,5007, providing a possible explanation for the 4.5~\micron\ detections and lack of flux at 3.6~\micron.  However, these objects lie within $\simeq 1$\arcsec\ from each other, with overlapping isophotes in the WFC3 images, and they are blended at the resolution of IRAC.\footnote[5]{The photometry at the positions of these two objects was performed separately, thus the detection significance of the combined IRAC fluxes from both is still $\sim$ 3 $\sigma$.}.  Furthermore, we find that each of these galaxies have photometric redshift probability distribution functions consistent within their 68\% confidence ranges (see \S~\ref{section:photoz}), with $z_\mathrm{phot}$ = 7.81$^{+0.44}_{-0.32}$ and 7.25$^{+0.27}_{-0.23}$, respectively.  Given their close proximity and similar redshifts (within the uncertainties) , these objects may be physically associated.  However, because we are unable to deblend the IRAC data, we assign all the 4.5~\micron\ flux to object ID 2056 because this object is brighter by 0.2~mag in \wfch\ and brighter by 0.4~mag in \wfcj.  Even in the case that they are are unassociated physically, ID 2056 would likely dominate the emission.

\section{Color--Selection of Galaxies at $6.3 < \MakeLowercase{z} \leq 8.6$}

\begin{figure}[t]
\epsscale{1.1}
\begin{center}
\plotone{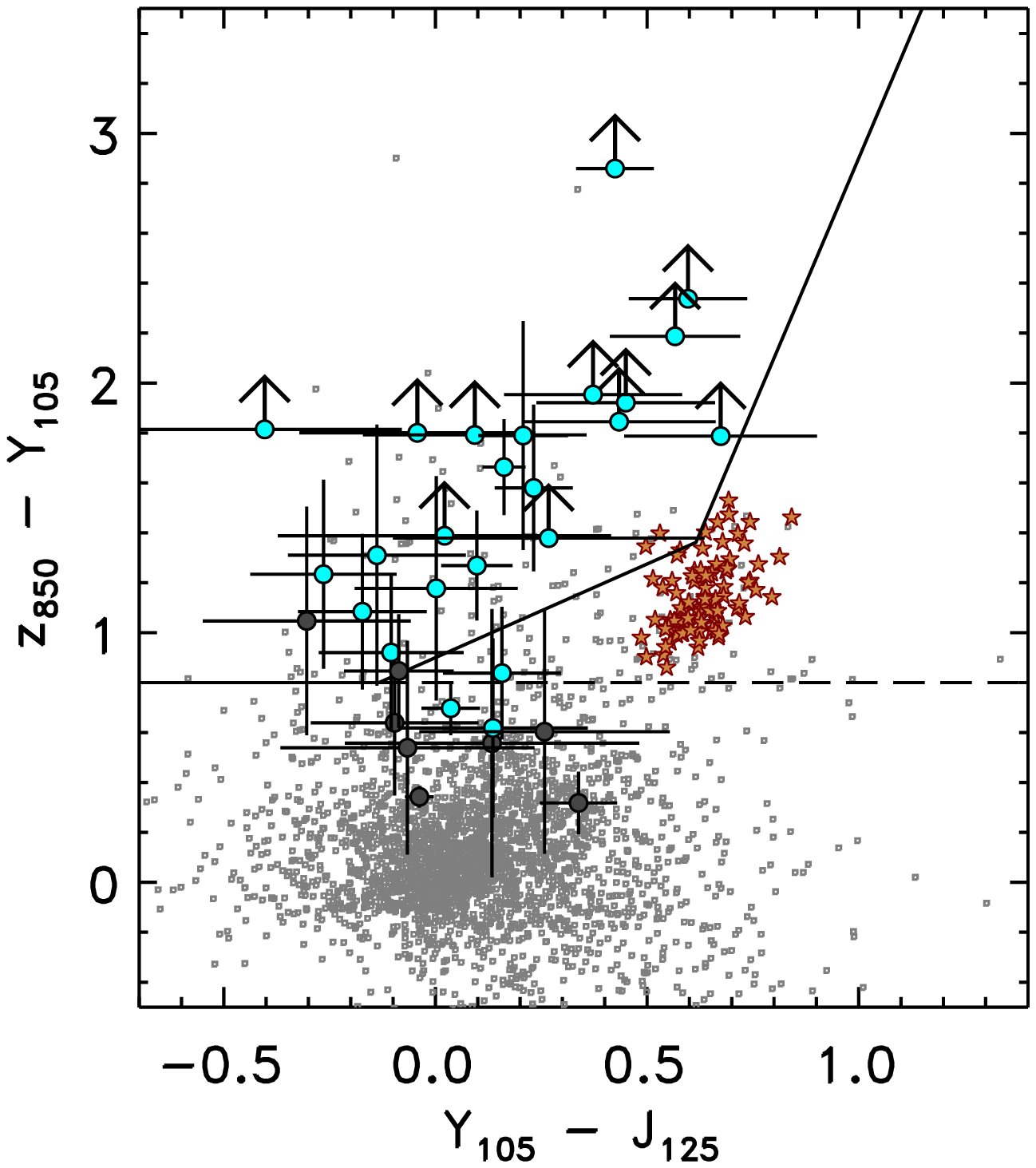}
\end{center}
\vspace{-8mm}
\caption{Color-color (\yb~- \jb~versus \zb~- \yb) plot showing how our photometric redshift selected objects compare to the selection criteria from various studies.  Small gray squares denote objects with $z_\mathrm{phot}$ $<$ 6.0.  Filled gray circles denote objects with 6.0 $<$ $z_\mathrm{phot}$ $<$ 6.3.  Filled cyan circles denote objects with 6.3 $<$ $z_\mathrm{phot}$ $<$ 7.5.  The arrows denote 1 $\sigma$ limits.  The solid lines show the selection criteria from \citet{oesch09}, while the dashed line shows the selection criterion from \citet{yan09}, both for LBGs at $z \sim$ 7.  Due to our photometric redshift analysis our sample includes galaxies outside of these windows.  Brown stars show the expected colors of galactic brown dwarfs.  Objects inside the LBG selection window which are not selected as high--redshift galaxies have significant optical flux, resulting in low--redshift photometric redshift fits.}\label{colcol1}
\end{figure}

\subsection{The Colors of Galaxies at $6.3 < \MakeLowercase{z} \leq 8.6$}

\fig{colcol1} shows the $\wfcy - \wfcj$ versus $\acsz - \wfcy$ color--color plot for galaxies in the HUDF, which is useful to isolate $z \sim$ 7 LBGs.  Galaxies with $z \gsim 6.3$ reside in the upper--left portion of the color-color space.  These objects have red $\acsz - \wfcy$ colors, indicative of the Ly$\alpha$ break redshifted to $\lambda \gsim 8900$~\AA.

We also show the color--selection criteria of \citet{oesch09} and \citet{yan09} used to select $z\sim 7$ galaxies in \fig{colcol1}, in order to contrast our selection method using the full photometric redshift probability distribution function with pure color--selected samples.  While these classical color--selection criteria identify the majority of objects in our sample, there are distinct differences.  Our sample selected using the full photometric redshift $P(z)$ includes galaxies excluded by these color--selection criteria.  Parenthetically, we find similar results if we use the photometric--redshift--derived samples from \citet{mclure09}.  Because both the criteria of Oesch et al.\ and Yan et al.\ require $\acsz - \wfcy \geq 0.8$~mag for object selection, they miss some objects with colors just below this threshold (owing to a combination of photometric uncertainties and color variations).  This is also expected as galaxies with intrinsically bluer $\wfcy - \wfcj$ colors enter these color--selection windows at lower redshifts than galaxies with redder $\wfcy - \wfcj$ colors.  This is illustrated by the fact that there is a source with $\wfcy - \wfcj = -0.3$~mag and $z_\mathrm{phot} < 6.3$ within the Oesch et al.\ color--selection ``wedge", whereas several galaxies with $z_\mathrm{phot} > 6.3$ with redder $\wfcy - \wfcj$ colors lie outside the color--selection windows.

Brown--dwarf stars are expected to have red $\acsz - \wfcy$ colors, mimicking the colors of high-redshift galaxies.  \fig{colcol1} shows the synthesized colors of brown--dwarf stars (using spectra of brown dwarfs from S. Leggett\footnote[6]{http://staff.gemini.edu/$\sim$sleggett/LTdata.html}, with the data from \citet{golomowski04},  \citet{knapp04} and \citet{chiu06}).  All the brown dwarfs in this sample would satisfy $\acsz - \wfcy > 0.8$~mag.  The Oesch et al.\ color--selection ``wedge" would exclude most of these objects.  However, all of these objects would be present in a sample using the criterion of Yan et al., although the authors argue their sample includes no unresolved sources.

\begin{figure}[t]
\epsscale{1.1}
\begin{center}
\plotone{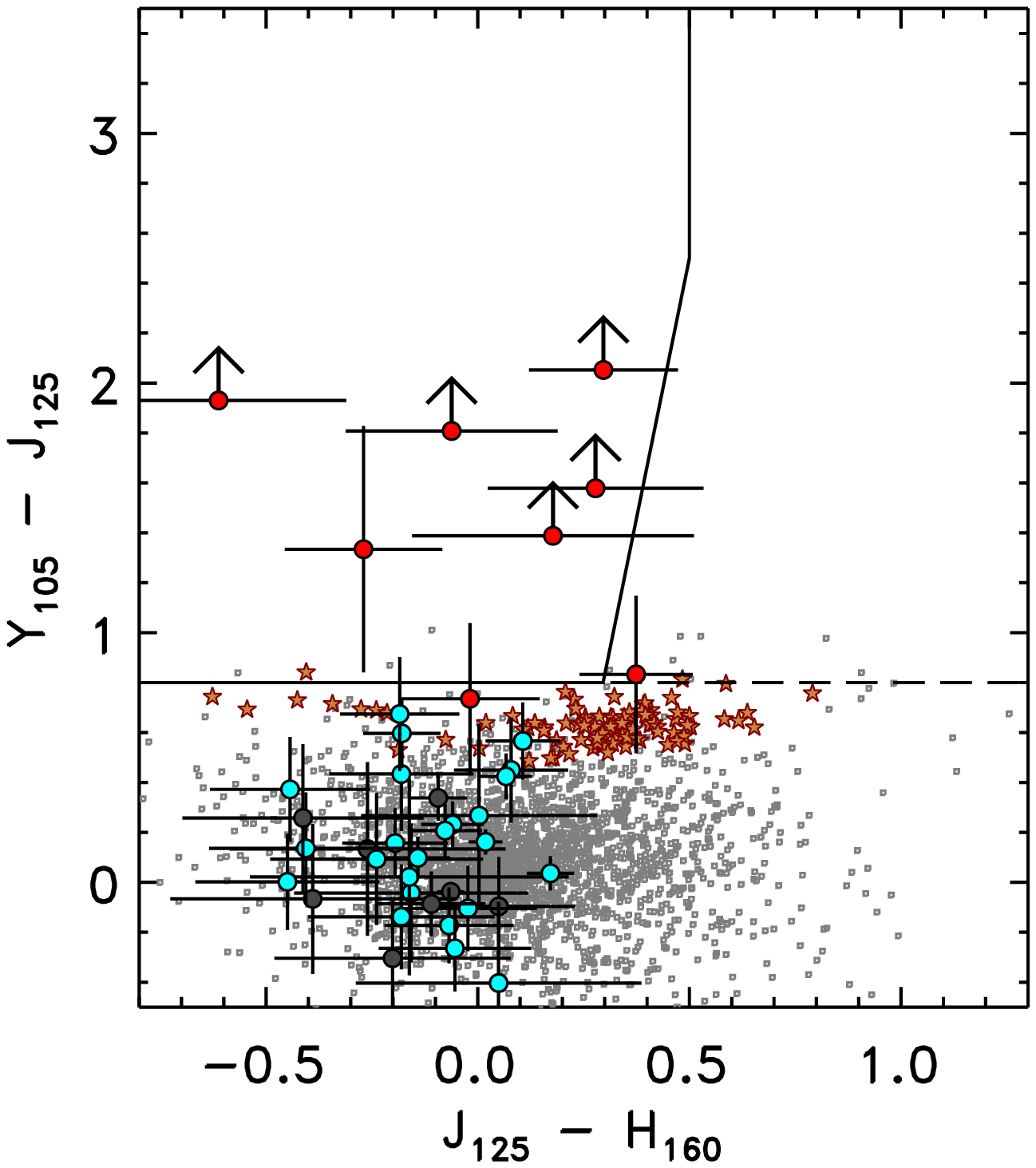}
\end{center}
\vspace{-8mm}
\caption{\jb~- \hb~versus \yb~- \jb~color-color plot for galaxies in our WFC3 sample.  Symbols are the same as in \fig{colcol1}, except here we also show galaxies with 7.5 $<$ $z_\mathrm{phot}$ $\leq$ 8.6 as red filled circles.  The solid lines show the selection criteria from \citet{bouwens09}, while the dashed line shows the selection criterion from \citet{yan09}, both for $z \sim$ 8 LBGs.}\label{colcol2}
\end{figure}

\fig{colcol2} shows the $\wfcj - \wfch$ versus $\wfcy - \wfcj$ color--color plot for galaxies in the HUDF, which is useful to isolate $z \sim$ 8 galaxies.  Those galaxies in our $z \sim$ 8 sample are indicated by red--filled circles, while galaxies in the lower--redshift $z \sim$ 7 sample are indicated as cyan--filled circles.  Similar to the $z \sim$ 7 sample in \fig{colcol1}, the $z \sim$ 8 sample resides largely in the upper--left portion of plot, and they have red $\wfcy$ -- $\wfcj$ colors, indicative of the Ly$\alpha$ break redshifted to $\lambda \gsim 11000$~\AA.  

\fig{colcol2} also shows the color--selection criteria of \citet{bouwens09} and \citet{yan09} to select $z\sim 8$ galaxies.  Of the eight galaxies in our $z \sim$ 8 sample, one with $\wfcy - \wfcj = 0.7$~mag (ID 819) would be missed by both color--selection criteria, and one additional galaxy (ID 2055) with $\wfcy - \wfcj = 0.8$~mag and $\wfcj$ -- $\wfch = 0.4$~mag lies outside the Bouwens et al.\ ``wedge".

As with \fig{colcol1} brown--dwarf stars provide contamination to $z\sim 8$ samples.  \fig{colcol2} shows that the expected $\wfcj - \wfch$ colors of brown dwarfs span a large range.  While most have $\wfcy - \wfcj < 0.8$~mag, the distribution lies near to this limit, implying that photometric uncertainties of brown dwarfs will mimic the colors of $z\sim 8$ galaxies.

\subsection{The Light Profiles of Galaxies at $6.3 < z \leq 8.6$}

\begin{figure}[t]
\epsscale{1.1}
\begin{center}
\plotone{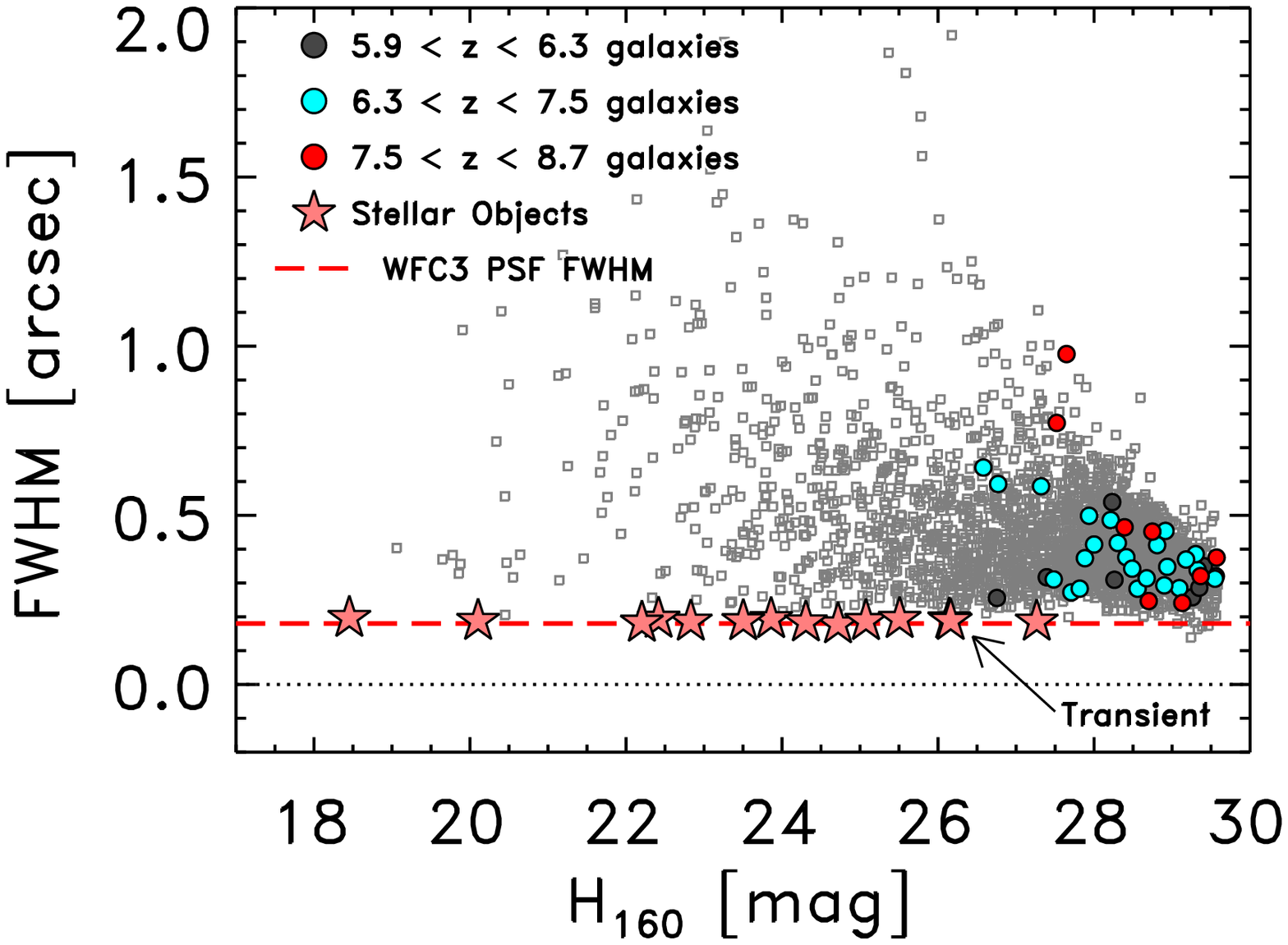}
\end{center}
\vspace{-5mm}
\caption{Because Galactic brown dwarfs can have similar colors to high redshift galaxies, we compare the measured FWHM of our sources to stars in the image.  Unresolved sources at the WFC3 resolution have FWHM $\leq$ 6 pixels ($0\farcs18$).  We visually inspected unresolved objects in the \hb\ image, and all appear very compact and consistent with being stellar.  We note that one of these objects is the transient first detected by \citet{oesch09}.  All of our candidate high-redshift galaxies are resolved, with FWHM $>$ $0\farcs2$.  We conclude our sample suffers no contamination by galactic brown dwarfs.}\label{fwhm}
\end{figure}

Owing to the fact that the expected $\acsz - \wfcy$, $\wfcy - \wfcj$, and $\wfcj - \wfch$ colors of brown--dwarf stars can mimic those of z $>$ 6 galaxies, we examined the size of the light profiles of the galaxies in our sample.  \fig{fwhm} shows the full--width at half--maximum (FWHM) of the galaxies in our catalog measured from their light profiles in the WFC3 F160W image.  The figure indicates FWHM = $0\farcs18$, the expected FWHM of WFC3 PSF in the F160W band.  We identified all bright stars in the WFC3 F160W image, and these all have a FWHM $\approx$ 6~pixels ($0\farcs18$ at $0\farcs03$~pix$^{-1}$).  We subsequently examined all objects in our full WFC3 catalog with a FWHM $<$ 6.6 pixels and found them to have stellar morphologies, consistent with point sources.  These sources are indicated as large filled pentagrams in \fig{fwhm}.  The temporally transient source identified originally in our sample of high--redshift galaxies is also indicated in the figure, and it is consistent with being a point source.  In contrast, the FWHM measured for the F160W light profiles of our high--redshift samples are resolved, and exceed the FWHM of point sources down to the magnitude limit of our catalog.  From this we conclude that the sources in our $z \sim$ 7 and $z \sim$ 8 samples are resolved galaxies, with no contamination from stellar sources (including brown dwarfs).

\subsection{Comparison to High--Redshift Sample Selection from the Literature}\label{section:comparison}

In this section we compare the galaxies in our $z \sim$ 7 sample and $z \sim$ 8 sample to others in the literature using these WFC3 HUDF data.  Table 1 provides a flag for sources in our sample that match the astrometric coordinates of sources in these other studies.  In the subsections below we provide qualitative comparisons.

\subsubsection{Comparison to Bouwens et al.\  \& Oesch et al.}
\citet{bouwens09} examined an early reduction of the HUDF WFC3 data, searching for ``\yb"--dropout galaxies at $7.5 \lesssim z \lesssim 8.5$ (see also \fig{colcol2}).  Using these criteria with additional criteria to reject low--redshift interlopers, they reported five candidate LBGs, all with \hb~$>$ 28 mag.  We recover all five of these $z \sim$ 8 candidate LBGs from \citet{bouwens09}, and we find that the photometric redshifts of these galaxies range from 7.6 $\leq$ $z_\mathrm{phot}$ $\leq$ 8.6.  The measured $\wfcj - \wfch$~colors of the matched objects are consistent within $<$ 0.2~mag.

\citet{oesch09} used the same reduction of the WFC3 HUDF data as Bouwens et al.\ and searched for ``\acsz''--dropout galaxies at $6.5 \lesssim z \lesssim 7.5$ (see also \fig{colcol1}).  Oesch et al.\ identified 16 \zb-dropout galaxies.  We match 14 of these galaxies, for which we derive a photometric redshift range from 6.3 $\leq$ z$_{phot}$ $\leq$ 7.4.  Of the two objects not in our catalog, one fails our magnitude limit for selection as it has $\wfcj = 29.6$~mag and $\wfch=30.3$~mag, and for the other we compute $z_\mathrm{phot} =$ 5.7.  The $\wfcj - \wfch$ colors we measure for these galaxies are consistent within $<$ 0.2~mag for 12 out of the 14 objects in common between our catalogs. For the remaining objects, we measure color discrepancies as high as 0.28 mag, which we attribute to variations in data reduction (see \S~\ref{section:data}).

\subsubsection{Bunker et al.}
\citet{bunker09} perform color--color selections similar to that employed by \citet{oesch09} and \citet{bouwens09}, selecting 11 \zb--dropout galaxies with $z \sim$ 7, and 7 \yb--dropout galaxies with $z \sim$ 8.  We match all 11 \zb-dropouts from Bunker et al.\, and we measure a photometric--redshift range of 6.4 $\leq$ $z_\mathrm{phot}$ $\leq$ 7.4.  We note, however, that in our catalog object zD4 of \citet{bunker09} is split into two objects in our sample, ID 2055 and 2056 (see \S 2.6).  We recover 4 of the 7 \yb-dropouts from Bunker et al., measuring a range of 7.5 $\leq$ $z_\mathrm{phot}$ $\leq$ 8.6.  We excluded 2 of the missing 3 sources selected by Bunker et al.\ because they are fainter than our $3.5\sigma$ magnitude limit in $\wfch$.  The remaining source (Bunker et al.\ ID YD5) has a photometric redshift of $z_\mathrm{phot} = 2.0^{+0.3}_{-0.6}$, and was therefore excluded from our sample.

\subsubsection{McLure et al.}
The selection of \citet{mclure09} is most similar to that used here.  McLure et al.\ used an independent reduction of the WFC3 HUDF data, and selected all WFC3 sources with no detected counterpart in the ACS F775W images.  We match 25 of the 35 objects identified by \citet{mclure09} at $z\geq 6.3$.  Of the 10 objects not in our catalog, 5 have 5.7 $\leq$ $z_\mathrm{phot}$ $\leq$ 6.2, one has $z_\mathrm{phot} = 1.3^{+0.3}_{-0.5}$, and 3 are fainter than our catalog magnitude limits.  The remaining object has $z_\mathrm{phot} = 6.6^{+0.2}_{-5.4}$, with the large redshift uncertainty giving $\mathcal{P}_6 <$ 0.6.

For the 25 objects in common between our catalogs, we find that our photometric redshifts are consistent within $\Delta$z $\approx$ 0.2 for 17 objects.  Although our \wfcj\ - \wfch\ colors are consistent with that measured by McLure et al.\ ($\lesssim$ 0.3~mag for all sources), small differences due to photometric aperture variations affect the redshift interpretation.  The catalog used by McLure et al.\ used $0\farcs6$--diameter circular apertures to measure colors, in contrast to the elliptical apertures used to construct our catalog.

\subsubsection{Yan et al.}
\citet{yan09} used an independent reduction of the WFC3 HUDF data to select \zb--dropout ($z\sim 7$), \yb--dropout ($z\sim 8$), and \jb--dropout ($z\sim 9$) galaxies.  They identify 20 \zb--dropouts, 15 \yb--dropouts, and 20 \jb--dropouts.  We match 16 of the 20 \zb-dropouts in our catalog, for which we measure a photometric redshift range of 6.4 $\leq$ $z_\mathrm{phot}$ $\leq$ 7.4.  We match 7 of the 15 \yb-dropouts, with 6/7 residing in our $z \sim$ 8 sample, and one additional having $z_\mathrm{phot}$ = 6.8 (included in our z $\sim$ 7 sample).  Of the 12 galaxies of Yan et al.\ unmatched in our samples, 8 are fainter than the magnitude limit required for our selection, including 1 galaxy that is split into two components in our catalog (both of which are then fainter than our magnitude limit).  For three others we derive photometric redshifts $z_\mathrm{phot} < 6.3$.  We excluded the remaining galaxy because it has overlapping isophotes with a bright nearby galaxy, which appears to contaminate its colors.  The additional galaxies in the sample of Yan et al.\ are fainter and less secure as Yan et al.\ require only a 3$\sigma$ detection in either the F125W or F160W--bands.  Additionally, Yan et al.\ did not restrict the $\wfcj - \wfch$ color of their \yb-dropout sample, which possibly suffers a higher contamination from galactic brown dwarfs (although Yan et al.\ exclude unresolved sources) and lower redshift galaxies with older stellar populations (see also \fig{colcol2}).  As detections were required in both the F125W and F160W --bands for our photometric redshift analysis, we have no sample to compare with the single--band--detected \jb-dropouts of Yan et al.

\section{Rest--Frame UV Properties of Galaxies at $6.3 < \MakeLowercase{z} \leq 8.6$}

\begin{figure*}[!ht]
\epsscale{0.8} 
\begin{center} 
\plotone{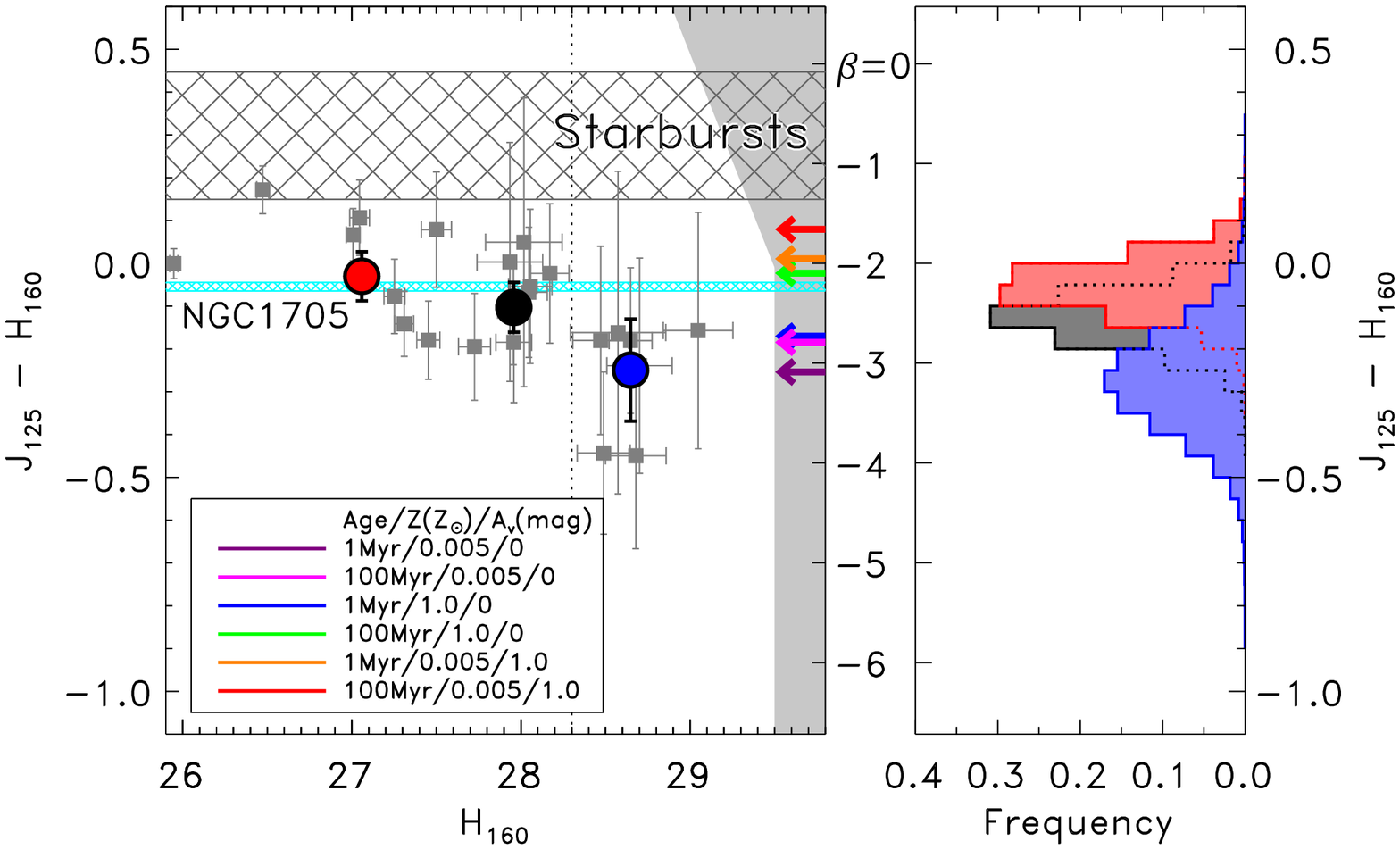} 
\end{center} 
\vspace{2mm} 
\caption{\hb\ versus $\wfcj - \wfch$ color-magnitude diagram for the $z \sim$ 7 sample.  The 6.3 $<$ $z_\mathrm{phot}$ $<$ 7.5 galaxies are plotted as gray squares, with the error bars representing the 1 $\sigma$ photometric uncertainties.  The gray shaded area corresponds to the region excluded by the magnitude limits of our sample.  The central vertical axis denotes the UV spectral slope $\beta$, which is a function of the color (see \S 4.2).  The black, red and blue large circles show the mean color for all objects, the bright subsample (\wfch\ $<$ 28.3) and the faint subsample (\wfch\ $\geq$ 28.3), respectively.  The dotted line denotes the magnitude split of \wfch\ $=$ 28.3.  The histograms in the right--hand panel represent the distribution of the means for all objects (black histogram), objects with \hb~$<$ 28.3 (red histogram), and objects with \hb~$\geq$ 28.3 (blue histogram) from the Monte Carlo bootstrap simulations.  The errors on the means were derived from the standard deviation of these distributions.  The gray cross-hatched region shows the color spanned by the \citet{kinney96} starburst galaxy templates (redshifted to $z = 7$).  We also show the Kinney template for NGC~1705, which is one of the bluest local starburst galaxies.  The colored arrows denote the colors of CB07 stellar population models, described in the figure legend, where the three numbers denote their age, metallicity (in solar units), and dust extinction (in A$_\mathrm{V}$), respectively. While many objects have measured colors bluer than NGC~1705, the average of the faint subsample is consistent within 2 $\sigma$.  The faint subsample is also consistent with all dust-free CB07 models.  This implies that these objects may have very low dust contents.  However, as the distribution of the CB07 models shows, the colors of stellar populations do not change much between 1--100 Myr and $Z$ = 0.005--1.0 $Z$\sol, so determinations of these properties are difficult, though it does appear as if these objects are $\lesssim$ 100 Myr and/or have $Z$ $<$ $Z$\sol.}\label{colmag7}
\end{figure*} 

\subsection{Rest--Frame UV Colors}
\fig{colmag7} shows the color--magnitude diagram of \hb\ versus $\wfcj - \wfch$ for our $z \sim$ 7 sample.  At these redshifts the $\wfcj - \wfch$ color probes rest--frame 1500--1900~\AA. This region is highly sensitive to massive star formation and the amount of dust attenuation.  Even though the photometric uncertainties on individual objects are large owing to their faint observed fluxes, the $z \sim$ 7 galaxies have blue $\wfcj - \wfch$ colors.  In few cases, these blue colors are consistent with the expected rest--frame UV colors of local starbursts \citep{kinney96}, but the majority of the galaxies are bluer than these local starbursts.  This is not a consequence of our selection, as we made no restriction on rest--frame UV color.  To verify this, in \fig{colmag7} we shade the region excluded by our magnitude limits (\wfcj\ $\leq$ 29.55 and \wfch\ $\leq$ 29.60).  This region would exclude galaxies only with $\wfcj - \wfch$ redder than local starbursts at \wfch\ $<$ 29.

To quantify the photometric scatter in our data, we ran simulations whereby we generated $\sim$ 5 $\times$ 10$^{4}$ mock galaxies and placed them in the WFC3 data.  These mock galaxies were generated with a uniform distribution of luminosities and redshifts over 6 $< z <$ 9.  We used models from the 2007 update to Bruzual \& Charlot (2003; hereafter CB07) for their intrinsic SEDs, with a constant star formation history, Salpeter initial mass function (IMF), age of $\sim$ 10$^{8}$ yr and $Z$ = 0.005 $Z$\sol.  They were attenuated by dust via the extinction curve of \citet{calzetti00} with $-$0.3 $<$ E(B-V) $<$ 0.5 (where the negative color excesses simulate extremely blue galaxies) as well as by the IGM \citep{madau95}.  The galaxy shapes were chosen to be exponential disks, with a log-normal radius distribution over the range of $0\farcs03$ $<$ r $<$ 1.0\arcs, with the mean recovered galaxy radius of 0.14\arcs.  These mock galaxies were then placed in the WFC3 data with the appropriate amount of noise using {\tt artdata} and we extracted their flux in much the same way as with our observations.  Upon investigation of these catalogs, we found that while the offset between the input color and measured color is small ($<$ 0.05 mag from \hb\ = 25--29 mag), the photometric scatter can be large, from $\sigma$ = 0.05 mag at \hb\ = 25, to $\sim$ 0.2 mag at \hb\ = 29.  This scatter is representative of the true photometric uncertainty on the colors of our objects.  We note that the scatter of $\sim$ 0.2 mag at m$_{H}$ = 29 is similar to what we obtained from our photometry in \S 2.

Given the large photometric scatter on individual galaxies, we compute the mean $\wfcj - \wfch$ color as a function of \wfch\ brightness, in subsamples of $\wfch \geq 28.3$~mag and $\wfch < 28.3$~mag (where $\wfch = 28.3$ is equivalent to M$_{UV}$ = $-$18.75 assuming $z = 7$ and the mean $\wfcj - \wfch$ color of the z $\sim$ 7 sample).  We compute the mean and uncertainty using Monte Carlo bootstrap simulations of $10^7$ realizations of the data.  In each simulation, we perturb randomly the colors of an object by its associated photometric scatter, and then compute the median $\wfcj - \wfch$ of the simulated objects.  Because these sample sizes are small (i.e. the \wfch\ $\geq$ 28.3 subsample at $z \sim$ 7 consists of 7 galaxies) we include the effects of Poisson noise.  We do this by including N + A$\times$$\sqrt{N}$ galaxies in the median for a given simulation, where N is the number of galaxies in a given sample, A is a random number drawn from a normal distribution with a mean of zero and a standard deviation of one, and the galaxies chosen are randomly self-sampled.

The mean and standard deviation of this distribution is thus an estimate of the true mean and uncertainty on the $\wfcj - \wfch$ distribution of the real sample.  The histograms in the right panel of \fig{colmag7} shows the results of our simulations for the $z \sim$ 7 sample, and Table 2 gives the measured mean and 68\% confidence range.  

\begin{figure*}[t]
\epsscale{0.8}
\plotone{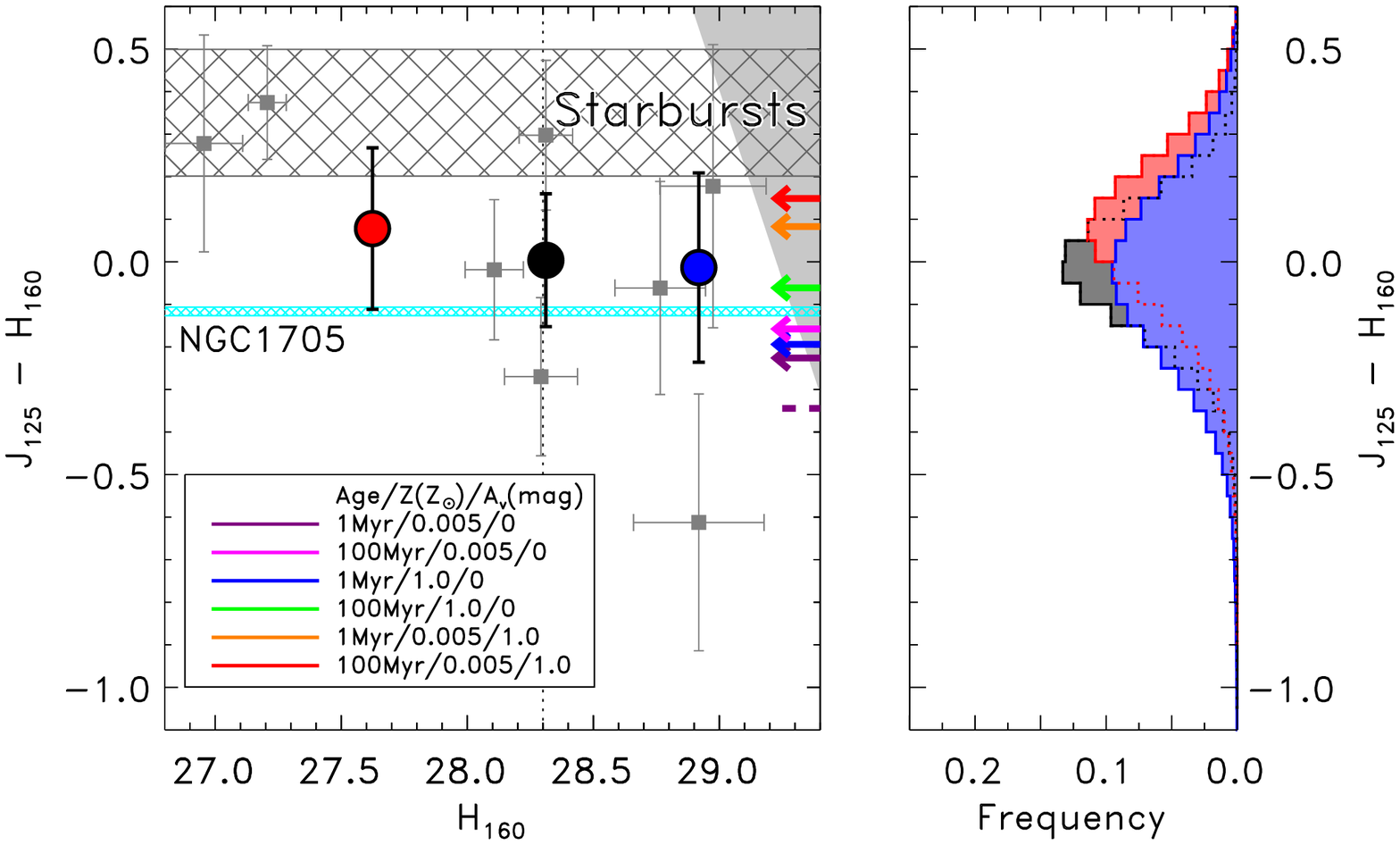}
\vspace{2mm}
\caption{\hb\ versus $\wfcj - \wfch$ color-magnitude diagram for the $z \sim$ 8 sample.  The $z_\mathrm{phot}$ $>$ 7.5 galaxy candidates are plotted with symbols defined in \fig{colmag7}, and models redshifted to z = 8.  As \lya~affects the F125W--band at $z \sim$ 8, we show an additional model track as a dashed line, which is the bluest possible model from CB07, with \lya~added in (this model has a rest--frame \lya\ EW $\sim$ 430 \AA; see \S 5.1).  Similar to $z \sim$ 7, these objects appear to be fairly dust free, especially the faint subsample, but determinations of their metallicities and ages are difficult, as the object's colors are consistent with a range of both parameters.  At these redshifts \lya\ lies in the F125W--band, thus we do not compute $\beta$.}\label{colmag8}
\end{figure*}

\fig{colmag8} shows the \hb\ versus $\wfcj - \wfch$ color--magnitude diagram for the $z \sim$ 8 sample.  We performed a similar calculation of the mean and 68\% confidence range for this sample, using the same Monte Carlo bootstrap method as for the $z \sim$ 7 sample discussed above.  At $z \sim$ 8 the \lya~emission line resides in the F125W-band, and we indicate the its effect on the CB07 model with the bluest colors in the figure.  In general, the objects in the $z \sim$ 8 sample have fainter \wfch\ magnitudes than the objects in the $z \sim$ 7 sample.  Along with the smaller sample size, the measurements of the mean colors of the $z \sim$ 8 sample have larger uncertainties. 

\subsection{UV Spectral Slope}

In order to facilitate the comparison of our results to previous studies, we use the WFC3 photometry to approximate the slope of the rest--frame UV continuum, $\beta$, of the galaxies in our $z \sim$ 7 sample, where $f_\lambda \propto \lambda^\beta$ \citep[e.g.,][]{calzetti94}.  Although the definition of $\beta$ is based on regions of the UV spectrum devoid of strong absorption features, it is common to estimate the UV spectral slope of distant galaxies using the broad--band measures of the rest--frame UV color \citep[e.g.][]{meurer99}.  Here we use the relation of \citet{bouwens09c} to estimate $\beta$ for galaxies in our $z \sim$ 7 sample,
\begin{equation}
\beta_\mathrm {phot} = 4.29 (J_{125} - H_{160}) - 2.0,
\end{equation}
where $\beta_\mathrm {phot}$ denotes a photometric estimate of $\beta$.  The formal definition of $\beta$ pertains only to wavelengths greater than that of \lya.  Therefore, we restrict our analysis to those objects in our $z \sim$ 7 sample with $z_\mathrm{phot}$ $<$ 7.5, where \lya~emission falls outside the F125W-band.  We show $\beta_\mathrm {phot}$ for our sample at $z_\mathrm{phot}$ $<$ 7.5 in \fig{colmag7} along the middle vertical axis, and our measurements of $\beta_\mathrm {phot}$ are tabulated in Table 2, from the colors from the Monte Carlo bootstrap simulations.

\subsection{Interpretation of the UV Colors and UV Slopes}

\fig{colmag7} compares our measurements of the mean color and its associated uncertainty for the $z \sim$ 7 sample to the empirical \citet{kinney96} local starburst template spectra.  We also compare our measurements to stellar--population synthesis models from CB07 spanning a range of age, metallicity and dust extinction.  Lastly, we also compare our measurements to the local extremely blue galaxy NGC~1705, also from \citet{kinney96}.  NGC~1705 is a blue compact dwarf galaxy with recent strong star-formation activity, and is unobscured with E(B-V) = 0.00 \citep{calzetti94}.  \citet{lee04} published metallicity measurements from 16 H\,{\sc ii} regions in NGC~1705, finding an average oxygen abundance of $\sim$ 0.35 $Z$\sol, making it comparable to Local Group dwarf irregular galaxies.  More recently, \citet{annibali09} find a good fit to the young stellar populations in NGC~1705 when assuming metallicities of only 0.02 $Z$\sol.  Thus, NGC~1705 is one of the more metal poor local star--forming galaxies.

Both the bright and faint $z \sim$ 7 subsamples are bluer than local starburst templates from \citet{kinney96}.  With our simulation results, we derive a 99.9987\% (4.4 $\sigma$) confidence that the color of an average $z \sim$ 7 galaxy is bluer than that of a local starburst galaxy.  The confidence that the bright and faint subsamples are individually bluer than local starbursts is slightly less given their smaller sample sizes, at 99.88 \% and 99.86 \% ($\sim$ 3.2 $\sigma$), respectively.  Rather, the colors of these $z \sim$ 7 galaxies are closer to the that of NGC~1705.  The confidences that these three samples are bluer than NGC~1705 are $\sim$ 80\%, 33\% and 95\%, respectively (i.e. consistent within 2 $\sigma$).

This is in contrast to studies of star--forming LBGs at lower redshifts, $2 < z < 6$ \citep[e.g.,][]{papovich01, ouchi04, reddy09, overzier09, bouwens09b}.  \citet{papovich01} showed that UV--selected galaxies at $z\sim$ 2--3 have rest--frame UV colors which span a range similar to that of the composite starbursts from Kinney et al., while these typical local templates are unable to reproduce the blue colors of the $z \sim$ 7 sample.  Therefore, we conclude that $z \sim$ 7 galaxies are dominated by stellar populations that have a combination of young ages, lower metallicity, and little dust extinction.  Additionally, in contrast to the color--magnitude relations observed in lower--redshift galaxies at $4 < z < 6$ \citep[e.g.,][]{papovich04,bouwens09b,overzier09}, with our relatively small sample size we observe only a weak ($<$ 2 $\sigma$) dependence between the UV--color and magnitude in the $z \sim$ 7 sample.

Galaxies at $z \sim$ 7 are uniformly blue, nearly independent of magnitude.  \fig{colmag7} shows the expected colors of six CB07 stellar population synthesis models, covering a range in age (1--100 Myr), metallicity ($0.005 - 1$~\zsol), and dust extinction ($A_{V} =$ 0--1 mag) formed with a constant star--formation history.  With only a single rest--frame UV color, it can be difficult to accurately constrain these properties.  For example, for a fixed age of 1~Myr (100~Myr), changing the metallicity from 0.005~\Zsol\ to $Z$\sol\ increases the $\wfcj - \wfch$ by 0.08 mag (0.16 mag).  Similarly, at fixed metallicity increasing the age from 1~Myr to 100~Myr increases the color by 0.07--0.14~mag, with a slight dependence on the metallicity.  The models with either very young ages ($<$ 100~Myr) or very low metallicity ($\sim$ 0.005~\Zsol) match the observed $\wfcj - \wfch$ colors of the faint $z \sim$ 7 subsample with $\wfch \geq 28.3$~mag.  This is not true for models with higher ages or higher metallicities.  If we include the effects of dust attenuation on any of these models the colors become considerably redder, extending them beyond the 1$\sigma$ range of the faint $\wfch \geq 28.3$~mag subsample, and also extending them beyond the $1\sigma$ range for the brighter $\wfch < 28.3$~mag for the model with age 100~Myr.  The conclusion here is strong.  \textit{The extremely blue colors of faint galaxies with $\wfch \geq 28.3$~mag in the $z \sim$ 7 sample implies that they have low dust contents.}  Their colors are also consistent with low metallicities and low stellar population ages, but further work is necessitated to place strong constraints on these properties (see \S 5).

We draw similar conclusions for the $z \sim$ 8 sample as for the $z \sim$ 7 sample.  However, due to the smaller sample size and increased distance to these galaxies, the conclusions have a lower significance, as the bin of all galaxies are bluer than local starbursts at only $\sim$ 1.7 sigma significance.  The mean color $\wfcj - \wfch = $-$0.01 \pm 0.22$~mag of the faint subsample is consistent with that at $z \sim$ 7 within $\sim$ 1 $\sigma$, and it is also consistent with the expected color of NGC~1705.
\begin{deluxetable*}{ccccccc}
\tabletypesize{\small}
\tablecaption{Average UV Continuum Properties}\label{tab:uv}
\tablewidth{0pt}
\tablehead{
\colhead{Redshift} & \colhead{Sample} & \colhead{$J_{125} - H_{160}$} & \colhead{$\beta_\mathrm {phot}$} & \colhead{Confidence bluer than}  & \colhead{Confidence bluer than}& \colhead{$\rho_\mathrm{UV}$}\\
\colhead{$ $} & \colhead{$ $} & \colhead{$ $} & \colhead{$ $} & \colhead{local starbursts} & \colhead{NGC1705} & \colhead{($\times$ 10$^{26}$)}\\
}
\startdata
z $\sim$ 7&All Objects&$-$0.10 $\pm$ 0.06&$-$2.44 $\pm$ 0.25&99.9987\%&79.8\%&0.44 $\pm$ 0.11\\
z $\sim$ 7&$H_{160} <$ 28.5&$-$0.03 $\pm$ 0.06&$-$2.13 $\pm$ 0.25&99.88\%&32.7\%&---\\
z $\sim$ 7&$H_{160} \geq$ 28.5&$-$0.25 $\pm$ 0.12&$-$3.07 $\pm$ 0.51&99.86\%&95.0\%&---\\
z $\sim$ 8&All Objects&\phantom{$-$}0.00 $\pm$ 0.16&---&91.1\%&20.3\%&0.14 $\pm$ 0.06\\
z $\sim$ 8&$H_{160} <$ 28.5&\phantom{$-$}0.08 $\pm$ 0.19&---&74.9\%&14.2\%&---\\
z $\sim$ 8&$H_{160} \geq$ 28.5&$-$0.01 $\pm$ 0.22&---&84.6\%&29.9\%&---\\
\enddata
\tablecomments{We do not measure $\beta_\mathrm {phot}$ for the z $\sim$ 8 galaxies, as \lya~(emission or absorption) will contaminate the measurement.  Additionally, the confidence ranges on the z $\sim$ 7 galaxy sample are more robust than at z $\sim$ 8, given the larger number of objects in the z $\sim$ 7 sample.  We derive the errors on these values via 10$^{7}$ bootstrap Monte Carlo simulations.  The specific luminosity density at 1500 \AA, $\rho_\mathrm{UV}$, has units of erg s$^{-1}$ Hz$^{-1}$ Mpc$^{-3}$.}
\end{deluxetable*}

\citet{schaerer02} have argued that nebular continuum emission from regions around massive star formation at very young ages could affect the observed colors of high--redshift galaxies at these redshifts.  While this continuum emission is not included in the CB07 stellar population synthesis models, we conclude here that its effects are mitigated in our sample (similar to conclusions reached by \citet{bouwens09c}).  The local starburst galaxy I Zw 18 has the lowest metallicity known of any galaxy, and it is dominated by actively forming young stars as evidenced by its large Wolf-Rayet stellar population.  Nevertheless, I Zw 18 shows no evidence for nebular continuum emission \citep{brown02}.  Furthermore, while this continuum emission is expected to redden the UV continuum of the galaxies, mimicking stellar populations with a combination of higher dust attenuation, older ages, or higher metallicity, the galaxies in our $z \sim$ 7 and $z \sim$ 8 samples have nearly maximally blue colors, implying that this red continuum does not dominate their integrated properties.  Lastly, the nebular continuum is significant only if a high fraction of hydrogen--ionizing photons are absorbed by the ISM, which may not be the case here as the dust extinction appears small \citep[see also][and \S~\ref{section:reion}]{bouwens09c}.  This may imply that the fainter galaxies have higher Lyman continuum escape fractions (see \S~\ref{section:reion}). 

\subsection{Comparison with Other Samples}
Several groups have studied the slope of the rest--frame UV continuum of LBGs \citep[e.g.,][]{steidel99, meurer99, ouchi04, stanway05, reddy06, reddy08, reddy09, hathi08, bouwens06, bouwens07, bouwens09b}, finding a consensus conclusion that the UV spectral slope evolves toward lower values (bluer colors) at progressively higher redshifts and fainter magnitudes.  \citet{bouwens09b} found that LBGs at at 2.5 $<$ z $<$ 6 shift systematically to bluer colors with increasing redshift, even when properly taking into account the LBG--selection criteria.  The slope of the UV continuum $\beta$ should be more sensitive to variations in the amount and configuration of dust attenuation in these galaxies, and rather less sensitive to differences in stellar ages, metal fraction, or the IMF \citep[e.g.,][]{calzetti94, meurer99}.  Therefore, this evolution in $\beta$ implies that the overall dust content in distant galaxies decreases with increasing redshift.  Additionally, \citet{bouwens09c} reported evidence for a shift of $\beta_\mathrm {phot}$ to steeper (i.e., bluer) values at lower UV luminosity, as high--redshift galaxies with L $\sim$ 0.15 L$^{*}_{z = 3}$ are systematically bluer by $\sim$ 0.3 in $\beta_\mathrm {phot}$ compared to L$^{*}_{z = 3}$ galaxies, where L$^{*}_{z = 3}$ refers to the characteristic luminosity at $z =$ 3 measured by \citet{steidel99}.  

For our full sample of galaxies with $z < 7.5$ we find a mean UV spectral slope of $\beta_\mathrm {phot}$ = $-$2.44 $\pm$ 0.25, dropping to $\beta_\mathrm {phot}$ = $-$3.07 $\pm$ 0.51 for the faint subsample with $\wfch \geq 28.3$~mag.  Evidence for a dependence of UV slope on the UV luminosity is only weakly observed in our data given the uncertainties.  These values for the spectral slopes are indicative of galaxies dominated by young stars with low dust attenuation \citep[e.g.,][]{calzetti94,kinney96}, and thus reinforces our conclusions from the \S 4.3.  In particular, the low values of $\beta_\mathrm {phot}$ for our faint subsample are consistent with the local starburst NGC~1705, which has near--zero dust attenuation.

\citet{bouwens09c} obtain similar UV spectral slopes in their study of the UV continuum slopes of their $z \sim$ 7 sample (with $-$20 $\leq$ M$_{1500}$  $\leq$ $-$18).  Bouwens et al.\ argue that this is evidence for very low metallicities ($Z$ $\leq$ 0.005 $Z$\sol, possibly as low as $Z$ = 0) as only these stellar populations reproduce the observed values for $\beta$.  The models studied by Bouwens et al.\ include effects of nebular continuum (see discussion in the previous section), which would \textit{redden} the UV slope, implying that the galaxies would need a very large escape fraction for UV photons ($f_{esc}$$\geq$ 0.3).  However, as we discuss in \S 4.3, the distribution of galaxy colors after properly accounting for the photometric uncertainties does not provide evidence for this effect.  Rather, the evidence suggests these objects have a make-up consistent with extremely blue local galaxies such as NGC~1705.

\vspace{5mm}
\section{The Stellar Populations of $6.3 < \MakeLowercase{z} \leq 8.6$ Galaxies}\label{section:sed}

\subsection{Stellar--Population Synthesis Model Fitting}

We compare the galaxy colors to stellar population synthesis models to further constrain the properties of these galaxies.  We fit all 31 galaxies in our $z \sim$ 7 and $z \sim$ 8 samples with a suite of CB07 models, and in all cases we use models with a Salpeter IMF (with the default CB07 mass range of 0.1 -- 100 $M$\sol).  Because the rest--frame UV colors of the galaxies are indicative of ongoing star formation we have fixed the star formation history (SFH) to be constant \citep[see also][]{labbe09b}.  Given the young ages we derive for the stellar populations, this is a reasonable approximation.  However, if we allow for exponentially declining SFHs, we would expect to find even younger stellar population ages \citep[see, e.g.,][]{papovich01, shapley01}, therefore this assumption yields conservative results.

During the fitting, we vary the age (where age represents the time elapsed since the onset of star formation) from 1 Myr to the time elapsed from $z =$ 20 (a reasonable estimate for the onset of galaxy formation) to the given redshift, and we apply dust extinction using the \citet{calzetti00} attenuation law over a range of 0 $\leq$ E(B-V) $\leq$ 0.4 (equivalent to 0 $\leq$ A$_\mathrm{V}$ $\leq$ 1.6 mag).  Given the blue colors of these objects from \S 4, we expect our maximum possible dust extinction to be sufficiently high.  We allow the metallicity to vary between the full range in CB07, from 0.005 -- 2.5 $Z$\sol.  We apply IGM attenuation via the prescription of \citet{madau95}.

We expect the galaxies in our samples to be dominated by intensely star--forming stellar populations.  We therefore include the effects of \lya~emission on the stellar population models \citep[for details see][]{finkelstein08, finkelstein09a}.  \lya\ emission may significantly affect the colors given that the observed EW of the line increases as $(1 + z)$ and the effective width of the filters decreases as $(1+z)^{-1}$.  To estimate the expected \lya\ line--flux, we use the number of ionizing photons provided for each CB07 model as a function of stellar population age.  Assuming Case~B recombination \citep{osterbrock89}, approximately 2/3rds of hydrogen--ionizing photons produce \lya~photons during recombination.  We sum the expected \lya~flux with the CB07 model spectrum.  This is done prior to the application of the dust and IGM attenuation in order to most accurately represent the processes in these galaxies.  \lya~lines in Ly$\alpha$ emitting galaxies typically show line profiles truncated on the blue-side presumably due to IGM absorption \citep[e.g.,][]{stern99, rhoads03}.  We model this effect by allowing the full IGM attenuation on one--half of the line flux.  This results in an effective IGM optical depth at \lya~of
\begin{equation}
\tau_\mathrm{eff} = -\ln(0.5) -\ln(1 + \exp[\tau_\mathrm{IGM} (\lambda_{Ly\alpha})]). 
\end{equation}

We tested the redshift dependence on our SED fits by allowing the redshift to be a free parameter, fitting the full ACS, WFC3 and IRAC photometry over a grid of 0 $< z \leq$ 11, with $\delta z = 0.05$.  We find that differences in stellar libraries and IGM treatment between our fitting method and that of EAZY result in a slightly higher redshift from this fitting, by a few tenths on average, with a range in the sample from 6.50 $\leq$ z $\leq$ 8.85.  At these correspondingly larger distances, the best--fit models are brighter, which the fitting process compensates for by making the models younger (in many cases maximally young; 1 Myr).  Younger stars are more luminous in the UV, thus the corresponding stellar mass is less.  When comparing the 68\% confidence ranges from this fitted redshift to the photometric redshift, we find that they are consistent in nearly all cases.  We thus adopt $z = z_\mathrm{phot}$ for the model fitting, as this will again yield conservative results (i.e., older and more massive).  However, we will examine how the uncertainties on this redshift affect our parameter uncertainties below.

We fit the model spectra, integrated with the photometric bandpasses (including bandpass transmission, telescope optics and detector efficiency) to the photometry from \acsz, \wfcy, \wfcj, \wfch, and IRAC [3.6] and [4.5] for each galaxy.  For each model we calculate $\chi^2$ using the photometric errors, including a systematic error of 0.05$\times f_\nu$, where $f_\nu$ is the flux measured in each bandpass. This systematic error accounts for possible zeropoint uncertainties, aperture corrections, and other uncertainties that scale multiplicatively with the flux. 

During the fitting, any data point with a wavelength below \lya~detected with $<$ 2 $\sigma$ significance was not included in the fit, unless a given model violated the 2 $\sigma$ upper limit, in which case the $\chi^2$ for that band is added in, using the 2 $\sigma$ upper limit as the flux, and the 1 $\sigma$ flux as the error.  At the photometric redshift, the flux in all models should be heavily attenuated by the IGM, so this essentially results in these points being excluded by the fit, which is preferable as there can be large variations in the IGM optical depth along various lines-of-sight.  

For the IRAC data most galaxies are undetected, and we allowed the models to fit the actual measured fluxes, as the photometric errors still provide constraints on the possible model parameters.  However, their contribution to the $\chi^2$ is minimal due to their large uncertainties, thus the IRAC fluxes for these galaxies for the most part do not provide strong constraints on the shape of the SED.  Rather, the blue rest--frame UV (\yb~-- \jb~and \jb~--~\hb) colors coupled with the constraint that the stellar population age be less than the time elapsed since $z =$ 20 places stronger constraints on the stellar population properties.

\begin{figure*}
\epsscale{1.0}
\vspace{2mm}
\plotone{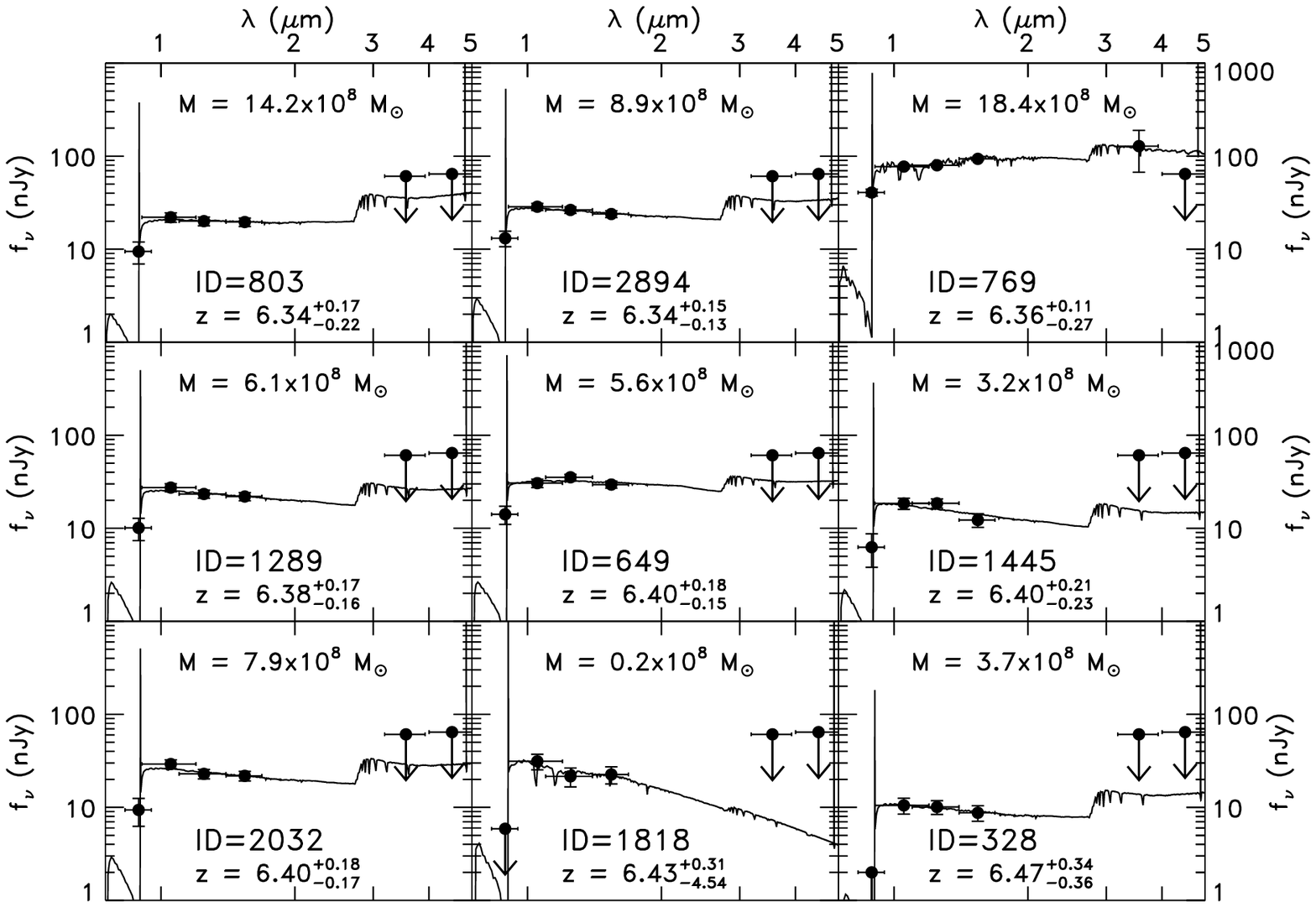}
\vspace{-4mm}
\hspace{3mm}
\plotone{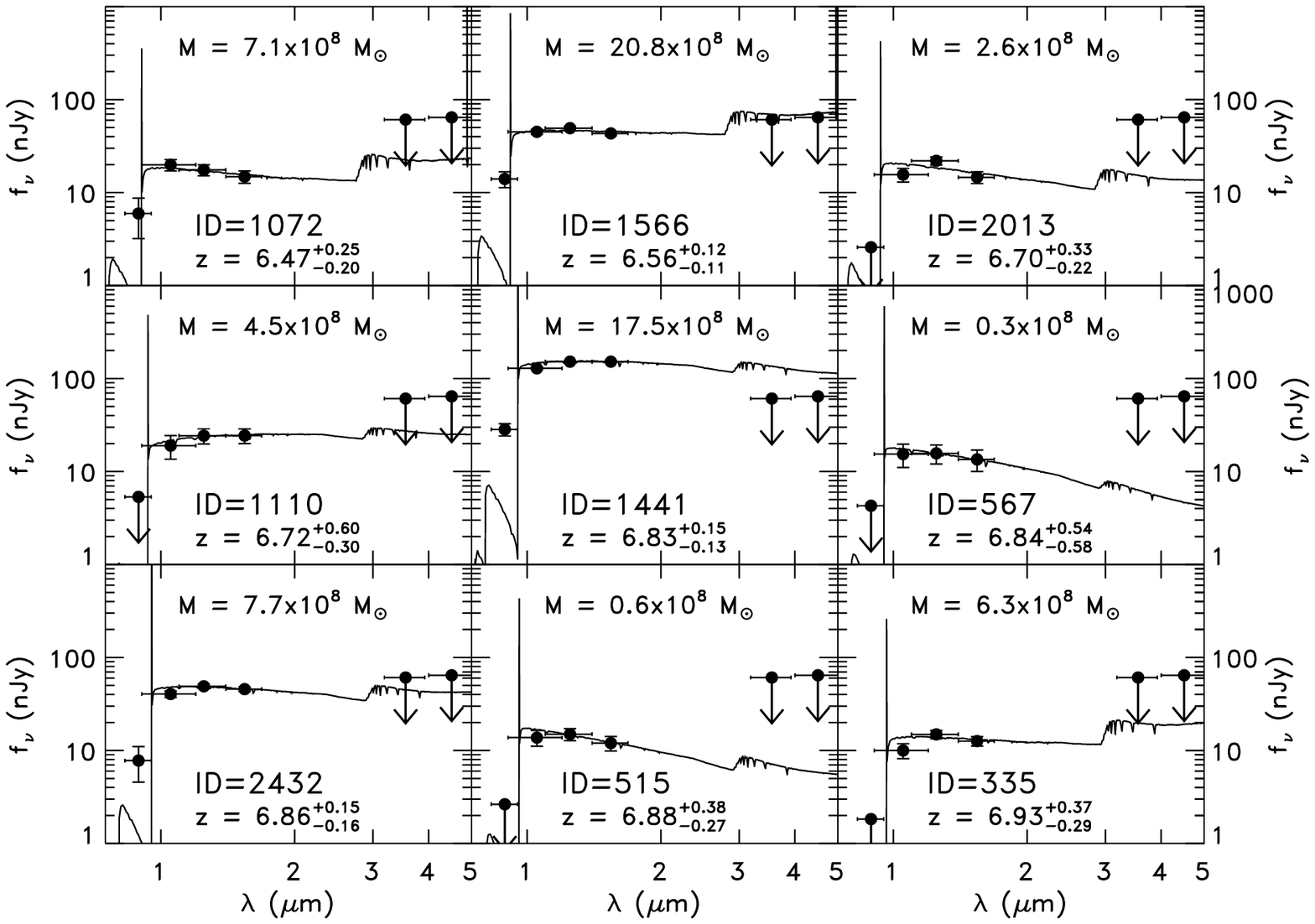}
\vspace{6mm}
\caption{The observed spectral energy distributions (SEDs) of our objects at 6.3 $<$ $z_\mathrm{phot}$ $<$ 7.5 with the best--fit model spectra overplotted, shown in order of increasing $z_\mathrm{phot}$.  The spike in the spectra represents \lya~emission as described in the text.  The best--fit results are tabulated in Table 3.  When a data point is detected at less than 2 $\sigma$ significance, we show the 1 $\sigma$ upper limits as arrows.  The blue colors of these objects typically result in best--fit spectra that are blue, implying very young ages.  We show the best--fit stellar mass in the top  of each panel.}\label{sed1}
\end{figure*}
\addtocounter{figure}{-1}

\begin{figure*}
\epsscale{1.0}
\plotone{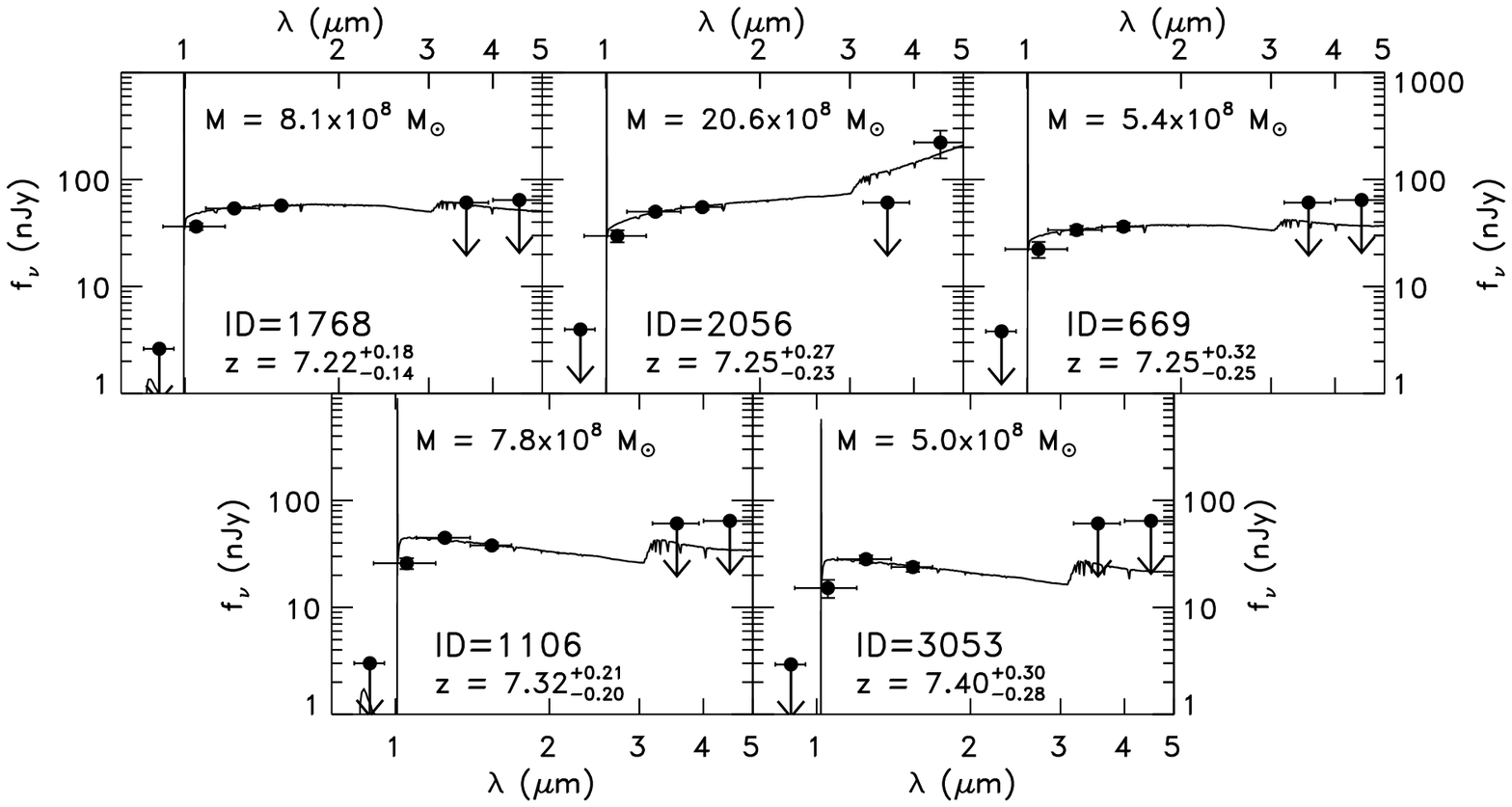}
\vspace{4mm}
\caption{Continued.}\label{sed1c}
\end{figure*}

\begin{figure*}
\epsscale{1.0}
\plotone{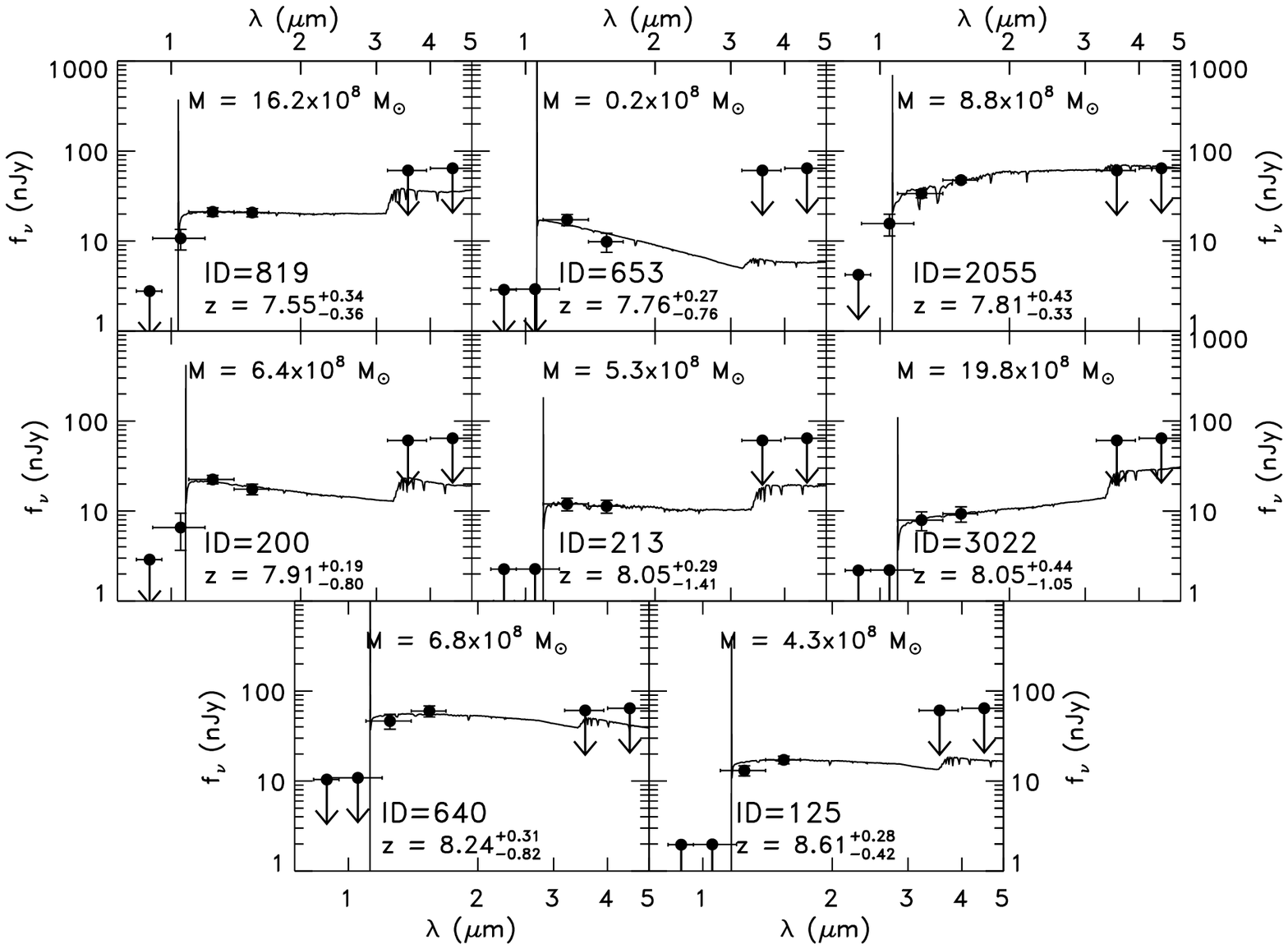}
\vspace{4mm}
\caption{Same as \fig{sed1}, for galaxies with $7.5 < z_\mathrm{phot}\leq 8.6$.}\label{sed2}
\end{figure*}

\begin{deluxetable*}{ccccccccccccc}
\tabletypesize{\small}
\tablecaption{Model Fitting Results}\label{tab:final}
\tablewidth{0pt}
\tablehead{
\colhead{ID} & \colhead{z$_{phot}$} & \colhead{Mass} & \colhead{Mass} & \colhead{Max Mass} & \colhead{Max Mass}& \colhead{Age} & \colhead{Age} & \colhead{A$_\mathrm{V}$} & \colhead{A$_\mathrm{V}$} & \colhead{Z} & \colhead{Z}\\
\colhead{$ $} & \colhead{$ $} & \colhead{Best--Fit} & \colhead{68\% Range} & \colhead{Best--Fit} & \colhead{68\% Range}& \colhead{Best--Fit} & \colhead{68\% Range} & \colhead{Best--Fit} & \colhead{68\% Range} & \colhead{Best--Fit} & \colhead{68\% Range}\\
\colhead{$ $} & \colhead{$ $} & \colhead{(10$^{8}$ M\sol)} & \colhead{(10$^{8}$ M\sol)} & \colhead{(10$^{8}$ M\sol)} & \colhead{(10$^{8}$ M\sol)} & \colhead{(Myr)} & \colhead{(Myr)} & \colhead{(mag)} & \colhead{(mag)} & \colhead{(Z\sol)} & \colhead{(Z\sol)}\\
}
\startdata
\phantom{0}803&6.34&14.2&\phantom{0}3.9 -- 25.6&\phantom{0}20.7&\phantom{0}4.6 -- 26.1&700&\phantom{0}50 -- 700&0.32&0.08 -- 0.89&0.005&0.005 -- 0.200\\
2894&6.34&\phantom{0}8.9&\phantom{0}2.6 -- 15.2&\phantom{0}15.2&\phantom{0}3.9 -- 16.6&400&\phantom{0}50 -- 700&0.24&0.00 -- 0.65&0.005&0.005 -- 0.020\\
\phantom{0}769&6.36&18.4&12.1 -- 44.8&\phantom{0}25.0&18.7 -- 74.5&100&\phantom{0}10 -- 100&0.49&0.49 -- 1.38&2.500&0.020 -- 2.500\\
1289&6.38&\phantom{0}6.1&\phantom{0}1.2 -- 12.2&\phantom{00}5.9&\phantom{0}2.3 -- 12.4&300&\phantom{0}30 -- 700&0.24&0.00 -- 0.65&0.005&0.005 -- 0.400\\
\phantom{0}649&6.40&\phantom{0}5.6&\phantom{0}2.9 -- 11.9&\phantom{0}24.3&\phantom{0}9.8 -- 33.4&\phantom{0}60&\phantom{0}20 -- 200&0.65&0.32 -- 0.89&0.005&0.005 -- 0.020\\
1445&6.40&\phantom{0}3.2&\phantom{0}0.3 -- \phantom{0}5.7&\phantom{00}5.0&\phantom{0}2.2 -- \phantom{0}6.5&400&\phantom{00}5 -- 500&0.00&0.00 -- 0.40&0.005&0.005 -- 0.005\\
2032&6.40&\phantom{0}7.9&\phantom{0}0.7 -- 12.3&\phantom{00}5.4&\phantom{0}2.3 -- 10.7&600&\phantom{00}5 -- 700&0.08&0.00 -- 0.57&0.005&0.005 -- 0.200\\
1818&6.43&\phantom{0}0.2&\phantom{0}0.2 -- \phantom{0}3.6&\phantom{00}2.5&\phantom{0}2.0 -- \phantom{0}8.0&\phantom{00}1&\phantom{00}1 -- \phantom{0}60&0.00&0.00 -- 0.81&2.500&0.005 -- 2.500\\
\phantom{0}328&6.47&\phantom{0}3.7&\phantom{0}0.2 -- \phantom{0}6.0&\phantom{00}2.9&\phantom{0}1.2 -- 11.8&700&\phantom{00}1 -- 700&0.00&0.00 -- 0.97&0.020&0.005 -- 0.200\\
1072&6.47&\phantom{0}7.1&\phantom{0}0.3 -- \phantom{0}7.7&\phantom{00}5.1&\phantom{0}2.1 -- \phantom{0}8.9&700&\phantom{00}2 -- 700&0.08&0.00 -- 0.65&0.005&0.005 -- 0.200\\
1566&6.56&20.8&10.1 -- 31.8&\phantom{0}37.8&16.8 -- 41.6&300&\phantom{0}80 -- 600&0.49&0.24 -- 0.73&0.005&0.005 -- 0.005\\
2013&6.70&\phantom{0}2.6&\phantom{0}0.4 -- \phantom{0}6.0&\phantom{00}5.9&\phantom{0}2.8 -- 10.1&200&\phantom{00}8 -- 400&0.08&0.00 -- 0.57&0.005&0.005 -- 0.005\\
1110&6.72&\phantom{0}4.5&\phantom{0}0.6 -- 10.8&\phantom{00}8.4&\phantom{0}3.2 -- 31.1&\phantom{0}20&\phantom{00}1 -- \phantom{0}30&0.97&0.16 -- 1.38&0.020&0.005 -- 2.500\\
1441&6.83&17.5&10.5 -- 29.4&\phantom{0}32.7&20.8 -- 43.2&\phantom{0}20&\phantom{0}20 -- \phantom{0}40&0.73&0.57 -- 0.89&0.020&0.020 -- 0.020\\
\phantom{0}567&6.84&\phantom{0}0.3&\phantom{0}0.2 -- \phantom{0}3.0&\phantom{00}1.8&\phantom{0}1.7 -- \phantom{0}8.4&\phantom{0}10&\phantom{00}1 -- \phantom{0}30&0.16&0.00 -- 0.89&0.020&0.005 -- 2.500\\
2432&6.86&\phantom{0}7.7&\phantom{0}2.4 -- 12.6&\phantom{0}22.9&\phantom{0}5.8 -- 26.9&\phantom{0}60&\phantom{00}8 -- 100&0.57&0.24 -- 0.81&0.005&0.005 -- 0.400\\
\phantom{0}515&6.88&\phantom{0}0.6&\phantom{0}0.2 -- \phantom{0}3.7&\phantom{00}1.4&\phantom{0}1.7 -- \phantom{0}6.2&\phantom{0}60&\phantom{00}1 -- 100&0.00&0.00 -- 0.73&0.005&0.005 -- 0.200\\
\phantom{0}335&6.93&\phantom{0}6.3&\phantom{0}0.6 -- 10.6&\phantom{0}10.0&\phantom{0}2.4 -- 18.6&400&\phantom{00}4 -- 600&0.32&0.00 -- 1.05&0.005&0.005 -- 0.020\\
1768&7.22&\phantom{0}8.1&\phantom{0}2.4 -- 14.8&\phantom{00}9.9&\phantom{0}9.5 -- 33.4&\phantom{0}10&\phantom{00}3 -- \phantom{0}30&0.97&0.57 -- 1.05&0.020&0.020 -- 2.500\\
2056&7.25&20.6&17.3 -- 95.6&\phantom{0}85.5&53.8 -- 135.3&\phantom{00}5&\phantom{00}4 -- 600&1.46&0.65 -- 1.62&0.005&0.005 -- 0.200\\
\phantom{0}669&7.25&\phantom{0}5.4&\phantom{0}2.5 -- 24.9&\phantom{0}11.7&\phantom{0}7.8 -- 52.8&\phantom{00}7&\phantom{00}1 -- 200&1.05&0.49 -- 1.46&0.020&0.005 -- 2.500\\
1106&7.32&\phantom{0}7.8&\phantom{0}1.8 -- 14.7&\phantom{0}14.7&\phantom{0}6.8 -- 22.0&200&\phantom{0}20 -- 400&0.16&0.00 -- 0.57&0.005&0.005 -- 0.020\\
3053&7.40&\phantom{0}5.0&\phantom{0}0.7 -- 11.1&\phantom{00}9.4&\phantom{0}3.8 -- 16.7&200&\phantom{00}8 -- 500&0.16&0.00 -- 0.81&0.005&0.005 -- 0.020\\
\phantom{0}819&7.55&16.2&\phantom{0}1.5 -- 21.8&\phantom{0}13.1&\phantom{0}4.3 -- 36.7&500&\phantom{00}2 -- 500&0.40&0.16 -- 1.30&0.005&0.005 -- 0.200\\
\phantom{0}653&7.76&\phantom{0}0.2&\phantom{0}0.2 -- \phantom{0}2.4&\phantom{00}2.0&\phantom{0}1.9 -- \phantom{0}2.7&\phantom{00}4&\phantom{00}1 -- 200&0.00&0.00 -- 0.24&0.005&0.005 -- 0.200\\
2055&7.81&\phantom{0}8.8&\phantom{0}6.6 -- 31.1&\phantom{0}43.8&25.2 -- 88.8&\phantom{00}2&\phantom{00}2 -- \phantom{0}10&1.38&1.21 -- 1.62&2.500&0.020 -- 2.500\\
\phantom{0}200&7.91&\phantom{0}6.4&\phantom{0}0.3 -- \phantom{0}9.4&\phantom{00}3.5&\phantom{0}2.8 -- 13.7&500&\phantom{00}1 -- 500&0.00&0.00 -- 0.81&0.005&0.005 -- 0.200\\
\phantom{0}213&8.05&\phantom{0}5.3&\phantom{0}0.2 -- 14.7&\phantom{0}10.5&\phantom{0}1.8 -- 31.5&400&\phantom{00}1 -- 400&0.16&0.00 -- 1.54&0.200&0.005 -- 2.500\\
3022&8.05&19.8&\phantom{0}0.3 -- 50.9&\phantom{0}20.6&\phantom{0}1.7 -- 66.1&400&\phantom{00}1 -- 400&0.89&0.16 -- 1.62&0.020&0.005 -- 2.500\\
\phantom{0}640&8.24&\phantom{0}6.8&\phantom{0}1.4 -- 12.8&\phantom{0}18.8&\phantom{0}7.4 -- 46.6&\phantom{0}20&\phantom{00}1 -- \phantom{0}20&0.65&0.16 -- 1.13&0.020&0.005 -- 2.500\\
\phantom{0}125&8.61&\phantom{0}4.3&\phantom{0}0.8 -- 20.6&\phantom{00}9.1&\phantom{0}3.5 -- 43.0&\phantom{0}40&\phantom{00}2 -- 300&0.73&0.24 -- 1.62&0.005&0.005 -- 2.500\\
\enddata
\tablecomments{We derive the 68\% confidence range on our parameters by taking the central 68\% of the probability distribution for each parameter from the 10$^{3}$ bootstrap Monte Carlo simulations, including the photometric uncertainties and the photometric redshift probability distribution function.  The maximum mass is the best-fit mass from the two--component models.}
\end{deluxetable*}

Although these objects are very blue in the rest--frame UV, we are unable to exclude the possibility that the galaxies contain generations of stars from previous star--formation episodes, lost in the ``glare'' of the UV--luminous population.  We allowed for the possibility that we were missing an older generation of stars by generating a second set of stellar--population synthesis models with two components.  The first model component is similar as described above, and is used to describe the UV emission from young stars, except that the star formation history is modeled as an instantaneous starburst.  The second stellar population is also modeled as an instantaneous starburst, with an age equal to the time elapsed between $z =$ 20 and $z_\mathrm{phot}$ derived for each galaxy.  Because of the limited number of data points we have for each galaxy, we considered only models where a fixed mass fraction of 90\% of the mass exists in this ``maximally old'' stellar population.

\subsection{Fitting Results}
Figures 8 and 9 display the best--fit models for the galaxies in our $z \sim$ 7 and $z \sim$ 8 samples, respectively, and Table~3 provides tabulated results.  The best--fit model parameters refer to those models with the minimum $\chi^2$ calculated between the object and model fluxes.  
We calculate the 68\% confidence ranges on each of the model parameters (age, dust, metallicity, and stellar mass) for our fits using a series of $10^3$ Monte Carlo bootstrap simulations including the photometric uncertainties and the probability distribution function of the photometric redshifts.  The resulting confidence range is the central 68\% of the best--fit results for each parameter for each object.  In each simulation we vary the input fluxes by an amount proportional to their errors, such that the parameter uncertainties reflect the photometric uncertainties.  Table 3 lists the 68\% confidence range on the parameters for each object.  The best--fit ages of these objects from the full sample range from 1 -- 720 Myr, with a median of $\sim$ 200 Myr.  The best--fit color excesses range from 0 $\leq$ A$_\mathrm{V}$ $\lesssim$ 1.5 mag (0 $\leq$ E(B-V) $\leq$ 0.36), with a median of A$_\mathrm{V}$ $\simeq$ 0.3.  For individual galaxies there is a relatively large 68\% confidence range on the model stellar--population age, dust color excess, and metallicity.  This is due to the fact that these parameters are degenerate (as they all result in reddening of the spectra), and the large 68\% range reflects this degeneracy.  In the absence of data measuring the strength of Balmer/4000~\AA--break, we are unable to improve these constraints.  In contrast, our model fitting provides relatively tighter constraints on the galaxy stellar masses.  

To quantify the effect that the redshift uncertainties have on our derived parameter uncertainties, we also performed a set of simulations assuming a redshift uncertainty of zero.  From both sets of simulations, we compute characteristic parameter errors by taking the mean of the difference between the best--fit value and the lower and upper 68\% confidence range.  When we compare the stellar population parameter uncertainties with and without redshift variation, we find that accounting for the redshift uncertainty increases the errors by $\sim$ 30\% in mass, 10\% in age, and 25\% in E(B-V).  We thus conclude that the effect of redshift uncertainties should be included in the error propogation when using photometrically estimated redshifts.  In the subsections that follow we discuss the constraints on the various stellar population parameters.

\subsubsection{Stellar Population Age, Dust Extinction and Metallicity}

The median best--fit age in our sample is $\sim$ 200 Myr, but all objects have 68\% confidence ranges which are consistent with ages $\leq$ 80 Myr, with 24/31 objects consistent with extremely young ages of $\leq$ 10 Myr.  However, the large spread in ages make it difficult to make strong conclusions.  Nevertheless, the colors of the majority of galaxies are consistent with model fits requiring young stellar-population ages.

The best--fit dust extinctions for galaxies in our sample range from 0 $\leq$ A$_\mathrm {v}$ $\lesssim$ 1.5 mag, with a median extinction of A$_\mathrm {v}$ $\sim$ 0.3 mag.  However, the 68\% confidence range on the extinction for 16/31 objects are consistent with zero dust extinction.  This is consistent with our conclusions in \S~4 based on the very blue colors of these objects.  

The fact that \textit{a few} galaxies show redder UV colors implies some dust attenuation, especially at brighter \wfch\ magnitudes.  This is consistent with dust--production rates in core--collapse supernovae (SNe).  \citet{todini01} show that $\sim$ 0.1--0.3 M\sol\ of dust form per star with initial masses of 12--35 M\sol, which end as SNe after $<$ 20 Myr.  Taking a typical stellar mass of $10^9$~\msol\ for galaxies in our $z \sim$ 7 and $z \sim$ 8 samples (see \S~\ref{section:sed}), we expect roughly $\sim$ 2 $\times$ 10$^{6}$ stars in this mass range will form (assuming a Salpeter IMF).  This implies the production of nearly 10$^{6}$ M\sol\ of dust from core--collapse supernovae.  Assuming this dust mixes with the galaxies ISM on a dynamical timescale ($\sim$ 30 Myr assuming a size of 1.5 kpc and a circular velocity of 50 km s$^{-1}$), it is highly plausible that this dust has a measurable effect on the $z \sim$ 7 and $z \sim$ 8 galaxy rest--frame UV colors.

\begin{figure}[t]
\epsscale{1.1}
\plotone{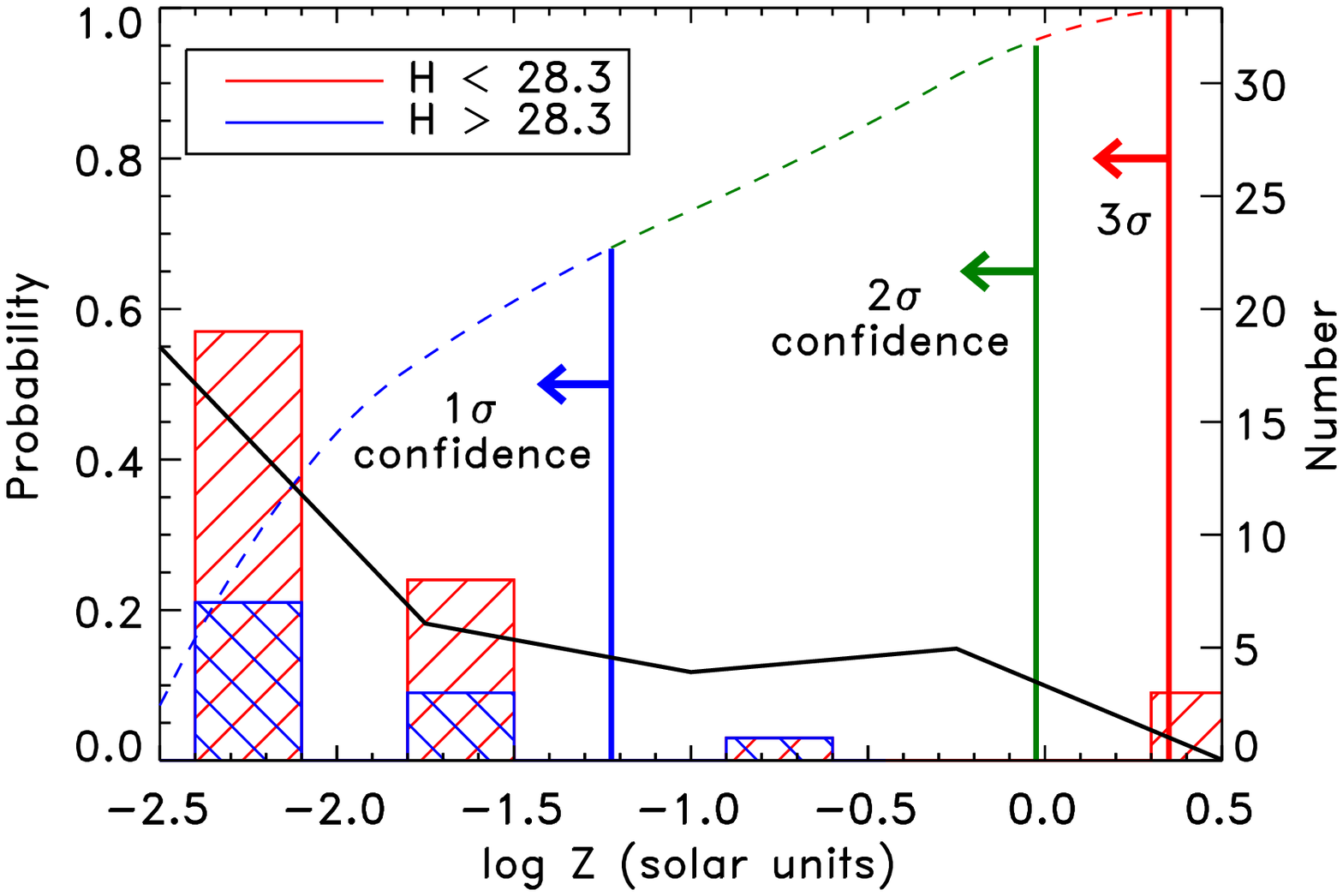}
\caption{The histogram shows the best--fit metallicites from our model fitting, with red and blue lined regions denoting \hb~$<$ 28.3 and \hb~$\geq$ 28.3, respectively.  The black line denotes the metallicity joint probability distribution for the whole sample.  The dashed line shows the cumulative probability, integrating from the peak in the probability distribution at $Z$ = 0.005 $Z$\sol.  This curve changes color from blue to green to red while in the 68, 95 and 99.5\% confidence ranges, respectively.  From this, we rule out solar metallicities at 95\% confidence for the sample, and constrain $Z$ $<$ 0.05 $Z$\sol~with 68\% confidence \citep[c.f.,][]{bouwens09c}.}\label{metfig}
\end{figure}

Surprisingly, our model fits place strong constraints on the metallicities of the galaxies in our samples, relative to the age and extinction.  The majority (19/31) of galaxies in our samples have best--fit metallicities $Z = 0.005$~\zsol.  A further 8 of the remaining 12 galaxies have best--fit metallicities of $Z = 0.02$~\zsol.  Additionally, all 31 galaxies have their entire 68\% confidence ranges at $Z$ $\leq$ 0.02 $Z$\sol.  This is consistent with our conclusions in \S~4 that the metallicites of the majority of galaxies at $z \sim$ 7 and 8 are a fraction of solar.  Galaxies at these redshifts do not have UV colors consistent with solar metallicities.  \fig{metfig} shows the distribution of best--fit metallicities for our samples, as well as the joint probability distribution function for metallicity, constructed from the probability distribution function for each object.  Only galaxies with brighter apparent magnitudes, $\wfch < 28.3$~mag have possibly solar (or super--solar) best--fit metallicities.  Given the model--parameter uncertainties, the joint probability distribution is more informative than the individual best--fits.  The probability distribution has a maximum at $Z$ = 0.005 \zsol\ with a wide distribution spanning the full parameter space.  \fig{metfig} also shows the cumulative probability distribution function.  We find that 68\% of the probability distribution function is constrained at $Z$ $\leq$ 0.05 $Z$\sol.  Furthermore, 95\% of the cumulative probability distribution function lies within metallicities $\leq$ $Z$\sol.  Therefore, we conclude that the majority of galaxies in our sample are best--fit by sub-solar metallicities, and in some cases are only a few percent of solar.

\subsubsection{Stellar Mass}

The stellar masses for each object are typically better constrained compared to the other model parameters.  In general these constraints result from the fact that the relatively young universe ($< 800$~Myr) limits the age of the stellar populations, and thus limits the amount of stellar mass in ``evolved'' stellar populations.  With this constraint, the rest--frame UV colors alone are sufficient to provide constraints on the total mass-to-light ratios and therefore the stellar masses.

As given in Table~3, the median 68\% confidence range on stellar mass spans factors of $\sim +2/-4.5$.  Thus, the upper limit on the masses is constrained to within a factor of two, while they may be a factor of 4.5 less massive.  \fig{massdist}~shows the distribution of best--fit stellar masses for the $z \sim$ 7 and $z \sim$ 8 populations.  We find average masses of slightly less than $\sim 10^{9}$~\msol\ for galaxies at $z \sim$ 7 and $z \sim$ 8, consistent with the stacking analysis of $z \sim$ 7 galaxies from \citet{labbe09}.  The figures also show the joint probability distribution for the stellar masses, constructed using the results for each individual galaxy.  The peak in the joint probability distributions is roughly the same at $z \sim$ 7 and $z \sim$ 8, with a maximum at $\approx 7\times 10^{8}$~M\sol, though $z \sim$ 8 LBGs are more likely to have M $\sim$ 10$^{7}$ M\sol.

\subsubsection{Maximal Stellar  Mass}

\begin{figure*}
\epsscale{1.1}
\begin{center}
\plottwo{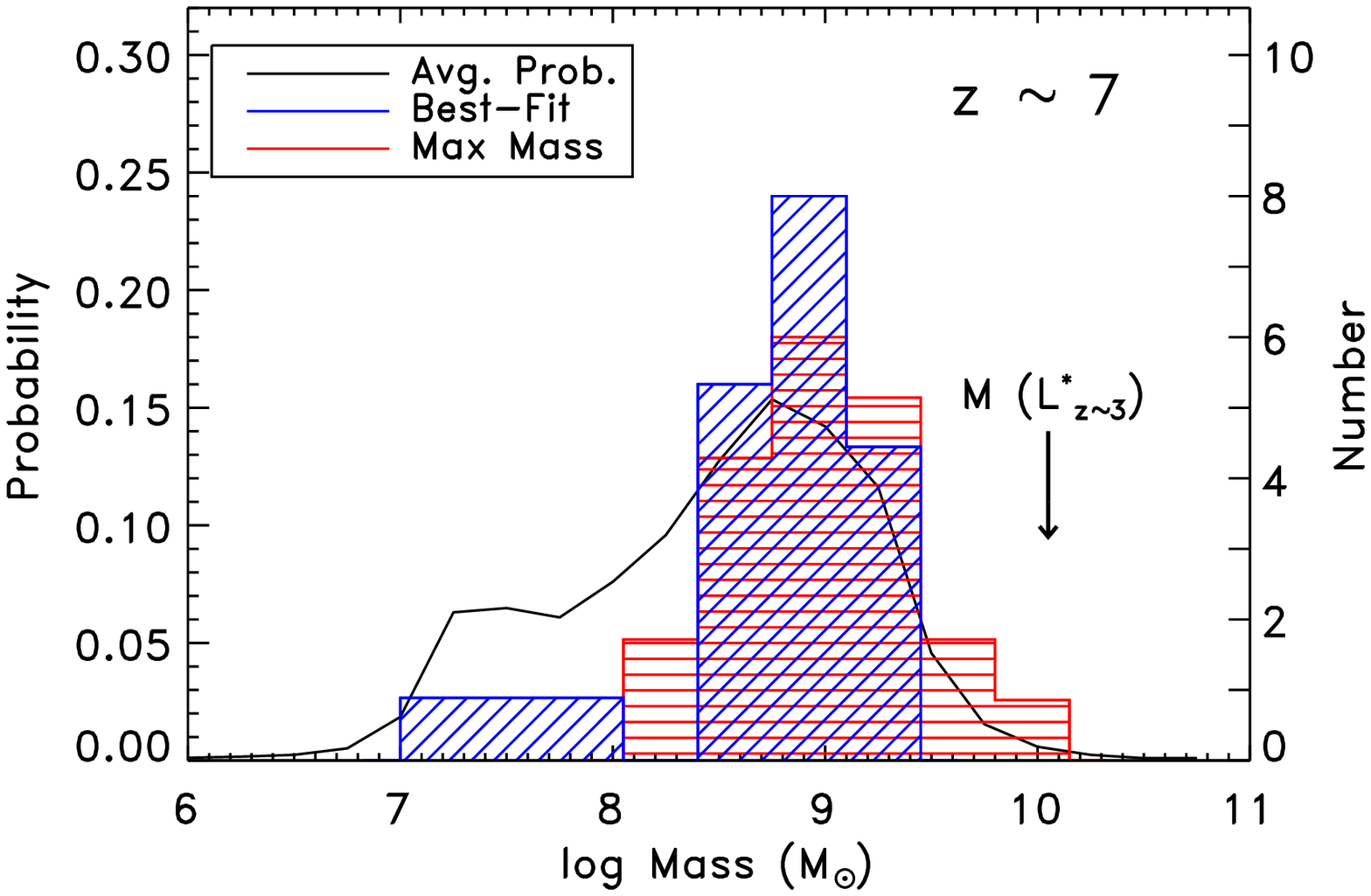}{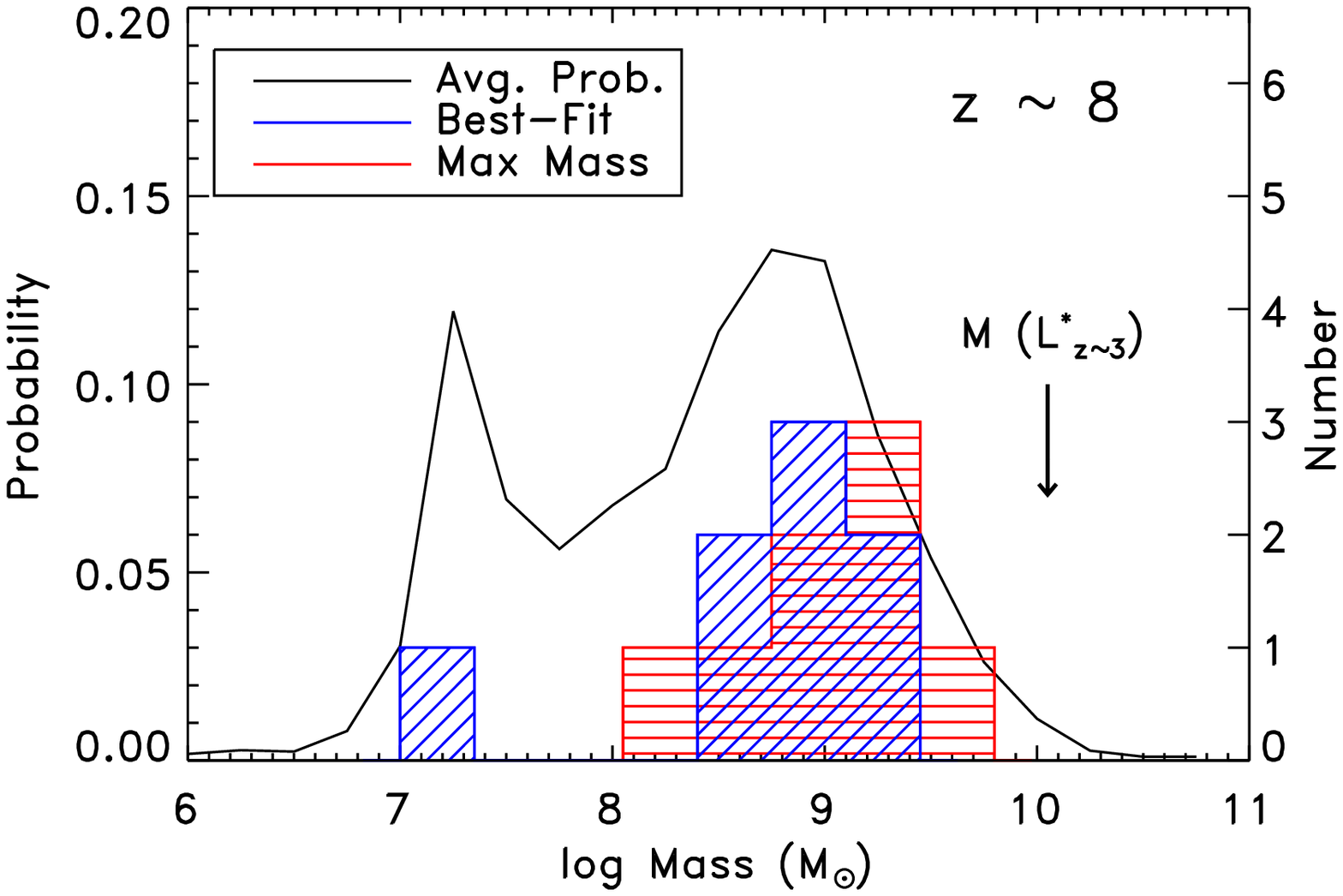}
\end{center}
\vspace{-5mm}
\caption{Left: The black curve shows the probability distribution of the stellar masses averaged over all 23 objects with 6.3 $<$ $z_\mathrm{phot}$ $<$ 7.5 in our sample.  In blue is the distribution of best--fit masses (with the numbers labeled on the right vertical axis).  All objects in this sample are best--fit with M $<$ 10$^{10}$ M\sol, with an interquartile range of $\sim$ 4 -- 9 $\times$ 10$^{8}$ M\sol, which is significantly less massive than L$^{\ast}$ LBGs at $z \sim$ 3.  The average mass distribution confirms this, showing an expected mass range from $\sim$ 7 $<$ log(M/M\sol) $<$ 10.  The red histogram shows the distribution of maximally allowable masses from the two--burst models, which peak at the same mass, but extend to log(M/M\sol) $\sim$ 10.0.  Thus, even if these galaxies formed 90\% of their mass at $z \sim$ 20, they would still be significantly less massive than L$^{\ast}$ galaxies at $z \sim$ 2--3, which have M (at L$^{\ast}$) $\sim$ 10$^{10}$ M\sol~\citep{shapley05, reddy06}.  Right: Same as the left, for galaxies with 7.5 $<$ $z_\mathrm{phot}$ $\leq$ 8.6.  We reach similar conclusions that these galaxies tend to be lower mass than $z \sim$ 3 LBGs.  There is some evidence for evolution from $z \sim$ 7 to $z \sim$ 8, as the latter are more likely to have M $\sim$ 10$^{7}$ M\sol.}\label{massdist}
\end{figure*}

\fig{massdist} also shows the distributions for the ``maximal'' stellar mass, defined as the stellar mass of our two--component models where 90\% of the stellar mass exists in a passively evolving population formed instantaneously at $z = 20$.  Surprisingly, we find that the maximal mass is shifted to a median value of slightly less than twice that for the best--fit single--population model.  Additionally, in many cases the upper end of the 68\% confidence range on the best--fit single--population model masses exceed the maximum mass value.  However, the most massive of either quantity is $<$ 10$^{10}$ M\sol~and they have similar distributions, thus in the following discussion, we use the best--fit two--burst mass as an indicator of the maximum mass.  This distribution is shifted only marginally to higher masses compared to the mass distribution from the single--component models discussed in the previous section.  This small shift is counterintuitive as our constraints stem primarily from photometry of the rest--frame UV, and contrasts with studies of lower redshift LBGs, which have maximal masses larger by up to an order of magnitude \citep[e.g.,][]{papovich01, erb06, pirzkal07, finkelstein09a}.  

To increase the maximal mass would require an older stellar population with a higher mass--to--light ratio.  However, this would require ages greater than the lookback time from $z = 20$ to the observed redshift.  Allowing the stellar mass in the older stellar population to increase would increase the emitted flux from this population in the F125W and F160W bands, increasing the $\chi^2$ and reducing the quality of fit.  To verify this, we also performed bootstrap Monte Carlo simulations on the two--burst models.  If the flux contributed from the old burst to the WFC3 bands was negligible, one would expect the 68\% confidence range on the maximal mass to be quite large.  However, as shown in Table 3, we find this to not be the case, as the maximal masses are quite well constrained, with a median uncertainty of a factor of $\pm$ 2.5
 
If the stellar mass fraction in old stars was above 90\%, this would also increase the total stellar mass. Allowing 99\% of the stellar mass to be in this component, the maximal stellar masses increase by a median factor of $\sim$ 3, not a factor of 10 as may be expected if the choice of mass ratio was driving our results.  However, we consider these masses unlikely as our model already adopts an extreme case where the vast majority of stellar mass formed at very high redshift.  If this is true for every galaxy, then we would expect a significant population of galaxies at $z \sim$ 7 and $z \sim$ 8 dominated by $\gsim$ 500~Myr stellar populations, which is not present.  Given the young ages we derive for the $z \sim$ 7 and $z \sim$ 8 samples (a few $\times$ 10 Myr) there would need to be a very high duty cycle of rejuvenated star formation every few $\times$ 10 Myr to reproduce the current observations.
 
We thus conclude that at $z \sim$ 7 and $z \sim$ 8, the relatively young age of the Universe since $z =$ 20 (400--700 Myr) constrains the age of any pre-existing stellar population.  This limits the contribution to the mass--to--light ratio to come solely from from earlier--type stars (earlier than A--type stars with $\gtrsim$ 3 M\sol), and limits the amplitude of the 4000 \AA~break.  The IRAC data available for the HUDF provides generally a weaker constraint on the stellar mass in these galaxies relative to the constraint from the age of the Universe \citep[c.f.,][]{labbe09}.
\begin{figure*}[t]
\epsscale{1.0}
\plotone{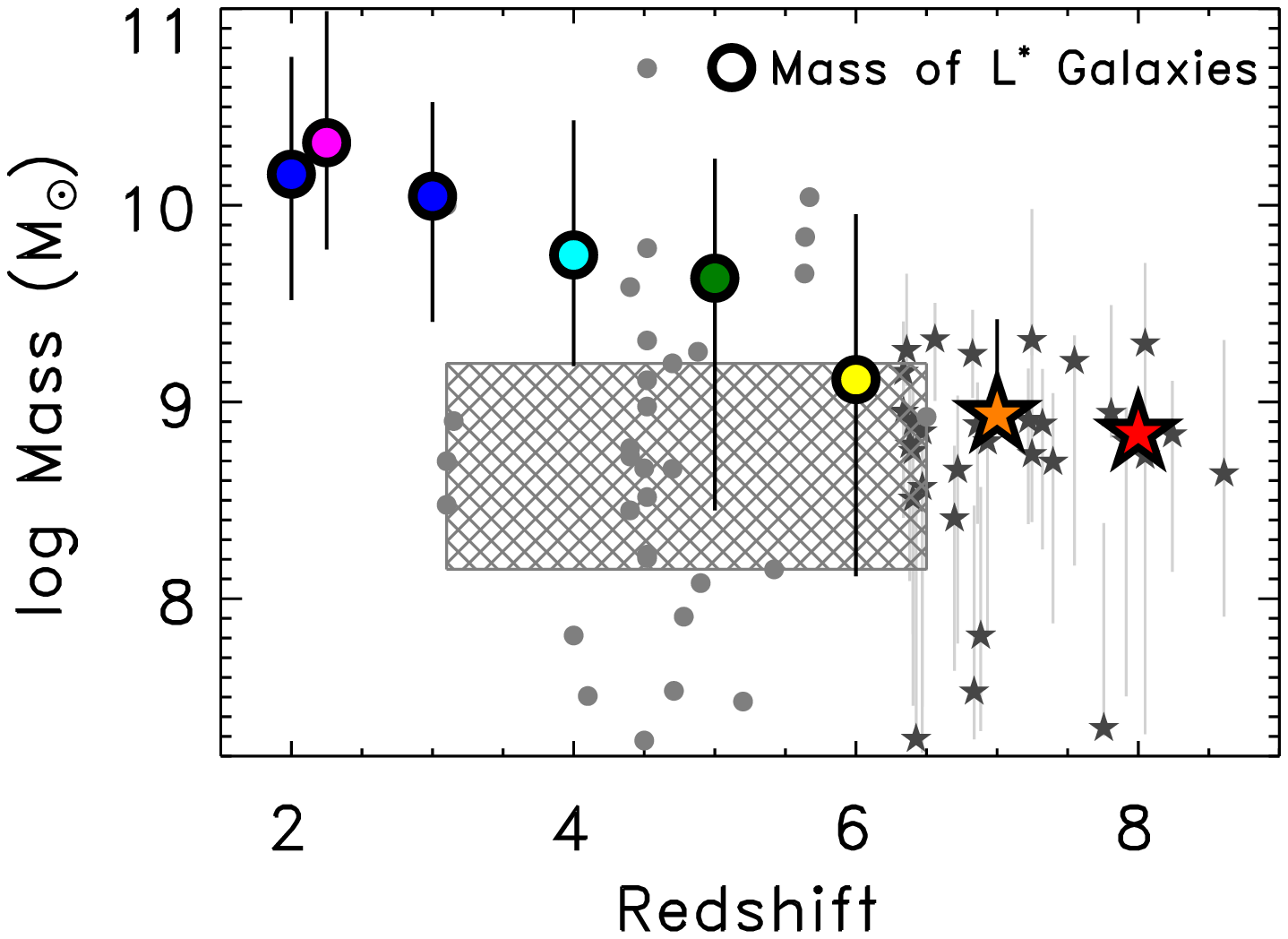}
\vspace{-7mm}
\caption{The stellar masses of L$^{\ast}$ (characteristic luminosity) LBGs versus redshift taken from the literature are shown as circles.  The masses were interpolated to the value at L$^{\ast}$ per the discussion in the text.   The remaining symbols are: blue--\citet{reddy06}; magenta--\citet{shapley05}; blue, green, yellow--\citet{stark09} at z = 4, 5 and 6.  The estimated mass at L$^{\ast}$ at $z = $7--8 from our results are shown by the large orange and red stars.  The displayed error bars represent the central 80\% range of the data, although the small numbers and tight mass ranges of our sample, particularly at $z \sim$ 8, result in small error bars.  For that reason, we show our individual galaxies as small stars with their uncertainties, which are typically on the order of a factor of ten.  At z $<$ 6, there was a slight decrease in mass with redshift, which is now extended by our new results.  The background gray circles denote stellar masses of \lya~emitting galaxies at 3.1 $\leq$ z $\leq$ 6.5 (\citet{chary05}, \citet{gawiser06b}, \citet{pirzkal07}, \citet{nilsson07a}, \citet{lai07}, \citet{finkelstein07}, \citet{lai08} and \citet{finkelstein09a}).  The gray hatched region denotes the interquartile range of the LAE masses.  The masses of the  z $>$ 6.3 LBGs studied here are more similar to those of LAEs at all redshifts than LBGs at any redshift $<$ 6.}\label{masses}
\end{figure*}

\subsection{Comparison of Stellar Masses to the Literature}

The interquartile range of stellar masses (encompassing the inner 50\%-tile) for our z $\sim$ 7 and 8 samples spans 4.5--8.9 $\times$ 10$^{8}$ M\sol.  These are lower than the stellar masses inferred for UV--selected galaxies at $2 < z < 6$.  At $z\sim 2-3$, the inferred stellar mass of ``$L^\ast$'' UV--selected galaxies are on the order of $10^{10}$~\msol\ \citep{papovich01, shapley01, erb06, reddy06}, and there are UV--selected galaxies with stellar masses exceeding $10^{11}$~\msol\ at these redshifts \citep{shapley05, erb06, forster09}.  The fact that the galaxies at $6.3 < z \leq 8.6$ in our sample have such low masses implies a rapid decrease in the stellar mass density of galaxies from lower redshift to this epoch.  

\citet{labbe09} recently studied the stellar populations of 14 objects at $z \sim$ 7 (all in common with our sample).  These authors fit models to the stacked WFC3 and IRAC fluxes, deriving a best--fit age of 300 Myr and a best--fit mass of 1.2$^{+0.3}_{-0.6}$ $\times$ 10$^{9}$ M\sol, consistent with our results.   However, at these redshifts nebular [O\,{\sc iii}] emission could contribute to the flux in the IRAC bands, enhancing the amplitude of the perceived Balmer break between the observed WFC3 and IRAC fluxes.

\fig{masses} compares our observations to a number of studies from the literature, from 2 $\leq$ z $\leq$ 7 \citep[][ all of which assumed a Salpeter IMF]{shapley05,reddy06,stark09}, and shows the evolution of the mass of an L$^{\ast}$ galaxy with redshift.  For \citet{shapley05} and \citet{reddy06}, we computed M$_{1500}$ for each object, where M$_{1500}$ is the absolute magnitude of an object at rest--frame 1500 \AA, and then computed the average mass in bins at M$_{1500}$ = -19.5, -20.5 and -21.5 mag, as well as the 80\% range in each bin.  We then interpolated these values to find the mass at L$^{\ast}$ using the luminosity function of \citet{reddy09} for these redshifts (we split the Reddy et al.\ sample into two bins at $z \sim$ 2 and $z \sim$ 3).  \citet{stark09} report the mean mass in bins of similar absolute magnitude, as well as the 80\% range, for LBGs at z = 4, 5 and 6.  We performed similar calculations to derive the mass at L$^{\ast}$, using the luminosity functions of \citet{bouwens07} at these redshifts.  We interpolate our observations here to the mass at L$^{\ast}$ using the luminosity function of \citet{oesch09}.

Comparing our objects to the literature in this consistent manner (i.e., at L$^{\ast}$, using the values of M$^{\ast}$ from \citet{oesch09} and \citet{bouwens09}) allows us to assess how the stellar masses of ``characteristic'' rest--frame UV-selected galaxies evolve with redshift.  From 2 $<$ z $<$ 6, the mass of a L$^{\ast}$ LBG drops by nearly an order of magnitude, from 10$^{10}$ to $\sim$ 10$^{9}$ M\sol.  Our observations at $z \sim$ 7--8 confirm this trend, finding that the masses of L$^{\ast}$ galaxies at z $>$ 6.5 are $\leq$ 10$^{9}$ M\sol.

Also shown in \fig{masses}~are the known stellar masses of galaxies selected on the basis of their Ly$\alpha$ emission (Ly$\alpha$ emitting galaxies, or LAEs) taken from the literature, along with their combined interquartile range.  The relationship between LAEs and LBGs is not yet clear, yet it may be likely that LAEs form the low--mass and low--luminosity end of the LBG population (e.g., Pentericci et al.\ 2009).  However, the masses of LAEs do not evolve much with time, evidence that they may be progenitors of larger galaxies at subsequent redshifts (e.g., Gawiser et al.\ 2006; Pirzkal et al.\ 2007; Finkelstein et al.\ 2009; Malhotra et al.\ 2010, in prep).  Comparing our results to previous studies of LBGs and LAEs, we find that the masses of galaxies in our $z \sim$ 7 and 8 samples more closely resemble LAEs (or sub--L$^{\ast}$ LBGs) at any redshift than $\gtrsim$ L$^{\ast}$ LBGs at z $<$ 6. 

Furthermore, LAEs are also similar to the $z \sim$ 7 and 8 galaxies in that they have blue colors (see \S~\ref{section:lya}).  \citet{shimasaku06} cite evidence that the incidence of strong \lya~emission in LBGs rises strongly from 3 $<$ z $<$ 6.  In \S~\ref{section:lya} we estimate that the minimum inferred rest--frame \lya~EW of the $z \sim$ 7--8 galaxies is $\sim$ 70 \AA, which meets the selection of a LAE.  At lower redshift, the typically intrinsically more luminous LBGs tend to be more evolved objects compared to LAEs.  We do not see many evolved objects at z $>$ 7, thus we conclude that the era of more--evolved and massive LBGs comes to an end at z $>$ 7.  At this early epoch, we are only seeing the progenitors later day, more evolved, galaxies.

\section{Lyman Alpha Emission}\label{section:lya}

Galaxies selected on the basis of their high Ly$\alpha$ emission generally correspond to younger, less massive galaxies with lower dust attenuation \citep[e.g.,][]{gawiser06b, pirzkal07, finkelstein09a}.  Among the $z \sim 3$ galaxy population, \citet{shapley03} showed roughly 25--30\% of the predominantly fainter LBG population exhibited \lya~in emission with sufficient strength to be selected as a LAE via narrowband surveys (with rest--frame equivalent width $\geq$ 20 \AA).  However, narrowband selected LAEs appear to be different than the typically brighter LBGs.  LAEs at each redshift have on average similar ages and masses, while LBGs are intrinsically more massive, and appear to evolve towards higher masses with decreasing redshift.  LAEs may become the dominant population at z $>$ 6, as the incidence of \lya~emission in LBGs appears to increase with increasing redshift \citep{shimasaku06}.  It is unclear if the increase in \lya~emission incidence among LBGs is due to decreasing dust obscuration \citep[e.g.,][]{bouwens09b}, or if higher--redshift galaxies have higher specific star formation rates, and thus would contain more massive, young stars.  In either case, we may expect \lya~emission to play a prominent role in the SEDs of LBGs at high redshifts (e.g., z $\geq$ 6).

\subsection{Effect on Best--Fits}

Although we have included the effects of \lya~emission in our model fits, we have no {\it a priori} knowledge that these galaxies in fact exhibit \lya~in emission.  However, simply allowing a model to have intrinsic \lya\ emission does not ensure that these photons will escape.  The amount of \lya\ is strongly model dependent.  We have restricted our models to constant star formation histories thus there will always be some intrinsic amount of \lya.  
As discussed above, the $z \sim$ 7--8 galaxies appear to contain little dust, thus any intrinsic \lya\ emission is not strongly internally attenuated.  We note that the escape fraction of \lya\ can also depend on the gas as well.  For example, at such high redshifts, these galaxies may be accreting large amounts of gas from the IGM.  If the covering fraction of the gas around the star--forming regions is large, the \lya\ emission could be scattered into a diffuse halo, which would likely not add significant flux to our broadband observations.  However, recent theoretical work \citep[e.g.,][]{dekel09, brooks09, keres09} shows that gas from the IGM likely accretes along a few discrete filaments, thus unless we happen to be observing along those filaments, the star--forming regions would not be obscured, and the \lya~escape fraction will be high.
\begin{figure*}[t]
\epsscale{1.1}
\begin{center}
\plottwo{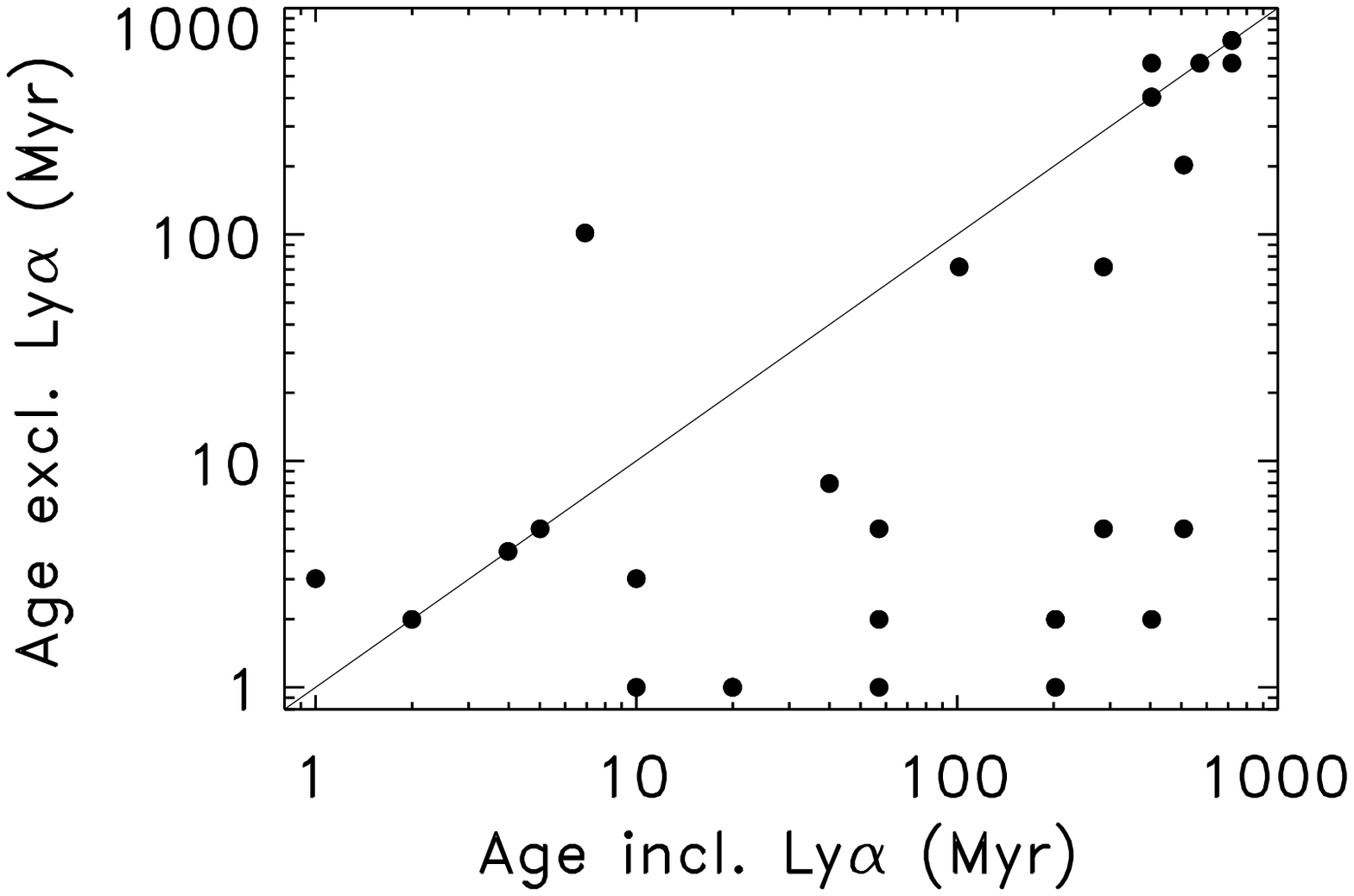}{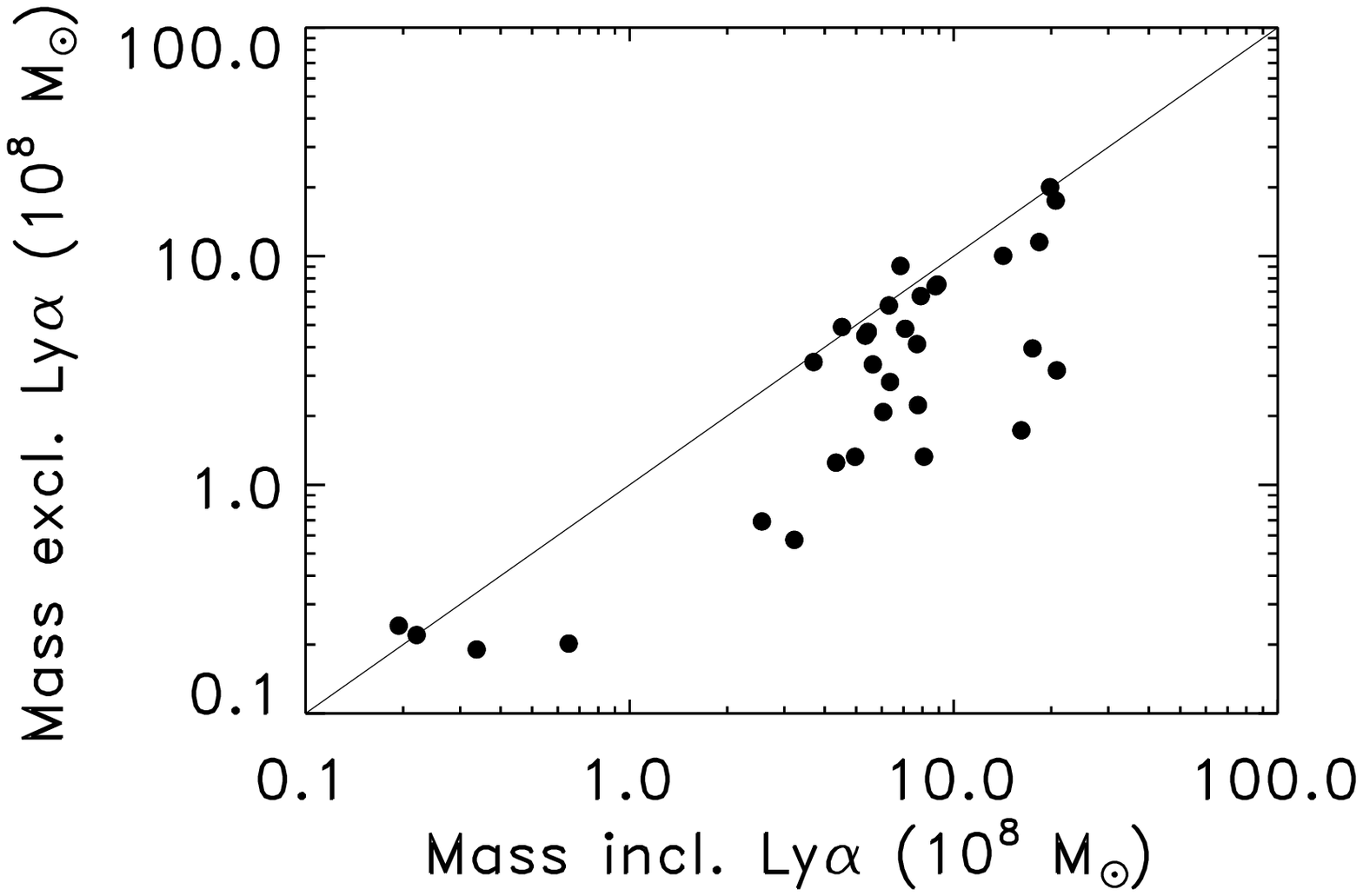}
\end{center}
\vspace{-5mm}
\caption{A comparison of model fitting results with and without \lya~emission in the models.  The left panel shows the best--fit ages.  Without \lya~emission the models tend to fit the objects with younger ages in most cases.  This is understandable, as without a \lya~emission line, the model has to fit the entirety of the \yb~-- \jb~or \jb~-- \hb~color with young stars, therefore reducing the age to match the observed blue colors.  The right panel shows the same effect with the masses, with \lya--free models underestimating the stellar mass.  The diagonal line shows the one--to--one correlation in both plots.  The figure is illustrative, and error bars are excluded for clarity.}\label{lyacomp}
\end{figure*}

Because our observations do not demand \lya\ emission, we quantify the effect \lya~has on our derived model fitting parameters by fitting the observed SEDs to models with and without emission lines.  \fig{lyacomp}~shows the best--fit age and stellar mass results for models with and without \lya~emission.  The models without \lya~emission yield younger ages and lower stellar masses than when \lya~emission is present.  This is understandable when looking an at object's SED.  For example, at z $\sim$ 6.5, an object would have \lya~in the F105W--band.  \lya~contributes to the amplitude of the color, making it bluer.  Removing this strong emission, the model is forced to decrease (make bluer) the \yb~-- \jb~color, producing lower best--fit ages, as shown in the left--hand panel of \fig{lyacomp}.  Since a younger stellar population is more luminous, the model thus requires less mass to fit the observations, as seen in the right--hand panel of \fig{lyacomp}, showing that the masses are also less in nearly all cases without \lya~emission.  Including emission lines in the models are thus necessary when dealing with star--forming galaxies with low dust contents \citep[see also][]{schaerer09}.  We conclude that our fits including \lya\ emission are likely to represent more closely the physical properties of our sample, and that these results are more conservative as they result in higher masses and older ages.

\begin{figure}[b]
\epsscale{1.2}
\plotone{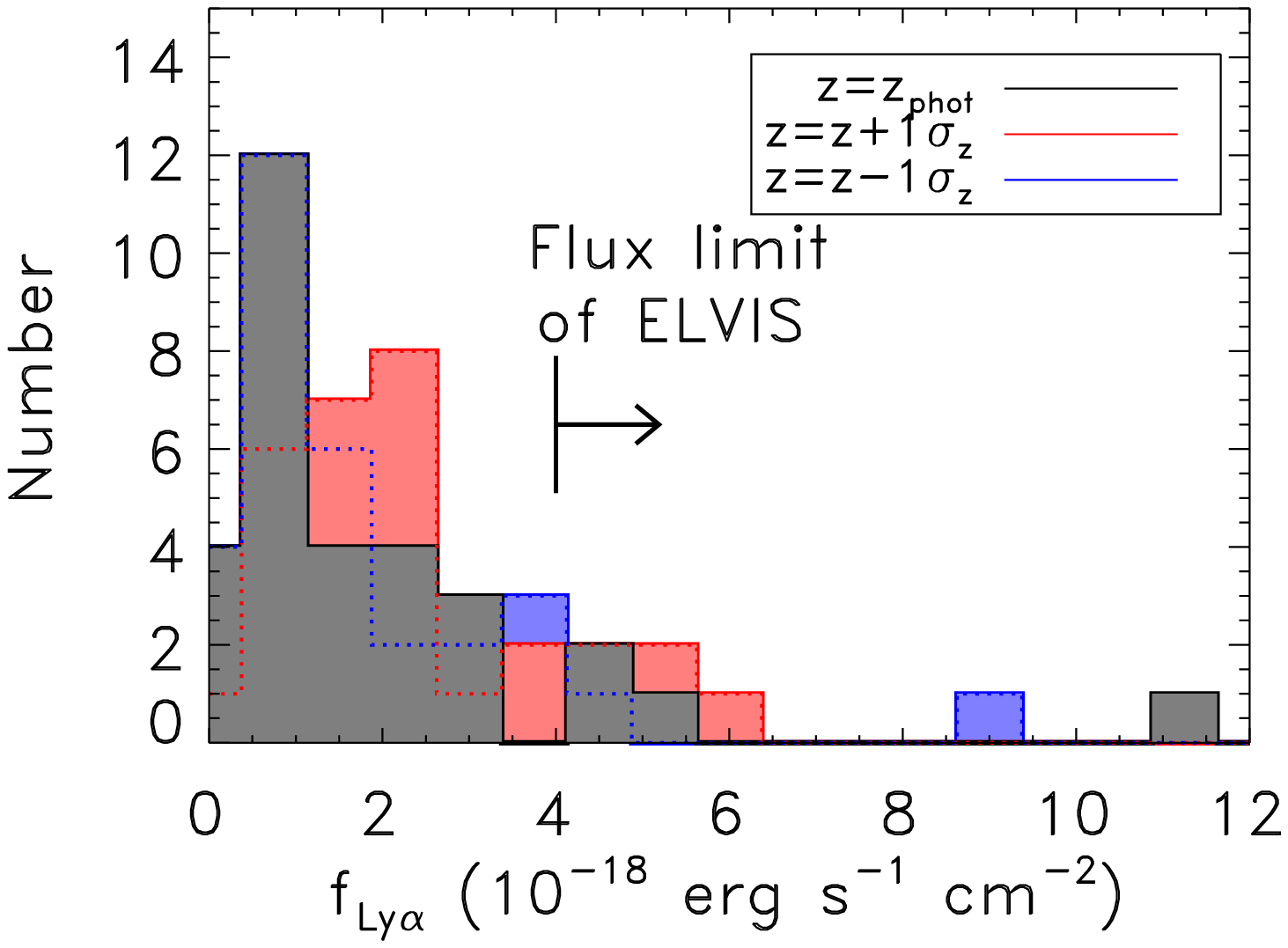}
\caption{The distribution of estimated \lya~line fluxes from the best--fit models for the $z \sim$ 7--8 galaxies is shown in gray.  The rest--frame \lya~EWs are very high ($\geq$ 70 \AA), but given the faint continua of these objects, the actual \lya~line fluxes are relatively weak.  All but four objects have f$_{Ly\alpha}$ $<$ 4 $\times$ 10$^{-18}$ erg s$^{-1}$ cm$^{-2}$, at or beyond the faint edge of current and planned high-redshift LAE searches \citep[e.g., ELVIS;][]{nilsson07b}.  Examining how these results change given the uncertainty in photometric redshift, we find that if our redshifts are underestimated, the implied line fluxes could be higher, with 8 objects having f$_{Ly\alpha}$ $>$ 4 $\times$ 10$^{-18}$ erg s$^{-1}$ cm$^{-2}$.}\label{lyaflux}
\end{figure}

\subsection{Implications for Future Surveys}
Many ground-based searches for z $>$ 7 galaxies focus on discovering them via the \lya~emission line.  The image depths necessary to detect the observed NIR continuum light are unfeasible with most ground-based observatories, but by choosing to search for \lya~emission at redshifts between strong atmospheric emission lines, a number of studies are underway which are expected to yield important results \citep[e.g.,][Tilvi et al.\ 2009 in prep]{nilsson07b,hibon09, sobral09}.

\fig{lyaflux}~shows the distribution of inferred \lya~line fluxes from all 35 candidate galaxies in our sample as derived from the best--fit stellar population models in \S 5.  As discussed above, we expect these objects to have strong \lya~emission as they appear to be young and unextincted.  The expected \lya~line flux for a typical object in our $z \sim$ 7--8 samples is $\sim$ 1--2 $\times$ 10$^{-18}$ erg s$^{-1}$ cm$^{-2}$, although 4 objects have inferred \lya~line fluxes brighter than 4 $\times$ 10$^{-18}$ erg s$^{-1}$ cm$^{-2}$.  We find that 8/31 objects can have f$_{Ly\alpha}$ ranging from 4--15 $\times$ 10$^{-18}$ erg s$^{-1}$ cm$^{-2}$ when considering our photometric redshift uncertainties.  In contrast, \citet{hibon09} reach a limiting depth of 8.3 $\times$ 10$^{-18}$ erg s$^{-1}$ cm$^{-2}$.  The ELVIS survey is expected to achieve sensitivities of 4.0 $\times$ 10$^{-18}$ erg s$^{-1}$ cm$^{-2}$ \citep{nilsson07b}.  These current and future surveys would not be expected to detect many objects similar to those in our sample, which would require a $\sim$ four--fold increase in limiting sensitivity.  Nonetheless, our sample was selected from deep WFC3 observations of the HUDF, thus narrowband surveys searching a wider area will likely have a much greater chance of discovering brighter and rarer objects.

\section{Joint Constraints on Cosmic Reionization and Galaxy Properties}\label{section:reion}
The presence of galaxies at $z \sim$ 7 and $z \sim$ 8 allows us to test how these galaxies contribute to cosmic reionization, which is believed to be in progress during this time \citep[e.g.,][]{becker01, fan06, dunkley09}.  To estimate the contribution of the galaxies in our $z \sim$ 7 and $z \sim$ 8 samples to the budget of hydrogen ionizing photons, we compute the specific luminosity density of each object at rest--frame 1500~\AA.  For each object, we interpolate the WFC3 \wfcy, \wfcj, and \wfch\ photometry and apply the luminosity distance to derive the specific luminosity at 1500~\AA.  We then sum the individual luminosities of each galaxy and divide by the comoving volume element over $6.3 < z < 7.5$ and $7.5 < z \leq 8.6$ for the $z \sim$ 7 and $z \sim$ 8 samples, respectively.  We make no corrections for survey incompleteness or for galaxies fainter than the magnitude limit of our catalog, although we note that either of these corrections would \textit{increase} the luminosity densities we calculate.  
Therefore, our luminosity densities are conservative lower limits.

\begin{figure}[t]
\epsscale{1.2}
\plotone{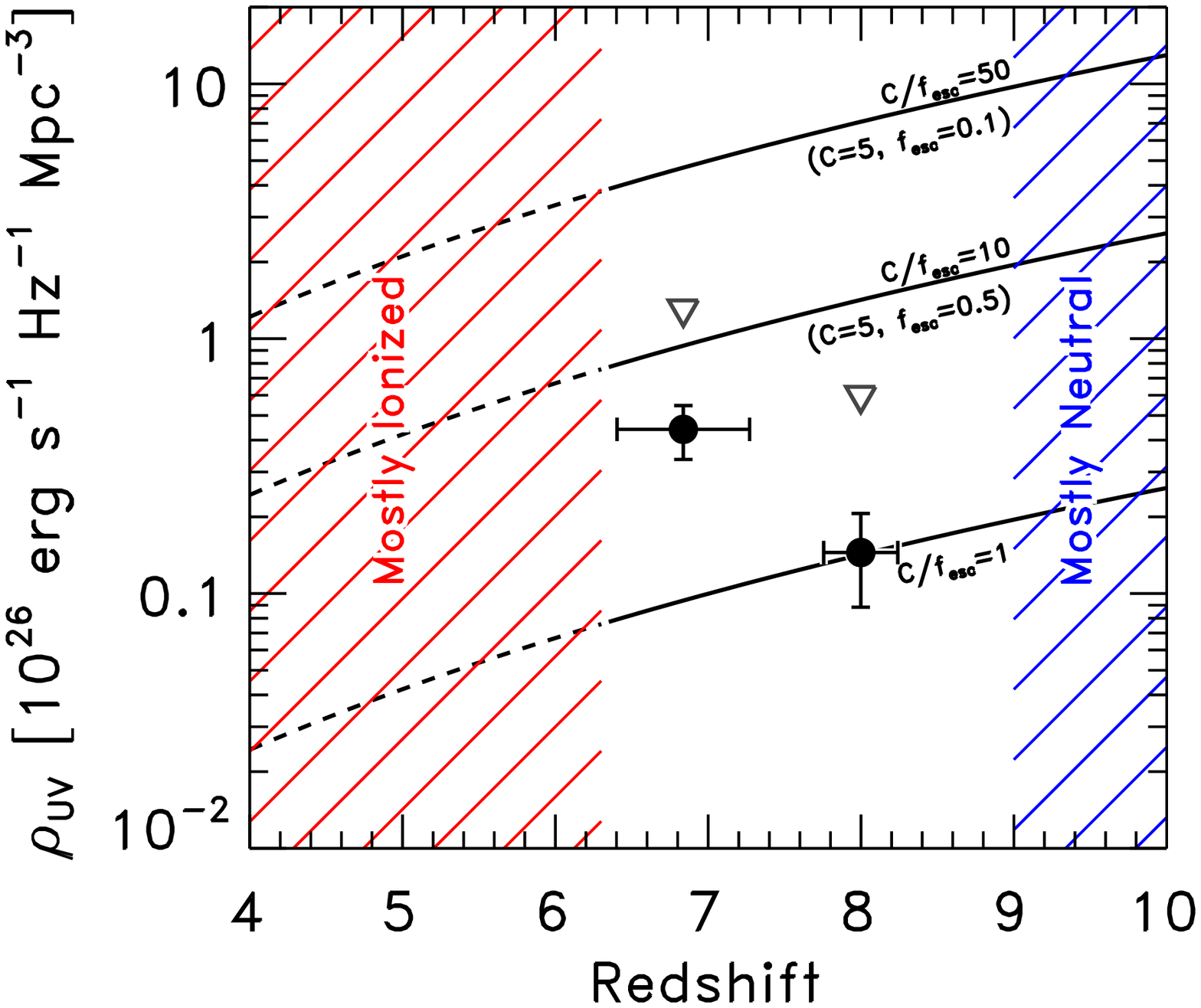}
\caption{The circles represent the specific luminosity density of our sample, uncorrected for dust or incompleteness, split into two bins in redshift.  We denote the regions where we believe the Universe is mostly ionized and mostly neutral.  We are now beginning to probe into the region where reionization is likely to be beginning.  The curves show the critical specific luminosity density needed to reionize the Universe for ratios of the clumping factor C over the escape fraction f$_{esc}$.  If our points lie above a given curve, that indicates that the Universe would be fully reionized by that redshift.  The downward triangles denote the total specific luminosity density for our samples if we account for fainter galaxies down to M$_{1500} = -15$ with the luminosity function of \citet{oesch09}.  For large values of $f_{esc}$($\sim$ 0.5) galaxies at these redshifts are able to reioinize the universe.  Given the evidence for low dust extinction, especially from the fainter galaxies in our sample, high values of $f_{esc}$ may be likely.}\label{reion}
\end{figure}

\fig{reion}~shows the rest--frame 1500~\AA\ specific luminosity density we derive for our samples at $z \sim$ 7 and $z \sim$ 8.  We estimate errors on the luminosity densities by performing a bootstrap simulation whereby we generate galaxy samples by randomly selecting galaxies from the true sample with replacements including the effects of Poisson uncertainties and the photometric uncertainties.

We compare the luminosity densities we derive to models for the required number of ionizing photons to keep the Universe ionized at a given redshift \citep{madau99}.  Following Madau et al.\ (1999) and \citet{pawlik09}, we find the required luminosity density at 1500~\AA\ needed to maintain an ionized universe is
\begin{equation}\label{eqn:reion}
\rho_\mathrm{UV} = 1.25 \times 10^{25}\,\, \epsilon_{53}\,^{-1} \left( \frac{1+z}{8} \right)^3  \left( \frac{\Omega_b h^2_{70}}{0.0463} \right)^2 \frac{C}{f_\mathrm{esc}},
\end{equation}
 in units of erg s$^{-1}$ Hz$^{-1}$ Mpc$^{-3}$, where $\Omega_b$ is the cosmic baryon density and $h_{70}$ is the Hubble parameter in units of $h=0.7$, and the constant 0.0463 is our assumed value of $\Omega_b$ ($\times$ $h_{70}^2$) from {\it Wilkinson Microwave Anisotropy Probe} Year 5 data \citep{komatsu09}.  The variable $\epsilon_{53}$ is the number of Lyman continuum photons per unit of forming stellar mass is units of 10$^{53}$ photons s$^{-1}$ (M\sol~yr$^{-1}$)$^{-1}$.  To determine this value, we used CB07 models with a Salpeter IMF, a constant SFH and $Z$ = 0.2 $Z$\sol, which is within the 95\% confidence range on the metallicities of our objects from \S 5, and gives $\epsilon_{53}$ = 1.2.  Varying the metallicity from $Z$\sol -- 0.02 $Z$\sol changes $\epsilon_{53}$ to 0.9 -- 1.4, respectively.  We note that if the high--mass end of the IMF is biased towards more massive stars, more photons would be emitted at 900 \AA~relative to this equation, thus our assumptions are conservative.

The equation above depends on two variables, the first being the ``clumping'' factor of neutral hydrogen, $C \equiv \langle \rho^2_\mathrm{HI} \rangle \langle \rho_\mathrm{HI} \rangle^{-2}$.  The second variable $f_\mathrm{esc}$ is the average escape fraction of hydrogen--ionizing photons.  While Eqn. 5 uses a single value for $f_{esc}$ this is the average over all galaxies, and certainly there is a distribution, possibly with a luminosity dependence.  Neither $C$ nor $f_{esc}$ are well understood theoretically or empirically.  While \citet{madau99} argued for a high \ion{H}{1} clumping factor $C \approx 30$, recent simulations and radiative transfer calculations imply a lower average clumping of $C \approx 5-6$ \citep{pawlik09,finlator09}, due to photo-heating of the IGM by star formation.  The escape fraction of ionizing photons has been measured at lower redshifts, with $f_\mathrm{esc} < 0.1$ at $z = 0$, $f_\mathrm{esc} = 0.08$ at $z\sim 1$ \citep{siana07}, and measurements of $f_\mathrm{esc}$ = 0.15--0.6 at $z \sim 3$ \citep{steidel01, shapley06}.  The escape fraction at z $>$ 6 is unknown, but current models require $f_{esc}$ $>$ 0.2 \citep{pawlik09, finlator09}.

\fig{reion} shows several curves from equation~\ref{eqn:reion} for various combinations of $C / f_\mathrm{esc}$.  Based on this figure we conclude that galaxies at $z \sim$ 7 and $z \sim$ 8 have sufficient hydrogen--ionizing photons to maintain cosmic reionization under certain conditions.  Even with no corrections to our measured luminosity densities, the galaxies in our sample would reionize the Universe for $C / f_\mathrm{esc} = 1$ to $\approx 10$.  This ratio of $C$ to $f_\mathrm{esc}$ already encompasses theoretically motivated values.  For example, taking $C$ = 5 from the recent radiative--transfer calculations of \citet{finlator09} implies that for $f_\mathrm{esc} = 0.5$, the galaxies we {\it observe} would nearly reionize their local volume by $z\sim 6.5-7$.

Although we have considered only the UV luminosities of the galaxies in our sample, we can estimate the contribution from fainter galaxies using the $z \sim 7$ UV luminosity function of \citet{oesch09} with our $z \sim$ 7 sample, and shifting it to match our $z \sim$ 8 sample.  We calculate a correction factor as the ratio between the integral of the Oesch et al.\ luminosity function down to M$_{1500} = -15$ by the integral of our observed luminosity function at $z \sim$ 7 and 8.  The implied corrections are a factor of 3 at $z \sim$ 7 and 4 at $z \sim$ 8.  These are indicated by the downward triangles in \fig{reion}.  These points represent plausible upper limits on the luminosity density (barring additional corrections from incompleteness and effective volume).  Achieving higher luminosity densities from these galaxies would require substantial corrections from galaxies fainter than our magnitude limits \citep[c.f.][]{yan09}.  Allowing for the maximal hydrogen--ionizing photon escape fraction, $f_\mathrm{esc} = 1$, the clumpiness \textit{must} be $C \lesssim 20$ to match the measured UV luminosity density and still maintain a partially ionized IGM.  For more plausible escape fractions, $f_\mathrm{esc} \approx 0.1 - 0.5$, the clumping factor must be closer to $C = $3--5.

The low levels of dust extinction inferred from the very blue rest--frame UV colors of our sample implies that the escape fraction for hydrogen--ionizing photons may be quite large (though see \S 6.1).  Indeed, although our above calculations have assumed no dependence of $f_{esc}$ on the UV luminosity (and thus the stellar mass as the stellar mass correlates with UV luminosity at these redshifts -- see Tables 1 and 3), this may be the case, as we showed above that the fainter galaxies may be slightly bluer.  Therefore, it may be that the fainter galaxies have higher ionizing escape fractions, $f_\mathrm{esc} > 0.5$, resulting from combination of low metallicity and minimal extinction in these galaxies.  As galaxies develop higher metallicity and dust content, it may be they achieve lower escape fractions, consistent with measurements at $z = 3$ \citep{shapley06}, but by then the universe is well ionized.

\subsection{Comparison to Recent Studies}
Several recent studies have also made estimates of the contribution of galaxies at $z \sim 7-10$ to the UV luminosity density required to reionize the Universe.  \citet{labbe09} used average WFC3 and IRAC colors from stacked data for their sample to estimate the average SFR and stellar mass.  Assuming the stellar mass in their galaxies formed at $7 < z < 11$, they arrive at a result consistent with the one we report here, assuming similar values for the \ion{H}{1} clumpiness and UV escape fraction.  Similar results are achieved by \citet{bunker09} and \citet{yan09}, if we adopt our values of $C$ and $f_\mathrm{esc}$ (and we note that the extremely steep slope of the UV luminosity function derived by Yan et al.\ is able to reionize completely the Universe at $z \gg 7$).  Therefore, the following (general) conclusion is fairly robust, as it has been derived from several studies using different techniques.  \textit{The galaxies we identify at $z \sim 7$ and $z \sim 8$ produce a sufficient UV luminosity density for reionization.}  The specific details are still unclear (in particular the unknown values for the UV escape fractions), and will require more observational studies of high--redshift galaxies.  Based on our study, and those in the literature, we predict that the galaxies at $z \sim 7-8$ have $f_\mathrm{esc} \approx 0.1-0.5$ in order for reionization to proceed.

\section{Conclusions}

We have analyzed new, very deep data from WFC3 on board \textit{HST} of the Hubble Ultra Deep Field, obtaining a sample of 31 galaxies at $6.3 < z_\mathrm{phot} \leq 8.6$.  We have examined the colors, physical properties, and Ly$\alpha$--emission characteristics of these galaxies using a slate of empirical galaxy templates.  We find that the rest--frame UV colors of these galaxies are very blue, bluer than the expected colors of local starburst galaxies ($>4\sigma$ significance).  This represents strong evolution from $z\sim 3$, where typical star--forming galaxies have colors consistent with these local starbursts.  Taking into account a full simulation of the galaxy photometric uncertainties, we find little evidence that the galaxies in our $z\sim 7$ and 8 samples host exotic stellar populations with primitive metallicities.  Rather these objects appear similar to very blue (nearly unextincted) local starburst galaxies such as NGC~1705.

We fit the measured colors of the galaxies in the $z \sim 7$ and 8 samples to stellar population models, using the updated code of Charlot \& Bruzual (2007).  We find that although the constraints on the ages and dust extinction are marginal, the data are consistent with the stellar populations in these galaxies having low ages ($\sim$100~Myr), little--to--no dust extinction, and low metallicity ($<$ 0.05 $Z$\sol~at $\sim$ 1 $\sigma$ confidence).

We derive tighter constraints on the objects' stellar masses, ranging from $10^8$--$10^9$~$M_\odot$ for $z \sim 7$ and 8 galaxies.  These results extend the evolutionary trend that the stellar mass of L$^{\ast}$ galaxies decreases with increasing redshift.  We examine the maximal allowable stellar masses in these objects by fitting them to a two--burst model, which includes an old ``burst" of star formation with 90\% of the total stellar mass, and with an age equal to the time from $z = 20$ to the photometric redshift of each galaxy.  The young age of the universe at these high redshifts (look back time of $t = 750$~Myr at $z=7$) places strong constraints on the amount of old stars in these objects, much more so than the upper limits derived from the deep IRAC data.  With this age constraint, even the maximal masses for the $z\sim 7$ and 8 galaxies are typically only a few times $10^9$~M\sol.  These are larger than their best--fit masses using single stellar populations, but they are significantly less than those derived for ``$L^\ast$'' star--forming galaxies at $z < 6$.

We investigate the effect of Ly$\alpha$ emission on the interpretation of the colors of the galaxies in the $z \sim 7$ and 8 samples.  As may be expected for these relatively unextincted, star--forming objects, Ly$\alpha$ can be strong.  If this emission is not included in the models, the best--fit results systematically underestimate both the age and stellar mass, as the models become bluer to account for the observed rest--frame UV colors.  We estimate the Ly$\alpha$ emission in these objects from the best--fit stellar population models, finding that while all objects have  rest--frame EW $>$70~\AA, the implied line fluxes for typical $z\sim 7$ and 8 galaxies are below the sensitivity of current and planned Ly$\alpha$ searches by factors of a few at these redshifts.

Additionally, the derived stellar masses for the $z \sim 7$ and 8 galaxies are comparable to those of galaxies at $3 < z < 7$ selected on the basis of their Ly$\alpha$ emission, which tend to be less evolved than typical lower--redshift LBGs.  We conclude that evolved galaxies at $z \gtrsim 7$ are very rare, and that the galaxies in our samples represent the progenitors of lower redshift galaxies in their first throes of star formation. 

We calculate the UV luminosity density of the galaxies in our $z \sim 7$ and 8 galaxies, and compare it to the necessary value for cosmic reionization.  We find that the galaxies in our sample provide sufficient UV luminosity density (within factors of a few) of that required for reionization, even without making corrections for incompleteness.  Estimating the contribution from galaxies below our detection limit, star--forming galaxies at $z \sim 7$ and 8 can ionize the the IGM if they have UV escape fractions $f_\mathrm{esc} \approx 0.3-0.5$.  These high escape fractions may be likely given the the low dust extinction and metal content in these galaxies.

The recent installment of WFC3 on board \textit{HST} now allows for the study of faint objects ($\sim 29$~mag) in deep near--IR imaging.  Even with the small amount of data available for these galaxies we are able to derive useful constraints on their properties, and their effect on the IGM.  Future infrared data will allow better constraints on the physical properties of these objects to link them both to lower redshift galaxies, and to galaxies yet to be discovered at higher redshifts.

\acknowledgements
We thank Helmut Katzgraber, Keely Finkelstein, Seth Cohen, Sangeeta Malhotra, Romeel Dav\'e, Kristian Finlator, Garth Illingworth, Rychard Bouwens, Ivo Labb{\'e} and Daniel Schaerer for insightful questions and useful conversations.  We also thank the anonymous referee for their constructive comments.  S. L. F. is supported by the Texas A\&M University Department of Physics and Astronomy.  Support for N. A. R. was provided by NASA through Hubble Fellowship grant HST-HF-01223.01 awarded by the Space Telescope Science Institute, which is operated by the Association of Universities for Research in Astronomy, Inc., for NASA, under contract NAS 5-26555.

\end{document}